\documentclass[a4paper,10.5pt]{article}

\usepackage{stmaryrd}
\usepackage{bm}
\usepackage{cite}
\usepackage{color}
\usepackage{dsfont}
\usepackage{nicefrac}
\usepackage{multirow}
\usepackage{graphics}
\usepackage{graphicx}
\usepackage{epsfig}
\usepackage{subfigure}
\usepackage{pgf}
\usepackage{amsmath,amssymb,amsfonts}
\usepackage{wasysym}
\usepackage{yhmath}
\usepackage{booktabs}
\usepackage{tikz-cd}

\begin{document}

\title{Fiber bundle topology optimization for mass and heat transfer in laminar flow}

\author{Yongbo Deng\footnote{yongbo.deng@kit.edu (Y. Deng)}, Jan G. Korvink\footnote{jan.korvink@kit.edu (J. G. Korvink)} \\
Institute of Microstructure Technology (IMT), \\
Karlsruhe Institute of Technology (KIT), \\
Hermann-von-Helmholtzplatz 1, \\
76344 Eggenstein-Leopoldshafen, Germany.}

\maketitle

\abstract{This paper presents fiber bundle topology optimization for mass and heat transfer in surface and volume flow in the laminar region, to optimize the matching between the pattern of a surface structure and the implicit 2-manifold on which the pattern is defined. The fiber bundle concept is used to describe the pattern of the surface structure together with the implicit 2-manifold as an ensemble defined on the preset base manifold. Topology optimization of the surface structure for mass and heat transfer in surface and volume flow is then implemented on the variable curved surface expressed as the implicit 2-manifold, which is defined on the preset base manifold by using a differentiable homeomorphism. For the surface flow, fiber bundle topology optimization for mass and heat transfer is implemented based on the porous medium model by using the material distribution method, where the material density is used to interpolate the impermeability of the porous medium filled on the implicit 2-manifold. For the volume flow, fiber bundle topology optimization for mass and heat transfer is implemented based on the mixed boundary condition interpolated by the material density used to represent the pattern of thin walls defined on the implicit 2-manifold, which is embedded in the three dimensional domain occupied by the volume flow. For both of the surface and volume flow, two sets of design variables are defined for the pattern of the surface structure and the implicit 2-manifold. The fiber bundle topology optimization problems are analyzed by using the continuous adjoint method to derive the gradient information of the design objectives and constraints, and they are then solved by using the gradient based iterative procedures. In the numerical results, the effects of variable amplitude of the implicit 2-manifold, Reynolds number, P\'{e}clet number, and pressure drop or dissipation power of the fluid flow are investigated to demonstrate the extended design freedom and design space of fiber bundle topology optimization for mass and heat transfer in surface and volume flow.

}

\section{Introduction} \label{sec:IntroductionManifold}

Topology optimization is a robust method used to determine the structural configuration, which corresponds to the material distribution in a structure \cite{BendsoeAndSigmund2003}. In contrast to designing devices by tuning a handful of structural parameters in size and shape optimization, topology optimization utilizes the full-parameter space to design a structure based on the user-desired performance, and it is more flexible and robust, because of its low dependence on the initial guess of the optimization procedure. Therefore, topology optimization is a more powerful tool to optimize structures with material distribution represented by design variables.

Optimization of structural topology was investigated as early as 1904 for trusses \cite{Michell1904}. Topology optimization was originated from the structural optimization problem in elasticity and compliance mechanisms \cite{Bendsoe1988,Sigmund2001,Sigmund1997,Saxena2005}.
It was then extended to multiple physical problems, such as acoustics, electromagnetics, fluidics, optics and thermal dynamics \cite{Borrvall2003,Gerborg-Hansen2006,Nomura2007,Sigmund2008,Duhring2008,Akl2008}.
Several approaches, such as the evolutionary techniques \cite{Steven2000}, the evolutionary structural optimization method \cite{HuangSMOESO2010,NabakiHuangJSE2018}, the homogenization method \cite{Bendsoe1988,Allaire2002}, the material distribution or variable density method \cite{Rozvany2001,Bendsoe1999}, the level set method \cite{Wang2003,Allaire2004,Liu2008,Xing2009}, the method of moving morphable components \cite{GuoJAM2014,GuoCMAME2016}, the feature driven method with signed distance function \cite{ZhouCMAME2016} and the phase field method \cite{TakezawaJCP2010}, have been developed to implement topology optimization. The material distribution method has the advances of rapid convergence and low dependence of the initial guess. It is thereby chosen to implement the research in this paper.

Topology optimization for fluid problems was pioneered by using the evolutionary techniques \cite{Steven2000}. The first attempt of material distribution method based topology optimization for the Stokes flow was performed in 2003 \cite{Borrvall2003}. This method was further investigated for the Stokes flow \cite{Guillaume2004,Aage2008} and the Darcy-Stokes flow \cite{Guest2006,Wiker2007}, where an artificial friction force proportional to the fluid velocity was added to the Stokes equations in order to implement topology optimization based on the porous medium model proposed in Ref. \cite{Borrvall2003}.
This optimization model was then extended to the Navier-Stokes flow with low and moderate Reynolds numbers \cite{Gersborg-Hansen2005,Olessen2006,Guest2006,Aage2008,Deng2010}, and the non-Newtonian flow \cite{Pingen2010}. Topology optimization for fluid problems primarily focused on the steady flow without body forces \cite{BendsoeAndSigmund2003,Borrvall2003,Gersborg-Hansen2005,Olessen2006,Aage2008,Guest2006,Zhou2008,Pingen2010,Challis20092}. However, unsteady flow exists widespread. Topology optimization was extended to unsteady Navier-Stokes flow to reveal the related dynamic effects on the optimal topology \cite{Kreissl2011,Deng2011}. 
External body forces that relate with the fluid inertia, such as the gravity, centrifugal force and Coriolis force, usually exist in fluid flow. Topology optimization of the steady and unsteady Navier-Stokes flow with body forces was implemented by penalizing the body force based an interpolation function of the design variable and using the level set method, respectively \cite{Deng2013SMOBF,Deng2013CMAMEBF}. 
Transport of fluids at high velocity, which lead to turbulence, is common in industry. Topology optimization for turbulent flow with high Reynolds number was developed based on the finite volume discretized Reynolds-averaged Navier–Stokes equations coupled with either one- or two-equation turbulence closure models \cite{Dilgen2018CMAMETurbulent}, the Spalart-Allmaras models \cite{Yoon2016CMAMETurbulent,Sa2016CMAMETurbulent} and the data-driven model \cite{Hammond2022SMOTurbulent}, respectively. Based on the development of topology optimization for steady and unsteady flow, topology optimization of microfluidic devices has been performed, such as topology optimization of micromixers \cite{Andreasen2009IJNMFMixer,Deng2012BMMixer,Li2023CPLMixer,Guo2018MicromachinesMixer}, microvalves \cite{Deng2010IEEEMEMSValve,Abdelwahed2020ANAValve,Bohm2022MNValve,Liu2022EOValve} and micropumps \cite{Deng2011TRANSDUCERSPump,Tanaka2016TJSMEPump,Zhang2016SMOPump,Sun2023JAMSTPump}. 

Except the above researches for volume flow, topology optimization for surface flow has been developed to find the patterns of the related surface structures \cite{DengJCPTOOPSurfaceFlow2022,DengCJMEFBTOOPSurfaceFlow2024}. Surface flow can greatly decrease the computational cost in the numerical design of the related fluidic structures. The fluid flow in the channels attached on the walls of equipments can be described as the surface flow on the curved surfaces corresponding to the outer shapes of the equipments. The streamsurfaces corresponding to the outer shapes of fluidic structures with complete-slip boundaries can be described as surface flow separated from volume flow, where the complete-slip boundaries can be approximated and achieved by chemically coating or physically structuring solid surfaces to derive the extreme hydrophobicity \cite{KwonLangmuir2009}, using the optimal control method to manipulate the boundary velocity of flow \cite{SritharanSIAM1998}, and producing vapor layers between the solid and liquid phases based on the Leidenfrost phenomenon \cite{ThimblebyPhysicsEducation1989}, etc. To topologically optimize the patterns of surface structures, researches were implemented for stiffness and multi-material structures \cite{VermaakSMO2014,SigmundJMPS1997,GaoZhangIJNME2011,
LuoAiaaJ2012,WangCMAME2004,ZhouWangSMO2007,PanagiotisVogiatzis2018}, layouts of shell structures \cite{KrogOlhoffComputersStructures1996,AnsolaComputersStructures2002,HassaniSMO2013, LochnerAldingerSchumacher2014,ClausenActaMechanicaSinica2017,DienemannStructMultidiscOptim2017, YanChengWangJSV2018}, electrode patterns of electroosmosis \cite{DengIJHMT2018}, fluid-structure and fluid-particle interaction \cite{YoonFSIIJNME2010,LundgaardFSISMO2018,AndreasenFSISMO2019}, energy absorption \cite{AuligUlm2012}, cohesion \cite{MauteIJNME2017}, actuation \cite{MauteSMO2005} and wettability control \cite{DengMicrotextureCMAME2018,DengMicrotextureAMM2019,DengMicrotextureSMO2020}, etc.; topology optimization implemented on 2-manifolds was developed with the applications in elasticity, wettability control, heat transfer and electromagnetics \cite{HuoJAM2022,ZhangFengSMO2022,DengCMAME2020}; and fiber bundle topology optimization was developed for wettability control at fluid/solid interfaces \cite{DengMicrotextureSMO2020}. Recently, topology optimization for surface flow extended the design space of fluidic structures to 2-manifolds, where the 2-manifold represents the topological space locally homeomorphous to an Euclidean space \cite{DengJCPTOOPSurfaceFlow2022}. Fiber bundle topology optimization for surface flow was developed to match the pattern of the surface flow and the implicit 2-manifold on which the pattern is defined \cite{DengCJMEFBTOOPSurfaceFlow2024}. Then, the design space and design freedom were further extended for flow problems by including the design domain for the pattern of the surface flow into the design space, where the design domain is the implicit 2-manifold defined on the base manifold of the fiber bundle.

Mass and heat transfer are two widespread phenomena in fluid flow. Because of the scaling effect, microflow is usually in the region of laminar flow where the convection is weak and the diffusion dominates the mass and heat transfer processes. This causes the relatively low efficiency of mass and heat transfer. Therefore, enhancing the efficiency of mass and heat transfer in microflow becomes one of the eternal topics in the development of microfluidic devices for industry \cite{NguyenJMM2005,YaoRSER2015,MirallesDIAG2013}. Topology optimization is one of the most popular approaches used to enhance the efficiency of mass and heat transfer in microflow, where the microfluidic structures have been optimized to strengthen the convection \cite{HoghojDTU2023,Tawk2011NHTPB,Marck2013NHTPB}. With regards to mass transfer, topology optimization has been implemented for micromixers and microreactors \cite{Andreasen2009IJNMFMixer,Deng2012BMMixer,Li2023CPLMixer,Guo2018MicromachinesMixer,OkkelsECCOMAS2006,
SchaepperBiotechBioeng2011,WangCEJ2023,BhattacharjeeRCE2022,ChenJDST2021}; 
with regards to heat transfer, topology optimization has been implemented for heat sinks and heat exchangers \cite{FawazEnergy2022,ZhangHMT2008,RogieSMO2023,PietropaoliSMO2019,AlexandersenIJNMF2013,LohanSMO2017,
ZhangSMO2019,JooIJHMT2017,YanSCTS2023,HoghojIJHMT2020,LiIJHMT2019}. Those researches were implemented in three-dimensional (3D) domains or on the reduced two-dimensional (2D) planes. With the manufacturability permitted by the currently developed 3D-printing technologies, topology optimization on 2-manifolds can bring new design space for the mass and heat transfer problems; fiber bundle topology optimization can further extend the design freedom by optimizing the matching between the pattern of a surface structure and the implicit 2-manifold on which the surface pattern is defined. Therefore, this paper develops fiber bundle topology optimization for mass and heat transfer in surface and volume flow.

Fiber bundle is a concept of differential geometry \cite{ChernDifferentialGeometry1999}. It is composed of the base manifold and the fiber defined on the base manifold. The pattern of the surface structure together with its definition domain corresponds to the fiber of the fiber bundle. If there exists a 2-manifold homeomorphous to the fiber, it can be set as the base manifold of the fiber bundle. In computation, the existence of the base manifold can be ensured by presetting a fixed geometrical surface as the base manifold, then the fiber can be found on the preset base manifold. This means that the definition domain of the pattern is an implicit 2-manifold defined on the preset base manifold. The reason for this paper to use the concept of fiber bundle is to describe the topology of a surface structure as an ensemble instead of three separated components. Therefore, the task of fiber bundle topology optimization for mass and heat transfer in surface and volume flow is to find the optimized matching between the pattern of the surface structure and the implicit 2-manifold defined on the preset base manifold, to achieve the desired mass and heat transfer performance.

For surface flow, the material distribution method is used to determine the pattern of the surface structure, where the implicit 2-manifold used to define the surface structure is described on the base manifold. Then, two sets of design variables are required for the pattern of the surface structure and the implicit 2-manifold, respectively. For the material distribution method, the porous medium model has been developed in topology optimization for fluid flow \cite{Borrvall2003,Gersborg-Hansen2006,Kreissl2011,Deng2011}. In this model, the porous medium was filled in the 2D/3D design domains. Correspondingly, the artificial Darcy friction was introduced into the force terms of the Stokes equations and the Navier-Stokes equations. The impermeability of the porous medium was evolved in the topology optimization procedure to derive the fluidic structures. Inspired by the porous medium model, topology optimization and fiber bundle topology optimization for surface flow have been implemented by filling the porous medium onto 2-manifolds, where the artificial Darcy friction is added to the surface Navier-Stokes equations \cite{DengJCPTOOPSurfaceFlow2022,DengCJMEFBTOOPSurfaceFlow2024}. This paper inherits this porous medium model to implement fiber bundle topology optimization for mass and heat transfer in the surface flow.

For volume flow, the material density derived from the design variable is used to interpolate the no-jump and no-slip boundary conditions to determine the pattern of the thin-wall structure embedded in the volume flow of a mass and heat transfer problem, where the implicit 2-manifold used to define the no-jump and no-slip boundary conditions is described on the preset base manifold. Then, two sets of design variables are required for the pattern of the thin-wall structure and the implicit 2-manifold, respectively. To interpolate two different types of boundary conditions, a mixed boundary condition interpolated by the material density has been developed for electroosmotic flow \cite{DengIJHMT2018}. It was then extended to implement topology optimization on a 2-manifold, where the mixed boundary condition is constructed and defined on the 2-manifold \cite{DengCMAME2020}. The mixed boundary condition degenerates into two different boundary conditions when the material density is iteratively evolved into an approximated binary distribution. Then, a mixed form of the no-jump and no-slip boundary conditions defined on the implicit 2-manifold can be inspired and introduced for the Navier-Stokes equations used to describe the volume flow in the mass and heat transfer problem. Therefore, this paper uses the mixed boundary condition interpolated by the material density to implement fiber bundle topology optimization for mass and heat transfer in the volume flow.

The remained sections of this paper are organized as follows.
In Sections \ref{sec:MethodologyFiberBundleTOOPTransferSurfaceFlows} and \ref{sec:MethodologyFiberBundleTOOPTransferBulkFlows}, the methodology of fiber bundle topology optimization for mass and heat transfer is presented for the surface and volume flow, including the introduction of the physical model and design variables, the fiber bundle topology optimization problems, adjoint analysis, numerical implementation, and numerical results and discussion.
In Sections \ref{sec:Conclusions} and \ref{sec:Acknowledgements}, conclusions and acknowledgments of this paper are provided. In Section \ref{sec:AppendixMHM}, appendix is provided for Sections \ref{sec:MethodologyFiberBundleTOOPTransferSurfaceFlows} and \ref{sec:MethodologyFiberBundleTOOPTransferBulkFlows}. In this paper, the incompressible Newtonian fluid is considered; all the mathematical descriptions are implemented in the Cartesian systems; the column form is defaulted for a vector; and the convention that the gradient of a vector function has the gradient of the components as column vectors is used.

\section{Mass and heat transfer in surface flow}\label{sec:MethodologyFiberBundleTOOPTransferSurfaceFlows}

In this section, fiber bundle topology optimization is described for mass and heat transfer in the surface flow to find the optimized matching between the pattern of the surface flow and the implicit 2-manifold on which the surface flow is defined. The implicit 2-manifold is homeomorphically defined on a preset base manifold. The mass and heat transfer processes are described by the surface Navier-Stokes equations, the surface convection-diffusion equation and the surface convective heat-transfer equation defined on the implicit 2-manifold.

\subsection{Physical model and material interpolation} \label{sec:PorousMediumModelSurfaceNSEqus}

The porous medium model is utilized in fiber bundle topology optimization for mass and heat transfer in the surface flow. In this model, the porous medium is filled onto the implicit 2-manifold. The artificial Darcy friction is added to the surface Navier-Stokes equations used to describe the surface flow. The artificial Darcy friction, derived based on the constitutive law of the porous medium, is assumed to be proportional to the fluid velocity \cite{Borrvall2003,Gersborg-Hansen2006}:
\begin{equation}\label{equ:ArtificialDarcyFriction}
  \mathbf{b}_\Gamma = - \alpha \mathbf{u},~\forall \mathbf{x}_\Gamma \in \Gamma
\end{equation}
where $\alpha$ is the impermeability; $\Gamma$ is the implicit 2-manifold; and $\mathbf{x}_\Gamma$ denotes a point on $\Gamma$. When the porosity is zero, the porous medium corresponds to a solid material with infinite impermeability and zero fluid velocity caused by the infinite friction force; when the porosity is infinite, it corresponds to the structural void for the transport of the fluid with zero impermeability. Therefore, the impermeability can be described as
\begin{equation}\label{equ:ImpermeabilityInterp}
\left\{
\begin{split}
\alpha|_{\mathbf{x}_{\Gamma} \in \Gamma_D} & =
\left\{
\begin{split}
 & +\infty,~\gamma_p = 0 \\
 & 0      ,~\gamma_p = 1
\end{split}
\right. \\
\alpha|_{\mathbf{x}_{\Gamma} \in \Gamma_F} & = 0, ~ \gamma_p = 1 \\
\end{split}
\right.
\end{equation}
where $\gamma_p\in\left\{0,1\right\}$ is a binary distribution defined on $\Gamma$, with $0$ and $1$ representing the solid and fluid phases, respectively; $\Gamma_D$ is the design domain for the pattern of the surface flow and $\Gamma_F$ is the fluid domain with the material density enforced to be $\gamma_p = 1$, with $\Gamma_D$ and $\Gamma_F$ satisfying $\Gamma_D \cup \Gamma_F = \Gamma$ and $\Gamma_D \cap \Gamma_F = \emptyset$. Especially, $\Gamma$ is the design domain, when there is no enforced fluid domain, i.e. $\Gamma_F = \emptyset$ and $\Gamma = \Gamma_D$. Equivalently, the design domain can also be specified by using an indicator defined as
\begin{equation}\label{equ:IndicatorFuncDesignDom}
  f_{id,\Gamma} \left( \mathbf{x}_\Gamma \right) = \left\{\begin{split}
  & 1, ~ \forall \mathbf{x}_\Gamma \in \Gamma_D \\
  & 0, ~ \forall \mathbf{x}_\Gamma \in \Gamma \setminus \Gamma_D
  \end{split}\right.
\end{equation}
where $f_{id,\Gamma}$ is the indicator function. 

To avoid the numerical difficulty on solving a binary optimization problem, the binary variable on the design domain is relaxed to vary continuously in $\left[0,1\right]$. The relaxed binary variable is referred to as the design variable of the surface structure and it is used to derive the material density for the material interpolation of the impermeability. Based on the description of the impermeability in Eq. \ref{equ:ImpermeabilityInterp}, the material interpolation of the impermeability can be implemented by using the convex and $q$-parameterized scheme \cite{Borrvall2003}:
\begin{equation}\label{equ:InterpolationForImpermeability}
\begin{split}
\alpha \left( \gamma_p \right) = \left\{ \begin{split}
& \alpha_f + \left( \alpha_s - \alpha_f \right) q { 1 - \gamma_p \over q + \gamma_p }, ~ \forall \mathbf{x}_\Gamma \in \Gamma_D \\
& 0, ~ \forall \mathbf{x}_\Gamma \in \Gamma \setminus \Gamma_D \\
\end{split}\right.
\end{split}
\end{equation}
where $\gamma_p$ is renamed as the material density; $\alpha_s$ and $\alpha_f$ are the impermeability of the solid and fluid phases, respectively; $q$ is the parameter used to tune the convexity of this interpolation. For the fluid phase, the impermeability is zero, i.e. $\alpha_f = 0$. For the solid phase, $\alpha_s$ should be infinite theoretically; numerically, a finite value much larger than the fluid density $\rho$ is chosen for $\alpha_s$, to ensure the stability of the numerical implementation and approximate the solid phase with acceptable accuracy. 

\subsection{Design variables}\label{sec:DesignVariablePattern}

In fiber bundle topology optimization for mass and heat transfer in the surface flow, two sets of design variables are required to be sequentially defined for the implicit 2-manifold and the pattern of the surface flow.

\subsubsection{Design variable for implicit 2-manifold}\label{subsec:DesignVariableImplicitManifold}

To describe the implicit 2-manifold, the design variable that takes continuous values in $\left[0,1\right]$ is defined on the base manifold. This design variable is used to describe the distribution of the relative displacement between the implicit 2-manifold and the base manifold. The relative displacement is in the normal direction of the base manifold and the implicit 2-manifold is defined based on this normal displacement.

\begin{figure}[!htbp]
  \centering
  \subfigure[$d_m$]
  {\includegraphics[width=0.3\textwidth]{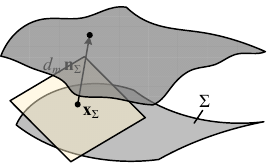}}\hspace{4em}
  \subfigure[$d_f$]
  {\includegraphics[width=0.3\textwidth]{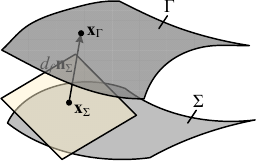}}
  \caption{Sketches for the design variable $d_m$ and the filtered design variable $d_f$ of the implicit 2-manifold $\Gamma$ defined on the base manifold $\Sigma$.}\label{fig:SketchPDEFilterOptimalMatchingManifold}
\end{figure}

To ensure the smoothness of the implicit 2-manifold and the well-posedness of the solution, a surface-PDE filter is imposed on the design variable of the implicit 2-manifold \cite{DengCMAME2020}:
      \begin{equation}\label{equ:PDEFilterzmBaseStructureMHM}
      \begin{split}
      & \left\{
        \begin{split}
          & - \mathrm{div}_\Sigma \left( r_m^2 \nabla_\Sigma d_f \right) + d_f = A_d \left( d_m - {1\over2} \right), ~ \forall \mathbf{x}_\Sigma \in \Sigma \\
          & \mathbf{n}_{\boldsymbol\tau_\Sigma} \cdot \nabla_\Sigma d_f = 0, ~ \forall \mathbf{x}_\Sigma \in \partial \Sigma \\
        \end{split}\right. \\
      \end{split}
      \end{equation}
where $d_m = d_m \left( \mathbf{x}_\Sigma \right)$ is the design variable of the implicit 2-manifold; $d_f = d_f \left( \mathbf{x}_\Sigma \right)$ is the filtered design variable and it is the normal displacement used to describe the implicit 2-manifold; $r_m$ is the filter radius, and it is constant; $\Sigma$ is the base manifold used to define the implicit 2-manifold, and $d_m$ and $d_f$ are defined on $\Sigma$; $\mathbf{x}_\Sigma$ denotes a point on $\Sigma$; $\nabla_\Sigma$ and $\mathrm{div}_\Sigma$ are the tangential gradient operator and tangential divergence operator defined on $\Sigma$, respectively; $\mathbf{n}_{\boldsymbol\tau_\Sigma} = \mathbf{n}_\Sigma\times\boldsymbol\tau_\Sigma$ is the unit outer conormal vector normal to $\partial\Sigma$ and tangent to $\Sigma$ at $\partial\Sigma$, with $\mathbf{n}_\Sigma$ and $\boldsymbol\tau_\Sigma$ representing the unit normal vector on $\Sigma$ and the unit tangential vector at $\partial\Sigma$, respectively; $A_d$ is the variable amplitude of the implicit 2-manifold, i.e. the parameter used to control the amplitude of the normal displacement, and it is nonnegative ($A_d \geq 0$). Because $d_m$ is valued in $\left[0,1\right]$, $d_f$ is valued in $\left[ -A_d/2, A_d/2 \right]$ and its magnitude is less than $A_d/2$. The design variable of the implicit 2-manifold and its filtered counterpart are sketched in Fig. \ref{fig:SketchPDEFilterOptimalMatchingManifold}.

After the filter operation, the implicit 2-manifold can be described by the filtered design variable:
\begin{equation}\label{equ:NormalDisplacementDistributionMHM}
  \Gamma = \left\{ \mathbf{x}_\Gamma \left| ~ \mathbf{x}_\Gamma = d_f \mathbf{n}_\Sigma + \mathbf{x}_\Sigma,~\forall \mathbf{x}_\Sigma \in \Sigma \right. \right\}
\end{equation}
where $\Gamma$ is the implicit 2-manifold and $\mathbf{x}_\Gamma$ denotes a point on $\Gamma$. A similar relation is held for the design domain of the pattern of the surface flow, i.e.
\begin{equation}\label{equ:NormalDisplacementDistributionMHMDesignDom}
  \Gamma_D = \left\{ \mathbf{x}_\Gamma \left| ~ \mathbf{x}_\Gamma = d_f \mathbf{n}_\Sigma + \mathbf{x}_\Sigma,~\forall \mathbf{x}_\Sigma \in \Sigma_D \right. \right\}
\end{equation}
where $\Sigma_D\subset\Sigma$ is the base manifold of $\Gamma_D$. From Eq. \ref{equ:NormalDisplacementDistributionMHM}, a differentiable homeomorphism can be determined corresponding to the bijection $d_f: \Sigma \mapsto \Gamma$ with $\mathbf{x}_\Gamma = d_f \mathbf{n}_\Sigma + \mathbf{x}_\Sigma $ at $\forall \mathbf{x}_\Sigma \in \Sigma$. Therefore, $\mathcal{H}\left(\Gamma\right)$ is homeomorphous to $\mathcal{H}\left(\Sigma\right)$. The Jacobian matrix of the homeomorphism in Eq. \ref{equ:NormalDisplacementDistributionMHM} for the implicit 2-manifold in the curvilinear coordinate system of the base manifold can be transformed into
\begin{equation}\label{equ:TransformedJacobian}
  \mathbf{T}_\Gamma = {\partial \mathbf{x}_\Gamma \over \partial \mathbf{x}_\Sigma} = \nabla_\Sigma d_f \mathbf{n}_\Sigma^\mathrm{T} + d_f \nabla_\Sigma \mathbf{n}_\Sigma + \mathbf{I}, ~ \forall \mathbf{x}_\Sigma \in \Sigma
\end{equation}
with $\left| \mathbf{T}_\Gamma \right|$ representing its determinant, where $\mathbf{I}$ is the unit tensor.

The variational formulation of the surface-PDE filter in Eq. \ref{equ:PDEFilterzmBaseStructureMHM} is considered in the first order Sobolev space defined on $\Sigma$. It can be derived based on the Galerkin method:
\begin{equation}\label{equ:VariationalFormulationPDEFilterBaseManifoldMHM}
\left\{\begin{split}
     & \mathrm{Find}~d_f \in\mathcal{H}\left(\Sigma\right) ~ \mathrm{for}~d_m \in \mathcal{L}^2\left(\Sigma\right) ~ \mathrm{and} ~ \forall \tilde{d}_f \in \mathcal{H}\left(\Sigma\right), \\
     & \mathrm{such~that} ~ \int_\Sigma r_m^2 \nabla_\Sigma d_f \cdot \nabla_\Sigma \tilde{d}_f + d_f\tilde{d}_f - A_d \left( d_m - {1\over2} \right) \tilde{d}_f \,\mathrm{d}\Sigma = 0
\end{split}\right.
\end{equation}
where $\tilde{d}_f$ is the test function of $d_f$; $\mathcal{H}\left(\Sigma\right)$ represents the first order Sobolev space defined on $\Sigma$; and $\mathcal{L}^2\left(\Sigma\right)$ represents the second order Lebesque space defined on $\Sigma$.

\subsubsection{Design variable for pattern of surface flow}\label{subsec:DesignVariablePattern}

\begin{figure}[!htbp]
  \centering
  \subfigure[$\gamma$]
  {\includegraphics[width=0.3\textwidth]{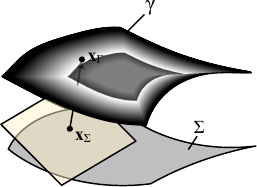}}\hspace{1em}
  \subfigure[$\gamma_f$]
  {\includegraphics[width=0.3\textwidth]{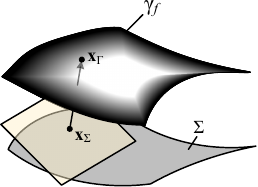}}\hspace{1em}
  \subfigure[$\gamma_p$]
  {\includegraphics[width=0.3\textwidth]{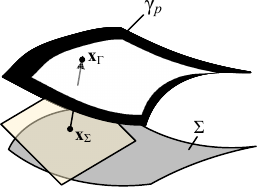}}
  \caption{Sketches for the design variable $\gamma$, the filtered design variable $\gamma_f$ and the material density $\gamma_p$ of the pattern of the surface flow.}\label{fig:SketchPDEFilterOptimalMatchingPattern}
\end{figure}

The pattern of the surface flow is represented by the material density defined on the implicit 2-manifold. The material density in Eqs. \ref{equ:ImpermeabilityInterp} and \ref{equ:InterpolationForImpermeability} is obtained by sequentially implementing the surface-PDE filter and the threshold projection on the design variable defined on the implicit 2-manifold, as sketched in Fig. \ref{fig:SketchPDEFilterOptimalMatchingPattern}. This design variable is also valued continuously in $\left[0,1\right]$. Here, the combination of the surface-PDE filter and the threshold projection can remove the gray regions and control the minimum length scale in the derived pattern.

The surface-PDE filter for the design variable of the pattern is implemented by solving the following surface PDE \cite{DengCMAME2020}:
      \begin{equation}\label{equ:PDEFilterGammaFilberMHM} 
        \left\{\begin{split}
        - \mathrm{div}_\Gamma \left( r_f^2 \nabla_\Gamma \gamma_f \right) + \gamma_f & = \gamma, ~\forall \mathbf{x}_\Gamma \in \Gamma_D \\
        \mathbf{n}_{\boldsymbol\tau_\Gamma} \cdot \nabla_\Gamma \gamma_f & = 0, ~\forall \mathbf{x}_\Gamma \in \partial\Gamma_D \\
        \end{split}\right.
      \end{equation}
where $\gamma$ is the design variable; $\gamma_f$ is the filtered design variable; $r_f$ is the filter radius, and it is constant; $\nabla_\Gamma$ and $\mathrm{div}_\Gamma$ are the tangential gradient operator and tangential divergence operator defined on the implicit 2-manifold $\Gamma$, respectively; $\mathbf{n}_{\boldsymbol\tau_\Gamma} = \mathbf{n}_\Gamma\times\boldsymbol\tau_\Gamma$ is the unit outer conormal vector normal to $\partial\Gamma$ and tangent to $\Gamma$ at $\partial\Gamma$, with $\mathbf{n}_\Gamma$ and $\boldsymbol\tau_\Gamma$ representing the unit normal vector on $\Gamma$ and the unit tangential vector at $\partial\Gamma$, respectively. The threshold projection of the filtered design variable is implemented as \cite{WangStructMultidiscipOptim2011,GuestIntJNumerMethodsEng2004}
      \begin{equation}\label{equ:ProjectionGammaFilberMHM}
        \gamma_p = { \tanh\left(\beta \xi\right) + \tanh\left(\beta \left(\gamma_f-\xi\right)\right) \over \tanh\left(\beta \xi\right) + \tanh\left(\beta \left(1-\xi\right)\right)}, ~ \forall \mathbf{x}_\Gamma \in \Gamma_D
      \end{equation}
where $\beta$ and $\xi$ are the parameters of the threshold projection, with values chosen based on numerical experiments \cite{GuestIntJNumerMethodsEng2004}.

The variational formulation of the surface-PDE filter is considered in the first order Sobolev space defined on $\Gamma_D$. It can be derived based on the Galerkin method:
\begin{equation}\label{equ:VariationalFormulationPDEFilterMHM} 
\left\{\begin{split}
     & \mathrm{Find}~\gamma_f \in\mathcal{H}\left(\Gamma_D\right) ~ \mathrm{for}~\gamma \in \mathcal{L}^2\left(\Gamma_D\right) ~ \mathrm{and} ~ \forall \tilde{\gamma}_f \in \mathcal{H}\left(\Gamma_D\right), \\
     & \mathrm{such~that} ~ \int_{\Gamma_D} r_f^2 \nabla_\Gamma \gamma_f \cdot \nabla_\Gamma \tilde{\gamma}_f + \gamma_f\tilde{\gamma}_f - \gamma \tilde{\gamma}_f \,\mathrm{d}\Gamma = 0
\end{split}\right.
\end{equation}
where $\tilde{\gamma}_f$ is the test function of $\gamma_f$; $\mathcal{H}\left(\Gamma_D\right)$ represents the first order Sobolev space defined on $\Gamma_D$; and $\mathcal{L}^2\left(\Gamma_D\right)$ represents the second order Lebesque space defined on $\Gamma_D$.

\subsubsection{Coupling of design variables}\label{subsec:CouplingDesignVariables}

The design variable introduced in Section \ref{subsec:DesignVariablePattern} for the pattern of the surface flow is defined on the implicit 2-manifold introduced in Section \ref{subsec:DesignVariableImplicitManifold}. Their coupling relation can be derived by transforming the tangential gradient operator $\nabla_\Gamma$, the tangential divergence operator $\mathrm{div}_\Gamma$ and the unit normal $\mathbf{n}_\Gamma$ into the forms defined on the base manifold $\Sigma$. 
The tangential gradient operator $\nabla_\Gamma$ can be transformed into
\begin{equation}\label{equ:ConcludeTangentialOperatorRelationMHM}
\begin{split}
  \nabla_\Gamma = \: & \mathbf{P} \nabla_{\mathbf{x}_\Xi} \\
  = \: & \nabla_{\mathbf{x}_\Xi} - \left( \mathbf{n}_\Gamma \cdot \nabla_{\mathbf{x}_\Xi} \right) \mathbf{n}_\Gamma \\
  = \: & \mathbf{T}_\Gamma^{-1} \nabla_{\mathbf{x}_\Omega} - \left[ \mathbf{n}_\Gamma \cdot \left(\mathbf{T}_\Gamma^{-1} \nabla_{\mathbf{x}_\Omega} \right) \right] \mathbf{n}_\Gamma \\
  = \: & \mathbf{T}_\Gamma^{-1} \left( \nabla_\Sigma + \nabla_\Sigma^{\perp} \right) - \left\{ \mathbf{n}_\Gamma \cdot \left[\mathbf{T}_\Gamma^{-1} \left( \nabla_\Sigma + \nabla_\Sigma^{\perp} \right)\right] \right\} \mathbf{n}_\Gamma \\
  = \: & \left\{ \mathbf{T}_\Gamma^{-1} \nabla_\Sigma - \left[ \mathbf{n}_\Gamma \cdot \left( \mathbf{T}_\Gamma^{-1} \nabla_\Sigma \right) \right] \mathbf{n}_\Gamma \right\} + \left\{ \mathbf{T}_\Gamma^{-1} \nabla_\Sigma^{\perp} - \left[ \mathbf{n}_\Gamma \cdot \left(\mathbf{T}_\Gamma^{-1} \nabla_\Sigma^{\perp} \right) \right] \mathbf{n}_\Gamma \right\}, \\
\end{split}
\end{equation}
where $\mathbf{P} = \mathbf{P}\left( \mathbf{x}_\Gamma \right) = \mathbf{I} - \mathbf{n}_\Gamma \mathbf{n}_\Gamma^\mathrm{T}$ is the normal projector in the tangential space of $\Gamma$ at $\mathbf{x}_\Gamma$; $\nabla_{\mathbf{x}_\Xi}$ and $\nabla_{\mathbf{x}_\Omega}$ are the gradient operators in the 3D Euclidean domains $\Xi$ and $\Omega$ imbedded with $\Gamma$ and $\Sigma$ in the extended Cartesian systems of $\mathbf{x}_\Gamma$ and $\mathbf{x}_\Sigma$, respectively, and they satisfy $\nabla_{\mathbf{x}_\Xi} = \mathbf{T}_\Gamma^{-1} \nabla_{\mathbf{x}_\Omega}$ at $\mathbf{x}_\Gamma \in \Gamma$ with $\forall \mathbf{x}_\Sigma \in \Sigma$; $\nabla_\Sigma^{\perp}$ is the normal component of $\nabla_{\mathbf{x}_\Sigma}$ on $\Sigma$. Because the variables on $\Sigma$ are defined intrinsically, $\nabla_\Sigma^{\perp}$ is nonexistent. Therefore, the tangential gradient operator in Eq. \ref{equ:ConcludeTangentialOperatorRelationMHM} can be transformed into
\begin{equation}\label{equ:TangentialOperatorRelationMHM}
  \nabla_\Gamma = \mathbf{T}_\Gamma^{-1} \nabla_\Sigma - \left[ \mathbf{n}_\Gamma \cdot \left( \mathbf{T}_\Gamma^{-1} \nabla_\Sigma \right) \right] \mathbf{n}_\Gamma.
\end{equation}
By implementing the dot product with $\mathbf{n}_\Gamma$ at the both sides of Eq. \ref{equ:TangentialOperatorRelationMHM}, $\mathbf{n}_\Gamma \cdot \nabla_\Gamma = 0$ can be retained for a differentiable function defined on $\Gamma$. Therefore, the transformed form of $\nabla_\Gamma$ in Eq. \ref{equ:TangentialOperatorRelationMHM} is self-consistent. 

\begin{figure}[!htbp]
  \centering
  \includegraphics[width=0.3\textwidth]{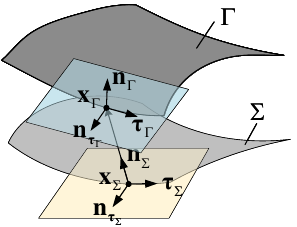}
  \caption{Sketch for relation among the unit tangential vector $\boldsymbol\tau_\Gamma$ at $\partial\Gamma$, the unit normal vector $\mathbf{n}_\Sigma$ on $\Sigma$ and the tangential gradient $\nabla_\Sigma d_f$.}\label{fig:DemonForTangentialVectorGammaMHM}
\end{figure}

From the relation in Eq. \ref{equ:NormalDisplacementDistributionMHM}, the following geometrical relation among the unit normal vectors on $\Sigma$ and $\Gamma$ and the gradient of $d_f$ can be derived as sketched in Fig. \ref{fig:DemonForTangentialVectorGammaMHM}:
\begin{equation}\label{equ:GeometricalRelationUnitaryNormalMHM}
  \mathbf{n}_\Gamma \parallel \left( \mathbf{n}_\Sigma - \nabla_\Sigma d_f \right), ~ \forall \mathbf{x}_\Sigma \in \Sigma
\end{equation}
where $\cdot \parallel \cdot$ represents the parallel relation of two vectors. Therefore, the unit normal vector on $\Gamma$ can be derived by normalizing $\left( \mathbf{n}_\Sigma - \nabla_\Sigma d_f \right)$ at $\forall \mathbf{x}_\Sigma \in \Sigma$, i.e. 
\begin{equation}\label{equ:UnitaryNormalGammaMHM}
  \mathbf{n}_\Gamma^{\left( d_f \right)} = { \mathbf{n}_\Sigma - \nabla_\Sigma d_f \over \left\| \mathbf{n}_\Sigma - \nabla_\Sigma d_f \right\|_2},
\end{equation}
where $\left\| \cdot \right\|_2$ is the 2-norm of a vector. In Eq. \ref{equ:UnitaryNormalGammaMHM}, the transformed unit normal vector is distinguished from the original form by using the filtered design variable $d_f$ as the superscript, and this identification method is adopted in the following for the other transformed operators and variables. 

The tangential gradient operator $\nabla_\Gamma$ is sequentially transformed into
\begin{equation}\label{equ:TransformedTangentialOperatorMHM} 
\begin{split}
  \nabla_\Gamma^{\left( d_f \right)} g = \mathbf{T}_\Gamma^{-1} \nabla_\Sigma g - \left[ \mathbf{n}_\Gamma^{\left( d_f \right)} \cdot \left( \mathbf{T}_\Gamma^{-1} \nabla_\Sigma g \right) \right] \mathbf{n}_\Gamma^{\left( d_f \right)}, ~ \forall g \in \mathcal{H}\left( \Sigma \right).
  \end{split}
\end{equation}
Based on the transformed tangential gradient operator, the tangential divergence operator $\mathrm{div}_\Gamma$ can be transformed into 
\begin{equation}\label{equ:TransformedDivergenceOperatorMHM}
\begin{split}
  \mathrm{div}_\Gamma^{\left( d_f \right)} \mathbf{g} =\: & \mathrm{tr}\left( \nabla_\Gamma^{\left( d_f \right)} \mathbf{g} \right) \\
  =\: & \mathrm{tr} \left( \mathbf{T}_\Gamma^{-1} \nabla_\Sigma \mathbf{g} - \left[ \mathbf{n}_\Gamma^{\left( d_f \right)} \cdot \left( \mathbf{T}_\Gamma^{-1} \nabla_\Sigma \mathbf{g} \right) \right] \mathbf{n}_\Gamma^{\left( d_f \right)} \right), ~ \forall \mathbf{g} \in \left( \mathcal{H}\left( \Sigma \right) \right)^3
\end{split}
\end{equation}
where $\mathrm{tr}$ is the operator used to extract the trace of a tensor.


Based on Eq. \ref{equ:FirstOrderVariOfVectorNormMHM} in the appendix in Section \ref{sec:AdjointAnalysisDesignObjectiveSurfaceCDMHM}, the first order variational of $\mathbf{n}_\Gamma^{\left( d_f \right)}$ to $d_f$ can be derived as
\begin{equation}\label{equ:1stVariofNormalGammaTodf}
  \mathbf{n}_\Gamma^{\left( d_f, \tilde{d}_f \right)} = - { \nabla_\Sigma \tilde{d}_f \over \left\| \mathbf{n}_\Sigma - \nabla_\Sigma d_f \right\|_2 } + { \mathbf{n}_\Gamma^{\left( d_f \right)} \cdot \nabla_\Sigma \tilde{d}_f \over \left\| \mathbf{n}_\Sigma - \nabla_\Sigma d_f \right\|_2} \mathbf{n}_\Gamma^{\left( d_f \right)}, ~ \forall \tilde{d}_f \in \mathcal{H}\left( \Sigma \right).
\end{equation}
Because the tangential gradient operator $\nabla_\Gamma$ depends on $d_f$, its first-order variational to $d_f$ can be derived as 
\begin{equation}\label{equ:FirstOrderVarisForTangentialOperatorsMHM}
\begin{split}
& \nabla_\Gamma^{\left(d_f, \tilde{d}_f\right)} g = \left( { \partial \mathbf{T}_\Gamma^{-1} \over \partial d_f } \tilde{d}_f + { \partial \mathbf{T}_\Gamma^{-1} \over \partial \nabla_\Sigma d_f } \cdot \nabla_\Sigma \tilde{d}_f \right) \nabla_\Sigma g - \left[ \mathbf{n}_\Gamma^{\left( d_f, \tilde{d}_f \right)} \cdot \left( \mathbf{T}_\Gamma^{-1} \nabla_\Sigma g \right) \right] \mathbf{n}_\Gamma^{\left( d_f \right)} \\
& ~~~~~~~~~~~~~~~~ - \left\{ \mathbf{n}_\Gamma^{\left( d_f \right)} \cdot \left[\left( { \partial \mathbf{T}_\Gamma^{-1} \over \partial d_f } \tilde{d}_f + { \partial \mathbf{T}_\Gamma^{-1} \over \partial \nabla_\Sigma d_f } \cdot \nabla_\Sigma \tilde{d}_f \right) \nabla_\Sigma g \right] \right\} \mathbf{n}_\Gamma^{\left( d_f \right)} \\
& ~~~~~~~~~~~~~~~~ - \left[ \mathbf{n}_\Gamma^{\left( d_f \right)} \cdot \left(\mathbf{T}_\Gamma^{-1} \nabla_\Sigma g \right) \right] \mathbf{n}_\Gamma^{\left( d_f, \tilde{d}_f \right)}, ~ \forall g \in \mathcal{H}\left( \Sigma \right) ~ \mathrm{and} ~ \forall \tilde{d}_f \in \mathcal{H}\left( \Sigma \right).
\end{split}
\end{equation}
Similarly, the first-order variational of $\mathrm{div}_\Gamma$ to $d_f$ can be derived as
\begin{equation}\label{equ:FirstOrderVarisForDivergenceOperatorsMHM}
\begin{split}
& \mathrm{div}_\Gamma^{\left(d_f,\tilde{d}_f\right)} \mathbf{g} = ~ \mathrm{tr} \Bigg( \left( { \partial \mathbf{T}_\Gamma^{-1} \over \partial d_f } \tilde{d}_f + { \partial \mathbf{T}_\Gamma^{-1} \over \partial \nabla_\Sigma d_f } \cdot \nabla_\Sigma \tilde{d}_f \right) \nabla_\Sigma \mathbf{g} - \left[ \mathbf{n}_\Gamma^{\left( d_f, \tilde{d}_f \right)} \cdot \left(\mathbf{T}_\Gamma^{-1} \nabla_\Sigma \mathbf{g} \right) \right] \mathbf{n}_\Gamma^{\left( d_f \right)} \\
& ~~~~~~~~~~~~~~~~~~ - \left\{ \mathbf{n}_\Gamma^{\left( d_f \right)} \cdot \left[\left( { \partial \mathbf{T}_\Gamma^{-1} \over \partial d_f } \tilde{d}_f + { \partial \mathbf{T}_\Gamma^{-1} \over \partial \nabla_\Sigma d_f } \cdot \nabla_\Sigma \tilde{d}_f \right) \nabla_\Sigma \mathbf{g} \right] \right\} \mathbf{n}_\Gamma^{\left( d_f \right)} \\
& ~~~~~~~~~~~~~~~~~~ - \left[ \mathbf{n}_\Gamma^{\left( d_f \right)} \cdot \left( \mathbf{T}_\Gamma^{-1} \nabla_\Sigma \mathbf{g} \right) \right] \mathbf{n}_\Gamma^{\left( d_f, \tilde{d}_f \right)} \Bigg), ~ \forall \mathbf{g} \in \left(\mathcal{H} \left( \Sigma \right)\right)^3 ~ \mathrm{and} ~ \forall \tilde{d}_f \in \mathcal{H}\left( \Sigma \right).
\end{split}
\end{equation}

Because $d_f$ is a differentiable homeomorphism, it can induce the Riemannian metric for $\Gamma$. Then, the differential on the base manifold and implicit 2-manifold satisfies
\begin{equation}\label{equ:DiffRiemannianMHM}
\left\{\begin{split}
  & \mathrm{d}\Gamma = \left| \mathbf{T}_\Gamma \right| \left\| \mathbf{T}_\Gamma \mathbf{n}_\Gamma^{\left( d_f \right)} \right\|_2^{-1} \mathrm{d}\Sigma \\
  & \mathrm{d}l_{\partial\Gamma} = \left\| \boldsymbol{\tau}_\Gamma \right\|_2 \left\| \mathbf{T}_\Gamma^{-1} \boldsymbol{\tau}_\Gamma \right\|_2^{-1} \mathrm{d}l_{\partial\Sigma} \\
\end{split}\right.,
\end{equation}
where $\mathrm{d}l_{\partial\Gamma}$ and $\mathrm{d}l_{\partial\Sigma}$ are the differential of the boundary curves of $\Gamma$ and $\Sigma$, respectively. In Eq. \ref{equ:DiffRiemannianMHM}, the unit tangential vector $\boldsymbol\tau_\Gamma$ at $\partial\Gamma$ sketched in Fig. \ref{fig:DemonForTangentialVectorGammaMHM} satisfies
\begin{equation}\label{equ:TauGammaMHM}
\left. \begin{split}
& \mathbf{n}_{{\boldsymbol\tau}_{\Sigma}} \parallel \left( \mathbf{n}_\Sigma \times \nabla_\Sigma d_f \right) \\
& \mathbf{n}_{\Gamma} \parallel \left( \mathbf{n}_\Sigma - \nabla_\Sigma d_f\right) \\
& \mathbf{n}_{{\boldsymbol\tau}_{\Gamma}} = \mathbf{n}_{\Gamma} \times {\boldsymbol\tau}_{\Gamma} \\
& {\boldsymbol\tau}_{\Sigma} \parallel \nabla_\Sigma d_f \\
\end{split}\right\}
  \Rightarrow \boldsymbol\tau_\Gamma \parallel \left[ \left( \mathbf{n}_\Sigma \times \nabla_\Sigma d_f \right) \times \left( \mathbf{n}_\Sigma - \nabla_\Sigma d_f\right) \right].
\end{equation}
Therefore, the second equation in Eq. \ref{equ:DiffRiemannianMHM} can be transformed into
\begin{equation}\label{equ:TransformedDiffRiemannianCurveMHM}
  \mathrm{d}l_{\partial\Gamma} = \left\| \left( \mathbf{n}_\Sigma \times \nabla_\Sigma d_f \right) \times \left( \mathbf{n}_\Sigma - \nabla_\Sigma d_f\right) \right\|_2 \left\| \mathbf{T}_\Gamma^{-1} \left[ \left( \mathbf{n}_\Sigma \times \nabla_\Sigma d_f \right) \times \left( \mathbf{n}_\Sigma - \nabla_\Sigma d_f\right)\right] \right\|_2^{-1} \,\mathrm{d}l_{\partial\Sigma}.
\end{equation}
In the following parts of this paper, $M^{\left( d_f \right)}$ and $L^{\left( d_f \right)}$ are defined as follows for Eqs. \ref{equ:DiffRiemannianMHM} and \ref{equ:TransformedDiffRiemannianCurveMHM}, i.e.
\begin{equation}\label{equ:MetricMLMHM}
  \left\{\begin{split}
  & M^{\left( d_f \right)} \doteq \left| \mathbf{T}_\Gamma \right| \left\| \mathbf{T}_\Gamma \mathbf{n}_\Gamma^{\left( d_f \right)} \right\|_2^{-1} \\
  & L^{\left( d_f \right)} \doteq \left\| \left( \mathbf{n}_\Sigma \times \nabla_\Sigma d_f \right) \times \left( \mathbf{n}_\Sigma - \nabla_\Sigma d_f\right) \right\|_2 \left\| \mathbf{T}_\Gamma^{-1} \left[ \left( \mathbf{n}_\Sigma \times \nabla_\Sigma d_f \right) \times \left( \mathbf{n}_\Sigma - \nabla_\Sigma d_f\right)\right] \right\|_2^{-1} \\
  \end{split}\right..
\end{equation}

Based on the transformed tangential gradient operator in Eq. \ref{equ:TransformedTangentialOperatorMHM}, the homeomorphism between $\mathcal{H}\left(\Gamma\right)$ and $\mathcal{H}\left(\Sigma\right)$ described in Eq. \ref{equ:NormalDisplacementDistributionMHM} and the homeomorphism between $\mathcal{H}\left(\Gamma_D\right)$ and $\mathcal{H}\left(\Sigma_D\right)$ described in Eq. \ref{equ:NormalDisplacementDistributionMHMDesignDom}, the coupling relation between the two sets of design variables can be derived by instituting Eq. \ref{equ:TransformedTangentialOperatorMHM} into Eq. \ref{equ:VariationalFormulationPDEFilterMHM}:
\begin{equation}\label{equ:CoupledVariationalPDEFilter} 
\left\{\begin{split}
     & \mathrm{Find}~\gamma_f \in\mathcal{H}\left(\Sigma_D\right) ~ \mathrm{for} ~ \gamma \in \mathcal{L}^2\left(\Sigma_D\right) ~ \mathrm{and} ~ \forall \tilde{\gamma}_f \in \mathcal{H}\left(\Sigma_D\right), \\
     & \mathrm{such~that} ~ \int_{\Sigma_D} \left( r_f^2 \nabla_\Gamma^{\left( d_f \right)} \gamma_f \cdot \nabla_\Gamma^{\left( d_f \right)} \tilde{\gamma}_f + \gamma_f\tilde{\gamma}_f - \gamma \tilde{\gamma}_f \right) M^{\left( d_f \right)} \,\mathrm{d}\Sigma = 0
\end{split}\right.
\end{equation}
where the tangential gradient operator $\nabla_\Gamma$ on $\Gamma$ is replaced by its transformed form $\nabla_\Gamma^{\left( d_f \right)}$ in Eq. \ref{equ:TransformedTangentialOperatorMHM}; and $\mathcal{H}\left(\Sigma_D\right)$ and $\mathcal{L}^2\left(\Sigma_D\right)$ defined on $\Sigma_D$ are the homeomorphous counterparts of $\mathcal{H}\left(\Gamma_D\right)$ and $\mathcal{L}^2\left(\Gamma_D\right)$, respectively.

\subsubsection{Fiber bundle of surface structure} \label{subsec:FiberBundleMassHeatTransfer}

\begin{figure}[!htbp]
  \centering
  \begin{tikzpicture}[node distance=0.2\columnwidth, auto]
  \node (P) {$\Gamma \times \left[0,1\right]$};
  \node (C) [right of=P] {$\Sigma \times \left( \Gamma \times \left[0,1\right] \right)$};
  \node (Ai) [below of=P] {$\Sigma$};
  \draw[->, thick] (C) to node {$proj_1$} (Ai);
  \draw[->, thick] (P) to node {$\varphi_2$} (C);
  \draw[->, thick] (P) to node [swap] {$\varphi_1^{-1}$} (Ai);
  \end{tikzpicture}
  \caption{Diagram for the fiber bundle composed of the base manifold, the implicit 2-manifold and the pattern of the surface structure.}\label{fig:DiagramFiberBundle1}
\end{figure}
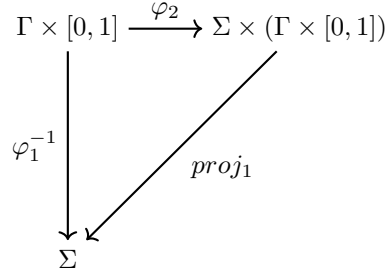

The fiber bundle of the surface structure for mass and heat transfer in the surface flow is composed of the base manifold together with the implicit 2-manifold and the pattern of the surface structure, where $\Sigma$ is the base manifold and $\Gamma \times \left[0,1\right]$ is the fiber, respectively. It can be expressed as
\begin{equation}\label{equ:FiberBundleMHM}
  \left( \Sigma \times \left(\Gamma \times \left[0,1\right]\right), \Sigma, proj_1, \Gamma \times \left[0, 1\right] \right),
\end{equation}
where $proj_1$ is the natural projection $proj_1: \Sigma \times \left(\Gamma \times \left[0,1\right]\right) \mapsto \Sigma$; $\varphi_1$ is the homeomorphous map $\varphi_1 : \Sigma \mapsto \Gamma \times \left[0,1\right]$; $\varphi_2$ is the homeomorphous map $\varphi_2 : \Gamma \times \left[0,1\right] \mapsto \Sigma \times \left(\Gamma \times \left[0,1\right]\right)$; and $proj_1$, $\varphi_1$ and $\varphi_2$ satisfy
\begin{equation}
  \left\{\begin{split}
    & proj_1\left(\mathbf{x}_\Sigma,\left(\mathbf{x}_\Gamma, \gamma_p \right)\right) = proj_1\left(\mathbf{x}_\Sigma, \left( d_f\left(\mathbf{x}_\Sigma \right), \gamma_p \right)\right) = \mathbf{x}_\Sigma, ~ \forall \mathbf{x}_\Sigma\in\Sigma \\
    & \varphi_1\left( \mathbf{x}_\Sigma \right) = \left( \mathbf{x}_\Gamma, \gamma_p \right) = \left( d_f\left( \mathbf{x}_\Sigma \right), \gamma_p \right), ~ \forall \mathbf{x}_\Sigma \in \Sigma \\
    & \varphi_2\left( \mathbf{x}_\Gamma, \gamma_p \right) = \left( \mathbf{x}_\Sigma, \left( \mathbf{x}_\Gamma, \gamma_p \right) \right) = \left( \mathbf{x}_\Sigma, \left( d_f \left(\mathbf{x}_\Sigma\right), \gamma_p \right) \right), ~ \forall \left( \mathbf{x}_\Gamma, \gamma_p \right) \in \Gamma \times \left[ 0, 1 \right] \\
    \end{split}\right..
  \end{equation}
The diagram of the fiber bundle in Eq. \ref{equ:FiberBundleMHM} is shown in Fig. \ref{fig:DiagramFiberBundle1}. 

\subsection{Mass transfer problem} \label{sec:MassTransferProblem}

The mass transfer process in the surface flow can be described by the surface Navier-Stokes equations and the surface convection-diffusion equation.

\subsubsection{Surface Navier-Stokes equations} \label{sec:SurfaceNSEqus}

The governing equations for the motion of a Newtonian surface fluid can be formulated intrinsically on a 2-manifold of codimension one in an Euclidian space.
Based on the conservation laws of momentum and mass, the surface Navier-Stokes equations can be derived to describe the incompressible surface flow \cite{ArroyoPRE2009,BrennerElsevier2013,RahimiSoftMatter2013}:
\begin{equation}\label{equ:NSequOnManifoldsMHM}
\begin{split}
\left.\begin{split}
\rho \left( \mathbf{u} \cdot \nabla_\Gamma \right) \mathbf{u} - \mathbf{P} \mathrm{div}_\Gamma \left[ \eta \left( \nabla_\Gamma \mathbf{u} + \nabla_\Gamma \mathbf{u}^\mathrm{T} \right) \right] + \nabla_\Gamma p & = - \alpha \mathbf{u} \\
- \mathrm{div}_\Gamma \mathbf{u} & = 0 \\
\mathbf{u} \cdot \mathbf{n}_\Gamma & = 0 \\
\end{split}\right\}~\forall \mathbf{x}_\Gamma \in \Gamma,
\end{split}
\end{equation}
where $\mathbf{u}$ is the fluid velocity; $p$ is the fluid pressure; $\rho$ is the fluid density; $\eta$ is the dynamic viscosity; and $\mathbf{u} \cdot \mathbf{n}_\Gamma = 0$ is the tangential constraint of the fluid velocity. The tangential constraint is imposed, because the fluid spatially flows on the 2-manifold $\Gamma$ and the fluid velocity is a vector in the tangential space of $\Gamma$.

To solve the mass and heat transfer problem, the fluid velocity and stress in the surface Navier-Stokes equations are required to be specified at the boundaries of the 2-manifold $\Gamma$:
\begin{equation}\label{equ:BNDConditionSurfaceNSEqu}
\left\{
\begin{split}
& \mathbf{u} = \mathbf{u}_{l_{v,\Gamma}}, ~ \forall \mathbf{x}_\Gamma \in l_{v,\Gamma} ~~~ \left( \mathrm{Inlet~boundary~condition} \right) \\
& \mathbf{u} = \mathbf{0}, ~ \forall \mathbf{x}_\Gamma \in l_{v_0,\Gamma} ~~~ \left( \mathrm{Wall~boundary~condition} \right) \\
& \left[ - \eta \left( \nabla_\Gamma \mathbf{u} + \nabla_\Gamma \mathbf{u}^\mathrm{T} \right) + p \right] \mathbf{n}_{\boldsymbol\tau_\Gamma} = \mathbf{0}, ~\forall \mathbf{x}_\Gamma \in l_{s,\Gamma} ~~~ \left( \mathrm{Outlet~boundary~condition} \right) \\
\end{split}
\right.,
\end{equation}
where $\mathbf{u}_{l_{v,\Gamma}}$ is a known distribution of the fluid velocity at the inlet boundary curve of $\Gamma$, depending on the specified fluid velocity $\mathbf{u}_{l_{v,\Sigma}}$ at $l_{v,\Sigma}$ representing a boundary curve of $\Sigma$; 
the no-slip boundary condition with velocity equal to $\mathbf{0}$ is imposed on the wall boundary curve $l_{v_0,\Gamma}$ of $\Gamma$; and $l_{s,\Gamma}$ is the boundary curve imposed with the outlet boundary condition.

The variational formulation of the surface Navier-Stokes equations is considered in the functional spaces without containing the tangential constraint of the fluid velocity. The tangential constraint of the fluid velocity is imposed by using the Lagrangian multiplier \cite{FriesIJNMF2018,ReutherPOF2018}. Based on the Galerkin method, the variational formulation of the surface Navier-Stokes equations can be derived as
\begin{equation}\label{equ:VariationalFormulationSurfaceNavierStokesEqus}
\left\{\begin{split}
  & \mathrm{Find} \left\{\begin{split}
    & \mathbf{u}\in\left(\mathcal{H}\left(\Gamma\right)\right)^3~\mathrm{with} ~ \left\{ \begin{split}
    & \mathbf{u} = \mathbf{u}_{l_{v,\Gamma}}~ \mathrm{at} ~ \forall \mathbf{x}_\Gamma \in l_{v,\Gamma} \\
    & \mathbf{u} = \mathbf{0}~ \mathrm{at} ~ \forall \mathbf{x}_\Gamma \in l_{v_0,\Gamma} \\
    \end{split}\right.\\
  & p \in \mathcal{H}\left(\Gamma\right) \\
  & \lambda\in\mathcal{L}^2\left(\Gamma\right)~\mathrm{with}~ \lambda=0~ \mathrm{at} ~ \forall \mathbf{x}_\Gamma \in l_{v,\Gamma} \cup l_{v_0,\Gamma} \\
  \end{split}\right.\\
  & \mathrm{for} \left\{\begin{split}
  & \forall \tilde{\mathbf{u}} \in\left(\mathcal{H}\left(\Gamma\right)\right)^3 \\
  & \forall \tilde{p} \in \mathcal{H}\left(\Gamma\right) \\
  & \forall \tilde{\lambda} \in \mathcal{L}^2\left(\Gamma\right)
  \end{split}\right., ~ \mathrm{such~that} \\
  &\int_\Gamma \rho \left( \mathbf{u} \cdot \nabla_\Gamma \right) \mathbf{u} \cdot \tilde{\mathbf{u}} + {\eta\over2} \left( \nabla_\Gamma \mathbf{u} + \nabla_\Gamma \mathbf{u}^\mathrm{T} \right) : \left( \nabla_\Gamma \tilde{\mathbf{u}} + \nabla_\Gamma \tilde{\mathbf{u}}^\mathrm{T} \right) - p\,\mathrm{div}_\Gamma \tilde{\mathbf{u}} - \tilde{p} \,\mathrm{div}_\Gamma \mathbf{u} \\
  & + \alpha \mathbf{u} \cdot \tilde{\mathbf{u}} + \lambda \tilde{\mathbf{u}} \cdot \mathbf{n}_\Gamma + \tilde{\lambda} \mathbf{u} \cdot \mathbf{n}_\Gamma \,\mathrm{d}\Gamma - \sum_{E_\Gamma\in\mathcal{E}_\Gamma} \int_{E_\Gamma} \tau_{BP,\Gamma} \nabla_\Gamma p \cdot \nabla_\Gamma \tilde{p} \,\mathrm{d}\Gamma = 0 \\
\end{split}\right.
\end{equation}
where $\lambda$ is the Lagrange multiplier used to impose the tangential constraint of the fluid velocity; $\tilde{\mathbf{u}}$, $\tilde{p}$ and $\tilde{\lambda}$ are the test functions of $\mathbf{u}$, $p$ and $\lambda$, respectively; the Brezzi-Pitk\"{a}ranta stabilization term
\begin{equation*}
  - \sum_{E_\Gamma\in\mathcal{E}_\Gamma} \int_{E_\Gamma} \tau_{BP,\Gamma} \nabla_\Gamma p \cdot \nabla_\Gamma \tilde{p} \,\mathrm{d}\Gamma
\end{equation*}
with $\tau_{BP,\Gamma}$ representing the stabilization parameter is imposed on the variational formulation, in order to use linear finite elements to solve both the fluid velocity and pressure \cite{DoneaWiley2003}; $\mathcal{E}_\Gamma$ is an elementization of $\Gamma$; and $E_\Gamma$ is an element of the elementization $\mathcal{E}_\Gamma$. The stabilization parameter is chosen as \cite{DoneaWiley2003}
\begin{equation}\label{equ:NSSurfaceStabilizationTermCD}
 \tau_{BP,\Gamma} = {h_{E_\Gamma}^2 \over 12\eta},
\end{equation}
where $h_{E_\Gamma}$ is the size of the element $E_\Gamma$. Because the element $E_\Gamma$ and the elementization $\mathcal{E}_\Gamma$ for the 2-manifold $\Gamma$ are implicitly defined on the base 2-manifold $\Sigma$, they are derived from the design variable for the implicit 2-manifold $\Gamma$ and the explicit elementization $\mathcal{E}_\Sigma$ of the base 2-manifold $\Sigma$. Based on Eqs. \ref{equ:DiffRiemannianMHM} and \ref{equ:MetricMLMHM}, $h_{E_\Gamma}^2$ in $\tau_{BP,\Gamma}$ can be approximated by the area of $E_\Gamma$. Then, it can be transformed into
\begin{equation}\label{equ:ElementAreaTransformationMHM}
\begin{split}
  h_{E_\Gamma}^2 \approx \, & \int_{E_\Gamma} 1 \, \mathrm{d}\Gamma \\
  = \, & \int_{E_\Sigma} M^{\left( d_f \right)} \, \mathrm{d}\Sigma \\
  \approx \, & h_{E_\Sigma}^2 \int_{E_\Sigma} M^{\left( d_f \right)} \, \mathrm{d}\Sigma \bigg/ \int_{E_\Sigma} 1 \, \mathrm{d}\Sigma \\
  = \, & h_{E_\Sigma}^2 \bar{M}^{\left( d_f \right)}_{E_\Sigma},
\end{split}
\end{equation}
where $h_{E_\Sigma}$ is the size of the element $E_\Sigma$ representing an element of the explicit elementization $\mathcal{E}_\Sigma$ of $\Sigma$; $\bar{M}^{\left( d_f \right)}_{E_\Sigma}$ is the average value of $M^{\left( d_f \right)}$ in the element $E_\Sigma$; and the area of $E_\Sigma$ can be approximated as $h_{E_\Sigma}^2$, i.e. $\int_{E_\Sigma} 1 \, \mathrm{d}\Sigma \approx h_{E_\Sigma}^2$. Because the elementization satisfies $h_{E_\Sigma}^2 \ll \left| \Sigma \right|$ with $\left| \Sigma \right|$ representing the area of $\Sigma$, $\bar{M}^{\left( d_f \right)}_{E_\Sigma}$ can be well approximated by the value of $M^{\left( d_f \right)}$ at $\forall \mathbf{x}_\Sigma \in E_\Sigma$, i.e.
\begin{equation}\label{equ:ApproximationMetricAverageMHM}
  \bar{M}^{\left( d_f \right)}_{E_\Sigma} \approx M^{\left( d_f \right)}, ~ \forall \mathbf{x}_\Sigma \in E_\Sigma.
\end{equation}
Therefore, the stabilization parameter $\tau_{BP,\Gamma}$ in Eq. \ref{equ:NSSurfaceStabilizationTermCD} can be transformed into
\begin{equation}\label{equ:TransformedNSSurfaceStabilizationTermCD}
  \tau_{BP,\Gamma}^{\left( d_f \right)} = {h_{E_\Sigma}^2 \over 12\eta} M^{\left( d_f \right)}. 
\end{equation}


Because $l_{v,\Gamma}$ and $l_{v_0,\Gamma}$ are homeomorphous to $l_{v,\Sigma}$ and $l_{v_0,\Sigma}$, respectively, $l_{v,\Sigma}$ and $l_{v_0,\Sigma}$ are fixed, and $\mathbf{u}_{l_{v,\Gamma}}$ is homeomorphous to $\mathbf{u}_{l_{v,\Sigma}}$, the variational formulation in Eq. \ref{equ:VariationalFormulationSurfaceNavierStokesEqus} can be transformed into the following form defined on $\Sigma$ based on the coupling relations in Section \ref{subsec:CouplingDesignVariables}:
\begin{equation}\label{equ:TransformedVariationalFormulationSurfaceNSEqusCD}
\left\{\begin{split}
  & \mathrm{Find} \left\{\begin{split}
    & \mathbf{u}\in\left(\mathcal{H}\left(\Sigma\right)\right)^3~\mathrm{with}~ \left\{ \begin{split}
    & \mathbf{u} = \mathbf{u}_{l_{v,\Sigma}}~ \mathrm{at} ~ \forall \mathbf{x}_\Sigma \in l_{v,\Sigma} \\
    & \mathbf{u} = \mathbf{0}~ \mathrm{at} ~ \forall \mathbf{x}_\Sigma \in l_{v_0,\Sigma}\\
    \end{split}\right.\\
  & p \in \mathcal{H}\left(\Sigma\right) \\
  & \lambda\in\mathcal{L}^2\left(\Sigma\right)~\mathrm{with}~ \lambda=0~ \mathrm{at} ~ \forall \mathbf{x}_\Sigma \in l_{v,\Sigma} \cup l_{v_0,\Sigma}\\
  \end{split}\right. \\
  & \mathrm{for} \left\{\begin{split}
  & \forall \tilde{\mathbf{u}} \in\left(\mathcal{H}\left(\Sigma\right)\right)^3 \\
  & \forall \tilde{p} \in \mathcal{H}\left(\Sigma\right) \\
  & \forall \tilde{\lambda} \in \mathcal{L}^2\left(\Sigma\right)
  \end{split}\right.,~ \mathrm{such~that}\\
  & \int_\Sigma \bigg[ \rho \left( \mathbf{u} \cdot \nabla_\Gamma^{\left(d_f\right)} \right) \mathbf{u} \cdot \tilde{\mathbf{u}} + {\eta\over2} \left( \nabla_\Gamma^{\left(d_f\right)} \mathbf{u} + \nabla_\Gamma^{\left(d_f\right)} \mathbf{u}^\mathrm{T} \right) : \left( \nabla_\Gamma^{\left(d_f\right)} \tilde{\mathbf{u}} + \nabla_\Gamma^{\left(d_f\right)} \tilde{\mathbf{u}}^\mathrm{T} \right) \\
  & - p \, \mathrm{div}_\Gamma^{\left( d_f \right)} \tilde{\mathbf{u}} - \tilde{p} \, \mathrm{div}_\Gamma^{\left( d_f \right)} \mathbf{u} + \alpha \mathbf{u} \cdot \tilde{\mathbf{u}} + \lambda \tilde{\mathbf{u}} \cdot \mathbf{n}_\Gamma^{\left( d_f \right)} + \tilde{\lambda} \mathbf{u} \cdot \mathbf{n}_\Gamma^{\left( d_f \right)} \bigg] M^{\left( d_f \right)} \,\mathrm{d}\Sigma \\
  & - \sum_{E_\Sigma\in\mathcal{E}_\Sigma} \int_{E_\Sigma} \tau_{BP,\Gamma}^{\left( d_f \right)} \nabla_\Gamma^{\left( d_f \right)} p \cdot \nabla_\Gamma^{\left( d_f \right)} \tilde{p} M^{\left( d_f \right)} \,\mathrm{d}\Sigma = 0.
\end{split}\right.
\end{equation}

\subsubsection{Surface convection-diffusion equation} \label{sec:SurfaceCDEqu}

The mass transfer process in the surface flow can be described by the surface convection-diffusion equation defined on the same implicit 2-manifold as the surface Navier-Stokes equations in Section \ref{sec:SurfaceNSEqus}. 
Based on the conservation law of mass transfer, the surface convection-diffusion equation can be derived to describe the mass transfer process in the surface flow \cite{DziukActaNumerica2013}:
\begin{equation}\label{equ:CDequOnManifoldsMHM}
\begin{split}
\mathbf{u} \cdot \nabla_\Gamma c - \mathrm{div}_\Gamma \left( D \nabla_\Gamma c \right) & = 0, ~\forall \mathbf{x}_\Gamma \in \Gamma
\end{split}
\end{equation}
where $c$ is the mass concentration in the surface flow; and $D$ is the diffusion coefficient. For the surface convection-diffusion equation, the distribution of the concentration is known at the inlet boundary curve $l_{v,\Gamma}$; and the remained part of the boundary curve is insulative:
\begin{equation}\label{equ:ConcentrationBoundaryCondition}
\left\{
\begin{split}
  & c = c_0, ~ \forall \mathbf{x}_\Gamma \in l_{v,\Gamma} \\
  & \nabla_\Gamma c \cdot \mathbf{n}_{\boldsymbol\tau_\Gamma} = 0, ~ \forall \mathbf{x}_\Gamma \in l_{v_0,\Gamma} \cup l_{s,\Gamma}
\end{split}\right.
\end{equation}
where $c_0$ is the known distribution of the concentration.

Based on the Galerkin method, the variational formulation of the surface convection-diffusion equation is considered in the first order Sobolev space defined on the implicit 2-manifold $\Gamma$:
\begin{equation}\label{equ:VariationalFormulationSurfaceConvecDiffusEqu}
\left\{\begin{split}
  & \mathrm{Find}~ c\in\mathcal{H}\left(\Gamma\right)~\mathrm{with} ~ c = c_0~ \mathrm{at} ~ \forall \mathbf{x}_\Gamma \in l_{v,\Gamma}, ~ \mathrm{for} ~ \forall \tilde{c} \in \mathcal{H}\left(\Gamma\right), ~ \mathrm{such~that} \\
  & \int_\Gamma \left( \mathbf{u} \cdot \nabla_\Gamma c \right) \tilde{c} + D \nabla_\Gamma c \cdot \nabla_\Gamma \tilde{c} \,\mathrm{d}\Gamma + \sum_{E_\Gamma\in\mathcal{E}_\Gamma} \int_{E_\Gamma} \tau_{PG,\Gamma} \left( \mathbf{u} \cdot \nabla_\Gamma c \right) \left( \mathbf{u} \cdot \nabla_\Gamma \tilde{c} \right) \,\mathrm{d}\Gamma = 0 \\
\end{split}\right.
\end{equation}
where $\tilde{c}$ is the test function of $c$; the Petrov-Galerkin stabilization term
\begin{equation}
  \sum_{E_\Gamma\in\mathcal{E}_\Gamma} \int_{E_\Gamma} \tau_{PG,\Gamma} \left( \mathbf{u} \cdot \nabla_\Gamma c \right) \left( \mathbf{u} \cdot \nabla_\Gamma \tilde{c} \right) \,\mathrm{d}\Gamma
\end{equation}
with $\tau_{PG,\Gamma}$ representing the stabilization parameter is imposed on the variational formulation, in order to use the linear finite elements to solve the distribution of the concentration \cite{DoneaWiley2003}. The stabilization parameter is chosen as \cite{DoneaWiley2003}
\begin{equation}\label{equ:CDSurfaceStabilizationTermCD}
 \tau_{PG,\Gamma} = \left( {4 \over h_{E_\Gamma}^2 D} + {2 \left\| \mathbf{u} \right\|_2 \over h_{E_\Gamma} } \right)^{-1}.
\end{equation}
Based on Eqs. \ref{equ:DiffRiemannianMHM}, \ref{equ:ElementAreaTransformationMHM} and \ref{equ:ApproximationMetricAverageMHM}, $\tau_{PG,\Gamma}$ can be transformed into
\begin{equation}\label{equ:TransformedCDSurfaceStabilizationTermCD}
  \tau_{PG,\Gamma}^{\left( d_f \right)} =\left( {4 \over h_{E_\Sigma}^2 M^{\left( d_f \right)} D} + {2 \left\| \mathbf{u} \right\|_2 \over h_{E_\Sigma} \left(M^{\left( d_f \right)}\right)^{1\over2} } \right)^{-1}. 
\end{equation}

Based on the homeomorphisms and the coupling relations in Section \ref{subsec:CouplingDesignVariables}, the variational formulation in Eq. \ref{equ:VariationalFormulationSurfaceConvecDiffusEqu} can be transformed into the form defined on $\Sigma$:
\begin{equation}\label{equ:TransformedVariationalFormulationSurfaceCDEqu}
\left\{\begin{split}
& \mathrm{Find}~ c\in\mathcal{H}\left(\Sigma\right)~\mathrm{with} ~ c = c_0~ \mathrm{at} ~ \forall \mathbf{x}_\Sigma \in l_{v,\Sigma}, ~ \mathrm{for} ~ \forall \tilde{c} \in \mathcal{H}\left(\Sigma\right), \\
& \mathrm{such~that} ~ \int_\Sigma \left[ \left( \mathbf{u} \cdot \nabla_\Gamma^{\left(d_f\right)} c \right) \tilde{c} + D \nabla_\Gamma^{\left(d_f\right)} c \cdot \nabla_\Gamma^{\left(d_f\right)} \tilde{c} \right] M^{\left( d_f \right)} \,\mathrm{d}\Sigma \\
& + \sum_{E_\Sigma\in\mathcal{E}_\Sigma} \int_{E_\Sigma} \tau_{PG,\Gamma}^{\left( d_f \right)} \left( \mathbf{u} \cdot \nabla_\Gamma^{\left( d_f \right)} c \right) \left( \mathbf{u} \cdot \nabla_\Gamma^{\left( d_f \right)} \tilde{c} \right) M^{\left( d_f \right)} \,\mathrm{d}\Sigma = 0. \\
\end{split}\right.
\end{equation}

\subsubsection{Design objective for mass transfer problem} \label{sec:DesignObjectiveSurfaceNSCD}

For mass transfer in the surface flow, the desired performance of the surface structure can be set to achieve the anticipated distribution of the concentration at the outlet. The mass transfer performance can be measured by the deviation between the obtained and anticipated distribution of the concentration. Therefore, the design objective of fiber bundle topology optimization for mass transfer in the surface flow is considered as
\begin{equation}\label{equ:DesignObjectiveSurfaceCD}
  J_c = \int_{l_{s,\Gamma}} \left( c - \bar{c} \right)^2 \,\mathrm{d}l_{\partial\Gamma} \bigg/ \int_{l_{v,\Gamma}} \left( c_0 - \bar{c} \right)^2 \,\mathrm{d}l_{\partial\Gamma},
\end{equation}
where $\bar{c}$ is the anticipated distribution of the concentration at the outlet and it is linearly mapped onto the inlet for the reference value of the concentration deviation. Based on the coupling relations in Section \ref{subsec:CouplingDesignVariables}, the design objective in Eq. \ref{equ:DesignObjectiveSurfaceCD} can be transformed into
\begin{equation}\label{equ:FurtherTransformedDesignObjectiveCD}
\begin{split}
  J_c^{\left( d_f \right)} = & \int_{l_{s,\Sigma}} \left( c - \bar{c} \right)^2 L^{\left( d_f \right)} \,\mathrm{d}l_{\partial\Sigma} \bigg/ \int_{l_{v,\Sigma}} \left( c_0 - \bar{c} \right)^2 L^{\left( d_f \right)} \,\mathrm{d}l_{\partial\Sigma}. \\
\end{split}
\end{equation}

\subsubsection{Constraint of pressure drop} \label{sec:PressureConstraintSurfaceNSCD}

To ensure the patency of the surface structure for mass transfer in the surface flow, the constraint of the pressure drop between the inlet and outlet can be imposed as
\begin{equation}\label{equ:PressureConstraintSurfaceNSCD}
  \left| \Delta P \left/ \: \Delta P_0 - 1 \right. \right| \leq 1\times10^{-3},
\end{equation}
where $\Delta P_0$ is the specified reference value of the pressure drop and $\Delta P$ is the pressure drop between the inlet and outlet:
\begin{equation}\label{equ:PressureDropSurfaceNSCD}
  \Delta P = \int_{l_{v,\Gamma}} p \,\mathrm{d}l_{\partial\Gamma} - \int_{l_{s,\Gamma}} p \,\mathrm{d}l_{\partial\Gamma}.
\end{equation}
Based on the relation in Eq. \ref{equ:TauGammaMHM} sketched in Fig. \ref{fig:DemonForTangentialVectorGammaMHM}, Eq. \ref{equ:PressureConstraintSurfaceNSCD} can be transformed into
\begin{equation}\label{equ:TransformedPressureConstraintSurfaceNSCD}
\begin{split}
  \Delta P^{\left( d_f \right)} = & \int_{l_{v,\Sigma}} p L^{\left( d_f \right)} \,\mathrm{d}l_{\partial\Sigma} - \int_{l_{s,\Sigma}} p L^{\left( d_f \right)} \,\mathrm{d}l_{\partial\Sigma}. \\
\end{split}
\end{equation}

\subsubsection{Fiber bundle topology optimization problem} \label{sec:MatchingOptimizationSurfaceNSEqus}

Based on the above introduction, the fiber bundle topology optimization problem for mass transfer in the surface flow can be constructed to find the optimized matching between the pattern of the surface flow and the implicit 2-manifold, i.e. to optimize the fiber bundle defined in Eq. \ref{equ:FiberBundleMHM}:
\begin{equation}\label{equ:VarProToopSurfaceNSCD}
\left\{\begin{split}
  & \mathrm{Find} \left\{\begin{split}
  & \gamma: \Gamma \mapsto \left[0,1\right] \\
  & d_m: \Sigma \mapsto \left[0,1\right]\end{split}\right.~ \mathrm{for} ~
  \left(\Sigma \times \left(\Gamma \times \left[0,1\right]\right), \Sigma, proj_1, \Gamma \times \left[0,1\right] \right), \\
  & \mathrm{to} ~ \mathrm{minimize}~{J_c \over J_{c,0}}~ \mathrm{with} ~ J_c = \int_{l_{s,\Gamma}} \left( c - \bar{c} \right)^2 \,\mathrm{d}l_{\partial\Gamma} \bigg/ \int_{l_{v,\Gamma}} \left( c_0 - \bar{c} \right)^2 \,\mathrm{d}l_{\partial\Gamma}, \\
  & \mathrm{constrained} ~ \mathrm{by} \\
  & \left\{\begin{split}
  & \begin{split}
  & \left\{\begin{split}
    & \left\{\begin{split}
       & \rho \left( \mathbf{u} \cdot \nabla_\Gamma \right) \mathbf{u} - \mathbf{P} \mathrm{div}_\Gamma \left[ \eta \left( \nabla_\Gamma \mathbf{u} + \nabla_\Gamma \mathbf{u}^\mathrm{T} \right) \right] + \nabla_\Gamma p = - \alpha \mathbf{u}, ~ \forall \mathbf{x}_\Gamma \in \Gamma \\
       & - \mathrm{div}_\Gamma \mathbf{u} = 0, ~ \forall \mathbf{x}_\Gamma \in \Gamma \\
       & \mathbf{u} \cdot \mathbf{n}_\Gamma = 0, ~ \forall \mathbf{x}_\Gamma \in \Gamma \\
    \end{split}\right. \\
    & \alpha \left( \gamma_p \right) = \alpha_f + \left( \alpha_s - \alpha_f \right) q { 1 - \gamma_p \over q + \gamma_p } \\
    \end{split}\right.
    \end{split} \\
  & \mathbf{u} \cdot \nabla_\Gamma c - \mathrm{div}_\Gamma \left( D \nabla_\Gamma c \right) = 0, ~\forall \mathbf{x}_\Gamma \in \Gamma \\
  & \left\{\begin{split}
    & \left\{\begin{split}
        & - \mathrm{div}_\Gamma \left( r_f^2 \nabla_\Gamma \gamma_f \right) + \gamma_f = \gamma,~\forall \mathbf{x}_\Gamma \in \Gamma_D \\
        & \mathbf{n}_{\boldsymbol\tau_\Gamma} \cdot \nabla_\Gamma \gamma_f = 0,~\forall \mathbf{x}_\Gamma \in \partial\Gamma_D \\
    \end{split}\right. \\
    & \gamma_p = { \tanh\left(\beta \xi\right) + \tanh\left(\beta \left(\gamma_f-\xi\right)\right) \over \tanh\left(\beta \xi\right) + \tanh\left(\beta \left(1-\xi\right)\right)} \\
    \end{split}\right. \\
  & \left\{
        \begin{split}
          & - \mathrm{div}_\Sigma \left( r_m^2 \nabla_\Sigma d_f \right) + d_f = A_d \left( d_m - {1\over2} \right), ~ \forall \mathbf{x}_\Sigma \in \Sigma \\
          & \mathbf{n}_{\boldsymbol\tau_\Sigma} \cdot \nabla_\Sigma d_f = 0, ~ \forall \mathbf{x}_\Sigma \in \partial \Sigma \\
        \end{split}\right. \\
  & \Gamma = \left\{ \mathbf{x}_\Gamma : \mathbf{x}_\Gamma = d_f \mathbf{n}_\Sigma + \mathbf{x}_\Sigma,~\forall \mathbf{x}_\Sigma \in \Sigma \right\} \\
  & \left| \Delta P \left/ \: \Delta P_0 - 1 \right. \right| \leq 1\times10^{-3}, ~\mathrm{with} ~ \Delta P = \int_{l_{v,\Gamma}} p \,\mathrm{d}l_{\partial\Gamma} - \int_{l_{s,\Gamma}} p \,\mathrm{d}l_{\partial\Gamma} \\
\end{split}\right., \\
\end{split}\right.
\end{equation}
where $J_{c,0}$ is the reference value of the design objective corresponding to the initial distribution of the design variables.

The coupling relations among the variables, functions, tangential divergence operator and tangential gradient operator in Eq. \ref{equ:VarProToopSurfaceNSCD} are illustrated by the arrow chart described as
\[\begin{array}{cccccccc}
 \textcolor{blue}{d_m} & \xrightarrow{\mathrm{Eq.~}\ref{equ:PDEFilterzmBaseStructureMHM}} & d_f & & & \\
 & & \bigg\downarrow\vcenter{\rlap{\scriptsize{Eq.~\ref{equ:TransformedTangentialOperatorMHM}}}} & & \\
 & & \left\{ \mathrm{div}_\Gamma, \nabla_\Gamma, \mathbf{n}_\Gamma \right\} & \xrightarrow{\mathrm{Eqs.~}\ref{equ:NSequOnManifoldsMHM}~\&~\ref{equ:CDequOnManifoldsMHM}} & \left\{ \mathbf{u},~p,~\lambda,~c \right\} & \xrightarrow{\mathrm{Eqs.~}\ref{equ:DesignObjectiveSurfaceCD} ~\&~ \ref{equ:PressureDropSurfaceNSCD}} & \left\{ \textcolor[rgb]{0.50,0.00,0.00}{J_c}, ~ \textcolor[rgb]{0.50,0.00,0.00}{\Delta P} \right\} \\
 & & \bigg\downarrow\vcenter{\rlap{\scriptsize{Eq.~\ref{equ:PDEFilterGammaFilberMHM}}}} & &  \bigg\uparrow\vcenter{\rlap{\scriptsize{Eq.~\ref{equ:NSequOnManifoldsMHM}}}}   \\
 \textcolor{blue}{\gamma} & \xrightarrow{\mathrm{Eq.~}\ref{equ:PDEFilterGammaFilberMHM}} & \gamma_f & \xrightarrow{\mathrm{Eq.~}\ref{equ:ProjectionGammaFilberMHM}} & \textcolor[rgb]{0.50,0.00,0.00}{\gamma_p} & & \\
\end{array}\]
where the design variables $d_m$ and $\gamma$, marked in blue, are the inputs; the design objective $J_c$, the pressure drop $\Delta P$ and the material density $\gamma_p$, marked in red, are the outputs.

\subsubsection{Adjoint analysis} \label{sec:AdjointAnalysisSurfaceNSEqusCD}

The fiber bundle topology optimization problem in Eq. \ref{equ:VarProToopSurfaceNSCD} can be solved by using a gradient-based iterative procedure, where the adjoint sensitivities are used to determine the relevant gradient information. The adjoint analysis is implemented for the design objective and pressure drop to derive the adjoint sensitivities. The details for the adjoint analysis have been provided in the appendix in Sections \ref{sec:AdjointAnalysisDesignObjectiveSurfaceCDMHM} and \ref{sec:AdjointAnalysisPressureConstraintMHM}.

Based on the transformed design objective in Eq. \ref{equ:FurtherTransformedDesignObjectiveCD} and transformed pressure drop in Eq. \ref{equ:TransformedPressureConstraintSurfaceNSCD}, the adjoint analysis of the fiber bundle topology optimization problem can be implemented on the functional spaces defined on the base manifold. Based on the continuous adjoint method \cite{HinzeSpringer2009}, the adjoint sensitivity of the design objective $J_c$ is derived as
\begin{equation}\label{equ:AdjSensitivityCDGaDmMHM}
\begin{split}
\delta J_c = - \int_{\Sigma_D} \gamma_{fa} \tilde{\gamma} M^{\left( d_f \right)} \,\mathrm{d}\Sigma - \int_\Sigma A_d d_{fa} \tilde{d}_m \,\mathrm{d}\Sigma, ~ \forall \tilde{\gamma} \in \mathcal{L}^2\left(\Sigma_D\right), ~ \forall \tilde{d}_m \in \mathcal{L}^2\left(\Sigma\right)
\end{split}
\end{equation}
where $\gamma_{fa}$ and $d_{fa}$ are the adjoint variables of the filtered design variables $\gamma_f$ and $d_f$, respectively. The adjoint variables can be derived from the adjoint equations in the variational formulations. The variational formulation for the adjoint equation of the surface convection-diffusion equation is derived as
\begin{equation}\label{equ:WeakAdjEquSCDEquMHM}
\left\{\begin{split}
  & \mathrm{Find} ~ c_a \in \mathcal{H}\left(\Sigma\right)~\mathrm{with}~ c_a=0 ~ \mathrm{at} ~ \forall \mathbf{x}_\Sigma \in l_{v,\Sigma}, ~ \mathrm{for} ~\forall \tilde{c}_a \in \mathcal{H} \left(\Sigma\right) \\
  & \mathrm{such~that} ~ \int_{l_{s,\Sigma}} 2 \left( c - \bar{c} \right) L^{\left( d_f \right)} \tilde{c}_a \,\mathrm{d}l_{\partial\Sigma} \bigg/ \int_{l_{v,\Sigma}} \left( c_0 - \bar{c} \right)^2 L^{\left( d_f \right)} \,\mathrm{d}l_{\partial\Sigma} \\
  & + \int_\Sigma \left[ \left( \mathbf{u} \cdot \nabla_\Gamma^{\left(d_f\right)} \tilde{c}_a \right) c_a + D \nabla_\Gamma^{\left(d_f\right)} \tilde{c}_a \cdot \nabla_\Gamma^{\left(d_f\right)} c_a \right] M^{\left( d_f \right)} \,\mathrm{d}\Sigma \\
  & + \sum_{E_\Sigma\in\mathcal{E}_\Sigma} \int_{E_\Sigma} \tau_{PG,\Gamma}^{\left( d_f \right)} \left( \mathbf{u} \cdot \nabla_\Gamma^{\left( d_f \right)} \tilde{c}_a \right) \left( \mathbf{u} \cdot \nabla_\Gamma^{\left( d_f \right)} c_a \right) M^{\left( d_f \right)} \,\mathrm{d}\Sigma = 0 \\
\end{split}\right.
\end{equation}
where $c_a$ is the adjoint variable of $c$ and $\tilde{c}_a$ is the test function of $c_a$. The variational formulation for the adjoint equations of the surface Naiver-Stokes equations is derived as
\begin{equation}\label{equ:AdjSurfaceNavierStokesEqusJObjectiveMHM} 
\left\{\begin{split}
  & \mathrm{Find} \left\{\begin{split}
  & \mathbf{u}_a \in\left(\mathcal{H}\left(\Sigma\right)\right)^3~\mathrm{with}~ \mathbf{u}_a = \mathbf{0}~ \mathrm{at} ~ {\forall \mathbf{x} \in l_{v,\Sigma} \cup l_{v_0,\Sigma} } \\
  & p_a \in \mathcal{H}\left(\Sigma\right) \\
  & \lambda_a \in \mathcal{L}^2\left(\Sigma\right)~\mathrm{with}~ \lambda_a = 0~ \mathrm{at} ~ \forall \mathbf{x} \in l_{v,\Sigma} \cup l_{v_0,\Sigma} \\
  \end{split}\right.\\
  & \mathrm{for} \left\{\begin{split} 
  & \forall \tilde{\mathbf{u}}_a \in\left(\mathcal{H}\left(\Sigma\right)\right)^3 \\
  & \forall \tilde{p}_a \in \mathcal{H}\left(\Sigma\right) \\
  & \forall \tilde{\lambda}_a \in \mathcal{L}^2\left(\Sigma\right)
  \end{split}\right.,~\mathrm{such~that} \\
  & \int_\Sigma \bigg[ \rho \left( \tilde{\mathbf{u}}_a \cdot \nabla_\Gamma^{\left(d_f\right)} \right) \mathbf{u} \cdot \mathbf{u}_a + \rho \left( \mathbf{u} \cdot \nabla_\Gamma^{\left(d_f\right)} \right) \tilde{\mathbf{u}}_a \cdot \mathbf{u}_a + {\eta\over2} \left( \nabla_\Gamma^{\left(d_f\right)} \tilde{\mathbf{u}}_a + \nabla_\Gamma^{\left(d_f\right)} \tilde{\mathbf{u}}_a^\mathrm{T} \right) \\
  & : \left( \nabla_\Gamma^{\left(d_f\right)} \mathbf{u}_a + \nabla_\Gamma^{\left(d_f\right)} \mathbf{u}_a^\mathrm{T} \right) - \tilde{p}_a \mathrm{div}_\Gamma^{\left( d_f \right)} \mathbf{u}_a - p_a \mathrm{div}_\Gamma^{\left( d_f \right)} \tilde{\mathbf{u}}_a + \alpha \tilde{\mathbf{u}}_a \cdot \mathbf{u}_a + \left( \tilde{\lambda}_a \mathbf{u}_a + \lambda_a \tilde{\mathbf{u}}_a \right) \\
  & \cdot \mathbf{n}_\Gamma^{\left( d_f \right)} + \left( \tilde{\mathbf{u}}_a \cdot \nabla_\Gamma^{\left(d_f\right)} c \right) c_a \bigg] M^{\left( d_f \right)} \,\mathrm{d}\Sigma + \sum_{E_\Sigma\in\mathcal{E}_\Sigma} \int_{E_\Sigma} \bigg[ - \tau_{BP,\Gamma}^{\left( d_f \right)} \nabla_\Gamma^{\left( d_f \right)} \tilde{p}_a \cdot \nabla_\Gamma^{\left( d_f \right)} p_a \\
  & + \tau_{PG,\Gamma}^{\left( d_f, \tilde{\mathbf{u}}_a \right)}\left( \mathbf{u} \cdot \nabla_\Gamma^{\left( d_f \right)} c \right) \left( \mathbf{u} \cdot \nabla_\Gamma^{\left( d_f \right)} c_a \right) + \tau_{PG,\Gamma}^{\left( d_f \right)} \left( \tilde{\mathbf{u}}_a \cdot \nabla_\Gamma^{\left( d_f \right)} c \right) \left( \mathbf{u} \cdot \nabla_\Gamma^{\left( d_f \right)} c_a \right) \\
  & + \tau_{PG,\Gamma}^{\left( d_f \right)} \left( \mathbf{u} \cdot \nabla_\Gamma^{\left( d_f \right)} c \right) \left( \tilde{\mathbf{u}}_a \cdot \nabla_\Gamma^{\left( d_f \right)} c_a \right) \bigg] M^{\left( d_f \right)} \,\mathrm{d}\Sigma = 0 \\
\end{split}\right.
\end{equation}
where $\mathbf{u}_a$, $p_a$ and $\lambda_a$ are the adjoint variables of $\mathbf{u}$, $p$ and $\lambda$, respectively; $\tilde{\mathbf{u}}_a$, $\tilde{p}_a$ and $\tilde{\lambda}_a$ are the test functions of $\mathbf{u}_a$, $p_a$ and $\lambda_a$, respectively; and $\tau_{PG,\Gamma}^{\left( d_f, \tilde{\mathbf{u}} \right)}$ is the first-order variational of $\tau_{PG,\Gamma}^{\left( d_f \right)}$ to $\mathbf{u}$, and it is expressed as
\begin{equation}\label{equ:1stVariTauPGtoU}
  \tau_{PG,\Gamma}^{\left( d_f, \tilde{\mathbf{u}} \right)} = - \left( {4 \over h_{E_\Sigma}^2 M^{\left( d_f \right)} D} + {2 \left\| \mathbf{u} \right\|_2 \over h_{E_\Sigma} \left(M^{\left( d_f \right)}\right)^{1\over2} } \right)^{-2} {2 \mathbf{u} \cdot \tilde{\mathbf{u}} \over h_{E_\Sigma} \left(M^{\left( d_f \right)}\right)^{1\over2} \left\| \mathbf{u} \right\|_2 }, ~ \forall \tilde{\mathbf{u}} \in \left( \mathcal{H}\left( \Sigma \right) \right)^3.
\end{equation}
The variational formulations for the adjoint equations of the surface-PDE filters for $\gamma$ and $d_m$ are derived as
\begin{equation}\label{equ:AdjPDEFilterJObjectiveGaMHM}  
\left\{\begin{split}
  & \mathrm{Find}~\gamma_{fa}\in\mathcal{H}\left(\Sigma_D\right) ~\mathrm{for}~ \forall \tilde{\gamma}_{fa} \in \mathcal{H}\left(\Sigma_D\right),~\mathrm{such~that} \\
  & \int_{\Sigma_D} \left( {\partial\alpha \over \partial \gamma_p} {\partial \gamma_p \over \partial \gamma_f} \mathbf{u} \cdot \mathbf{u}_a \tilde{\gamma}_{fa} + r_f^2 \nabla_\Gamma^{\left( d_f \right)} \tilde{\gamma}_{fa} \cdot \nabla_\Gamma^{\left( d_f \right)} \gamma_{fa} + \tilde{\gamma}_{fa} \gamma_{fa} \right) M^{\left( d_f \right)} \,\mathrm{d}\Sigma = 0
\end{split}\right.
\end{equation}
and
\begin{equation}\label{equ:AdjPDEFilterJObjectiveDmMHM} 
\left\{\begin{split}
  & \mathrm{Find}~d_{fa}\in\mathcal{H}\left(\Sigma\right)~\mathrm{for}~\forall \tilde{d}_{fa} \in \mathcal{H}\left(\Sigma\right),~\mathrm{such~that} \\
  & \int_{l_{s,\Sigma}} \left( c - \bar{c} \right)^2 L^{\left( d_f, \tilde{d}_{fa} \right)} \,\mathrm{d}l_{\partial\Sigma} \bigg/ \int_{l_{v,\Sigma}} \left( c_0 - \bar{c} \right)^2 L^{\left( d_f \right)} \,\mathrm{d}l_{\partial\Sigma} - \int_{l_{s,\Sigma}} \left( c - \bar{c} \right)^2 L^{\left( d_f \right)} \, \mathrm{d}l_{\partial\Sigma} \\
  & \int_{l_{v,\Sigma}} \left( c_0 - \bar{c} \right)^2 L^{\left( d_f, \tilde{d}_{fa} \right)} \,\mathrm{d}l_{\partial\Sigma} \bigg/ \left( \int_{l_{v,\Sigma}} \left( c_0 - \bar{c} \right)^2 L^{\left( d_f \right)} \,\mathrm{d}l_{\partial\Sigma} \right)^2 + \int_\Sigma \bigg[ \rho \left( \mathbf{u} \cdot \nabla_\Gamma^{\left(d_f, \tilde{d}_{fa} \right)} \right) \mathbf{u} \\
  & \cdot \mathbf{u}_a + {\eta\over2} \left( \nabla_\Gamma^{\left(d_f, \tilde{d}_{fa}\right)} \mathbf{u} + \nabla_\Gamma^{\left(d_f, \tilde{d}_{fa}\right)} \mathbf{u}^\mathrm{T} \right) : \left( \nabla_\Gamma^{\left(d_f\right)} \mathbf{u}_a + \nabla_\Gamma^{\left(d_f\right)} \mathbf{u}_a^\mathrm{T} \right) + {\eta\over2} \left( \nabla_\Gamma^{\left(d_f\right)} \mathbf{u} + \nabla_\Gamma^{\left(d_f\right)} \mathbf{u}^\mathrm{T} \right) \\
  & : \left( \nabla_\Gamma^{\left(d_f, \tilde{d}_{fa}\right)} \mathbf{u}_a + \nabla_\Gamma^{\left(d_f, \tilde{d}_{fa}\right)} \mathbf{u}_a^\mathrm{T} \right) - p \, \mathrm{div}_\Gamma^{\left( d_f, \tilde{d}_{fa} \right)} \mathbf{u}_a - p_a \mathrm{div}_\Gamma^{\left( d_f, \tilde{d}_{fa} \right)} \mathbf{u} + \left( \lambda \mathbf{u}_a + \lambda_a \mathbf{u} \right) \\
  & \cdot \mathbf{n}_\Gamma^{\left( d_f, \tilde{d}_{fa} \right)} + \left( \mathbf{u} \cdot \nabla_\Gamma^{\left(d_f, \tilde{d}_{fa}\right)} c \right) c_a + D \nabla_\Gamma^{\left(d_f, \tilde{d}_{fa} \right)} c \cdot \nabla_\Gamma^{\left(d_f\right)} c_a + D \nabla_\Gamma^{\left(d_f\right)} c \cdot \nabla_\Gamma^{\left(d_f, \tilde{d}_{fa}\right)} c_a \\
  & + f_{id,\Gamma} r_f^2 \left( \nabla_\Gamma^{\left( d_f, \tilde{d}_{fa} \right)} \gamma_f \cdot \nabla_\Gamma^{\left( d_f \right)} \gamma_{fa} + \nabla_\Gamma^{\left( d_f \right)} \gamma_f \cdot \nabla_\Gamma^{\left( d_f, \tilde{d}_{fa} \right)} \gamma_{fa} \right) \bigg] M^{\left( d_f \right)} + \bigg[ \rho \left( \mathbf{u} \cdot \nabla_\Gamma^{\left(d_f\right)} \right) \mathbf{u} \\
  & \cdot \mathbf{u}_a + {\eta\over2} \left( \nabla_\Gamma^{\left(d_f\right)} \mathbf{u} + \nabla_\Gamma^{\left(d_f\right)} \mathbf{u}^\mathrm{T} \right) : \bigg( \nabla_\Gamma^{\left(d_f\right)} \mathbf{u}_a 
  + \nabla_\Gamma^{\left(d_f\right)} \mathbf{u}_a^\mathrm{T} \bigg) - p \, \mathrm{div}_\Gamma^{\left( d_f \right)} \mathbf{u}_a - p_a \mathrm{div}_\Gamma^{\left( d_f \right)} \mathbf{u} \\
  & + \alpha \mathbf{u} \cdot \mathbf{u}_a + \lambda \mathbf{u}_a \cdot \mathbf{n}_\Gamma^{\left( d_f \right)} + \lambda_a \mathbf{u} \cdot \mathbf{n}_\Gamma^{\left( d_f \right)} + \left( \mathbf{u} \cdot \nabla_\Gamma^{\left(d_f\right)} c \right) c_a + D \nabla_\Gamma^{\left(d_f\right)} c \cdot \nabla_\Gamma^{\left(d_f\right)} c_a \\
  & + f_{id,\Gamma} \left( r_f^2 \nabla_\Gamma^{\left( d_f \right)} \gamma_f \cdot \nabla_\Gamma^{\left( d_f \right)} \gamma_{fa} + \gamma_f \gamma_{fa} - \gamma \gamma_{fa} \right) \bigg] M^{\left( d_f, \tilde{d}_{fa} \right)} \\
  & + r_m^2 \nabla_\Sigma \tilde{d}_{fa} \cdot \nabla_\Sigma d_{fa} + \tilde{d}_{fa} d_{fa} \,\mathrm{d}\Sigma \\
  & + \sum_{E_\Sigma\in\mathcal{E}_\Sigma} \int_{E_\Sigma} \bigg[ \bigg( - \tau_{BP,\Gamma}^{\left( d_f, \tilde{d}_f \right)} \nabla_\Gamma^{\left( d_f \right)} p \cdot \nabla_\Gamma^{\left( d_f \right)} p_a - \tau_{BP,\Gamma}^{\left( d_f \right)} \nabla_\Gamma^{\left( d_f, \tilde{d}_{fa} \right)} p \cdot \nabla_\Gamma^{\left( d_f \right)} p_a - \tau_{BP,\Gamma}^{\left( d_f \right)} \nabla_\Gamma^{\left( d_f \right)} p \\
  & \cdot \nabla_\Gamma^{\left( d_f, \tilde{d}_{fa} \right)} p_a \bigg) + \tau_{PG,\Gamma}^{\left( d_f, \tilde{d}_{fa} \right)} \left( \mathbf{u} \cdot \nabla_\Gamma^{\left( d_f \right)} c \right) \left( \mathbf{u} \cdot \nabla_\Gamma^{\left( d_f \right)} c_a \right) + \tau_{PG,\Gamma}^{\left( d_f \right)} \left( \mathbf{u} \cdot \nabla_\Gamma^{\left( d_f, \tilde{d}_{fa} \right)} c \right) \\
  & \left( \mathbf{u} \cdot \nabla_\Gamma^{\left( d_f \right)} c_a \right) + \tau_{PG,\Gamma}^{\left( d_f \right)} \left( \mathbf{u} \cdot \nabla_\Gamma^{\left( d_f \right)} c \right) \left( \mathbf{u} \cdot \nabla_\Gamma^{\left( d_f, \tilde{d}_{fa} \right)} c_a \right) \bigg] M^{\left( d_f \right)} + \bigg[ - \tau_{BP,\Gamma}^{\left( d_f \right)} \nabla_\Gamma^{\left( d_f \right)} p \\
  & \cdot \nabla_\Gamma^{\left( d_f \right)} p_a + \tau_{PG,\Gamma}^{\left( d_f \right)} \left( \mathbf{u} \cdot \nabla_\Gamma^{\left( d_f \right)} c \right) \left( \mathbf{u} \cdot \nabla_\Gamma^{\left( d_f \right)} c_a \right) \bigg] M^{\left( d_f, \tilde{d}_{fa} \right)} \,\mathrm{d}\Sigma = 0 \\
\end{split}\right.
\end{equation}
where $\tilde{\gamma}_{fa}$ and $\tilde{d}_{fa}$ are the test functions of $\gamma_{fa}$ and $d_{fa}$, respectively; $\tau_{BP,\Gamma}^{\left( d_f, \tilde{d}_f \right)}$ and $\tau_{PG,\Gamma}^{\left( d_f, \tilde{d}_f \right)}$ are the first-order variationals of $\tau_{BP,\Gamma}^{\left( d_f \right)}$ and $\tau_{PG,\Gamma}^{\left( d_f \right)}$ to $d_f$, respectively, and they are expressed as
\begin{equation}\label{equ:1stVariTauBPtodf}
  \tau_{BP,\Gamma}^{\left( d_f, \tilde{d}_f \right)} = {h_{E_\Sigma}^2 \over 12\eta} M^{\left( d_f, \tilde{d}_f \right)}, ~ \forall \tilde{d}_f \in \mathcal{H}\left( \Sigma \right)
\end{equation}
and
\begin{equation}\label{equ:1stVariTauPGtodf}
\begin{split}
  \tau_{PG,\Gamma}^{\left( d_f, \tilde{d}_f \right)} = \, & \left( {4 \over h_{E_\Sigma}^2 M^{\left( d_f \right)} D} + {2 \left\| \mathbf{u} \right\|_2 \over h_{E_\Sigma} \left(M^{\left( d_f \right)}\right)^{1\over2} } \right)^{-2} \\
  & \left( {4 \over h_{E_\Sigma}^2 \left(M^{\left( d_f \right)}\right)^2 D} + { \left\| \mathbf{u} \right\|_2 \over h_{E_\Sigma} \left( M^{\left( d_f \right)} \right)^{3\over2} } \right) M^{\left( d_f, \tilde{d}_f \right)}, ~ \forall \tilde{d}_f \in \mathcal{H} \left(\Sigma\right);
\end{split}
\end{equation}
$M^{\left( d_f, \tilde{d}_f \right)}$ and $L^{\left( d_f, \tilde{d}_f \right)}$ are the first-order variationals of $M^{\left( d_f \right)}$ and $L^{\left( d_f \right)}$ to $d_f$ derived based on Eq. \ref{equ:TransformedJacobian} and Eq. \ref{equ:FirstOrderVariOfVectorNormMHM} in the appendix, and they are expressed as
\begin{equation}\label{equ:1stVariofRiemannMetricMTodf}
\begin{split}
  M^{\left( d_f, \tilde{d}_f \right)} = \, & \left( {\partial \left| \mathbf{T}_\Gamma \right| \over \partial d_f} \tilde{d}_f + {\partial \left| \mathbf{T}_\Gamma \right| \over \partial \nabla_\Sigma d_f} \cdot \nabla_\Sigma \tilde{d}_f \right) \left\| \mathbf{T}_\Gamma \mathbf{n}_\Gamma^{\left( d_f \right)} \right\|_2^{-1} - \left| \mathbf{T}_\Gamma \right| \left\| \mathbf{T}_\Gamma \mathbf{n}_\Gamma^{\left( d_f \right)} \right\|_2^{-3} \\
  & \left[ {\partial \mathbf{T}_\Gamma \over \partial d_f } \mathbf{n}_\Gamma^{\left( d_f \right)} \tilde{d}_f + \left( {\partial \mathbf{T}_\Gamma \over \partial \nabla_\Sigma d_f } \cdot \nabla_\Sigma \tilde{d}_f \right) \mathbf{n}_\Gamma^{\left( d_f \right)} + \mathbf{T}_\Gamma \mathbf{n}_\Gamma^{\left( d_f, \tilde{d}_f \right)} \right] \\
  & \cdot \left( \mathbf{T}_\Gamma \mathbf{n}_\Gamma^{\left( d_f \right)} \right), ~ \forall \tilde{d}_f \in \mathcal{H}\left( \Sigma \right)
\end{split}
\end{equation}
and
\begin{equation}\label{equ:1stVariofRiemannMetricLTodf}
\begin{split}
  L^{\left( d_f, \tilde{d}_f \right)} = \, & { \left( \mathbf{n}_\Sigma \times \nabla_\Sigma d_f \right) \times \left( \mathbf{n}_\Sigma - \nabla_\Sigma d_f\right) \over \left\| \left( \mathbf{n}_\Sigma \times \nabla_\Sigma d_f \right) \times \left( \mathbf{n}_\Sigma - \nabla_\Sigma d_f\right) \right\|_2 } \cdot \left[ {\partial \left[ \left( \mathbf{n}_\Sigma \times \nabla_\Sigma d_f \right) \times \left( \mathbf{n}_\Sigma - \nabla_\Sigma d_f\right) \right] \over \nabla_\Sigma d_f } \cdot \nabla_\Sigma \tilde{d}_f \right] \\
  & \left\| \mathbf{T}_\Gamma^{-1} \left[ \left( \mathbf{n}_\Sigma \times \nabla_\Sigma d_f \right) \times \left( \mathbf{n}_\Sigma - \nabla_\Sigma d_f\right)\right] \right\|_2^{-1} + \left\| \left( \mathbf{n}_\Sigma \times \nabla_\Sigma d_f \right) \times \left( \mathbf{n}_\Sigma - \nabla_\Sigma d_f\right) \right\|_2 \\
  & \left\| \mathbf{T}_\Gamma^{-1} \left[ \left( \mathbf{n}_\Sigma \times \nabla_\Sigma d_f \right) \times \left( \mathbf{n}_\Sigma - \nabla_\Sigma d_f\right)\right] \right\|_2^{-3} \Bigg\{ \left( {\partial \mathbf{T}_\Gamma^{-1} \over \partial d_f } \tilde{d}_f + {\partial \mathbf{T}_\Gamma^{-1} \over \partial \nabla_\Sigma d_f } \cdot \nabla_\Sigma \tilde{d}_f \right) \\
  & \left[ \left( \mathbf{n}_\Sigma \times \nabla_\Sigma d_f \right) \times \left( \mathbf{n}_\Sigma - \nabla_\Sigma d_f\right)\right] + \mathbf{T}_\Gamma^{-1} \bigg[ {\partial \left[ \left( \mathbf{n}_\Sigma \times \nabla_\Sigma d_f \right) \times \left( \mathbf{n}_\Sigma - \nabla_\Sigma d_f\right)\right] \over \partial \nabla_\Sigma d_f } \\
  & \cdot \nabla_\Sigma \tilde{d}_f \bigg] \Bigg\} \cdot \left\{ \mathbf{T}_\Gamma^{-1} \left[ \left( \mathbf{n}_\Sigma \times \nabla_\Sigma d_f \right) \times \left( \mathbf{n}_\Sigma - \nabla_\Sigma d_f\right)\right] \right\}, ~ \forall \tilde{d}_f \in \mathcal{H} \left( \Sigma \right). \\
\end{split}
\end{equation}

For the constraint of the pressure drop, the adjoint sensitivity of the pressure drop $\Delta P$ is derived as
\begin{equation}\label{equ:AdjSensitivityGaDmPressureConstrMHM} 
\begin{split}
\delta \Delta P = - \int_{\Sigma_D} \gamma_{fa} \tilde{\gamma} M^{\left( d_f \right)} \,\mathrm{d}\Sigma - \int_\Sigma A_d d_{fa} \tilde{d}_m \,\mathrm{d}\Sigma,~ \forall \tilde{\gamma} \in \mathcal{L}^2\left(\Sigma_D\right), ~ \forall \tilde{d}_m \in \mathcal{L}^2\left(\Sigma\right).
\end{split}
\end{equation}
In Eq. \ref{equ:AdjSensitivityGaDmPressureConstrMHM}, the adjoint variables $\gamma_{fa}$ and $d_{fa}$ are derived by sequentially solving the variational formulation for the adjoint equations of the surface Navier-Stokes equations
\begin{equation}\label{equ:AdjEquSurfaceNSMHMPressureDrop}
\left\{\begin{split}
  & \mathrm{Find} \left\{\begin{split}
  & \mathbf{u}_a \in\left(\mathcal{H}\left(\Sigma\right)\right)^3~\mathrm{with}~ \mathbf{u}_a = \mathbf{0}~ \mathrm{at} ~  {\forall \mathbf{x} \in l_{v,\Sigma} \cup l_{v_0,\Sigma} } \\
  & p_a \in \mathcal{H}\left(\Sigma\right) \\
  & \lambda_a \in \mathcal{L}^2\left(\Sigma\right)~\mathrm{with}~ \lambda_a = 0~ \mathrm{at} ~ \forall \mathbf{x} \in l_{v,\Sigma} \cup l_{v_0,\Sigma} \\
  \end{split}\right. \\
  & \mathrm{for} \left\{\begin{split}
  & \forall \tilde{\mathbf{u}}_a \in\left(\mathcal{H}\left(\Sigma\right)\right)^3 \\
  & \forall \tilde{p}_a \in \mathcal{H}\left(\Sigma\right) \\
  & \forall \tilde{\lambda}_a \in \mathcal{L}^2\left(\Sigma\right)
  \end{split}\right.,~\mathrm{such~that} \\
  & \int_{l_{v,\Sigma}} \tilde{p}_a L^{\left( d_f \right)} \,\mathrm{d}l_{\partial\Sigma} - \int_{l_{s,\Sigma}} \tilde{p}_a L^{\left( d_f \right)} \,\mathrm{d}l_{\partial\Sigma} + \int_\Sigma \bigg[ \rho \left( \tilde{\mathbf{u}}_a \cdot \nabla_\Gamma^{\left(d_f\right)} \right) \mathbf{u} \cdot \mathbf{u}_a + \rho \left( \mathbf{u} \cdot \nabla_\Gamma^{\left(d_f\right)} \right) \tilde{\mathbf{u}}_a \\
  & \cdot \mathbf{u}_a + {\eta\over2} \left( \nabla_\Gamma^{\left(d_f\right)} \tilde{\mathbf{u}}_a + \nabla_\Gamma^{\left(d_f\right)} \tilde{\mathbf{u}}_a^\mathrm{T} \right) : \left( \nabla_\Gamma^{\left(d_f\right)} \mathbf{u}_a + \nabla_\Gamma^{\left(d_f\right)} \mathbf{u}_a^\mathrm{T} \right) - \tilde{p}_a \, \mathrm{div}_\Gamma^{\left( d_f \right)} \mathbf{u}_a \\
  & - p_a \mathrm{div}_\Gamma^{\left( d_f \right)} \tilde{\mathbf{u}}_a + \alpha \tilde{\mathbf{u}}_a \cdot \mathbf{u}_a + \tilde{\lambda}_a \mathbf{u}_a \cdot \mathbf{n}_\Gamma^{\left( d_f \right)} + \lambda_a \tilde{\mathbf{u}}_a \cdot \mathbf{n}_\Gamma^{\left( d_f \right)} \bigg] M^{\left( d_f \right)} \,\mathrm{d}\Sigma \\
  & - \sum_{E_\Sigma\in\mathcal{E}_\Sigma} \int_{E_\Sigma} \tau_{BP,\Gamma}^{\left( d_f \right)} \nabla_\Gamma^{\left( d_f \right)} \tilde{p}_a \cdot \nabla_\Gamma^{\left( d_f \right)} p_a M^{\left( d_f \right)} \,\mathrm{d}\Sigma = 0 \\  
\end{split}\right.
\end{equation}
and the variational formulations for the adjoint equations of the surface-PDE filters
\begin{equation}\label{equ:AdjPDEFilterPressureDropGaMHM} 
\left\{\begin{split}
  & \mathrm{Find}~\gamma_{fa}\in\mathcal{H}\left(\Sigma_D\right) ~\mathrm{for}~ \forall \tilde{\gamma}_{fa} \in \mathcal{H}\left(\Sigma_D\right),~\mathrm{such~that} \\
  & \int_{\Sigma_D} \left( {\partial\alpha \over \partial \gamma_p} {\partial \gamma_p \over \partial \gamma_f} \mathbf{u} \cdot \mathbf{u}_a \tilde{\gamma}_{fa} + r_f^2 \nabla_\Gamma^{\left( d_f \right)} \tilde{\gamma}_{fa} \cdot \nabla_\Gamma^{\left( d_f \right)} \gamma_{fa} + \tilde{\gamma}_{fa} \gamma_{fa} \right) M^{\left( d_f \right)} \,\mathrm{d}\Sigma = 0 
\end{split}\right.
\end{equation}
and
\begin{equation}\label{equ:AdjPDEFilterJPressureDropDmMHM} 
\left\{\begin{split}
  & \mathrm{Find}~d_{fa}\in\mathcal{H}\left(\Sigma\right)~\mathrm{for}~\forall \tilde{d}_{fa} \in \mathcal{H}\left(\Sigma\right),~\mathrm{such~that} \\
  & \int_{l_{v,\Sigma}} p L^{\left( d_f, \tilde{d}_{fa} \right)} \,\mathrm{d}l_{\partial\Sigma} - \int_{l_{s,\Sigma}} p L^{\left( d_f, \tilde{d}_{fa} \right)} \,\mathrm{d}l_{\partial\Sigma} + \int_\Sigma \bigg[ \rho \left( \mathbf{u} \cdot \nabla_\Gamma^{\left(d_f, \tilde{d}_{fa} \right)} \right) \mathbf{u} \cdot \mathbf{u}_a \\
  & + {\eta\over2} \left( \nabla_\Gamma^{\left(d_f, \tilde{d}_{fa}\right)} \mathbf{u} + \nabla_\Gamma^{\left(d_f, \tilde{d}_{fa}\right)} \mathbf{u}^\mathrm{T} \right) : \left( \nabla_\Gamma^{\left(d_f\right)} \mathbf{u}_a + \nabla_\Gamma^{\left(d_f\right)} \mathbf{u}_a^\mathrm{T} \right) + {\eta\over2} \bigg( \nabla_\Gamma^{\left(d_f\right)} \mathbf{u} \\
  & + \nabla_\Gamma^{\left(d_f\right)} \mathbf{u}^\mathrm{T} \bigg) : \left( \nabla_\Gamma^{\left(d_f, \tilde{d}_{fa}\right)} \mathbf{u}_a + \nabla_\Gamma^{\left(d_f, \tilde{d}_{fa} \right)} \mathbf{u}_a^\mathrm{T} \right) - p \, \mathrm{div}_\Gamma^{\left( d_f, \tilde{d}_{fa}\right)} \mathbf{u}_a - p_a \mathrm{div}_\Gamma^{\left( d_f, \tilde{d}_{fa} \right)} \mathbf{u} \\
  & + \lambda \mathbf{u}_a \cdot \mathbf{n}_\Gamma^{\left( d_f, \tilde{d}_{fa} \right)} + \lambda_a \mathbf{u} \cdot \mathbf{n}_\Gamma^{\left( d_f, \tilde{d}_{fa} \right)} + f_{id,\Gamma} r_f^2 \bigg( \nabla_\Gamma^{\left( d_f, \tilde{d}_{fa} \right)} \gamma_f \cdot \nabla_\Gamma^{\left( d_f \right)} \gamma_{fa} + \nabla_\Gamma^{\left( d_f \right)} \gamma_f \\
  & \cdot \nabla_\Gamma^{\left( d_f, \tilde{d}_{fa} \right)} \gamma_{fa} \bigg) \bigg] M^{\left( d_f \right)} + \bigg[ \rho \left( \mathbf{u} \cdot \nabla_\Gamma^{\left(d_f\right)} \right) \mathbf{u} \cdot \mathbf{u}_a + {\eta\over2} \left( \nabla_\Gamma^{\left(d_f\right)} \mathbf{u} + \nabla_\Gamma^{\left(d_f\right)} \mathbf{u}^\mathrm{T} \right) : \bigg( \nabla_\Gamma^{\left(d_f\right)} \mathbf{u}_a \\
  & + \nabla_\Gamma^{\left(d_f\right)} \mathbf{u}_a^\mathrm{T} \bigg) - p \, \mathrm{div}_\Gamma^{\left( d_f \right)} \mathbf{u}_a - p_a \mathrm{div}_\Gamma^{\left( d_f \right)} \mathbf{u} + \alpha \mathbf{u} \cdot \mathbf{u}_a + \lambda \mathbf{u}_a \cdot \mathbf{n}_\Gamma^{\left( d_f \right)} + \lambda_a \mathbf{u} \cdot \mathbf{n}_\Gamma^{\left( d_f \right)} \\
  & + f_{id,\Gamma} \left( r_f^2 \nabla_\Gamma^{\left( d_f \right)} \gamma_f \cdot \nabla_\Gamma^{\left( d_f \right)} \gamma_{fa} + \gamma_f \gamma_{fa} - \gamma \gamma_{fa} \right) \bigg] M^{\left( d_f, \tilde{d}_{fa} \right)}  + r_m^2 \nabla_\Sigma \tilde{d}_{fa} \cdot \nabla_\Sigma d_{fa} \\
  & + \tilde{d}_{fa} d_{fa} \,\mathrm{d}\Sigma - \sum_{E_\Sigma\in\mathcal{E}_\Sigma} \int_{E_\Sigma} \bigg( \tau_{BP,\Gamma}^{\left( d_f, \tilde{d}_{fa} \right)} \nabla_\Gamma^{\left( d_f \right)} p \cdot \nabla_\Gamma^{\left( d_f \right)} p_a + \tau_{BP,\Gamma}^{\left( d_f \right)} \nabla_\Gamma^{\left( d_f \right)} p \\
  & \cdot \nabla_\Gamma^{\left( d_f, \tilde{d}_{fa} \right)} p_a \bigg) M^{\left( d_f \right)} + \tau_{BP,\Gamma}^{\left( d_f \right)} \nabla_\Gamma^{\left( d_f \right)} p \cdot \nabla_\Gamma^{\left( d_f \right)} p_a M^{\left( d_f, \tilde{d}_{fa} \right)} \,\mathrm{d}\Sigma = 0.
\end{split}\right.
\end{equation}

After the derivation of the adjoint sensitivities in Eqs. \ref{equ:AdjSensitivityCDGaDmMHM} and \ref{equ:AdjSensitivityGaDmPressureConstrMHM}, the design variables $\gamma$ and $d_m$ can be evolved iteratively to determine the fiber bundle of the surface structure for mass transfer in the surface flow.

\subsection{Heat transfer problem} \label{sec:HeatTransferProblem}

The heat transfer process in the surface flow can be described by the surface Navier-Stokes equations and the surface convective heat-transfer equation.

\subsubsection{Surface Navier-Stokes equations} \label{sec:SurfaceNSEqusCHM}

The surface Navier-Stokes equations used to describe the motion of the surface fluid is the same as that introduced in Section \ref{sec:SurfaceNSEqus}. The difference is on the choice of the stabilization term in the variational formulation of the surface Navier-Stokes equations to numerically solve the fluid velocity and pressure by using linear finite elements. For heat transfer, the variational formulation of the surface Navier-Stokes equations can be derived as
\begin{equation}\label{equ:VariationalFormulationSurfaceNavierStokesEqusCHM}
\left\{\begin{split}
  & \mathrm{Find} \left\{\begin{split}
    & \mathbf{u}\in\left(\mathcal{H}\left(\Gamma\right)\right)^3~\mathrm{with} ~ \left\{ \begin{split}
    & \mathbf{u} = \mathbf{u}_{l_{v,\Gamma}}~ \mathrm{at} ~ \forall \mathbf{x}_\Gamma \in l_{v,\Gamma} \\
    & \mathbf{u} = \mathbf{0}~ \mathrm{at} ~ \forall \mathbf{x}_\Gamma \in l_{v_0,\Gamma} \\
    \end{split}\right.\\
  & p \in \mathcal{H}\left(\Gamma\right) \\
  & \lambda\in\mathcal{L}^2\left(\Gamma\right)~\mathrm{with}~ \lambda=0~ \mathrm{at} ~ \forall \mathbf{x}_\Gamma \in l_{v,\Gamma} \cup l_{v_0,\Gamma} \\
  \end{split}\right. \\
  & \mathrm{for} \left\{\begin{split}
  & \forall \tilde{\mathbf{u}} \in\left(\mathcal{H}\left(\Gamma\right)\right)^3 \\
  & \forall \tilde{p} \in \mathcal{H}\left(\Gamma\right) \\
  & \forall \tilde{\lambda} \in \mathcal{L}^2\left(\Gamma\right)
  \end{split}\right., ~ \mathrm{such~that} \\
  &\int_\Gamma \rho \left( \mathbf{u} \cdot \nabla_\Gamma \right) \mathbf{u} \cdot \tilde{\mathbf{u}} + {\eta\over2} \left( \nabla_\Gamma \mathbf{u} + \nabla_\Gamma \mathbf{u}^\mathrm{T} \right) : \left( \nabla_\Gamma \tilde{\mathbf{u}} + \nabla_\Gamma \tilde{\mathbf{u}}^\mathrm{T} \right) - p\,\mathrm{div}_\Gamma \tilde{\mathbf{u}} - \tilde{p} \,\mathrm{div}_\Gamma \mathbf{u} \\
  & + \alpha \mathbf{u} \cdot \tilde{\mathbf{u}} + \lambda \tilde{\mathbf{u}} \cdot \mathbf{n}_\Gamma + \tilde{\lambda} \mathbf{u} \cdot \mathbf{n}_\Gamma \,\mathrm{d}\Gamma - \sum_{E_\Gamma\in\mathcal{E}_\Gamma} \int_{E_\Gamma} \tau_{LS\mathbf{u},\Gamma} \left( \rho \mathbf{u} \cdot \nabla_\Gamma \mathbf{u} + \nabla_\Gamma p + \alpha \mathbf{u} \right) \\
  & \cdot \left( \rho \mathbf{u} \cdot \nabla_\Gamma \tilde{\mathbf{u}} + \nabla_\Gamma \tilde{p} \right) + \tau_{LSp,\Gamma} \left( \rho \mathrm{div}_\Gamma \mathbf{u} \right) \left( \mathrm{div}_\Gamma \tilde{\mathbf{u}} \right) \,\mathrm{d}\Gamma = 0 \\
\end{split}\right.
\end{equation}
where the general least square stabilization term is imposed as
\begin{equation}
-\sum_{E_\Gamma\in\mathcal{E}_\Gamma} \int_{E_\Gamma} \tau_{LS\mathbf{u},\Gamma} \left( \rho \mathbf{u} \cdot \nabla_\Gamma \mathbf{u} + \nabla_\Gamma p + \alpha \mathbf{u} \right) \cdot \left( \rho \mathbf{u} \cdot \nabla_\Gamma \tilde{\mathbf{u}} + \nabla_\Gamma \tilde{p} \right) + \tau_{LSp,\Gamma} \left( \rho \mathrm{div}_\Gamma \mathbf{u} \right) \left( \mathrm{div}_\Gamma \tilde{\mathbf{u}} \right) \: \mathrm{d}\Gamma
\end{equation}
with $\tau_{LS\mathbf{u},\Gamma}$ and $\tau_{LSp,\Gamma}$ representing the stabilization parameters. The stabilization parameters are set as \cite{DoneaWiley2003}
\begin{equation}\label{equ:NSSurfaceStabilizationTermCHM}
\left\{\begin{split}
 & \tau_{LS\mathbf{u},\Gamma} = \min \left( {h_{E_\Gamma} \over 2 \rho \left\| \mathbf{u} \right\|_2}, { h_{E_\Gamma}^2 \over 12 \eta } \right) \\
 & \tau_{LSp,\Gamma} = \left\{\begin{split}
 & { 1 \over 2} h_{E_\Gamma} \left\| \mathbf{u} \right\|_2, ~ \mathbf{u}^2 < \epsilon_{eps}^{1\over2} \\
 & { 1 \over 2} h_{E_\Gamma}, ~ \mathbf{u}^2 \geq \epsilon_{eps}^{1\over2}
 \end{split}\right.
 \end{split}\right.
\end{equation}
where $\epsilon_{eps}$ with the value of $2.2\times10^{-16}$ is the floating point precision. Based on Eqs. \ref{equ:DiffRiemannianMHM}, \ref{equ:ElementAreaTransformationMHM} and \ref{equ:ApproximationMetricAverageMHM}, the stabilization parameters in Eq. \ref{equ:NSSurfaceStabilizationTermCHM} can be transformed into
\begin{equation}\label{equ:TransformedNSSurfaceStabilizationTermCHM}
\left\{\begin{split}
 & \tau_{LS\mathbf{u},\Gamma}^{\left( d_f \right)} = \min \left( {h_{E_\Sigma} \over 2 \rho \left\| \mathbf{u} \right\|_2 } \left(M^{\left( d_f \right)}\right)^{1\over2}, { h_{E_\Sigma}^2 \over 12 \eta } M^{\left( d_f \right)} \right) \\
 & \tau_{LSp,\Gamma}^{\left( d_f \right)} = \left\{\begin{split}
 & { 1 \over 2} h_{E_\Sigma} \left\| \mathbf{u} \right\|_2 \left(M^{\left( d_f \right)}\right)^{1\over2}, ~ \mathbf{u}^2 < \epsilon_{eps}^{1\over2} \\
 & { 1 \over 2} h_{E_\Sigma} \left(M^{\left( d_f \right)}\right)^{1\over2}, ~ \mathbf{u}^2 \geq \epsilon_{eps}^{1\over2}
 \end{split}\right.
 \end{split}\right..
\end{equation}

Based on the coupling relations in Section \ref{subsec:CouplingDesignVariables}, the variational formulation in Eq. \ref{equ:VariationalFormulationSurfaceNavierStokesEqusCHM} can be transformed into the form defined on the base manifold $\Sigma$:
\begin{equation}\label{equ:TransformedVariationalFormulationSurfaceNSEqusHM}
\left\{\begin{split}
  & \mathrm{Find} \left\{\begin{split}
    & \mathbf{u}\in\left(\mathcal{H}\left(\Sigma\right)\right)^3~\mathrm{with}~ \left\{ \begin{split}
    & \mathbf{u} = \mathbf{u}_{l_{v,\Sigma}}~ \mathrm{at} ~ \forall \mathbf{x}_\Sigma \in l_{v,\Sigma} \\
    & \mathbf{u} = \mathbf{0}~ \mathrm{at} ~ \forall \mathbf{x}_\Sigma \in l_{v_0,\Sigma}\\
    \end{split}\right.\\
  & p \in \mathcal{H}\left(\Sigma\right) \\
  & \lambda\in\mathcal{L}^2\left(\Sigma\right)~\mathrm{with}~ \lambda=0~ \mathrm{at} ~ \forall \mathbf{x}_\Sigma \in l_{v,\Sigma} \cup l_{v_0,\Sigma}\\
  \end{split}\right. \\
  & \mathrm{for} \left\{\begin{split}
  & \forall \tilde{\mathbf{u}} \in\left(\mathcal{H}\left(\Sigma\right)\right)^3 \\
  & \forall \tilde{p} \in \mathcal{H}\left(\Sigma\right) \\
  & \forall \tilde{\lambda} \in \mathcal{L}^2\left(\Sigma\right)
  \end{split}\right.,~ \mathrm{such~that}\\
  & \int_\Sigma \bigg[ \rho \left( \mathbf{u} \cdot \nabla_\Gamma^{\left(d_f\right)} \right) \mathbf{u} \cdot \tilde{\mathbf{u}} + {\eta\over2} \left( \nabla_\Gamma^{\left(d_f\right)} \mathbf{u} + \nabla_\Gamma^{\left(d_f\right)} \mathbf{u}^\mathrm{T} \right) : \left( \nabla_\Gamma^{\left(d_f\right)} \tilde{\mathbf{u}} + \nabla_\Gamma^{\left(d_f\right)} \tilde{\mathbf{u}}^\mathrm{T} \right) \\
  & - p \, \mathrm{div}_\Gamma^{\left( d_f \right)} \tilde{\mathbf{u}} - \tilde{p} \, \mathrm{div}_\Gamma^{\left( d_f \right)} \mathbf{u} + \alpha \mathbf{u} \cdot \tilde{\mathbf{u}} + \lambda \tilde{\mathbf{u}} \cdot \mathbf{n}_\Gamma^{\left( d_f \right)} + \tilde{\lambda} \mathbf{u} \cdot \mathbf{n}_\Gamma^{\left( d_f \right)} \bigg] M^{\left( d_f \right)} \,\mathrm{d}\Sigma \\
  & - \sum_{E_\Sigma\in\mathcal{E}_\Sigma} \int_{E_\Sigma} \bigg[ \tau_{LS\mathbf{u},\Gamma}^{\left( d_f \right)} \left( \rho \mathbf{u} \cdot \nabla_\Gamma^{\left( d_f \right)} \mathbf{u} + \nabla_\Gamma^{\left( d_f \right)} p + \alpha \mathbf{u} \right) \cdot \bigg( \rho \mathbf{u} \cdot \nabla_\Gamma^{\left( d_f \right)} \tilde{\mathbf{u}} \\
  & + \nabla_\Gamma^{\left( d_f \right)} \tilde{p} \bigg) + \tau_{LSp,\Gamma}^{\left( d_f \right)} \left( \rho \mathrm{div}_\Gamma^{\left( d_f \right)} \mathbf{u} \right) \left( \mathrm{div}_\Gamma^{\left( d_f \right)} \tilde{\mathbf{u}} \right) \bigg] M^{\left( d_f \right)} \,\mathrm{d}\Sigma = 0.
\end{split}\right.
\end{equation}

\subsubsection{Surface convective heat-transfer equation} \label{sec:SurfaceCHMEqu}

The heat transfer process in the surface flow can be described by the surface convective heat-transfer equation defined on the implicit 2-manifold. 
Based on the conservation law of energy, the surface convective heat-transfer equation can be derived to describe the heat transfer phenomenon in the surface flow \cite{DziukActaNumerica2013}:
\begin{equation}\label{equ:CHMequOnManifolds}
\begin{split}
\rho C_p \mathbf{u} \cdot \nabla_\Gamma T - \mathrm{div}_\Gamma \left( k \nabla_\Gamma T \right) & = Q, ~\forall \mathbf{x}_\Gamma \in \Gamma
\end{split}
\end{equation}
where $T$ is the temperature; $C_p$ is the specific heat capacity; $k$ is the coefficient of heat conductivity; and $Q$ is the power of the heat source. For the surface convective heat-transfer equation, the inlet boundary is heat sink, i.e. the temperature is known at $l_{v,\Gamma}$; and the remained part of the boundary curve is insulative:
\begin{equation}\label{equ:TemperatureBoundaryCondition}
\left\{
\begin{split}
  & T = T_0, ~ \forall \mathbf{x}_\Gamma \in l_{v,\Gamma} \\
  & \nabla_\Gamma T \cdot \mathbf{n}_{\boldsymbol\tau_\Gamma} = 0, ~ \forall \mathbf{x}_\Gamma \in l_{v_0,\Gamma} \cup l_{s,\Gamma}
\end{split}\right.
\end{equation}
where $T_0$ is the known distribution of the temperature.

On the material interpolation in fiber bundle topology optimization for heat transfer in the surface flow, it is implemented for the specific heat capacity and heat conductivity together with the impermeability of the porous medium in Eq. \ref{equ:InterpolationForImpermeability}. The the specific heat capacity and heat conductivity exist in both the solid and fluid phases. In the solid phase, the convective heat-transfer degenerates into conductive heat transfer, because the fluid velocity is nearly zero in the solid phase
of an optimization result; and in the fluid phase, the convective heat-transfer is remained. Therefore, the material interpolations for the specific heat capacity and coefficient of heat conductivity are implemented by using the convex and $q$-parameterized scheme as
\begin{equation}\label{equ:MaterialIntepolationHeatConductivity}
\left\{\begin{split}
  & C_p \left( \gamma_p \right) = C_{pf} + \left( C_{ps} - C_{pf} \right) q { 1 - \gamma_p \over q + \gamma_p } \\
  & k \left( \gamma_p \right) = k_f + \left( k_s - k_f \right) q { 1 - \gamma_p \over q + \gamma_p } \\
\end{split}\right.,
\end{equation}
where $C_{ps}$ and $C_{pf}$ are the specific heat capacity of the solid and fluid phases, respectively; and $k_s$ and $k_f$ are the coefficients of the heat conductivity of the solid and fluid phases, respectively.

Based on the Galerkin method, the variational formulation of the surface convective heat-transfer equation is considered in the first order Sobolev space defined on the implicit 2-manifold $\Gamma$:
\begin{equation}\label{equ:VariationalFormulationSurfaceCHMEqu}
\left\{\begin{split}
  & \mathrm{Find}~ T\in\mathcal{H}\left(\Gamma\right)~\mathrm{with} ~ T = T_0~ \mathrm{at} ~ \forall \mathbf{x}_\Gamma \in l_{v,\Gamma}, ~ \mathrm{for} ~ \forall \tilde{T} \in \mathcal{H}\left(\Gamma\right), \\
  & \mathrm{such~that} ~ \int_\Gamma \left( \rho C_p \mathbf{u} \cdot \nabla_\Gamma T - Q \right) \tilde{T} + k \nabla_\Gamma T \cdot \nabla_\Gamma \tilde{T} \,\mathrm{d}\Gamma \\
  & + \sum_{E_\Gamma\in\mathcal{E}_\Gamma} \int_{E_\Gamma} \tau_{LST,\Gamma} \left( \rho C_p \mathbf{u} \cdot \nabla_\Gamma T - Q \right) \left( \rho C_p \mathbf{u} \cdot \nabla_\Gamma \tilde{T} \right) \,\mathrm{d}\Gamma = 0 \\
\end{split}\right.
\end{equation}
where $\tilde{T}$ is the test function of $T$; and the general least square stabilization term
\begin{equation}
  \sum_{E_\Gamma\in\mathcal{E}_\Gamma} \int_{E_\Gamma} \tau_{LST,\Gamma} \left( \rho C_p \mathbf{u} \cdot \nabla_\Gamma T - Q \right) \left( \rho C_p \mathbf{u} \cdot \nabla_\Gamma \tilde{T} \right) \,\mathrm{d}\Gamma
\end{equation}
with $\tau_{LST,\Gamma}$ representing the stabilization parameter is imposed on the variational formulation, in order to use linear finite elements to solve Eq. \ref{equ:CHMequOnManifolds} \cite{DoneaWiley2003}. The stabilization parameter is expressed as \cite{DoneaWiley2003}
\begin{equation}\label{equ:CHMSurfaceStabilizationTermCHM}
 \tau_{LST,\Gamma} = \min \left( {h_{E_\Gamma} \over 2 \rho C_p \left\|\mathbf{u}\right\|_2 }, { h_{E_\Gamma}^2 \over 12k } \right).
\end{equation}
Based on Eqs. \ref{equ:DiffRiemannianMHM}, \ref{equ:ElementAreaTransformationMHM} and \ref{equ:ApproximationMetricAverageMHM}, $\tau_{LST,\Gamma}$ can be transformed into
\begin{equation}\label{equ:TransformedCHMSurfaceStabilizationTermCHM}
  \tau_{LST,\Gamma}^{\left( d_f \right)} = \min \left( {h_{E_\Sigma} \over 2 \rho C_p \left\| \mathbf{u} \right\|_2 } \left(M^{\left( d_f \right)}\right)^{1\over2}, { h_{E_\Sigma}^2 \over 12k } M^{\left( d_f \right)} \right).
\end{equation}

Based on the homeomorphisms and the coupling relations in Section \ref{subsec:CouplingDesignVariables}, the variational formulation in Eq. \ref{equ:VariationalFormulationSurfaceConvecDiffusEqu} can be transformed into the form defined on the base manifold $\Sigma$:
\begin{equation}\label{equ:TransformedVariationalFormulationSurfaceCHMEqu}
\left\{\begin{split}
& \mathrm{Find}~ c\in\mathcal{H}\left(\Sigma\right)~\mathrm{with} ~ c = c_0~ \mathrm{at} ~ \forall \mathbf{x}_\Sigma \in l_{v,\Sigma}, ~ \mathrm{for} ~ \forall \tilde{c} \in \mathcal{H}\left(\Sigma\right), \\
& \mathrm{such~that} ~ \int_\Sigma \left[ \left( \rho C_p \mathbf{u} \cdot \nabla_\Gamma^{\left(d_f\right)} T - Q \right) \tilde{T} + k \nabla_\Gamma^{\left(d_f\right)} T \cdot \nabla_\Gamma^{\left(d_f\right)} \tilde{T} \right] M^{\left( d_f \right)} \,\mathrm{d}\Sigma \\
& + \sum_{E_\Sigma\in\mathcal{E}_\Sigma} \int_{E_\Sigma} \tau_{LST,\Gamma}^{\left( d_f \right)} \left( \rho C_p \mathbf{u} \cdot \nabla_\Gamma^{\left( d_f \right)} T - Q \right) \left( \rho C_p \mathbf{u} \cdot \nabla_\Gamma^{\left( d_f \right)} \tilde{T} \right) M^{\left( d_f \right)} \,\mathrm{d}\Sigma = 0.
\end{split}\right.
\end{equation}

\subsubsection{Design objective for heat transfer problem} \label{sec:DesignObjectiveSurfaceNSCHM}

For heat transfer in the surface flow, the desired performance of the surface structure can be set to achieve the minimized thermal compliance. The thermal compliance can be measured by the integration of the square of the temperature gradient in the computational domain. Therefore, the design objective of fiber bundle topology optimization for heat transfer in the surface flow is considered as
\begin{equation}\label{equ:DesignObjectiveSurfaceCHM}
  J_T = \int_\Gamma f_{id,\Gamma} k \nabla_\Gamma T \cdot \nabla_\Gamma T \,\mathrm{d}\Gamma,
\end{equation}
where $J_T$ is the thermal compliance and $f_{id,\Gamma}$ is the indicator function defined in Eq. \ref{equ:IndicatorFuncDesignDom}. Based on the coupling relations in Section \ref{subsec:CouplingDesignVariables}, the design objective in Eq. \ref{equ:DesignObjectiveSurfaceCHM} can be transformed into
\begin{equation}\label{equ:TransformedDesignObjectiveSurfaceCHM}
\begin{split}
  J_T^{\left( d_f \right)} = & \int_\Sigma f_{id,\Gamma} k \nabla_\Gamma^{\left( d_f \right)} T \cdot \nabla_\Gamma^{\left( d_f \right)} T M^{\left( d_f \right)} \,\mathrm{d}\Sigma.
\end{split}
\end{equation}

\subsubsection{Constraint of dissipation power} \label{sec:DissipationConstraintSurfaceNSCHM}

To ensure the patency of the surface structure for heat transfer in the surface flow, a constraint of the power of the viscous dissipation is imposed and it is expressed as
\begin{equation}\label{equ:DissipationConstraintSurfaceNSCHM}
  \left| \Phi \left/ \: \Phi_0 - 1 \right. \right| \leq 1\times10^{-3},
\end{equation}
where $\Phi_0$ is the specified reference value of the dissipation power and $\Phi$ is the dissipation power in the surface flow:
\begin{equation}\label{equ:DissipationSurfaceNSCHM}
  \Phi = \int_\Gamma {\eta\over2} \left( \nabla_\Gamma \mathbf{u} + \nabla_\Gamma \mathbf{u}^\mathrm{T} \right) : \left( \nabla_\Gamma \mathbf{u} + \nabla_\Gamma \mathbf{u}^\mathrm{T} \right) + \alpha \mathbf{u}^2 \,\mathrm{d}\Gamma.
\end{equation}
Based on the coupling relations in Section \ref{subsec:CouplingDesignVariables}, Eq. \ref{equ:DissipationSurfaceNSCHM} can be transformed into
\begin{equation}\label{equ:TransformedDissipationSurfaceNSCHM}
\begin{split}
  \Phi^{\left( d_f \right)} = \int_\Sigma \left[ {\eta\over2} \left( \nabla_\Gamma^{\left( d_f \right)} \mathbf{u} + \nabla_\Gamma^{\left( d_f \right)} \mathbf{u}^\mathrm{T} \right) : \left( \nabla_\Gamma^{\left( d_f \right)} \mathbf{u} + \nabla_\Gamma^{\left( d_f \right)} \mathbf{u}^\mathrm{T} \right) + \alpha \mathbf{u}^2 \right] M^{\left( d_f \right)} \,\mathrm{d}\Sigma. \\
\end{split}
\end{equation}

\subsubsection{Area constraint of surface structure} \label{sec:AreaConstraintSurfaceNSCHM}

To regularize the fiber bundle topology optimization problem for heat transfer in the surface flow, the area constraint of the surface structure can be imposed as
\begin{equation}\label{equ:AreaConstraintSurfaceNSCHM}
  \left| s \left/\: s_0 \right. - 1 \right| \leq 1\times10^{-3},
\end{equation}
where $s$ is the area fraction of the surface structure and $s_0$ is the specified area fraction. The area fraction of the surface structure is defined as
\begin{equation}\label{equ:AreaSurfaceNSCHM}
  s = {1\over\left|\Gamma_D\right|} \int_{\Gamma_D} \gamma_p \,\mathrm{d}\Gamma = {1\over\left|\Gamma_D\right|} \int_\Gamma f_{id,\Gamma} \gamma_p \,\mathrm{d}\Gamma,
\end{equation}
where $\left|\Gamma_D\right|$ is the area of the implicit 2-manifold and it is expressed as
\begin{equation}\label{equ:AreaImplicitManifoldSurfaceNSCHM}
\left|\Gamma_D\right| = \int_{\Gamma_D} 1 \,\mathrm{d}\Gamma = \int_\Gamma f_{id,\Gamma} \,\mathrm{d}\Gamma.
\end{equation}
Based on the coupling relations in Section \ref{subsec:CouplingDesignVariables}, Eqs. \ref{equ:AreaSurfaceNSCHM} and \ref{equ:AreaImplicitManifoldSurfaceNSCHM} can be transformed into
\begin{equation}\label{equ:TransformedAreaSurfaceNSCHM}
s^{\left( d_f \right)} = {1\over\left|\Gamma_D\right|^{\left( d_f \right)}} \int_{\Sigma_D} \gamma_p M^{\left( d_f \right)} \,\mathrm{d}\Sigma = {1\over\left|\Gamma_D\right|^{\left( d_f \right)}} \int_\Sigma f_{id,\Gamma} \gamma_p M^{\left( d_f \right)} \,\mathrm{d}\Sigma
\end{equation}
and
\begin{equation}\label{equ:TransformedAreaImplicitManifoldSurfaceNSCHM}
\left|\Gamma_D\right|^{\left( d_f \right)} = \int_{\Sigma_D} M^{\left( d_f \right)} \,\mathrm{d}\Sigma = \int_\Sigma f_{id,\Gamma} M^{\left( d_f \right)} \,\mathrm{d}\Sigma.
\end{equation}

\subsubsection{Fiber bundle topology optimization problem} \label{sec:MatchingOptimizationSurfaceNSEqusHM}

Based on the above introduction, the fiber bundle topology optimization problem for heat transfer in the surface flow can be constructed to optimize the fiber bundle in Eq. \ref{equ:FiberBundleMHM}:
\begin{equation}\label{equ:VarProToopSurfaceNSCHM} 
\left\{\begin{split}
  & \mathrm{Find} \left\{\begin{split}
  & \gamma: \Gamma \mapsto \left[0,1\right] \\
  & d_m: \Sigma \mapsto \left[0,1\right]\end{split}\right.~ \mathrm{for} ~
  \left(\Sigma \times \left(\Gamma \times \left[0,1\right]\right), \Sigma, proj_1, \Gamma \times \left[0,1\right] \right), \\
  & \mathrm{to} ~ \mathrm{minimize}~{J_T \over J_{T,0}}~ \mathrm{with} ~ J_T = \int_\Gamma f_{id,\Gamma} k \nabla_\Gamma T \cdot \nabla_\Gamma T \,\mathrm{d}\Gamma, \\
  & \mathrm{constrained} ~ \mathrm{by} \\
  & \left\{\begin{split}
  & \begin{split}
  & \left\{\begin{split}
    & \left\{\begin{split}
       & \rho \left( \mathbf{u} \cdot \nabla_\Gamma \right) \mathbf{u} - \mathbf{P} \mathrm{div}_\Gamma \left[ \eta \left( \nabla_\Gamma \mathbf{u} + \nabla_\Gamma \mathbf{u}^\mathrm{T} \right) \right] + \nabla_\Gamma p = - \alpha \mathbf{u}, ~ \forall \mathbf{x}_\Gamma \in \Gamma \\
       & - \mathrm{div}_\Gamma \mathbf{u} = 0, ~ \forall \mathbf{x}_\Gamma \in \Gamma \\
       & \mathbf{u} \cdot \mathbf{n}_\Gamma = 0, ~ \forall \mathbf{x}_\Gamma \in \Gamma \\
    \end{split}\right. \\
    & \alpha \left( \gamma_p \right) = \alpha_f + \left( \alpha_s - \alpha_f \right) q { 1 - \gamma_p \over q + \gamma_p } \\
    \end{split}\right.
    \end{split} \\
  & \left\{\begin{split}
  & \rho C_p \mathbf{u} \cdot \nabla_\Gamma T - \mathrm{div}_\Gamma \left( k \nabla_\Gamma T \right) = Q, ~\forall \mathbf{x}_\Gamma \in \Gamma \\
  & \left\{\begin{split}
  & C_p \left( \gamma_p \right) = C_{pf} + \left( C_{ps} - C_{pf} \right) q { 1 - \gamma_p \over q + \gamma_p } \\
  & k \left( \gamma_p \right) = k_f + \left( k_s - k_f \right) q { 1 - \gamma_p \over q + \gamma_p } \\
  \end{split}\right.\\
  \end{split}\right.\\
  & \left\{\begin{split}
    & \left\{\begin{split}
        & - \mathrm{div}_\Gamma \left( r_f^2 \nabla_\Gamma \gamma_f \right) + \gamma_f = \gamma,~\forall \mathbf{x}_\Gamma \in \Gamma_D \\
        & \mathbf{n}_{\boldsymbol\tau_\Gamma} \cdot \nabla_\Gamma \gamma_f = 0,~\forall \mathbf{x}_\Gamma \in \partial\Gamma_D \\
    \end{split}\right. \\
    & \gamma_p = { \tanh\left(\beta \xi\right) + \tanh\left(\beta \left(\gamma_f-\xi\right)\right) \over \tanh\left(\beta \xi\right) + \tanh\left(\beta \left(1-\xi\right)\right)} \\
  \end{split}\right. \\
  & \left\{
        \begin{split}
          & - \mathrm{div}_\Sigma \left( r_m^2 \nabla_\Sigma d_f \right) + d_f = A_d \left( d_m - {1\over2} \right), ~ \forall \mathbf{x}_\Sigma \in \Sigma \\
          & \mathbf{n}_{\boldsymbol\tau_\Sigma} \cdot \nabla_\Sigma d_f = 0, ~ \forall \mathbf{x}_\Sigma \in \partial \Sigma \\
        \end{split}\right. \\
  & \Gamma = \left\{ \mathbf{x}_\Gamma : \mathbf{x}_\Gamma = d_f \mathbf{n}_\Sigma + \mathbf{x}_\Sigma,~\forall \mathbf{x}_\Sigma \in \Sigma \right\} \\
  & \left| \Phi \left/ \: \Phi_0 - 1 \right. \right| \leq 1\times10^{-3}, ~\mathrm{with} ~ \Phi = \int_\Gamma {\eta\over2} \left( \nabla_\Gamma \mathbf{u} + \nabla_\Gamma \mathbf{u}^\mathrm{T} \right) : \left( \nabla_\Gamma \mathbf{u} + \nabla_\Gamma \mathbf{u}^\mathrm{T} \right) + \alpha \mathbf{u}^2 \,\mathrm{d}\Gamma \\
  & \left| s \left/\: s_0 \right. - 1 \right| \leq 1\times10^{-3}, ~\mathrm{with} ~ s = {1\over\left|\Gamma_D\right|} \int_{\Gamma_D} \gamma_p \,\mathrm{d}\Gamma ~ \mathrm{and} ~ \left|\Gamma_D\right| = \int_{\Gamma_D} 1 \,\mathrm{d}\Gamma
\end{split}\right.
\end{split}\right.
\end{equation}
where $J_{T,0}$ is the reference value of the design objective corresponding to the initial distribution of the design variables.

The coupling relations among the variables, functions, tangential divergence operator and tangential gradient operator in Eq. \ref{equ:VarProToopSurfaceNSCHM} are illustrated by the arrow chart described as
\[\begin{array}{cccccccc}
 \textcolor{blue}{d_m} & \xrightarrow{\mathrm{Eq.~}\ref{equ:PDEFilterzmBaseStructureMHM}} & d_f & & & \\
 & & \bigg\downarrow\vcenter{\rlap{\scriptsize{Eq.~\ref{equ:TransformedTangentialOperatorMHM}}}} & & \\
 & & \left\{ \mathrm{div}_\Gamma, \nabla_\Gamma, \mathbf{n}_\Gamma \right\} & \xrightarrow{\mathrm{Eqs.~}\ref{equ:NSequOnManifoldsMHM}~\&~\ref{equ:CHMequOnManifolds}} & \left\{ \mathbf{u},~p,~\lambda,~T \right\} & \xrightarrow{\mathrm{Eqs.~}\ref{equ:DesignObjectiveSurfaceCHM}, ~ \ref{equ:DissipationSurfaceNSCHM} ~\&~ \ref{equ:AreaSurfaceNSCHM}} & \left\{ \textcolor[rgb]{0.50,0.00,0.00}{J_T}, ~ \textcolor[rgb]{0.50,0.00,0.00}{\Phi}, ~ \textcolor[rgb]{0.50,0.00,0.00}{s} \right\} \\
 & & \bigg\downarrow\vcenter{\rlap{\scriptsize{Eq.~\ref{equ:PDEFilterGammaFilberMHM}}}} & &  \bigg\uparrow\vcenter{\rlap{\scriptsize{Eq.~\ref{equ:NSequOnManifoldsMHM}}}}   \\
 \textcolor{blue}{\gamma} & \xrightarrow{\mathrm{Eq.~}\ref{equ:PDEFilterGammaFilberMHM}} & \gamma_f & \xrightarrow{\mathrm{Eq.~}\ref{equ:ProjectionGammaFilberMHM}} & \textcolor[rgb]{0.50,0.00,0.00}{\gamma_p} & & \\
\end{array}\]
where the design variables $d_m$ and $\gamma$, marked in blue, are the inputs; the design objective $J_T$, the dissipation power $\Phi$ and the material density $\gamma_p$, marked in red, are the outputs.

\subsubsection{Adjoint analysis} \label{sec:AdjointAnalysisSurfaceNSEqusCHM}

To solve the fiber bundle topology optimization problem in Eq. \ref{equ:VarProToopSurfaceNSCHM} by using a gradient based iterative procedure, the adjoint analysis is implemented for the design objective, dissipation power and area fraction of the surface structure to derive the adjoint sensitivities. The details for the adjoint analysis have been provided in the appendix in Sections \ref{sec:AdjointAnalysisDesignObjectiveSurfaceCHTMHM} and \ref{sec:AdjointAnalysisDissipationConstraintMHM}.

Based on the transformed design objective in Eq. \ref{equ:TransformedDesignObjectiveSurfaceCHM} and transformed dissipation power in Eq. \ref{equ:TransformedDissipationSurfaceNSCHM}, the adjoint analysis of the fiber bundle topology optimization problem can be implemented on the functional spaces defined on the base manifold. Based on the continuous adjoint method \cite{HinzeSpringer2009}, the adjoint sensitivity of the design objective $J_T$ is derived as
\begin{equation}\label{equ:AdjSensitivityCDGaDmObjCHM}
\begin{split}
\delta J_T = - \int_{\Sigma_D} \gamma_{fa} \tilde{\gamma} M^{\left( d_f \right)}\,\mathrm{d}\Sigma - \int_\Sigma A_d d_{fa} \tilde{d}_m \,\mathrm{d}\Sigma, ~ \forall \tilde{\gamma} \in \mathcal{L}^2\left(\Sigma_D\right), ~ \forall \tilde{d}_m \in \mathcal{L}^2\left(\Sigma\right).
\end{split}
\end{equation}
The adjoint variables in Eq. \ref{equ:AdjSensitivityCDGaDmObjCHM} can be derived from the adjoint equations in the variational formulations. The variational formulation for the adjoint equation of the surface convective heat-transfer equation is derived as
\begin{equation}\label{equ:WeakAdjEquSCHMEquMHMObj}
\left\{\begin{split}
  & \mathrm{Find} ~ T_a \in \mathcal{H}\left(\Sigma\right)~\mathrm{with}~ T_a=0 ~ \mathrm{at} ~ \forall \mathbf{x}_\Sigma \in l_{v,\Sigma}, ~ \mathrm{for} ~ \forall \tilde{T}_a \in \mathcal{H} \left(\Sigma\right),~\mathrm{such~that} \\
  & \int_\Sigma \left[ 2 f_{id,\Gamma} k \nabla_\Gamma^{\left( d_f \right)} T \cdot \nabla_\Gamma^{\left( d_f \right)} \tilde{T}_a + \left( \rho C_p \mathbf{u} \cdot \nabla_\Gamma^{\left(d_f\right)} \tilde{T}_a \right) T_a + k \nabla_\Gamma^{\left(d_f\right)} T_a \cdot \nabla_\Gamma^{\left(d_f\right)} \tilde{T}_a \right] \\
  & M^{\left( d_f \right)} \,\mathrm{d}\Sigma + \sum_{E_\Sigma\in\mathcal{E}_\Sigma} \int_{E_\Sigma} \tau_{LST,\Gamma}^{\left( d_f \right)} \left( \rho C_p \mathbf{u} \cdot \nabla_\Gamma^{\left( d_f \right)} T_a \right) \left( \rho C_p \mathbf{u} \cdot \nabla_\Gamma^{\left( d_f \right)} \tilde{T}_a \right) M^{\left( d_f \right)} \,\mathrm{d}\Sigma = 0 \\
\end{split}\right.
\end{equation}
where $T_a$ is the adjoint variable of $T$ and $\tilde{T}_a$ is the test function of $T_a$.
The variational formulation for the adjoint equations of the surface Naiver-Stokes equations is derived as
\begin{equation}\label{equ:AdjSurfaceNavierStokesEqusJObjectiveCHMObj} 
\left\{\begin{split}
  & \mathrm{Find} \left\{\begin{split}
  & \mathbf{u}_a \in\left(\mathcal{H}\left(\Sigma\right)\right)^3~\mathrm{with}~ \mathbf{u}_a = \mathbf{0}~ \mathrm{at} ~ {\forall \mathbf{x} \in l_{v,\Sigma} \cup l_{v_0,\Sigma} } \\
  & p_a \in \mathcal{H}\left(\Sigma\right) \\
  & \lambda_a \in \mathcal{L}^2\left(\Sigma\right)~\mathrm{with}~ \lambda_a = 0~ \mathrm{at} ~ \forall \mathbf{x} \in l_{v,\Sigma} \cup l_{v_0,\Sigma} \\
  \end{split}\right. \\
  & \mathrm{for} \left\{\begin{split}
  & \forall \tilde{\mathbf{u}}_a \in\left(\mathcal{H}\left(\Sigma\right)\right)^3 \\
  & \forall \tilde{p}_a \in \mathcal{H}\left(\Sigma\right) \\
  & \forall \tilde{\lambda}_a \in \mathcal{L}^2\left(\Sigma\right)
  \end{split}\right.,~\mathrm{such~that} \\
  & \int_\Sigma \bigg[ \rho \left( \tilde{\mathbf{u}}_a \cdot \nabla_\Gamma^{\left(d_f\right)} \mathbf{u} + \mathbf{u} \cdot \nabla_\Gamma^{\left(d_f\right)} \tilde{\mathbf{u}}_a \right) \cdot \mathbf{u}_a + {\eta\over2} \left( \nabla_\Gamma^{\left(d_f\right)} \tilde{\mathbf{u}}_a + \nabla_\Gamma^{\left(d_f\right)} \tilde{\mathbf{u}}_a^\mathrm{T} \right) : \left( \nabla_\Gamma^{\left(d_f\right)} \mathbf{u}_a + \nabla_\Gamma^{\left(d_f\right)} \mathbf{u}_a^\mathrm{T} \right) \\
  & - \tilde{p}_a \mathrm{div}_\Gamma^{\left( d_f \right)} \mathbf{u}_a - p_a \mathrm{div}_\Gamma^{\left( d_f \right)} \tilde{\mathbf{u}}_a + \alpha \tilde{\mathbf{u}}_a \cdot \mathbf{u}_a + \tilde{\lambda}_a  \mathbf{u}_a \cdot \mathbf{n}_\Gamma^{\left( d_f \right)} + \lambda_a \tilde{\mathbf{u}}_a \cdot \mathbf{n}_\Gamma^{\left( d_f \right)} + \rho C_p \tilde{\mathbf{u}}_a \cdot \nabla_\Gamma^{\left(d_f\right)} T T_a \bigg] \\
  & M^{\left( d_f \right)} \,\mathrm{d}\Sigma - \sum_{E_\Sigma\in\mathcal{E}_\Sigma} \int_{E_\Sigma} \bigg\{ \tau_{LS\mathbf{u},\Gamma}^{\left( d_f, \tilde{\mathbf{u}}_a \right)} \left( \rho \mathbf{u} \cdot \nabla_\Gamma^{\left( d_f \right)} \mathbf{u} + \nabla_\Gamma^{\left( d_f \right)} p + \alpha \mathbf{u} \right) \cdot \left( \rho \mathbf{u} \cdot \nabla_\Gamma^{\left( d_f \right)} \mathbf{u}_a + \nabla_\Gamma^{\left( d_f \right)} p_a \right) \\
  & + \tau_{LS\mathbf{u},\Gamma}^{\left( d_f \right)} \left[ \rho \left( \tilde{\mathbf{u}}_a \cdot \nabla_\Gamma^{\left( d_f \right)} \mathbf{u} + \mathbf{u} \cdot \nabla_\Gamma^{\left( d_f \right)} \tilde{\mathbf{u}}_a \right) + \nabla_\Gamma^{\left( d_f \right)} \tilde{p}_a + \alpha \tilde{\mathbf{u}}_a \right] \cdot \left( \rho \mathbf{u} \cdot \nabla_\Gamma^{\left( d_f \right)} \mathbf{u}_a + \nabla_\Gamma^{\left( d_f \right)} p_a \right) \\
  & + \tau_{LS\mathbf{u},\Gamma}^{\left( d_f \right)} \left( \rho \mathbf{u} \cdot \nabla_\Gamma^{\left( d_f \right)} \mathbf{u} + \nabla_\Gamma^{\left( d_f \right)} p + \alpha \mathbf{u} \right) \cdot \left( \rho \tilde{\mathbf{u}}_a \cdot \nabla_\Gamma^{\left( d_f \right)} \mathbf{u}_a \right) + \tau_{LSp,\Gamma}^{\left( d_f, \tilde{\mathbf{u}}_a \right)} \left( \rho \mathrm{div}_\Gamma^{\left( d_f \right)} \mathbf{u} \right) \\
  & \left( \mathrm{div}_\Gamma^{\left( d_f \right)} \mathbf{u}_a \right) + \tau_{LSp,\Gamma}^{\left( d_f \right)} \left( \rho \mathrm{div}_\Gamma^{\left( d_f \right)} \tilde{\mathbf{u}}_a \right) \left( \mathrm{div}_\Gamma^{\left( d_f \right)} \mathbf{u}_a \right) - \tau_{LST,\Gamma}^{\left( d_f, \tilde{\mathbf{u}}_a \right)} \left( \rho C_p \mathbf{u} \cdot \nabla_\Gamma^{\left( d_f \right)} T - Q \right) \\
  & \left( \rho C_p \mathbf{u} \cdot \nabla_\Gamma^{\left( d_f \right)} T_a \right) - \tau_{LST,\Gamma}^{\left( d_f \right)} \left( \rho C_p \tilde{\mathbf{u}}_a \cdot \nabla_\Gamma^{\left( d_f \right)} T \right) \left( \rho C_p \mathbf{u} \cdot \nabla_\Gamma^{\left( d_f \right)} T_a \right) - \tau_{LST,\Gamma}^{\left( d_f \right)} \\
  & \left( \rho C_p \mathbf{u} \cdot \nabla_\Gamma^{\left( d_f \right)} T - Q \right) \left( \rho C_p \tilde{\mathbf{u}}_a \cdot \nabla_\Gamma^{\left( d_f \right)} T_a \right) \bigg\} M^{\left( d_f \right)} \,\mathrm{d}\Sigma = 0
\end{split}\right.
\end{equation}
where $\tau_{LS\mathbf{u},\Gamma}^{\left( d_f, \tilde{\mathbf{u}} \right)}$, $\tau_{LSp,\Gamma}^{\left( d_f, \tilde{\mathbf{u}} \right)}$ and $\tau_{LST,\Gamma}^{\left( d_f, \tilde{\mathbf{u}} \right)}$ are the first-order variationals of $\tau_{LS\mathbf{u},\Gamma}^{\left( d_f \right)}$, $\tau_{LSp,\Gamma}^{\left( d_f \right)}$ and $\tau_{LST,\Gamma}^{\left( d_f \right)}$ to $\mathbf{u}$, respectively, and they are expressed as
\begin{equation}\label{equ:TransformedNSSurfaceStabilizationTerm1stVariUnormCHM}
\begin{split}
& \left.\begin{split}
 & \tau_{LS\mathbf{u},\Gamma}^{\left( d_f, \tilde{\mathbf{u}} \right)} = \left\{\begin{split}
   & - {h_{E_\Sigma} \mathbf{u} \cdot \tilde{\mathbf{u}} \over 2 \rho \left\| \mathbf{u} \right\|_2^3 } \left(M^{\left( d_f \right)}\right)^{1\over2}, ~ {h_{E_\Sigma} \over 2 \rho \left\| \mathbf{u} \right\|_2 } \left(M^{\left( d_f \right)}\right)^{1\over2} < { h_{E_\Sigma}^2 \over 12 \eta } M^{\left( d_f \right)} \\
   & 0, ~ {h_{E_\Sigma} \over 2 \rho \left\| \mathbf{u} \right\|_2 } \left(M^{\left( d_f \right)}\right)^{1\over2} \geq { h_{E_\Sigma}^2 \over 12 \eta } M^{\left( d_f \right)}
  \end{split}\right. \\
 & \tau_{LSp,\Gamma}^{\left( d_f, \tilde{\mathbf{u}} \right)} = \left\{\begin{split}
 & { h_{E_\Sigma} \mathbf{u} \cdot \tilde{\mathbf{u}} \over 2 \left\| \mathbf{u} \right\|_2 } \left(M^{\left( d_f \right)}\right)^{1\over2}, ~ \mathbf{u}^2 < \epsilon_{eps}^{1\over2} \\
 & 0, ~ \mathbf{u}^2 \geq \epsilon_{eps}^{1\over2}
 \end{split}\right. \\
 & \tau_{LST,\Gamma}^{\left( d_f, \tilde{\mathbf{u}} \right)} = \left\{\begin{split} 
  & - {h_{E_\Sigma} \mathbf{u} \cdot \tilde{\mathbf{u}} \over 2 \rho C_p \left\| \mathbf{u} \right\|_2^3 } \left(M^{\left( d_f \right)}\right)^{1\over2}, ~ {h_{E_\Sigma} \over 2 \rho C_p \left\| \mathbf{u} \right\|_2 } \left(M^{\left( d_f \right)}\right)^{1\over2} < { h_{E_\Sigma}^2 \over 12k } M^{\left( d_f \right)} \\
  & 0, ~ {h_{E_\Sigma} \over 2 \rho C_p \left\| \mathbf{u} \right\|_2 } \left(M^{\left( d_f \right)}\right)^{1\over2} \geq { h_{E_\Sigma}^2 \over 12k } M^{\left( d_f \right)}
  \end{split}\right.
 \end{split}\right\} \\
 & \forall \tilde{\mathbf{u}} \in \left(\mathcal{H}\left( \Sigma \right)\right)^3.
\end{split}
\end{equation}
The variational formulations for the adjoint equations of the surface-PDE filters for $\gamma$ and $d_m$ are derived as
\begin{equation}\label{equ:AdjPDEFilterJObjectiveGaCHMObj}  
\left\{\begin{split}
  & \mathrm{Find}~\gamma_{fa}\in\mathcal{H}\left(\Sigma_D\right) ~\mathrm{for}~ \forall \tilde{\gamma}_{fa} \in \mathcal{H}\left(\Sigma_D\right),~\mathrm{such~that} \\
  & \int_{\Sigma_D} \bigg\{ \bigg[ {\partial \alpha \over \partial \gamma_p} {\partial \gamma_p \over \partial \gamma_f} \mathbf{u} \cdot \mathbf{u}_a + \rho {\partial C_p \over \partial \gamma_p} {\partial \gamma_p \over \partial \gamma_f} \mathbf{u} \cdot \nabla_\Gamma^{\left(d_f\right)} T T_a + {\partial k \over \partial \gamma_p} {\partial \gamma_p \over \partial \gamma_f} \bigg( f_{id,\Gamma} \nabla_\Gamma^{\left(d_f\right)} T \\
  & \cdot \nabla_\Gamma^{\left(d_f\right)} T + \nabla_\Gamma^{\left(d_f\right)} T \cdot \nabla_\Gamma^{\left(d_f\right)} T_a \bigg) \bigg] \tilde{\gamma}_{fa} + r_f^2 \nabla_\Gamma^{\left( d_f \right)} \gamma_{fa} \cdot \nabla_\Gamma^{\left( d_f \right)} \tilde{\gamma}_{fa} + \gamma_{fa} \tilde{\gamma}_{fa} \bigg\} M^{\left( d_f \right)} \,\mathrm{d}\Sigma \\
  & - \sum_{E_\Sigma\in\mathcal{E}_\Sigma} \int_{{E_\Sigma}\cap\Sigma_D} \Bigg\{ \tau_{LS\mathbf{u},\Gamma}^{\left( d_f \right)} {\partial \alpha \over \partial \gamma_p} {\partial \gamma_p \over \partial \gamma_f} \mathbf{u} \cdot \left( \rho \mathbf{u} \cdot \nabla_\Gamma^{\left( d_f \right)} \mathbf{u}_a + \nabla_\Gamma^{\left( d_f \right)} p_a \right) - \Bigg( {\partial \tau_{LST,\Gamma}^{\left( d_f \right)} \over \partial C_p} {\partial C_p \over \partial \gamma_p} \\
  & + {\partial \tau_{LST,\Gamma}^{\left( d_f \right)} \over \partial k} {\partial k \over \partial \gamma_p} \Bigg) {\partial \gamma_p \over \partial \gamma_f} \left( \rho C_p \mathbf{u} \cdot \nabla_\Gamma^{\left( d_f \right)} T - Q \right) \left( \rho C_p \mathbf{u} \cdot \nabla_\Gamma^{\left( d_f \right)} T_a \right) - \tau_{LST,\Gamma}^{\left( d_f \right)} \rho {\partial C_p \over \partial \gamma_p} {\partial \gamma_p \over \partial \gamma_f} \\
  & \bigg[ \left( \mathbf{u} \cdot \nabla_\Gamma^{\left( d_f \right)} T \right) \left( \rho C_p \mathbf{u} \cdot \nabla_\Gamma^{\left( d_f \right)} T_a \right) + \left( \rho C_p \mathbf{u} \cdot \nabla_\Gamma^{\left( d_f \right)} T - Q \right) \\
  & \left( \mathbf{u} \cdot \nabla_\Gamma^{\left( d_f \right)} T_a \right) \bigg] \Bigg\} \tilde{\gamma}_{fa} M^{\left( d_f \right)} \,\mathrm{d}\Sigma = 0 \\
\end{split}\right.
\end{equation}
and
\begin{equation}\label{equ:AdjPDEFilterJObjectiveDmCHMObj} 
\left\{\begin{split}
  & \mathrm{Find}~d_{fa}\in\mathcal{H}\left(\Sigma\right) ~\mathrm{for}~ \forall \tilde{d}_{fa} \in \mathcal{H}\left(\Sigma\right),~\mathrm{such~that} \\
  & \int_\Sigma \bigg[ 2 f_{id,\Gamma} k \nabla_\Gamma^{\left( d_f, \tilde{d}_{fa} \right)} T \cdot \nabla_\Gamma^{\left( d_f \right)} T + \rho \left( \mathbf{u} \cdot \nabla_\Gamma^{\left(d_f, \tilde{d}_{fa} \right)} \right) \mathbf{u} \cdot \mathbf{u}_a \\
  & + {\eta\over2} \left( \nabla_\Gamma^{\left(d_f, \tilde{d}_{fa}\right)} \mathbf{u} + \nabla_\Gamma^{\left(d_f, \tilde{d}_{fa}\right)} \mathbf{u}^\mathrm{T} \right) : \left( \nabla_\Gamma^{\left(d_f\right)} \mathbf{u}_a + \nabla_\Gamma^{\left(d_f\right)} \mathbf{u}_a^\mathrm{T} \right) + {\eta\over2} \left( \nabla_\Gamma^{\left(d_f\right)} \mathbf{u} + \nabla_\Gamma^{\left(d_f\right)} \mathbf{u}^\mathrm{T} \right) \\
  & : \left( \nabla_\Gamma^{\left(d_f, \tilde{d}_{fa} \right)} \mathbf{u}_a + \nabla_\Gamma^{\left(d_f, \tilde{d}_{fa} \right)} \mathbf{u}_a^\mathrm{T} \right) - p \, \mathrm{div}_\Gamma^{\left( d_f, \tilde{d}_{fa} \right)} \mathbf{u}_a - p_a \mathrm{div}_\Gamma^{\left( d_f, \tilde{d}_{fa} \right)} \mathbf{u} + \lambda \mathbf{u}_a \cdot \mathbf{n}_\Gamma^{\left( d_f, \tilde{d}_{fa} \right)} + \lambda_a \\
  & \mathbf{u} \cdot \mathbf{n}_\Gamma^{\left( d_f, \tilde{d}_{fa} \right)} + \left( \rho C_p \mathbf{u} \cdot \nabla_\Gamma^{\left(d_f, \tilde{d}_{fa} \right)} T \right) T_a + k \nabla_\Gamma^{\left(d_f, \tilde{d}_{fa} \right)} T \cdot \nabla_\Gamma^{\left(d_f\right)} T_a + k \nabla_\Gamma^{\left(d_f\right)} T \cdot \nabla_\Gamma^{\left(d_f, \tilde{d}_{fa} \right)} T_a \\
  & + f_{id,\Gamma} r_f^2 \left( \nabla_\Gamma^{\left( d_f, \tilde{d}_{fa} \right)} \gamma_f \cdot \nabla_\Gamma^{\left( d_f \right)} \gamma_{fa} + \nabla_\Gamma^{\left( d_f \right)} \gamma_f \cdot \nabla_\Gamma^{\left( d_f, \tilde{d}_{fa} \right)} \gamma_{fa} \right) \bigg] M^{\left( d_f \right)} + \bigg[ f_{id,\Gamma} k \nabla_\Gamma^{\left( d_f \right)} T \\
  & \cdot \nabla_\Gamma^{\left( d_f \right)} T + \rho \left( \mathbf{u} \cdot \nabla_\Gamma^{\left(d_f\right)} \right) \mathbf{u} \cdot \mathbf{u}_a + {\eta\over2} \left( \nabla_\Gamma^{\left(d_f\right)} \mathbf{u} + \nabla_\Gamma^{\left(d_f\right)} \mathbf{u}^\mathrm{T} \right) : \left( \nabla_\Gamma^{\left(d_f\right)} \mathbf{u}_a + \nabla_\Gamma^{\left(d_f\right)} \mathbf{u}_a^\mathrm{T} \right) \\
  & - p \, \mathrm{div}_\Gamma^{\left( d_f \right)} \mathbf{u}_a - p_a \, \mathrm{div}_\Gamma^{\left( d_f \right)} \mathbf{u} + \alpha \mathbf{u} \cdot \mathbf{u}_a + \lambda \mathbf{u}_a \cdot \mathbf{n}_\Gamma^{\left( d_f \right)} + \lambda_a \mathbf{u} \cdot \mathbf{n}_\Gamma^{\left( d_f \right)} + \left( \rho C_p \mathbf{u} \cdot \nabla_\Gamma^{\left(d_f\right)} T - Q \right) \\
  & T_a + k \nabla_\Gamma^{\left(d_f\right)} T \cdot \nabla_\Gamma^{\left(d_f\right)} T_a + f_{id,\Gamma} \left( r_f^2 \nabla_\Gamma^{\left( d_f \right)} \gamma_f \cdot \nabla_\Gamma^{\left( d_f \right)} \gamma_{fa} + \gamma_f \gamma_{fa} - \gamma \gamma_{fa} \right) \bigg] M^{\left( d_f, \tilde{d}_{fa} \right)} \\
  & + r_m^2 \nabla_\Sigma \tilde{d}_{fa} \cdot \nabla_\Sigma d_{fa} + \tilde{d}_{fa} d_{fa} \,\mathrm{d}\Sigma - \sum_{E_\Sigma\in\mathcal{E}_\Sigma} \int_{E_\Sigma} \bigg[ \tau_{LS\mathbf{u},\Gamma}^{\left( d_f, \tilde{d}_{fa} \right)} \left( \rho \mathbf{u} \cdot \nabla_\Gamma^{\left( d_f \right)} \mathbf{u} + \nabla_\Gamma^{\left( d_f \right)} p + \alpha \mathbf{u} \right) \\
  & \cdot \left( \rho \mathbf{u} \cdot \nabla_\Gamma^{\left( d_f \right)} \mathbf{u}_a + \nabla_\Gamma^{\left( d_f \right)} p_a \right) + \tau_{LS\mathbf{u},\Gamma}^{\left( d_f \right)} \left( \rho \mathbf{u} \cdot \nabla_\Gamma^{\left( d_f, \tilde{d}_{fa} \right)} \mathbf{u} + \nabla_\Gamma^{\left( d_f, \tilde{d}_{fa} \right)} p \right) \cdot \bigg( \rho \mathbf{u} \cdot \nabla_\Gamma^{\left( d_f \right)} \mathbf{u}_a \\
  & + \nabla_\Gamma^{\left( d_f \right)} p_a \bigg) + \tau_{LS\mathbf{u},\Gamma}^{\left( d_f \right)} \left( \rho \mathbf{u} \cdot \nabla_\Gamma^{\left( d_f \right)} \mathbf{u} + \nabla_\Gamma^{\left( d_f \right)} p + \alpha \mathbf{u} \right) \cdot \left( \rho \mathbf{u} \cdot \nabla_\Gamma^{\left( d_f, \tilde{d}_{fa} \right)} \mathbf{u}_a + \nabla_\Gamma^{\left( d_f, \tilde{d}_{fa} \right)} p_a \right) \\
  & + \tau_{LSp,\Gamma}^{\left( d_f, \tilde{d}_{fa} \right)} \left( \rho \mathrm{div}_\Gamma^{\left( d_f \right)} \mathbf{u} \right) \left( \mathrm{div}_\Gamma^{\left( d_f \right)} \mathbf{u}_a \right) + \tau_{LSp,\Gamma}^{\left( d_f \right)} \left( \rho \mathrm{div}_\Gamma^{\left( d_f, \tilde{d}_{fa} \right)} \mathbf{u} \right) \left( \mathrm{div}_\Gamma^{\left( d_f \right)} \mathbf{u}_a \right) + \tau_{LSp,\Gamma}^{\left( d_f \right)} \\
  & \left( \rho \mathrm{div}_\Gamma^{\left( d_f \right)} \mathbf{u} \right) \left( \mathrm{div}_\Gamma^{\left( d_f, \tilde{d}_{fa} \right)} \mathbf{u}_a \right) - \tau_{LST,\Gamma}^{\left( d_f, \tilde{d}_{fa} \right)} \left( \rho C_p \mathbf{u} \cdot \nabla_\Gamma^{\left( d_f \right)} T - Q \right) \left( \rho C_p \mathbf{u} \cdot \nabla_\Gamma^{\left( d_f \right)} T_a \right) \\
  & - \tau_{LST,\Gamma}^{\left( d_f \right)} \left( \rho C_p \mathbf{u} \cdot \nabla_\Gamma^{\left( d_f, \tilde{d}_{fa} \right)} T \right) \left( \rho C_p \mathbf{u} \cdot \nabla_\Gamma^{\left( d_f \right)} T_a \right) - \tau_{LST,\Gamma}^{\left( d_f \right)} \left( \rho C_p \mathbf{u} \cdot \nabla_\Gamma^{\left( d_f \right)} T - Q \right) \\
  & \left( \rho C_p \mathbf{u} \cdot \nabla_\Gamma^{\left( d_f, \tilde{d}_{fa} \right)} T_a \right) \bigg] M^{\left( d_f \right)} + \bigg[ \tau_{LS\mathbf{u},\Gamma}^{\left( d_f \right)} \left( \rho \mathbf{u} \cdot \nabla_\Gamma^{\left( d_f \right)} \mathbf{u} + \nabla_\Gamma^{\left( d_f \right)} p + \alpha \mathbf{u} \right) \\
  & \cdot \left( \rho \mathbf{u} \cdot \nabla_\Gamma^{\left( d_f \right)} \mathbf{u}_a + \nabla_\Gamma^{\left( d_f \right)} p_a \right) + \tau_{LSp,\Gamma}^{\left( d_f \right)} \left( \rho \mathrm{div}_\Gamma^{\left( d_f \right)} \mathbf{u} \right) \left( \mathrm{div}_\Gamma^{\left( d_f \right)} \mathbf{u}_a \right) - \tau_{LST,\Gamma}^{\left( d_f \right)} \\
  & \left( \rho C_p \mathbf{u} \cdot \nabla_\Gamma^{\left( d_f \right)} T - Q \right) \left( \rho C_p \mathbf{u} \cdot \nabla_\Gamma^{\left( d_f \right)} T_a \right) \bigg] M^{\left( d_f, \tilde{d}_{fa} \right)} \,\mathrm{d}\Sigma = 0
\end{split}\right.
\end{equation}
where $\tau_{LS\mathbf{u},\Gamma}^{\left( d_f, \tilde{d}_f \right)}$, $\tau_{LSp,\Gamma}^{\left( d_f, \tilde{d}_f \right)}$ and $\tau_{LST,\Gamma}^{\left( d_f, \tilde{d}_f \right)}$ are the first-order variationals of $\tau_{LS\mathbf{u},\Gamma}^{\left( d_f \right)}$, $\tau_{LSp,\Gamma}^{\left( d_f \right)}$ and $\tau_{LST,\Gamma}^{\left( d_f \right)}$ to $d_f$, respectively, and they are derived as
\begin{equation}\label{equ:TransformedNSSurfaceStabilizationTerm1stVaridfCHM}
\begin{split}
& \left.\begin{split}
 & \tau_{LS\mathbf{u},\Gamma}^{\left( d_f, \tilde{d}_f \right)} = \left\{\begin{split}
   & {h_{E_\Sigma} \over 4 \rho \left\| \mathbf{u} \right\|_2 \left(M^{\left( d_f \right)}\right)^{1\over2} } M^{\left( d_f, \tilde{d}_f \right)}, ~ {h_{E_\Sigma} \over 2 \rho \left\| \mathbf{u} \right\|_2 } \left(M^{\left( d_f \right)}\right)^{1\over2} < { h_{E_\Sigma}^2 \over 12 \eta } M^{\left( d_f \right)} \\
   & { h_{E_\Sigma}^2 \over 12 \eta } M^{\left( d_f, \tilde{d}_f \right)}, ~ {h_{E_\Sigma} \over 2 \rho \left\| \mathbf{u} \right\|_2 } \left(M^{\left( d_f \right)}\right)^{1\over2} \geq { h_{E_\Sigma}^2 \over 12 \eta } M^{\left( d_f \right)}
  \end{split}\right. \\
 & \tau_{LSp,\Gamma}^{\left( d_f, \tilde{d}_f \right)} = \left\{\begin{split}
 & { h_{E_\Sigma} \left\| \mathbf{u} \right\|_2 \over 4 \left(M^{\left( d_f \right)}\right)^{1\over2}} M^{\left( d_f, \tilde{d}_f \right)}, ~ \mathbf{u}^2 < \epsilon_{eps}^{1\over2} \\
 & { h_{E_\Sigma} \over 4 \left(M^{\left( d_f \right)}\right)^{1\over2}} M^{\left( d_f, \tilde{d}_f \right)}, ~ \mathbf{u}^2 \geq \epsilon_{eps}^{1\over2}
 \end{split}\right. \\
 & \tau_{LST,\Gamma}^{\left( d_f, \tilde{d}_f \right)} = \left\{\begin{split} 
  & {h_{E_\Sigma} \over 4 \rho C_p \left\| \mathbf{u} \right\|_2 \left(M^{\left( d_f \right)}\right)^{1\over2} } M^{\left( d_f, \tilde{d}_f \right)}, ~ {h_{E_\Sigma} \over 2 \rho C_p \left\| \mathbf{u} \right\|_2 } \left(M^{\left( d_f \right)}\right)^{1\over2} < { h_{E_\Sigma}^2 \over 12k } M^{\left( d_f \right)} \\
  & { h_{E_\Sigma}^2 \over 12k } M^{\left( d_f, \tilde{d}_f \right)}, ~ {h_{E_\Sigma} \over 2 \rho C_p \left\| \mathbf{u} \right\|_2 } \left(M^{\left( d_f \right)}\right)^{1\over2} \geq { h_{E_\Sigma}^2 \over 12k } M^{\left( d_f \right)}
  \end{split}\right.
 \end{split}\right\} \\
& \forall \tilde{d}_f \in \mathcal{H}\left( \Sigma \right).
\end{split}
\end{equation}

For the constraint of the dissipation power, the adjoint sensitivity of the dissipation power $\Phi$ is derived as
\begin{equation}\label{equ:AdjSensitivityGaDmDissipationConstrCHM}  
\begin{split}
\delta \Phi = - \int_{\Sigma_D} \gamma_{fa} \tilde{\gamma} M^{\left( d_f \right)} \,\mathrm{d}\Sigma - \int_\Sigma A_d d_{fa} \tilde{d}_m \,\mathrm{d}\Sigma,~ \forall \tilde{\gamma} \in \mathcal{L}^2\left(\Sigma_D\right),~ \forall \tilde{d}_m \in \mathcal{L}^2\left(\Sigma\right).
\end{split}
\end{equation}
In Eq. \ref{equ:AdjSensitivityGaDmDissipationConstrCHM}, the adjoint variables $\gamma_{fa}$ and $d_{fa}$ are derived by sequentially solving the variational formulation for the adjoint equations of the surface Navier-Stokes equations
\begin{equation}\label{equ:AdjEquSurfaceNSMHMDissipationPower}
\left\{\begin{split}
  & \mathrm{Find} \left\{\begin{split}
  & \mathbf{u}_a \in\left(\mathcal{H}\left(\Sigma\right)\right)^3~\mathrm{with}~ \mathbf{u}_a = \mathbf{0}~ \mathrm{at} ~ {\forall \mathbf{x} \in l_{v,\Sigma} \cup l_{v_0,\Sigma} } \\
  & p_a \in \mathcal{H}\left(\Sigma\right) \\
  & \lambda_a \in \mathcal{L}^2\left(\Sigma\right)~\mathrm{with}~ \lambda_a = 0~ \mathrm{at} ~ \forall \mathbf{x} \in l_{v,\Sigma} \cup l_{v_0,\Sigma} \\
  \end{split}\right. \\
  & \mathrm{for} \left\{\begin{split}
  & \forall \tilde{\mathbf{u}}_a \in\left(\mathcal{H}\left(\Sigma\right)\right)^3 \\
  & \forall \tilde{p}_a \in \mathcal{H}\left(\Sigma\right) \\
  & \tilde{\lambda}_a \in \mathcal{L}^2\left(\Sigma\right)
  \end{split}\right.,~\mathrm{such~that} \\
  & \int_\Sigma \bigg\{ \eta \left( \nabla_\Gamma^{\left( d_f \right)} \tilde{\mathbf{u}}_a + \nabla_\Gamma^{\left( d_f \right)} \tilde{\mathbf{u}}_a^\mathrm{T} \right) : \left( \nabla_\Gamma^{\left( d_f \right)} \mathbf{u} + \nabla_\Gamma^{\left( d_f \right)} \mathbf{u}^\mathrm{T} \right) + 2 \alpha \mathbf{u} \cdot \tilde{\mathbf{u}}_a + \rho \bigg[ \left( \tilde{\mathbf{u}}_a \cdot \nabla_\Gamma^{\left(d_f\right)} \right) \mathbf{u} \\
  & + \left( \mathbf{u} \cdot \nabla_\Gamma^{\left(d_f\right)} \right) \tilde{\mathbf{u}}_a \bigg] \cdot \mathbf{u}_a + {\eta\over2} \left( \nabla_\Gamma^{\left(d_f\right)} \tilde{\mathbf{u}}_a + \nabla_\Gamma^{\left(d_f\right)} \tilde{\mathbf{u}}_a^\mathrm{T} \right) : \left( \nabla_\Gamma^{\left(d_f\right)} \mathbf{u}_a + \nabla_\Gamma^{\left(d_f\right)} \mathbf{u}_a^\mathrm{T} \right) - \tilde{p}_a \\
  & \mathrm{div}_\Gamma^{\left( d_f \right)} \mathbf{u}_a - p_a \mathrm{div}_\Gamma^{\left( d_f \right)} \tilde{\mathbf{u}}_a + \alpha \tilde{\mathbf{u}}_a \cdot \mathbf{u}_a + \tilde{\lambda}_a \mathbf{u}_a \cdot \mathbf{n}_\Gamma^{\left( d_f \right)} + \lambda_a \tilde{\mathbf{u}}_a \cdot \mathbf{n}_\Gamma^{\left( d_f \right)} \bigg\} M^{\left( d_f \right)} \,\mathrm{d}\Sigma \\
  & - \sum_{E_\Sigma\in\mathcal{E}_\Sigma} \int_{E_\Sigma} \bigg\{ \tau_{LS\mathbf{u},\Gamma}^{\left( d_f, \tilde{\mathbf{u}}_a \right)} \left( \rho \mathbf{u} \cdot \nabla_\Gamma^{\left( d_f \right)} \mathbf{u} + \nabla_\Gamma^{\left( d_f \right)} p + \alpha \mathbf{u} \right) \cdot \left( \rho \mathbf{u} \cdot \nabla_\Gamma^{\left( d_f \right)} \mathbf{u}_a + \nabla_\Gamma^{\left( d_f \right)} p_a \right) \\
  & + \tau_{LS\mathbf{u},\Gamma}^{\left( d_f \right)} \bigg[ \rho \left( \tilde{\mathbf{u}}_a \cdot \nabla_\Gamma^{\left( d_f \right)} \mathbf{u} + \mathbf{u} \cdot \nabla_\Gamma^{\left( d_f \right)} \tilde{\mathbf{u}}_a \right) + \nabla_\Gamma^{\left( d_f \right)} \tilde{p}_a + \alpha \tilde{\mathbf{u}}_a \bigg] \cdot \left( \rho \mathbf{u} \cdot \nabla_\Gamma^{\left( d_f \right)} \mathbf{u}_a + \nabla_\Gamma^{\left( d_f \right)} p_a \right) \\
  & + \tau_{LS\mathbf{u},\Gamma}^{\left( d_f \right)} \left( \rho \mathbf{u} \cdot \nabla_\Gamma^{\left( d_f \right)} \mathbf{u} + \nabla_\Gamma^{\left( d_f \right)} p + \alpha \mathbf{u} \right) \cdot \left( \rho \tilde{\mathbf{u}}_a \cdot \nabla_\Gamma^{\left( d_f \right)} \mathbf{u}_a \right) + \tau_{LSp,\Gamma}^{\left( d_f, \tilde{\mathbf{u}}_a \right)} \left( \rho \mathrm{div}_\Gamma^{\left( d_f \right)} \mathbf{u} \right) \\
  & \left( \mathrm{div}_\Gamma^{\left( d_f \right)} \mathbf{u}_a \right) + \tau_{LSp,\Gamma}^{\left( d_f \right)} \left( \rho \mathrm{div}_\Gamma^{\left( d_f \right)} \tilde{\mathbf{u}}_a \right) \left( \mathrm{div}_\Gamma^{\left( d_f \right)} \mathbf{u}_a \right) \bigg\} M^{\left( d_f \right)} \,\mathrm{d}\Sigma = 0 \\  
\end{split}\right.
\end{equation}
and the variational formulations for the adjoint equations of the surface-PDE filters
\begin{equation}\label{equ:AdjPDEFilterDissipationPowerGaMHM} 
\left\{\begin{split}
  & \mathrm{Find}~\gamma_{fa}\in\mathcal{H}\left(\Sigma_D\right)~\mathrm{for}~\forall \tilde{\gamma}_{fa} \in \mathcal{H}\left(\Sigma_D\right),~\mathrm{such~that} \\
  & \int_{\Sigma_D} \left[ \left( {\partial \alpha \over \partial \gamma_p} {\partial \gamma_p \over \partial \gamma_f} \mathbf{u}^2 + {\partial \alpha \over \partial \gamma_p} {\partial \gamma_p \over \partial \gamma_f} \mathbf{u} \cdot \mathbf{u}_a \right) \tilde{\gamma}_{fa} + r_f^2 \nabla_\Gamma^{\left( d_f \right)} \tilde{\gamma}_{fa} \cdot \nabla_\Gamma^{\left( d_f \right)} \gamma_{fa} + \tilde{\gamma}_{fa} \gamma_{fa} \right] M^{\left( d_f \right)} \,\mathrm{d}\Sigma \\
  & - \sum_{E_\Sigma\in\mathcal{E}_\Sigma} \int_{E_\Sigma\cap\Sigma_D} \tau_{LS\mathbf{u},\Gamma}^{\left( d_f \right)} {\partial \alpha \over \partial \gamma_p} {\partial \gamma_p \over \partial \gamma_f} \mathbf{u} \cdot \left( \rho \mathbf{u} \cdot \nabla_\Gamma^{\left( d_f \right)} \mathbf{u}_a + \nabla_\Gamma^{\left( d_f \right)} p_a \right) \tilde{\gamma}_{fa} M^{\left( d_f \right)} \,\mathrm{d}\Sigma = 0
\end{split}\right.
\end{equation}
and
\begin{equation}\label{equ:AdjPDEFilterJDissipationPowerDmMHM} 
\left\{\begin{split}
  & \mathrm{Find}~d_{fa}\in\mathcal{H}\left(\Sigma\right) ~\mathrm{for}~ \forall \tilde{d}_{fa} \in \mathcal{H}\left(\Sigma\right),~\mathrm{such~that} \\
  & \int_\Sigma \bigg[ \eta \left( \nabla_\Gamma^{\left( d_f, \tilde{d}_{fa} \right)} \mathbf{u} + \nabla_\Gamma^{\left( d_f, \tilde{d}_{fa} \right)} \mathbf{u}^\mathrm{T} \right) : \left( \nabla_\Gamma^{\left( d_f \right)} \mathbf{u} + \nabla_\Gamma^{\left( d_f \right)} \mathbf{u}^\mathrm{T} \right) + \rho \mathbf{u} \cdot \nabla_\Gamma^{\left(d_f, \tilde{d}_{fa}\right)} \mathbf{u} \cdot \mathbf{u}_a \\
  & + {\eta\over2} \left( \nabla_\Gamma^{\left(d_f, \tilde{d}_{fa}\right)} \mathbf{u} + \nabla_\Gamma^{\left(d_f, \tilde{d}_{fa}\right)} \mathbf{u}^\mathrm{T} \right) : \left( \nabla_\Gamma^{\left(d_f\right)} \mathbf{u}_a + \nabla_\Gamma^{\left(d_f\right)} \mathbf{u}_a^\mathrm{T} \right) + {\eta\over2} \left( \nabla_\Gamma^{\left(d_f\right)} \mathbf{u} + \nabla_\Gamma^{\left(d_f\right)} \mathbf{u}^\mathrm{T} \right) \\
  & : \left( \nabla_\Gamma^{\left(d_f, \tilde{d}_{fa}\right)} \mathbf{u}_a + \nabla_\Gamma^{\left(d_f, \tilde{d}_{fa} \right)} \mathbf{u}_a^\mathrm{T} \right) - p \, \mathrm{div}_\Gamma^{\left( d_f, \tilde{d}_{fa} \right)} \mathbf{u}_a - p_a \mathrm{div}_\Gamma^{\left( d_f, \tilde{d}_{fa} \right)} \mathbf{u} + \lambda \mathbf{u}_a \cdot \mathbf{n}_\Gamma^{\left( d_f, \tilde{d}_{fa} \right)} \\
  & + \lambda_a \mathbf{u} \cdot \mathbf{n}_\Gamma^{\left( d_f, \tilde{d}_{fa} \right)} + f_{id,\Gamma} r_f^2 \left( \nabla_\Gamma^{\left( d_f, \tilde{d}_{fa} \right)} \gamma_f \cdot \nabla_\Gamma^{\left( d_f \right)} \gamma_{fa} + \nabla_\Gamma^{\left( d_f \right)} \gamma_f \cdot \nabla_\Gamma^{\left( d_f, \tilde{d}_{fa} \right)} \gamma_{fa} \right) \bigg] M^{\left( d_f \right)} \\
  & + \bigg[ {\eta\over2} \left( \nabla_\Gamma^{\left( d_f \right)} \mathbf{u} + \nabla_\Gamma^{\left( d_f \right)} \mathbf{u}^\mathrm{T} \right) : \left( \nabla_\Gamma^{\left( d_f \right)} \mathbf{u} + \nabla_\Gamma^{\left( d_f \right)} \mathbf{u}^\mathrm{T} \right) + \alpha \mathbf{u}^2 + \rho \left( \mathbf{u} \cdot \nabla_\Gamma^{\left(d_f\right)} \right) \mathbf{u} \cdot \mathbf{u}_a \\
  & + {\eta\over2} \left( \nabla_\Gamma^{\left(d_f\right)} \mathbf{u} + \nabla_\Gamma^{\left(d_f\right)} \mathbf{u}^\mathrm{T} \right) : \left( \nabla_\Gamma^{\left(d_f\right)} \mathbf{u}_a + \nabla_\Gamma^{\left(d_f\right)} \mathbf{u}_a^\mathrm{T} \right) - p \, \mathrm{div}_\Gamma^{\left( d_f \right)} \mathbf{u}_a - p_a \mathrm{div}_\Gamma^{\left( d_f \right)} \mathbf{u} \\
  & + \alpha \mathbf{u} \cdot \mathbf{u}_a + \lambda \mathbf{u}_a \cdot \mathbf{n}_\Gamma^{\left( d_f \right)} + \lambda_a \mathbf{u} \cdot \mathbf{n}_\Gamma^{\left( d_f \right)} + f_{id,\Gamma} \bigg( r_f^2 \nabla_\Gamma^{\left( d_f \right)} \gamma_f \cdot \nabla_\Gamma^{\left( d_f \right)} \gamma_{fa} + \gamma_f \gamma_{fa} \\
  & - \gamma \gamma_{fa} \bigg) \bigg] M^{\left( d_f, \tilde{d}_{fa} \right)} + r_m^2 \nabla_\Sigma d_{fa} \cdot \nabla_\Sigma \tilde{d}_{fa} + d_{fa} \tilde{d}_{fa} \,\mathrm{d}\Sigma \\
  & - \sum_{E_\Sigma\in\mathcal{E}_\Sigma} \int_{E_\Sigma} \bigg[ \tau_{LS\mathbf{u},\Gamma}^{\left( d_f, \tilde{d}_{fa} \right)} \left( \rho \mathbf{u} \cdot \nabla_\Gamma^{\left( d_f \right)} \mathbf{u} + \nabla_\Gamma^{\left( d_f \right)} p + \alpha \mathbf{u} \right) \cdot \left( \rho \mathbf{u} \cdot \nabla_\Gamma^{\left( d_f \right)} \mathbf{u}_a + \nabla_\Gamma^{\left( d_f \right)} p_a \right) \\
  & + \tau_{LS\mathbf{u},\Gamma}^{\left( d_f \right)} \left( \rho \mathbf{u} \cdot \nabla_\Gamma^{\left( d_f, \tilde{d}_{fa} \right)} \mathbf{u} + \nabla_\Gamma^{\left( d_f, \tilde{d}_{fa} \right)} p \right) \cdot \left( \rho \mathbf{u} \cdot \nabla_\Gamma^{\left( d_f \right)} \mathbf{u}_a + \nabla_\Gamma^{\left( d_f \right)} p_a \right) + \tau_{LS\mathbf{u},\Gamma}^{\left( d_f \right)} \\
  & \left( \rho \mathbf{u} \cdot \nabla_\Gamma^{\left( d_f \right)} \mathbf{u} + \nabla_\Gamma^{\left( d_f \right)} p + \alpha \mathbf{u} \right) \cdot \left( \rho \mathbf{u} \cdot \nabla_\Gamma^{\left( d_f, \tilde{d}_{fa} \right)} \mathbf{u}_a + \nabla_\Gamma^{\left( d_f, \tilde{d}_{fa} \right)} p_a \right) + \tau_{LSp,\Gamma}^{\left( d_f, \tilde{d}_{fa} \right)} \\
  & \left( \rho \mathrm{div}_\Gamma^{\left( d_f \right)} \mathbf{u} \right) \left( \mathrm{div}_\Gamma^{\left( d_f \right)} \mathbf{u}_a \right) + \tau_{LSp,\Gamma}^{\left( d_f \right)} \left( \rho \mathrm{div}_\Gamma^{\left( d_f, \tilde{d}_{fa} \right)} \mathbf{u} \right) \left( \mathrm{div}_\Gamma^{\left( d_f \right)} \mathbf{u}_a \right) + \tau_{LSp,\Gamma}^{\left( d_f \right)} \left( \rho \mathrm{div}_\Gamma^{\left( d_f \right)} \mathbf{u} \right) \\
  & \left( \mathrm{div}_\Gamma^{\left( d_f, \tilde{d}_{fa} \right)} \mathbf{u}_a \right) \bigg] M^{\left( d_f \right)} + \bigg[ \tau_{LS\mathbf{u},\Gamma}^{\left( d_f \right)} \left( \rho \mathbf{u} \cdot \nabla_\Gamma^{\left( d_f \right)} \mathbf{u} + \nabla_\Gamma^{\left( d_f \right)} p + \alpha \mathbf{u} \right) \cdot \bigg( \rho \mathbf{u} \cdot \nabla_\Gamma^{\left( d_f \right)} \mathbf{u}_a \\
  & + \nabla_\Gamma^{\left( d_f \right)} p_a \bigg) + \tau_{LSp,\Gamma}^{\left( d_f \right)} \left( \rho \mathrm{div}_\Gamma^{\left( d_f \right)} \mathbf{u} \right) \left( \mathrm{div}_\Gamma^{\left( d_f \right)} \mathbf{u}_a \right) \bigg] M^{\left( d_f, \tilde{d}_{fa} \right)} \,\mathrm{d}\Sigma = 0.
\end{split}\right.
\end{equation}

For the constraint of the area fraction, the adjoint sensitivity of the area fraction $s$ is derived from that of $s\left|\Gamma_D\right|$ and $\left|\Gamma_D\right|$ as
\begin{equation}\label{equ:AdjSensitivityGaDmAreaFracConstrCHM}
  \delta s = \delta \left( {s\left|\Gamma_D\right| \over \left|\Gamma_D\right|} \right) = {1 \over \left|\Gamma_D\right|} \delta \left( s\left|\Gamma_D\right| \right) - {s \over \left|\Gamma_D\right|} \delta \left|\Gamma_D\right|,
\end{equation}
where the adjoint sensitivities $\delta \left( s\left|\Gamma_D\right| \right)$ and $\delta \left|\Gamma_D\right|$ can be derived based on the adjoint analysis of $s^{\left(d_f\right)}\left|\Gamma_D\right|^{\left(d_f\right)} = \int_{\Sigma_D} \gamma_p M^{\left(d_f\right)} \:\mathrm{d}\Sigma$ and $\left|\Gamma_D\right|^{\left(d_f\right)} = \int_{\Sigma_D} M^{\left(d_f\right)} \:\mathrm{d}\Sigma$. The adjoint sensitivity of $s\left|\Gamma_D\right|$ is
\begin{equation}\label{equ:AdjSensitivityGaDmAreaFluidStrucConstrCHM}  
\begin{split}
\delta \left(s\left|\Gamma_D\right|\right) = - \int_{\Sigma_D} \gamma_{fa} \tilde{\gamma} M^{\left( d_f \right)} \,\mathrm{d}\Sigma - \int_\Sigma A_d d_{fa} \tilde{d}_m \,\mathrm{d}\Sigma,~ \forall \tilde{\gamma} \in \mathcal{L}^2\left(\Sigma_D\right),~ \forall \tilde{d}_m \in \mathcal{L}^2\left(\Sigma\right).
\end{split}
\end{equation}
In Eq. \ref{equ:AdjSensitivityGaDmAreaFluidStrucConstrCHM}, the adjoint variables $\gamma_{fa}$ and $d_{fa}$ are derived by sequentially solving the variational formulations for the adjoint equations of the surface-PDE filters:
\begin{equation}\label{AdjPDEFilterAreaFluidStrucGaMHM}
\left\{\begin{split}
  & \mathrm{Find} ~ \gamma_{fa} \mathcal{H}\left(\Sigma_D\right) ~ \mathrm{for}~\forall \tilde{\gamma}_{fa} \mathcal{H}\left(\Sigma_D\right) ,~\mathrm{such~that} \\
  & \int_{\Sigma_D} \left( {\partial \gamma_p \over \partial \gamma_f} \tilde{\gamma}_{fa} + r_f^2 \nabla_\Gamma^{\left( d_f \right)} \gamma_{fa} \cdot \nabla_\Gamma^{\left( d_f \right)} \tilde{\gamma}_{fa} + \gamma_{fa} \tilde{\gamma}_{fa} \right) M^{\left( d_f \right)} \,\mathrm{d}\Sigma = 0 \\  
\end{split}\right.
\end{equation}
and
\begin{equation}\label{equ:AdjPDEFilterAreaFluidStrucDfMHM}  
\left\{\begin{split}
  & \mathrm{Find} ~ d_{fa} \mathcal{H}\left(\Sigma\right) ~ \mathrm{for}~\forall \tilde{d}_{fa} \mathcal{H}\left(\Sigma\right) ,~\mathrm{such~that} \\
  & \int_\Sigma f_{id,\Gamma} r_f^2 \left( \nabla_\Gamma^{\left( d_f, \tilde{d}_{fa} \right)} \gamma_f \cdot \nabla_\Gamma^{\left( d_f \right)} \gamma_{fa} + \nabla_\Gamma^{\left( d_f \right)} \gamma_f \cdot \nabla_\Gamma^{\left( d_f, \tilde{d}_{fa} \right)} \gamma_{fa} \right) M^{\left( d_f \right)} \\
  & + f_{id,\Gamma} \left( \gamma_p + r_f^2 \nabla_\Gamma^{\left( d_f \right)} \gamma_f \cdot \nabla_\Gamma^{\left( d_f \right)} \gamma_{fa} + \gamma_f \gamma_{fa} - \gamma \gamma_{fa} \right) M^{\left( d_f, \tilde{d}_{fa} \right)} \\
  & + r_m^2 \nabla_\Sigma d_{fa} \cdot \nabla_\Sigma \tilde{d}_{fa} + d_{fa} \tilde{d}_{fa} \,\mathrm{d}\Sigma = 0.
\end{split}\right.
\end{equation}
The adjoint sensitivity of $\left|\Gamma_D\right|$ is
\begin{equation}\label{equ:AdjSensitivityGaDmAreaImplitManifConstrCHM}  
\begin{split}
\delta \left|\Gamma_D\right| = \int_\Sigma - A_d d_{fa} \tilde{d}_m \,\mathrm{d}\Sigma,~ \forall \tilde{d}_m \in \mathcal{L}^2\left(\Sigma\right).
\end{split}
\end{equation}
In Eq. \ref{equ:AdjSensitivityGaDmAreaImplitManifConstrCHM}, the adjoint variable $d_{fa}$ is derived by solving the variational formulation for the adjoint equation of the surface-PDE filter for the implicit 2-manifold:
\begin{equation}\label{equ:AdjPDEFilterAreaImplitManifDfMHM} 
\left\{\begin{split}
  & \mathrm{Find} ~ d_{fa} \mathcal{H}\left(\Sigma\right) ~ \mathrm{for}~\forall \tilde{d}_{fa} \mathcal{H}\left(\Sigma\right) ,~\mathrm{such~that} \\
  & \int_\Sigma f_{id,\Gamma} M^{\left( d_f, \tilde{d}_{fa} \right)} + r_m^2 \nabla_\Sigma d_{fa} \cdot \nabla_\Sigma \tilde{d}_{fa} + d_{fa} \tilde{d}_{fa} \,\mathrm{d}\Sigma = 0.
\end{split}\right.
\end{equation}
Then, the adjoint sensitivity of the area fraction in Eq. \ref{equ:AdjSensitivityGaDmAreaFracConstrCHM} can be derived based on Eqs. \ref{equ:AdjSensitivityGaDmAreaFluidStrucConstrCHM} and \ref{equ:AdjSensitivityGaDmAreaImplitManifConstrCHM}.

After the derivation of the adjoint sensitivities in Eqs. \ref{equ:AdjSensitivityCDGaDmObjCHM}, \ref{equ:AdjSensitivityGaDmDissipationConstrCHM} and \ref{equ:AdjSensitivityGaDmAreaFracConstrCHM}, the design variables $\gamma$ and $d_m$ can be evolved iteratively to determine the fiber bundle of the surface structure for heat transfer in the surface flow.

\subsection{Numerical implementation} \label{sec:NumericalImplementationSurfaceNSEqus}

The fiber bundle topology optimization problems in Eqs. \ref{equ:VarProToopSurfaceNSCD} and \ref{equ:VarProToopSurfaceNSCHM} can be solved by using the iterative algorithms described in Tabs. \ref{tab:IterativeProcedureSurfaceFlowMM} and \ref{tab:IterativeProcedureSurfaceFlowHM}. The surface finite element method is utilized to solve the variational formulations of the relevant PDEs and adjoint equations.
On the details for the surface finite element solution, one can refer to Ref. \cite{DziukActaNumerica2013}. Especially, when the surface finite element method is used to solve the surface flow problem on the implicit 2-manifold filled with the porous medium, the Lagrange multiplier method is used to enforce the tangential constraint of the fluid velocity \cite{FriesIJNMF2018,ReutherPOF2018}. To avoid the numerical singularity caused by the null value, the 2-norm of a vector function $\mathbf{f}$ as the factor of the denominator are approximated as $ \left( \mathbf{f}^2 + \epsilon_{eps} \right)^{1/2} $, e.g. the 2-norm of $ \mathbf{n}_\Sigma - \nabla_\Sigma d_f $ in Eq. \ref{equ:UnitaryNormalGammaMHM} and the 2-norm of fluid velocity $ \mathbf{u} $ in Eqs. \ref{equ:TransformedNSSurfaceStabilizationTermCHM} and \ref{equ:TransformedCHMSurfaceStabilizationTermCHM} are numerically approximated as $ \left[ \left(\mathbf{n}_\Sigma - \nabla_\Sigma d_f\right)^2 + \epsilon_{eps} \right]^{1/2}$ and $\left( \mathbf{u}^2 + \epsilon_{eps} \right)^{1/2}$, respectively.

\begin{table}[!htbp]
\centering
\begin{tabular}{l}
  \hline
  \textbf{Algorithm 1}: iterative solution of Eq. \ref{equ:VarProToopSurfaceNSCD} \\
  \hline
  Set $\mathbf{u}_{l_{v,\Sigma}}$, $\rho$, $\eta$, $D$, $c_0$, $\bar{c}$, $A_d$ and $\Delta P_0$;\\
  Set $\left\{
  \begin{array}{l}
    \gamma \leftarrow 1 \\
    d_m \leftarrow 1/2
  \end{array}
  \right.$, $\left\{
  \begin{array}{l}
    r_f = \pi/30 \\
    r_m = \pi/4 \\
  \end{array}
  \right.$, $\left\{
  \begin{array}{l}
    n_{\max} \leftarrow 230 \\
    n_i \leftarrow 1
  \end{array}
  \right.$, $\left\{
  \begin{array}{l}
    n_{upd} \leftarrow 20 \\
    n_{1st} \leftarrow 10
  \end{array}
  \right.$, \\
  ~~~~~$\left\{
  \begin{array}{l}
    \xi \leftarrow 0.5 \\
    \beta \leftarrow 1
  \end{array}
  \right.$, $\left\{
  \begin{array}{l}
    \alpha_f \leftarrow 0 \\
    \alpha_s \leftarrow 10^4 \rho \\
  \end{array}
  \right.$, $q \leftarrow 1\times10^{-2}$; \\
  \hline
  \textbf{for} $n_i = 1:n_{\max}$ \\
          \hspace{1em} Solve $d_f$ from Eq. \ref{equ:PDEFilterzmBaseStructureMHM}, and solve $\gamma_f$ from Eq. \ref{equ:PDEFilterGammaFilberMHM}; \\
          \hspace{1em} Project $\gamma_f$ to derive $\gamma_p$ based on Eq. \ref{equ:ProjectionGammaFilberMHM}; \\
          \hspace{1em} Solve $\mathbf{u}$, $p$ and $\lambda$ from Eq. \ref{equ:TransformedVariationalFormulationSurfaceNSEqusCD}, and evaluate $\Delta P_{n_i}$; \\
          \hspace{1em} Solve $c$ from Eq. \ref{equ:TransformedVariationalFormulationSurfaceCDEqu}, and evaluate $J_{c,n_i} / J_{c,0}$; \\
          \hspace{1em} Solve $c_a$ from Eq. \ref{equ:WeakAdjEquSCDEquMHM}; \\
          \hspace{1em} Solve $\mathbf{u}_a$, $p_a$ and $\lambda_a$ from Eq. \ref{equ:AdjSurfaceNavierStokesEqusJObjectiveMHM}; \\
          \hspace{1em} Solve $\gamma_{fa}$ from Eq. \ref{equ:AdjPDEFilterJObjectiveGaMHM}, and solve $d_{fa}$ from Eq. \ref{equ:AdjPDEFilterJObjectiveDmMHM}; \\
          \hspace{1em} Evaluate $\delta J_{c,n_i}$ from Eq. \ref{equ:AdjSensitivityCDGaDmMHM}; \\
          \hspace{1em} Solve $\mathbf{u}_a$, $p_a$ and $\lambda_a$ from Eq. \ref{equ:AdjEquSurfaceNSMHMPressureDrop}; \\
          \hspace{1em} Solve $\gamma_{fa}$ from Eq. \ref{equ:AdjPDEFilterPressureDropGaMHM}, and solve $d_{fa}$ from Eq. \ref{equ:AdjPDEFilterJObjectiveDmMHM}; \\
          \hspace{1em} Evaluate $\delta \Delta P_{n_i}$ from Eq. \ref{equ:AdjSensitivityGaDmPressureConstrMHM}; \\          
          \hspace{1em} Update $\gamma$ and $d_m$ based on $\delta J_{c,n_i}$ and $C_{P,n_i} \delta \Delta P_{n_i}$ by using MMA; \\
          \hspace{1em} \textbf{if} $\left(n_i == n_{upd} + n_{1st} \right) \vee \left(\left(n_i>n_{upd} + n_{1st}\right)\wedge\left(\mathrm{mod}\left(n_i - n_{upd} - n_{1st},n_{upd}\right)==0\right)\right)$ \\
          \hspace{2em} $\beta \leftarrow 2\beta$; \\
          \hspace{1em} \textbf{end} \textbf{if} \\
          \hspace{1em} \textbf{if} $ \left( n_i==n_{\max} \right) \vee \left\{
          \begin{array}{l}
            \beta == 2^{10} \\
            {1\over5}\sum_{m=0}^4 \left| J_{c,n_i-m} - J_{c,n_i-\left(m+1\right)} \right|\big/J_{c,0} \leq 1\times10^{-3} \\
            \left|\Delta P_{n_i} / \Delta P_0-1\right| \leq 1\times10^{-3}
          \end{array}
          \right.$ \\
          \hspace{2em} break; \\
          \hspace{1em} \textbf{end} \textbf{if} \\
          \hspace{1em} $n_i \leftarrow n_i+1$ \\
  \textbf{end for} \\
  \hline
\end{tabular}
\caption{Pseudocodes used to solve the fiber bundle topology optimization problem for mass transfer in the surface flow. In the iterative solution loop, $n_i$ is the loop-index; $n_{\max}$ is the maximal value of $n_i$; $J_{c,n_i}$, $J_{c,n_i-m}$ and $J_{c,n_i-\left(m+1\right)}$ are the values of $J_c$ in the $\left(n_i\right)$-th, $\left(n_i-m\right)$-th and $\left(n_i-\left(m+1\right)\right)$-th iterations, respectively; $\Delta P_{n_i}$ is the value of $\Delta P$ in the $\left(n_i\right)$-th iteration; $C_{c,n_i}$ is the scaling factor for the adjoint sensitivity of the pressure drop and it is iteratively updated based on Eq. \ref{equ:ScalingFactorsForConstraintsMHM1}; and $\mathrm{mod}$ is the operator used to take the remainder.}\label{tab:IterativeProcedureSurfaceFlowMM}
\end{table}

\begin{table}[!htbp]
\centering
\begin{tabular}{l}
  \hline
  \textbf{Algorithm 2}: iterative solution of Eq. \ref{equ:VarProToopSurfaceNSCHM} \\
  \hline
  Set $\mathbf{u}_{l_{v,\Sigma}}$, $\rho$, $\eta$, $T_0$, $A_d$, $\Phi_0$ and $s_0$;\\
  Set $\left\{
  \begin{array}{l}
    \gamma \leftarrow 1/2 \\
    d_m \leftarrow 1/2
  \end{array}
  \right.$, $\left\{
  \begin{array}{l}
    r_f = \pi/30 \\
    r_m = \pi/4 \\
  \end{array}
  \right.$, $\left\{
  \begin{array}{l}
    n_i \leftarrow 1 \\
    n_{2^{10}} \leftarrow 0
  \end{array}
  \right.$, $\left\{
  \begin{array}{l}
    n_{upd} \leftarrow 20 \\
    n_{1st} \leftarrow 10
  \end{array}
  \right.$, $\left\{
  \begin{array}{l}
    \xi \leftarrow 0.5 \\
    \beta \leftarrow 1 \\
  \end{array}
  \right.$, $\left\{
  \begin{array}{l}
    \beta' \leftarrow 0 \\
    \gamma'_p \leftarrow 0 \\
  \end{array}
  \right.$, \\
  ~~~~ $\left\{
  \begin{array}{l}
    \alpha_f \leftarrow 0 \\
    \alpha_s \leftarrow 10^4 \rho \\
  \end{array}
  \right.$, $\left\{
  \begin{array}{l}
    C_{pf} \leftarrow 1\times10^{0} \\
    C_{ps} \leftarrow 1\times10^{-1} \\
  \end{array}
  \right.$, $\left\{
  \begin{array}{l}
    k_f \leftarrow 1\times10^{-1} \\
    k_s \leftarrow 1\times10^{0} \\
  \end{array}
  \right.$, $q \leftarrow 1\times10^{-2}$, $\epsilon_{\gamma_p} \leftarrow 5\times10^{-2}$; \\
  \hline
  \textbf{while} $\beta \leq 2^{10}$ \\
          \hspace{1em} Solve $d_f$ from Eq. \ref{equ:PDEFilterzmBaseStructureMHM}, and solve $\gamma_f$ from Eq. \ref{equ:PDEFilterGammaFilberMHM}; \\
          \hspace{1em} \textbf{if} $ \left(n_i \geq n_{upd} + n_{1st}\right)\wedge\left(\mathrm{mod}\left(n_i-n_{upd} - n_{1st},n_{upd}\right)==1\right) $ \\
          \hspace{2em} Compute $\gamma_p$ from $\gamma_f$ based on Eq. \ref{equ:ProjectionGammaFilberMHM}; \\
          \hspace{2em} \textbf{if} $\beta < 2^5$ \\
          \hspace{3em} $n_{2^5} \leftarrow 0$; $\beta' \leftarrow 2\beta$; \\
          \hspace{3em} Compute $\gamma'_p$ from $\gamma_f$ based on Eq. \ref{equ:ProjectionGammaFilberMHM} with $\gamma_p$ and $\beta$ replaced to be $\gamma'_p$ and $\beta'$; \\
          \hspace{3em} \textbf{while} $\left\| \gamma'_p - \gamma_p \right\|_\infty \geq \epsilon_{\gamma_p}$ \\
          \hspace{4em} $\beta' \leftarrow \left( \beta' + \beta \right) / 2 $; \\
          \hspace{4em} Compute $\gamma'_p$ from $\gamma_f$ based on Eq. \ref{equ:ProjectionGammaFilberMHM} with $\gamma_p$ and $\beta$ replaced to be $\gamma'_p$ and $\beta'$; \\
          \hspace{3em} \textbf{end while} \\
          \hspace{3em} $\beta \leftarrow \beta'$; \\
          \hspace{2em} \textbf{else} \\
          \hspace{3em} \textbf{if} $ n_{2^5} == 1 $ \\
          \hspace{4em} $\beta \leftarrow 2\beta$; \\
          \hspace{3em} \textbf{elseif} $ n_{2^5} == 0 $\\
          \hspace{4em} $\beta \leftarrow 2^5$; $n_{2^5} \leftarrow 1$; \\
          \hspace{3em} \textbf{end if} \\
          \hspace{2em} \textbf{end if} \\
          \hspace{2em} \textbf{if} $\beta == 2^{10}$ \\
          \hspace{3em} $n_{2^{10}} \leftarrow n_{2^{10}} + 1$; \\
          \hspace{2em} \textbf{end} \textbf{if} \\
          \hspace{2em} \textbf{if} $\big( \left(\beta == 2^{10}\right) \wedge \left({1\over5}\sum_{m=0}^4 \left| J_{T,n_i-m} - J_{T,n_i-\left(m+1\right)} \right|\big/J_{T,0} \leq 1\times10^{-3} \right)$ \\
          \hspace{3em} $\wedge \left( \left|\Phi_{n_i}/\Phi_0-1\right| \leq 1\times10^{-3} \right) \wedge \left( \left|s_{n_i}/s_0-1\right| \leq 1\times10^{-3} \right) \big) \vee \left( n_{2^{10}} == n_{upd} \right) $ \\
          \hspace{3em} break; \\
          \hspace{2em} \textbf{end} \textbf{if} \\
          \hspace{1em} \textbf{end} \textbf{if} \\
          \hspace{1em} Project $\gamma_f$ to derive $\gamma_p$ based on Eq. \ref{equ:ProjectionGammaFilberMHM}; \\
          \hspace{1em} Solve $\mathbf{u}$, $p$ and $\lambda$ from Eq. \ref{equ:TransformedVariationalFormulationSurfaceNSEqusHM}, and evaluate $\Phi_{n_i}$; \\
          \hspace{1em} Solve $T$ from Eq. \ref{equ:TransformedVariationalFormulationSurfaceCHMEqu}, and evaluate $J_{T,n_i} / J_{T,0}$; \\
          \hspace{1em} Solve $T_a$ from Eq. \ref{equ:WeakAdjEquSCHMEquMHMObj}, and solve $\mathbf{u}_a$, $p_a$ and $\lambda_a$ from Eq. \ref{equ:AdjSurfaceNavierStokesEqusJObjectiveCHMObj}; \\
          \hspace{1em} Solve $\gamma_{fa}$ from Eq. \ref{equ:AdjPDEFilterJObjectiveGaCHMObj}, and solve $d_{fa}$ from Eq. \ref{equ:AdjPDEFilterJObjectiveDmCHMObj}; \\
          \hspace{1em} Evaluate $\delta J_{T,n_i}$ from Eq. \ref{equ:AdjSensitivityCDGaDmObjCHM}; \\
          \hspace{1em} Solve $\mathbf{u}_a$, $p_a$ and $\lambda_a$ from Eq. \ref{equ:AdjEquSurfaceNSMHMDissipationPower}; \\
          \hspace{1em} Solve $\gamma_{fa}$ from Eq. \ref{equ:AdjPDEFilterDissipationPowerGaMHM}, and solve $d_{fa}$ from Eq. \ref{equ:AdjPDEFilterJDissipationPowerDmMHM}; \\
          \hspace{1em} Evaluate $\delta \Phi_{n_i}$ from Eq. \ref{equ:AdjSensitivityGaDmDissipationConstrCHM}; \\
          \hspace{1em} Solve $\gamma_{fa}$ from Eq. \ref{AdjPDEFilterAreaFluidStrucGaMHM}, and solve $d_{fa}$ from Eq. \ref{equ:AdjPDEFilterAreaFluidStrucDfMHM}; \\
          \hspace{1em} Evaluate $\delta \left(s\left|\Gamma\right|\right)_{n_i}$ from Eq. \ref{equ:AdjSensitivityGaDmAreaFluidStrucConstrCHM}; \\
          \hspace{1em} Solve $d_{fa}$ from Eq. \ref{equ:AdjPDEFilterAreaImplitManifDfMHM}; \\
          \hspace{1em} Evaluate $\delta \left|\Gamma\right|_{n_i}$ from Eq. \ref{equ:AdjSensitivityGaDmAreaImplitManifConstrCHM}; \\ 
          \hspace{1em} Evaluate $\delta s_{n_i}$ from $\delta \left(s\left|\Gamma\right|\right)_{n_i}$ and $\delta \left|\Gamma\right|_{n_i}$ from Eq. \ref{equ:AdjSensitivityGaDmAreaFracConstrCHM}; \\          
          \hspace{1em} Update $\gamma$ and $d_m$ based on $\delta J_{T,n_i}$, $C_{\Phi,n_i}\delta \Phi_{n_i}$ and $C_{s,n_i}\delta s_{n_i}$ by using MMA; \\
          \hspace{1em} $n_i \leftarrow n_i+1$; \\
  \textbf{end while} \\
  \hline
\end{tabular}
\caption{Pseudocodes used to solve the fiber bundle topology optimization problem for heat transfer in the surface flow. In the iterative solution loop, $\left\|\cdot\right\|_\infty$ represents the $\infty$-norm of a function and its discrete counterpart is the maximal component of the absolute of the discretized function; $J_{T,n_i}$, $J_{T,n_i-m}$ and $J_{T,n_i-\left(m+1\right)}$ are the values of $J_T$ in the $\left(n_i\right)$-th, $\left(n_i-m\right)$-th and $\left(n_i-\left(m+1\right)\right)$-th iterations, respectively; $\Phi_{n_i}$ and $s_{n_i}$ are the values of $\Phi$ and $s$ in the $\left(n_i\right)$-th iteration; $C_{T,n_i}$ is the scaling factor for the adjoint sensitivities of the dissipation power and area fraction, and they are iteratively updated based on Eq. \ref{equ:ScalingFactorsForConstraintsMHM11}.}\label{tab:IterativeProcedureSurfaceFlowHM}
\end{table}

In the algorithm for the iterative solution of the fiber bundle topology optimization problem in Eq. \ref{equ:VarProToopSurfaceNSCD}, the projection parameter $\beta$ with the initial value of $1$ is doubled after the beginning $30$ iterations and then doubled after every $20$ iterations; the loops of the algorithms are stopped when the maximal iteration numbers are reached, or if $\beta$ reaching $2^{10}$, the averaged variations of the design objectives in continuous 5 iterations and the residuals of the constraints are simultaneously satisfied. In the algorithm for the iterative solution of the fiber bundle topology optimization problem in Eq. \ref{equ:VarProToopSurfaceNSCHM}, the doubling operation of the projection parameter $\beta$ before it reaching $2^5$ can cause significant oscillations of the values of the dissipation power. Therefore, the constraint of the dissipation power can not be well satisfied in the iterative procedure. A bisection method is used to update $\beta$ and control the amplitude variations of the material density, when $\beta$ is less than $2^5$. The doubling operation is enabled again, when $\beta$ reaches $2^5$. In the algorithms in Tabs. \ref{tab:IterativeProcedureSurfaceFlowMM} and \ref{tab:IterativeProcedureSurfaceFlowHM}, the design variables are updated by using the method of moving asymptotes (MMA) \cite{SvanbergIntJNumerMethodsEng1987}.

In the algorithms, the constraints in the $\left(n_i\right)$-th iteration are equivalently set as
\begin{equation}\label{equ:EquivalentOperationOfConstraintsMHT1}
C_{P,n_i} \left|\Delta P_{n_i} / \Delta P_0-1\right| \leq 1\times10^{-3} C_{P,n_i} 
\end{equation}
and
\begin{equation}\label{equ:EquivalentOperationOfConstraintsMHT11}
\left\{\begin{split}
& C_{\Phi,n_i} \left|\Phi_{n_i} / \Phi_0-1\right| \leq 1\times10^{-3} C_{\Phi,n_i} \\
& C_{s,n_i} \left|s_{n_i} \left/\: s_0 \right. - 1 \right| \leq 1\times10^{-3} C_{s,n_i}
\end{split}\right.
\end{equation}
to scale the adjoint sensitivities of the constraints as
\begin{equation}\label{equ:AdjSensiesOfEquivalentConstraintsMHT1}
C_{P,n_i} \delta \Delta P_{n_i}
\end{equation}
and
\begin{equation}\label{equ:AdjSensiesOfEquivalentConstraintsMHT11}
\left\{\begin{split}
& C_{\Phi,n_i} \delta \Phi_{n_i} \\
& C_{s,n_i} \delta s_{n_i}
\end{split}\right.,
\end{equation}
where $C_{P,n_i}$, $C_{\Phi,n_i}$ and $C_{s,n_i}$ are the scaling factors. The equivalent operations in Eqs. \ref{equ:EquivalentOperationOfConstraintsMHT1} and \ref{equ:EquivalentOperationOfConstraintsMHT11} can ensure the robust satisfication of the constraints in the iterative procedures, by keeping the adjoint sensitivities of the constraints possess the same magnitude as that of the design objectives. The scaling factors in the $\left( n_i \right)$-th iteration are set as
\begin{equation}\label{equ:ScalingFactorsForConstraintsMHM1}
C_{P,n_i} = {\Delta P_0 \over J_{c,0}} \left\|{\Delta J_{c,n_i} \over \Delta \boldsymbol{\Upsilon}}\right\|_2 \left/ \, \left\|{\Delta \Delta P_{n_i} \over \Delta \boldsymbol{\Upsilon}}\right\|_2 \right. 
\end{equation}
and
\begin{equation}\label{equ:ScalingFactorsForConstraintsMHM11}
\left\{\begin{split}
& C_{\Phi,n_i} = {\Phi_0 \over J_{T,0}} \left\|{\Delta J_{T,n_i} \over \Delta \boldsymbol{\Upsilon}}\right\|_2 \left/ \, \left\|{\Delta \Phi_{n_i} \over \Delta \boldsymbol{\Upsilon}}\right\|_2 \right. \\
& C_{s,n_i} = {s_0 \over J_{T,0}} \left\|{\Delta J_{T,n_i} \over \Delta \boldsymbol{\Upsilon}}\right\|_2 \left/ \, \left\|{\Delta s_{n_i} \over \Delta \boldsymbol{\Upsilon}}\right\|_2 \right.
\end{split}\right.,
\end{equation}
where $$\left\{{\Delta J_{c,n_i} \left/ \, \Delta \boldsymbol{\Upsilon}\right.}, {\Delta \Delta P_{n_i} \left/ \, \Delta \boldsymbol{\Upsilon}\right.}\right\},$$ $$\left\{{\Delta J_{T,n_i} \left/ \, \Delta \boldsymbol{\Upsilon}\right.}, {\Delta \Phi_{n_i} \left/ \, \Delta \boldsymbol{\Upsilon}\right.}\right\},$$ and $$\left\{{\Delta J_{T,n_i} \left/ \, \Delta \boldsymbol{\Upsilon}\right.}, {\Delta s_{n_i} \left/ \, \Delta \boldsymbol{\Upsilon}\right.}\right\}$$ are the discretized counterparts of $$\left\{{\delta J_{c,n_i} \left/ \, \delta \gamma\right.}, {\delta \Delta P_{n_i} \left/ \, \delta \gamma\right.}\right\},$$ $$\left\{{\delta J_{T,n_i} \left/ \, \delta \gamma\right.}, {\delta \Phi_{n_i} \left/ \, \delta \gamma\right.}\right\},$$ and $$\left\{{\delta J_{T,n_i} \left/ \, \delta \gamma\right.}, {\delta s_{n_i} \left/ \, \delta \gamma\right.}\right\}.$$

\begin{figure}[!htbp]
  \centering
  \includegraphics[width=0.6\textwidth]{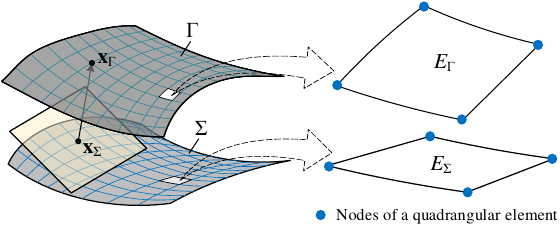}
  \caption{Sketch for the meshes of the quadrangular-element based discretization of the base manifold $\Sigma$ and the mapped meshes on the implicit 2-manifold $\Gamma$.}\label{fig:ElementNodesMHM}
\end{figure}

Linear quadrangular elements are used to interpolate the design variable of the surface structure and that of the implicit 2-manifold, and solve the variational formulations of the governing equations, the surface-PDE filters and the adjoint equations. The meshes of the quadrangular-element based discretization of the base manifold have been sketched in Fig. \ref{fig:ElementNodesMHM}, including the mapped meshes on the implicit 2-manifold.

\subsection{Results and discussion} \label{sec:NumericalExamplesMatchinOptSurfaceFlow}

\begin{figure}[!htbp]
  \centering
  \subfigure[]
  {\includegraphics[width=0.32\textwidth]{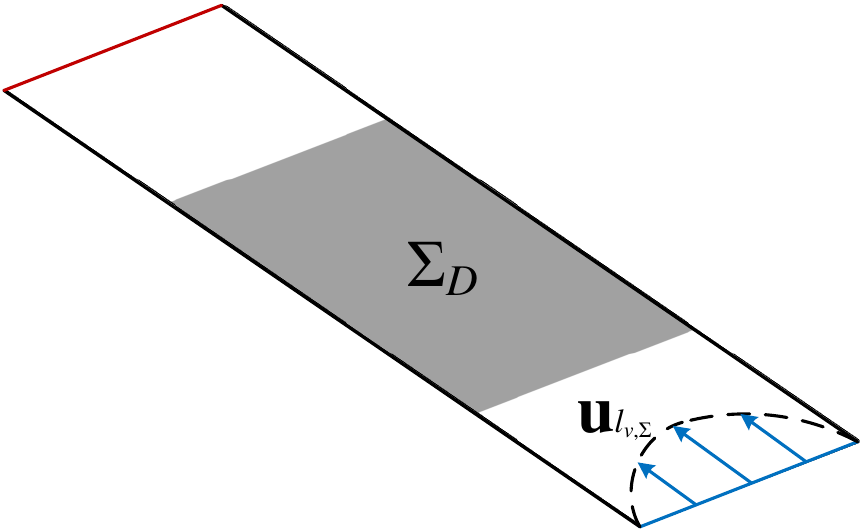}}
  \subfigure[]
  {\includegraphics[width=0.32\textwidth]{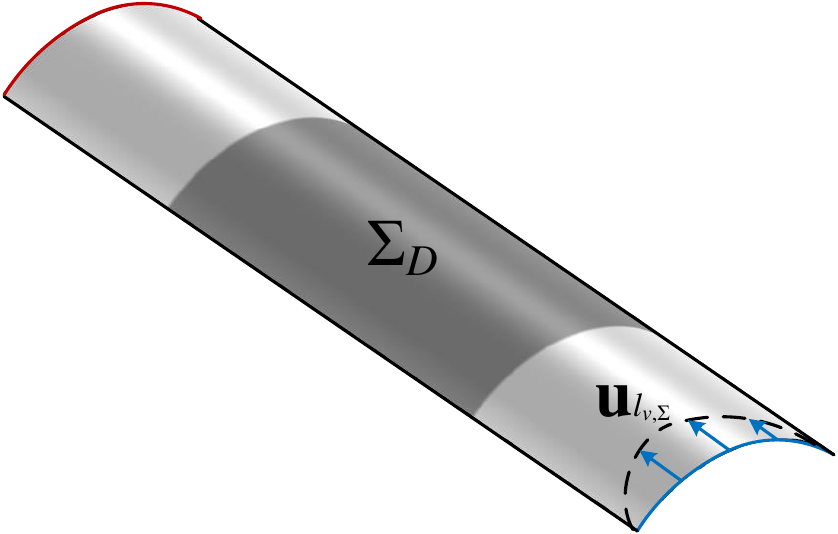}}
  \subfigure[]
  {\includegraphics[width=0.32\textwidth]{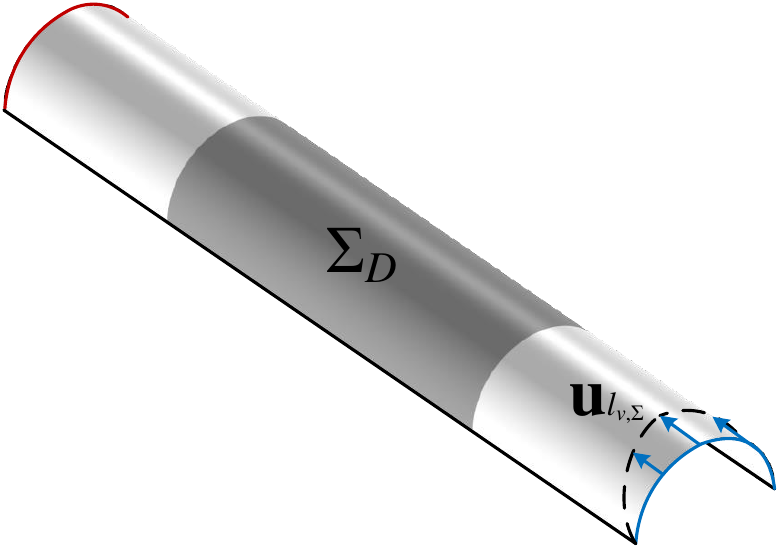}}
  \subfigure[]
  {\includegraphics[width=0.32\textwidth]{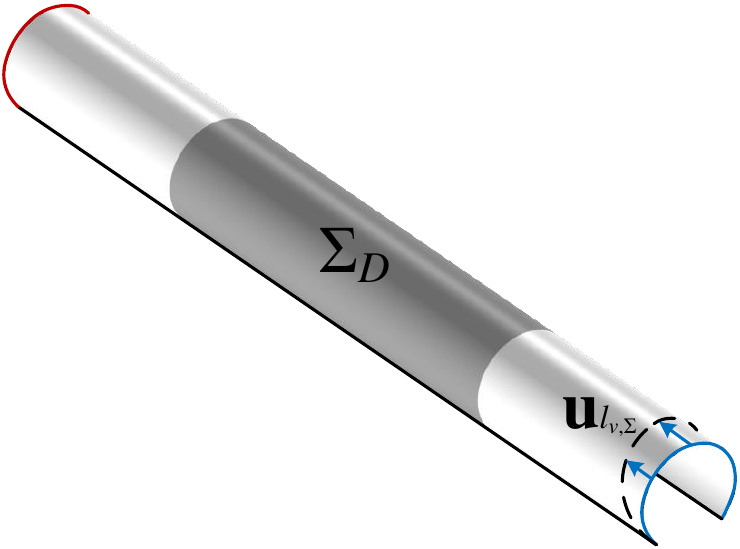}}
  \subfigure[]
  {\includegraphics[width=0.32\textwidth]{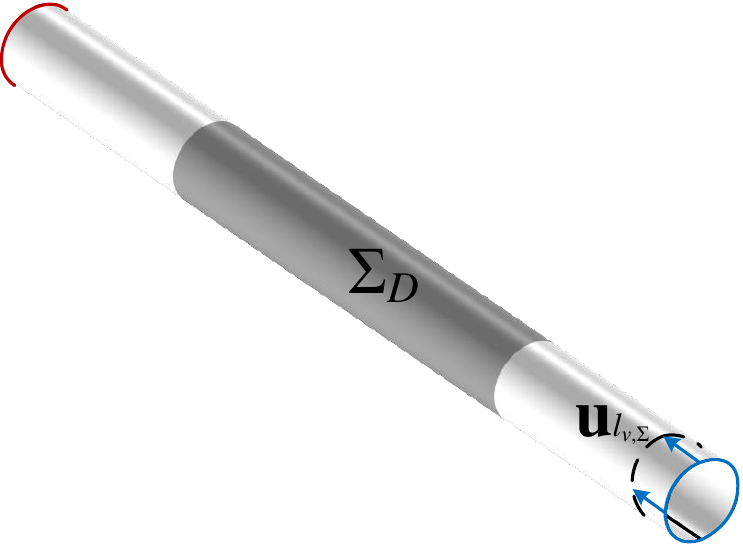}}
  \includegraphics[width=0.5\textwidth]{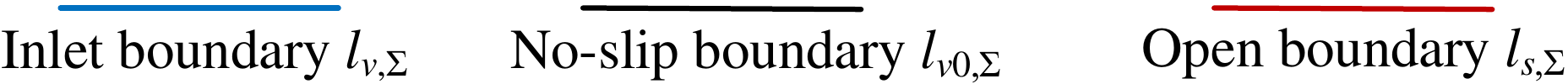}
  \caption{Design domains of fiber bundle topology optimization for mass and heat transfer in the surface flow.}\label{fig:MassHeatTransferManifoldsDesignDomain}
\end{figure}

For fiber bundle topology optimization for mass and heat transfer in the surface flow, the design domains are shown in Fig. \ref{fig:MassHeatTransferManifoldsDesignDomain}, including a series of curved surfaces obtained by deforming a flat surface into cylindrical surfaces, where the lengths of the inlet, outlet and wall boundaries are remained to be unchanged during the deformation. The flat surface is a rectangle with the lengths of the inlet and outlet equal to $\pi$, the length of computational domain equal to $4\pi$ and the length of design domain equal to $2\pi$. In the numerical computation, the design domains are discretized by using the curved quadrilateral element with the side length of $\pi/60$. The filter radii of the design variables for the implicit 2-manifolds and patterns of the surface flow are set to be $\pi/30$ and $\pi/4$, respectively.

\subsubsection{Mass transfer in surface flow} \label{sec:ResultsDiscussionMassTransferSurface}

\begin{figure}[!htbp]
  \centering
  \subfigure[]
  {\includegraphics[width=1\textwidth]{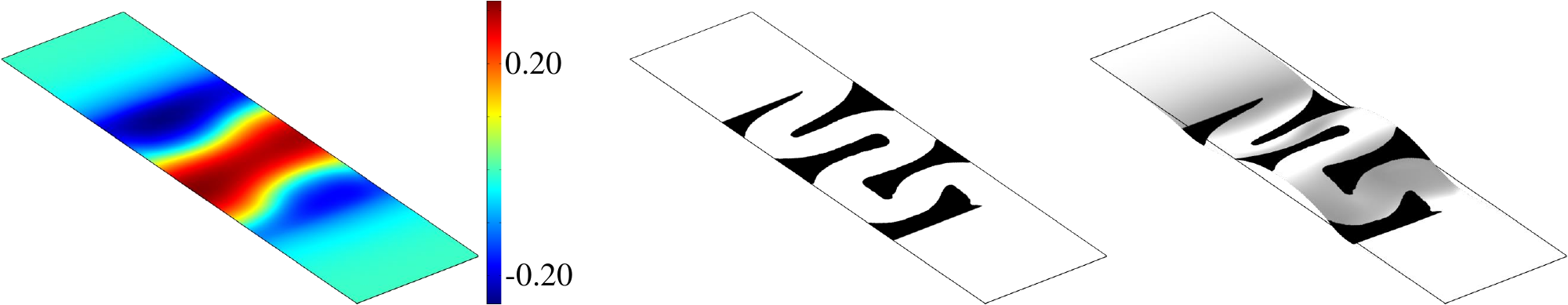}}
  \subfigure[]
  {\includegraphics[width=1\textwidth]{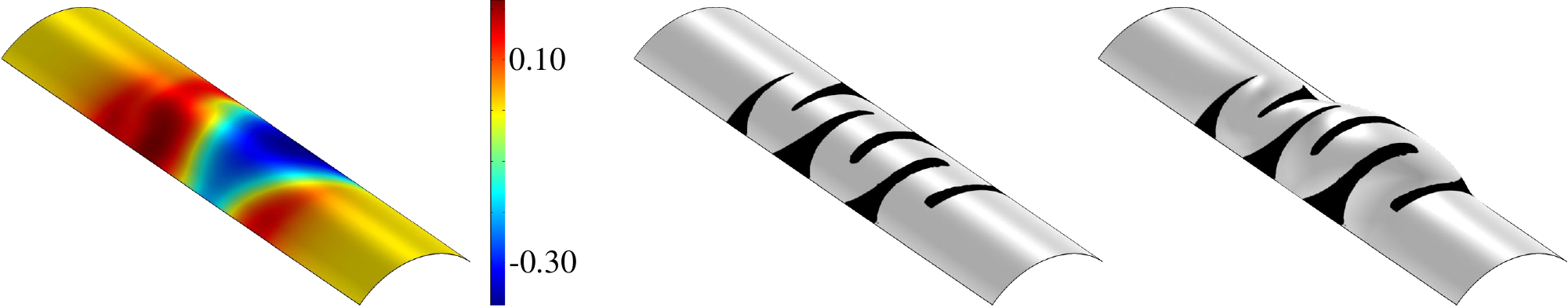}}
  \subfigure[]
  {\includegraphics[width=1\textwidth]{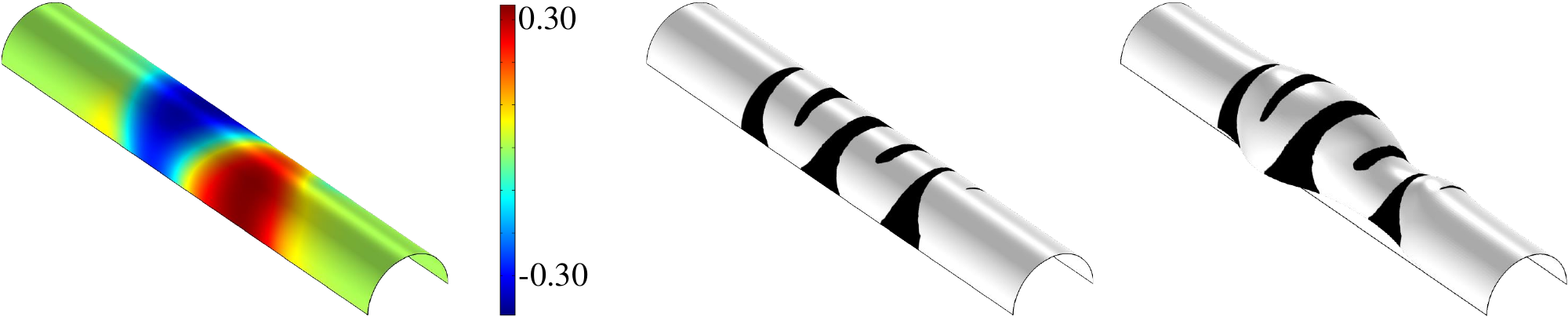}}
  \subfigure[]
  {\includegraphics[width=1\textwidth]{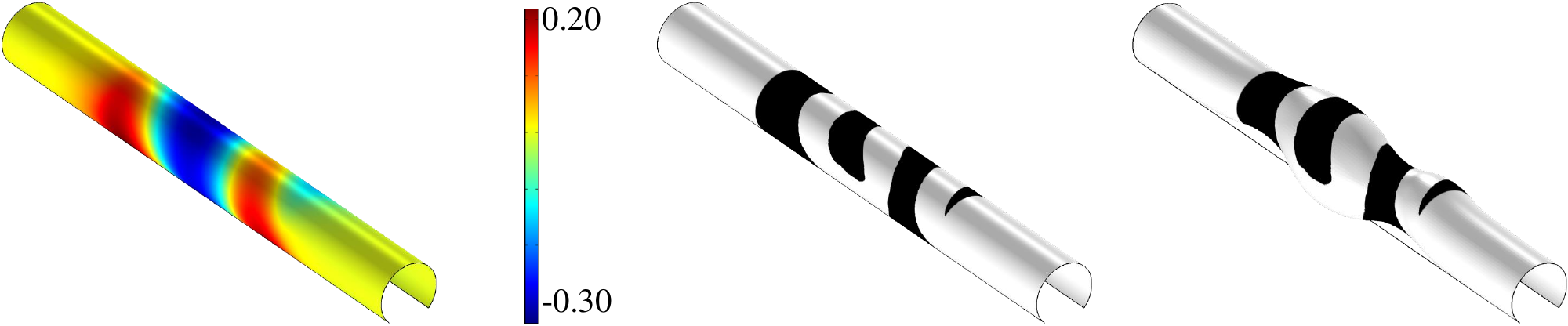}}
  \subfigure[]
  {\includegraphics[width=1\textwidth]{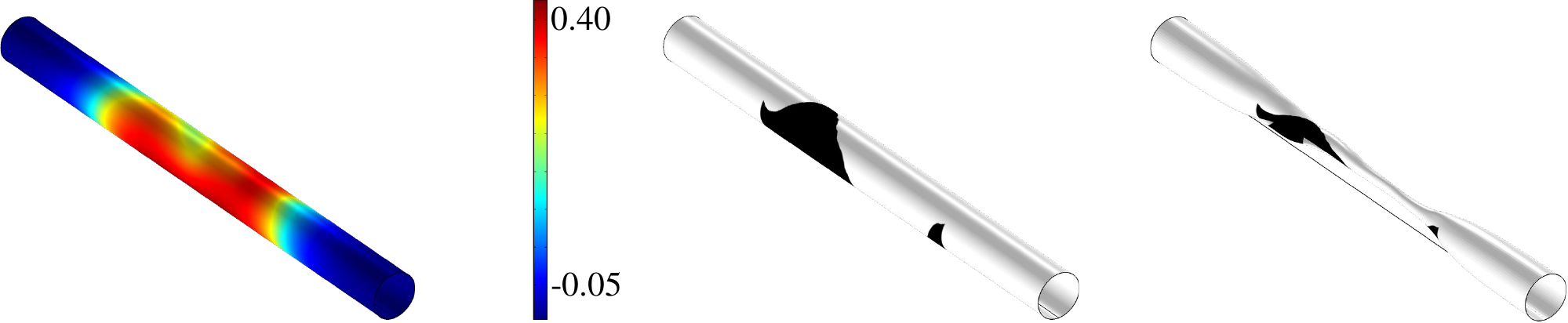}}
  \caption{Distribution of the normal displacement for the implicit 2-manifolds and the material density for the surface patterns obtained for mass transfer in the surface flow on the design domains sketched in Fig. \ref{fig:MassHeatTransferManifoldsDesignDomain}, including the fiber bundles composed of the base 2-manifold and the surface patterns defined on the implicit 2-manifolds.}\label{fig:MassTransferManifoldsAd=10e-1}
\end{figure}

\begin{figure}[!htbp]
  \centering
  \includegraphics[width=0.7\textwidth]{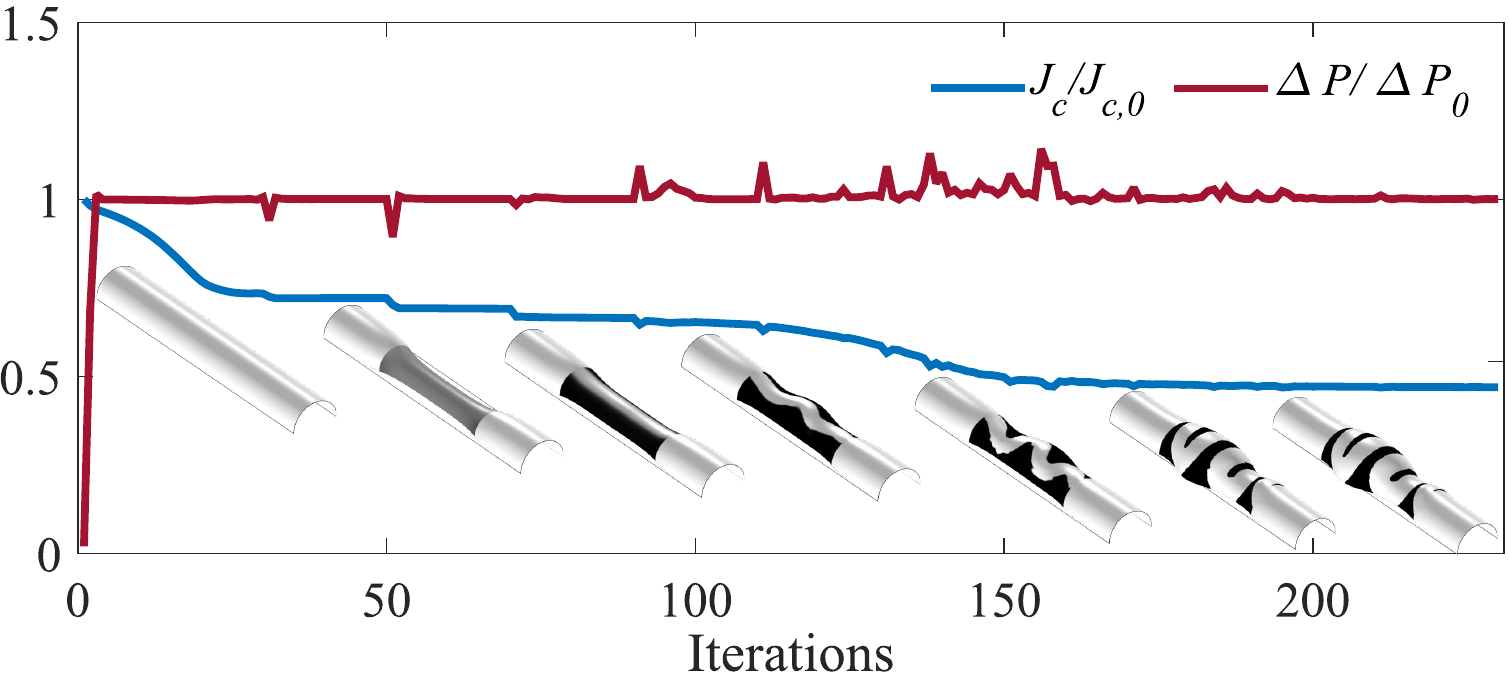}
  \caption{Convergent histories of the design objective and constraint of the pressure drop for mass transfer in the surface flow on the design domain sketched in Fig. \ref{fig:MassHeatTransferManifoldsDesignDomain}c, including the snapshots for the evolution of the fiber bundle during the iterative solution of the optimization problem.}\label{fig:MassTransferManifold3ConvergentHistories}
\end{figure}

\begin{figure}[!htbp]
  \centering
  \subfigure[Velocity]
  {\includegraphics[width=0.32\textwidth]{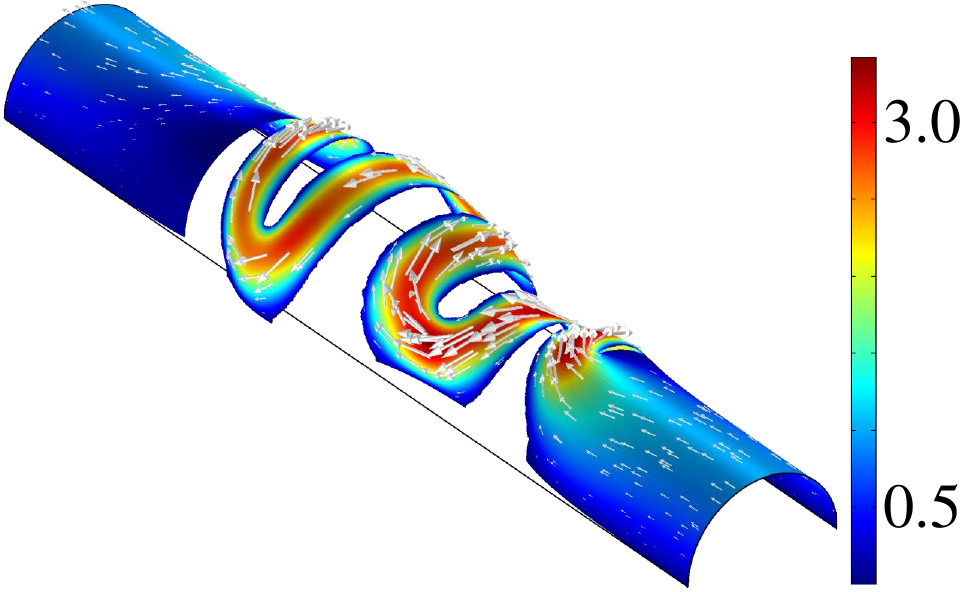}}
  \subfigure[Pressure]
  {\includegraphics[width=0.32\textwidth]{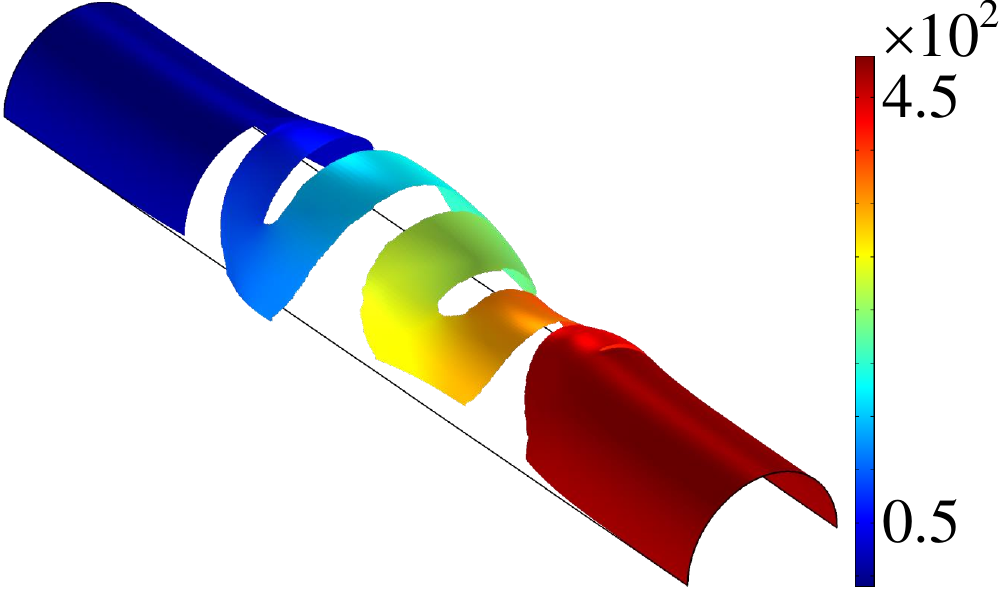}}
  \subfigure[Concentration]
  {\includegraphics[width=0.32\textwidth]{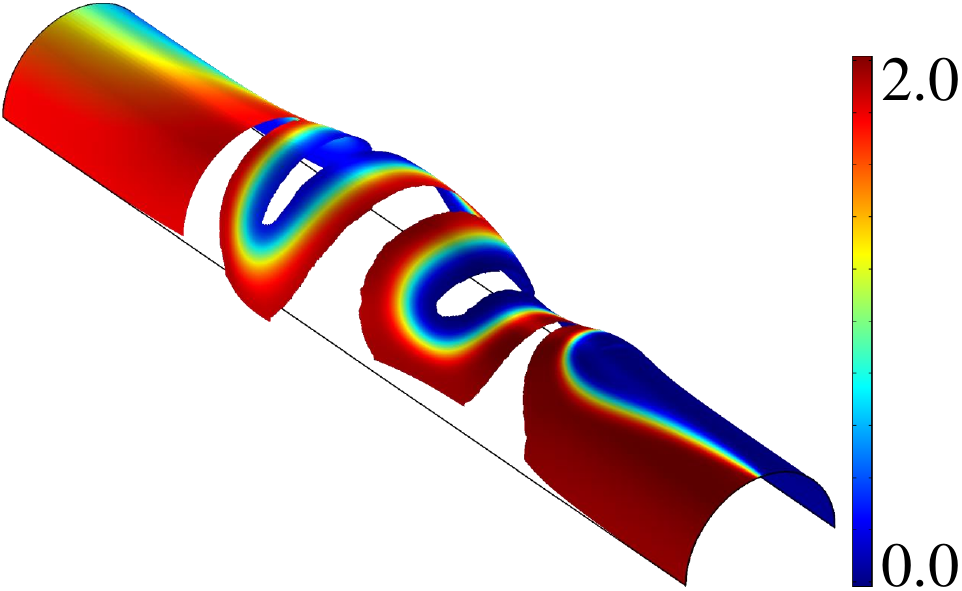}}
  \caption{Distribution of the velocity, pressure and concentration on the fiber bundle obtained for mass transfer in the surface flow on the design domain sketched in Fig. \ref{fig:MassHeatTransferManifoldsDesignDomain}c.}\label{fig:MassTransferManifold3VelocityPressureConcentration}
\end{figure}

In fiber bundle topology optimization for mass transfer in the surface flow, the fluid density and viscosity are set as the unitary, the diffusion coefficient is set as $D = 5\times10^{-3}$, the optimization parameters in the material interpolation are set as $\alpha_s = 1\times10^4$ and $q=1\times10^{-2}$, the pressure drop is set as $\Delta P_0 = 1.5\times10^3$, the inlet velocity is set as the parabolic distribution with the maximal value of $1$, and the concentration distribution at the inlet is set by using the step function with the mid-value at the central point of the inlet boundary and the maximal and minimal values of $2$ and $0$. Correspondingly, the anticipated distribution of the concentration at the outlet is $\bar{c}=1$. By setting the variable magnitude of the implicit 2-manifold as $1.0$, the fiber bundle topology optimization problem in Eq. \ref{equ:VarProToopSurfaceNSCD} is solved on the design domains shown in Fig. \ref{fig:MassHeatTransferManifoldsDesignDomain}. The distribution of the filtered design variables for the implicit 2-manifolds and the material density for the patterns are obtained as shown in Fig. \ref{fig:MassTransferManifoldsAd=10e-1}, where the fiber bundles composed of the implicit 2-manifolds and the surface patterns are included. Convergent histories of the design objective and constraint of the pressure drop are plotted in Fig. \ref{fig:MassTransferManifold3ConvergentHistories} for fiber bundle topology optimization on the design domain in Fig. \ref{fig:MassHeatTransferManifoldsDesignDomain}c, including snapshots for the evolution of the fiber bundle during the iterative solution of the optimization problem. From the monotonicity of the objective values and satisfication of the constraint of the pressure drop, the robustness of the iterative solution can be confirmed for fiber bundle topology optimization for mass transfer in the surface flow. The distribution of the velocity, pressure and concentration are provided in Fig. \ref{fig:MassTransferManifold3VelocityPressureConcentration} for fiber bundle topology optimization on the design domain in Fig. \ref{fig:MassHeatTransferManifoldsDesignDomain}c, where the zig-zag shaped curved-channel is obtained for the surface flow to enhance the convection and mixing length and hence the mass-transfer efficiency is improved.

\begin{figure}[!htbp]
  \centering
  \includegraphics[width=0.7\textwidth]{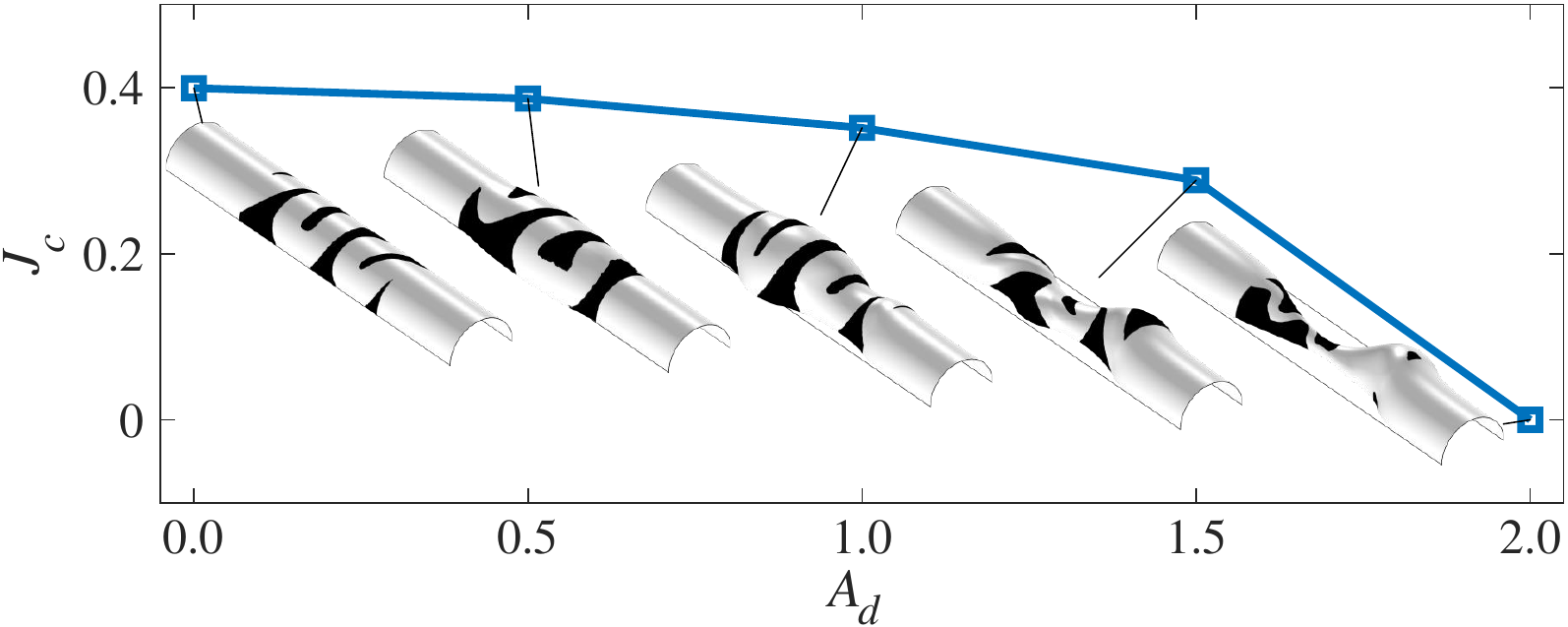}
  \caption{Objective values for the variable magnitudes of $A_d = \left\{0.0,0.5,1.0,1.5,2.0\right\}$ and the obtained fiber bundles for mass transfer in the surface flow.}\label{fig:MassTransferManifold3AdPlot}
\end{figure}

The variable magnitude of the implicit 2-manifold can determine the design space of the fiber bundle topology optimization problem. Therefore, the fiber bundle topology optimization problem is solved for the variable magnitudes of $A_d = \left\{0.0,0.5,1.0,1.5,2.0\right\}$, respectively, where the other parameters are remained to be unchanged. The values of the design objective are obtained and plotted in Fig. \ref{fig:MassTransferManifold3AdPlot} including the obtained fiber bundles. Fig. \ref{fig:MassTransferManifold3AdPlot} shows that the objective values decrease along with increasing the variable magnitude. This is because that larger variable magnitude permits more flexible evolution of the fiber bundle to improve the mixing performance of the surface structure. Especially, the objective value is less than $0.05$ when the variable magnitude is $A_d=2.0$. This means that the complete mixing is achieved by the surface structure corresponding to the fiber bundle with the variable magnitude of $A_d=2.0$.

\begin{figure}[!htbp]
  \centering
  \includegraphics[width=0.7\textwidth]{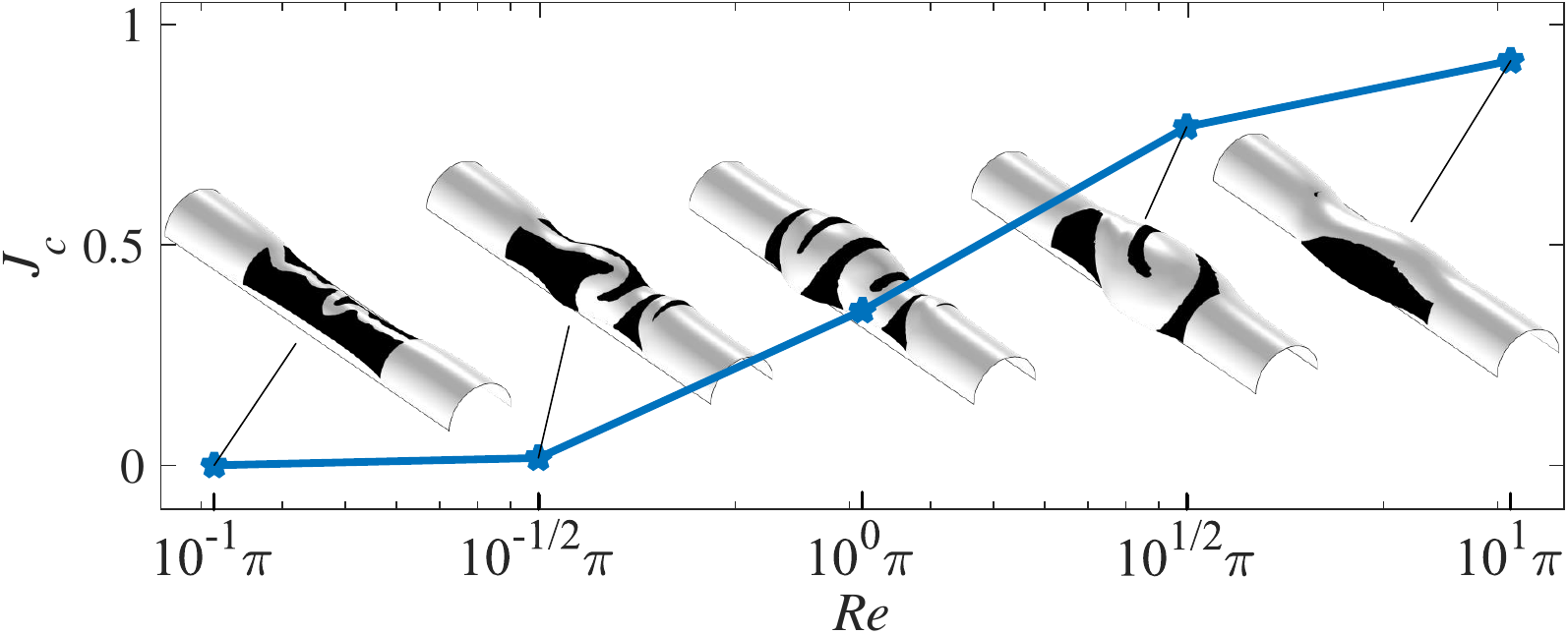}
  \caption{Objective values for the Reynolds numbers of $Re = \left\{10^{-1}\pi, 10^{-1/2}\pi, 10^0\pi, 10^{1/2}\pi, 10^1\pi \right\}$ and the obtained fiber bundles for mass transfer in the surface flow.}\label{fig:MassTransferManifold3Re}
\end{figure}

Reynolds number can be used to characterize the relative dominance of the convection and viscosity in the surface flow. It can be defined as 
\begin{equation}\label{equ:ReFBTOOPMassSurfaceFlow}
  Re = { \rho \sup_{\forall \mathbf{x}_\Gamma \in l_{v,\Gamma}} \left\|\mathbf{u}_{l_{v,\Gamma}}\right\|_2 \left| l_{v,\Gamma} \right| \over \eta}
\end{equation}
where $\sup_{\forall \mathbf{x}_\Gamma \in l_{v,\Gamma}} \left\|\mathbf{u}_{l_{v,\Gamma}}\right\|_2$ is the maximal value of the inlet velocity and $\left| l_{v,\Gamma} \right|$ is the length of the inlet boundary. Then, the fiber bundle topology optimization problem is solved for the Reynolds numbers of $Re = \left\{10^{-1}\pi, 10^{-1/2}\pi, 10^0\pi, 10^{1/2}\pi, 10^1\pi \right\}$, respectively, where the other parameters are remained to be unchanged. The values of the design objective are obtained and plotted in Fig. \ref{fig:MassTransferManifold3Re} including the obtained fiber bundles. Fig. \ref{fig:MassTransferManifold3Re} shows that the objective values increase along with increasing the Reynolds number. This is because that larger Reynolds number means larger fluid velocity and shorter mixing time during the fluid flowing through the surface structure. At low Reynolds number, relatively thin surface structures are obtained to satisfy the constraint of the pressure drop. They become thick along with the increase of the Reynolds number, because higher Reynolds number increase the pressure drop and the surface structure needs to be thick enough to decrease the pressure drop and satisfy the corresponding constraint.

\begin{figure}[!htbp]
  \centering
  \includegraphics[width=0.7\textwidth]{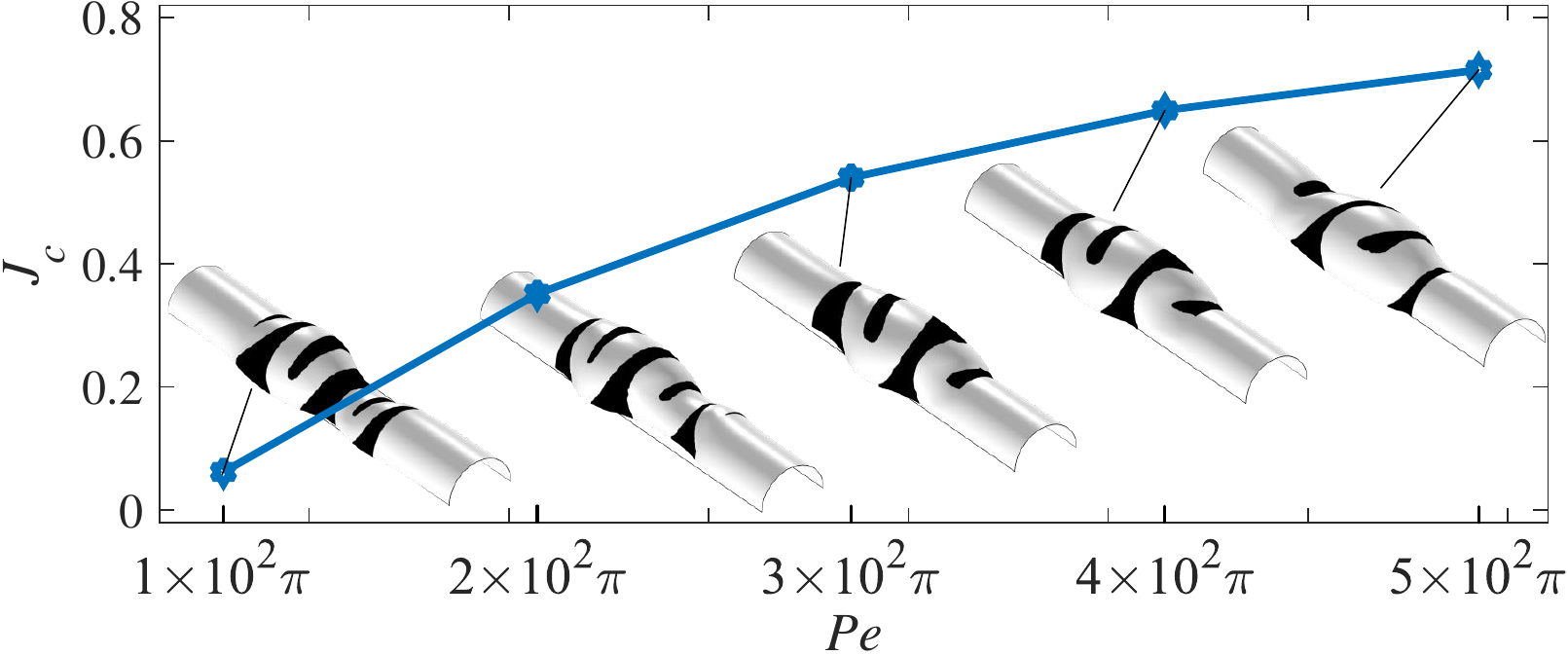}
  \caption{Objective values for the P\'{e}clet numbers of $Pe = \left\{ 1\times10^2\pi, 2\times10^2\pi, 3\times10^2\pi, 4\times10^2\pi, 5\times10^2\pi \right\}$ and the obtained fiber bundles for mass transfer in the surface flow.}\label{fig:MassTransferManifold3Pe}
\end{figure}

P\'{e}clet number can be used to characterize the relative dominance of the convection and diffusion in the mass transfer problems. It can be defined as 
\begin{equation}\label{equ:PeFBTOOPMassSurfaceFlow}
  Pe = { \sup_{\forall \mathbf{x}_\Gamma \in l_{v,\Gamma}} \left\|\mathbf{u}_{l_{v,\Gamma}}\right\|_2 \left| l_{v,\Gamma} \right| \over D}.
\end{equation}
Then, the fiber bundle topology optimization problem is solved for the P\'{e}clet numbers of $Pe = \left\{ 1\times10^2\pi, 2\times10^2\pi, 3\times10^2\pi, 4\times10^2\pi, 5\times10^2\pi \right\}$, respectively, where the other parameters are remained to be unchanged. The values of the design objective are obtained and plotted in Fig. \ref{fig:MassTransferManifold3Pe} including the obtained fiber bundles. Fig. \ref{fig:MassTransferManifold3Pe} shows that the objective values increase along with increasing the P\'{e}clet number. This is because that larger P\'{e}clet number means the dominance of the convection becomeing stronger and the mixing time being equivalently reduced in the surface structure.

\begin{figure}[!htbp]
  \centering
  \subfigure[$A_d = 1.0$]
  {\includegraphics[width=0.7\textwidth]{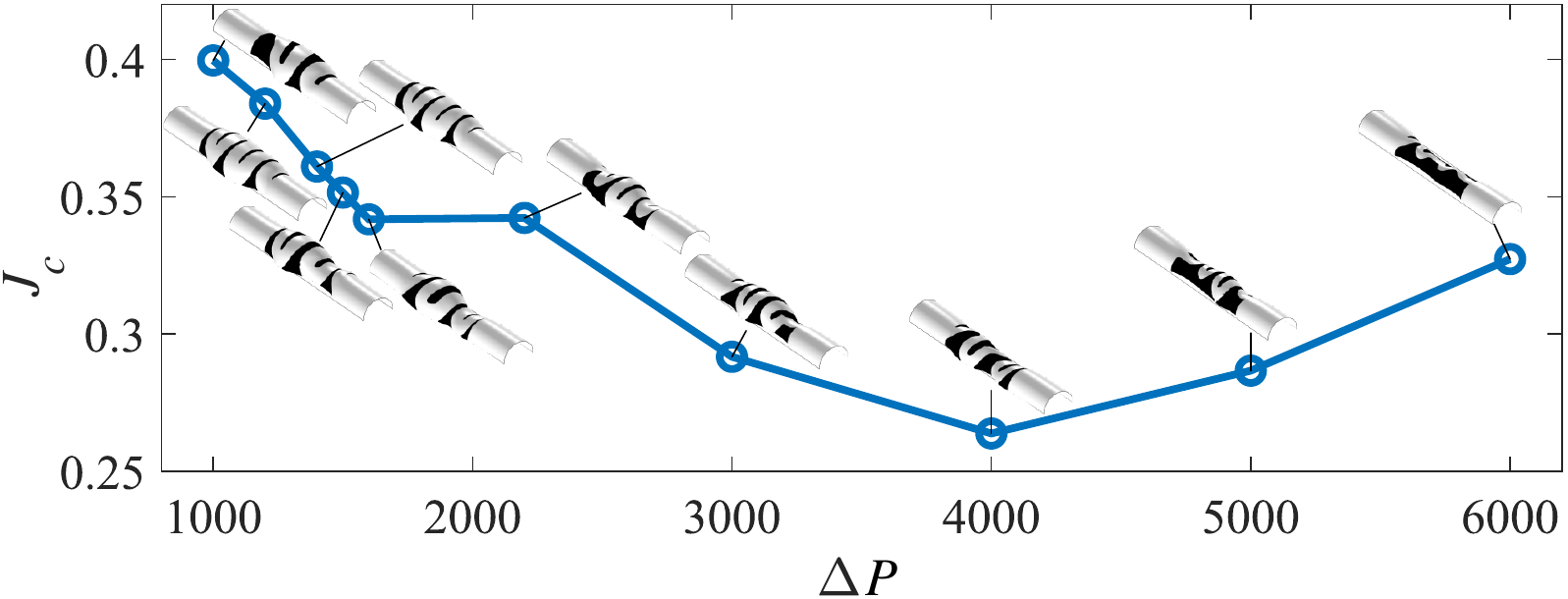}}
  \subfigure[$A_d = 1.5$]
  {\includegraphics[width=0.7\textwidth]{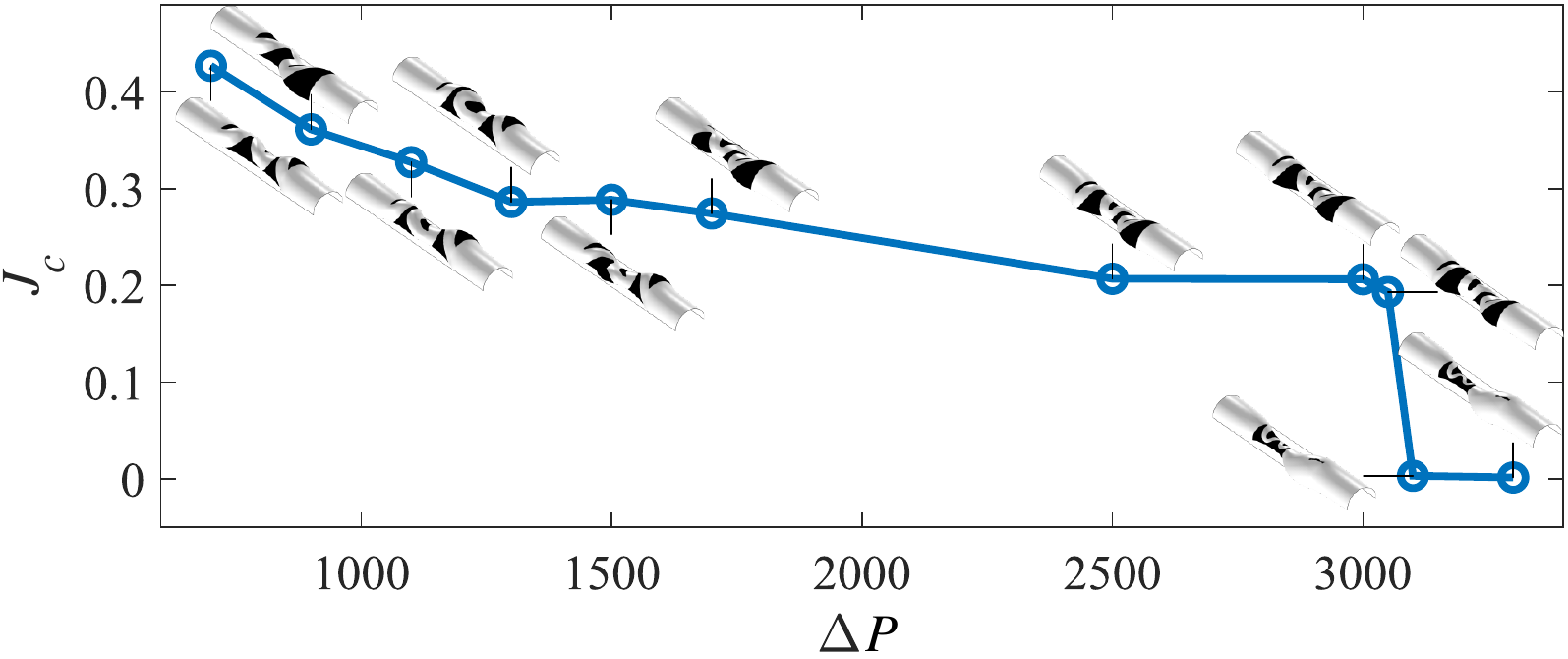}}
  \subfigure[$A_d = 2.0$]
  {\includegraphics[width=0.7\textwidth]{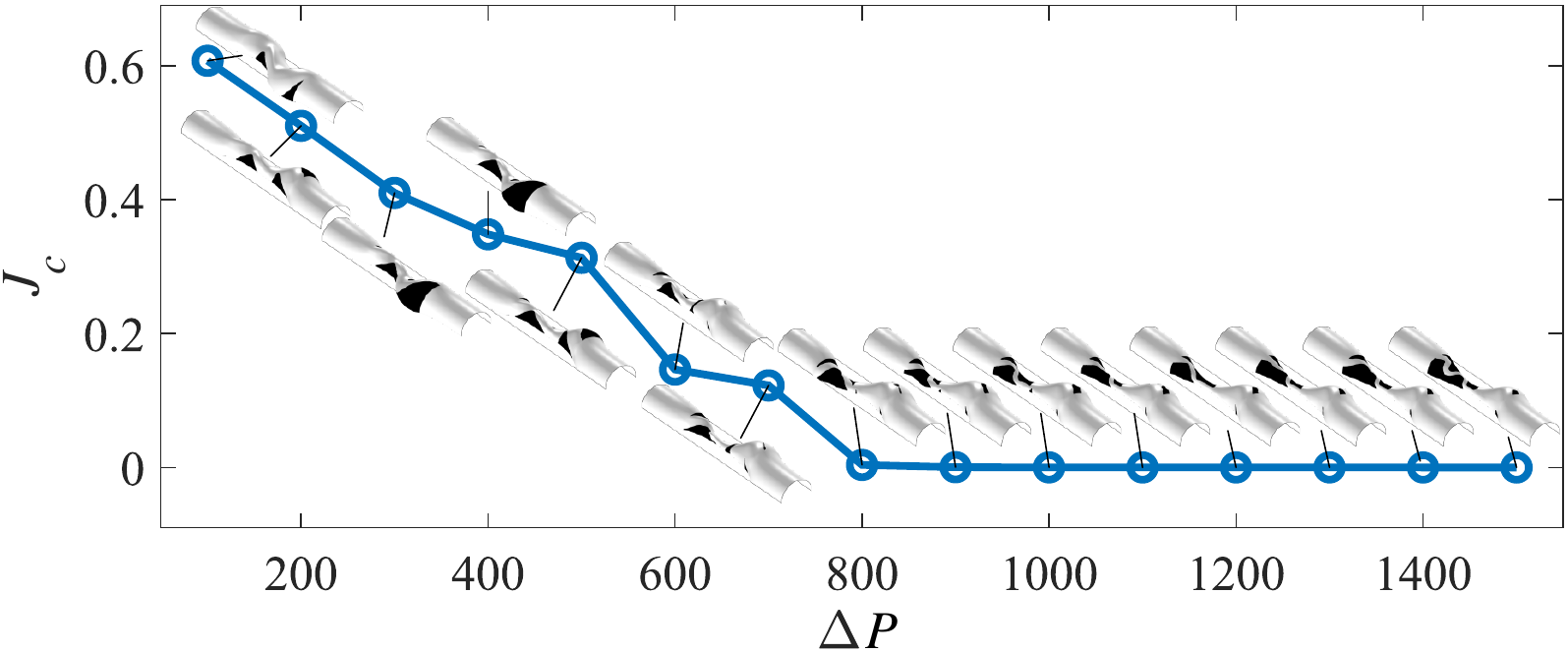}}
  \caption{Objective values and the obtained fiber bundles for different pressure drop for mass transfer in the surface flow.}\label{fig:MassTransferManifoldAdPressure}
\end{figure}

\begin{figure}[!htbp]
  \centering
  \subfigure[$A_d = 1.5, \Delta P = 3100$]
  {\includegraphics[width=0.3\textwidth]{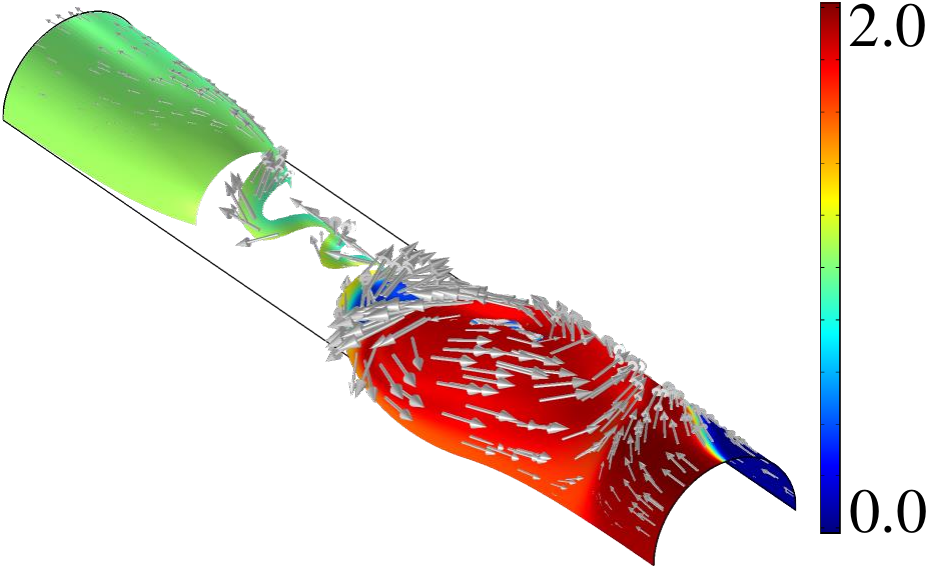}}\hspace{3em}
  \subfigure[$A_d = 2.0, \Delta P = 800$]
  {\includegraphics[width=0.3\textwidth]{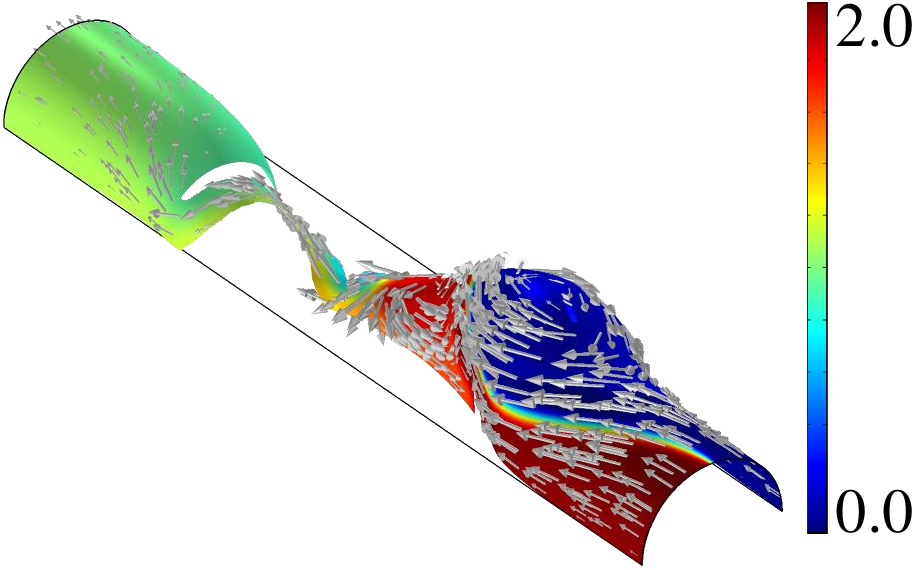}}
  \caption{Concentration distribution in the vortex based mixing modes on the topologically optimized fiber bundles.}\label{fig:MassTransferManifoldAdCriticalPressure}
\end{figure}

The constraint of the pressure drop can be used to ensure the smoothness of the surface flow. Based on the results in Fig. \ref{fig:MassTransferManifold3AdPlot}, it is investigated for the variable magnitudes of $A_d = \left\{ 1.0, 1.5, 2.0 \right\}$ on the design domain in Fig. \ref{fig:MassHeatTransferManifoldsDesignDomain}c. For the variable magnitude of $A_d = 1.0$, the fiber bundle topology optimization problem is solved for different values of the pressure drop and the objective values are plotted in Fig. \ref{fig:MassTransferManifoldAdPressure}a including the obtained fiber bundles. In Fig. \ref{fig:MassTransferManifoldAdPressure}a, there is a transition of the monotonicity of the objective values at the pressure drop of $4\times10^3$. The transition shows that reasonable increase of the pressure drop can improve the mixing performance of the surface structures. This is because that reasonable increase of the pressure drop can enhance the convection of the surface flow and mixing efficiency can be improved by the enhanced convection. However, the pressure drop can not be increased consistently, because too large pressure drop causes the thin surface structures and large fluid velocity and this equivalently increases the mixing length. When the variable magnitude is increased to $A_d = 1.5$ and $A_d = 2.0$, the transition of the monotonicity of the objective values disappears as shown in Figs. \ref{fig:MassTransferManifoldAdPressure}b and \ref{fig:MassTransferManifoldAdPressure}c. This is because that the mixing mode is changed after a critical value of the pressure drop. When the pressure drop arrives at the critical value, the vortex based mixing modes present as shown in Fig. \ref{fig:MassTransferManifoldAdCriticalPressure}, where vortexes are generated on the implicit 2-manifolds, the convection is enhanced effectively and the mixed fluid around the interfaces of two solutions is extracted by a thin surface channel connected to the interface of the vortexes. The vortex based mixing modes can achieve complete mixing as shown in Fig. \ref{fig:MassTransferManifoldAdCriticalPressure}. Therefore, the increase of the pressure drop can achieve complete mixing by inducing vortex based mixing modes, when the variable magnitude is preset to be large enough for the fiber bundle topology optimization problem.

To confirm the optimality, the results in Fig. \ref{fig:MassTransferManifold3Pe} are cross-compared by computing the objective values for the obtained fiber bundles at different P\'{e}clet numbers, where the Reynolds number is remained as $10^0\pi$. The computed objective values are listed in Tab. \ref{tab:MassTransferManifold3Optimality}. From the comparison of the objective values in every row of Tab. \ref{tab:MassTransferManifold3Optimality}, the optimized performance of the obtained fiber bundles can be confirmed.

\begin{table}[!htbp]
\centering
\begin{tabular}{l|ccccc}
  \toprule
        & \includegraphics[width=0.14\textwidth]{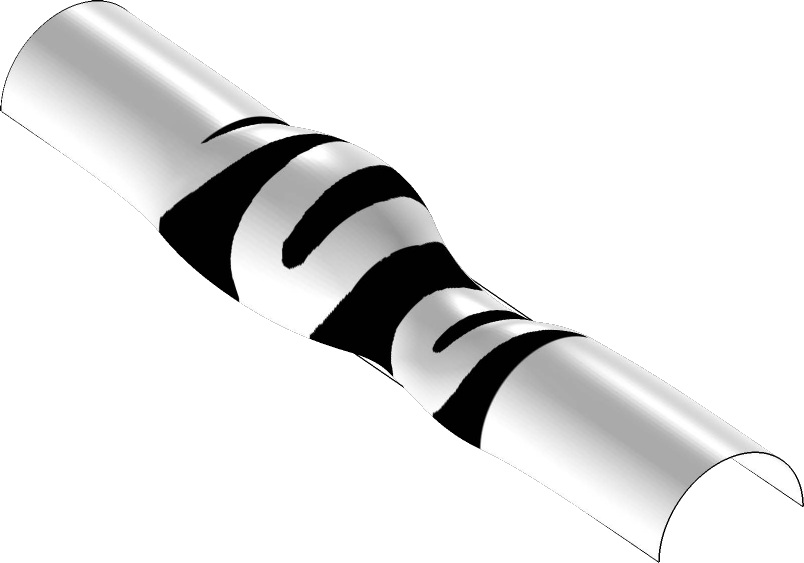}
        & \includegraphics[width=0.14\textwidth]{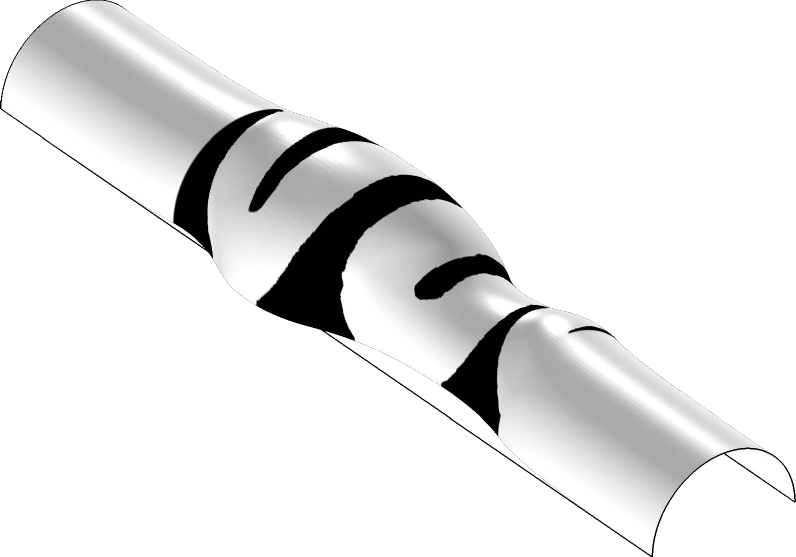}
        & \includegraphics[width=0.14\textwidth]{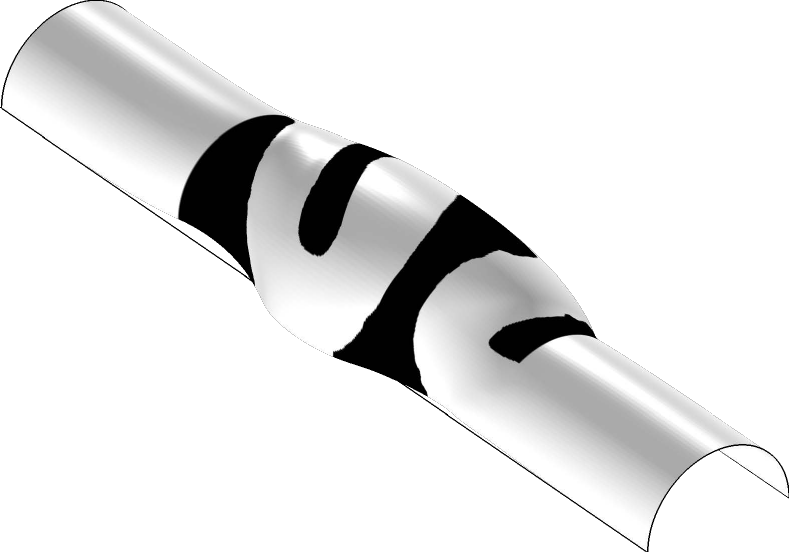}
        & \includegraphics[width=0.14\textwidth]{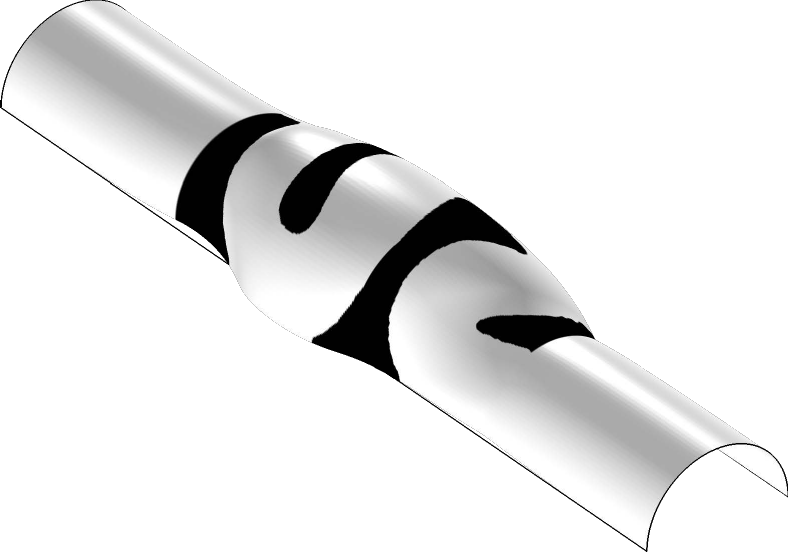}
        & \includegraphics[width=0.14\textwidth]{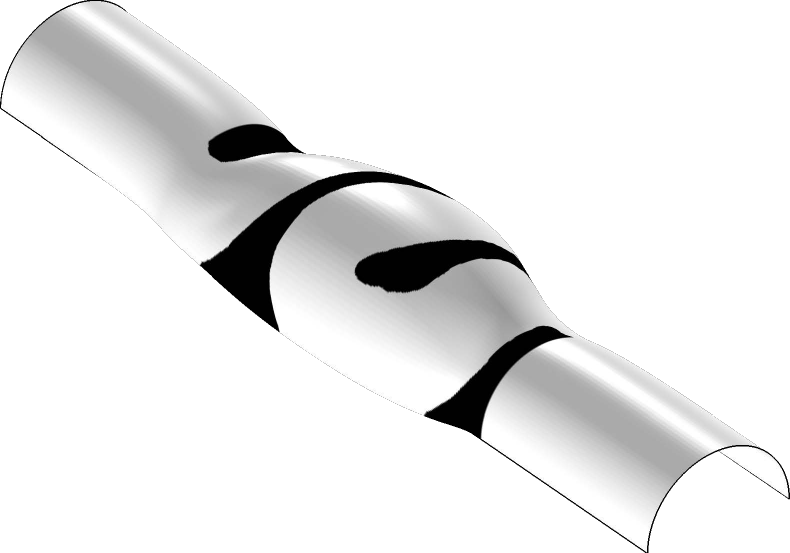} \\
   & $Pe = 1\times10^2\pi$ & $Pe = 2\times10^2\pi$ & $Pe = 3\times10^2\pi$ & $Pe = 4\times10^2\pi$ & $Pe = 5\times10^2\pi$ \\     
  \midrule
  $Pe = 1\times10^2\pi$ & $\mathbf{0.0621}$ & $0.6243$ & $0.6491$ & $0.6548$ & $0.6419$ \\
  \midrule
  $Pe = 2\times10^2\pi$ & $0.7640$ & $\mathbf{0.3515}$ & $0.7763$ & $0.7789$ & $0.7700$ \\
  \midrule
  $Pe = 3\times10^2\pi$ & $0.8156$ & $0.8125$ & $\mathbf{0.5402}$ & $0.8269$ & $0.8197$ \\
  \midrule
  $Pe = 4\times10^2\pi$ & $0.8448$ & $0.8421$ & $0.8531$ & $\mathbf{0.6495}$ & $0.8482$ \\
  \midrule
  $Pe = 5\times10^2\pi$ & $0.8643$ & $0.8619$ & $0.8716$ & $0.8727$ & $\mathbf{0.7153}$ \\
  \bottomrule
\end{tabular}
\caption{Objective values for the obtained fiber bundle in Fig. \ref{fig:MassTransferManifold3Pe} at different P\'{e}clet numbers, where the Reynolds number is remained as $10^0\pi$.}\label{tab:MassTransferManifold3Optimality}
\end{table}

\subsubsection{Heat transfer in surface flow} \label{sec:ResultsDiscussionHeatTransferSurface}

\begin{figure}[!htbp]
  \centering
  \subfigure[]
  {\includegraphics[width=1\textwidth]{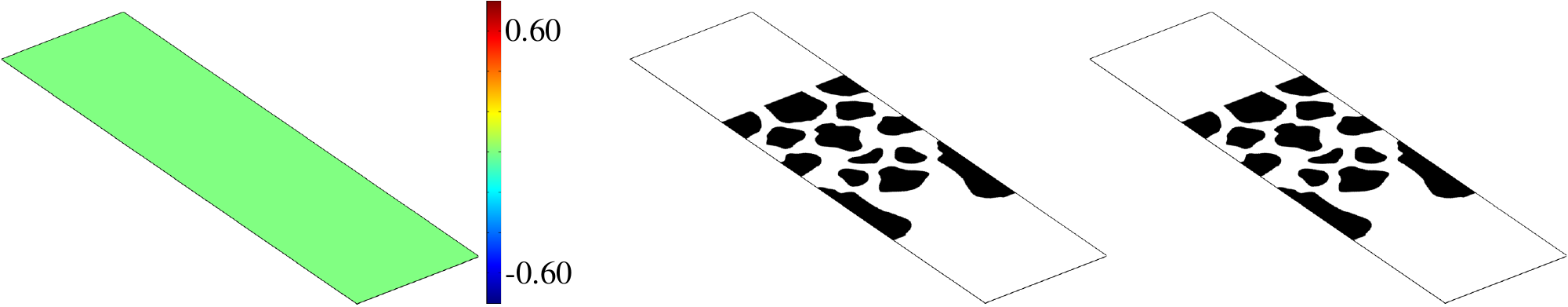}}
  \subfigure[]
  {\includegraphics[width=1\textwidth]{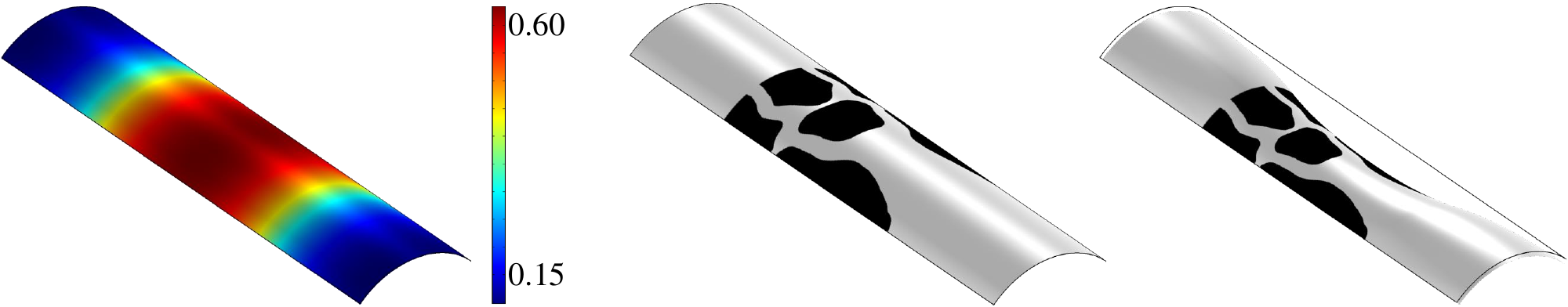}}
  \subfigure[]
  {\includegraphics[width=1\textwidth]{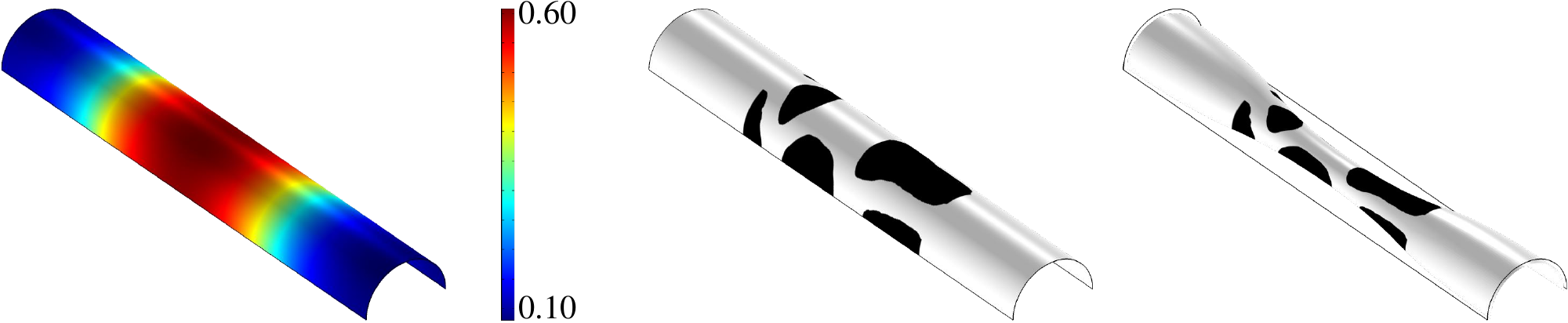}}
  \subfigure[]
  {\includegraphics[width=1\textwidth]{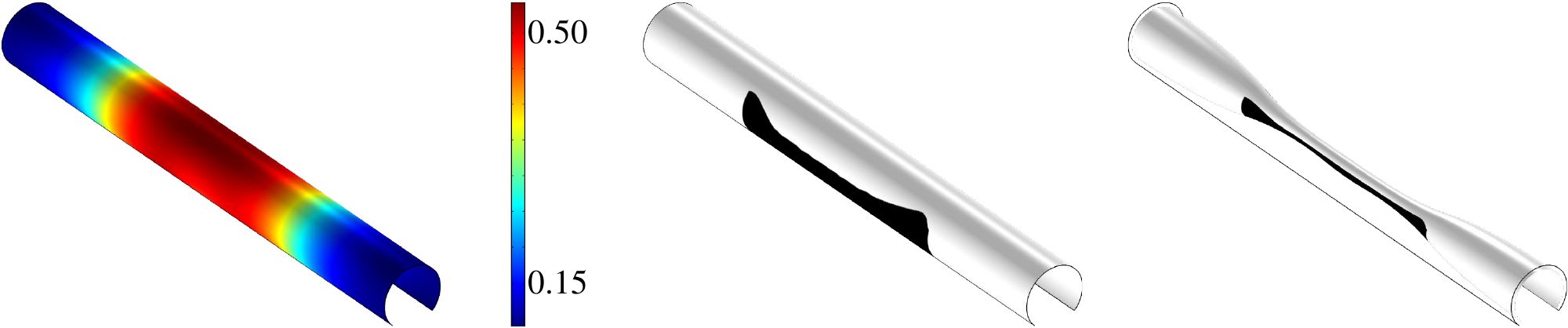}}
  \subfigure[]
  {\includegraphics[width=1\textwidth]{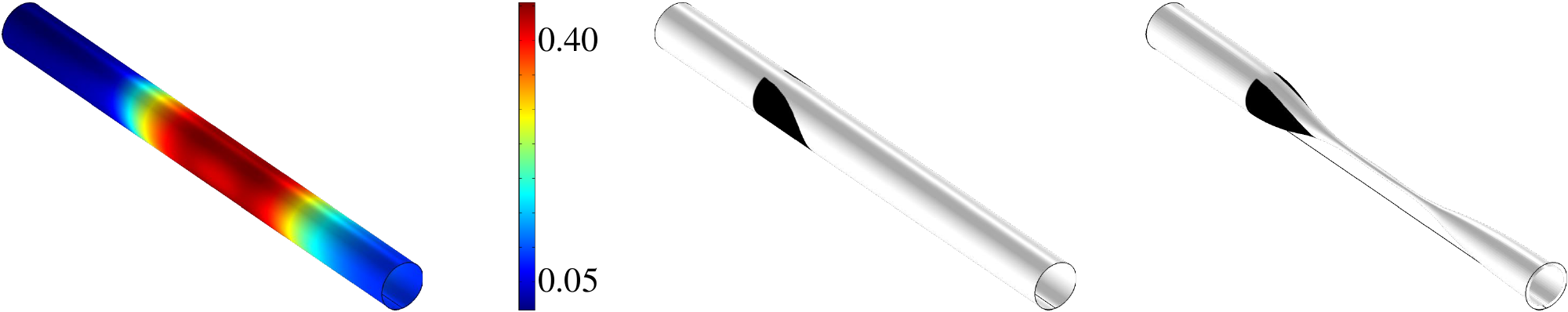}}
  \caption{Distribution of the normal displacement for the implicit 2-manifolds and the material density of the surface patterns for heat transfer in the surface flow on the design domains sketched in Fig. \ref{fig:MassHeatTransferManifoldsDesignDomain}, including the fiber bundles composed of the implicit 2-manifolds and the surface patterns.}\label{fig:HeatTransferManifoldsAd=15e-1}
\end{figure}

\begin{figure}[!htbp]
  \centering
  \includegraphics[width=0.7\textwidth]{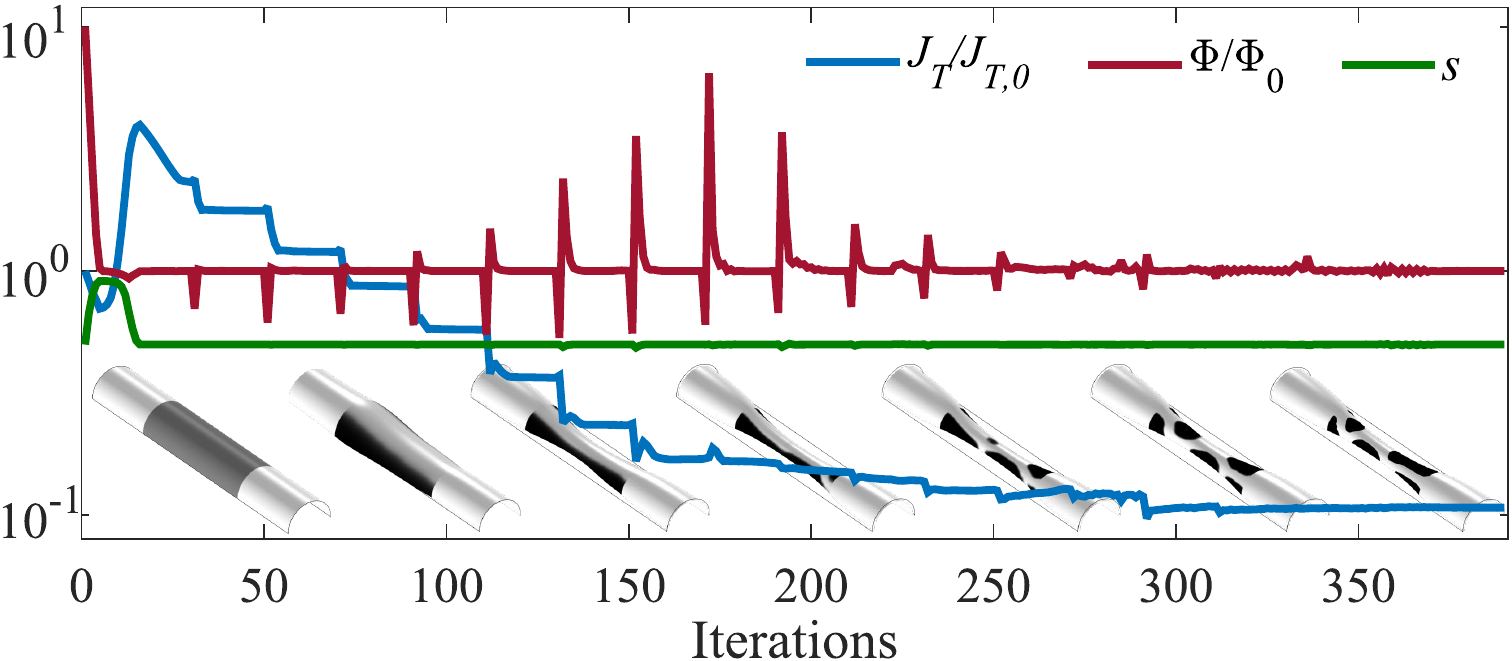}
  \caption{Convergent histories of the design objective and constraints of the dissipation power and area fraction for heat transfer in the surface flow on the design domain sketched in Fig. \ref{fig:MassHeatTransferManifoldsDesignDomain}c, including snapshots for the evolution of the fiber bundle during the iterative solution of the optimization problem.}\label{fig:HeatTransferManifold3ConvergentHistories}
\end{figure}

\begin{figure}[!htbp]
  \centering
  \subfigure[Velocity]
  {\includegraphics[width=0.32\textwidth]{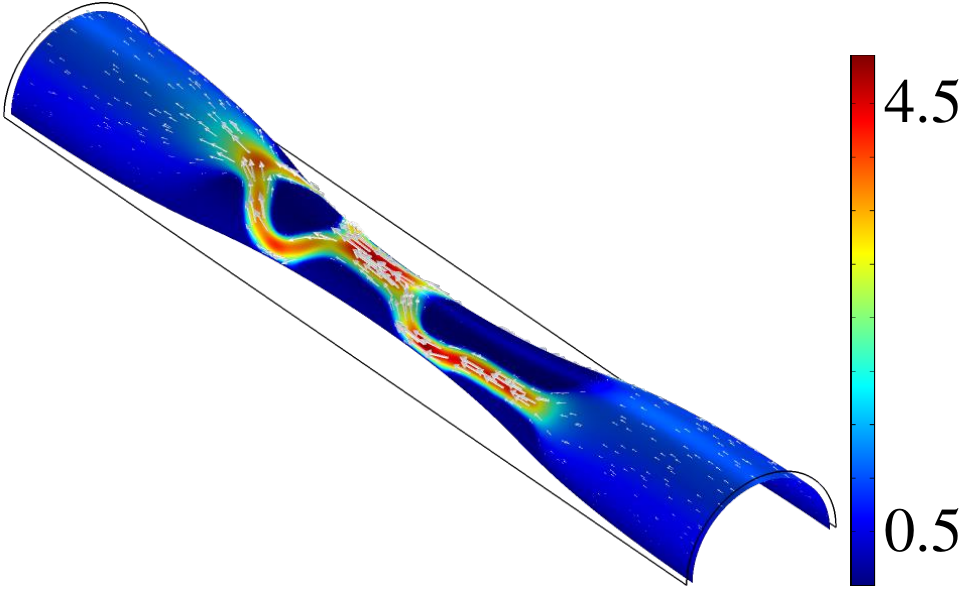}}
  \subfigure[Pressure]
  {\includegraphics[width=0.32\textwidth]{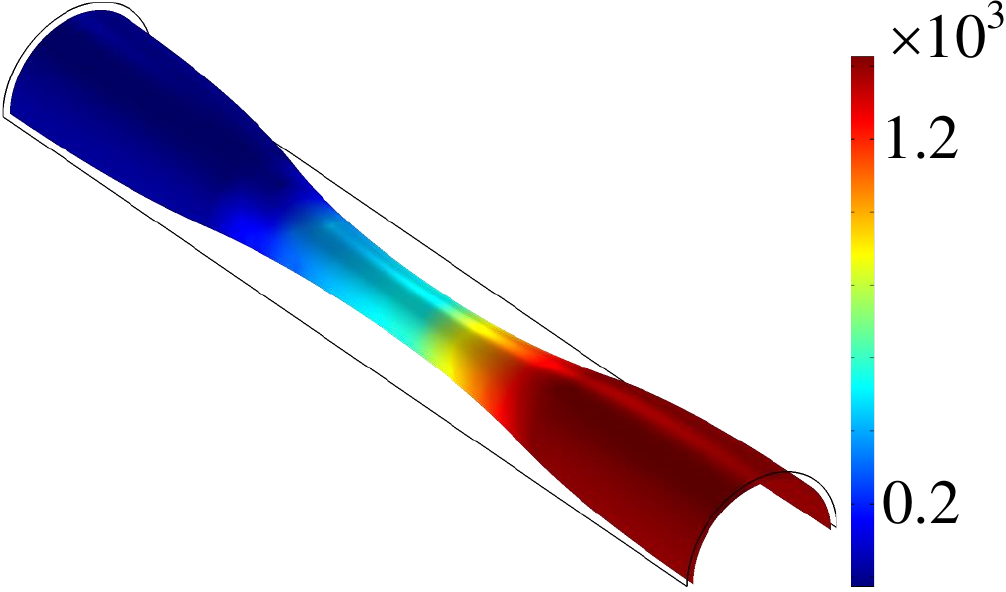}}
  \subfigure[Temperature]
  {\includegraphics[width=0.32\textwidth]{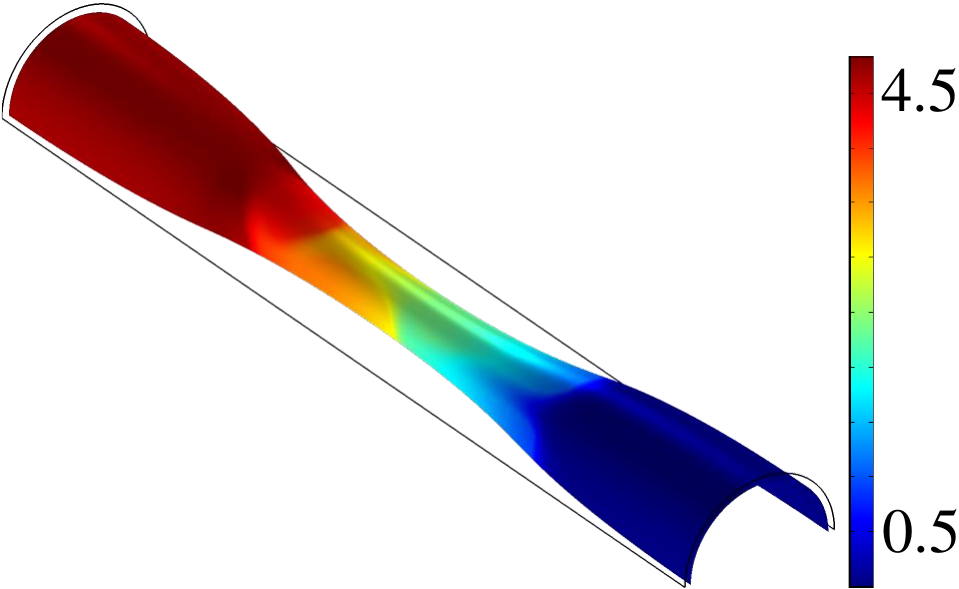}}
  \caption{Distribution of the velocity, pressure and temperature for the fiber bundle obtained on the design domain sketched in Fig. \ref{fig:MassHeatTransferManifoldsDesignDomain}c.}\label{fig:HeatTransferManifold3VelocityPressureTemperature}
\end{figure}

In fiber bundle topology optimization for heat transfer in the surface flow, the fluid density and viscosity are also considered as unitary. The material coefficients and optimization parameters are set as that in Tab. \ref{tab:IterativeProcedureSurfaceFlowHM}. The inlet velocity is set as the parabolic distribution with the maximal value of $1$, and the temperature distribution at the inlet is set to be homogeneous with the unitary value. By setting the variable magnitude as $1.5$, the fiber bundle topology optimization problem in Eq. \ref{equ:VarProToopSurfaceNSCHM} is solved on the design domains in Fig. \ref{fig:MassHeatTransferManifoldsDesignDomain}. The distribution of the filtered design variables for the implicit 2-manifolds and the material density for the patterns are obtained as shown in Fig. \ref{fig:HeatTransferManifoldsAd=15e-1}, where the fiber bundles composed of the implicit 2-manifolds and the surface patterns are included. Convergent histories of the optimization objective and constraints of the dissipation power and area fraction are plotted in Fig. \ref{fig:HeatTransferManifold3ConvergentHistories} for fiber bundle topology optimization on the design domain in Fig. \ref{fig:HeatTransferManifoldsAd=15e-1}c, including snapshots for the evolution of the fiber bundle during the iterative solution of the optimization problem. From the monotonicity of the objective values and satisfication of the constraints of the dissipation power and area fraction, the robustness of the iterative procedure can be confirmed for fiber bundle topology optimization for heat transfer in the surface flow. The distribution of the velocity, pressure and temperature are provided in Fig. \ref{fig:HeatTransferManifold3VelocityPressureTemperature} for fiber bundle topology optimization on the design domain in Fig. \ref{fig:MassHeatTransferManifoldsDesignDomain}c, where the splitting-merging shaped curved-channel is obtained for the surface flow to enhance the convection, enlarge the diffusion length, decrease the surface gradient of the temperature distribution and the efficiency of heat transfer is thereby improved.

\begin{figure}[!htbp]
  \centering
  \includegraphics[width=0.7\textwidth]{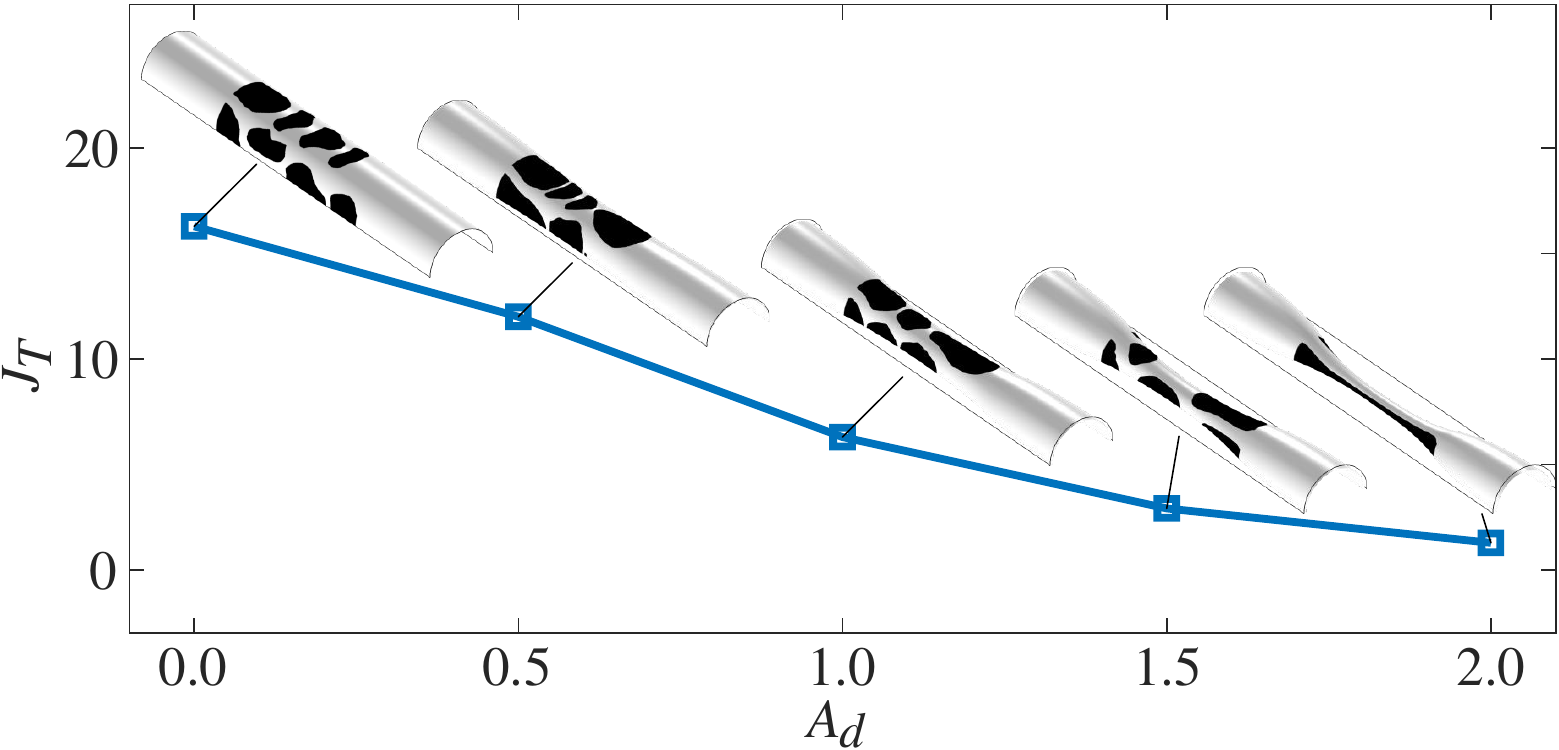}
  \caption{Objective values for the variable magnitudes of $A_d = \left\{0.0,0.5,1.0,1.5,2.0\right\}$ and the obtained fiber bundles for heat transfer in the surface flow.}\label{fig:HeatTransferManifold3AdPlot}
\end{figure}

The variable magnitudes of $A_d = \left\{0.0,0.5,1.0,1.5,2.0\right\}$ are investigated for fiber bundle topology optimization of heat transfer in the surface flow, where the other parameters are remained to be unchanged. The values of the design objective are obtained and plotted in Fig. \ref{fig:HeatTransferManifold3AdPlot} including the obtained fiber bundles. Because larger variable magnitude permits more flexible evolution of the fiber bundle to improve the heat transfer performance of the surface structure, the objective values in Fig. \ref{fig:HeatTransferManifold3AdPlot} decrease along with increasing the variable magnitude.

\begin{figure}[!htbp]
  \centering
  \includegraphics[width=0.7\textwidth]{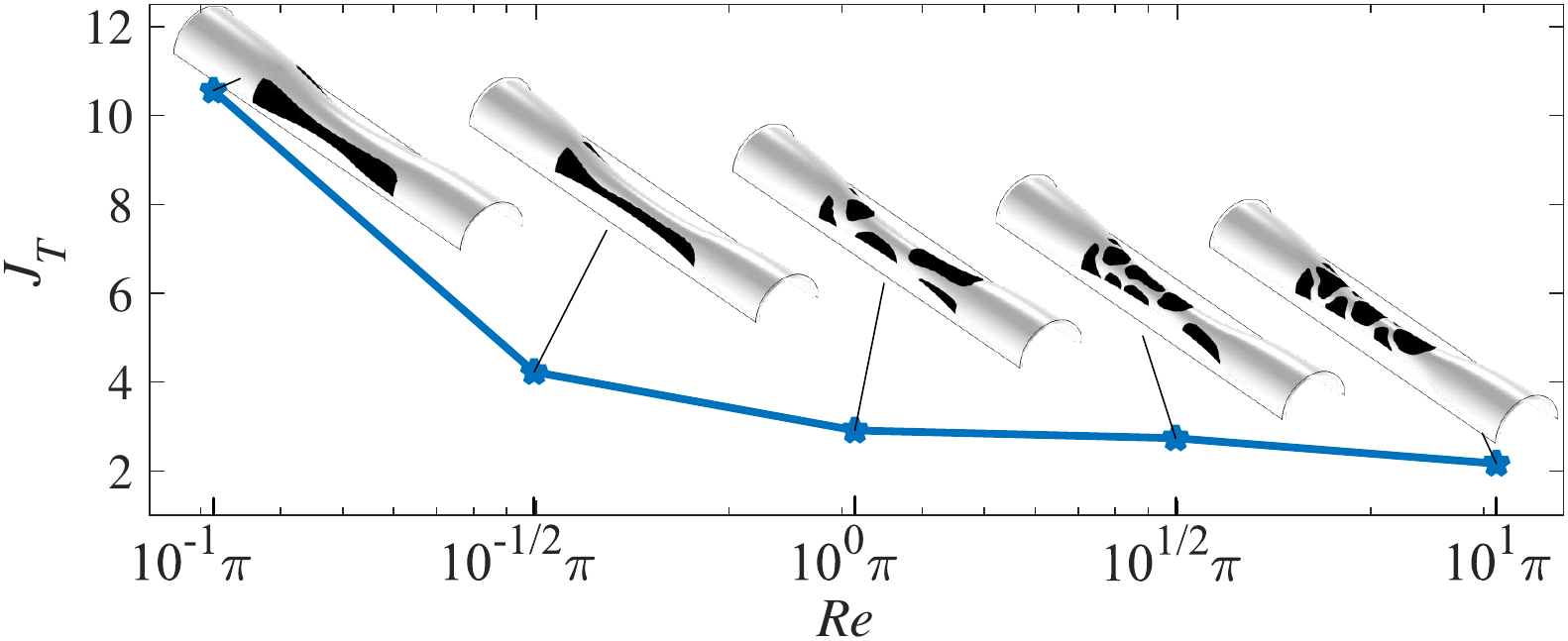}
  \caption{Objective values for the Reynolds numbers of $Re = \left\{10^{-1}\pi, 10^{-1/2}\pi, 10^0\pi, 10^{1/2}\pi, 10^1\pi \right\}$ and the obtained fiber bundles for heat transfer in the surface flow.}\label{fig:HeatTransferManifold3Re}
\end{figure}

To investigate the effect of the Reynolds number defined in Eq. \ref{equ:ReFBTOOPMassSurfaceFlow}, the fiber bundle topology optimization problem for heat transfer in the surface flow is solved for the Reynolds numbers of $Re = \left\{10^{-1}\pi, 10^{-1/2}\pi, 10^0\pi, 10^{1/2}\pi, 10^1\pi \right\}$, where the other parameters are remained to be unchanged. The values of the design objective are obtained and plotted in Fig. \ref{fig:HeatTransferManifold3Re} including the obtained fiber bundles. To remain the P\'{e}clet number to be unchanged in the investigation of the Reynolds number, the dynamic viscosity of the fluid is changed, where high Reynolds number corresponds to low dynamic viscosity. Because the specified value of the dissipation power is remained to be unchanged, splitting-merging shaped curved-channel and decrease of the width of the channel can help to ensure the satisfication of the constraint of the dissipation power for the surface flow with high Reynolds number and low dynamic viscosity. This can simultaneously improve the heat transfer performance by decreasing the surface gradient of the temperature distribution. Therefore, the objective values decrease along with increasing the Reynolds number in Fig. \ref{fig:HeatTransferManifold3Re}.

\begin{figure}[!htbp]
  \centering
  \includegraphics[width=0.7\textwidth]{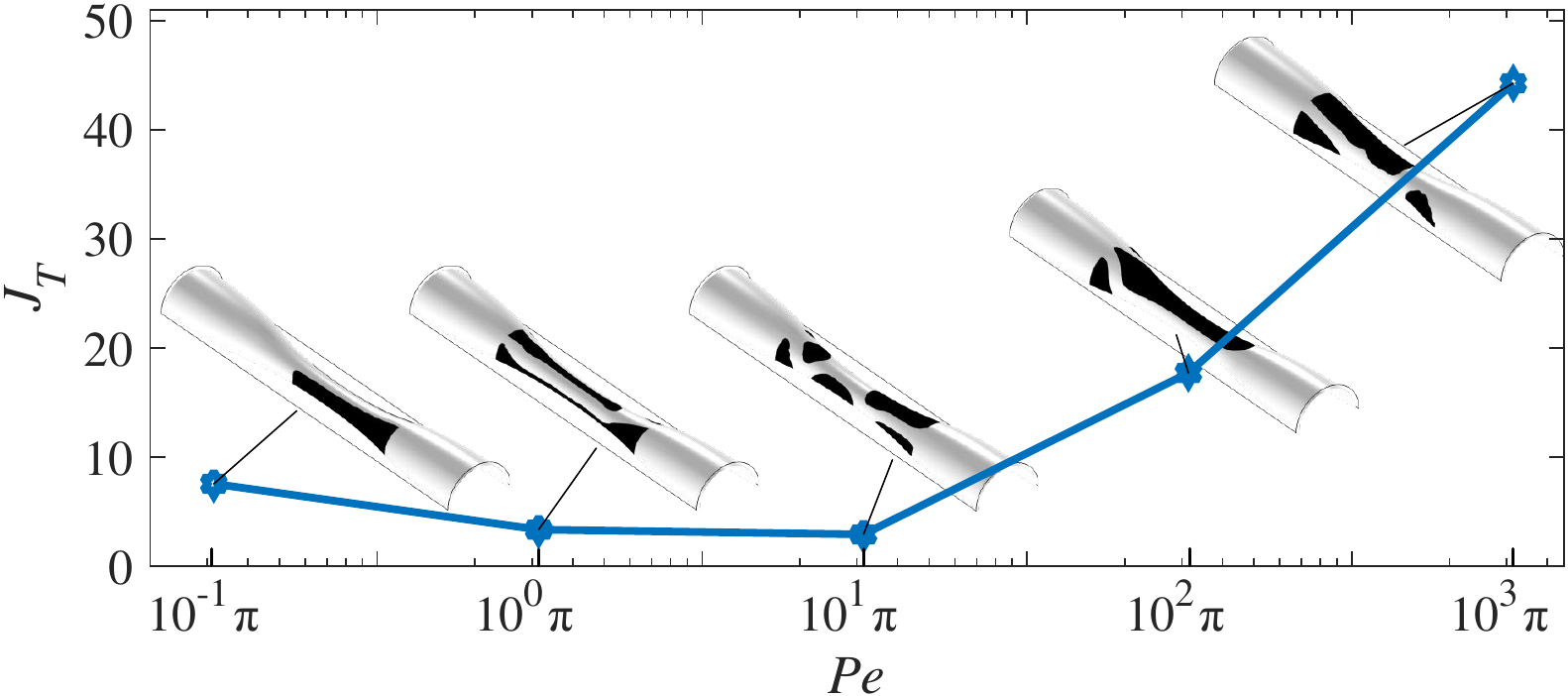}
  \caption{Objective values for the P\'{e}clet numbers of $Pe = \left\{ 10^{-1}\pi, 10^0\pi, 10^1\pi, 10^2\pi, 10^3\pi \right\}$ and the obtained fiber bundles for heat transfer in the surface flow.}\label{fig:HeatTransferManifold3Pe}
\end{figure}

In the convective heat-transfer problem, P\'{e}clet number is used to characterize the relative dominance of the convection and thermal conductivity. It can be defined as 
\begin{equation}\label{equ:PeFBTOOPHeatSurfaceFlow}
  Pe = { \rho C_{pf} \sup_{\forall \mathbf{x}_\Gamma \in l_{v,\Gamma}} \left\|\mathbf{u}_{l_{v,\Gamma}}\right\|_2 \left| l_{v,\Gamma} \right| \over k_f}.
\end{equation}
Then, the fiber bundle topology optimization problem is solved for the P\'{e}clet numbers of $Pe = \left\{ 10^{-1}\pi, 10^0\pi, 10^1\pi, 10^2\pi, 10^3\pi \right\}$, where the other parameters are remained to be unchanged. The values of the design objective are obtained and plotted in Fig. \ref{fig:HeatTransferManifold3Pe} including the obtained fiber bundles. To remain the Reynolds number to be unchanged in the investigation of the P\'{e}clet number, the thermal conductivity of the fluid is changed, where high P\'{e}clet number corresponds to small coefficient of the thermal conductivity. Because both thermal conductivity and convection are helpful for the improvement of the heat transfer performance, there is a joint point of those two factors when the P\'{e}clet number is changed. Therefore, the obtained fiber bundle at the P\'{e}clet number of $10^1\pi$ has the lowest value of the design objective. This is achieved by the splitting-merging shaped curved-channel together with the thermal conductivity of the fluid, where the convection of the surface flow is enhanced by the splitting-merging shape of the obtained curved-channel.

\begin{figure}[!htbp]
  \centering
  \includegraphics[width=0.71\textwidth]{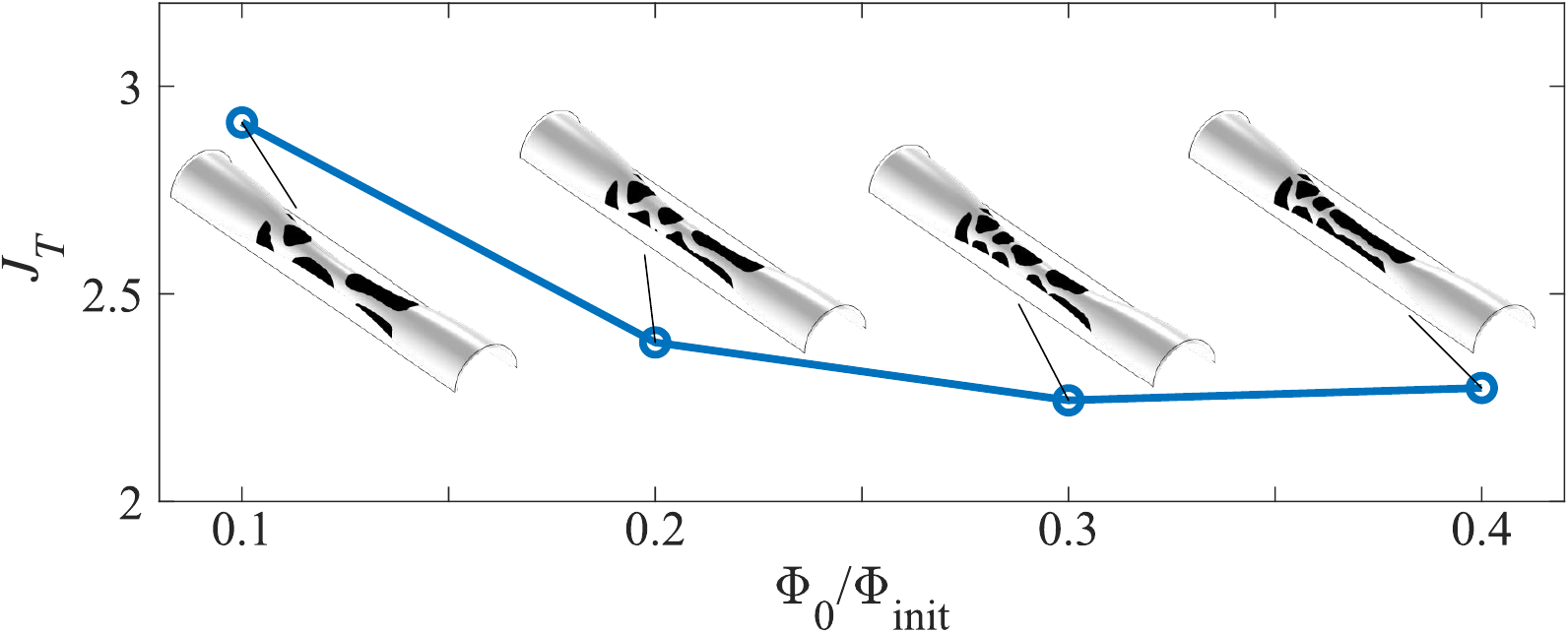}
  \caption{Objective values for different dissipation power and the obtained fiber bundles for heat transfer in the surface flow.}\label{fig:HeatTransferManifold3Df}
\end{figure}

The constraint of the dissipation power can be used to ensure the smoothness of the surface flow. Based on the results in Fig. \ref{fig:HeatTransferManifold3Df}, it is investigated for the variable magnitude of $A_d = 1.5$ on the design domain in Fig. \ref{fig:MassHeatTransferManifoldsDesignDomain}c. The fiber bundle topology optimization problem is solved for different values of the specified dissipation power and the objective values are plotted in Fig. \ref{fig:HeatTransferManifold3Df} including the obtained fiber bundles. When the dissipation power is specified with a large value, more blocks are presented and the width of the fluid channel becomes thin to enlarge the gradient of the fluid velocity and satisfy the constraint of the dissipation power. Then, the convection of the surface flow is enhanced and the heat transfer performance is improved. Therefore, the objective values in Fig. \ref{fig:HeatTransferManifold3Df} decrease along with the increase of the specified values of the dissipation power.

To confirm the optimality, the results in Fig. \ref{fig:HeatTransferManifold3Pe} are cross-compared by computing the objective values for the obtained fiber bundles at different P\'{e}clet numbers, where the Reynolds number is remained as $10^0\pi$. The computed objective values are listed in Tab. \ref{tab:HeatTransferManifold3OptimalityPe}. From the comparison of the objective values in every row of Tab. \ref{tab:HeatTransferManifold3OptimalityPe}, the optimized performance of the obtained fiber bundles can be confirmed.

\begin{table}[!htbp]
\centering
\begin{tabular}{l|ccccc}
  \toprule
        & \includegraphics[width=0.14\textwidth]{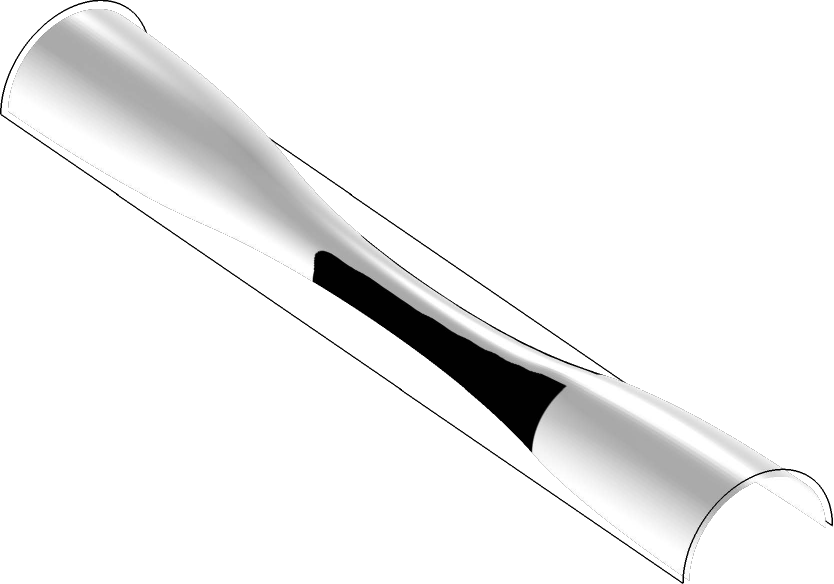}
        & \includegraphics[width=0.14\textwidth]{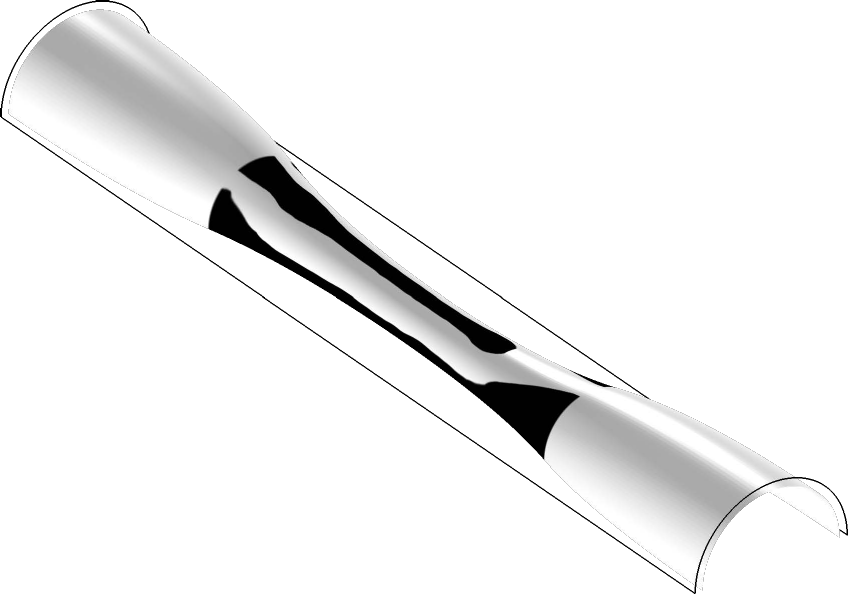}
        & \includegraphics[width=0.14\textwidth]{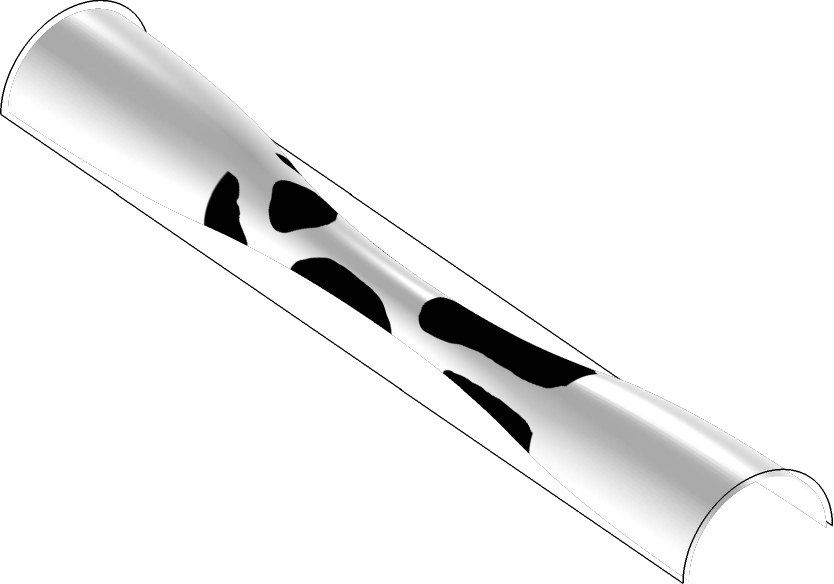}
        & \includegraphics[width=0.14\textwidth]{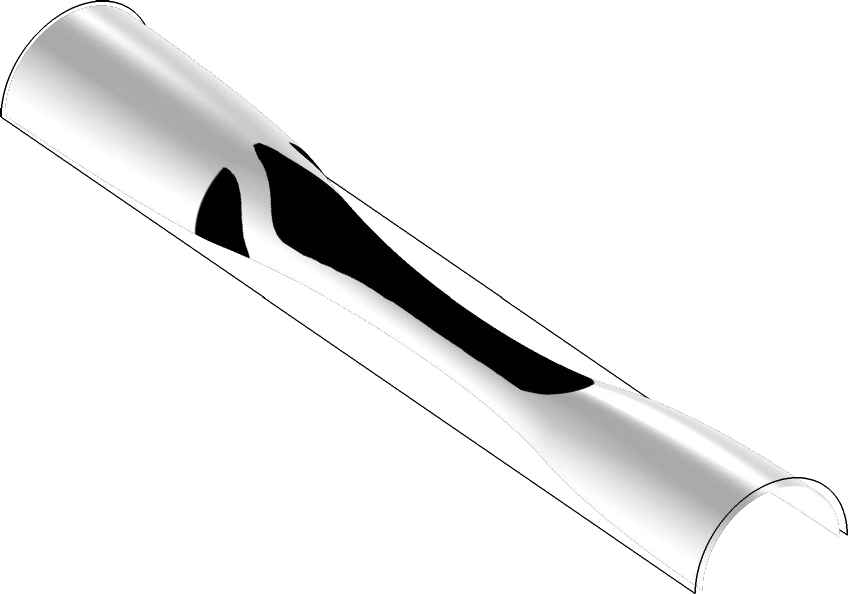}
        & \includegraphics[width=0.14\textwidth]{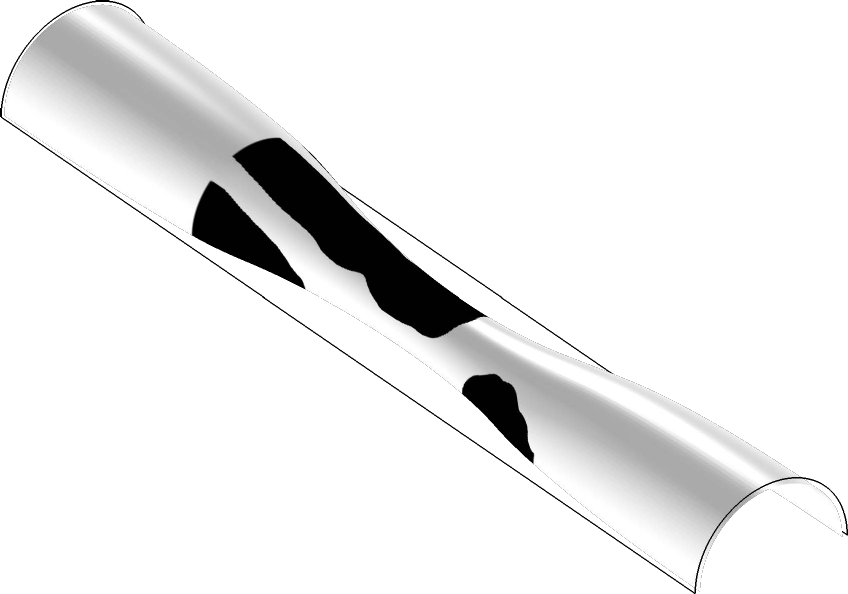} \\
   & $Pe = 10^{-1}\pi$ & $Pe = 10^0\pi$ & $Pe = 10^1\pi$ & $Pe = 10^2\pi$ & $Pe = 10^3\pi$ \\
  \midrule
  $Pe = 10^{-1}\pi$ & $\mathbf{7.5505}$ & $7.7949$ & $7.9965$ & $8.4704$ & $9.5489$ \\
  \midrule
  $Pe = 10^{0}\pi$ & $3.7587$ & $\mathbf{3.3512}$ & $3.9927$ & $5.5317$ & $6.8946$ \\
  \midrule
  $Pe = 10^1\pi$ & $3.7922$ & $3.5995$ & $\mathbf{2.9127}$ & $4.9974$ & $6.1530$ \\
  \midrule
  $Pe = 10^2\pi$ & $25.2534$ & $18.8232$ & $19.3612$ & $\mathbf{17.7171}$ & $18.3672$ \\
  \midrule
  $Pe = 10^3\pi$ & $62.8847$ & $51.7062$ & $75.3778$ & $98.3435$ & $\mathbf{44.2847}$ \\
  \bottomrule
\end{tabular}
\caption{Objective values for the obtained fiber bundles in Fig. \ref{fig:HeatTransferManifold3Pe} at different P\'{e}clet numbers, where the Reynolds number is remained as $10^0\pi$.}\label{tab:HeatTransferManifold3OptimalityPe}
\end{table}

\section{Mass and heat transfer in volume flow}\label{sec:MethodologyFiberBundleTOOPTransferBulkFlows}

For the volume flow, fiber bundle topology optimization for mass and heat transfer can be implemented for the thin walls embedded in the corresponding 3D domain. When the thickness of the thin walls are much less than the chracteristic size of the 3D domain, the thin walls can be approximated as curved surfaces imposed with no-slip boundary conditions. To find the optimized matching between the thin-wall pattern and the surface defined with the thin-wall pattern, fiber bundle topology optimization can be implemented. 

\subsection{Description of deformed 3D domain} \label{sec:DescriptionDeformedDomain}

\begin{figure}[!htbp]
  \centering
  \includegraphics[width=0.8\textwidth]{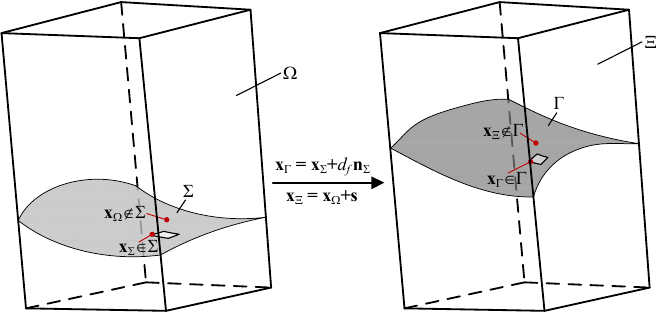}
  \caption{Sketch for the implicit 2-manifold induced deformation of the original domain, where $\Xi$ is the deformed domain, $\Omega$ is the original domain and $\mathbf{x}_\Omega$ and $\mathbf{x}_\Xi$ are sketched by the points not localized on $\Sigma$ and $\Gamma$.}\label{fig:SketchDeformedDomBulk}
\end{figure}

A surface used to define the thin-wall pattern can be described as the implicit 2-manifold defined on the preset base 2-manifold imbedded into the 3D domain. The implicit 2-manifold induces the deformation of the 3D domain. The governing equations of the mass and heat transfer processes are defined on the deformed 3D domain. The homeomorphism relation between the implicit 2-manifold and the base 2-manifold are expressed as that in Eq. \ref{equ:NormalDisplacementDistributionMHM}, where $\Sigma$ and $\Gamma$ continuously represents the base 2-manifold and the implicit 2-manifold, respectively. The design variable of the implicit 2-manifold and its surface-PDE filter are the same as that defined in Section \ref{subsec:DesignVariableImplicitManifold}. The deformed 3D domain is sketched in Fig. \ref{fig:SketchDeformedDomBulk}, and it can be described as
\begin{equation}\label{equ:HomeomorphismOmegaBulkFlowMHT}
  \Xi = \left\{ \mathbf{x}_\Xi \left| \, \begin{split}
  & \mathbf{x}_\Xi = \mathbf{x}_\Omega + \mathbf{s}, ~ \forall \mathbf{x}_\Omega \in \Omega \\
  & \mathbf{x}_\Xi = \mathbf{x}_\Gamma, ~ \forall \mathbf{x}_\Gamma \in \Gamma \\
  & \mathbf{x}_\Omega = \mathbf{x}_\Sigma, ~ \forall \mathbf{x}_\Sigma \in \Sigma \\
  & \mathbf{x}_\Gamma = \mathbf{x}_\Sigma + d_f \mathbf{n}_\Sigma, ~ \forall \mathbf{x}_\Sigma \in \Sigma
  \end{split}\right. \right\}
\end{equation}
where $\Omega$ is the 3D domain and it is open; $\mathbf{s}$ is the displacement in $\Omega$ caused by $d_f$ defined on $\Sigma$; $\Xi$ is the deformed counterpart of $\Omega$, with the deformation corresponding to the distribution of the displacement $\mathbf{s}$; $\mathbf{x}_\Omega$ is the Cartesian coordinate in $\Omega$; and $\mathbf{x}_\Xi$ is the harmonic coordinate in $\Xi$. In the deformed domain, the gradient operator $\nabla_{\mathbf{x}_\Xi}$, the divergence operator $\mathrm{div}_{\mathbf{x}_\Xi}$ and the unit outer normal $\mathbf{n}_{\partial\Xi}$ at $\partial\Xi$ can be defined in the coordinate system of $\mathbf{x}_\Xi$. The displacement $\mathbf{s}$ in Eq. \ref{equ:HomeomorphismOmegaBulkFlowMHT} can be described by the Laplace's equation:
\begin{equation}\label{equ:HarmonicCoordinateEquMHT}
\left\{
\begin{split}
& \mathrm{div}_{\mathbf{x}_\Omega} \left( \nabla_{\mathbf{x}_\Omega} \mathbf{s} \right) = \mathbf{0}, ~ \forall \mathbf{x}_\Omega \in \Omega \\
& \mathbf{s} = \mathbf{0}, ~ \forall \mathbf{x}_\Omega \in \Sigma_{v,\Omega} \cup \Sigma_{s,\Omega} \\
& \mathbf{s} = d_f \mathbf{n}_\Sigma, ~ \forall \mathbf{x}_\Omega \in \Sigma \\
& \mathbf{n}_{\partial\Omega} \cdot \nabla_{\mathbf{x}_\Omega} \mathbf{s} = \mathbf{0}, ~ \forall \mathbf{x}_\Omega \in \Sigma_{v_0,\Omega} \\
\end{split}
\right.
\end{equation}
where $\nabla_{\mathbf{x}_\Omega}$ and $\mathrm{div}_{\mathbf{x}_\Omega}$ are the gradient and divergence operators in $\Omega$, respectively; $\mathbf{n}_{\partial\Omega}$ is the unit outer normal vector at $\partial \Omega$; $\Sigma_{v,\Omega}$, $\Sigma_{v_0,\Omega}$ and $\Sigma_{s,\Omega}$ are boundary parts of $\partial \Omega$ corresponding to the inlet, wall and outlet of the deformed domain $\Xi$, respectively. Then, the Jacobian matrix for the deformed domain can be derived as
\begin{equation}\label{equ:TransformedJacobianOmegaMHT} 
  \mathbf{T}_\Xi = {\partial \mathbf{x}_\Xi \over \partial \mathbf{x}_\Omega } = \nabla_{\mathbf{x}_\Omega} \mathbf{s} + \mathbf{I}, ~ \forall \mathbf{x}_\Omega \in \Omega
\end{equation}
with $\left| \mathbf{T}_\Xi \right|$ representing its determinant.

The variational formulation of the Laplace's equation in Eq. \ref{equ:HarmonicCoordinateEquMHT} can be derived as
\begin{equation}\label{equ:WeakFormLaplacianBulkFlowMHT}
\left\{\begin{split}
& \mathrm{Find} 
\left\{\begin{split}
& \mathbf{s} \in \left(\mathcal{H}\left(\Omega\right)\right)^3 \\
& \boldsymbol{\lambda}_\mathbf{s} \in \left(\mathcal{H}^{1\over2}\left(\Sigma\right)\right)^3 \\
\end{split}\right. \\
& \mathrm{with} ~ \mathbf{s} = \mathbf{0} ~ \mathrm{at} ~ \forall \mathbf{x}_\Omega \in \Sigma_{v,\Omega} \cup \Sigma_{s,\Omega}, \\
& \mathrm{for} 
\left\{\begin{split}
& \forall \tilde{\mathbf{s}} \in \left(\mathcal{H}\left(\Omega\right)\right)^3 \\
& \forall \tilde{\boldsymbol{\lambda}}_\mathbf{s} \in \left(\mathcal{H}^{-{1\over2}}\left(\Sigma\right)\right)^3\\
\end{split}\right., ~ \mathrm{such~that} \\
& \int_\Omega - \nabla_{\mathbf{x}_\Omega} \mathbf{s} : \nabla_{\mathbf{x}_\Omega} \tilde{\mathbf{s}} \,\mathrm{d}\Omega + \int_\Sigma \left( \mathbf{s} - d_f \mathbf{n}_\Sigma \right) \cdot \tilde{\boldsymbol{\lambda}}_\mathbf{s} + \boldsymbol{\lambda}_\mathbf{s} \cdot \tilde{\mathbf{s}} \,\mathrm{d}\Sigma = 0
\end{split}\right.
\end{equation}
where $\tilde{\mathbf{s}}$ is the test function of $\mathbf{s}$; the Lagrangian multiplier is used to impose the Dirichlet boundary condition of the displacement on $\Sigma$, and $\boldsymbol{\lambda}_\mathbf{s}$ is the Lagrangian multiplier with $\tilde{\boldsymbol{\lambda}}_\mathbf{s}$ representing its test function; $\mathcal{H}\left(\Omega\right)$ is the first order Sobolev space defined on $\Omega$; $\mathcal{H}^{1\over2}\left(\Sigma\right)$ is the trace space defined on $\Sigma$; and $\mathcal{H}^{-{1\over2}}\left(\Sigma\right)$ is the dual space of $\mathcal{H}^{1\over2}\left(\Sigma\right)$.

\subsection{Description of thin-wall pattern and fiber bundle} \label{sec:DescriptionWallPattern}

The thin-wall pattern is defined on the implicit 2-manifold which is defined on the base manifold. The definition of the design variables of the implicit 2-manifold and the thin-wall pattern are the same as that have been presented in Sections \ref{subsec:DesignVariableImplicitManifold} and \ref{subsec:DesignVariablePattern}. Therefore, the related coupling relations and definition of the fiber bundle are the same as that in Sections \ref{subsec:CouplingDesignVariables} and \ref{subsec:FiberBundleMassHeatTransfer}.

\subsection{Coupling of design variables}\label{subsec:CouplingDesignVariablesBulkFlowMHT}

Based on Eqs. \ref{equ:HomeomorphismOmegaBulkFlowMHT} and \ref{equ:TransformedJacobianOmegaMHT}, the gradient operator in the deformed domain $\Xi$ can be transformed into
\begin{equation}\label{equ:BulkTransformedGradDeformedDomMHT}
  \nabla_{\mathbf{x}_\Xi}^{\left(\mathbf{s}\right)} = \mathbf{T}_\Xi^{-1} \nabla_{\mathbf{x}_\Omega},
\end{equation}
where $\nabla_{\mathbf{x}_\Xi}$ is the gradient operator in the deformed domain $\Xi$ and $\nabla_{\mathbf{x}_\Xi}^{\left(\mathbf{s}\right)}$ is the transformed counterpart of $\nabla_{\mathbf{x}_\Xi}$. Based on the transformed gradient operator, the divergence operator in the deformed domain can be transformed into 
\begin{equation}\label{equ:BulkTransformedDivergenceOperatorMHT} 
\begin{split}
  \mathrm{div}_{\mathbf{x}_\Xi}^{\left( \mathbf{s} \right)} \mathbf{g} = \mathrm{tr}\left( \nabla_{\mathbf{x}_\Xi}^{\left( \mathbf{s} \right)} \mathbf{g} \right) = \mathrm{tr} \left( \mathbf{T}_\Xi^{-1} \nabla_{\mathbf{x}_\Omega} \mathbf{g} \right), ~ \forall \mathbf{g} \in \left( \mathcal{H}\left( \Omega \right) \right)^3
\end{split}
\end{equation}
where $\mathrm{div}_{\mathbf{x}_\Xi}$ is the divergence operator in the deformed domain $\Xi$ and $\mathrm{div}_{\mathbf{x}_\Xi}^{\left(\mathbf{s}\right)}$ is the transformed counterpart of $\mathrm{div}_{\mathbf{x}_\Xi}$; and $\mathrm{tr}$ is the operator used to extract the trace of a tensor. The unit outer normal vector at the boundary of the deformed domain can be transformed based on the following parallel relation:
\begin{equation}\label{equ:ParallelRelationForNormalXi}
  \mathbf{n}_{\partial\Xi} \parallel \left[ \mathbf{n}_{\partial\Omega} - \nabla_{\partial\Omega} \left( \mathbf{s} \cdot \mathbf{n}_{\partial\Omega} \right) \right],
\end{equation}
where $\nabla_{\partial\Omega}$ is the tangential gradient operator at $\partial\Omega$. Because of $\mathbf{s} = \mathbf{0}$ at $\forall \mathbf{x}_\Omega \in \Sigma_{v,\Omega} \cup \Sigma_{s,\Omega}$ and $\mathbf{n}_{\partial\Omega} \cdot \nabla_{\mathbf{x}_\Omega} \mathbf{s} = \mathbf{0}$ at $\forall \mathbf{x}_\Omega \in \Sigma_{v_0,\Omega}$, the right side of Eq. \ref{equ:ParallelRelationForNormalXi} satisfies
\begin{equation}\label{equ:DeformedOutwardNormal}
\begin{split}
\mathbf{n}_{\partial\Omega} - \nabla_{\partial\Omega} \left( \mathbf{s} \cdot \mathbf{n}_{\partial\Omega} \right) = \: & \mathbf{n}_{\partial\Omega} - \left\{ \nabla_{\mathbf{x}_\Omega} \left( \mathbf{s} \cdot \mathbf{n}_{\partial\Omega} \right) - \left[ \mathbf{n}_{\partial\Omega} \cdot \nabla_{\mathbf{x}_\Omega} \left( \mathbf{s} \cdot \mathbf{n}_{\partial\Omega} \right) \right] \mathbf{n}_{\partial\Omega} \right\} \\
= \: & \mathbf{n}_{\partial\Omega} - \left\{ \mathbf{s} \cdot \nabla_{\mathbf{x}_\Omega} \mathbf{n}_{\partial\Omega} - \left[ \mathbf{n}_{\partial\Omega} \cdot \left( \mathbf{s} \cdot \nabla_{\mathbf{x}_\Omega} \mathbf{n}_{\partial\Omega} \right) \right] \mathbf{n}_{\partial\Omega} \right\} \\
= \: & \mathbf{n}_{\partial\Omega} - \left\{ \mathbf{s} \cdot \nabla_{\partial\Omega} \mathbf{n}_{\partial\Omega} - \left[ \mathbf{n}_{\partial\Omega} \cdot \left( \mathbf{s} \cdot \nabla_{\partial\Omega} \mathbf{n}_{\partial\Omega} \right) \right] \mathbf{n}_{\partial\Omega} \right\}. \\
\end{split}
\end{equation}
Therefore, $\mathbf{n}_{\partial\Xi}$ can be transformed into
\begin{equation}\label{equ:DeformedOutwardUnitaryNormal}
\begin{split}
\mathbf{n}_{\partial\Xi}^{\left(\mathbf{s}\right)} = { \mathbf{n}_{\partial\Omega} - \left\{ \mathbf{s} \cdot \nabla_{\partial\Omega} \mathbf{n}_{\partial\Omega} - \left[ \mathbf{n}_{\partial\Omega} \cdot \left( \mathbf{s} \cdot \nabla_{\partial\Omega} \mathbf{n}_{\partial\Omega} \right) \right] \mathbf{n}_{\partial\Omega} \right\} \over \left\| \mathbf{n}_{\partial\Omega} - \left\{ \mathbf{s} \cdot \nabla_{\partial\Omega} \mathbf{n}_{\partial\Omega} - \left[ \mathbf{n}_{\partial\Omega} \cdot \left( \mathbf{s} \cdot \nabla_{\partial\Omega} \mathbf{n}_{\partial\Omega} \right) \right] \mathbf{n}_{\partial\Omega} \right\} \right\|_2 }.
\end{split}
\end{equation}
From Eq. \ref{equ:DeformedOutwardNormal}, $\mathbf{n}_{\partial\Xi} = \mathbf{n}_{\partial\Omega}$ is always satisfied at $\Sigma_{v,\Omega} \cup \Sigma_{s,\Omega}$, and it is satisfied at $\Sigma_{v_0,\Omega}$ when $\Sigma_{v_0,\Omega}$ is locally or piecewisely flat.

In Eqs. \ref{equ:BulkTransformedGradDeformedDomMHT}, \ref{equ:BulkTransformedDivergenceOperatorMHT} and \ref{equ:DeformedOutwardUnitaryNormal}, the transformed counterparts of the gradient operator, divergence operator and unit outer normal are distinguished from the original forms by using $\mathbf{s}$ as the superscript. This identification method is used in the following for the other transformed operators and variables.

Because the transformed gradient operator $\nabla_{\mathbf{x}_\Xi}$ depends on $\mathbf{s}$, its first-order variational to $\mathbf{s}$ can be derived as 
\begin{equation}\label{equ:BulkFirstOrderVarisForGradOperatorsMHT}
\begin{split}
& \nabla_{\mathbf{x}_\Xi}^{\left(\mathbf{s}, \tilde{\mathbf{s}}\right)} g = \left( {\partial \mathbf{T}_\Xi^{-1} \over \partial \nabla_{\mathbf{x}_\Omega} \mathbf{s}} : \nabla_{\mathbf{x}_\Omega} \tilde{\mathbf{s}} \right) \nabla_{\mathbf{x}_\Omega} g, ~ \forall g \in \mathcal{H}\left( \Omega \right) ~\mathrm{and}~ \forall \tilde{\mathbf{s}} \in \left(\mathcal{H}\left( \Omega \right)\right)^3;
\end{split}
\end{equation}
similarly, the first-order variational of $\mathrm{div}_{\mathbf{x}_\Xi}$ to $\mathbf{s}$ can be derived as
\begin{equation}\label{equ:BulkFirstOrderVarisForDivergenceOperatorsMHT}
\begin{split}
  & \mathrm{div}_{\mathbf{x}_\Xi}^{\left(\mathbf{s},\tilde{\mathbf{s}} \right)} \mathbf{g} = \mathrm{tr} \left( \left( {\partial \mathbf{T}_\Xi^{-1} \over \partial \nabla_{\mathbf{x}_\Omega} \mathbf{s}} : \nabla_{\mathbf{x}_\Omega} \tilde{\mathbf{s}} \right) \nabla_\Sigma \mathbf{g} \right), ~ \forall \mathbf{g} \in \left(\mathcal{H} \left( \Omega \right)\right)^3 ~ \mathrm{and} ~ \forall \tilde{\mathbf{s}} \in \left(\mathcal{H}\left( \Omega \right)\right)^3;
\end{split}
\end{equation}
and the first-order variational of $\mathbf{n}_{\partial\Xi}^{\left(\mathbf{s}\right)}$ to $\mathbf{s}$ can be derived as
\begin{equation}\label{equ:BulkFirstOrderVariDeformedOutwardUnitaryNormal}
\begin{split}
\mathbf{n}_{\partial\Xi}^{\left(\mathbf{s}, \tilde{\mathbf{s}}\right)} = \: &  - { \tilde{\mathbf{s}} \cdot \nabla_{\partial\Omega} \mathbf{n}_{\partial\Omega} - \left[ \mathbf{n}_{\partial\Omega} \cdot \left( \tilde{\mathbf{s}} \cdot \nabla_{\partial\Omega} \mathbf{n}_{\partial\Omega} \right) \right] \mathbf{n}_{\partial\Omega} \over \left\| \mathbf{n}_{\partial\Omega} - \left\{ \mathbf{s} \cdot \nabla_{\partial\Omega} \mathbf{n}_{\partial\Omega} - \left[ \mathbf{n}_{\partial\Omega} \cdot \left( \mathbf{s} \cdot \nabla_{\partial\Omega} \mathbf{n}_{\partial\Omega} \right) \right] \mathbf{n}_{\partial\Omega} \right\} \right\|_2 } \\
& + { \mathbf{n}_{\partial\Xi}^{\left(\mathbf{s}\right)} \cdot \left\{ \tilde{\mathbf{s}} \cdot \nabla_{\partial\Omega} \mathbf{n}_{\partial\Omega} - \left[ \mathbf{n}_{\partial\Omega} \cdot \left( \tilde{\mathbf{s}} \cdot \nabla_{\partial\Omega} \mathbf{n}_{\partial\Omega} \right) \right] \mathbf{n}_{\partial\Omega} \right\} \over \left\| \mathbf{n}_{\partial\Omega} - \left\{ \mathbf{s} \cdot \nabla_{\partial\Omega} \mathbf{n}_{\partial\Omega} - \left[ \mathbf{n}_{\partial\Omega} \cdot \left( \mathbf{s} \cdot \nabla_{\partial\Omega} \mathbf{n}_{\partial\Omega} \right) \right] \mathbf{n}_{\partial\Omega} \right\} \right\|_2 } \mathbf{n}_{\partial\Xi}^{\left(\mathbf{s}\right)} , ~ \forall \tilde{\mathbf{s}} \in \left(\mathcal{H}\left( \Omega \right)\right)^3.
\end{split}
\end{equation}

Because $\mathbf{s}$ is a differentiable homeomorphism, it can induce a Riemannian metric. Then, the differentials on the deformed domain and the original domain satisfy
\begin{equation}\label{equ:DiffRiemannianBulkMHM}
\left\{\begin{split}
  & \mathrm{d}\Xi = \left| \mathbf{T}_\Xi \right| \mathrm{d}\Omega \\
  & \mathrm{d}\Gamma_{\partial\Xi} = \left| \mathbf{T}_\Xi \right| \left\| \mathbf{T}_\Xi \mathbf{n}_{\partial\Xi}^{\left( \mathbf{s} \right)} \right\|_2^{-1} \mathrm{d}\Sigma_{\partial\Omega} \\
\end{split}\right.,
\end{equation}
where $\mathrm{d}\Xi$, $\mathrm{d}\Gamma_{\partial\Xi}$ and $\mathrm{d}\Sigma_{\partial\Omega}$ are the differentials of $\Xi$, $\partial\Xi$ and $\partial\Omega$, respectively. $K^{\left( \mathbf{s} \right)}$ and $M^{\left( \mathbf{s} \right)}$ are used to represent the Riemannian metric in Eq. \ref{equ:DiffRiemannianBulkMHM}, i.e.
\begin{equation}\label{equ:MetricMLBulkMHM}
  \left\{\begin{split}
  & K^{\left( \mathbf{s} \right)} \doteq \left| \mathbf{T}_\Xi \right| \\
  & M^{\left( \mathbf{s} \right)} \doteq \left| \mathbf{T}_\Xi \right| \left\| \mathbf{T}_\Xi \mathbf{n}_{\partial\Xi}^{\left( \mathbf{s} \right)} \right\|_2^{-1} \\
  \end{split}\right..
\end{equation}

Based on the above relations, the related variables and operators are coupled as illustrated by the arrow chart described as
\[\begin{array}{cccccccc}
 d_m & \xrightarrow{\mathrm{Eq.~}\ref{equ:PDEFilterzmBaseStructureMHM}} & d_f & \xrightarrow{\mathrm{Eq.~}\ref{equ:HarmonicCoordinateEquMHT}} & \mathbf{s} & \\
 & & \bigg\downarrow\vcenter{\rlap{\scriptsize{Eq.~\ref{equ:TransformedTangentialOperatorMHM}}}} & &  \bigg\downarrow\vcenter{\rlap{\scriptsize{Eqs.~\ref{equ:TransformedJacobianOmegaMHT}, \ref{equ:BulkTransformedGradDeformedDomMHT}~\&~\ref{equ:BulkTransformedDivergenceOperatorMHT}}}} \\
 & & \left\{ \mathrm{div}_\Gamma, \nabla_\Gamma, \mathbf{n}_\Gamma \right\} & & \left\{\mathbf{T}_\Xi, \mathrm{div}_{\mathbf{x}_\Xi}, \nabla_{\mathbf{x}_\Xi} \right\} & & \\
 & & \bigg\downarrow\vcenter{\rlap{\scriptsize{Eq.~\ref{equ:PDEFilterGammaFilberMHM}}}} & & \bigg\downarrow\vcenter{\rlap{\scriptsize{Eq.~\ref{equ:MetricMLBulkMHM}}}} \\
 \gamma & \xrightarrow{\mathrm{Eq.~}\ref{equ:PDEFilterGammaFilberMHM}} & \gamma_f & & \left\{ K^{\left( \mathbf{s} \right)}, M^{\left( \mathbf{s} \right)} \right\}. & & \\
\end{array}\]

\subsection{Mass transfer in volume flow}\label{sec:MassTransferBulkFlows}

The mass transfer process in the volume flow can be described by the Navier-Stokes equations and the convection-diffusion equation.

\subsubsection{Navier-Stokes equations for volume flow} \label{sec:NavierStokesEquBulkFlows}

Based on the description of the deformed 3D domain in Eq. \ref{equ:HomeomorphismOmegaBulkFlowMHT}, the Navier-Stokes equations used to describe the volume flow can be derived as
\begin{equation}\label{equ:NavierStokesEquBulkFlowMassTransferMHT}
\left.\begin{split}
\rho \mathbf{u} \cdot \nabla_{\mathbf{x}_\Xi} \mathbf{u} - \mathrm{div}_{\mathbf{x}_\Xi} \left[ \eta \left( \nabla_{\mathbf{x}_\Xi} \mathbf{u} + \nabla_{\mathbf{x}_\Xi} \mathbf{u}^\mathrm{T} \right) \right] + \nabla_{\mathbf{x}_\Xi} p = \: & \mathbf{0}\\
- \mathrm{div}_{\mathbf{x}_\Xi} \mathbf{u} = \: & 0 
\end{split}\right\} ~ \forall \mathbf{x}_\Xi \in \Xi.
\end{equation}
The material interpolation is implemented on the no-jump part and the no-slip part of the implicit 2-manifold, i.e.
\begin{equation}\label{equ:MaterialInterpolationMassTransferBulkFlow}
   \left\llbracket \left[ \eta \left( \nabla_{\mathbf{x}_\Xi} \mathbf{u} + \nabla_{\mathbf{x}_\Xi} \mathbf{u}^\mathrm{T} \right) - p \right] \mathbf{n}_{\partial\Xi} \right\rrbracket + \alpha \left(\gamma_p\right) \mathbf{u} = \mathbf{0}, ~ \forall \mathbf{x}_\Gamma \in \Gamma
\end{equation}
where $\mathbf{n}_{\partial\Xi}$ is the unit outer normal vector at $\partial\Xi$; and $\alpha$ is the material interpolation and it is expressed as
\begin{equation}\label{equ:AlphaMassTransferBulkFlow}
  \alpha \left(\gamma_p\right) = \alpha_{\max} q {1-\gamma_p \over q + \gamma_p}
\end{equation}
with $\alpha_{\max}$ representing the maximal value of $\alpha$. Theoretically, $\alpha_{max}$ should be infinite; numerically, it is valued to be finite and large enough to approximate the infinity and ensure the numerical stability. In Eq. \ref{equ:MaterialInterpolationMassTransferBulkFlow}, the design variable of the thin-wall pattern is defined on $\Gamma$ instead of $\Gamma_D$ in Section \ref{sec:PorousMediumModelSurfaceNSEqus}. Therefore, the surface-PDE filter of the design variable of the thin-wall pattern is implemented as
      \begin{equation}\label{equ:PDEFilterGammaFilberMHMVolumeFlow} 
        \left\{\begin{split}
        - \mathrm{div}_\Gamma \left( r_f^2 \nabla_\Gamma \gamma_f \right) + \gamma_f & = \gamma, ~\forall \mathbf{x}_\Gamma \in \Gamma \\
        \mathbf{n}_{\boldsymbol\tau_\Gamma} \cdot \nabla_\Gamma \gamma_f & = 0, ~\forall \mathbf{x}_\Gamma \in \partial\Gamma \\
        \end{split}\right..
      \end{equation}
The variational formulation of the surface-PDE filter in Eq. \ref{equ:PDEFilterGammaFilberMHMVolumeFlow} is considered in the first order Sobolev space defined on $\Gamma$. It can be derived based on the Galerkin method as
\begin{equation}\label{equ:VariationalFormulationPDEFilterMHMVolumeFlow} 
\left\{\begin{split}
     & \mathrm{Find}~\gamma_f \in\mathcal{H}\left(\Gamma\right) ~ \mathrm{for}~\gamma \in \mathcal{L}^2\left(\Gamma\right) ~ \mathrm{and} ~ \forall \tilde{\gamma}_f \in \mathcal{H}\left(\Gamma\right), \\
     & \mathrm{such~that} ~ \int_{\Gamma} r_f^2 \nabla_\Gamma \gamma_f \cdot \nabla_\Gamma \tilde{\gamma}_f + \gamma_f\tilde{\gamma}_f - \gamma \tilde{\gamma}_f \,\mathrm{d}\Gamma = 0.
\end{split}\right.
\end{equation}
Based on the transformed tangential gradient operator in Eq. \ref{equ:TransformedTangentialOperatorMHM} and the homeomorphism between $\mathcal{H}\left(\Gamma\right)$ and $\mathcal{H}\left(\Sigma\right)$ described in Eq. \ref{equ:NormalDisplacementDistributionMHM}, Eq. \ref{equ:VariationalFormulationPDEFilterMHMVolumeFlow} can be transformed into
\begin{equation}\label{equ:CoupledVariationalPDEFilterVolumeFlow}
\left\{\begin{split}
     & \mathrm{Find}~\gamma_f \in\mathcal{H}\left(\Sigma\right) ~ \mathrm{for} ~ \gamma \in \mathcal{L}^2\left(\Sigma\right) ~ \mathrm{and} ~ \forall \tilde{\gamma}_f \in \mathcal{H}\left(\Sigma\right), \\
     & \mathrm{such~that} ~ \int_\Sigma \left( r_f^2 \nabla_\Gamma^{\left( d_f \right)} \gamma_f \cdot \nabla_\Gamma^{\left( d_f \right)} \tilde{\gamma}_f + \gamma_f\tilde{\gamma}_f - \gamma \tilde{\gamma}_f \right) M^{\left( d_f \right)} \,\mathrm{d}\Sigma = 0.
\end{split}\right.
\end{equation}

The boundary conditions for the Navier-Stokes equations in Eq. \ref{equ:NavierStokesEquBulkFlowMassTransferMHT} include the Dirichlet boundary conditions with known fluid velocity at the inlet and zero velocity at the walls and the Neumann boundary condition with zero stress at the outlet:
\begin{equation}\label{equ:BoundaryCondBulkNSEquBulkFlowMHT}
  \left\{\begin{split}
  & \mathbf{u} = \mathbf{u}_{\Gamma_{v,\Xi}}, ~ \forall \mathbf{x}_\Xi \in \Gamma_{v,\Xi} ~~ \left( \mathrm{Inlet ~ boundary ~ condition} \right) \\
  & \mathbf{u} = \mathbf{0}, ~ \forall \mathbf{x}_\Xi \in \Gamma_{v_0,\Xi} ~~ \left( \mathrm{Wall ~ boundary ~ condition} \right) \\
  & \left[ \eta \left( \nabla_{\mathbf{x}_\Xi} \mathbf{u} + \nabla_{\mathbf{x}_\Xi} \mathbf{u}^\mathrm{T} \right) + p \right] \mathbf{n}_{\partial\Xi} = \mathbf{0}, ~ \forall \mathbf{x}_\Xi \in \Gamma_{s,\Xi} ~~ \left( \mathrm{Outlet ~ boundary ~ condition} \right)
  \end{split}\right.
\end{equation}
where $\Gamma_{v,\Xi}$, $\Gamma_{v_0,\Xi}$, and $\Gamma_{s,\Xi}$ are the inlet, wall and outlet boundaries, respectively, and they correspond to the boundary parts of $\Sigma_{v,\Omega}$, $\Sigma_{v_0,\Omega}$, and $\Sigma_{s,\Omega}$ included in $\partial\Omega$, respectively; and $\mathbf{u}_{\Gamma_{v,\Xi}}$ is the known velocity at the inlet and $\mathbf{u}_{\Sigma_{v,\Omega}}$ is its counterpart defined on $\Sigma_{v,\Omega}$.

Based on the Galerkin method, the variational formulation of the Navier-Stokes equations in Eq. \ref{equ:NavierStokesEquBulkFlowMassTransferMHT} is considered in the first order Sobolev space defined on the deformed domain $\Xi$:
\begin{equation}\label{equ:VariationalFormulationNavierStokesEquBulkFlowMHT}
\left\{\begin{split}
  & \mathrm{Find} \left\{\begin{split}
    & \mathbf{u}\in\left(\mathcal{H}\left(\Xi\right)\right)^3~\mathrm{with} ~ \left\{ \begin{split}
    & \mathbf{u} = \mathbf{u}_{\Gamma_{v,\Xi}}~ \mathrm{at} ~ \forall \mathbf{x}_\Xi \in \Gamma_{v,\Xi} \\
    & \mathbf{u} = \mathbf{0}~ \mathrm{at} ~ \forall \mathbf{x}_\Xi \in \Gamma_{v_0,\Xi} \\
    \end{split}\right.\\
  & p \in \mathcal{H}\left(\Xi\right) \\
  \end{split}\right.\\
  & \mathrm{for} \left\{\begin{split}
  & \forall \tilde{\mathbf{u}} \in\left(\mathcal{H}\left(\Xi\right)\right)^3 \\
  & \forall \tilde{p} \in \mathcal{H}\left(\Xi\right) \\
  \end{split}\right., ~ \mathrm{such~that} \\
  &\int_\Xi \rho \left( \mathbf{u} \cdot \nabla_{\mathbf{x}_\Xi} \right) \mathbf{u} \cdot \tilde{\mathbf{u}} + {\eta\over2} \left( \nabla_{\mathbf{x}_\Xi} \mathbf{u} + \nabla_{\mathbf{x}_\Xi} \mathbf{u}^\mathrm{T} \right) : \left( \nabla_{\mathbf{x}_\Xi} \tilde{\mathbf{u}} + \nabla_{\mathbf{x}_\Xi} \tilde{\mathbf{u}}^\mathrm{T} \right) - p\,\mathrm{div}_{\mathbf{x}_\Xi} \tilde{\mathbf{u}} \\
  & - \tilde{p} \,\mathrm{div}_{\mathbf{x}_\Xi} \mathbf{u} \,\mathrm{d}\Xi - \sum_{E_\Xi\in\mathcal{E}_\Xi} \int_{E_\Xi} \tau_{BP,\Xi} \nabla_{\mathbf{x}_\Xi} p \cdot \nabla_{\mathbf{x}_\Xi} \tilde{p} \,\mathrm{d}\Xi + \int_\Gamma \alpha \mathbf{u} \cdot \tilde{\mathbf{u}} \, \mathrm{d}\Gamma = 0 \\
\end{split}\right.
\end{equation}
where $\mathcal{H}\left(\Xi\right)$ is the first order Sobolev space defined on $\Xi$; the Brezzi-Pitk\"{a}ranta stabilization term
\begin{equation}
  - \sum_{E_\Xi\in\mathcal{E}_\Xi} \int_{E_\Xi} \tau_{BP,\Xi} \nabla_{\mathbf{x}_\Xi} p \cdot \nabla_{\mathbf{x}_\Xi} \tilde{p} \,\mathrm{d}\Xi
\end{equation}
is imposed on the variational formulation, in order to use linear finite elements to solve both the fluid velocity and pressure \cite{DoneaWiley2003}; $\mathcal{E}_\Xi$ is an elementization of $\Xi$; and $E_\Xi$ is an element of the elementization $\mathcal{E}_\Xi$. The stabilization parameter is chosen as \cite{DoneaWiley2003}
\begin{equation}\label{equ:NSBulkStabilizationTermCD}
 \tau_{BP,\Xi} = {h_{E_\Xi}^2 \over 12\eta},
\end{equation}
where $h_{E_\Xi}$ is the size of the element $E_\Xi$. Because the element $E_\Xi$ and the elementization $\mathcal{E}_\Xi$ for the deformed domain $\Xi$ are implicitly defined on $\Omega$, they are derived from the design variable for the implicit 2-manifold $\Gamma$ and the explicit elementization $\mathcal{E}_\Omega$ of $\Omega$. Based on Eqs. \ref{equ:HomeomorphismOmegaBulkFlowMHT} and \ref{equ:TransformedJacobianOmegaMHT}, $h_{E_\Xi}^2$ in $\tau_{BP,\Xi}$ can be approximated based on the volume of $E_\Xi$. Then, it can be transformed into
\begin{equation}\label{equ:ElementVolumeTransformationMHM} 
\begin{split}
  h_{E_\Xi}^3 \approx \, & \int_{E_\Xi} 1 \, \mathrm{d}\Xi \\
  = \, & \int_{E_\Omega} K^{\left( \mathbf{s} \right)} \, \mathrm{d}\Omega \\
  \approx \, & h_{E_\Omega}^3 \int_{E_\Omega} K^{\left( \mathbf{s} \right)} \, \mathrm{d}\Omega \bigg/ \int_{E_\Omega} 1 \, \mathrm{d}\Omega \\
  = \, & h_{E_\Omega}^3 \bar{K}^{\left( \mathbf{s} \right)}_{E_\Omega},
\end{split}
\end{equation}
where $h_{E_\Omega}$ is the size of the element $E_\Omega$ representing an element of the explicit elementization $\mathcal{E}_\Omega$ of $\Omega$; $\bar{K}^{\left( \mathbf{s} \right)}_{E_\Omega}$ is the average value of $K^{\left( \mathbf{s} \right)}$ in the element $E_\Omega$; and the volume of $E_\Omega$ can be approximated as $h_{E_\Omega}^3$, i.e. $\int_{E_\Omega} 1 \, \mathrm{d}\Omega \approx h_{E_\Omega}^3$. Because the elementization satisfies $h_{E_\Omega}^3 \ll \left| \Omega \right|$ with $\left| \Omega \right|$ representing the volume of $\Omega$, $\bar{K}^{\left( \mathbf{s} \right)}_{E_\Omega}$ can be well approximated by the value of $K^{\left( \mathbf{s} \right)}$ at $\forall \mathbf{x}_\Omega \in E_\Omega$, i.e.
\begin{equation}\label{equ:ApproximationBulkMetricAverageMHM}
  \bar{K}^{\left( \mathbf{s} \right)}_{E_\Omega} \approx K^{\left( \mathbf{s} \right)}, ~ \forall \mathbf{x}_\Omega \in E_\Omega.
\end{equation} 
Therefore, the stabilization parameter $\tau_{BP,\Xi}$ in Eq. \ref{equ:NSBulkStabilizationTermCD} can be transformed into
\begin{equation}\label{equ:TransformedNSBulkStabilizationTermCD}
  \tau_{BP,\Xi}^{\left( \mathbf{s} \right)} = {h_{E_\Omega}^2 \over 12\eta} \left( K^{\left( \mathbf{s} \right)} \right)^{2\over3}.
\end{equation}

Because of $\mathbf{s} = \mathbf{0}$ at $\Sigma_{v,\Omega} \cup \Sigma_{s,\Omega}$, $\Gamma_{v,\Xi}$, $\Gamma_{s,\Xi}$, $\mathbf{u}_{\Gamma_{v,\Xi}}$ and $\mathbf{n}_{\partial\Xi}$ on $\Gamma_{v,\Xi}\cup\Gamma_{s,\Xi}$ coincide with $\Sigma_{v,\Omega}$, $\Sigma_{s,\Omega}$, $\mathbf{u}_{\Sigma_{v,\Omega}}$ and $\mathbf{n}_{\partial\Omega}$ on $\Sigma_{v,\Omega}\cup\Sigma_{s,\Omega}$, respectively. Then, based on the relations in Sections \ref{sec:DesignVariablePattern} and \ref{sec:DescriptionDeformedDomain}, the variational formulation in Eq. \ref{equ:VariationalFormulationNavierStokesEquBulkFlowMHT} can be transformed into the form defined on $\Omega$:
\begin{equation}\label{equ:TransformedVariationalFormulationNavierStokesEquBulkFlowMHT}
\left\{\begin{split}
  & \mathrm{Find} \left\{\begin{split}
    & \mathbf{u}\in\left(\mathcal{H}\left(\Omega\right)\right)^3~\mathrm{with} ~ \left\{ \begin{split}
    & \mathbf{u} = \mathbf{u}_{\Gamma_{v,\Omega}}~ \mathrm{at} ~ \forall \mathbf{x}_\Omega \in \Sigma_{v,\Omega} \\
    & \mathbf{u} = \mathbf{0}~ \mathrm{at} ~ \forall \mathbf{x}_\Omega \in \Sigma_{v_0,\Omega} \\
    \end{split}\right.\\
  & p \in \mathcal{H}\left(\Omega\right) \\
  \end{split}\right.\\
  & \mathrm{for} \left\{\begin{split}
  & \forall \tilde{\mathbf{u}} \in\left(\mathcal{H}\left(\Omega\right)\right)^3 \\
  & \forall \tilde{p} \in \mathcal{H}\left(\Omega\right) \\
  \end{split}\right., ~ \mathrm{such~that} \\
  &\int_\Omega \Big[ \rho \left( \mathbf{u} \cdot \nabla_{\mathbf{x}_\Xi}^{\left(\mathbf{s}\right)} \right) \mathbf{u} \cdot \tilde{\mathbf{u}} + {\eta\over2} \left( \nabla_{\mathbf{x}_\Xi}^{\left(\mathbf{s}\right)} \mathbf{u} + \nabla_{\mathbf{x}_\Xi}^{\left(\mathbf{s}\right)} \mathbf{u}^\mathrm{T} \right) : \left( \nabla_{\mathbf{x}_\Xi}^{\left(\mathbf{s}\right)} \tilde{\mathbf{u}} + \nabla_{\mathbf{x}_\Xi}^{\left(\mathbf{s}\right)} \tilde{\mathbf{u}}^\mathrm{T} \right) - p\,\mathrm{div}_{\mathbf{x}_\Xi}^{\left(\mathbf{s}\right)} \tilde{\mathbf{u}} \\
  & - \tilde{p} \,\mathrm{div}_{\mathbf{x}_\Xi}^{\left(\mathbf{s}\right)} \mathbf{u} \Big] K^{\left( \mathbf{s} \right)} \,\mathrm{d}\Omega - \sum_{E_\Omega\in\mathcal{E}_\Omega} \int_{E_\Omega} \tau_{BP,\Xi}^{\left(\mathbf{s}\right)} \nabla_{\mathbf{x}_\Xi}^{\left(\mathbf{s}\right)} p \cdot \nabla_{\mathbf{x}_\Xi}^{\left(\mathbf{s}\right)} \tilde{p} K^{\left( \mathbf{s} \right)} \,\mathrm{d}\Omega \\
  & + \int_\Sigma \alpha \mathbf{u} \cdot \tilde{\mathbf{u}} M^{\left( d_f \right)} \, \mathrm{d}\Sigma = 0.
\end{split}\right.
\end{equation}

\subsubsection{Convection-diffusion equation for volume flow} \label{sec:ConvectionDiffusionEquBulkFlows}

The mass transfer process in the volume flow can be described by the convection-diffusion equation defined on the deformed domain:
\begin{equation}\label{equ:BulkConvectionDiffusionEqu}
  \mathbf{u} \cdot \nabla_{\mathbf{x}_\Xi} c + \nabla_{\mathbf{x}_\Xi} \cdot \left( D \nabla_{\mathbf{x}_\Xi} c \right) = 0, ~ \forall \mathbf{x}_\Xi \in \Xi.
\end{equation}
The boundary conditions for the convection-diffusion equation in Eq. \ref{equ:BulkConvectionDiffusionEqu} include the Dirichlet boundary condition at the inlet with know distribution of concentration and the Neumann boundary condition at the walls and outlet with insulation:
\begin{equation}\label{equ:BoundaryCondBulkCDEquMT}
  \left\{\begin{split}
  & c = c_0, ~ \forall \mathbf{x}_\Xi \in \Gamma_{v,\Xi} \\
  & \mathbf{n}_{\partial\Xi} \cdot \nabla_{\mathbf{x}_\Xi} c = 0, ~ \forall \mathbf{x}_\Xi \in \Gamma_{v_0,\Xi} \cup \Gamma_{s,\Xi} 
  \end{split}\right..
\end{equation}

Based on the Galerkin method, the variational formulation of the convection-diffusion equation is considered in the first order Sobolev space defined on the deformed domain $\Xi$:
\begin{equation}\label{equ:VariationalFormulationBulkConvecDiffusEqu}
\left\{\begin{split}
  & \mathrm{Find}~ c\in\mathcal{H}\left(\Xi\right)~\mathrm{with} ~ c = c_0~ \mathrm{at} ~ \forall \mathbf{x}_\Xi \in \Gamma_{v,\Xi}, ~ \mathrm{for} ~ \forall \tilde{c} \in \mathcal{H}\left(\Xi\right), ~ \mathrm{such~that} \\
  & \int_\Xi \left( \mathbf{u} \cdot \nabla_{\mathbf{x}_\Xi} c \right) \tilde{c} + D \nabla_{\mathbf{x}_\Xi} c \cdot \nabla_{\mathbf{x}_\Xi} \tilde{c} \,\mathrm{d}\Xi + \sum_{E_\Xi\in\mathcal{E}_\Xi} \int_{E_\Xi} \tau_{PG,\Xi} \left( \mathbf{u} \cdot \nabla_{\mathbf{x}_\Xi} c \right) \left( \mathbf{u} \cdot \nabla_{\mathbf{x}_\Xi} \tilde{c} \right) \,\mathrm{d}\Xi = 0 \\
\end{split}\right.
\end{equation}
where the Petrov-Galerkin stabilization term
\begin{equation}
  \sum_{E_\Xi\in\mathcal{E}_\Xi} \int_{E_\Xi} \tau_{PG,\Xi} \left( \mathbf{u} \cdot \nabla_{\mathbf{x}_\Xi} c \right) \left( \mathbf{u} \cdot \nabla_{\mathbf{x}_\Xi} \tilde{c} \right) \,\mathrm{d}\Xi
\end{equation}
with $\tau_{PG,\Xi}$ representing the stabilization parameter is imposed on the variational formulation, in order to use linear finite elements to solve the distribution of the concentration \cite{DoneaWiley2003}. The stabilization parameter is chosen as \cite{DoneaWiley2003}
\begin{equation}\label{equ:CDBulkStabilizationTermCD}
 \tau_{PG,\Xi} = \left( {4 \over h_{E_\Xi}^2 D} + {2 \left\| \mathbf{u} \right\|_2 \over h_{E_\Xi} } \right)^{-1}.
\end{equation}
Based on Eqs. \ref{equ:ElementVolumeTransformationMHM} and \ref{equ:ApproximationBulkMetricAverageMHM}, $\tau_{PG,\Xi}$ can be transformed into
\begin{equation}\label{equ:TransformedCDBulkStabilizationTermCD}
  \tau_{PG,\Xi}^{\left( \mathbf{s} \right)} =\left( {4 \over h_{E_\Omega}^2 \left( K^{\left( \mathbf{s} \right)} \right)^{2\over3} D} + {2 \left\| \mathbf{u} \right\|_2 \over h_{E_\Omega} \left( K^{\left( \mathbf{s} \right)} \right)^{1\over3} } \right)^{-1}.
\end{equation}

Based on the coupling relations in Section \ref{subsec:CouplingDesignVariablesBulkFlowMHT}, the variational formulation in Eq. \ref{equ:VariationalFormulationBulkConvecDiffusEqu} can be transformed into the form defined on $\Omega$:
\begin{equation}\label{equ:TransformedVariationalFormulationBulkCDEqu}
\left\{\begin{split}
  & \mathrm{Find}~ c\in\mathcal{H}\left(\Omega\right)~\mathrm{with} ~ c = c_0~ \mathrm{at} ~ \forall \mathbf{x}_\Omega \in \Sigma_{v,\Omega}, ~ \mathrm{for} ~ \forall \tilde{c} \in \mathcal{H}\left(\Omega\right), \\
  & \mathrm{such~that} ~ \int_\Omega \left[ \left( \mathbf{u} \cdot \nabla_{\mathbf{x}_\Xi}^{\left(\mathbf{s}\right)} c \right) \tilde{c} + D \nabla_{\mathbf{x}_\Xi}^{\left(\mathbf{s}\right)} c \cdot \nabla_{\mathbf{x}_\Xi}^{\left(\mathbf{s}\right)} \tilde{c} \right] K^{\left(\mathbf{s}\right)} \,\mathrm{d}\Omega \\
  & + \sum_{E_\Omega\in\mathcal{E}_\Omega} \int_{E_\Omega} \tau_{PG,\Xi}^{\left(\mathbf{s}\right)} \left( \mathbf{u} \cdot \nabla_{\mathbf{x}_\Xi}^{\left(\mathbf{s}\right)} c \right) \left( \mathbf{u} \cdot \nabla_{\mathbf{x}_\Xi}^{\left(\mathbf{s}\right)} \tilde{c} \right) K^{\left(\mathbf{s}\right)} \,\mathrm{d}\Omega = 0.
\end{split}\right.
\end{equation}

\subsubsection{Design objective and constraint of pressure drop} \label{sec:DesignObjectiveConstraintBulkNSCDMHT}

For the mass transfer problem in the volume flow, the desired performance of the fluid structure can also be set to achieve the anticipated distribution of the concentration at the outlet, and it can be measured by the deviation between the obtained and anticipated distribution of the concentration. Therefore, the design objective is considered as 
\begin{equation}\label{equ:DesignObjectiveBulkCDMHT}
  J_c = \int_{\Gamma_{s,\Xi}} \left( c - \bar{c} \right)^2 \,\mathrm{d}\Gamma_{\partial\Xi} \bigg/ \int_{\Gamma_{v,\Xi}} \left( c_0 - \bar{c} \right)^2 \,\mathrm{d}\Gamma_{\partial\Xi},
\end{equation}
where $\bar{c}$ is the anticipated distribution of the concentration at the outlet and it is linearly mapped onto the inlet for the reference value of the concentration deviation. 
Because of $\mathbf{s} = \mathbf{0}$ at $\Sigma_{v,\Omega} \cup \Sigma_{s,\Omega}$, $\Gamma_{v,\Xi}$ and $\Gamma_{s,\Xi}$ coincide with $\Sigma_{v,\Omega}$ and $\Sigma_{s,\Omega}$, respectively.
Therefore, the design objective in Eq. \ref{equ:DesignObjectiveBulkCDMHT} can be transformed into
\begin{equation}\label{equ:FurtherTransformedDesignObjectiveBulkCD}
\begin{split}
  J_c^{\left( \mathbf{s} \right)} = \: & \int_{\Sigma_{s,\Omega}} \left( c - \bar{c} \right)^2 M^{\left( \mathbf{s} \right)} \,\mathrm{d}\Sigma_{\partial\Omega} \bigg/ \int_{\Sigma_{v,\Omega}} \left( c_0 - \bar{c} \right)^2 M^{\left( \mathbf{s} \right)} \,\mathrm{d}\Sigma_{\partial\Omega} \\
   = \: & \int_{\Sigma_{s,\Omega}} \left( c - \bar{c} \right)^2 \,\mathrm{d}\Sigma_{\partial\Omega} \bigg/ \int_{\Sigma_{v,\Omega}} \left( c_0 - \bar{c} \right)^2 \,\mathrm{d}\Sigma_{\partial\Omega}.
\end{split}
\end{equation}

A constraint of the pressure drop between the inlet and outlet is imposed to ensure the patency of the fluid structure for mass transfer in the volume flow:
\begin{equation}\label{equ:PressureConstraintBulkNSCD}
  \left| \Delta P \left/ \: \Delta P_0 - 1 \right. \right| \leq 1\times10^{-3},
\end{equation}
where the pressure drop between the inlet and outlet is
\begin{equation}\label{equ:PressureDropBulkNSCD}
  \Delta P = \int_{\Gamma_{v,\Xi}} p \,\mathrm{d}\Gamma_{\partial\Xi} - \int_{\Gamma_{s,\Xi}} p \,\mathrm{d}\Gamma_{\partial\Xi}.
\end{equation}
Based on Eqs. \ref{equ:DiffRiemannianBulkMHM} and \ref{equ:MetricMLBulkMHM}, the pressure drop in Eq. \ref{equ:PressureDropBulkNSCD} can be transformed into
\begin{equation}\label{equ:TransformedPressureConstraintSurfaceNSCD}
\begin{split}
  \Delta P^{\left(\mathbf{s}\right)} = \: & \int_{\Sigma_{v,\Omega}} p M^{\left( \mathbf{s} \right)} \,\mathrm{d}\Sigma_{\partial\Omega} - \int_{\Sigma_{s,\Omega}} p M^{\left( \mathbf{s} \right)} \,\mathrm{d}\Sigma_{\partial\Omega} \\
   = \: & \int_{\Sigma_{v,\Omega}} p \,\mathrm{d}\Sigma_{\partial\Omega} - \int_{\Sigma_{s,\Omega}} p \,\mathrm{d}\Sigma_{\partial\Omega}. \\
\end{split}
\end{equation}

\subsubsection{Fiber bundle topology optimization problem}\label{sec:FiberBundleToopMTBulkFlows}

Based on the above introduction, the topology optimization problem for mass transfer in the volume flow can be constructed to optimize the fiber bundle in Eq. \ref{equ:FiberBundleMHM} for the thin-wall structure defined on the implicit 2-manifold:
\begin{equation}\label{equ:VarProToopBulkNSCDMHT}
\left\{\begin{split}
  & \mathrm{Find} \left\{\begin{split}
  & \gamma: \Gamma \mapsto \left[0,1\right] \\
  & d_m: \Sigma \mapsto \left[0,1\right]\end{split}\right.~ \mathrm{for} ~
  \left(\Sigma \times \left(\Gamma \times \left[0,1\right]\right), \Sigma, proj_1, \Gamma \times \left[0,1\right] \right), \\
  & \mathrm{to} ~ \mathrm{minimize}~{J_c \over J_{c,0}}~ \mathrm{with} ~ J_c = \int_{\Gamma_{s,\Xi}} \left( c - \bar{c} \right)^2 \,\mathrm{d}\Gamma_{\partial\Xi} \bigg/ \int_{\Gamma_{v,\Xi}} \left( c_0 - \bar{c} \right)^2 \,\mathrm{d}\Gamma_{\partial\Xi}, \\
  & \mathrm{constrained} ~ \mathrm{by} \\
  & \left\{\begin{split}
  & \begin{split}
  & \left\{\begin{split}
    & \left\{\begin{split}
       & \rho \mathbf{u} \cdot \nabla_{\mathbf{x}_\Xi} \mathbf{u} - \mathrm{div}_{\mathbf{x}_\Xi} \left[ \eta \left( \nabla_{\mathbf{x}_\Xi} \mathbf{u} + \nabla_{\mathbf{x}_\Xi} \mathbf{u}^\mathrm{T} \right) \right] + \nabla_{\mathbf{x}_\Xi} p = \mathbf{0}, ~ \forall \mathbf{x}_\Xi \in \Xi \\
       & - \mathrm{div}_{\mathbf{x}_\Xi} \mathbf{u} = 0, ~ \forall \mathbf{x}_\Xi \in \Xi \\
       & \left\llbracket \left[ \eta \left( \nabla_{\mathbf{x}_\Xi} \mathbf{u} + \nabla_{\mathbf{x}_\Xi} \mathbf{u}^\mathrm{T} \right) - p \right] \mathbf{n}_{\partial\Xi} \right\rrbracket + \alpha \left(\gamma_p\right) \mathbf{u} = \mathbf{0}, ~ \forall \mathbf{x}_\Gamma \in \Gamma \\
    \end{split}\right. \\
    & \alpha \left(\gamma_p\right) = \alpha_{\max} q {1-\gamma_p \over q + \gamma_p} \\
    \end{split}\right.
    \end{split} \\
  & \mathbf{u} \cdot \nabla_{\mathbf{x}_\Xi} c + \nabla_{\mathbf{x}_\Xi} \cdot \left( D \nabla_{\mathbf{x}_\Xi} c \right) = 0, ~ \forall \mathbf{x}_\Xi \in \Xi \\
  & \left\{\begin{split}
  & \left\{\begin{split}
        & - \mathrm{div}_\Gamma \left( r_f^2 \nabla_\Gamma \gamma_f \right) + \gamma_f = \gamma,~\forall \mathbf{x}_\Gamma \in \Gamma \\
        & \mathbf{n}_{\boldsymbol\tau_\Gamma} \cdot \nabla_\Gamma \gamma_f = 0,~\forall \mathbf{x}_\Gamma \in \partial\Gamma \\
    \end{split}\right. \\
    & \gamma_p = { \tanh\left(\beta \xi\right) + \tanh\left(\beta \left(\gamma_f-\xi\right)\right) \over \tanh\left(\beta \xi\right) + \tanh\left(\beta \left(1-\xi\right)\right)} \\
    \end{split}\right. \\
  & \left\{
        \begin{split}
          & - \mathrm{div}_\Sigma \left( r_m^2 \nabla_\Sigma d_f \right) + d_f = A_d \left( d_m - {1\over2} \right), ~ \forall \mathbf{x}_\Sigma \in \Sigma \\
          & \mathbf{n}_{\boldsymbol\tau_\Sigma} \cdot \nabla_\Sigma d_f = 0, ~ \forall \mathbf{x}_\Sigma \in \partial \Sigma \\
        \end{split}\right. \\
  & \left\{\begin{split}
         & \mathrm{div}_{\mathbf{x}_\Omega} \left( \nabla_{\mathbf{x}_\Omega} \mathbf{s} \right) = \mathbf{0}, ~ \forall \mathbf{x}_\Omega \in \Omega \\
         & \mathbf{s} = \mathbf{0}, ~ \forall \mathbf{x}_\Omega \in \Sigma_{v,\Omega} \cup \Sigma_{s,\Omega} \\
         & \mathbf{s} = d_f \mathbf{n}_\Sigma, ~ \forall \mathbf{x}_\Omega \in \Sigma \\
         & \mathbf{n}_{\partial\Omega} \cdot \nabla_{\mathbf{x}_\Omega} \mathbf{s} = \mathbf{0}, ~ \forall \mathbf{x}_\Omega \in \Sigma_{v_0,\Omega} \\
  \end{split}\right. \\
  & \Xi = \left\{ \mathbf{x}_\Xi \left| \, \begin{split}
  \mathbf{x}_\Xi = \: & \mathbf{x}_\Omega + \mathbf{s}, ~ \forall \mathbf{x}_\Omega \in \Omega \\
  \mathbf{x}_\Xi = \: & \mathbf{x}_\Gamma, ~ \forall \mathbf{x}_\Gamma \in \Gamma \\
  \mathbf{x}_\Omega = \: & \mathbf{x}_\Sigma, ~ \forall \mathbf{x}_\Sigma \in \Sigma \\
  \mathbf{x}_\Gamma = \: & \mathbf{x}_\Sigma + d_f \mathbf{n}_\Sigma 
  \end{split}\right. \right\} \\
  & \left| \Delta P \left/ \: \Delta P_0 - 1 \right. \right| \leq 1\times10^{-3}, ~\mathrm{with} ~ \Delta P = \int_{\Gamma_{v,\Xi}} p \,\mathrm{d}\Gamma_{\partial\Xi} - \int_{\Gamma_{s,\Xi}} p \,\mathrm{d}\Gamma_{\partial\Xi}. \\
\end{split}\right.
\end{split}\right.
\end{equation}

The coupling relations among the variables, functions, and differential operators in Eq. \ref{equ:VarProToopBulkNSCDMHT} are illustrated by the arrow chart described as
\[\begin{array}{cccccccc}
 \textcolor{blue}{d_m} & \xrightarrow{\mathrm{Eq.~}\ref{equ:PDEFilterzmBaseStructureMHM}} & d_f & \xrightarrow{\mathrm{Eqs.~}\ref{equ:HarmonicCoordinateEquMHT}~\&~\ref{equ:BulkTransformedGradDeformedDomMHT}} & \left\{ \mathbf{s}, \nabla_{\mathbf{x}_\Xi}, \mathrm{div}_{\mathbf{x}_\Xi}, \mathbf{n}_{\partial\Xi} \right\} & \\
 & & \bigg\downarrow\vcenter{\rlap{\scriptsize{Eq.~\ref{equ:TransformedTangentialOperatorMHM}}}} & & \bigg\downarrow\vcenter{\rlap{\scriptsize{Eqs.~\ref{equ:BulkConvectionDiffusionEqu}
 ~\&~\ref{equ:NavierStokesEquBulkFlowMassTransferMHT}}}} \\
 & & \left\{ \nabla_\Gamma, \mathrm{div}_\Gamma, \mathbf{n}_\Gamma \right\} & & \left\{ \mathbf{u},~p,~c \right\} & \xrightarrow{\mathrm{Eqs.~}\ref{equ:DesignObjectiveBulkCDMHT} ~\&~ \ref{equ:PressureDropBulkNSCD}} & \left\{ \textcolor[rgb]{0.50,0.00,0.00}{J_c}, ~ \textcolor[rgb]{0.50,0.00,0.00}{\Delta P} \right\} \\
 & & \bigg\downarrow\vcenter{\rlap{\scriptsize{Eq.~\ref{equ:PDEFilterGammaFilberMHM}}}} & &  \bigg\uparrow\vcenter{\rlap{\scriptsize{Eq.~\ref{equ:MaterialInterpolationMassTransferBulkFlow}}}}   \\
 \textcolor{blue}{\gamma} & \xrightarrow{\mathrm{Eq.~}\ref{equ:PDEFilterGammaFilberMHM}} & \gamma_f & \xrightarrow{\mathrm{Eq.~}\ref{equ:ProjectionGammaFilberMHM}} & \textcolor[rgb]{0.50,0.00,0.00}{\gamma_p} & & \\
\end{array}\]
where the design variables $d_m$ and $\gamma$, marked in blue, are the inputs; the design objective $J_c$, the pressure drop $\Delta P$ and the material density $\gamma_p$, marked in red, are the outputs.

\subsubsection{Adjoint analysis} \label{sec:AdjointAnalysisBulkNSCDEqusMHT}

To solve the fiber bundle topology optimization problem in Eq. \ref{equ:VarProToopBulkNSCDMHT}, the adjoint analysis is implemented for the design objective and constraint of the pressure drop to derive the adjoint sensitivities. The details for the adjoint analysis are provided in the appendix in Sections \ref{sec:AdjointAnalysisDesignObjectiveBulkCHTMHM} and \ref{sec:AdjointAnalysisPressureDropBulkCHTMHMBulk}.

Based on the transformed design objective in Eq. \ref{equ:FurtherTransformedDesignObjectiveBulkCD} and transformed pressure drop in Eq. \ref{equ:TransformedPressureConstraintSurfaceNSCD}, the adjoint analysis of the fiber bundle topology optimization problem can be implemented on the functional spaces defined on the original domain $\Omega$. Based on the continuous adjoint method \cite{HinzeSpringer2009}, the adjoint sensitivity of the design objective $J_c$ is derived as
\begin{equation}\label{equ:AdjSensitivityNSCDGaDmObjBulkMHT}
\begin{split}
\delta J_c = \int_\Sigma - \gamma_{fa} \tilde{\gamma} M^{\left( d_f \right)} - A_d d_{fa} \tilde{d}_m \,\mathrm{d}\Sigma,~ \forall \left( \tilde{\gamma}, \tilde{d}_m \right) \in \left(\mathcal{L}^2\left(\Sigma\right)\right)^2.
\end{split}
\end{equation}
The adjoint variables in Eq. \ref{equ:AdjSensitivityNSCDGaDmObjBulkMHT} can be derived by sequentially solving the following adjoint equations in variational formulations. The variational formulation for the adjoint equation of the convection-diffusion equation is derived as
\begin{equation}\label{equ:WeakAdjEquCDEquBulkMTCa}
\left\{\begin{split}
  & \mathrm{Find} ~ c_a \in \mathcal{H}\left(\Omega\right)~\mathrm{with}~ c_a=0 ~ \mathrm{at} ~ \forall \mathbf{x}_\Omega \in \Sigma_{v,\Omega}, ~ \mathrm{for} ~\forall \tilde{c}_a \in \mathcal{H} \left(\Omega\right) \\
  & \mathrm{such~that} ~ \int_{\Sigma_{s,\Omega}} 2 \left( c - \bar{c} \right) \tilde{c}_a \,\mathrm{d}\Sigma_{\partial\Omega} \bigg/ \int_{\Sigma_{v,\Omega}} \left( c_0 - \bar{c} \right)^2 \,\mathrm{d}\Sigma_{\partial\Omega} \\
  & + \int_\Omega \left[ \left( \mathbf{u} \cdot \nabla_{\mathbf{x}_\Xi}^{\left(\mathbf{s}\right)} \tilde{c}_a \right) c_a + D \nabla_{\mathbf{x}_\Xi}^{\left(\mathbf{s}\right)} \tilde{c}_a \cdot \nabla_{\mathbf{x}_\Xi}^{\left(\mathbf{s}\right)} c_a \right] K^{\left(\mathbf{s}\right)} \,\mathrm{d}\Omega \\
  & + \sum_{E_\Omega\in\mathcal{E}_\Omega} \int_{E_\Omega} \tau_{PG,\Xi}^{\left(\mathbf{s}\right)} \left( \mathbf{u} \cdot \nabla_{\mathbf{x}_\Xi}^{\left(\mathbf{s}\right)} \tilde{c}_a \right) \left( \mathbf{u} \cdot \nabla_{\mathbf{x}_\Xi}^{\left(\mathbf{s}\right)} c_a \right) K^{\left(\mathbf{s}\right)} \,\mathrm{d}\Omega = 0.
\end{split}\right.
\end{equation}
The variational formulation for the adjoint equations of the Naiver-Stokes equations is derived as
\begin{equation}\label{equ:AdjBulkNavierStokesEqusJObjectiveMTUaPa}
\left\{\begin{split}
  & \mathrm{Find} \left\{\begin{split}
  & \mathbf{u}_a \in\left(\mathcal{H}\left(\Omega\right)\right)^3~\mathrm{with}~ \mathbf{u}_a = \mathbf{0}~ \mathrm{at} ~ {\forall \mathbf{x}_\Omega \in \Sigma_{v,\Omega} \cup \Sigma_{v_0,\Omega} } \\
  & p_a \in \mathcal{H}\left(\Omega\right) \\
  \end{split}\right.\\
  & \mathrm{for} \left\{\begin{split} 
  & \forall \tilde{\mathbf{u}}_a \in\left(\mathcal{H}\left(\Omega\right)\right)^3 \\
  & \forall \tilde{p}_a \in \mathcal{H}\left(\Omega\right) \\
  \end{split}\right.,~\mathrm{such~that} \\
  & \int_\Omega \Big[ \rho \left( \tilde{\mathbf{u}}_a \cdot \nabla_{\mathbf{x}_\Xi}^{\left(\mathbf{s}\right)} \right) \mathbf{u} \cdot \mathbf{u}_a + \rho \left( \mathbf{u} \cdot \nabla_{\mathbf{x}_\Xi}^{\left(\mathbf{s}\right)} \right) \tilde{\mathbf{u}}_a \cdot \mathbf{u}_a + {\eta\over2} \left( \nabla_{\mathbf{x}_\Xi}^{\left(\mathbf{s}\right)} \tilde{\mathbf{u}}_a + \nabla_{\mathbf{x}_\Xi}^{\left(\mathbf{s}\right)} \tilde{\mathbf{u}}_a^\mathrm{T} \right) \\
  & : \left( \nabla_{\mathbf{x}_\Xi}^{\left(\mathbf{s}\right)} \mathbf{u}_a + \nabla_{\mathbf{x}_\Xi}^{\left(\mathbf{s}\right)} \mathbf{u}_a^\mathrm{T} \right) - \tilde{p}_a \mathrm{div}_{\mathbf{x}_\Xi}^{\left(\mathbf{s}\right)} \mathbf{u}_a - p_a \mathrm{div}_{\mathbf{x}_\Xi}^{\left(\mathbf{s}\right)} \tilde{\mathbf{u}}_a + \left( \tilde{\mathbf{u}}_a \cdot \nabla_{\mathbf{x}_\Xi}^{\left(\mathbf{s}\right)} c \right) c_a \Big] K^{\left( \mathbf{s} \right)} \,\mathrm{d}\Omega \\
  & + \sum_{E_\Omega\in\mathcal{E}_\Omega} \int_{E_\Omega} \Big[ - \tau_{BP,\Xi}^{\left(\mathbf{s}\right)} \nabla_{\mathbf{x}_\Xi}^{\left(\mathbf{s}\right)} \tilde{p}_a \cdot \nabla_{\mathbf{x}_\Xi}^{\left(\mathbf{s}\right)} p_a + \tau_{PG,\Xi}^{\left(\mathbf{s}, \tilde{\mathbf{u}}_a\right)} \left( \mathbf{u} \cdot \nabla_{\mathbf{x}_\Xi}^{\left(\mathbf{s}\right)} c \right) \left( \mathbf{u} \cdot \nabla_{\mathbf{x}_\Xi}^{\left(\mathbf{s}\right)} c_a \right) \\
  & + \tau_{PG,\Xi}^{\left(\mathbf{s}\right)} \left( \tilde{\mathbf{u}}_a \cdot \nabla_{\mathbf{x}_\Xi}^{\left(\mathbf{s}\right)} c \right) \left( \mathbf{u} \cdot \nabla_{\mathbf{x}_\Xi}^{\left(\mathbf{s}\right)} c_a \right) + \tau_{PG,\Xi}^{\left(\mathbf{s}\right)} \left( \mathbf{u} \cdot \nabla_{\mathbf{x}_\Xi}^{\left(\mathbf{s}\right)} c \right) \left( \tilde{\mathbf{u}}_a \cdot \nabla_{\mathbf{x}_\Xi}^{\left(\mathbf{s}\right)} c_a \right) \Big] \\
  & K^{\left(\mathbf{s}\right)} \,\mathrm{d}\Omega + \int_\Sigma \alpha \tilde{\mathbf{u}}_a \cdot \mathbf{u}_a M^{\left( d_f \right)} \, \mathrm{d}\Sigma = 0
\end{split}\right.
\end{equation}
where $\tau_{PG,\Xi}^{\left( \mathbf{s}, \tilde{\mathbf{u}} \right)}$ is the first-order variational of $\tau_{PG,\Xi}^{\left( \mathbf{s} \right)}$ to $\mathbf{u}$, and it is expressed as
\begin{equation}\label{equ:1stVariTauPGtoUBulk}
  \tau_{PG,\Xi}^{\left( \mathbf{s}, \tilde{\mathbf{u}} \right)} = - \left( {4 \over h_{E_\Omega}^2 \left( K^{\left( \mathbf{s} \right)} \right)^{2\over3} D} + {2 \left\| \mathbf{u} \right\|_2 \over h_{E_\Omega} \left( K^{\left( \mathbf{s} \right)} \right)^{1\over3} } \right)^{-2} {2 \mathbf{u} \cdot \tilde{\mathbf{u}} \over h_{E_\Omega} \left( K^{\left( \mathbf{s} \right)} \right)^{1\over3} \left\| \mathbf{u} \right\|_2 }, ~ \forall \tilde{\mathbf{u}} \in \left( \mathcal{H}\left( \Omega \right) \right)^3.
\end{equation}
The variational formulation for the adjoint equation of the Laplace's equation for $\mathbf{s}$ is derived as
\begin{equation}\label{equ:WeakAdjEquHarmonicEquBulkMTSa} 
\left\{\begin{split}
  & \mathrm{Find} \left\{\begin{split}
  & \mathbf{s}_a \in \left(\mathcal{H}\left(\Omega\right)\right)^3~\mathrm{with}~ \mathbf{s}_a = \mathbf{0} ~ \mathrm{at} ~ \forall \mathbf{x}_\Omega \in \Sigma_{v,\Omega} \cup \Sigma_{s,\Omega} \\
  & \boldsymbol{\lambda}_{\mathbf{s}a} \in \left(\mathcal{H}^{-{1\over2}}\left(\Sigma\right)\right)^3 \\
  \end{split}\right.,\\
  & \mathrm{for} \left\{\begin{split} 
  & \forall \tilde{\mathbf{s}}_a \in\left(\mathcal{H}\left(\Omega\right)\right)^3 \\
  & \forall \tilde{\boldsymbol{\lambda}}_{\mathbf{s}a} \in \left(\mathcal{H}^{{1\over2}}\left(\Sigma\right)\right)^3 \\
  \end{split}\right.,~\mathrm{such~that} \\
  & \int_\Omega \Big[ \rho \left( \mathbf{u} \cdot \nabla_{\mathbf{x}_\Xi}^{\left(\mathbf{s},\tilde{\mathbf{s}}_a\right)} \right) \mathbf{u} \cdot \mathbf{u}_a + {\eta\over2} \left( \nabla_{\mathbf{x}_\Xi}^{\left(\mathbf{s}, \tilde{\mathbf{s}}_a\right)} \mathbf{u} + \nabla_{\mathbf{x}_\Xi}^{\left(\mathbf{s}, \tilde{\mathbf{s}}_a\right)} \mathbf{u}^\mathrm{T} \right) : \left( \nabla_{\mathbf{x}_\Xi}^{\left(\mathbf{s}\right)} \mathbf{u}_a + \nabla_{\mathbf{x}_\Xi}^{\left(\mathbf{s}\right)} \mathbf{u}_a^\mathrm{T} \right) \\
  & + {\eta\over2} \left( \nabla_{\mathbf{x}_\Xi}^{\left(\mathbf{s}\right)} \mathbf{u} + \nabla_{\mathbf{x}_\Xi}^{\left(\mathbf{s}\right)} \mathbf{u}^\mathrm{T} \right) : \left( \nabla_{\mathbf{x}_\Xi}^{\left(\mathbf{s}, \tilde{\mathbf{s}}_a\right)} \mathbf{u}_a + \nabla_{\mathbf{x}_\Xi}^{\left(\mathbf{s}, \tilde{\mathbf{s}}_a\right)} \mathbf{u}_a^\mathrm{T} \right) - p\,\mathrm{div}_{\mathbf{x}_\Xi}^{\left(\mathbf{s}, \tilde{\mathbf{s}}_a\right)} \mathbf{u}_a - p_a \mathrm{div}_{\mathbf{x}_\Xi}^{\left(\mathbf{s}, \tilde{\mathbf{s}}_a\right)} \mathbf{u} \\
  & + \left( \mathbf{u} \cdot \nabla_{\mathbf{x}_\Xi}^{\left(\mathbf{s},\tilde{\mathbf{s}}_a\right)} c \right) c_a + D \left( \nabla_{\mathbf{x}_\Xi}^{\left(\mathbf{s},\tilde{\mathbf{s}}_a\right)} c \cdot \nabla_{\mathbf{x}_\Xi}^{\left(\mathbf{s}\right)} c_a + \nabla_{\mathbf{x}_\Xi}^{\left(\mathbf{s}\right)} c \cdot \nabla_{\mathbf{x}_\Xi}^{\left(\mathbf{s}, \tilde{\mathbf{s}}_a\right)} c_a \right) \Big] K^{\left( \mathbf{s} \right)} + \Big[ \rho \left( \mathbf{u} \cdot \nabla_{\mathbf{x}_\Xi}^{\left(\mathbf{s}\right)} \right) \mathbf{u} \\
  & \cdot \mathbf{u}_a + {\eta\over2} \left( \nabla_{\mathbf{x}_\Xi}^{\left(\mathbf{s}\right)} \mathbf{u} + \nabla_{\mathbf{x}_\Xi}^{\left(\mathbf{s}\right)} \mathbf{u}^\mathrm{T} \right) : \left( \nabla_{\mathbf{x}_\Xi}^{\left(\mathbf{s}\right)} \mathbf{u}_a + \nabla_{\mathbf{x}_\Xi}^{\left(\mathbf{s}\right)} \mathbf{u}_a^\mathrm{T} \right) - p\,\mathrm{div}_{\mathbf{x}_\Xi}^{\left(\mathbf{s}\right)} \mathbf{u}_a - p_a \mathrm{div}_{\mathbf{x}_\Xi}^{\left(\mathbf{s}\right)} \mathbf{u} + \left( \mathbf{u} \cdot \nabla_{\mathbf{x}_\Xi}^{\left(\mathbf{s}\right)} c \right) \\
  & c_a + D \nabla_{\mathbf{x}_\Xi}^{\left(\mathbf{s}\right)} c \cdot \nabla_{\mathbf{x}_\Xi}^{\left(\mathbf{s}\right)} c_a \Big] K^{\left( \mathbf{s}, \tilde{\mathbf{s}}_a \right)} - \nabla_{\mathbf{x}_\Omega} \tilde{\mathbf{s}}_a : \nabla_{\mathbf{x}_\Omega} \mathbf{s}_a \,\mathrm{d}\Omega + \sum_{E_\Omega\in\mathcal{E}_\Omega} \int_{E_\Omega} \Big[ - \tau_{BP,\Xi}^{\left(\mathbf{s}, \tilde{\mathbf{s}}_a \right)} \nabla_{\mathbf{x}_\Xi}^{\left(\mathbf{s}\right)} p \cdot \nabla_{\mathbf{x}_\Xi}^{\left(\mathbf{s}\right)} p_a \\
  & - \tau_{BP,\Xi}^{\left(\mathbf{s}\right)} \nabla_{\mathbf{x}_\Xi}^{\left(\mathbf{s}, \tilde{\mathbf{s}}_a\right)} p \cdot \nabla_{\mathbf{x}_\Xi}^{\left(\mathbf{s}\right)} p_a - \tau_{BP,\Xi}^{\left(\mathbf{s}\right)} \nabla_{\mathbf{x}_\Xi}^{\left(\mathbf{s}\right)} p \cdot \nabla_{\mathbf{x}_\Xi}^{\left(\mathbf{s}, \tilde{\mathbf{s}}_a\right)} p_a + \tau_{PG,\Xi}^{\left(\mathbf{s}, \tilde{\mathbf{s}}_a\right)} \left( \mathbf{u} \cdot \nabla_{\mathbf{x}_\Xi}^{\left(\mathbf{s}\right)} c \right) \left( \mathbf{u} \cdot \nabla_{\mathbf{x}_\Xi}^{\left(\mathbf{s}\right)} c_a \right) \\
  & + \tau_{PG,\Xi}^{\left(\mathbf{s}\right)} \left( \mathbf{u} \cdot \nabla_{\mathbf{x}_\Xi}^{\left(\mathbf{s}, \tilde{\mathbf{s}}_a \right)} c \right) \left( \mathbf{u} \cdot \nabla_{\mathbf{x}_\Xi}^{\left(\mathbf{s}\right)} c_a \right) + \tau_{PG,\Xi}^{\left(\mathbf{s}\right)} \left( \mathbf{u} \cdot \nabla_{\mathbf{x}_\Xi}^{\left(\mathbf{s}\right)} c \right) \left( \mathbf{u} \cdot \nabla_{\mathbf{x}_\Xi}^{\left(\mathbf{s}, \tilde{\mathbf{s}}_a \right)} c_a \right) \Big] K^{\left(\mathbf{s}\right)} \\
  & + \Big[ - \tau_{BP,\Xi}^{\left(\mathbf{s}\right)} \nabla_{\mathbf{x}_\Xi}^{\left(\mathbf{s}\right)} p \cdot \nabla_{\mathbf{x}_\Xi}^{\left(\mathbf{s}\right)} p_a + \tau_{PG,\Xi}^{\left(\mathbf{s}\right)} \left( \mathbf{u} \cdot \nabla_{\mathbf{x}_\Xi}^{\left(\mathbf{s}\right)} c \right) \left( \mathbf{u} \cdot \nabla_{\mathbf{x}_\Xi}^{\left(\mathbf{s}\right)} c_a \right) \Big] K^{\left(\mathbf{s},\tilde{\mathbf{s}}_a \right)}  \,\mathrm{d}\Omega \\
  & + \int_\Sigma \tilde{\mathbf{s}}_a \cdot \boldsymbol{\lambda}_{\mathbf{s}a} + \tilde{\boldsymbol{\lambda}}_{\mathbf{s}a} \cdot \mathbf{s}_a \,\mathrm{d}\Sigma = 0 \\
\end{split}\right.
\end{equation}  
where $\mathbf{s}_a$ and $\boldsymbol{\lambda}_{\mathbf{s}a}$ are the adjoint variables of $\mathbf{s}$ and $\boldsymbol{\lambda}_\mathbf{s}$, respectively; $\tilde{\mathbf{s}}_a$ and $\tilde{\boldsymbol{\lambda}}_{\mathbf{s}a}$ are the test functions of $\mathbf{s}_a$ and $\boldsymbol{\lambda}_{\mathbf{s}a}$, respectively; $K^{\left( \mathbf{s}, \tilde{\mathbf{s}} \right)}$ is the first-order variational of $K^{\left( \mathbf{s} \right)}$ to $\mathbf{s}$ derived based on Eq. \ref{equ:FirstOrderVariOfVectorNormMHM} in the appendix, and it is expressed as
\begin{equation}\label{equ:1stVariofRiemannMetricTodf}
\begin{split}
  K^{\left( \mathbf{s}, \tilde{\mathbf{s}} \right)} = \: & {\partial \left| \mathbf{T}_\Xi \right| \over \partial \nabla_{\mathbf{x}_\Omega} \mathbf{s} } : \nabla_{\mathbf{x}_\Omega} \tilde{\mathbf{s}}, ~ \forall \tilde{\mathbf{s}} \in \left( \mathcal{H}\left( \Omega \right) \right)^3; \\
\end{split}
\end{equation}
$\tau_{BP,\Xi}^{\left( \mathbf{s}, \tilde{\mathbf{s}} \right)}$ and $\tau_{PG,\Xi}^{\left( \mathbf{s}, \tilde{\mathbf{s}} \right)}$ are the first-order variationals of $\tau_{BP,\Xi}^{\left( \mathbf{s} \right)}$ and $\tau_{PG,\Xi}^{\left( \mathbf{s} \right)}$ to $\mathbf{s}$, respectively, and they are expressed as
\begin{equation}\label{equ:1stVariBulkTauBPtodf}
  \tau_{BP,\Xi}^{\left( \mathbf{s}, \tilde{\mathbf{s}} \right)} = {h_{E_\Omega}^2 \over 18\eta} \left( K^{\left( \mathbf{s} \right)} \right)^{-{1\over3}} K^{\left( \mathbf{s}, \tilde{\mathbf{s}} \right)}, ~ \forall \tilde{\mathbf{s}} \in \left(\mathcal{H}\left( \Omega \right)\right)^3
\end{equation}
and
\begin{equation}\label{equ:1stVariTauPGtosBulk}
\begin{split}
  \tau_{PG,\Xi}^{\left( \mathbf{s}, \tilde{\mathbf{s}} \right)} = \, & \left( {4 \over h_{E_\Omega}^2 \left( K^{\left( \mathbf{s} \right)} \right)^{2\over3} D} + {2 \left\| \mathbf{u} \right\|_2 \over h_{E_\Omega} \left( K^{\left( \mathbf{s} \right)} \right)^{1\over3} } \right)^{-2} \\
  & \left( {8 \over 3 h_{E_\Omega}^2 \left(K^{\left( \mathbf{s} \right)}\right)^{5\over3} D} + { 2 \left\| \mathbf{u} \right\|_2 \over 3 h_{E_\Omega} \left( K^{\left( \mathbf{s} \right)} \right)^{4\over3} } \right) K^{\left( \mathbf{s}, \tilde{\mathbf{s}} \right)}, ~ \forall \tilde{\mathbf{s}} \in \left( \mathcal{H} \left(\Omega\right) \right)^3.
\end{split}
\end{equation}

The variational formulations for the adjoint equations of the surface-PDE filters for $\gamma$ and $d_m$ are derived as 
\begin{equation}\label{equ:AdjPDEFilterJObjectiveGafMHMGafa}  
\left\{\begin{split}
  & \mathrm{Find}~\gamma_{fa}\in\mathcal{H}\left(\Sigma\right) ~\mathrm{for}~ \forall \tilde{\gamma}_{fa} \in \mathcal{H}\left(\Sigma\right),~\mathrm{such~that} \\
  & \int_\Sigma \left( {\partial\alpha\over\partial\gamma_p} {\partial\gamma_p\over\partial\gamma_f} \mathbf{u} \cdot \mathbf{u}_a \tilde{\gamma}_{fa} + r_f^2 \nabla_\Gamma^{\left( d_f \right)} \tilde{\gamma}_{fa} \cdot \nabla_\Gamma^{\left( d_f \right)} \gamma_{fa} + \tilde{\gamma}_{fa} \gamma_{fa} \right) M^{\left( d_f \right)} \,\mathrm{d}\Sigma = 0
\end{split}\right.
\end{equation}
and
\begin{equation}\label{equ:AdjPDEFilterJObjectiveDmMHMDfa} 
\left\{\begin{split}
  & \mathrm{Find}~d_{fa}\in\mathcal{H}\left(\Sigma\right)~\mathrm{for}~\forall \tilde{d}_{fa} \in \mathcal{H}\left(\Sigma\right),~\mathrm{such~that} \\
  & \int_\Sigma r_f^2 \left( \nabla_\Gamma^{\left( d_f, \tilde{d}_{fa} \right)} \gamma_f \cdot \nabla_\Gamma^{\left( d_f \right)} \gamma_{fa} + \nabla_\Gamma^{\left( d_f \right)} \gamma_f \cdot \nabla_\Gamma^{\left( d_f, \tilde{d}_{fa} \right)} \gamma_{fa} \right) M^{\left( d_f \right)} \\
  & + \left( r_f^2 \nabla_\Gamma^{\left( d_f \right)} \gamma_f \cdot \nabla_\Gamma^{\left( d_f \right)} \gamma_{fa} + \gamma_f \gamma_{fa} - \gamma \gamma_{fa} + \alpha \mathbf{u} \cdot \mathbf{u}_a \right) M^{\left( d_f, \tilde{d}_{fa} \right)} \\
  & + r_m^2 \nabla_\Sigma \tilde{d}_{fa} \cdot \nabla_\Sigma d_{fa} + \tilde{d}_{fa} d_{fa} - \mathbf{n}_\Sigma \cdot \boldsymbol{\lambda}_{\mathbf{s}a} \tilde{d}_{fa} \,\mathrm{d}\Sigma = 0.
\end{split}\right.
\end{equation}

For the constraint of the pressure drop, the adjoint sensitivity of the pressure drop $\Delta P$ is derived as
\begin{equation}\label{equ:AdjSensitivityGaDmPressureConstrMHMBulk} 
\begin{split}
\delta \Delta P = \int_\Sigma - \gamma_{fa} \tilde{\gamma} M^{\left( d_f \right)} - A_d d_{fa} \tilde{d}_m \,\mathrm{d}\Sigma,~ \forall \left( \tilde{\gamma}, \tilde{d}_m \right) \in \left(\mathcal{L}^2\left(\Sigma\right)\right)^2.
\end{split}
\end{equation}
In Eq. \ref{equ:AdjSensitivityGaDmPressureConstrMHMBulk}, the adjoint variables $\gamma_{fa}$ and $d_{fa}$ are derived by sequentially solving the variational formulation for the adjoint equations of the Navier-Stokes equations
\begin{equation}\label{equ:AdjEquSurfaceNSMHMPressureDropBulk}
\left\{\begin{split}
  & \mathrm{Find} \left\{\begin{split}
  & \mathbf{u}_a \in\left(\mathcal{H}\left(\Omega\right)\right)^3~\mathrm{with}~ \mathbf{u}_a = \mathbf{0}~ \mathrm{at} ~  {\forall \mathbf{x}_\Omega \in \Sigma_{v,\Omega} \cup \Sigma_{v_0,\Omega} } \\
  & p_a \in \mathcal{H}\left(\Omega\right) \\
  \end{split}\right.\\
  & \mathrm{for} \left\{\begin{split}
  & \forall \tilde{\mathbf{u}}_a \in\left(\mathcal{H}\left(\Omega\right)\right)^3 \\
  & \forall \tilde{p}_a \in \mathcal{H}\left(\Omega\right) \\
  \end{split}\right.,~\mathrm{such~that} \\
  & \int_{\Sigma_{v,\Omega}} \tilde{p}_a \,\mathrm{d}\Sigma_{\partial\Omega} - \int_{\Sigma_{s,\Omega}} \tilde{p}_a \,\mathrm{d}\Sigma_{\partial\Omega} + \int_\Omega \Big[ \rho \left( \tilde{\mathbf{u}}_a \cdot \nabla_{\mathbf{x}_\Xi}^{\left(\mathbf{s}\right)} \right) \mathbf{u} \cdot \mathbf{u}_a + \rho \left( \mathbf{u} \cdot \nabla_{\mathbf{x}_\Xi}^{\left(\mathbf{s}\right)} \right) \tilde{\mathbf{u}}_a \cdot \mathbf{u}_a \\
  & + {\eta\over2} \left( \nabla_{\mathbf{x}_\Xi}^{\left(\mathbf{s}\right)} \tilde{\mathbf{u}}_a + \nabla_{\mathbf{x}_\Xi}^{\left(\mathbf{s}\right)} \tilde{\mathbf{u}}_a^\mathrm{T} \right) : \left( \nabla_{\mathbf{x}_\Xi}^{\left(\mathbf{s}\right)} \mathbf{u}_a + \nabla_{\mathbf{x}_\Xi}^{\left(\mathbf{s}\right)} \mathbf{u}_a^\mathrm{T} \right) - \tilde{p}_a\mathrm{div}_{\mathbf{x}_\Xi}^{\left(\mathbf{s}\right)} \mathbf{u}_a - p_a \mathrm{div}_{\mathbf{x}_\Xi}^{\left(\mathbf{s}\right)} \tilde{\mathbf{u}}_a \Big] K^{\left( \mathbf{s} \right)} \,\mathrm{d}\Omega \\
  & + \sum_{E_\Omega\in\mathcal{E}_\Omega} \int_{E_\Omega} - \tau_{BP,\Xi}^{\left(\mathbf{s}\right)} \nabla_{\mathbf{x}_\Xi}^{\left(\mathbf{s}\right)} \tilde{p}_a \cdot \nabla_{\mathbf{x}_\Xi}^{\left(\mathbf{s}\right)} p_a K^{\left( \mathbf{s} \right)} \,\mathrm{d}\Omega + \int_\Sigma \alpha \tilde{\mathbf{u}}_a \cdot \mathbf{u}_a M^{\left( d_f \right)} \, \mathrm{d}\Sigma = 0 \\  
\end{split}\right.,
\end{equation}
the variational formulation for the adjoint equation of the Laplace's equation
\begin{equation}\label{equ:WeakAdjEquHarmonicEquBulkMTSaPressureDrop} 
\left\{\begin{split}
  & \mathrm{Find} \left\{\begin{split}
  & \mathbf{s}_a \in \left(\mathcal{H}\left(\Omega\right)\right)^3~\mathrm{with}~ \mathbf{s}_a = \mathbf{0} ~ \mathrm{at} ~ \forall \mathbf{x}_\Omega \in \Sigma_{v,\Omega} \cup \Sigma_{s,\Omega} \\
  & \boldsymbol{\lambda}_{\mathbf{s}a} \in \left(\mathcal{H}^{-{1\over2}}\left(\Sigma\right)\right)^3 \\
  \end{split}\right.,\\
  & \mathrm{for} \left\{\begin{split} 
  & \forall \tilde{\mathbf{s}}_a \in\left(\mathcal{H}\left(\Omega\right)\right)^3 \\
  & \forall \tilde{\boldsymbol{\lambda}}_{\mathbf{s}a} \in \left(\mathcal{H}^{{1\over2}}\left(\Sigma\right)\right)^3 \\
  \end{split}\right.,~\mathrm{such~that} \\
  & \int_\Omega \Big[ \rho \left( \mathbf{u} \cdot \nabla_{\mathbf{x}_\Xi}^{\left(\mathbf{s}, \tilde{\mathbf{s}}_a\right)} \right) \mathbf{u} \cdot \mathbf{u}_a + {\eta\over2} \left( \nabla_{\mathbf{x}_\Xi}^{\left(\mathbf{s}, \tilde{\mathbf{s}}_a\right)} \mathbf{u} + \nabla_{\mathbf{x}_\Xi}^{\left(\mathbf{s}, \tilde{\mathbf{s}}_a\right)} \mathbf{u}^\mathrm{T} \right) : \left( \nabla_{\mathbf{x}_\Xi}^{\left(\mathbf{s}\right)} \mathbf{u}_a + \nabla_{\mathbf{x}_\Xi}^{\left(\mathbf{s}\right)} \mathbf{u}_a^\mathrm{T} \right) \\
  & + {\eta\over2} \left( \nabla_{\mathbf{x}_\Xi}^{\left(\mathbf{s}\right)} \mathbf{u} + \nabla_{\mathbf{x}_\Xi}^{\left(\mathbf{s}\right)} \mathbf{u}^\mathrm{T} \right) : \left( \nabla_{\mathbf{x}_\Xi}^{\left(\mathbf{s}, \tilde{\mathbf{s}}_a\right)} \mathbf{u}_a + \nabla_{\mathbf{x}_\Xi}^{\left(\mathbf{s}, \tilde{\mathbf{s}}_a\right)} \mathbf{u}_a^\mathrm{T} \right) - p\,\mathrm{div}_{\mathbf{x}_\Xi}^{\left(\mathbf{s}, \tilde{\mathbf{s}}_a\right)} \mathbf{u}_a - p_a \mathrm{div}_{\mathbf{x}_\Xi}^{\left(\mathbf{s}, \tilde{\mathbf{s}}_a\right)} \mathbf{u} \Big] K^{\left( \mathbf{s} \right)} \\
  & + \Big[ \rho \left( \mathbf{u} \cdot \nabla_{\mathbf{x}_\Xi}^{\left(\mathbf{s}\right)} \right) \mathbf{u} \cdot \mathbf{u}_a + {\eta\over2} \left( \nabla_{\mathbf{x}_\Xi}^{\left(\mathbf{s}\right)} \mathbf{u} + \nabla_{\mathbf{x}_\Xi}^{\left(\mathbf{s}\right)} \mathbf{u}^\mathrm{T} \right) : \left( \nabla_{\mathbf{x}_\Xi}^{\left(\mathbf{s}\right)} \mathbf{u}_a + \nabla_{\mathbf{x}_\Xi}^{\left(\mathbf{s}\right)} \mathbf{u}_a^\mathrm{T} \right) - p\,\mathrm{div}_{\mathbf{x}_\Xi}^{\left(\mathbf{s}\right)} \mathbf{u}_a \\
  & - p_a \mathrm{div}_{\mathbf{x}_\Xi}^{\left(\mathbf{s}\right)} \mathbf{u} \Big] K^{\left( \mathbf{s}, \tilde{\mathbf{s}}_a \right)} - \nabla_{\mathbf{x}_\Omega} \tilde{\mathbf{s}}_a : \nabla_{\mathbf{x}_\Omega} \mathbf{s}_a \,\mathrm{d}\Omega + \sum_{E_\Omega\in\mathcal{E}_\Omega} \int_{E_\Omega} \Big[ - \tau_{BP,\Xi}^{\left(\mathbf{s}, \tilde{\mathbf{s}}_a\right)} \nabla_{\mathbf{x}_\Xi}^{\left(\mathbf{s}\right)} p \cdot \nabla_{\mathbf{x}_\Xi}^{\left(\mathbf{s}\right)} p_a \\
  & - \tau_{BP,\Xi}^{\left(\mathbf{s}\right)} \nabla_{\mathbf{x}_\Xi}^{\left(\mathbf{s}, \tilde{\mathbf{s}}_a\right)} p \cdot \nabla_{\mathbf{x}_\Xi}^{\left(\mathbf{s}\right)} p_a - \tau_{BP,\Xi}^{\left(\mathbf{s}\right)} \nabla_{\mathbf{x}_\Xi}^{\left(\mathbf{s}\right)} p \cdot \nabla_{\mathbf{x}_\Xi}^{\left(\mathbf{s}, \tilde{\mathbf{s}}_a\right)} p_a \Big] K^{\left( \mathbf{s} \right)} - \tau_{BP,\Xi}^{\left(\mathbf{s}\right)} \nabla_{\mathbf{x}_\Xi}^{\left(\mathbf{s}\right)} p \\
  & \cdot \nabla_{\mathbf{x}_\Xi}^{\left(\mathbf{s}\right)} p_a K^{\left( \mathbf{s}, \tilde{\mathbf{s}}_a \right)} \,\mathrm{d}\Omega + \int_\Sigma \tilde{\mathbf{s}}_a \cdot \boldsymbol{\lambda}_{\mathbf{s}a} + \tilde{\boldsymbol{\lambda}}_{\mathbf{s}a} \cdot \mathbf{s}_a \,\mathrm{d}\Sigma = 0 \\
\end{split}\right.,
\end{equation}
and the variational formulations of the adjoint equations of the surface-PDE filters 
\begin{equation}\label{equ:AdjPDEFilterPressureDropGaMHMBulk} 
\left\{\begin{split}
  & \mathrm{Find}~\gamma_{fa}\in\mathcal{H}\left(\Sigma\right) ~\mathrm{for}~ \forall \tilde{\gamma}_{fa} \in \mathcal{H}\left(\Sigma\right),~\mathrm{such~that} \\
  & \int_\Sigma \left( {\partial\alpha\over\partial\gamma_p} {\partial\gamma_p\over\partial\gamma_f} \mathbf{u} \cdot \mathbf{u}_a \tilde{\gamma}_{fa} + r_f^2 \nabla_\Gamma^{\left( d_f \right)} \tilde{\gamma}_{fa} \cdot \nabla_\Gamma^{\left( d_f \right)} \gamma_{fa} + \tilde{\gamma}_{fa} \gamma_{fa} \right) M^{\left( d_f \right)} \,\mathrm{d}\Sigma = 0
\end{split}\right.
\end{equation}
and 
\begin{equation}\label{equ:AdjPDEFilterJPressureDropDmMHMBulk} 
\left\{\begin{split}
  & \mathrm{Find}~d_{fa}\in\mathcal{H}\left(\Sigma\right)~\mathrm{for}~\forall \tilde{d}_{fa} \in \mathcal{H}\left(\Sigma\right),~\mathrm{such~that} \\
  & \int_\Sigma r_f^2 \left( \nabla_\Gamma^{\left( d_f, \tilde{d}_{fa} \right)} \gamma_f \cdot \nabla_\Gamma^{\left( d_f \right)} \gamma_{fa} + \nabla_\Gamma^{\left( d_f \right)} \gamma_f \cdot \nabla_\Gamma^{\left( d_f, \tilde{d}_{fa} \right)} \gamma_{fa} \right) M^{\left( d_f \right)} \\
  & + \left( r_f^2 \nabla_\Gamma^{\left( d_f \right)} \gamma_f \cdot \nabla_\Gamma^{\left( d_f \right)} \gamma_{fa} + \gamma_f \gamma_{fa} - \gamma \gamma_{fa} + \alpha \mathbf{u} \cdot \mathbf{u}_a \right) M^{\left( d_f, \tilde{d}_{fa} \right)} \\
  & + r_m^2 \nabla_\Sigma \tilde{d}_{fa} \cdot \nabla_\Sigma d_{fa} + \tilde{d}_{fa} d_{fa} - \tilde{d}_{fa} \mathbf{n}_\Sigma \cdot \boldsymbol{\lambda}_{\mathbf{s}a} \,\mathrm{d}\Sigma = 0.
\end{split}\right.
\end{equation}

After the derivation of the adjoint sensitivities in Eqs. \ref{equ:AdjSensitivityNSCDGaDmObjBulkMHT} and \ref{equ:AdjSensitivityGaDmPressureConstrMHMBulk}, the design variables $\gamma$ and $d_m$ can be evolved iteratively to determine the fiber bundle of the thin walls for mass transfer in the volume flow.

\subsection{Heat transfer in volume flow}\label{sec:HeatTransferBulkFlows}

The mass transfer problem in the volume flow can be described by the Navier-Stokes equations and the convective heat-transfer equation.

\subsubsection{Navier-Stokes equations for volume flow} \label{sec:BulkNSEqusHT}

The governing equations for the motion of the fluid and the material interpolation on the implicit 2-manifold are the same as that introduced in Section \ref{sec:NavierStokesEquBulkFlows}. The difference is on the choice of the stabilization term in the variational formulation of the Navier-Stokes equations to numerically solve the fluid velocity and pressure by using linear finite elements. For the heat transfer problem, the variational formulation for the Navier-Stokes equations can be derived as
\begin{equation}\label{equ:VariationalFormulationBulkNavierStokesEqusHT}
\left\{\begin{split}
  & \mathrm{Find} \left\{\begin{split}
    & \mathbf{u}\in\left(\mathcal{H}\left(\Xi\right)\right)^3~\mathrm{with} ~ \left\{ \begin{split}
    & \mathbf{u} = \mathbf{u}_{\Gamma_{v,\Xi}}~ \mathrm{at} ~ \forall \mathbf{x}_\Xi \in \Gamma_{v,\Xi} \\
    & \mathbf{u} = \mathbf{0} ~ \mathrm{at} ~ \forall \mathbf{x}_\Xi \in \Gamma_{v_0,\Xi} \\
    \end{split}\right.\\
  & p \in \mathcal{H}\left(\Xi\right) \\
  \end{split}\right. \\
  & \mathrm{for} \left\{\begin{split}
  & \forall \tilde{\mathbf{u}} \in\left(\mathcal{H}\left(\Xi\right)\right)^3 \\
  & \forall \tilde{p} \in \mathcal{H}\left(\Xi\right) \\
  \end{split}\right., ~ \mathrm{such~that} \\
  &\int_\Xi \rho \left( \mathbf{u} \cdot \nabla_{\mathbf{x}_\Xi} \right) \mathbf{u} \cdot \tilde{\mathbf{u}} + {\eta\over2} \left( \nabla_{\mathbf{x}_\Xi} \mathbf{u} + \nabla_{\mathbf{x}_\Xi} \mathbf{u}^\mathrm{T} \right) : \left( \nabla_{\mathbf{x}_\Xi} \tilde{\mathbf{u}} + \nabla_{\mathbf{x}_\Xi} \tilde{\mathbf{u}}^\mathrm{T} \right) - p\,\mathrm{div}_{\mathbf{x}_\Xi} \tilde{\mathbf{u}} \\
  & - \tilde{p} \,\mathrm{div}_{\mathbf{x}_\Xi} \mathbf{u} \,\mathrm{d}\Xi - \sum_{E_\Xi\in\mathcal{E}_\Xi} \int_{E_\Xi} \tau_{LS\mathbf{u},\Xi} \left( \rho \mathbf{u} \cdot \nabla_{\mathbf{x}_\Xi} \mathbf{u} + \nabla_{\mathbf{x}_\Xi} p \right) \cdot ( \rho \mathbf{u} \cdot \nabla_{\mathbf{x}_\Xi} \tilde{\mathbf{u}} \\
  & + \nabla_{\mathbf{x}_\Xi} \tilde{p} ) + \tau_{LSp,\Xi} \left( \rho \mathrm{div}_{\mathbf{x}_\Xi} \mathbf{u} \right) \left( \mathrm{div}_{\mathbf{x}_\Xi} \tilde{\mathbf{u}} \right) \,\mathrm{d}\Xi + \int_\Gamma \alpha \mathbf{u} \cdot \tilde{\mathbf{u}} \, \mathrm{d}\Gamma = 0
\end{split}\right.
\end{equation}
where the general least square stabilization term is imposed as
\begin{equation}
\begin{split}
- \sum_{E_\Xi\in\mathcal{E}_\Xi} \int_{E_\Xi} \tau_{LS\mathbf{u},\Xi} \left( \rho \mathbf{u} \cdot \nabla_{\mathbf{x}_\Xi} \mathbf{u} + \nabla_{\mathbf{x}_\Xi} p \right) \cdot \left( \rho \mathbf{u} \cdot \nabla_{\mathbf{x}_\Xi} \tilde{\mathbf{u}} + \nabla_{\mathbf{x}_\Xi} \tilde{p} \right) + \tau_{LSp,\Xi} \left( \rho \mathrm{div}_{\mathbf{x}_\Xi} \mathbf{u} \right) \left( \mathrm{div}_{\mathbf{x}_\Xi} \tilde{\mathbf{u}} \right) \,\mathrm{d}\Xi
\end{split}
\end{equation}
with $\tau_{LS\mathbf{u},\Xi}$ and $\tau_{LSp,\Xi}$ representing the stabilization parameters. The stabilization parameters are set as \cite{DoneaWiley2003}
\begin{equation}\label{equ:NSBulkStabilizationTermHT}
\left\{\begin{split}
 & \tau_{LS\mathbf{u},\Xi} = \min \left( {h_{E_\Xi} \over 2 \rho \left\| \mathbf{u} \right\|_2}, { h_{E_\Xi}^2 \over 12 \eta } \right) \\
 & \tau_{LSp,\Xi} = \left\{\begin{split}
 & { 1 \over 2} h_{E_\Xi} \left\| \mathbf{u} \right\|_2, ~ \mathbf{u}^2 < \epsilon_{eps}^{1\over2} \\
 & { 1 \over 2} h_{E_\Xi}, ~ \mathbf{u}^2 \geq \epsilon_{eps}^{1\over2}
 \end{split}\right.
 \end{split}\right..
\end{equation}
Based on Eqs. \ref{equ:DiffRiemannianBulkMHM}, \ref{equ:ElementVolumeTransformationMHM} and \ref{equ:ApproximationBulkMetricAverageMHM}, the stabilization parameters in Eq. \ref{equ:NSBulkStabilizationTermHT} can be transformed into
\begin{equation}\label{equ:TransformedNSBulkStabilizationTermHT}
\left\{\begin{split}
 & \tau_{LS\mathbf{u},\Xi}^{\left( \mathbf{s} \right)} = \min \left( {h_{E_\Omega} \over 2 \rho \left\| \mathbf{u} \right\|_2 } \left(K^{\left( \mathbf{s} \right)}\right)^{1\over3}, { h_{E_\Omega}^2 \over 12 \eta } \left(K^{\left( \mathbf{s} \right)}\right)^{2\over3} \right) \\
 & \tau_{LSp,\Xi}^{\left( \mathbf{s} \right)} = \left\{\begin{split}
 & { 1 \over 2} h_{E_\Omega} \left\| \mathbf{u} \right\|_2 \left(K^{\left( \mathbf{s} \right)}\right)^{1\over3}, ~ \mathbf{u}^2 < \epsilon_{eps}^{1\over2} \\
 & { 1 \over 2} h_{E_\Omega} \left(K^{\left( \mathbf{s} \right)}\right)^{1\over3}, ~ \mathbf{u}^2 \geq \epsilon_{eps}^{1\over2}
 \end{split}\right.
 \end{split}\right..
\end{equation}

Based on the coupling relations in Section \ref{subsec:CouplingDesignVariablesBulkFlowMHT}, the variational formulation in Eq. \ref{equ:VariationalFormulationBulkNavierStokesEqusHT} can be transformed into the form defined on the original domain $\Omega$:
\begin{equation}\label{equ:TransformedVariationalFormulationBulkNSEqusHM}
\left\{\begin{split}
  & \mathrm{Find} \left\{\begin{split}
    & \mathbf{u}\in\left(\mathcal{H}\left(\Omega\right)\right)^3~\mathrm{with}~ \left\{ \begin{split}
    & \mathbf{u} = \mathbf{u}_{\Sigma_{v,\Omega}}~ \mathrm{at} ~ \forall \mathbf{x}_\Omega \in \Sigma_{v,\Omega} \\
    & \mathbf{u} = \mathbf{0}~ \mathrm{at} ~ \forall \mathbf{x}_\Omega \in \Sigma_{v_0,\Omega}\\
    \end{split}\right.\\
  & p \in \mathcal{H}\left(\Omega\right) \\
  \end{split}\right. \\
  & \mathrm{for} \left\{\begin{split}
  & \forall \tilde{\mathbf{u}} \in\left(\mathcal{H}\left(\Omega\right)\right)^3 \\
  & \forall \tilde{p} \in \mathcal{H}\left(\Omega\right) \\
  \end{split}\right.,~ \mathrm{such~that}\\
  &\int_\Omega \Big[ \rho \left( \mathbf{u} \cdot \nabla_{\mathbf{x}_\Xi}^{\left(\mathbf{s}\right)} \right) \mathbf{u} \cdot \tilde{\mathbf{u}} + {\eta\over2} \left( \nabla_{\mathbf{x}_\Xi}^{\left(\mathbf{s}\right)} \mathbf{u} + \nabla_{\mathbf{x}_\Xi}^{\left(\mathbf{s}\right)} \mathbf{u}^\mathrm{T} \right) : \left( \nabla_{\mathbf{x}_\Xi}^{\left(\mathbf{s}\right)} \tilde{\mathbf{u}} + \nabla_{\mathbf{x}_\Xi}^{\left(\mathbf{s}\right)} \tilde{\mathbf{u}}^\mathrm{T} \right) - p\,\mathrm{div}_{\mathbf{x}_\Xi}^{\left(\mathbf{s}\right)} \tilde{\mathbf{u}} \\
  & - \tilde{p} \, \mathrm{div}_{\mathbf{x}_\Xi}^{\left(\mathbf{s}\right)} \mathbf{u} \Big] K^{\left(\mathbf{s}\right)} \,\mathrm{d}\Omega - \sum_{E_\Omega\in\mathcal{E}_\Omega} \int_{E_\Omega} \Big[ \tau_{LS\mathbf{u},\Xi}^{\left(\mathbf{s}\right)} \left( \rho \mathbf{u} \cdot \nabla_{\mathbf{x}_\Xi}^{\left(\mathbf{s}\right)} \mathbf{u} + \nabla_{\mathbf{x}_\Xi}^{\left(\mathbf{s}\right)} p \right) \\
  & \cdot \left( \rho \mathbf{u} \cdot \nabla_{\mathbf{x}_\Xi}^{\left(\mathbf{s}\right)} \tilde{\mathbf{u}} + \nabla_{\mathbf{x}_\Xi}^{\left(\mathbf{s}\right)} \tilde{p} \right) + \tau_{LSp,\Xi}^{\left(\mathbf{s}\right)} \left( \rho \mathrm{div}_{\mathbf{x}_\Xi}^{\left(\mathbf{s}\right)} \mathbf{u} \right) \left( \mathrm{div}_{\mathbf{x}_\Xi}^{\left(\mathbf{s}\right)} \tilde{\mathbf{u}} \right) \Big] \\
  & \cdot K^{\left(\mathbf{s}\right)} \,\mathrm{d}\Omega + \int_\Sigma \alpha \mathbf{u} \cdot \tilde{\mathbf{u}} M^{\left(d_f\right)} \, \mathrm{d}\Sigma = 0.
\end{split}\right.
\end{equation}

\subsubsection{Convective heat-transfer equation for volume flow} \label{sec:BulkCHMEqu}

The heat transfer process in the volume flow can be described by the convective heat-transfer equation defined on the deformed domain. 
Based on the conservation law of energy, the convective heat-transfer equation can be derived to describe the heat transfer in the volume flow:
\begin{equation}\label{equ:BulkCHMequHT}
\begin{split}
\rho C_p \mathbf{u} \cdot \nabla_{\mathbf{x}_\Xi} T - \mathrm{div}_{\mathbf{x}_\Xi} \left( k \nabla_{\mathbf{x}_\Xi} T \right) & = Q, ~\forall \mathbf{x}_\Xi \in \Xi.
\end{split}
\end{equation}
On the material interpolation in fiber bundle topology optimization for heat transfer in the volume flow, it is implemented on the no-jump and no-slip boundary in Eq. \ref{equ:MaterialInterpolationMassTransferBulkFlow}, instead of the heat conductivity.
For the convective heat-transfer equation, the inlet boundary is heat sink, i.e. the temperature is known at $\Gamma_{v,\Xi}$; and the remained part of the boundary curve is insulative:
\begin{equation}\label{equ:TemperatureBoundaryConditionBulk}
\left\{
\begin{split}
  & T = T_0, ~ \forall \mathbf{x}_\Xi \in \Gamma_{v,\Xi} \\
  & \nabla_{\mathbf{x}_\Xi} T \cdot \mathbf{n}_{\partial\Xi} = 0, ~ \forall \mathbf{x}_\Xi \in \Gamma_{v_0,\Xi} \cup \Gamma_{s,\Xi}
\end{split}\right..
\end{equation}

Based on the Galerkin method, the variational formulation of the convective heat-transfer equation is considered in the first order Sobolev space defined on the deformed domain $\Xi$:
\begin{equation}\label{equ:VariationalFormulationBulkCHMEqu}
\left\{\begin{split}
  & \mathrm{Find}~ T\in\mathcal{H}\left(\Xi\right)~\mathrm{with} ~ T = T_0~ \mathrm{at} ~ \forall \mathbf{x}_\Xi \in \Gamma_{v,\Xi}, ~ \mathrm{for} ~ \forall \tilde{T} \in \mathcal{H}\left(\Xi\right), \\
  & \mathrm{such~that} ~ \int_\Xi \left( \rho C_p \mathbf{u} \cdot \nabla_{\mathbf{x}_\Xi} T - Q \right) \tilde{T} + k \nabla_{\mathbf{x}_\Xi} T \cdot \nabla_{\mathbf{x}_\Xi} \tilde{T} \,\mathrm{d}\Xi \\
  & + \sum_{E_\Xi\in\mathcal{E}_\Xi} \int_{E_\Xi} \tau_{LST,\Xi} \left( \rho C_p \mathbf{u} \cdot \nabla_{\mathbf{x}_\Xi} T - Q \right) \left( \rho C_p \mathbf{u} \cdot \nabla_{\mathbf{x}_\Xi} \tilde{T} \right) \,\mathrm{d}\Xi = 0 \\
\end{split}\right.
\end{equation}
where the general least square stabilization term
\begin{equation}
  \sum_{E_\Xi\in\mathcal{E}_\Xi} \int_{E_\Xi} \tau_{LST,\Xi} \left( \rho C_p \mathbf{u} \cdot \nabla_{\mathbf{x}_\Xi} T - Q \right) \left( \rho C_p \mathbf{u} \cdot \nabla_{\mathbf{x}_\Xi} \tilde{T} \right) \,\mathrm{d}\Xi
\end{equation}
with $\tau_{LST,\Xi}$ representing the stabilization parameter is imposed on the variational formulation, in order to use linear finite elements to solve the distribution of the temperature \cite{DoneaWiley2003}. The stabilization parameter is expressed as \cite{DoneaWiley2003}
\begin{equation}\label{equ:CHMBulkStabilizationTermCHM}
 \tau_{LST,\Xi} = \min \left( {h_{E_\Xi} \over 2 \rho C_p \left\|\mathbf{u}\right\|_2 }, { h_{E_\Xi}^2 \over 12k } \right).
\end{equation}
Based on Eqs. \ref{equ:DiffRiemannianBulkMHM}, \ref{equ:ElementVolumeTransformationMHM} and \ref{equ:ApproximationBulkMetricAverageMHM}, $\tau_{LST,\Xi}$ can be transformed into
\begin{equation}\label{equ:TransformedCHMBulkStabilizationTermCHM}
  \tau_{LST,\Xi}^{\left( \mathbf{s} \right)} = \min \left( {h_{E_\Omega} \over 2 \rho C_p \left\| \mathbf{u} \right\|_2 } \left(K^{\left( \mathbf{s} \right)}\right)^{1\over3}, { h_{E_\Omega}^2 \over 12k } \left(K^{\left( \mathbf{s} \right)}\right)^{2\over3} \right).
\end{equation}

Based on the coupling relations in Section \ref{subsec:CouplingDesignVariablesBulkFlowMHT}, the variational formulation in Eq. \ref{equ:VariationalFormulationBulkCHMEqu} can be transformed into the form defined on the original domain $\Omega$:
\begin{equation}\label{equ:TransformedVariationalFormulationBulkCHMEqu}
\left\{\begin{split}
  & \mathrm{Find}~ c\in\mathcal{H}\left(\Omega\right)~\mathrm{with} ~ c = c_0~ \mathrm{at} ~ \forall \mathbf{x}_\Omega \in \Sigma_{v,\Omega}, ~ \mathrm{for} ~ \forall \tilde{c} \in \mathcal{H}\left(\Omega\right), \\
  & \mathrm{such~that} ~ \int_\Omega \left[ \left( \rho C_p \mathbf{u} \cdot \nabla_{\mathbf{x}_\Xi}^{\left(\mathbf{s}\right)} T - Q \right) \tilde{T} + k \nabla_{\mathbf{x}_\Xi}^{\left(\mathbf{s}\right)} T \cdot \nabla_{\mathbf{x}_\Xi}^{\left(\mathbf{s}\right)} \tilde{T} \right] K^{\left(\mathbf{s}\right)} \,\mathrm{d}\Omega \\
  & + \sum_{E_\Omega\in\mathcal{E}_\Omega} \int_{E_\Omega} \tau_{LST,\Xi}^{\left(\mathbf{s}\right)} \left( \rho C_p \mathbf{u} \cdot \nabla_{\mathbf{x}_\Xi}^{\left(\mathbf{s}\right)} T - Q \right) \left( \rho C_p \mathbf{u} \cdot \nabla_{\mathbf{x}_\Xi}^{\left(\mathbf{s}\right)} \tilde{T} \right) K^{\left(\mathbf{s}\right)} \,\mathrm{d}\Omega = 0.
\end{split}\right.
\end{equation}

\subsubsection{Design objective and constraint of pressure drop} \label{sec:DesignObjectiveConstraintPressureDropBulkNSCHM}

For the heat transfer problem in the volume flow, the desired performance of the surface structure can be set to achieve the minimized thermal compliance. The thermal compliance can be measured by the integration of the square of the temperature gradient on the design domain. Therefore, the design objective of fiber bundle topology optimization for heat transfer is considered as
\begin{equation}\label{equ:DesignObjectiveBulkCHM} 
  J_T = \int_\Xi f_{id,\Xi} k \nabla_{\mathbf{x}_\Xi} T \cdot \nabla_{\mathbf{x}_\Xi} T \,\mathrm{d}\Xi,
\end{equation}
where $f_{id,\Xi} = f_{id,\Xi}\left(\mathbf{x}_\Xi\right)$ is the indicator function used to specify the computational domain of the design objective, i.e. $f_{id,\Xi}$ is valued as $1$ in the computational domain of the design objective, or else it is valued as $0$. Based on the coupling relations in Section \ref{subsec:CouplingDesignVariablesBulkFlowMHT}, the design objective in Eq. \ref{equ:DesignObjectiveBulkCHM} can be transformed into the following form:
\begin{equation}\label{equ:TransformedDesignObjectiveBulkCHM}
\begin{split}
  J_T^{\left( \mathbf{s} \right)} = & \int_\Omega f_{id,\Xi}^{\left( \mathbf{s} \right)} k \nabla_{\mathbf{x}_\Xi}^{\left( \mathbf{s} \right)} T \cdot \nabla_{\mathbf{x}_\Xi}^{\left( \mathbf{s} \right)} T K^{\left( \mathbf{s} \right)} \,\mathrm{d}\Omega,
\end{split}
\end{equation}
where $f_{id,\Xi}^{\left( \mathbf{s} \right)}$ defined on $\Omega$ is the homeomorphism of the indicator function $f_{id,\Xi}$ defined on $\Xi$.

The thermal compliance is constrained by the specified pressure drop, which is the same as that described by Eqs. \ref{equ:PressureConstraintBulkNSCD}, \ref{equ:PressureDropBulkNSCD} and \ref{equ:TransformedPressureConstraintSurfaceNSCD} in Section \ref{sec:DesignObjectiveConstraintBulkNSCDMHT}.

\subsubsection{Fiber bundle topology optimization problem}\label{sec:FiberBundleToopHTBulkFlows}

Based on the above introduction, the topology optimization problem for heat transfer in the volume flow can be constructed to optimize the fiber bundle in Eq. \ref{equ:FiberBundleMHM} for the thin-wall pattern defined on the implicit 2-manifold:
\begin{equation}\label{equ:VarProToopBulkNSCHTMHT}
\left\{\begin{split}
  & \mathrm{Find} \left\{\begin{split}
  & \gamma: \Gamma \mapsto \left[0,1\right] \\
  & d_m: \Sigma \mapsto \left[0,1\right]\end{split}\right.~ \mathrm{for} ~
  \left(\Sigma \times \left(\Gamma \times \left[0,1\right]\right), \Sigma, proj_1, \Gamma \times \left[0,1\right] \right), \\
  & \mathrm{to} ~ \mathrm{minimize}~{J_T \over J_{T,0}}~ \mathrm{with} ~ J_T = \int_\Xi f_{id,\Xi} k \nabla_{\mathbf{x}_\Xi} T \cdot \nabla_{\mathbf{x}_\Xi} T \,\mathrm{d}\Xi, \\
  & \mathrm{constrained} ~ \mathrm{by} \\
  & \left\{\begin{split}
  & \left\{\begin{split}
  & \left\{\begin{split}
  & \left\{\begin{split}
       & \rho \mathbf{u} \cdot \nabla_{\mathbf{x}_\Xi} \mathbf{u} - \mathrm{div}_{\mathbf{x}_\Xi} \left[ \eta \left( \nabla_{\mathbf{x}_\Xi} \mathbf{u} + \nabla_{\mathbf{x}_\Xi} \mathbf{u}^\mathrm{T} \right) \right] + \nabla_{\mathbf{x}_\Xi} p = \mathbf{0}, ~ \forall \mathbf{x}_\Xi \in \Xi \\
       & - \mathrm{div}_{\mathbf{x}_\Xi} \mathbf{u} = 0, ~ \forall \mathbf{x}_\Xi \in \Xi \\
       & \left\llbracket \left[ \eta \left( \nabla_{\mathbf{x}_\Xi} \mathbf{u} + \nabla_{\mathbf{x}_\Xi} \mathbf{u}^\mathrm{T} \right) - p \right] \mathbf{n}_{\partial\Xi} \right\rrbracket + \alpha \left(\gamma_p\right) \mathbf{u} = \mathbf{0}, ~ \forall \mathbf{x}_\Gamma \in \Gamma \\
    \end{split}\right.\\
  & \rho C_p \mathbf{u} \cdot \nabla_{\mathbf{x}_\Xi} T - \mathrm{div}_{\mathbf{x}_\Xi} \left( k \nabla_{\mathbf{x}_\Xi} T \right) = Q, ~\forall \mathbf{x}_\Xi \in \Xi \\
  \end{split}\right. \\
  & \alpha \left(\gamma_p\right) = \alpha_{\max} q {1-\gamma_p \over q + \gamma_p} \\
  \end{split}\right. \\
  & \left\{\begin{split}
  & \left\{\begin{split}
        & - \mathrm{div}_\Gamma \left( r_f^2 \nabla_\Gamma \gamma_f \right) + \gamma_f = \gamma,~\forall \mathbf{x}_\Gamma \in \Gamma \\
        & \mathbf{n}_{\boldsymbol\tau_\Gamma} \cdot \nabla_\Gamma \gamma_f = 0,~\forall \mathbf{x}_\Gamma \in \partial\Gamma \\
    \end{split}\right. \\
  & \gamma_p = { \tanh\left(\beta \xi\right) + \tanh\left(\beta \left(\gamma_f-\xi\right)\right) \over \tanh\left(\beta \xi\right) + \tanh\left(\beta \left(1-\xi\right)\right)} \\
  \end{split}\right.\\
  & \left\{
        \begin{split}
          & - \mathrm{div}_\Sigma \left( r_m^2 \nabla_\Sigma d_f \right) + d_f = A_d \left( d_m - {1\over2} \right), ~ \forall \mathbf{x}_\Sigma \in \Sigma \\
          & \mathbf{n}_{\boldsymbol\tau_\Sigma} \cdot \nabla_\Sigma d_f = 0, ~ \forall \mathbf{x}_\Sigma \in \partial \Sigma \\
        \end{split}\right. \\
  & \left\{\begin{split}
         & \mathrm{div}_{\mathbf{x}_\Omega} \left( \nabla_{\mathbf{x}_\Omega} \mathbf{s} \right) = \mathbf{0}, ~ \forall \mathbf{x}_\Omega \in \Omega \\
         & \mathbf{s} = \mathbf{0}, ~ \forall \mathbf{x}_\Omega \in \Sigma_{v,\Omega} \cup \Sigma_{s,\Omega} \\
         & \mathbf{s} = d_f \mathbf{n}_\Sigma, ~ \forall \mathbf{x}_\Omega \in \Sigma \\
         & \mathbf{n}_{\partial\Omega} \cdot \nabla_{\mathbf{x}_\Omega} \mathbf{s} = \mathbf{0}, ~ \forall \mathbf{x}_\Omega \in \Sigma_{v_0,\Omega} \\
  \end{split}\right. \\
  & \Xi = \left\{ \mathbf{x}_\Xi \left| \, \begin{split}
  \mathbf{x}_\Xi = \: & \mathbf{x}_\Omega + \mathbf{s}, ~ \forall \mathbf{x}_\Omega \in \Omega \\
  \mathbf{x}_\Xi = \: & \mathbf{x}_\Gamma, ~ \forall \mathbf{x}_\Gamma \in \Gamma \\
  \mathbf{x}_\Omega = \: & \mathbf{x}_\Sigma, ~ \forall \mathbf{x}_\Sigma \in \Sigma \\
  \mathbf{x}_\Gamma = \: & \mathbf{x}_\Sigma + d_f \mathbf{n}_\Sigma 
  \end{split}\right. \right\} \\
  & \left| \Delta P \left/ \: \Delta P_0 - 1 \right. \right| \leq 1\times10^{-3}, ~\mathrm{with} ~ \Delta P = \int_{\Gamma_{v,\Xi}} p \,\mathrm{d}\Gamma_{\partial\Xi} - \int_{\Gamma_{s,\Xi}} p \,\mathrm{d}\Gamma_{\partial\Xi}. \\
\end{split}\right.
\end{split}\right.
\end{equation}

The coupling relations among the variables, functions, and differential operators in Eq. \ref{equ:VarProToopBulkNSCHTMHT} are illustrated by the arrow chart described as
\[\begin{array}{cccccccc}
 \textcolor{blue}{d_m} & \xrightarrow{\mathrm{Eq.~}\ref{equ:PDEFilterzmBaseStructureMHM}} & d_f & \xrightarrow{\mathrm{Eqs.~}\ref{equ:HarmonicCoordinateEquMHT}~\&~\ref{equ:BulkTransformedGradDeformedDomMHT}} & \left\{ \mathbf{s}, \nabla_{\mathbf{x}_\Xi}, \mathrm{div}_{\mathbf{x}_\Xi}, \mathbf{n}_{\partial\Xi} \right\} & \\
 & & \bigg\downarrow\vcenter{\rlap{\scriptsize{Eq.~\ref{equ:TransformedTangentialOperatorMHM}}}} & & \bigg\downarrow\vcenter{\rlap{\scriptsize{Eqs.~\ref{equ:NavierStokesEquBulkFlowMassTransferMHT}
 ~\&~\ref{equ:BulkCHMequHT}}}} \\
 & & \left\{ \nabla_\Gamma, \mathrm{div}_\Gamma, \mathbf{n}_\Gamma \right\} & & \left\{ \mathbf{u},~p,~T \right\} & \xrightarrow{\mathrm{Eqs.~}\ref{equ:DesignObjectiveBulkCHM} ~\&~ \ref{equ:PressureDropBulkNSCD}} & \left\{ \textcolor[rgb]{0.50,0.00,0.00}{J_T}, ~ \textcolor[rgb]{0.50,0.00,0.00}{\Delta P} \right\} \\
 & & \bigg\downarrow\vcenter{\rlap{\scriptsize{Eq.~\ref{equ:PDEFilterGammaFilberMHM}}}} & &  \bigg\uparrow\vcenter{\rlap{\scriptsize{Eq.~\ref{equ:MaterialInterpolationMassTransferBulkFlow}}}}   \\
 \textcolor{blue}{\gamma} & \xrightarrow{\mathrm{Eq.~}\ref{equ:PDEFilterGammaFilberMHM}} & \gamma_f & \xrightarrow{\mathrm{Eq.~}\ref{equ:ProjectionGammaFilberMHM}} & \textcolor[rgb]{0.50,0.00,0.00}{\gamma_p} & & \\
\end{array}\]
where the design variables $d_m$ and $\gamma$, marked in blue, are the inputs; the design objective $J_T$, the pressure drop $\Delta P$, and the material density $\gamma_p$, marked in red, are the outputs.

\subsubsection{Adjoint analysis} \label{sec:AdjointAnalysisBulkNSHTEqusMHT}

To solve the fiber bundle topology optimization problem in Eq. \ref{equ:VarProToopBulkNSCHTMHT}, the adjoint analysis is implemented for the design objective and constraint of the pressure drop to derive the adjoint sensitivities. The details for the adjoint analysis are provided in the appendix in Sections \ref{sec:AdjointAnalysisDesignObjectiveBulkCHT} and \ref{sec:AdjointAnalysisDissipationConstraintBulkMHM}.

Based on the transformed design objective in Eq. \ref{equ:TransformedDesignObjectiveBulkCHM} and transformed pressure drop in Eq. \ref{equ:TransformedPressureConstraintSurfaceNSCD}, the adjoint analysis of the fiber bundle topology optimization problem can be implemented in the functional spaces defined on the original domain $\Omega$. Based on the continuous adjoint method \cite{HinzeSpringer2009}, the adjoint sensitivity of the design objective $J_T$ is derived as
\begin{equation}\label{equ:AdjSensitivityNSCHTGaDmObjBulkMHT}
\begin{split}
\delta J_T = \int_\Sigma - \gamma_{fa} \tilde{\gamma} M^{\left( d_f \right)} - A_d d_{fa} \tilde{d}_m \,\mathrm{d}\Sigma,~ \forall \left( \tilde{\gamma}, \tilde{d}_m \right) \in \left(\mathcal{L}^2\left(\Sigma\right)\right)^2.
\end{split}
\end{equation}
The adjoint variables in Eq. \ref{equ:AdjSensitivityNSCHTGaDmObjBulkMHT} can be derived by sequentially solving the following adjoint equations in variational formulations. The variational formulation for the adjoint equation of the convective heat-transfer equation is derived as
\begin{equation}\label{equ:WeakAdjEquHTEquBulkMTTa}
\left\{\begin{split}
  & \mathrm{Find} ~ T_a \in \mathcal{H}\left(\Omega\right)~\mathrm{with}~ T_a=0 ~ \mathrm{at} ~ \forall \mathbf{x}_\Omega \in \Sigma_{v,\Omega}, ~ \mathrm{for} ~\forall \tilde{T}_a \in \mathcal{H} \left(\Omega\right),~\mathrm{such~that} \\
  & \int_\Omega \left[ 2 f_{id,\Xi}^{\left( \mathbf{s} \right)} k \nabla_{\mathbf{x}_\Xi}^{\left( \mathbf{s} \right)} \tilde{T}_a \cdot \nabla_{\mathbf{x}_\Xi}^{\left( \mathbf{s} \right)} T + \left( \rho C_p \mathbf{u} \cdot \nabla_{\mathbf{x}_\Xi}^{\left(\mathbf{s}\right)} \tilde{T}_a \right) T_a + k \nabla_{\mathbf{x}_\Xi}^{\left(\mathbf{s}\right)} \tilde{T}_a \cdot \nabla_{\mathbf{x}_\Xi}^{\left(\mathbf{s}\right)} T_a \right] K^{\left(\mathbf{s}\right)} \,\mathrm{d}\Omega \\
  & + \sum_{E_\Omega\in\mathcal{E}_\Omega} \int_{E_\Omega} \tau_{LST,\Xi}^{\left(\mathbf{s}\right)} \left( \rho C_p \mathbf{u} \cdot \nabla_{\mathbf{x}_\Xi}^{\left(\mathbf{s}\right)} \tilde{T}_a \right) \left( \rho C_p \mathbf{u} \cdot \nabla_{\mathbf{x}_\Xi}^{\left(\mathbf{s}\right)} T_a \right) K^{\left(\mathbf{s}\right)} \,\mathrm{d}\Omega = 0.
\end{split}\right.
\end{equation}
The variational formulation for the adjoint equations of the Naiver-Stokes equations is derived as
\begin{equation}\label{equ:AdjBulkNavierStokesEqusJObjectiveHTUaPa} 
\left\{\begin{split}
  & \mathrm{Find} \left\{\begin{split}
  & \mathbf{u}_a \in\left(\mathcal{H}\left(\Omega\right)\right)^3~\mathrm{with}~ \mathbf{u}_a = \mathbf{0}~ \mathrm{at} ~ {\forall \mathbf{x}_\Omega \in \Sigma_{v,\Omega} \cup \Sigma_{v_0,\Omega} } \\
  & p_a \in \mathcal{H}\left(\Omega\right) \\
  \end{split}\right.\\
  & \mathrm{for} \left\{\begin{split} 
  & \forall \tilde{\mathbf{u}}_a \in\left(\mathcal{H}\left(\Omega\right)\right)^3 \\
  & \forall \tilde{p}_a \in \mathcal{H}\left(\Omega\right) \\
  \end{split}\right.,~\mathrm{such~that} \\
  & \int_\Omega \Big[ \rho \left( \tilde{\mathbf{u}}_a \cdot \nabla_{\mathbf{x}_\Xi}^{\left(\mathbf{s}\right)} \right) \mathbf{u} \cdot \mathbf{u}_a + \rho \left( \mathbf{u} \cdot \nabla_{\mathbf{x}_\Xi}^{\left(\mathbf{s}\right)} \right) \tilde{\mathbf{u}}_a \cdot \mathbf{u}_a + {\eta\over2} \left( \nabla_{\mathbf{x}_\Xi}^{\left(\mathbf{s}\right)} \tilde{\mathbf{u}}_a + \nabla_{\mathbf{x}_\Xi}^{\left(\mathbf{s}\right)} \tilde{\mathbf{u}}_a^\mathrm{T} \right) \\
  & : \left( \nabla_{\mathbf{x}_\Xi}^{\left(\mathbf{s}\right)} \mathbf{u}_a + \nabla_{\mathbf{x}_\Xi}^{\left(\mathbf{s}\right)} \mathbf{u}_a^\mathrm{T} \right) - \tilde{p}_a\mathrm{div}_{\mathbf{x}_\Xi}^{\left(\mathbf{s}\right)} \mathbf{u}_a - p_a \mathrm{div}_{\mathbf{x}_\Xi}^{\left(\mathbf{s}\right)} \tilde{\mathbf{u}}_a + \left( \rho C_p \tilde{\mathbf{u}}_a \cdot \nabla_{\mathbf{x}_\Xi}^{\left(\mathbf{s}\right)} T \right) T_a \Big] K^{\left(\mathbf{s}\right)} \,\mathrm{d}\Omega \\
  & - \sum_{E_\Omega\in\mathcal{E}_\Omega} \int_{E_\Omega} \Big[ \tau_{LS\mathbf{u},\Xi}^{\left(\mathbf{s}, \tilde{\mathbf{u}}_a \right)} \left( \rho \mathbf{u} \cdot \nabla_{\mathbf{x}_\Xi}^{\left(\mathbf{s}\right)} \mathbf{u} + \nabla_{\mathbf{x}_\Xi}^{\left(\mathbf{s}\right)} p \right) \cdot \left( \rho \mathbf{u} \cdot \nabla_{\mathbf{x}_\Xi}^{\left(\mathbf{s}\right)} \mathbf{u}_a + \nabla_{\mathbf{x}_\Xi}^{\left(\mathbf{s}\right)} p_a \right) \\
  & + \tau_{LS\mathbf{u},\Xi}^{\left(\mathbf{s}\right)} \left( \rho \tilde{\mathbf{u}}_a \cdot \nabla_{\mathbf{x}_\Xi}^{\left(\mathbf{s} \right)} \mathbf{u} + \rho \mathbf{u} \cdot \nabla_{\mathbf{x}_\Xi}^{\left(\mathbf{s}\right)} \tilde{\mathbf{u}}_a + \nabla_{\mathbf{x}_\Xi}^{\left(\mathbf{s}\right)} \tilde{p}_a \right) \cdot \left( \rho \mathbf{u} \cdot \nabla_{\mathbf{x}_\Xi}^{\left(\mathbf{s}\right)} \mathbf{u}_a + \nabla_{\mathbf{x}_\Xi}^{\left(\mathbf{s}\right)} p_a \right) \\
  &  + \tau_{LS\mathbf{u},\Xi}^{\left(\mathbf{s}\right)} \left( \rho \mathbf{u} \cdot \nabla_{\mathbf{x}_\Xi}^{\left(\mathbf{s}\right)} \mathbf{u} + \nabla_{\mathbf{x}_\Xi}^{\left(\mathbf{s}\right)} p \right) \cdot \left( \rho \tilde{\mathbf{u}}_a \cdot \nabla_{\mathbf{x}_\Xi}^{\left(\mathbf{s}\right)} \mathbf{u}_a \right) + \tau_{LSp,\Xi}^{\left(\mathbf{s}, \tilde{\mathbf{u}}_a \right)} \left( \rho \mathrm{div}_{\mathbf{x}_\Xi}^{\left(\mathbf{s}\right)} \mathbf{u} \right) \left( \mathrm{div}_{\mathbf{x}_\Xi}^{\left(\mathbf{s}\right)} \mathbf{u}_a \right) \\
  & + \tau_{LSp,\Xi}^{\left(\mathbf{s}\right)} \left( \rho \mathrm{div}_{\mathbf{x}_\Xi}^{\left(\mathbf{s}\right)} \tilde{\mathbf{u}}_a \right) \left( \mathrm{div}_{\mathbf{x}_\Xi}^{\left(\mathbf{s}\right)} \mathbf{u}_a \right) - \tau_{LST,\Xi}^{\left(\mathbf{s}, \tilde{\mathbf{u}}_a \right)} \left( \rho C_p \mathbf{u} \cdot \nabla_{\mathbf{x}_\Xi}^{\left(\mathbf{s}\right)} T - Q \right) \left( \rho C_p \mathbf{u} \cdot \nabla_{\mathbf{x}_\Xi}^{\left(\mathbf{s}\right)} T_a \right) \\
  & - \tau_{LST,\Xi}^{\left(\mathbf{s}\right)} \left( \rho C_p \tilde{\mathbf{u}}_a \cdot \nabla_{\mathbf{x}_\Xi}^{\left(\mathbf{s}\right)} T \right) \left( \rho C_p \mathbf{u} \cdot \nabla_{\mathbf{x}_\Xi}^{\left(\mathbf{s}\right)} T_a \right) - \tau_{LST,\Xi}^{\left(\mathbf{s}\right)} \left( \rho C_p \mathbf{u} \cdot \nabla_{\mathbf{x}_\Xi}^{\left(\mathbf{s}\right)} T - Q \right) \\
  & \left( \rho C_p \tilde{\mathbf{u}}_a \cdot \nabla_{\mathbf{x}_\Xi}^{\left(\mathbf{s}\right)} T_a \right) \Big] K^{\left(\mathbf{s}\right)} \,\mathrm{d}\Omega + \int_\Sigma \alpha \tilde{\mathbf{u}}_a \cdot \mathbf{u}_a M^{\left(d_f\right)} \, \mathrm{d}\Sigma = 0
\end{split}\right.
\end{equation}
where $\tau_{LS\mathbf{u},\Xi}^{\left( \mathbf{s}, \tilde{\mathbf{u}} \right)}$, $\tau_{LSp,\Xi}^{\left( \mathbf{s}, \tilde{\mathbf{u}} \right)}$ and $\tau_{LST,\Xi}^{\left( \mathbf{s}, \tilde{\mathbf{u}} \right)}$ are the first-order variationals of $\tau_{LS\mathbf{u},\Xi}^{\left( \mathbf{s} \right)}$, $\tau_{LSp,\Xi}^{\left( \mathbf{s} \right)}$ and $\tau_{LST,\Xi}^{\left( \mathbf{s} \right)}$ to $\mathbf{u}$, respectively, and they are expressed as
\begin{equation}\label{equ:TransformedNSBulkStabilizationTerm1stVariUnormHT}
\begin{split}
& \left.\begin{split}
 & \tau_{LS\mathbf{u},\Xi}^{\left( \mathbf{s}, \tilde{\mathbf{u}} \right)} = \left\{\begin{split}
   & - {h_{E_\Omega} \mathbf{u} \cdot \tilde{\mathbf{u}} \over 2 \rho \left\| \mathbf{u} \right\|_2^3 } \left(K^{\left( \mathbf{s} \right)}\right)^{1\over3}, ~ {h_{E_\Omega} \over 2 \rho \left\| \mathbf{u} \right\|_2 } \left(K^{\left( \mathbf{s} \right)}\right)^{1\over3} < { h_{E_\Omega}^2 \over 12 \eta } \left(K^{\left( \mathbf{s} \right)}\right)^{2\over3} \\
   & 0, ~ {h_{E_\Omega} \over 2 \rho \left\| \mathbf{u} \right\|_2 } \left(K^{\left( \mathbf{s} \right)}\right)^{1\over3} \geq { h_{E_\Omega}^2 \over 12 \eta } \left(K^{\left( \mathbf{s} \right)}\right)^{2\over3}
  \end{split}\right. \\
 & \tau_{LSp,\Xi}^{\left( \mathbf{s}, \tilde{\mathbf{u}} \right)} = \left\{\begin{split}
 & { h_{E_\Omega} \mathbf{u} \cdot \tilde{\mathbf{u}} \over 2 \left\| \mathbf{u} \right\|_2 } \left(K^{\left( \mathbf{s} \right)}\right)^{1\over3}, ~ \mathbf{u}^2 < \epsilon_{eps}^{1\over2} \\
 & 0, ~ \mathbf{u}^2 \geq \epsilon_{eps}^{1\over2}
 \end{split}\right. \\
 & \tau_{LST,\Xi}^{\left( \mathbf{s}, \tilde{\mathbf{u}} \right)} = \left\{\begin{split} 
  & - {h_{E_\Sigma} \mathbf{u} \cdot \tilde{\mathbf{u}} \over 2 \rho C_p \left\| \mathbf{u} \right\|_2^3 } \left(K^{\left( \mathbf{s} \right)}\right)^{1\over3}, ~ {h_{E_\Omega} \over 2 \rho C_p \left\| \mathbf{u} \right\|_2 } \left(K^{\left( \mathbf{s} \right)}\right)^{1\over3} < { h_{E_\Omega}^2 \over 12k } \left(K^{\left( \mathbf{s} \right)}\right)^{2\over3} \\
  & 0, ~ {h_{E_\Omega} \over 2 \rho C_p \left\| \mathbf{u} \right\|_2 } \left(K^{\left( \mathbf{s} \right)}\right)^{1\over3} \geq { h_{E_\Omega}^2 \over 12k } \left(K^{\left( \mathbf{s} \right)}\right)^{2\over3}
  \end{split}\right.
 \end{split}\right\} \\
 & \forall \tilde{\mathbf{u}} \in \left( \mathcal{H} \left(\Omega\right) \right)^3.
\end{split}
\end{equation}
The variational formulation for the adjoint equation of the Laplace's equation for $\mathbf{s}$ is derived as
\begin{equation}\label{equ:WeakAdjEquHarmonicEquBulkHTSa} 
\left\{\begin{split}
  & \mathrm{Find} \left\{\begin{split}
  & \mathbf{s}_a \in \left(\mathcal{H}\left(\Omega\right)\right)^3~\mathrm{with}~ \mathbf{s}_a = \mathbf{0} ~ \mathrm{at} ~ \forall \mathbf{x}_\Omega \in \Sigma_{v,\Omega} \cup \Sigma_{s,\Omega} \\
  & \boldsymbol{\lambda}_{\mathbf{s}a} \in \left(\mathcal{H}^{-{1\over2}}\left(\Sigma\right)\right)^3 \\
  \end{split}\right.,\\
  & \mathrm{for} \left\{\begin{split} 
  & \forall \tilde{\mathbf{s}}_a \in\left(\mathcal{H}\left(\Omega\right)\right)^3 \\
  & \forall \tilde{\boldsymbol{\lambda}}_{\mathbf{s}a} \in \left(\mathcal{H}^{{1\over2}}\left(\Sigma\right)\right)^3 \\
  \end{split}\right.,~\mathrm{such~that} \\
  & \int_\Omega \Big[ f_{id,\Xi}^{\left( \mathbf{s}, \tilde{\mathbf{s}}_a \right)} k \nabla_{\mathbf{x}_\Xi}^{\left( \mathbf{s} \right)} T \cdot \nabla_{\mathbf{x}_\Xi}^{\left( \mathbf{s} \right)} T + 2f_{id,\Xi}^{\left( \mathbf{s} \right)} k \nabla_{\mathbf{x}_\Xi}^{\left( \mathbf{s}, \tilde{\mathbf{s}}_a \right)} T \cdot \nabla_{\mathbf{x}_\Xi}^{\left( \mathbf{s} \right)} T \\
  & + \rho \left( \mathbf{u} \cdot \nabla_{\mathbf{x}_\Xi}^{\left(\mathbf{s}, \tilde{\mathbf{s}}_a\right)} \right) \mathbf{u} \cdot \mathbf{u}_a + {\eta\over2} \left( \nabla_{\mathbf{x}_\Xi}^{\left(\mathbf{s}, \tilde{\mathbf{s}}_a\right)} \mathbf{u} + \nabla_{\mathbf{x}_\Xi}^{\left(\mathbf{s}, \tilde{\mathbf{s}}_a\right)} \mathbf{u}^\mathrm{T} \right) : \left( \nabla_{\mathbf{x}_\Xi}^{\left(\mathbf{s}\right)} \mathbf{u}_a + \nabla_{\mathbf{x}_\Xi}^{\left(\mathbf{s}\right)} \mathbf{u}_a^\mathrm{T} \right) \\
  & + {\eta\over2} \left( \nabla_{\mathbf{x}_\Xi}^{\left(\mathbf{s}\right)} \mathbf{u} + \nabla_{\mathbf{x}_\Xi}^{\left(\mathbf{s}\right)} \mathbf{u}^\mathrm{T} \right) : \left( \nabla_{\mathbf{x}_\Xi}^{\left(\mathbf{s}, \tilde{\mathbf{s}}_a\right)} \mathbf{u}_a + \nabla_{\mathbf{x}_\Xi}^{\left(\mathbf{s}, \tilde{\mathbf{s}}_a\right)} \mathbf{u}_a^\mathrm{T} \right) - p\,\mathrm{div}_{\mathbf{x}_\Xi}^{\left(\mathbf{s}, \tilde{\mathbf{s}}_a\right)} \mathbf{u}_a - p_a \mathrm{div}_{\mathbf{x}_\Xi}^{\left(\mathbf{s}, \tilde{\mathbf{s}}_a\right)} \mathbf{u} \\
  & + \left( \rho C_p \mathbf{u} \cdot \nabla_{\mathbf{x}_\Xi}^{\left(\mathbf{s}, \tilde{\mathbf{s}}_a\right)} T \right) T_a + k \nabla_{\mathbf{x}_\Xi}^{\left(\mathbf{s}, \tilde{\mathbf{s}}_a\right)} T \cdot \nabla_{\mathbf{x}_\Xi}^{\left(\mathbf{s}\right)} T_a + k \nabla_{\mathbf{x}_\Xi}^{\left(\mathbf{s}\right)} T \cdot \nabla_{\mathbf{x}_\Xi}^{\left(\mathbf{s}, \tilde{\mathbf{s}}_a\right)} T_a \Big] K^{\left(\mathbf{s}\right)} \\
  & + \Big[ f_{id,\Xi}^{\left( \mathbf{s} \right)} k \nabla_{\mathbf{x}_\Xi}^{\left( \mathbf{s} \right)} T \cdot \nabla_{\mathbf{x}_\Xi}^{\left( \mathbf{s} \right)} T + \rho \left( \mathbf{u} \cdot \nabla_{\mathbf{x}_\Xi}^{\left(\mathbf{s}\right)} \right) \mathbf{u} \cdot \mathbf{u}_a + {\eta\over2} \left( \nabla_{\mathbf{x}_\Xi}^{\left(\mathbf{s}\right)} \mathbf{u} + \nabla_{\mathbf{x}_\Xi}^{\left(\mathbf{s}\right)} \mathbf{u}^\mathrm{T} \right) \\
  & : \left( \nabla_{\mathbf{x}_\Xi}^{\left(\mathbf{s}\right)} \mathbf{u}_a + \nabla_{\mathbf{x}_\Xi}^{\left(\mathbf{s}\right)} \mathbf{u}_a^\mathrm{T} \right) - p\,\mathrm{div}_{\mathbf{x}_\Xi}^{\left(\mathbf{s}\right)} \mathbf{u}_a - p_a \mathrm{div}_{\mathbf{x}_\Xi}^{\left(\mathbf{s} \right)} \mathbf{u} + \left( \rho C_p \mathbf{u} \cdot \nabla_{\mathbf{x}_\Xi}^{\left(\mathbf{s}\right)} T - Q \right) T_a \\
  & + k \nabla_{\mathbf{x}_\Xi}^{\left(\mathbf{s}\right)} T \cdot \nabla_{\mathbf{x}_\Xi}^{\left(\mathbf{s}\right)} T_a \Big] K^{\left(\mathbf{s}, \tilde{\mathbf{s}}_a\right)} - \nabla_{\mathbf{x}_\Omega} \tilde{\mathbf{s}}_a : \nabla_{\mathbf{x}_\Omega} \mathbf{s}_a \,\mathrm{d}\Omega \\
  & - \sum_{E_\Omega\in\mathcal{E}_\Omega} \int_{E_\Omega} \Big[ \tau_{LS\mathbf{u},\Xi}^{\left(\mathbf{s}, \tilde{\mathbf{s}}_a\right)} \left( \rho \mathbf{u} \cdot \nabla_{\mathbf{x}_\Xi}^{\left(\mathbf{s}\right)} \mathbf{u} + \nabla_{\mathbf{x}_\Xi}^{\left(\mathbf{s}\right)} p \right) \cdot \left( \rho \mathbf{u} \cdot \nabla_{\mathbf{x}_\Xi}^{\left(\mathbf{s}\right)} \mathbf{u}_a + \nabla_{\mathbf{x}_\Xi}^{\left(\mathbf{s}\right)} p_a \right) \\
  & + \tau_{LS\mathbf{u},\Xi}^{\left(\mathbf{s}\right)} \left( \rho \mathbf{u} \cdot \nabla_{\mathbf{x}_\Xi}^{\left(\mathbf{s}, \tilde{\mathbf{s}}_a\right)} \mathbf{u} + \nabla_{\mathbf{x}_\Xi}^{\left(\mathbf{s}, \tilde{\mathbf{s}}_a\right)} p \right) \cdot \left( \rho \mathbf{u} \cdot \nabla_{\mathbf{x}_\Xi}^{\left(\mathbf{s}\right)} \mathbf{u}_a + \nabla_{\mathbf{x}_\Xi}^{\left(\mathbf{s}\right)} p_a \right) \\
  & + \tau_{LS\mathbf{u},\Xi}^{\left(\mathbf{s}\right)} \left( \rho \mathbf{u} \cdot \nabla_{\mathbf{x}_\Xi}^{\left(\mathbf{s}\right)} \mathbf{u} + \nabla_{\mathbf{x}_\Xi}^{\left(\mathbf{s}\right)} p \right) \cdot \left( \rho \mathbf{u} \cdot \nabla_{\mathbf{x}_\Xi}^{\left(\mathbf{s}, \tilde{\mathbf{s}}_a\right)} \mathbf{u}_a + \nabla_{\mathbf{x}_\Xi}^{\left(\mathbf{s}, \tilde{\mathbf{s}}_a\right)} p_a \right) \\
  & + \tau_{LSp,\Xi}^{\left(\mathbf{s}, \tilde{\mathbf{s}}_a\right)} \left( \rho \mathrm{div}_{\mathbf{x}_\Xi}^{\left(\mathbf{s}\right)} \mathbf{u} \right) \left( \mathrm{div}_{\mathbf{x}_\Xi}^{\left(\mathbf{s}\right)} \mathbf{u}_a \right) + \tau_{LSp,\Xi}^{\left(\mathbf{s}\right)} \left( \rho \mathrm{div}_{\mathbf{x}_\Xi}^{\left(\mathbf{s}, \tilde{\mathbf{s}}_a\right)} \mathbf{u} \right) \left( \mathrm{div}_{\mathbf{x}_\Xi}^{\left(\mathbf{s}\right)} \mathbf{u}_a \right) \\
  & + \tau_{LSp,\Xi}^{\left(\mathbf{s}\right)} \left( \rho \mathrm{div}_{\mathbf{x}_\Xi}^{\left(\mathbf{s}\right)} \mathbf{u} \right) \left( \mathrm{div}_{\mathbf{x}_\Xi}^{\left(\mathbf{s}, \tilde{\mathbf{s}}_a\right)} \mathbf{u}_a \right) - \tau_{LST,\Xi}^{\left(\mathbf{s}, \tilde{\mathbf{s}}_a\right)} \left( \rho C_p \mathbf{u} \cdot \nabla_{\mathbf{x}_\Xi}^{\left(\mathbf{s}\right)} T - Q \right) \\
  & \left( \rho C_p \mathbf{u} \cdot \nabla_{\mathbf{x}_\Xi}^{\left(\mathbf{s}\right)} T_a \right) - \tau_{LST,\Xi}^{\left(\mathbf{s}\right)} \left( \rho C_p \mathbf{u} \cdot \nabla_{\mathbf{x}_\Xi}^{\left(\mathbf{s}, \tilde{\mathbf{s}}_a\right)} T \right) \left( \rho C_p \mathbf{u} \cdot \nabla_{\mathbf{x}_\Xi}^{\left(\mathbf{s}\right)} T_a \right) \\
  & - \tau_{LST,\Xi}^{\left(\mathbf{s}\right)} \left( \rho C_p \mathbf{u} \cdot \nabla_{\mathbf{x}_\Xi}^{\left(\mathbf{s}\right)} T - Q \right) \left( \rho C_p \mathbf{u} \cdot \nabla_{\mathbf{x}_\Xi}^{\left(\mathbf{s}, \tilde{\mathbf{s}}_a \right)} T_a \right) \Big] K^{\left(\mathbf{s}\right)} \\
  & + \Big[ \tau_{LS\mathbf{u},\Xi}^{\left(\mathbf{s}\right)} \left( \rho \mathbf{u} \cdot \nabla_{\mathbf{x}_\Xi}^{\left(\mathbf{s}\right)} \mathbf{u} + \nabla_{\mathbf{x}_\Xi}^{\left(\mathbf{s}\right)} p \right) \cdot \left( \rho \mathbf{u} \cdot \nabla_{\mathbf{x}_\Xi}^{\left(\mathbf{s}\right)} \mathbf{u}_a + \nabla_{\mathbf{x}_\Xi}^{\left(\mathbf{s}\right)} p_a \right) \\
  & + \tau_{LSp,\Xi}^{\left(\mathbf{s}\right)} \left( \rho \mathrm{div}_{\mathbf{x}_\Xi}^{\left(\mathbf{s}\right)} \mathbf{u} \right) \left( \mathrm{div}_{\mathbf{x}_\Xi}^{\left(\mathbf{s}\right)} \mathbf{u}_a \right) - \tau_{LST,\Xi}^{\left(\mathbf{s}\right)} \left( \rho C_p \mathbf{u} \cdot \nabla_{\mathbf{x}_\Xi}^{\left(\mathbf{s}\right)} T - Q \right) \\
  & \left( \rho C_p \mathbf{u} \cdot \nabla_{\mathbf{x}_\Xi}^{\left(\mathbf{s}\right)} T_a \right) \Big] K^{\left(\mathbf{s}, \tilde{\mathbf{s}}_a\right)} \,\mathrm{d}\Omega + \int_\Sigma \tilde{\mathbf{s}}_a \cdot \boldsymbol{\lambda}_{\mathbf{s}a} + \tilde{\boldsymbol{\lambda}}_{\mathbf{s}a} \cdot \mathbf{s}_a \,\mathrm{d}\Sigma = 0
\end{split}\right.
\end{equation}
where $\tau_{LS\mathbf{u},\Xi}^{\left( \mathbf{s}, \tilde{\mathbf{s}} \right)}$, $\tau_{LSp,\Xi}^{\left( \mathbf{s}, \tilde{\mathbf{s}} \right)}$ and $\tau_{LST,\Xi}^{\left( \mathbf{s}, \tilde{\mathbf{s}} \right)}$ are the first-order variationals of $\tau_{LS\mathbf{u},\Xi}^{\left( \mathbf{s} \right)}$, $\tau_{LSp,\Xi}^{\left( \mathbf{s} \right)}$ and $\tau_{LST,\Xi}^{\left( \mathbf{s} \right)}$ to $\mathbf{s}$, respectively, and they are expressed as
\begin{equation}\label{equ:TransformedNSBulkStabilizationTerm1stVarisHT}
\begin{split}
& \left.\begin{split}
 & \tau_{LS\mathbf{u},\Xi}^{\left( \mathbf{s}, \tilde{\mathbf{s}} \right)} = \left\{\begin{split}
   & {h_{E_\Omega} \over 6 \rho \left\| \mathbf{u} \right\|_2 \left(K^{\left( \mathbf{s} \right)}\right)^{2\over3}} K^{\left( \mathbf{s}, \tilde{\mathbf{s}} \right)} , ~ {h_{E_\Omega} \over 2 \rho \left\| \mathbf{u} \right\|_2 } \left(K^{\left( \mathbf{s} \right)}\right)^{1\over3} < { h_{E_\Omega}^2 \over 12 \eta } \left(K^{\left( \mathbf{s} \right)}\right)^{2\over3} \\
   & { h_{E_\Omega}^2 \over 18 \eta \left(K^{\left( \mathbf{s} \right)}\right)^{1\over3} } K^{\left( \mathbf{s}, \tilde{\mathbf{s}} \right)} , ~ {h_{E_\Omega} \over 2 \rho \left\| \mathbf{u} \right\|_2 } \left(K^{\left( \mathbf{s} \right)}\right)^{1\over3} \geq { h_{E_\Omega}^2 \over 12 \eta } \left(K^{\left( \mathbf{s} \right)}\right)^{2\over3}
  \end{split}\right. \\
 & \tau_{LSp,\Xi}^{\left( \mathbf{s}, \tilde{\mathbf{s}} \right)} = \left\{\begin{split}
 & { 1 \over 6} h_{E_\Omega} \left\| \mathbf{u} \right\|_2 \left(K^{\left( \mathbf{s} \right)}\right)^{-{2\over3}} K^{\left( \mathbf{s}, \tilde{\mathbf{s}} \right)}, ~ \mathbf{u}^2 < \epsilon_{eps}^{1\over2} \\
 & { 1 \over 6} h_{E_\Omega} \left(K^{\left( \mathbf{s} \right)}\right)^{-{2\over3}} K^{\left( \mathbf{s}, \tilde{\mathbf{s}} \right)}, ~ \mathbf{u}^2 \geq \epsilon_{eps}^{1\over2}
 \end{split}\right. \\
 & \tau_{LST,\Xi}^{\left( \mathbf{s}, \tilde{\mathbf{s}} \right)} = \left\{\begin{split} 
  & {h_{E_\Omega} \over 6 \rho C_p \left\| \mathbf{u} \right\|_2 \left(K^{\left( \mathbf{s} \right)}\right)^{2\over3} } K^{\left( \mathbf{s}, \tilde{\mathbf{s}} \right)}, ~ {h_{E_\Omega} \over 2 \rho C_p \left\| \mathbf{u} \right\|_2 } \left(K^{\left( \mathbf{s} \right)}\right)^{1\over3} < { h_{E_\Omega}^2 \over 12k } \left(K^{\left( \mathbf{s} \right)}\right)^{2\over3} \\
  & { h_{E_\Omega}^2 \over 18 k \left(K^{\left( \mathbf{s} \right)}\right)^{1\over3} } K^{\left( \mathbf{s}, \tilde{\mathbf{s}} \right)}, ~ {h_{E_\Omega} \over 2 \rho C_p \left\| \mathbf{u} \right\|_2 } \left(K^{\left( \mathbf{s} \right)}\right)^{1\over3} \geq { h_{E_\Omega}^2 \over 12k } \left(K^{\left( \mathbf{s} \right)}\right)^{2\over3}
  \end{split}\right.
 \end{split}\right\} \\
 & \forall \tilde{\mathbf{s}} \in \left( \mathcal{H} \left(\Omega\right) \right)^3.
\end{split}
\end{equation}
The variational formulations for the adjoint equations of the surface-PDE filters for $\gamma$ and $d_m$ are derived as 
\begin{equation}\label{equ:AdjPDEFilterJObjectiveGafHTGafa}  
\left\{\begin{split}
  & \mathrm{Find}~\gamma_{fa}\in\mathcal{H}\left(\Sigma\right) ~\mathrm{for}~ \forall \tilde{\gamma}_{fa} \in \mathcal{H}\left(\Sigma\right),~\mathrm{such~that} \\
  & \int_\Sigma \left( {\partial\alpha\over\partial\gamma_p} {\partial\gamma_p\over\partial\gamma_f} \mathbf{u} \cdot \mathbf{u}_a \tilde{\gamma}_{fa} + r_f^2 \nabla_\Gamma^{\left( d_f \right)} \tilde{\gamma}_{fa} \cdot \nabla_\Gamma^{\left( d_f \right)} \gamma_{fa} + \tilde{\gamma}_{fa} \gamma_{fa} \right) M^{\left( d_f \right)} \,\mathrm{d}\Sigma = 0
\end{split}\right.
\end{equation}
and 
\begin{equation}\label{equ:AdjPDEFilterJObjectiveDmHTDfa} 
\left\{\begin{split}
  & \mathrm{Find}~d_{fa}\in\mathcal{H}\left(\Sigma\right)~\mathrm{for}~\forall \tilde{d}_{fa} \in \mathcal{H}\left(\Sigma\right),~\mathrm{such~that} \\
  & \int_\Sigma r_f^2 \left( \nabla_\Gamma^{\left( d_f, \tilde{d}_{fa} \right)} \gamma_f \cdot \nabla_\Gamma^{\left( d_f \right)} \gamma_{fa} + \nabla_\Gamma^{\left( d_f \right)} \gamma_f \cdot \nabla_\Gamma^{\left( d_f, \tilde{d}_{fa} \right)} \gamma_{fa} \right) M^{\left( d_f \right)} \\
  & + \left( r_f^2 \nabla_\Gamma^{\left( d_f \right)} \gamma_f \cdot \nabla_\Gamma^{\left( d_f \right)} \gamma_{fa} + \gamma_f \gamma_{fa} - \gamma \gamma_{fa} + \alpha \mathbf{u} \cdot \mathbf{u}_a \right) M^{\left( d_f, \tilde{d}_{fa} \right)} \\
  & + r_m^2 \nabla_\Sigma \tilde{d}_{fa} \cdot \nabla_\Sigma d_{fa} + \tilde{d}_{fa} d_{fa} - \mathbf{n}_\Sigma \cdot \boldsymbol{\lambda}_{\mathbf{s}a} \tilde{d}_{fa} \,\mathrm{d}\Sigma = 0.
\end{split}\right.
\end{equation}

For the constraint of the pressure drop, the adjoint sensitivity of the pressure drop $\Delta P$ is derived as
\begin{equation}\label{equ:AdjSensitivityGaDmDissipationConstrHTBulk}
\begin{split}
\delta \Delta P = \int_\Sigma - \gamma_{fa} \tilde{\gamma} M^{\left( d_f \right)} - A_d d_{fa} \tilde{d}_m \,\mathrm{d}\Sigma,~ \forall \left( \tilde{\gamma}, \tilde{d}_m \right) \in \left(\mathcal{L}^2\left(\Sigma\right)\right)^2
\end{split}
\end{equation}
where the adjoint variables $\gamma_{fa}$ and $d_{fa}$ are derived by sequentially solving the variational formulation for the adjoint equations of the Navier-Stokes equations
\begin{equation}\label{equ:AdjEquSurfaceNSMHMDissipationBulk}
\left\{\begin{split}
  & \mathrm{Find} \left\{\begin{split}
  & \mathbf{u}_a \in\left(\mathcal{H}\left(\Omega\right)\right)^3~\mathrm{with}~ \mathbf{u}_a = \mathbf{0}~ \mathrm{at} ~  {\forall \mathbf{x}_\Omega \in \Sigma_{v,\Omega} \cup \Sigma_{v_0,\Omega} } \\
  & p_a \in \mathcal{H}\left(\Omega\right) \\
  \end{split}\right.\\
  & \mathrm{for} \left\{\begin{split}
  & \forall \tilde{\mathbf{u}}_a \in\left(\mathcal{H}\left(\Omega\right)\right)^3 \\
  & \forall \tilde{p}_a \in \mathcal{H}\left(\Omega\right) \\
  \end{split}\right.,~\mathrm{such~that} \\
  & \int_{\Sigma_{v,\Omega}} \tilde{p}_a \,\mathrm{d}\Sigma_{\partial\Omega} - \int_{\Sigma_{s,\Omega}} \tilde{p}_a \,\mathrm{d}\Sigma_{\partial\Omega} + \int_\Omega \Big[ \rho \left( \tilde{\mathbf{u}}_a \cdot \nabla_{\mathbf{x}_\Xi}^{\left(\mathbf{s}\right)} \right) \mathbf{u} \cdot \mathbf{u}_a + \rho \left( \mathbf{u} \cdot \nabla_{\mathbf{x}_\Xi}^{\left(\mathbf{s}\right)} \right) \tilde{\mathbf{u}}_a \cdot \mathbf{u}_a \\
  & + {\eta\over2} \left( \nabla_{\mathbf{x}_\Xi}^{\left(\mathbf{s}\right)} \tilde{\mathbf{u}}_a + \nabla_{\mathbf{x}_\Xi}^{\left(\mathbf{s}\right)} \tilde{\mathbf{u}}_a^\mathrm{T} \right) : \left( \nabla_{\mathbf{x}_\Xi}^{\left(\mathbf{s}\right)} \mathbf{u}_a + \nabla_{\mathbf{x}_\Xi}^{\left(\mathbf{s}\right)} \mathbf{u}_a^\mathrm{T} \right) - \tilde{p}_a\,\mathrm{div}_{\mathbf{x}_\Xi}^{\left(\mathbf{s}\right)} \mathbf{u}_a - p_a \mathrm{div}_{\mathbf{x}_\Xi}^{\left(\mathbf{s}\right)} \tilde{\mathbf{u}}_a \Big] K^{\left(\mathbf{s}\right)} \,\mathrm{d}\Omega \\
  & - \sum_{E_\Omega\in\mathcal{E}_\Omega} \int_{E_\Omega} \Big[ \tau_{LS\mathbf{u},\Xi}^{\left(\mathbf{s}, \tilde{\mathbf{u}}_a \right)} \left( \rho \mathbf{u} \cdot \nabla_{\mathbf{x}_\Xi}^{\left(\mathbf{s}\right)} \mathbf{u} + \nabla_{\mathbf{x}_\Xi}^{\left(\mathbf{s}\right)} p \right) \cdot \left( \rho \mathbf{u} \cdot \nabla_{\mathbf{x}_\Xi}^{\left(\mathbf{s}\right)} \mathbf{u}_a + \nabla_{\mathbf{x}_\Xi}^{\left(\mathbf{s}\right)} p_a \right) \\
  & + \tau_{LS\mathbf{u},\Xi}^{\left(\mathbf{s}\right)} \left( \rho \tilde{\mathbf{u}}_a \cdot \nabla_{\mathbf{x}_\Xi}^{\left(\mathbf{s}\right)} \mathbf{u} + \rho \mathbf{u} \cdot \nabla_{\mathbf{x}_\Xi}^{\left(\mathbf{s}\right)} \tilde{\mathbf{u}}_a + \nabla_{\mathbf{x}_\Xi}^{\left(\mathbf{s}\right)} \tilde{p}_a \right) \cdot \Big( \rho \mathbf{u} \cdot \nabla_{\mathbf{x}_\Xi}^{\left(\mathbf{s}\right)} \mathbf{u}_a + \nabla_{\mathbf{x}_\Xi}^{\left(\mathbf{s}\right)} p_a \Big) \\
  & + \tau_{LS\mathbf{u},\Xi}^{\left(\mathbf{s}\right)} \left( \rho \mathbf{u} \cdot \nabla_{\mathbf{x}_\Xi}^{\left(\mathbf{s}\right)} \mathbf{u} + \nabla_{\mathbf{x}_\Xi}^{\left(\mathbf{s}\right)} p \right) \cdot \left( \rho \tilde{\mathbf{u}}_a \cdot \nabla_{\mathbf{x}_\Xi}^{\left(\mathbf{s}\right)} \mathbf{u}_a \right) + \tau_{LSp,\Xi}^{\left(\mathbf{s}, \tilde{\mathbf{u}}_a \right)} \left( \rho \mathrm{div}_{\mathbf{x}_\Xi}^{\left(\mathbf{s}\right)} \mathbf{u} \right) \left( \mathrm{div}_{\mathbf{x}_\Xi}^{\left(\mathbf{s}\right)} \mathbf{u}_a \right) \\
  & + \tau_{LSp,\Xi}^{\left(\mathbf{s}\right)} \left( \rho \mathrm{div}_{\mathbf{x}_\Xi}^{\left(\mathbf{s}\right)} \tilde{\mathbf{u}}_a \right) \left( \mathrm{div}_{\mathbf{x}_\Xi}^{\left(\mathbf{s}\right)} \mathbf{u}_a \right) \Big] K^{\left(\mathbf{s}\right)} \,\mathrm{d}\Omega + \int_\Sigma \alpha \tilde{\mathbf{u}}_a \cdot \mathbf{u}_a M^{\left(d_f\right)} \, \mathrm{d}\Sigma = 0, \\  
\end{split}\right.
\end{equation}
the variational formulation for the adjoint equation of the Laplace's equation 
\begin{equation}\label{equ:WeakAdjEquHarmonicEquBulkHTSaDissipation} 
\left\{\begin{split}
  & \mathrm{Find} \left\{\begin{split}
  & \mathbf{s}_a \in \left(\mathcal{H}\left(\Omega\right)\right)^3~\mathrm{with}~ \mathbf{s}_a = \mathbf{0} ~ \mathrm{at} ~ \forall \mathbf{x}_\Omega \in \Sigma_{v,\Omega} \cup \Sigma_{s,\Omega} \\
  & \boldsymbol{\lambda}_{\mathbf{s}a} \in \left(\mathcal{H}^{-{1\over2}}\left(\Sigma\right)\right)^3 \\
  \end{split}\right.,\\
  & \mathrm{for} \left\{\begin{split} 
  & \forall \tilde{\mathbf{s}}_a \in\left(\mathcal{H}\left(\Omega\right)\right)^3 \\
  & \forall \tilde{\boldsymbol{\lambda}}_{\mathbf{s}a} \in \left(\mathcal{H}^{{1\over2}}\left(\Sigma\right)\right)^3 \\
  \end{split}\right.,~\mathrm{such~that} \\
  & \int_\Omega \Big[ \rho \left( \mathbf{u} \cdot \nabla_{\mathbf{x}_\Xi}^{\left(\mathbf{s}, \tilde{\mathbf{s}}_a\right)} \right) \mathbf{u} \cdot \mathbf{u}_a + {\eta\over2} \Big( \nabla_{\mathbf{x}_\Xi}^{\left(\mathbf{s}, \tilde{\mathbf{s}}_a\right)} \mathbf{u} + \nabla_{\mathbf{x}_\Xi}^{\left(\mathbf{s}, \tilde{\mathbf{s}}_a\right)} \mathbf{u}^\mathrm{T} \Big) : \left( \nabla_{\mathbf{x}_\Xi}^{\left(\mathbf{s}\right)} \mathbf{u}_a + \nabla_{\mathbf{x}_\Xi}^{\left(\mathbf{s}\right)} \mathbf{u}_a^\mathrm{T} \right) \\
  & + {\eta\over2} \left( \nabla_{\mathbf{x}_\Xi}^{\left(\mathbf{s}\right)} \mathbf{u} + \nabla_{\mathbf{x}_\Xi}^{\left(\mathbf{s}\right)} \mathbf{u}^\mathrm{T} \right) : \left( \nabla_{\mathbf{x}_\Xi}^{\left(\mathbf{s}, \tilde{\mathbf{s}}_a\right)} \mathbf{u}_a + \nabla_{\mathbf{x}_\Xi}^{\left(\mathbf{s}, \tilde{\mathbf{s}}_a\right)} \mathbf{u}_a^\mathrm{T} \right) - p\,\mathrm{div}_{\mathbf{x}_\Xi}^{\left(\mathbf{s}, \tilde{\mathbf{s}}_a\right)} \mathbf{u}_a \\
  & - p_a \mathrm{div}_{\mathbf{x}_\Xi}^{\left(\mathbf{s}, \tilde{\mathbf{s}}_a\right)} \mathbf{u} \Big] K^{\left(\mathbf{s}\right)} + \Big[ {\eta\over2} \left( \nabla_{\mathbf{x}_\Xi}^{\left( \mathbf{s} \right)} \mathbf{u} + \nabla_{\mathbf{x}_\Xi}^{\left( \mathbf{s} \right)} \mathbf{u}^\mathrm{T} \right) : \left( \nabla_{\mathbf{x}_\Xi}^{\left( \mathbf{s} \right)} \mathbf{u} + \nabla_{\mathbf{x}_\Xi}^{\left( \mathbf{s} \right)} \mathbf{u}^\mathrm{T} \right) \\
  & + \rho \left( \mathbf{u} \cdot \nabla_{\mathbf{x}_\Xi}^{\left(\mathbf{s}\right)} \right) \mathbf{u} \cdot \mathbf{u}_a + {\eta\over2} \left( \nabla_{\mathbf{x}_\Xi}^{\left(\mathbf{s}\right)} \mathbf{u} + \nabla_{\mathbf{x}_\Xi}^{\left(\mathbf{s}\right)} \mathbf{u}^\mathrm{T} \right) : \left( \nabla_{\mathbf{x}_\Xi}^{\left(\mathbf{s}\right)} \mathbf{u}_a + \nabla_{\mathbf{x}_\Xi}^{\left(\mathbf{s}\right)} \mathbf{u}_a^\mathrm{T} \right) \\
  & - p\,\mathrm{div}_{\mathbf{x}_\Xi}^{\left(\mathbf{s}\right)} \mathbf{u}_a - p_a \mathrm{div}_{\mathbf{x}_\Xi}^{\left(\mathbf{s}\right)} \mathbf{u} \Big] K^{\left(\mathbf{s}, \tilde{\mathbf{s}}_a\right)} - \nabla_{\mathbf{x}_\Omega} \tilde{\mathbf{s}}_a : \nabla_{\mathbf{x}_\Omega} \mathbf{s}_a \,\mathrm{d}\Omega \\
  & - \sum_{E_\Omega\in\mathcal{E}_\Omega} \int_{E_\Omega} \Big[ \tau_{LS\mathbf{u},\Xi}^{\left(\mathbf{s}, \tilde{\mathbf{s}}_a\right)} \left( \rho \mathbf{u} \cdot \nabla_{\mathbf{x}_\Xi}^{\left(\mathbf{s}\right)} \mathbf{u} + \nabla_{\mathbf{x}_\Xi}^{\left(\mathbf{s}\right)} p \right) \cdot \left( \rho \mathbf{u} \cdot \nabla_{\mathbf{x}_\Xi}^{\left(\mathbf{s}\right)} \mathbf{u}_a + \nabla_{\mathbf{x}_\Xi}^{\left(\mathbf{s}\right)} p_a \right) + \tau_{LS\mathbf{u},\Xi}^{\left(\mathbf{s}\right)} \\
  & \Big( \rho \mathbf{u} \cdot \nabla_{\mathbf{x}_\Xi}^{\left(\mathbf{s}, \tilde{\mathbf{s}}_a\right)} \mathbf{u} + \nabla_{\mathbf{x}_\Xi}^{\left(\mathbf{s}, \tilde{\mathbf{s}}_a\right)} p \Big) \cdot \left( \rho \mathbf{u} \cdot \nabla_{\mathbf{x}_\Xi}^{\left(\mathbf{s}\right)} \mathbf{u}_a + \nabla_{\mathbf{x}_\Xi}^{\left(\mathbf{s}\right)} p_a \right) + \tau_{LS\mathbf{u},\Xi}^{\left(\mathbf{s}\right)} \left( \rho \mathbf{u} \cdot \nabla_{\mathbf{x}_\Xi}^{\left(\mathbf{s}\right)} \mathbf{u} + \nabla_{\mathbf{x}_\Xi}^{\left(\mathbf{s}\right)} p \right) \\
  & \cdot \Big( \rho \mathbf{u} \cdot \nabla_{\mathbf{x}_\Xi}^{\left(\mathbf{s}, \tilde{\mathbf{s}}_a\right)} \mathbf{u}_a + \nabla_{\mathbf{x}_\Xi}^{\left(\mathbf{s}, \tilde{\mathbf{s}}_a\right)} p_a \Big) + \tau_{LSp,\Xi}^{\left(\mathbf{s}, \tilde{\mathbf{s}}_a\right)} \left( \rho \mathrm{div}_{\mathbf{x}_\Xi}^{\left(\mathbf{s}\right)} \mathbf{u} \right) \left( \mathrm{div}_{\mathbf{x}_\Xi}^{\left(\mathbf{s}\right)} \mathbf{u}_a \right) + \tau_{LSp,\Xi}^{\left(\mathbf{s}\right)} \left( \rho \mathrm{div}_{\mathbf{x}_\Xi}^{\left(\mathbf{s}, \tilde{\mathbf{s}}_a\right)} \mathbf{u} \right) \\
  & \left( \mathrm{div}_{\mathbf{x}_\Xi}^{\left(\mathbf{s}\right)} \mathbf{u}_a \right) + \tau_{LSp,\Xi}^{\left(\mathbf{s}\right)} \left( \rho \mathrm{div}_{\mathbf{x}_\Xi}^{\left(\mathbf{s}\right)} \mathbf{u} \right) \left( \mathrm{div}_{\mathbf{x}_\Xi}^{\left(\mathbf{s}, \tilde{\mathbf{s}}_a\right)} \mathbf{u}_a \right) \Big] K^{\left(\mathbf{s}\right)} + \Big[ \tau_{LS\mathbf{u},\Xi}^{\left(\mathbf{s}\right)} \left( \rho \mathbf{u} \cdot \nabla_{\mathbf{x}_\Xi}^{\left(\mathbf{s}\right)} \mathbf{u} + \nabla_{\mathbf{x}_\Xi}^{\left(\mathbf{s}\right)} p \right) \\
  & \cdot \left( \rho \mathbf{u} \cdot \nabla_{\mathbf{x}_\Xi}^{\left(\mathbf{s}\right)} \mathbf{u}_a + \nabla_{\mathbf{x}_\Xi}^{\left(\mathbf{s}\right)} p_a \right) + \tau_{LSp,\Xi}^{\left(\mathbf{s}\right)} \left( \rho \mathrm{div}_{\mathbf{x}_\Xi}^{\left(\mathbf{s}\right)} \mathbf{u} \right) \left( \mathrm{div}_{\mathbf{x}_\Xi}^{\left(\mathbf{s}\right)} \mathbf{u}_a \right) \Big] K^{\left(\mathbf{s}, \tilde{\mathbf{s}}_a\right)} \,\mathrm{d}\Omega \\
  & + \int_\Sigma \tilde{\mathbf{s}}_a \cdot \boldsymbol{\lambda}_{\mathbf{s}a} + \tilde{\boldsymbol{\lambda}}_{\mathbf{s}a} \cdot \mathbf{s}_a \,\mathrm{d}\Sigma = 0, \\
\end{split}\right.
\end{equation}
and the variational formulations for the adjoint equations of the surface-PDE filters
\begin{equation}\label{equ:AdjPDEFilterDissipationGaHTBulk} 
\left\{\begin{split}
  & \mathrm{Find}~\gamma_{fa}\in\mathcal{H}\left(\Sigma\right) ~\mathrm{for}~ \forall \tilde{\gamma}_{fa} \in \mathcal{H}\left(\Sigma\right),~\mathrm{such~that} \\
  & \int_\Sigma \left( {\partial\alpha\over\partial\gamma_p} {\partial\gamma_p\over\partial\gamma_f} \mathbf{u} \cdot \mathbf{u}_a \tilde{\gamma}_{fa} + r_f^2 \nabla_\Gamma^{\left( d_f \right)} \tilde{\gamma}_{fa} \cdot \nabla_\Gamma^{\left( d_f \right)} \gamma_{fa} + \tilde{\gamma}_{fa} \gamma_{fa} \right) M^{\left( d_f \right)} \,\mathrm{d}\Sigma = 0
\end{split}\right.
\end{equation}
and 
\begin{equation}\label{equ:AdjPDEFilterJDissipationDmHTBulk} 
\left\{\begin{split}
  & \mathrm{Find}~d_{fa}\in\mathcal{H}\left(\Sigma\right)~\mathrm{for}~\forall \tilde{d}_{fa} \in \mathcal{H}\left(\Sigma\right),~\mathrm{such~that} \\
  & \int_\Sigma r_f^2 \left( \nabla_\Gamma^{\left( d_f, \tilde{d}_{fa} \right)} \gamma_f \cdot \nabla_\Gamma^{\left( d_f \right)} \gamma_{fa} + \nabla_\Gamma^{\left( d_f \right)} \gamma_f \cdot \nabla_\Gamma^{\left( d_f, \tilde{d}_{fa} \right)} \gamma_{fa} \right) M^{\left( d_f \right)} \\
  & + \left( r_f^2 \nabla_\Gamma^{\left( d_f \right)} \gamma_f \cdot \nabla_\Gamma^{\left( d_f \right)} \gamma_{fa} + \gamma_f \gamma_{fa} - \gamma \gamma_{fa} + \alpha \mathbf{u} \cdot \mathbf{u}_a \right) M^{\left( d_f, \tilde{d}_{fa} \right)} \\
  & + r_m^2 \nabla_\Sigma \tilde{d}_{fa} \cdot \nabla_\Sigma d_{fa} + \tilde{d}_{fa} d_{fa} - \tilde{d}_{fa} \mathbf{n}_\Sigma \cdot \boldsymbol{\lambda}_{\mathbf{s}a} \,\mathrm{d}\Sigma = 0.
\end{split}\right.
\end{equation}

After the derivation of the adjoint sensitivities in Eqs. \ref{equ:AdjSensitivityNSCHTGaDmObjBulkMHT} and \ref{equ:AdjSensitivityGaDmDissipationConstrHTBulk}, the design variables $\gamma$ and $d_m$ can be evolved iteratively to determine the fiber bundle of the thin-wall patterns for heat transfer in the volume flow.

\subsection{Numerical implementation}\label{sec:NumericalImplementationBulkNSEqus}

The fiber bundle topology optimization problems for the volume flow in Eqs. \ref{equ:VarProToopBulkNSCDMHT} and \ref{equ:VarProToopBulkNSCHTMHT} can be solved by using the iterative algorithms described in Tabs. \ref{tab:IterativeProcedureBulkFlowMM} and \ref{tab:IterativeProcedureBulkFlowHM}. The finite element method is utilized to solve the variational formulations of the relevant PDEs and adjoint equations.
On the details for the finite element solution, one can refer to Ref. \cite{ElmanFEMFlow2006}. In the iterative procedures of the algorithms, dealing with the numerical singularity caused by the null value, updating the projection parameter, updating the design variables, scaling the constraints and the corresponding adjoint sensitivities, and judging the convergence of the iterative loop are implemented by using the same approaches as that have been presented in Section \ref{sec:NumericalImplementationSurfaceNSEqus}.

\begin{table}[!htbp]
\centering
\begin{tabular}{l}
  \hline
  \textbf{Algorithm 3}: iterative solution of Eq. \ref{equ:VarProToopBulkNSCDMHT} \\
  \hline
  Set $\mathbf{u}_{\Gamma_{v,\Omega}}$, $\rho$, $\eta$, $D$, $c_0$, $\bar{c}$, $A_d$ and $\Delta P_0$;\\
  Set $\left\{
  \begin{array}{l}
    \gamma \leftarrow 1 \\
    d_m \leftarrow 1/2
  \end{array}
  \right.$, $\left\{
  \begin{array}{l}
    r_f = 1/10 \\
    r_m = 1/4 \\
  \end{array}
  \right.$, $\left\{
  \begin{array}{l}
    n_{\max} \leftarrow 230 \\
    n_i \leftarrow 1
  \end{array}
  \right.$, $\left\{
  \begin{array}{l}
    \xi \leftarrow 0.5 \\
    \beta \leftarrow 1
  \end{array}
  \right.$, $\left\{
  \begin{array}{l}
    \alpha_{\max} \leftarrow 10^5 \\
    q \leftarrow 1\times10^{-2}
  \end{array}
  \right.$; \\
  \hline
  \textbf{loop} \\
          \hspace{1em} Solve $d_f$ from Eq. \ref{equ:PDEFilterzmBaseStructureMHM}; \\
          \hspace{1em} Solve $\gamma_f$ from Eq. \ref{equ:PDEFilterGammaFilberMHM}; \\
          \hspace{1em} Project $\gamma_f$ to derive $\gamma_p$ based on Eq. \ref{equ:ProjectionGammaFilberMHM}; \\
          \hspace{1em} Solve $\mathbf{s}$ and $\boldsymbol{\lambda}_{\mathbf{s}}$ from Eq. \ref{equ:HarmonicCoordinateEquMHT}; \\
          \hspace{1em} Solve $\mathbf{u}$, $p$ and $\lambda$ from Eq. \ref{equ:VariationalFormulationNavierStokesEquBulkFlowMHT}, and evaluate $\Delta P_{n_i}$ from Eq. \ref{equ:PressureDropBulkNSCD}; \\
          \hspace{1em} Solve $c$ from Eq. \ref{equ:VariationalFormulationBulkConvecDiffusEqu}, and evaluate $J_{c,n_i} / J_{c,0}$ from Eq. \ref{equ:DesignObjectiveBulkCDMHT}; \\
          \hspace{1em} Solve $c_a$ from Eq. \ref{equ:WeakAdjEquCDEquBulkMTCa}; \\
          \hspace{1em} Solve $\mathbf{u}_a$, $p_a$ and $\lambda_a$ from Eq. \ref{equ:AdjBulkNavierStokesEqusJObjectiveMTUaPa}; \\
          \hspace{1em} Solve $\mathbf{s}_a$ and $\boldsymbol{\lambda}_{\mathbf{s}a}$ from Eq. \ref{equ:WeakAdjEquHarmonicEquBulkMTSa}; \\ 
          \hspace{1em} Solve $\gamma_{fa}$ from Eq. \ref{equ:AdjPDEFilterJObjectiveGafMHMGafa}; \\
          \hspace{1em} Solve $d_{fa}$ from Eq. \ref{equ:AdjPDEFilterJObjectiveDmMHMDfa}; \\
          \hspace{1em} Evaluate $\delta J_{c,n_i}$ from Eq. \ref{equ:AdjSensitivityNSCDGaDmObjBulkMHT}; \\
          \hspace{1em} Solve $\mathbf{u}_a$, $p_a$ and $\lambda_a$ from Eq. \ref{equ:AdjEquSurfaceNSMHMPressureDropBulk}; \\
          \hspace{1em} Solve $\mathbf{s}_a$ and $\boldsymbol{\lambda}_{\mathbf{s}a}$ from Eq. \ref{equ:WeakAdjEquHarmonicEquBulkMTSaPressureDrop}; \\
          \hspace{1em} Solve $\gamma_{fa}$ from Eq. \ref{equ:AdjPDEFilterPressureDropGaMHMBulk}; \\
          \hspace{1em} Solve $d_{fa}$ from Eq. \ref{equ:AdjPDEFilterJPressureDropDmMHMBulk}; \\
          \hspace{1em} Evaluate $\delta \Delta P_{n_i}$ from Eq. \ref{equ:AdjSensitivityGaDmPressureConstrMHMBulk}; \\          
          \hspace{1em} Update $\gamma$ and $d_m$ based on $\delta J_{c,n_i}$ and $C_{P,n_i} \delta \Delta P_{n_i}$ by using MMA; \\
          \hspace{1em} \textbf{if} $\left(n_i == 30 \right) \vee \left(\left(n_i>30\right)\wedge\left(\mathrm{mod}\left(n_i-30,20\right)==0\right)\right)$ \\
          \hspace{2em} $\beta \leftarrow 2\beta$; \\
          \hspace{1em} \textbf{end} \textbf{if} \\
          \hspace{1em} \textbf{if} $ \left( n_i==n_{\max} \right) \vee \left\{
          \begin{array}{l}
            \beta == 2^{10} \\
            {1\over5}\sum_{m=0}^4 \left| J_{c,n_i-m} - J_{c,n_i-\left(m+1\right)} \right|\big/J_{c,0} \leq 1\times10^{-3} \\
            \left|\Delta P_{n_i} / \Delta P_0-1\right| \leq 1\times10^{-3}
          \end{array}
          \right.$ \\
          \hspace{2em} break; \\
          \hspace{1em} \textbf{end} \textbf{if} \\
          \hspace{1em} $n_i \leftarrow n_i+1$ \\
  \textbf{end} \textbf{loop} \\
  \hline
\end{tabular}
\caption{Pseudocodes of the algorithm used to solve the fiber bundle topology optimization problem for mass transfer in the volume flow.}\label{tab:IterativeProcedureBulkFlowMM}
\end{table}

\begin{table}[!htbp]
\centering
\begin{tabular}{l}
  \hline
  \textbf{Algorithm 4}: iterative solution of Eq. \ref{equ:VarProToopBulkNSCHTMHT} \\
  \hline
  Set $\mathbf{u}_{l_{v,\Sigma}}$, $\rho$, $\eta$, $T_0$, $A_d$ and $\Delta P_0$;\\
  Set $\left\{
  \begin{array}{l}
    \gamma \leftarrow 1 \\
    d_m \leftarrow 1/2
  \end{array}
  \right.$, $\left\{
  \begin{array}{l}
    r_f = 1/10 \\
    r_m = 1/4 \\
  \end{array}
  \right.$, $\left\{
  \begin{array}{l}
    n_i \leftarrow 1 \\
    n_{2^{10}} \leftarrow 0
  \end{array}
  \right.$, $\left\{
  \begin{array}{l}
    n_{upd} \leftarrow 20 \\
    n_{1st} \leftarrow 10
  \end{array}
  \right.$, $\left\{
  \begin{array}{l}
    \xi \leftarrow 0.5 \\
    \beta \leftarrow 1 \\
  \end{array}
  \right.$, $\left\{
  \begin{array}{l}
    \beta' \leftarrow 0 \\
    \gamma'_p \leftarrow 0 \\
  \end{array}
  \right.$, \\
  ~~~~ $\left\{
  \begin{array}{l}
    C_p \leftarrow 1\times10^0 \\
    k \leftarrow 1\times10^0 \\
    Q \leftarrow 1\times10^0 \\
  \end{array}
  \right.$, $\left\{
  \begin{array}{l}
   \alpha_{\max} \leftarrow 1\times10^5 \\
    q \leftarrow 1\times10^{-2} \\
  \end{array}
  \right.$, $\epsilon_{\gamma_p} \leftarrow 1\times10^{-1}$; \\
  \hline
  \textbf{while} $\beta \leq 2^{10}$ \\
          \hspace{1em} Solve $d_f$ from Eq. \ref{equ:PDEFilterzmBaseStructureMHM}, and solve $\gamma_f$ from Eq. \ref{equ:PDEFilterGammaFilberMHM}; \\
          \hspace{1em} \textbf{if} $ \left(n_i \geq n_{upd} + n_{1st}\right)\wedge\left(\mathrm{mod}\left(n_i-n_{upd} - n_{1st},n_{upd}\right)==1\right) $ \\
          \hspace{2em} Compute $\gamma_p$ from $\gamma_f$ based on Eq. \ref{equ:ProjectionGammaFilberMHM}; \\
          \hspace{2em} \textbf{if} $\beta < 2^5$ \\
          \hspace{3em} $n_{2^5} \leftarrow 0$; $\beta' \leftarrow 2\beta$; \\
          \hspace{3em} Compute $\gamma'_p$ from $\gamma_f$ based on Eq. \ref{equ:ProjectionGammaFilberMHM} with $\gamma_p$ and $\beta$ replaced to be $\gamma'_p$ and $\beta'$; \\
          \hspace{3em} \textbf{while} $\left\| \gamma'_p - \gamma_p \right\|_\infty \geq \epsilon_{\gamma_p}$ \\
          \hspace{4em} $\beta' \leftarrow \left( \beta' + \beta \right) / 2 $; \\
          \hspace{4em} Compute $\gamma'_p$ from $\gamma_f$ based on Eq. \ref{equ:ProjectionGammaFilberMHM} with $\gamma_p$ and $\beta$ replaced to be $\gamma'_p$ and $\beta'$; \\
          \hspace{3em} \textbf{end while} \\
          \hspace{3em} $\beta \leftarrow \beta'$; \\
          \hspace{2em} \textbf{else} \\
          \hspace{3em} \textbf{if} $ n_{2^5} == 1 $ \\
          \hspace{4em} $\beta \leftarrow 2\beta$; \\
          \hspace{3em} \textbf{elseif} $ n_{2^5} == 0 $\\
          \hspace{4em} $\beta \leftarrow 2^5$; $n_{2^5} \leftarrow 1$; \\
          \hspace{3em} \textbf{end if} \\
          \hspace{2em} \textbf{end if} \\
          \hspace{2em} \textbf{if} $\beta == 2^{10}$ \\
          \hspace{3em} $n_{2^{10}} \leftarrow n_{2^{10}} + 1$; \\
          \hspace{2em} \textbf{end} \textbf{if} \\
          \hspace{2em} \textbf{if} $\big( \left(\beta == 2^{10}\right) \wedge \left({1\over5}\sum_{m=0}^4 \left| J_{T,n_i-m} - J_{T,n_i-\left(m+1\right)} \right|\big/J_{T,0} \leq 1\times10^{-3} \right)$ \\
          \hspace{3em} $\wedge \left( \left|\Phi_{n_i}/\Phi_0-1\right| \leq 1\times10^{-3} \right) \wedge \left( \left|s_{n_i}/s_0-1\right| \leq 1\times10^{-3} \right) \big) \vee \left( n_{2^{10}} == n_{upd} \right) $ \\
          \hspace{3em} break; \\
          \hspace{2em} \textbf{end} \textbf{if} \\
          \hspace{1em} \textbf{end} \textbf{if} \\
          \hspace{1em} Project $\gamma_f$ to derive $\gamma_p$ based on Eq. \ref{equ:ProjectionGammaFilberMHM}; \\
          \hspace{1em} Solve $\mathbf{s}$ and $\boldsymbol{\lambda}_{\mathbf{s}}$ from Eq. \ref{equ:HarmonicCoordinateEquMHT}; \\
          \hspace{1em} Solve $\mathbf{u}$, $p$ and $\lambda$ from Eq. \ref{equ:VariationalFormulationBulkNavierStokesEqusHT}, and evaluate $\Delta P_{n_i}$ from Eq. \ref{equ:TransformedPressureConstraintSurfaceNSCD}; \\
          \hspace{1em} Solve $T$ from Eq. \ref{equ:VariationalFormulationBulkCHMEqu}, and evaluate $J_{T,n_i} / J_{T,0}$ from Eq. \ref{equ:DesignObjectiveBulkCHM}; \\
          \hspace{1em} Solve $T_a$ from Eq. \ref{equ:WeakAdjEquHTEquBulkMTTa}, and solve $\mathbf{u}_a$, $p_a$ and $\lambda_a$ from Eq. \ref{equ:AdjBulkNavierStokesEqusJObjectiveHTUaPa}; \\
          \hspace{1em} Solve $\mathbf{s}_a$ and $\boldsymbol{\lambda}_{\mathbf{s}a}$ from Eq. \ref{equ:WeakAdjEquHarmonicEquBulkHTSa}; \\
          \hspace{1em} Solve $\gamma_{fa}$ from Eq. \ref{equ:AdjPDEFilterJObjectiveGafHTGafa}, and solve $d_{fa}$ from Eq. \ref{equ:AdjPDEFilterJObjectiveDmHTDfa}; \\
          \hspace{1em} Evaluate $\delta J_{T,n_i}$ from Eq. \ref{equ:AdjSensitivityNSCHTGaDmObjBulkMHT}; \\
          \hspace{1em} Solve $\mathbf{u}_a$, $p_a$ and $\lambda_a$ from Eq. \ref{equ:AdjEquSurfaceNSMHMDissipationBulk}; \\
          \hspace{1em} Solve $\mathbf{s}_a$ and $\boldsymbol{\lambda}_{\mathbf{s}a}$ from Eq. \ref{equ:WeakAdjEquHarmonicEquBulkHTSaDissipation}; \\
          \hspace{1em} Solve $\gamma_{fa}$ from Eq. \ref{equ:AdjPDEFilterDissipationGaHTBulk}, and solve $d_{fa}$ from Eq. \ref{equ:AdjPDEFilterJDissipationDmHTBulk}; \\
          \hspace{1em} Evaluate $\delta \Delta P_{n_i}$ from Eq. \ref{equ:AdjSensitivityGaDmDissipationConstrHTBulk}; \\
          \hspace{1em} Update $\gamma$ and $d_m$ based on $\delta J_{T,n_i}$ and $C_{P,n_i}\delta \Phi_{n_i}$ by using MMA; \\
          \hspace{1em} $n_i \leftarrow n_i+1$; \\
  \textbf{end while} \\
  \hline
\end{tabular}
\caption{Pseudocodes of the algorithm used to solve the fiber bundle topology optimization problem for heat transfer in the volume flow.}\label{tab:IterativeProcedureBulkFlowHM}
\end{table}

\begin{figure}[!htbp]
  \centering
  \includegraphics[width=0.55\textwidth]{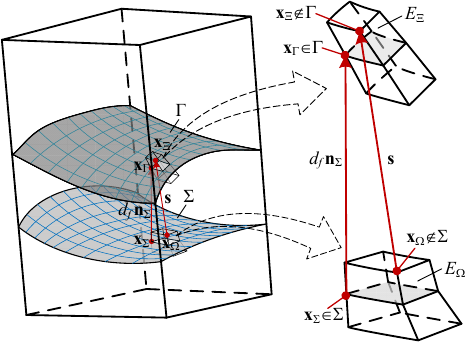}
  \caption{Sketch for the hexahedral elements of the discretizing meshes for the original domain and the mapping elements for the deformed domain, together with the quadrangular-element based discretization of the base manifold $\Sigma$ and the mapped meshes on the implicit 2-manifold $\Gamma$, where $\mathbf{x}_\Omega$ and $\mathbf{x}_\Xi$ are sketched by the points not localized on $\Sigma$ and $\Gamma$.}\label{fig:ElementNodesMHMBulk}
\end{figure}

Linear quadrangular elements are used to interpolate the design variables of the thin-wall pattern and that of the implicit 2-manifold, and solve the variational formulations of the related PDEs defined on the implicit 2-manifolds and the base manifolds; linear hexahedral elements are used to solve the variational formulations of the governing equations and their adjoint equations. The meshes of the quadrangular-element based discretization of the base manifolds and the mapped meshes on the implicit 2-manifolds are sketched in Fig. \ref{fig:ElementNodesMHM}. The meshes of the hexahedral-element based discretization of the original domain and the mapped meshes of the deformed domain are sketched in Fig. \ref{fig:ElementNodesMHMBulk}.

\subsection{Results and discussion}\label{sec:NumericalExamplesMatchinOptBulkFlow}

\begin{figure}[!htbp]
  \centering
  \includegraphics[width=0.6\textwidth]{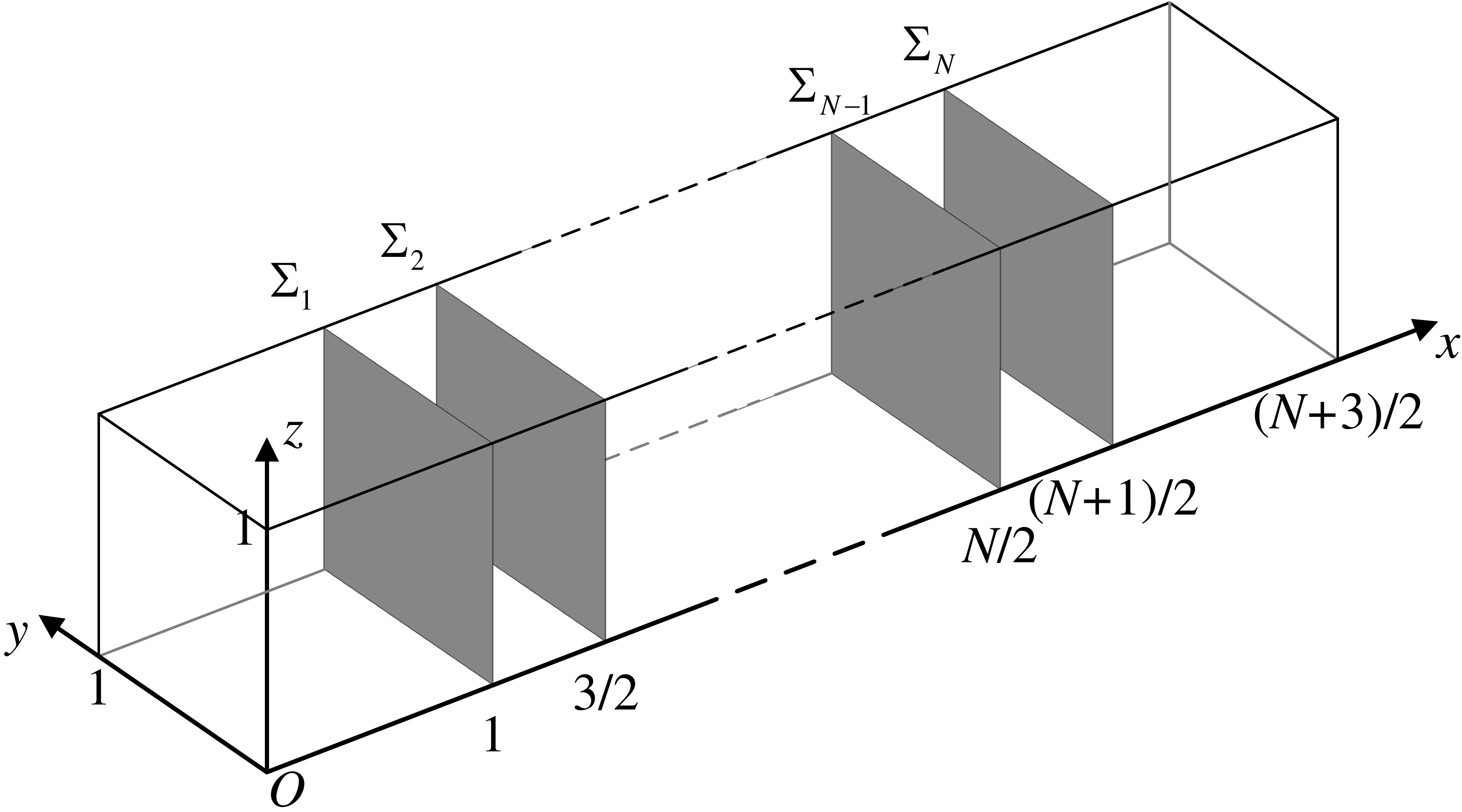}
  \caption{Design domains of fiber bundle topology optimization for mass and heat transfer in the volume flow. $\Sigma_n$ at $x=\left(n+1\right)/2$ with $n$ valued in $\left\{1,2,\cdots N\right\}$ is the cross-section of the straight channel; the design domain is the union of the cross sections; the inlet and outlet are the terminal faces at $x=0$ and $x=\left(N+3\right)/2$, respectively; and the remained outer faces are wall boundaries.}\label{fig:MassHeatTransferManifoldsDesignDomainBulkFlow}
\end{figure}

For fiber bundle topology optimization for mass and heat transfer in the volume flow, the fluid density and dynamic viscosity are considered as unitary. The design domain is sketched in Fig. \ref{fig:MassHeatTransferManifoldsDesignDomainBulkFlow}, including a series of equally spaced square cross-sections of a straight channel, i.e.
\begin{equation}\label{equ:DesignDomVolumeFlow}
  \Sigma = \bigcup_{n=1}^N \Sigma_n
\end{equation}
where $N$ is the total number of the cross-sections included in the design domain; and $\Sigma_n$ with $n$ valued from $\left\{1,2, \cdots N\right\}$ is a cross-section of the straight channel. The straight channel is localized in the Cartesian system of $O$-$xyz$ with the coordinate origin at $O$. The $n$-th cross-section is localized at $x=\left(n+1\right)/2$. The inlet and outlet are the terminal faces of the straight channel with $x=0$ and $x=\left(N+3\right)/2$, respectively. The remained outer surfaces are wall boundaries with zero velocity. The edge length of the square cross-sections is $1$, and the spacing distance of the cross-sections is $1/2$. In the numerical computation, the computational domain of the fluid channel is discretized by using regular hexahedral elements with the element size of $1/20$. The filter radii of the design variables of the implicit 2-manifold and the thin-wall pattern defined on the implicit 2-manifold are set to be $1/4$ and $1/10$, respectively.

\begin{figure}[!htbp]
  \centering
  \subfigure[$N=1$, $J_c = 0.4933$]
  {\includegraphics[width=0.32\textwidth]{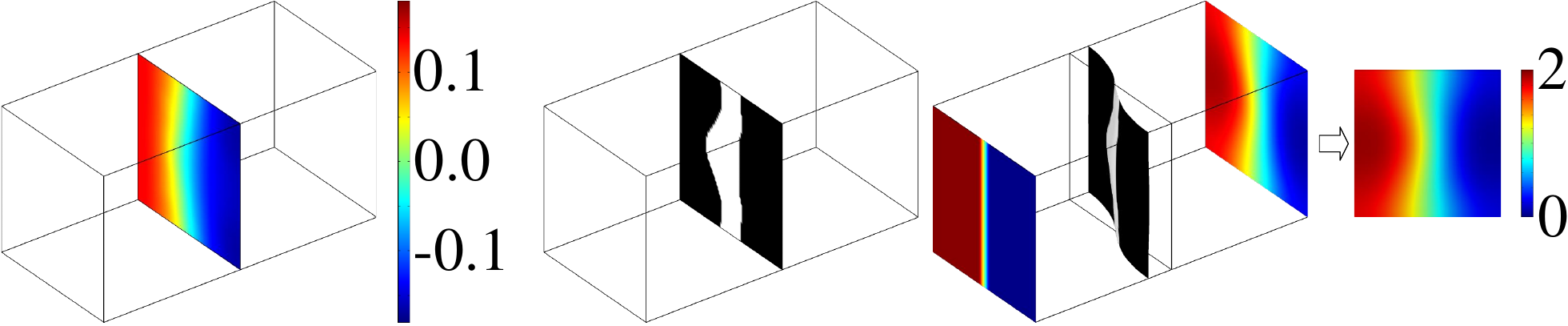}}\\
  \subfigure[$N=3$, $J_c = 0.3096$]
  {\includegraphics[width=0.4\textwidth]{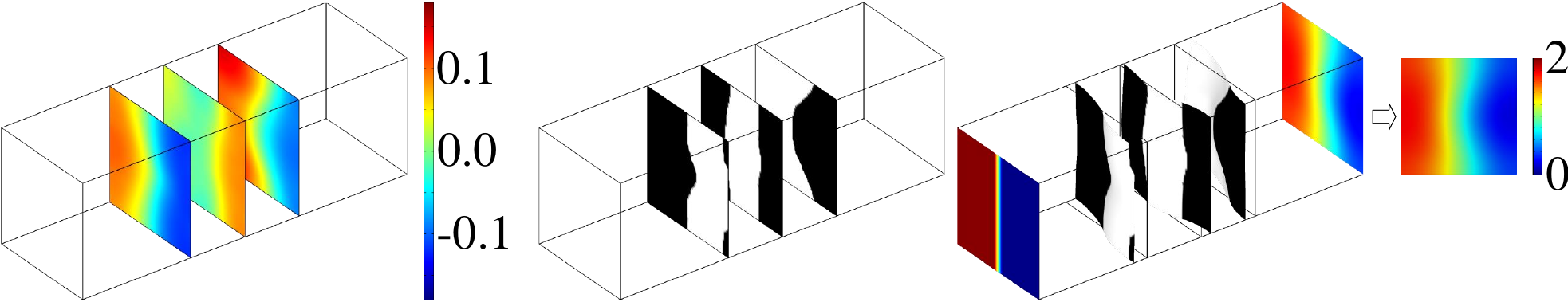}}\\
  \subfigure[$N=5$, $J_c = 0.1845$]
  {\includegraphics[width=0.5\textwidth]{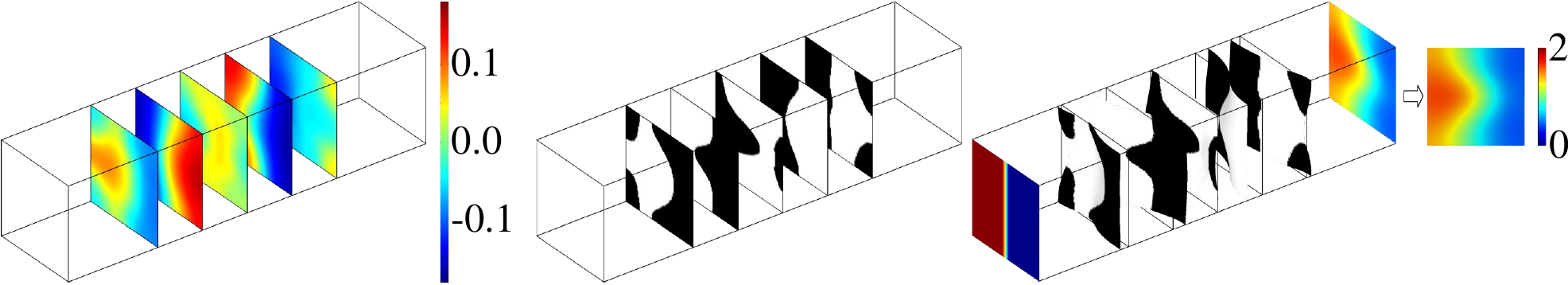}}\\
  \subfigure[$N=7$, $J_c = 0.1241$]
  {\includegraphics[width=0.6\textwidth]{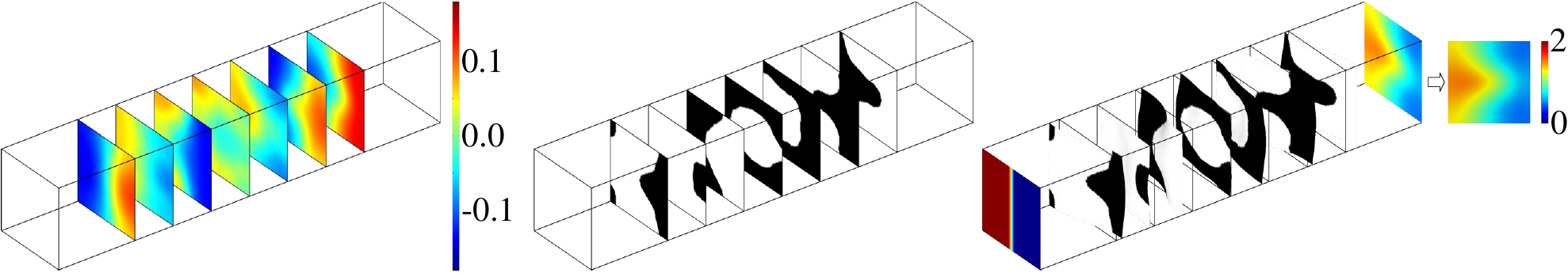}}\\
  \subfigure[$N=9$, $J_c = 0.0578$]
  {\includegraphics[width=0.7\textwidth]{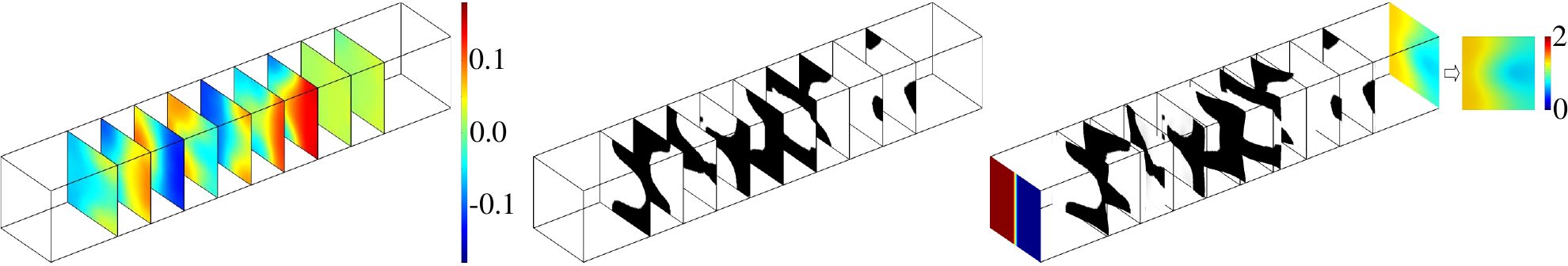}}\\
  \subfigure[$N=11$, $J_c = 0.0501$]
  {\includegraphics[width=0.8\textwidth]{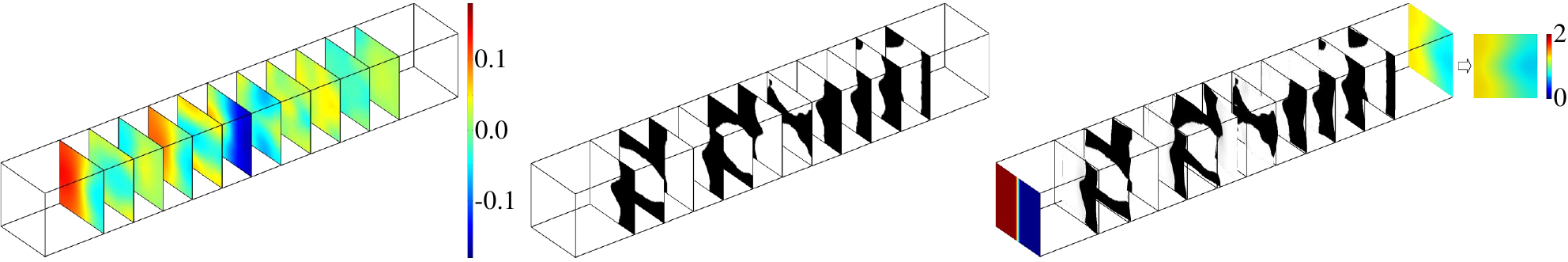}}\\
  \subfigure[$N=13$, $J_c = 0.0292$]
  {\includegraphics[width=0.9\textwidth]{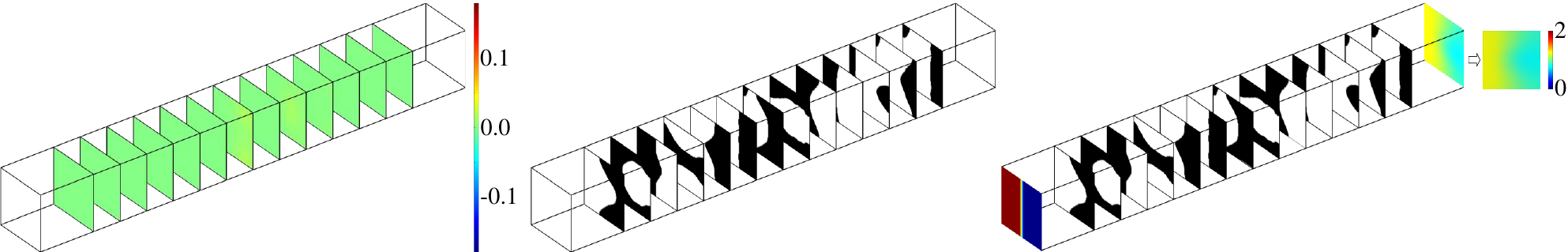}}
  \caption{Distribution of the filtered design variables for the implicit 2-manifolds and the material density of the thin-wall patterns obtained by solving the fiber bundle topology optimization problem for mass transfer in the volume flow on the design domains sketched in Fig. \ref{fig:MassHeatTransferManifoldsDesignDomainBulkFlow} with $N$ valued in $\left\{1,3,5,7,9,11,13\right\}$, where the first column is the distribution of the filtered design variables for the implicit 2-manifolds, the central column is the material density of the thin-wall patterns, and the third column is the deformed thin-wall patterns on the cross-sections including the concentration distribution at the inlet and outlet.}\label{fig:ImbeddedSurfaceCDVolumeFlows}
\end{figure}

\begin{figure}[!htbp]
  \centering
  \subfigure[$N=1$, $J_T = 0.5627$]
  {\includegraphics[width=0.32\textwidth]{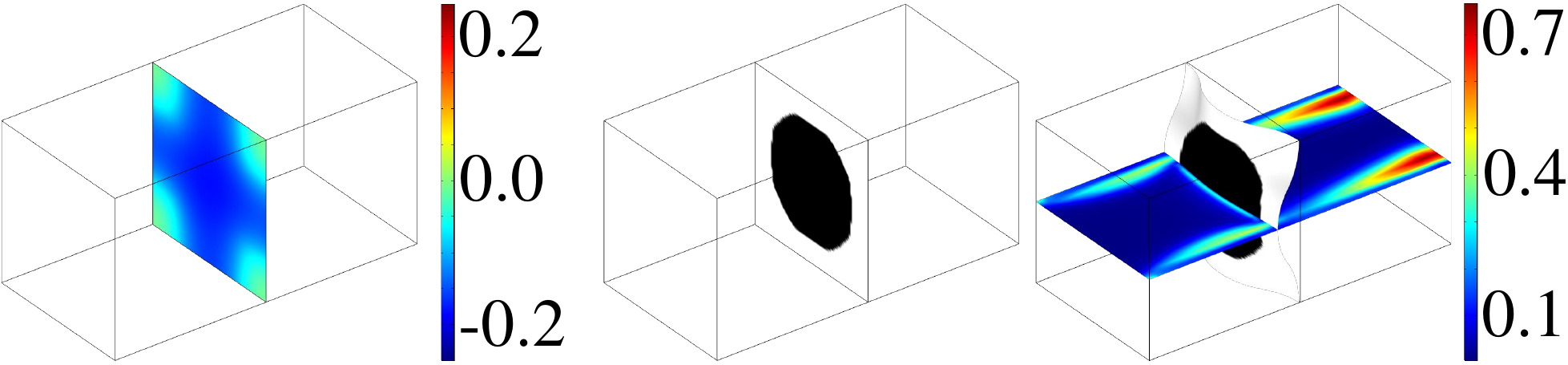}}\\
  \subfigure[$N=3$, $J_T = 0.7827$]
  {\includegraphics[width=0.4\textwidth]{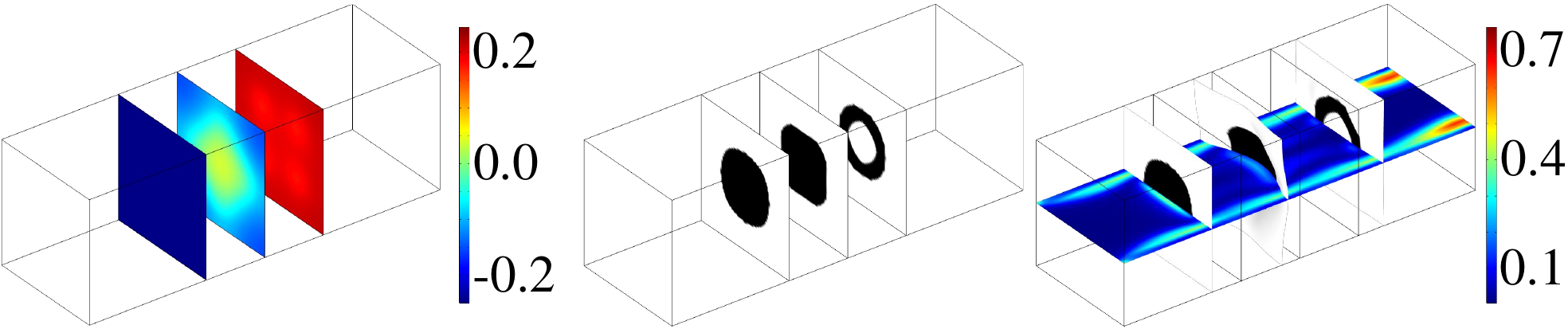}}\\
  \subfigure[$N=5$, $J_T = 1.0913$]
  {\includegraphics[width=0.5\textwidth]{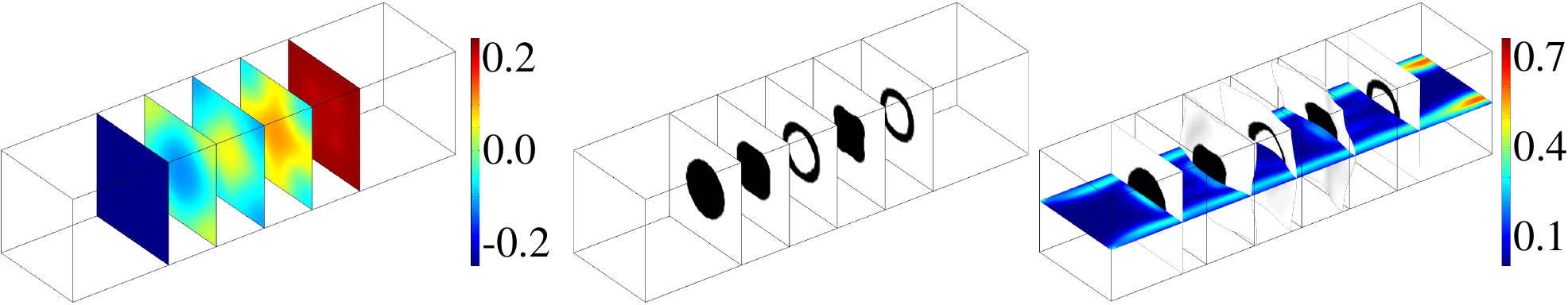}}\\
  \subfigure[$N=7$, $J_T = 1.4219$]
  {\includegraphics[width=0.6\textwidth]{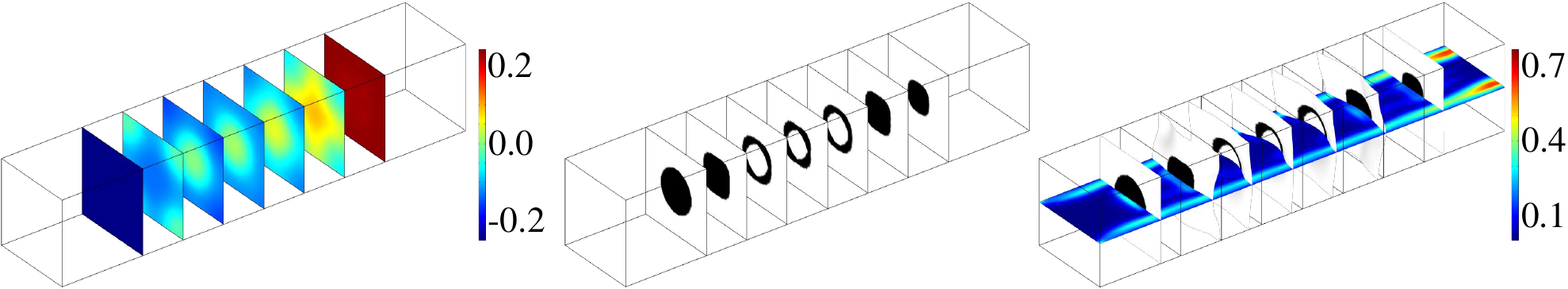}}\\
  \subfigure[$N=9$, $J_T = 1.7677$]
  {\includegraphics[width=0.7\textwidth]{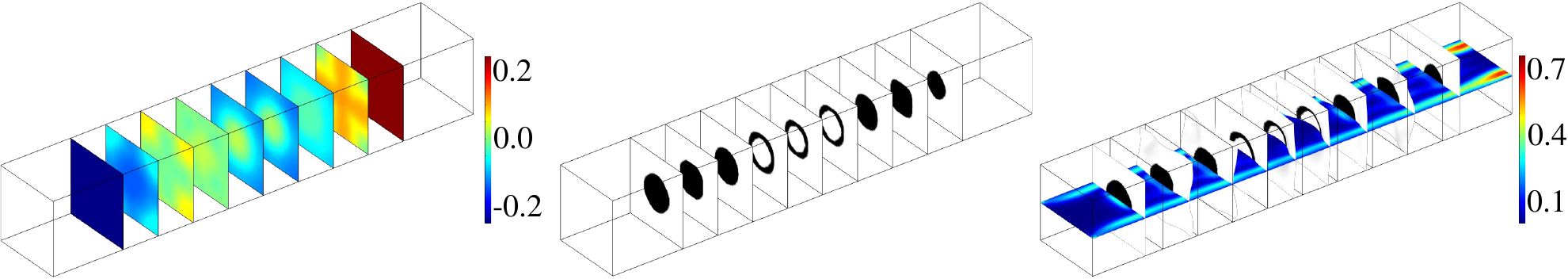}}\\
  \subfigure[$N=11$, $J_T = 2.0951$]
  {\includegraphics[width=0.8\textwidth]{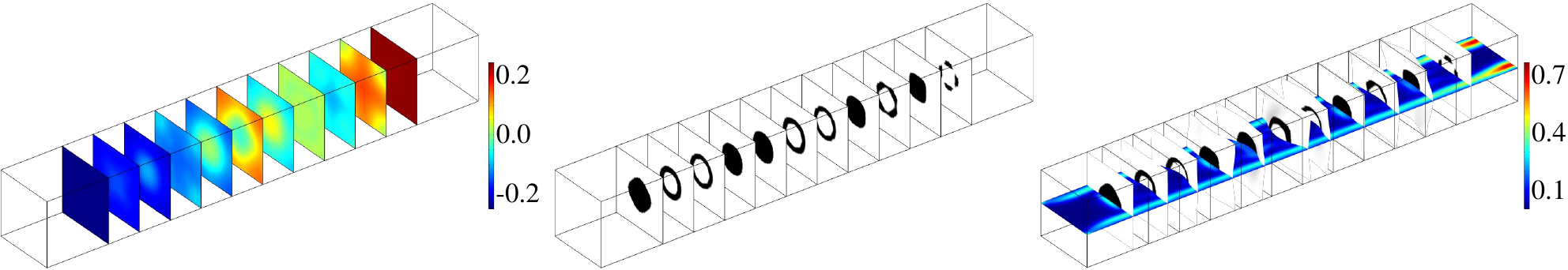}}\\
  \subfigure[$N=13$, $J_T = 2.4384$]
  {\includegraphics[width=0.9\textwidth]{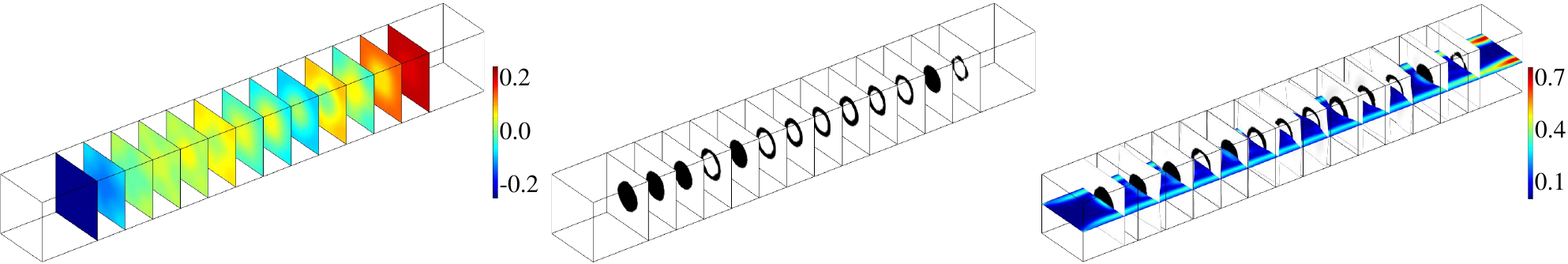}}
  \caption{Distribution of the filtered design variables for the implicit 2-manifolds and the material density of the thin-wall patterns obtained by solving the fiber bundle topology optimization problem for heat transfer in the volume flow on the design domains sketched in Fig. \ref{fig:MassHeatTransferManifoldsDesignDomainBulkFlow} with $N$ valued in $\left\{1,3,5,7,9,11,13\right\}$, where the first column is the distribution of the filtered design variables for the implicit 2-manifolds, the central column is the material density of the thin-wall patterns, and the third column is the deformed thin-wall patterns on the cross-sections including the thermal compliance distribution on the central cross-section.}\label{fig:ImbeddedSurfaceHTVolumeFlows}
\end{figure}

In fiber bundle topology optimization for mass and heat transfer in the volume flow, the inlet velocity is set as the parabolic distribution with $\mathbf{u}_{\Gamma_{v,\Omega}} = \left(2^4y\left(1-y\right)z\left(1-z\right), 0, 0 \right)$, corresponding to the Reynolds number of $Re=1\times10^0$; the coefficient of mass diffusion and the coefficient of heat conductivity are set as $D = 5\times10^{-3}$ and $k = 5\times10^{-3}$ corresponding to the P\'{e}clet number of $Pe = 2\times10^2$, respectively; the optimization parameters are set as $\alpha_{\max} = 1\times10^5$ and $q=1\times10^{-2}$; and the pressure drop is set as $\Delta P_0 = \left(N-2\right)/3\times10^2$. For mass transfer, the concentration distribution at the inlet is set by using the step function with the mid-value at $y=1/2$ on the inlet boundary and the minimal and maximal values of $0$ and $2$, respectively; correspondingly, the anticipation distribution of the concentration at the outlet is $\bar{c}=1$ for $\forall \mathbf{x}_\Xi \in \Gamma_{s,\Xi}$. For heat transfer, the known temperature at the inlet boundary is set with $T_0=0$ for $\forall \mathbf{x}_\Xi \in \Gamma_{v,\Xi}$ and the power of the heat source in the computational domain is set as $Q=1\times10^0$ for $\forall \mathbf{x}_\Xi \in \Xi$. By setting the variable magnitude of the implicit 2-manifold as $0.5$, the fiber bundle topology optimization problems in Eqs. \ref{equ:VarProToopBulkNSCDMHT} and \ref{equ:VarProToopBulkNSCHTMHT} are solved on the design domain in Fig. \ref{fig:MassHeatTransferManifoldsDesignDomainBulkFlow}. The distribution of the filtered design variables for the implicit 2-manifolds and the material density for the thin-wall patterns are obtained as shown in Figs. \ref{fig:ImbeddedSurfaceCDVolumeFlows} and \ref{fig:ImbeddedSurfaceHTVolumeFlows} for the cases with $N$ valued from $\left\{1,3,5,7,9,11,13\right\}$, where the specified reference value of the pressure drop linearly increases along with the increase of the length of the computational domain. For mass transfer, complete mixing is achieved in Fig. \ref{fig:ImbeddedSurfaceCDVolumeFlows}g. In Fig. \ref{fig:ImbeddedSurfaceCDVolumeFlows}, the implicit 2-manifold is prone to coincide with the base manifold and the deformation variable approaches to zero when the number of the cross-sections included in the design domain and the length of the computational domain are large enough. The thin-wall patterns on the cross-sections are hence the primary to achieve complete mixing and the implicit 2-manifolds are the supplemental, when the mixing length is large enough. For heat transfer, the thin-wall patterns on the cross-sections have the shapes of disc and ring as shown in Fig. \ref{fig:ImbeddedSurfaceHTVolumeFlows} and the obtained thin-wall structures are localized at the center of the cross-sections. Such thin-wall structures enhance the convection of the flow and strengthen the efficiency of heat transfer to minimize the thermal compliance. Fiber bundle topology optimization for mass and heat transfer in the volume flow is further investigated as follows for the cases with $N=7$.

\begin{figure}[!htbp]
  \centering
  \subfigure[]
  {\includegraphics[width=0.9\textwidth]{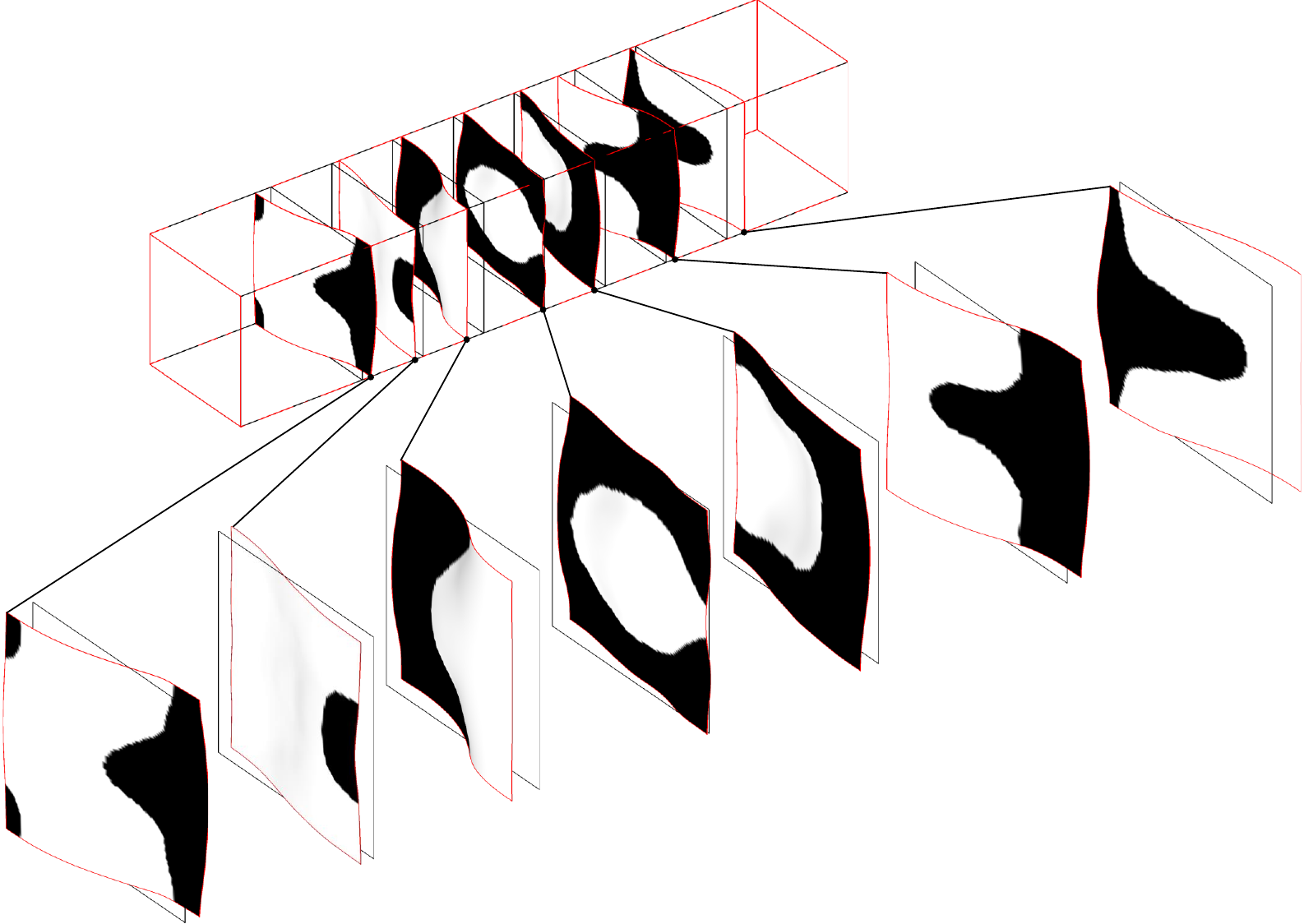}}
  \subfigure[]
  {\includegraphics[width=1\textwidth]{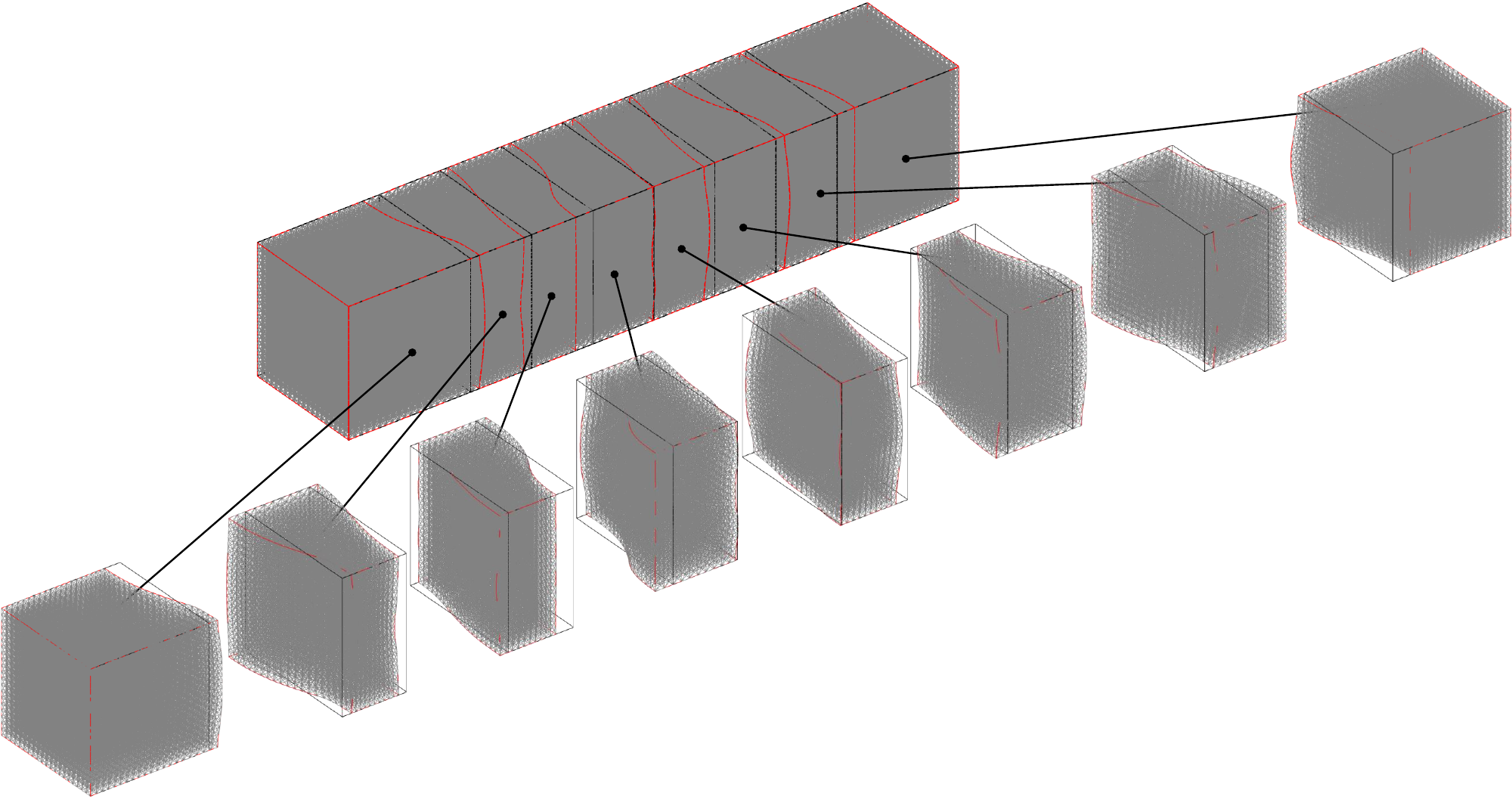}}
  \subfigure[]
  {\includegraphics[width=0.9\textwidth]{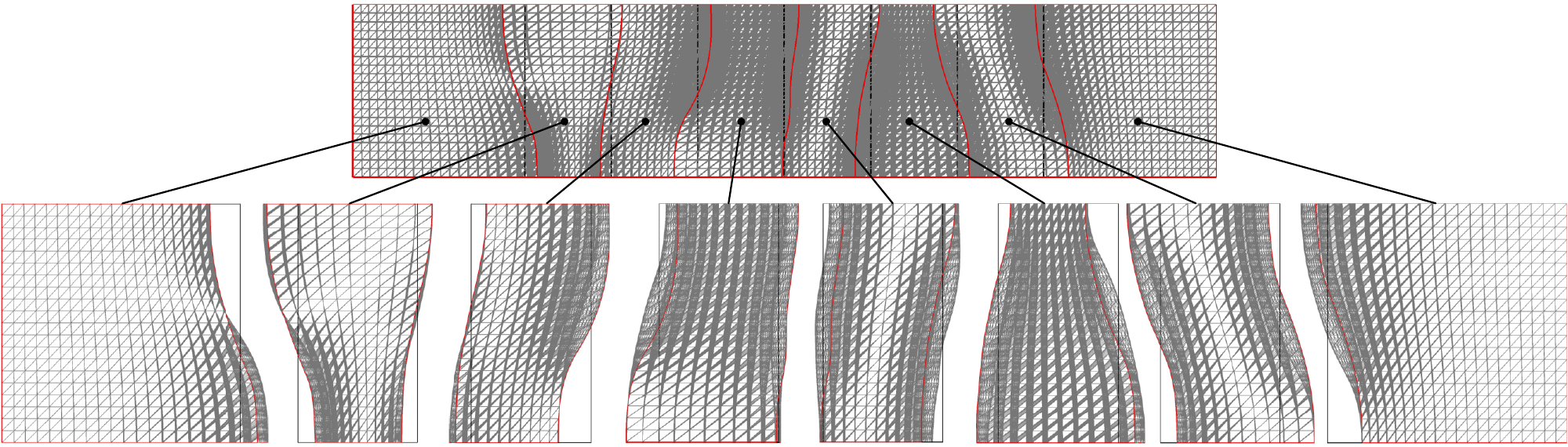}}
  \caption{Deformed meshes in the obtained fiber bundle for mass transfer in the volume flow, where the design domain is sketched in Fig. \ref{fig:MassHeatTransferManifoldsDesignDomainBulkFlow} with $N=7$. (a) Partial views of the obtained fiber bundle; (b) partial view of the deformed meshes; (c) top view of the deformed meshes.}\label{fig:MassTransferImmbeddedManifold73DPatternMesh}
\end{figure}

\begin{figure}[!htbp]
  \centering
  \subfigure[]
  {\includegraphics[width=0.9\textwidth]{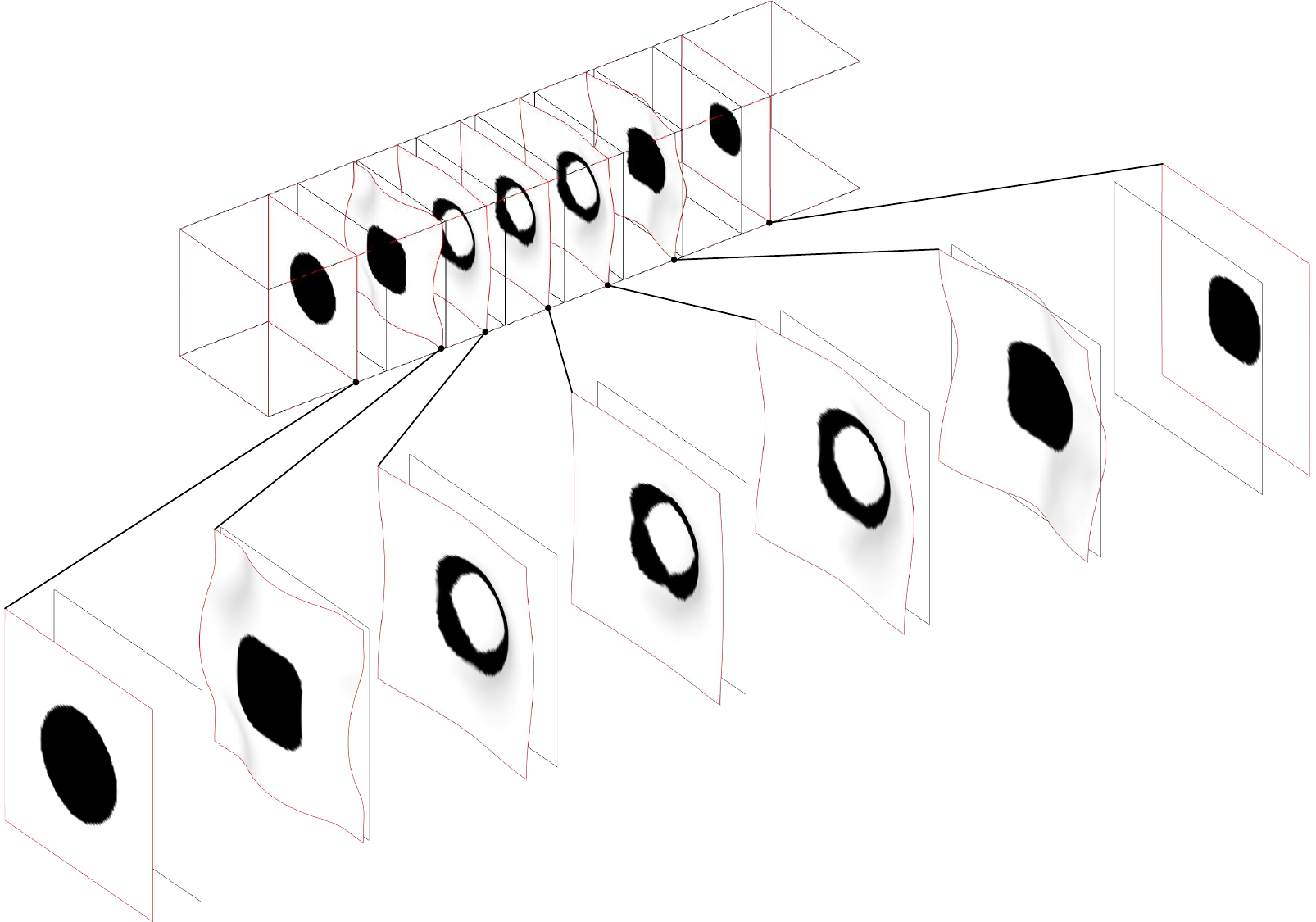}}
  \subfigure[]
  {\includegraphics[width=1\textwidth]{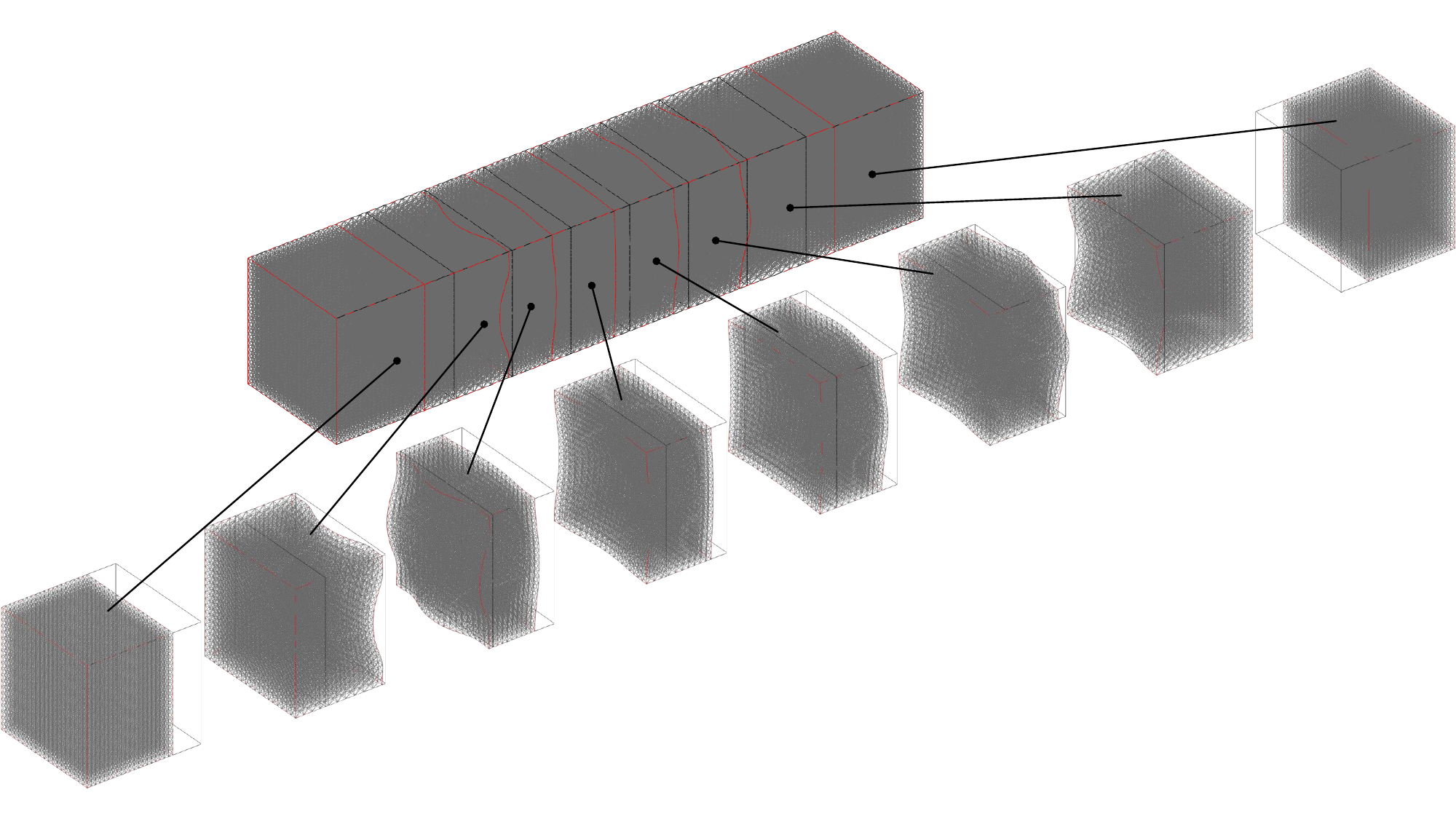}}
  \subfigure[]
  {\includegraphics[width=0.9\textwidth]{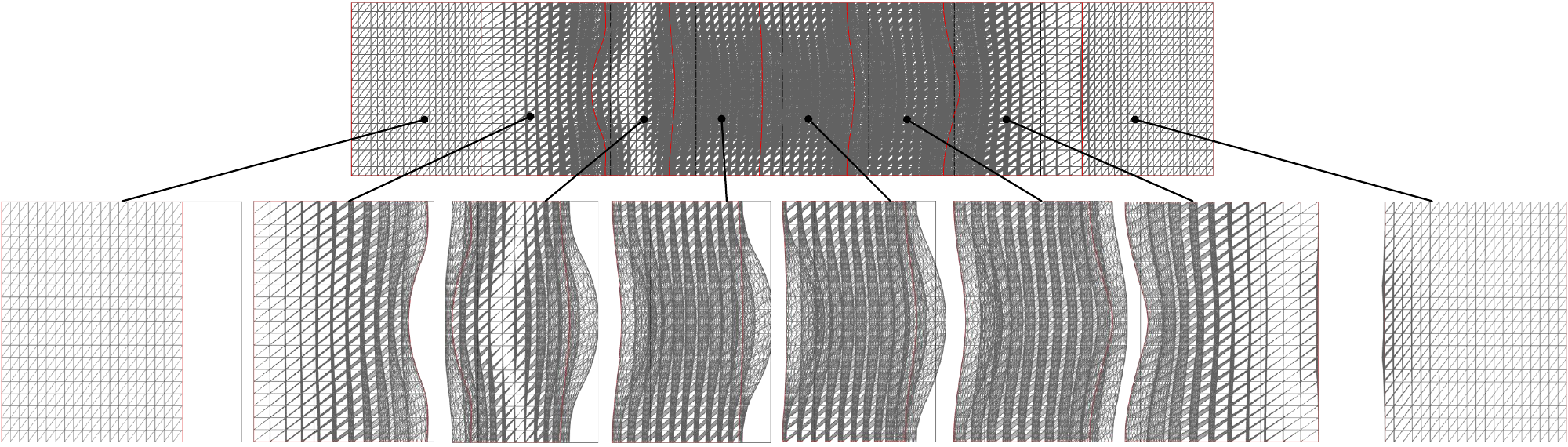}}
  \caption{Deformed meshes in the obtained fiber bundle for heat transfer in the volume flow, where the design domain is sketched in Fig. \ref{fig:MassHeatTransferManifoldsDesignDomainBulkFlow} with $N=7$. (a) Partial views of the obtained fiber bundle; (b) partial view of the deformed meshes; (c) top view of the deformed meshes.}\label{fig:HeatTransferImmbeddedManifold73DPatternMesh}
\end{figure}

\begin{figure}[!htbp]
  \centering
  \subfigure[]
  {\includegraphics[width=0.7\textwidth]{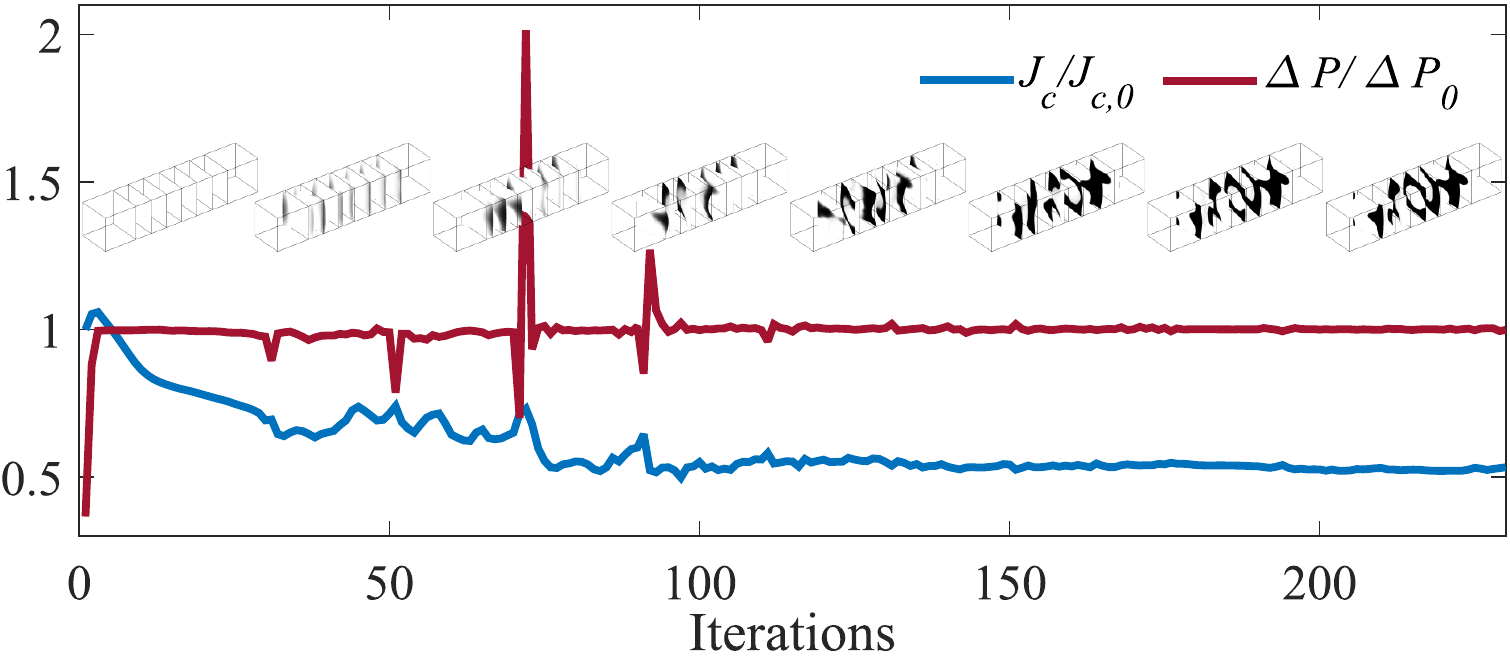}}
  \subfigure[]
  {\includegraphics[width=0.7\textwidth]{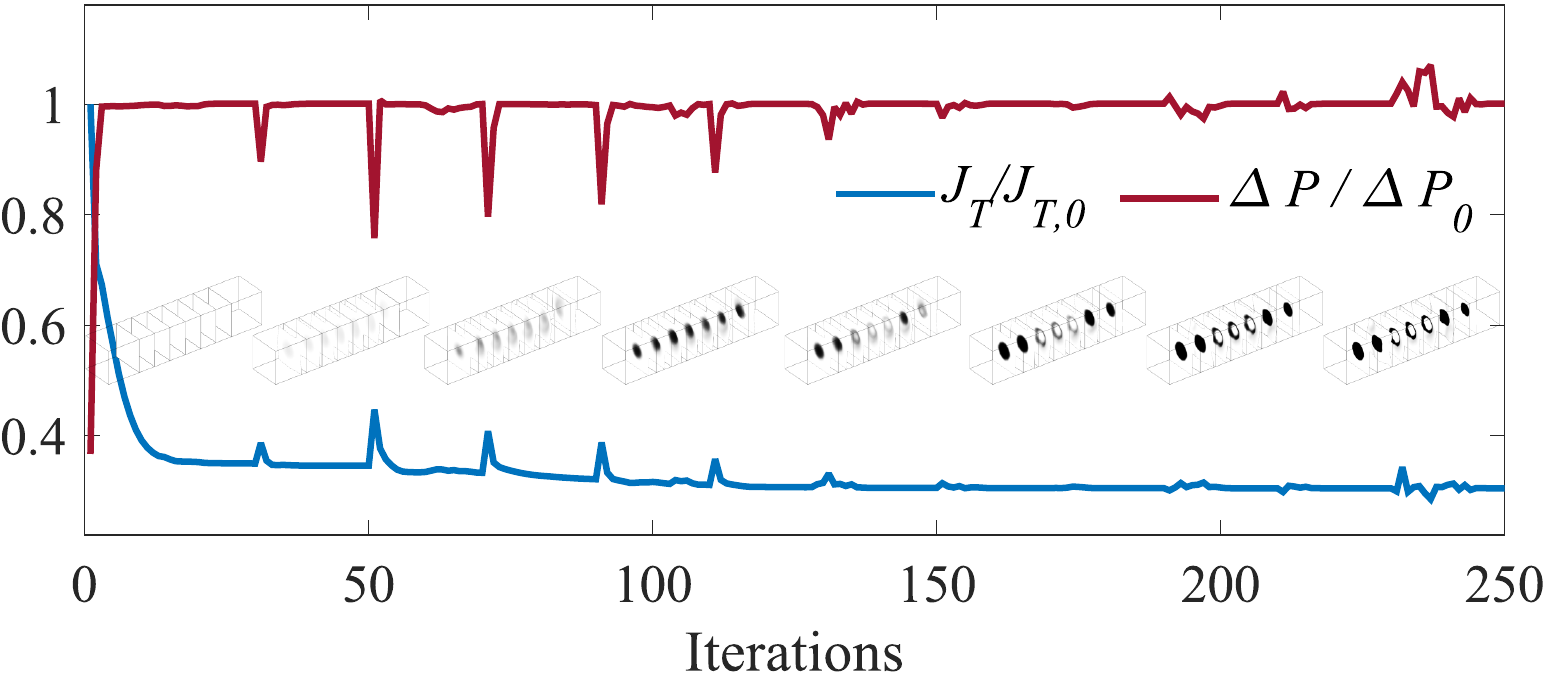}}
  \caption{Convergent histories of the design objective and constraint of the pressure drop in fiber bundle topology optimization for mass and heat transfer on the design domain sketched in Fig. \ref{fig:MassHeatTransferManifoldsDesignDomainBulkFlow} with $N=7$, where snapshots for the evolution of the fiber bundle during the iterative solution are included.}\label{fig:MassHeatTransferImmbeddedManifold7ConvergentHistories}
\end{figure}

\begin{figure}[!htbp]
  \centering
  \subfigure[]
  {\includegraphics[height=0.28\textwidth]{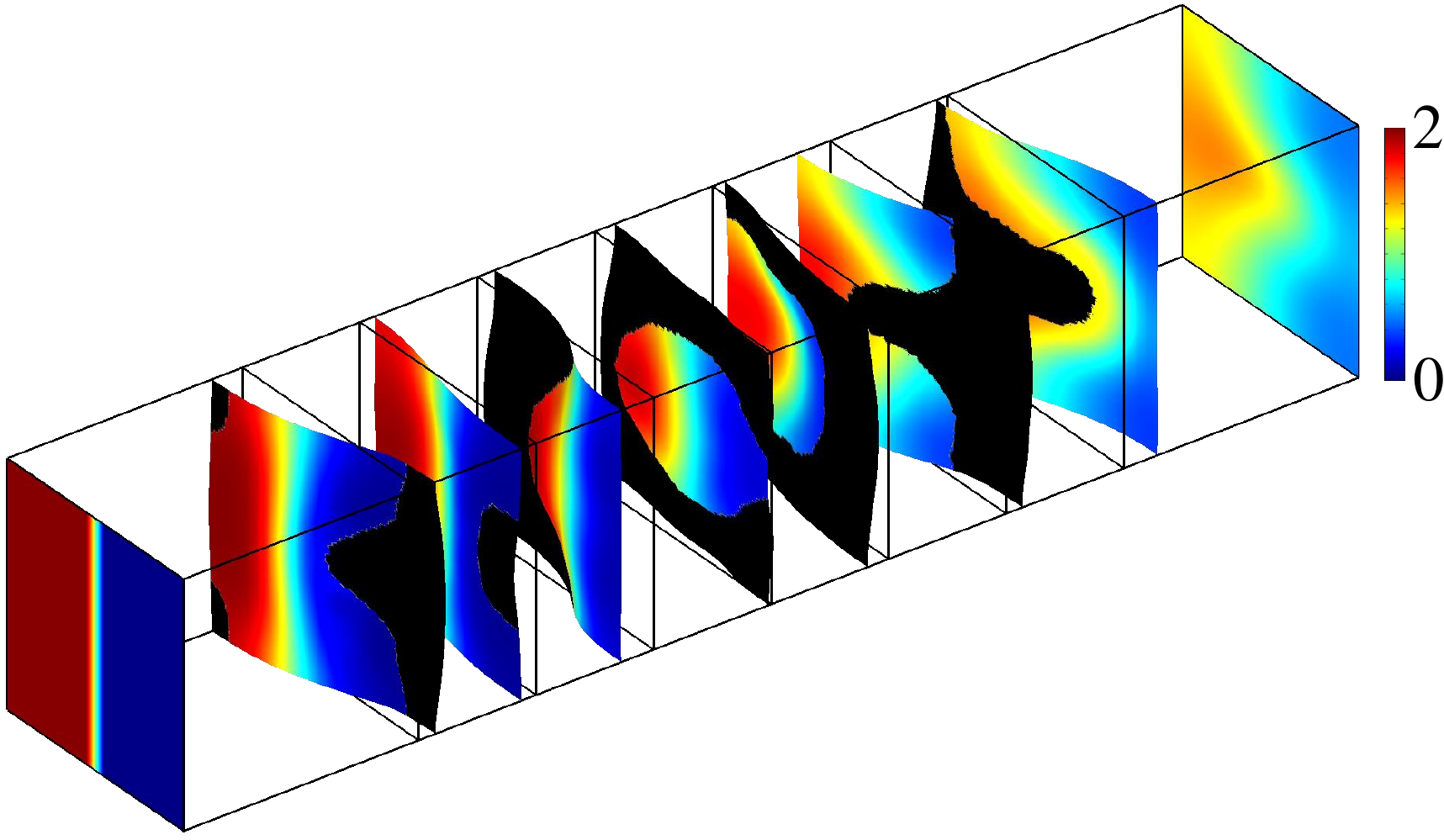}}\hspace{1em}
  \subfigure[]
  {\includegraphics[height=0.28\textwidth]{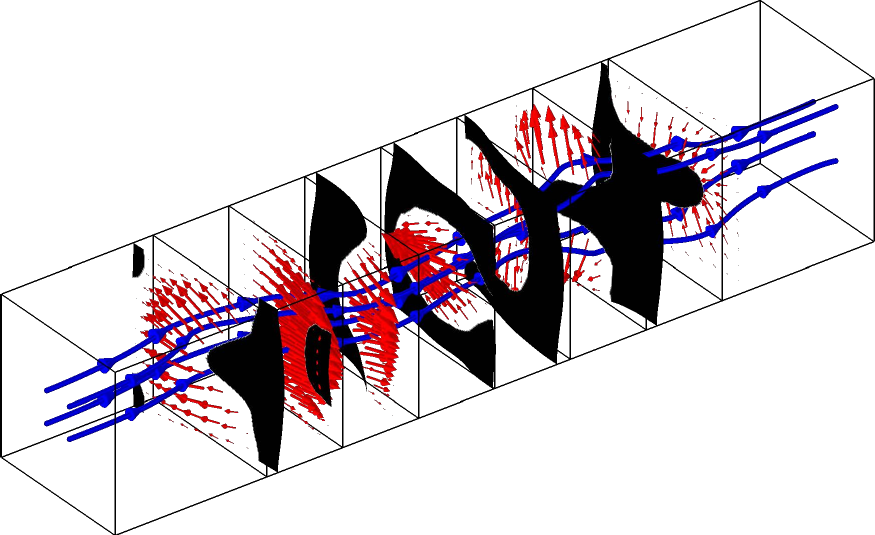}} \\
  \subfigure[]
  {\includegraphics[height=0.26\textwidth]{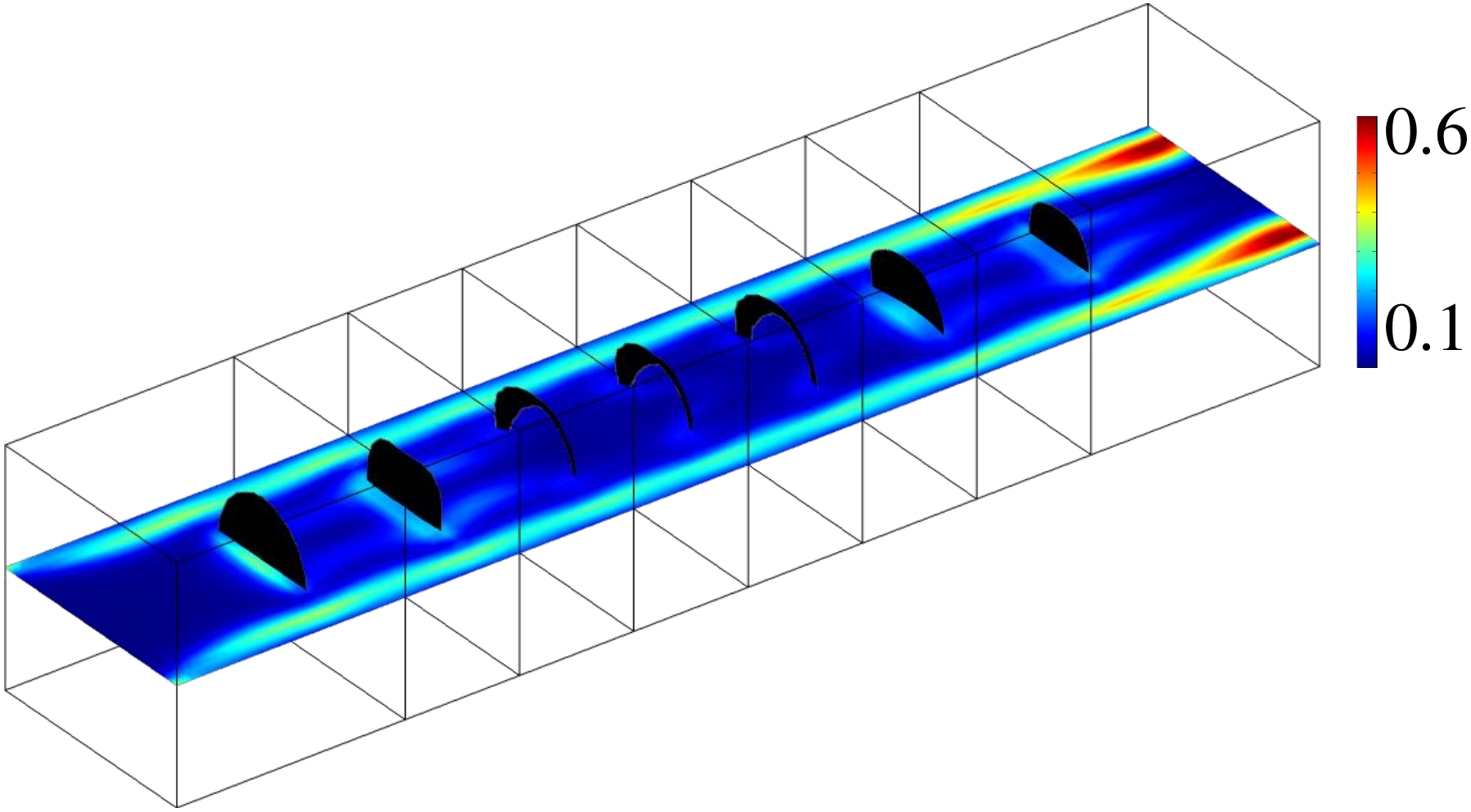}}\hspace{1em}
  \subfigure[]
  {\includegraphics[height=0.26\textwidth]{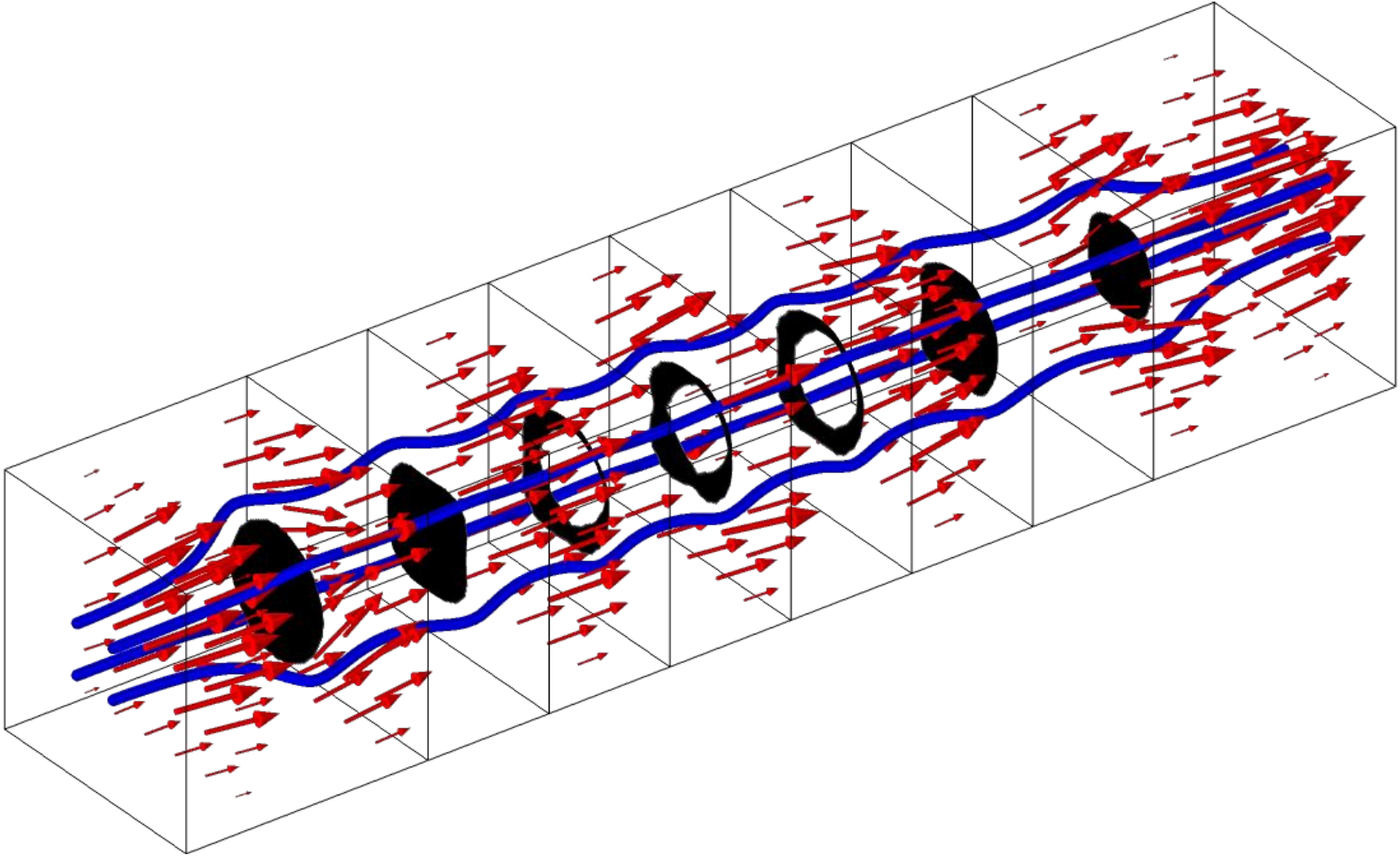}}
  \caption{Filed distribution in the obtained fiber bundles for mass and heat transfer on the design domain sketched in Fig. \ref{fig:MassHeatTransferManifoldsDesignDomainBulkFlow} with $N=7$. (a) Distribution of the concentration; (b) velocity of the secondary flow together with the streamlines in the straight channel; (c) distribution of the thermal compliance; (d) flow velocity together with the streamlines.}\label{fig:MassHeatTransferImmbeddedManifold7ConcenVArrowsStreamline}
\end{figure}

For the results in Figs. \ref{fig:ImbeddedSurfaceCDVolumeFlows}d and \ref{fig:ImbeddedSurfaceHTVolumeFlows}d, the detail views of the fiber bundle for the topologically optimized thin-wall patterns are presented in Figs. \ref{fig:MassTransferImmbeddedManifold73DPatternMesh}a and \ref{fig:HeatTransferImmbeddedManifold73DPatternMesh}a; and the detail views of the deformed meshes together and their top views are presented in Figs. \ref{fig:MassTransferImmbeddedManifold73DPatternMesh}b, \ref{fig:HeatTransferImmbeddedManifold73DPatternMesh}b, \ref{fig:MassTransferImmbeddedManifold73DPatternMesh}c and \ref{fig:HeatTransferImmbeddedManifold73DPatternMesh}c, where the mesh deformation is implicitly described by the Laplace's equation in Eq. \ref{equ:HarmonicCoordinateEquMHT} and it is caused by the relative displacement between the base manifold and the implicit 2-manifold. Convergent histories of the design objective and constraint of the pressure drop are plotted in Fig. \ref{fig:MassHeatTransferImmbeddedManifold7ConvergentHistories}, including snapshots for the evolution of the fiber bundle during the iterative solution of the optimization problem. From the monotonicity of the objective values and satisfication of the constraints of the pressure drop, the robustness of the iterative solutions can be confirmed for both the mass and heat transfer problems. Distribution of the concentration, thermal compliance and velocity together with streamlines in the straight channel are provided in Fig. \ref{fig:MassHeatTransferImmbeddedManifold7ConcenVArrowsStreamline}, where the obtained thin-wall patterns induce the secondary flow to enlarge the mixing length and strengthen the convection to enhance the heat transfer, and the mass and heat transfer efficiencies are thereby improved.

\begin{figure}[!htbp]
  \centering
  \subfigure[]
  {\includegraphics[width=0.7\textwidth]{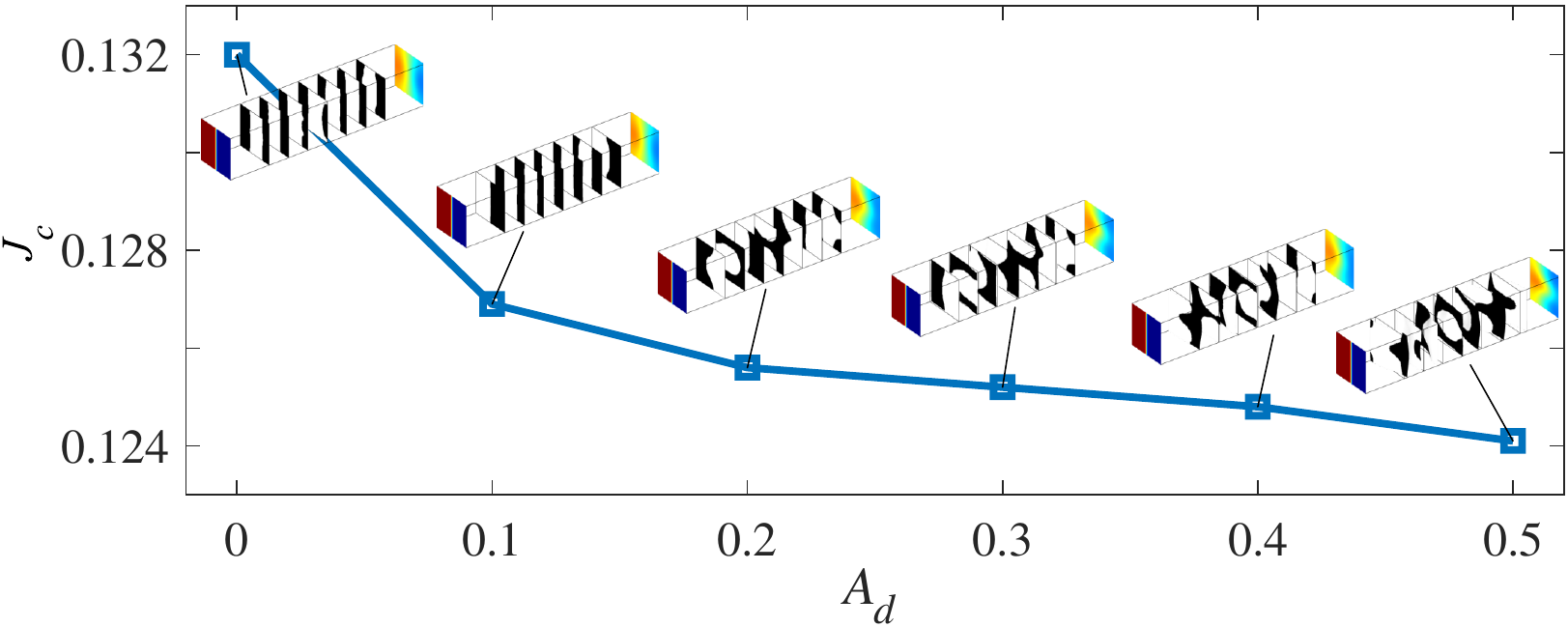}}
  \subfigure[]
  {\includegraphics[width=0.7\textwidth]{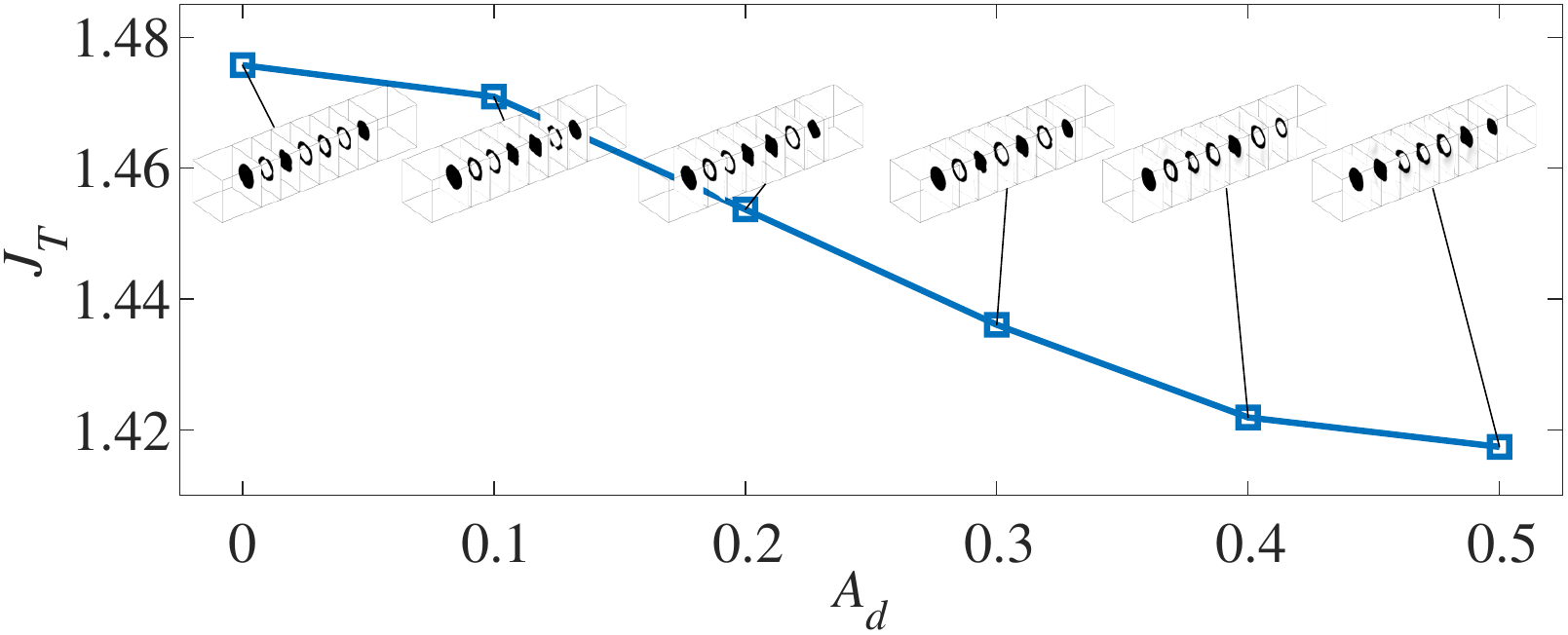}}
  \caption{Objective values and the obtained fiber bundles for the variable magnitudes of $A_d = \left\{0.0,0.1,0.2,0.3,0.4,0.5\right\}$ and the obtained fiber bundles on the design domain sketched in Fig. \ref{fig:MassHeatTransferManifoldsDesignDomainBulkFlow} with $N=7$, where the Reynolds number is $Re = 1\times10^0$ and the P\'{e}clet number is $Pe = 2\times10^2$.}\label{fig:MassHeatTransferImmbeddedManifold7Ad}
\end{figure}

\begin{figure}[!htbp]
  \centering
  \subfigure[]
  {\includegraphics[width=0.7\textwidth]{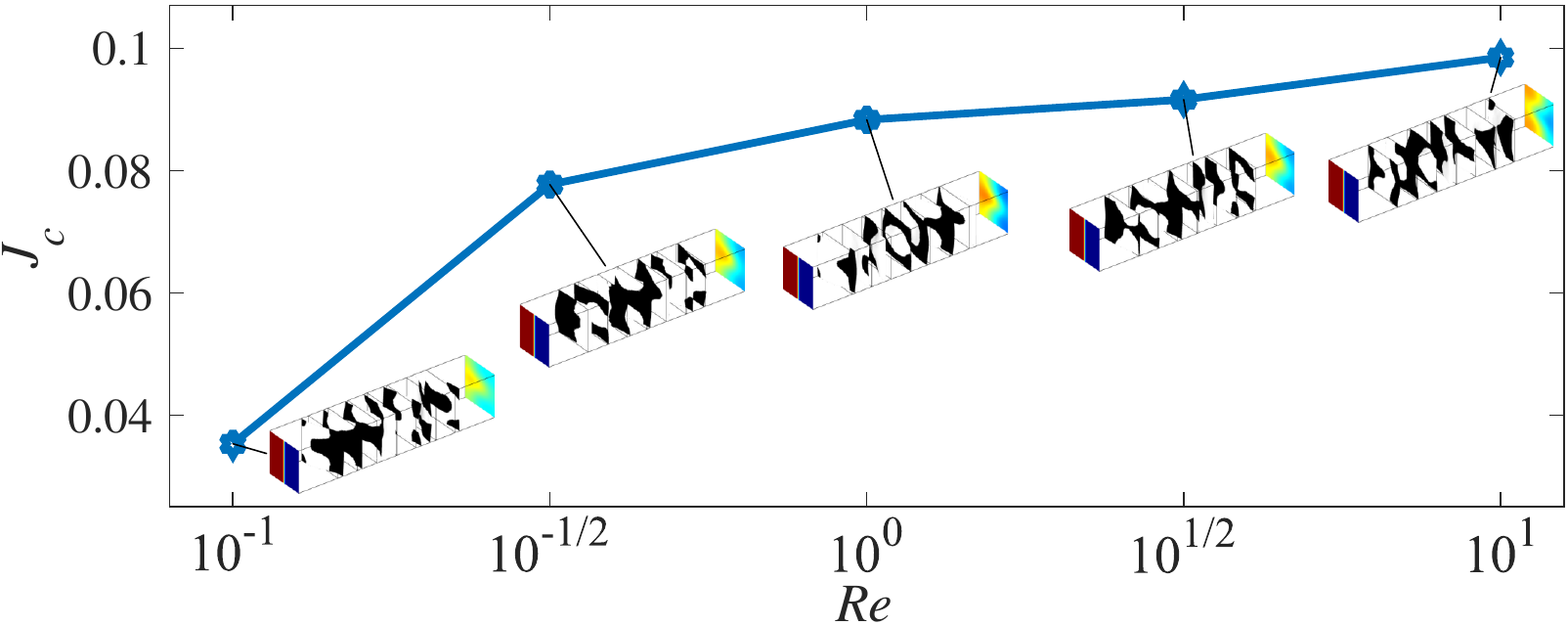}}
  \subfigure[]
  {\includegraphics[width=0.7\textwidth]{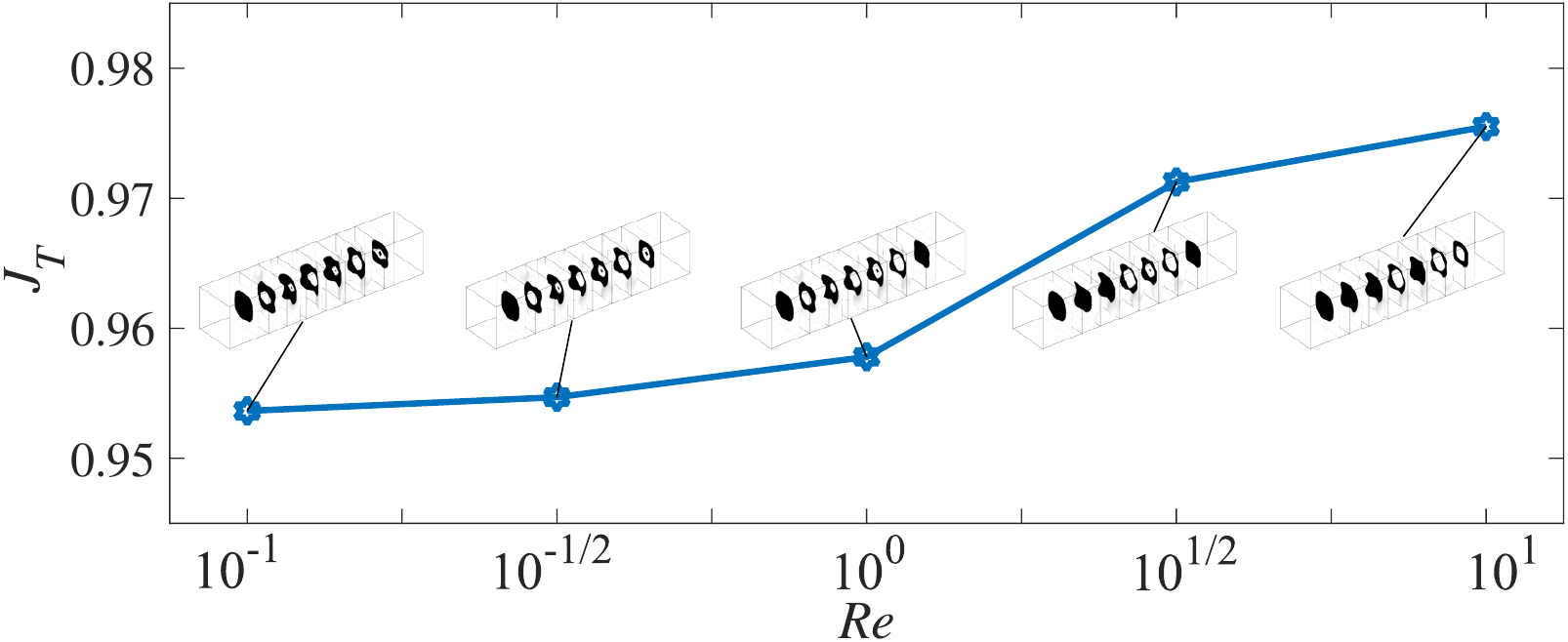}}
  \caption{Objective values and the obtained fiber bundles for the Reynolds numbers of $Re = \left\{10^{-1}, 10^{-1/2}, 10^0, 10^{1/2}, 10^1 \right\}$ and the obtained fiber bundles on the design domain sketched in Fig. \ref{fig:MassHeatTransferManifoldsDesignDomainBulkFlow} with $N=7$, where the variable magnitude is $A_d=0.5$ and the P\'{e}clet number is $Pe = 2\times10^2$.}\label{fig:MassHeatTransferImmbeddedManifold7Re}
\end{figure}

\begin{figure}[!htbp]
  \centering
  \subfigure[]
  {\includegraphics[width=0.7\textwidth]{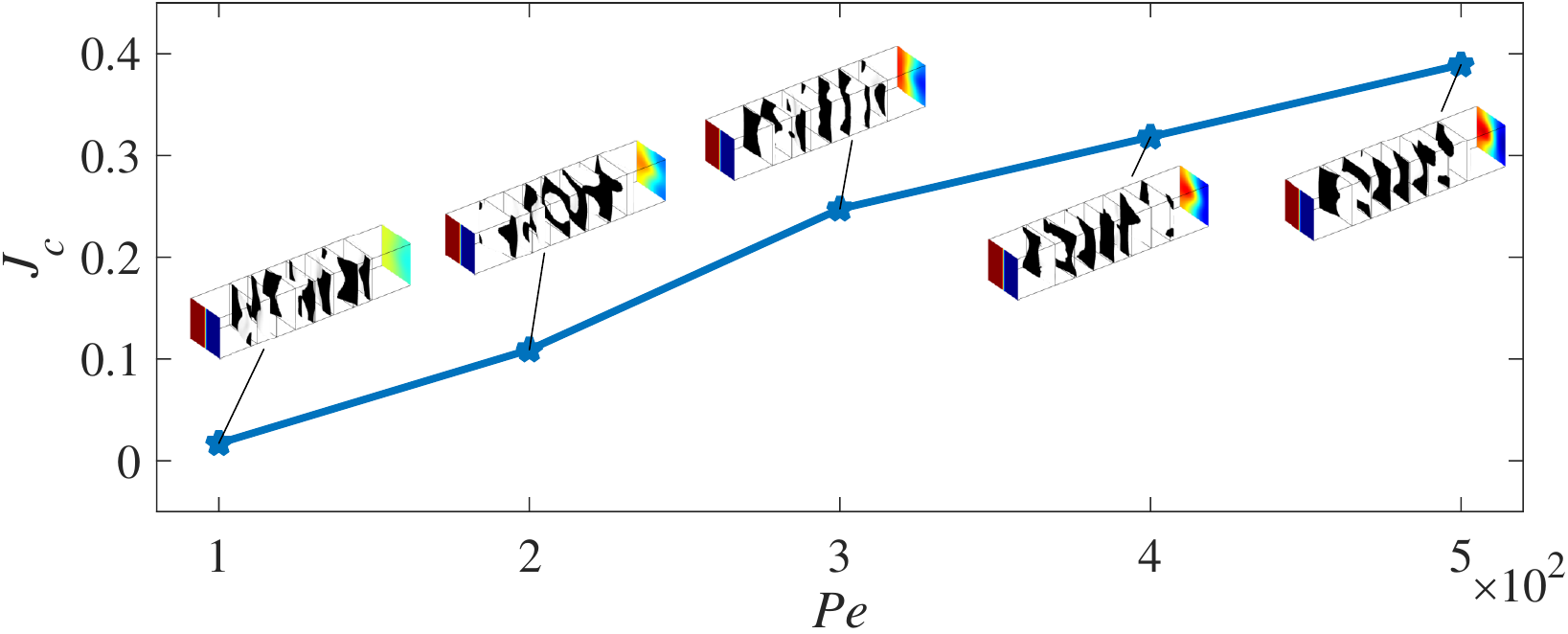}}
  \subfigure[]
  {\includegraphics[width=0.7\textwidth]{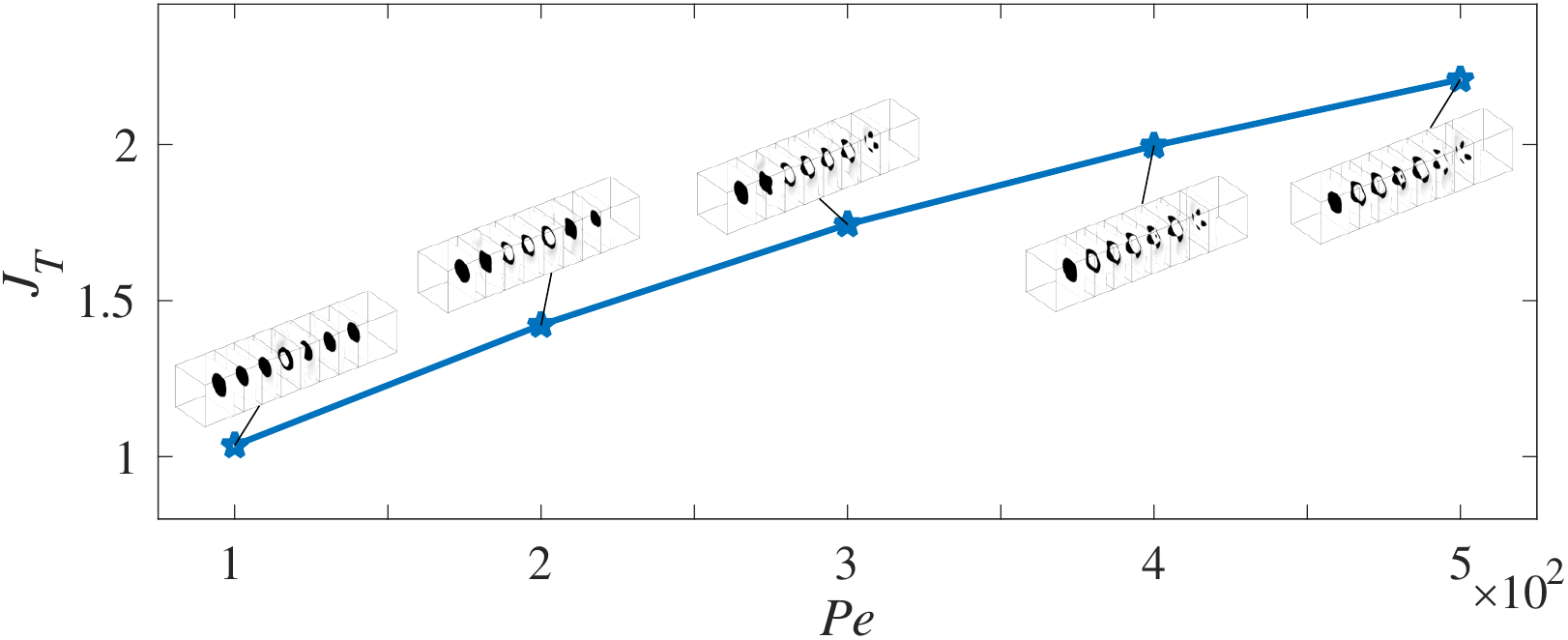}}
  \caption{Objective values and the obtained fiber bundles for the P\'{e}clet numbers of $Pe = \left\{ 1\times10^2, 2\times10^2, 3\times10^2, 4\times10^2, 5\times10^2 \right\}$ and the obtained fiber bundles on the design domain sketched in Fig. \ref{fig:MassHeatTransferManifoldsDesignDomainBulkFlow} with $N=7$, where the variable magnitude is $A_d=0.5$ and the Reynolds number is $Re = 1\times10^0$.}\label{fig:MassHeatTransferImmbeddedManifold7Pe}
\end{figure}

\begin{figure}[!htbp]
  \centering
  \subfigure[]
  {\includegraphics[width=0.7\textwidth]{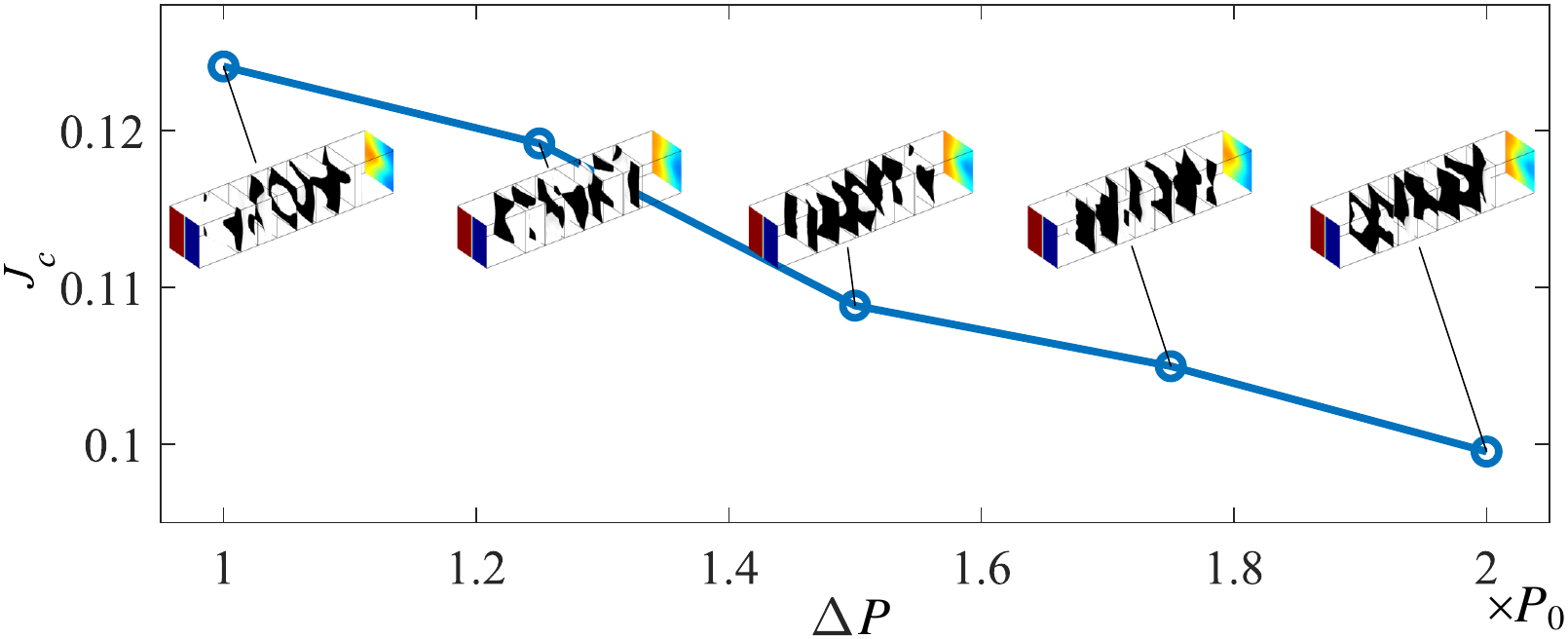}}
  \subfigure[]
  {\includegraphics[width=0.7\textwidth]{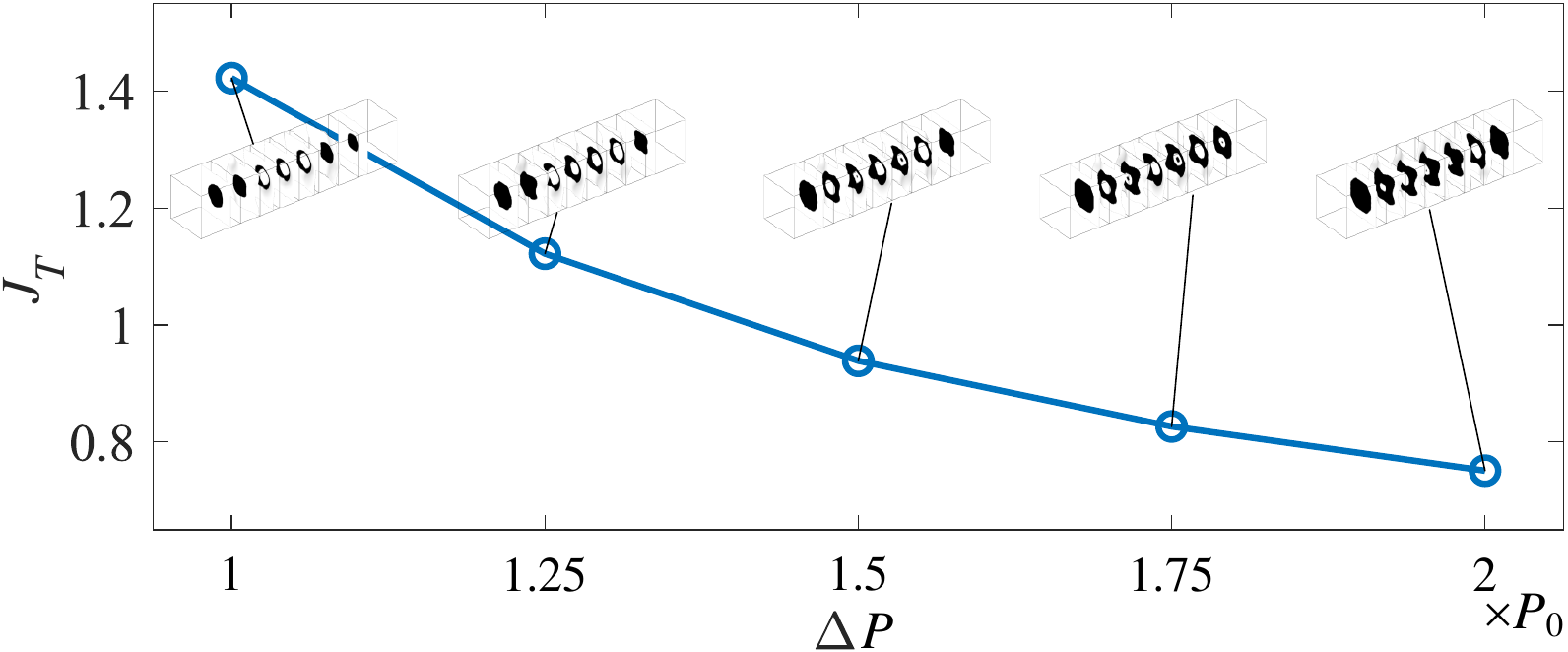}}
  \caption{Objective values and the obtained fiber bundles for different pressure drop on the design domain sketched in Fig. \ref{fig:MassHeatTransferManifoldsDesignDomainBulkFlow} with $N=7$, where the Reynolds number is $Re = 1\times10^0$ and the P\'{e}clet number is $Pe = 2\times10^2$.}\label{fig:MassHeatTransferImmbeddedManifold7DP}
\end{figure}

The effects of the variable magnitude, Reynolds number, P\'{e}clet number and reference pressure drop are further investigated for the fiber bundle topology optimization problem of mass and heat transfer in the volume flow. As shown in Fig. \ref{fig:MassHeatTransferImmbeddedManifold7Ad}, the mass and heat transfer performance achieved by the optimized matching between the thin-wall patterns and the implicit 2-manifolds can be improved by enlarging the variable magnitude, because larger variable magnitude enlarges the design space. The necessity of fiber bundle topology optimization is also demonstrated by the results in Fig. \ref{fig:MassHeatTransferImmbeddedManifold7Ad}. As shown in Figs. \ref{fig:MassHeatTransferImmbeddedManifold7Re} and \ref{fig:MassHeatTransferImmbeddedManifold7Pe}, the optimized matchings corresponding the lower Reynolds number and lower P\'{e}clet number can achieve more efficient mass and heat transfer, this is because of the higher diffusion efficiency. As shown in Fig. \ref{fig:MassHeatTransferImmbeddedManifold7DP}, the increase of the pressure drop can enhance the convection of the flow and thereby improve the mass and heat transfer performance of the obtained fiber bundles.

To confirm the optimality, the results in Fig. \ref{fig:MassHeatTransferImmbeddedManifold7Pe} are cross-compared by computing the objective values for the obtained fiber bundle at different P\'{e}clet numbers, where the Reynolds number is remained as $1\times10^0$. The computed objective values are listed in Tab. \ref{tab:MassHeatTransferManifold3VolumeFlowOptimality}. From the comparison of the objective values in every row of Tabs. \ref{tab:MassHeatTransferManifold3VolumeFlowOptimality}a and \ref{tab:MassHeatTransferManifold3VolumeFlowOptimality}b, the optimized performance of the obtained fiber bundles can be confirmed.

\begin{table}[!htbp]
\centering
\subtable[]
{\begin{tabular}{l|ccccc}
  \toprule
        & \includegraphics[width=0.14\textwidth]{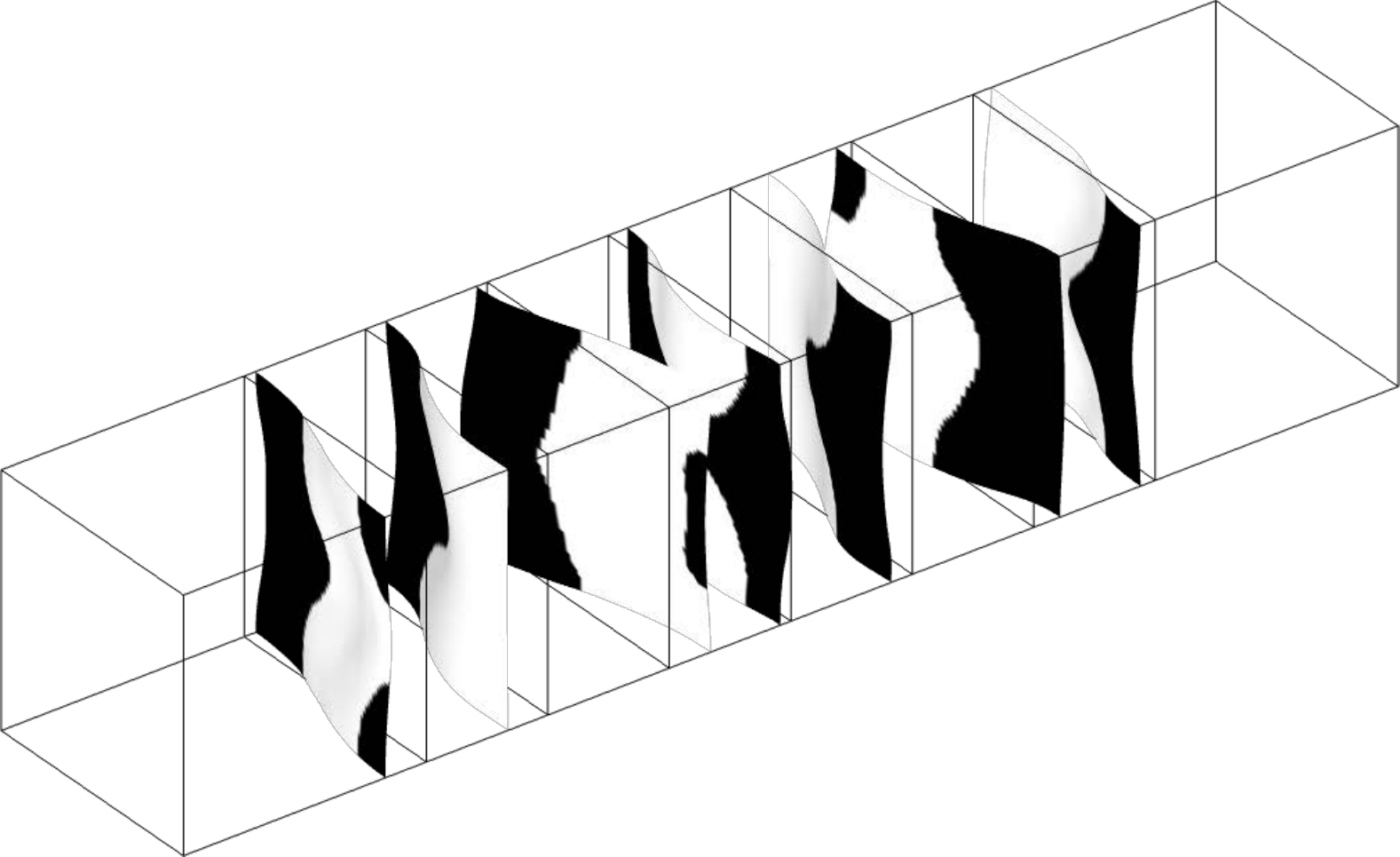}
        & \includegraphics[width=0.14\textwidth]{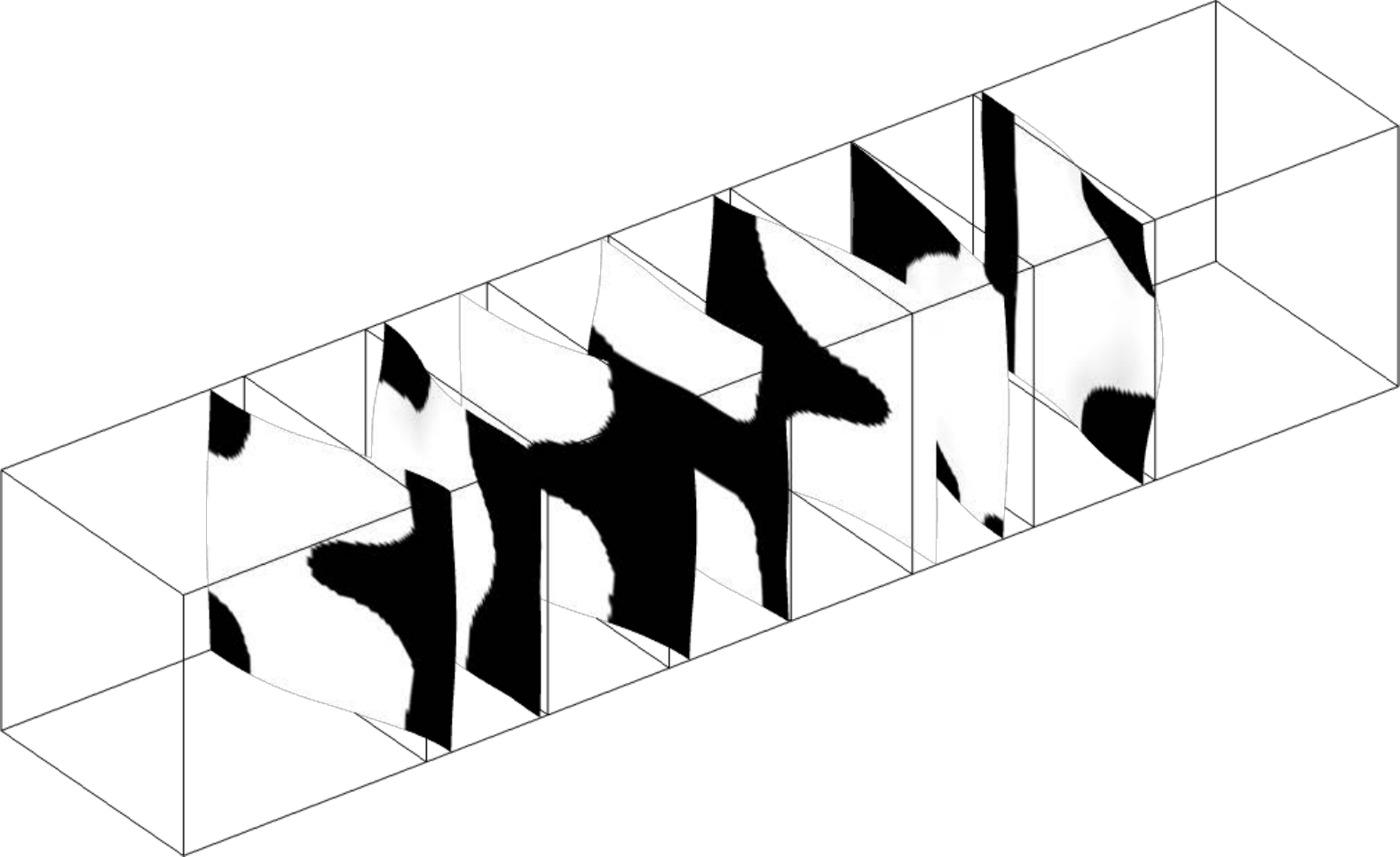}
        & \includegraphics[width=0.14\textwidth]{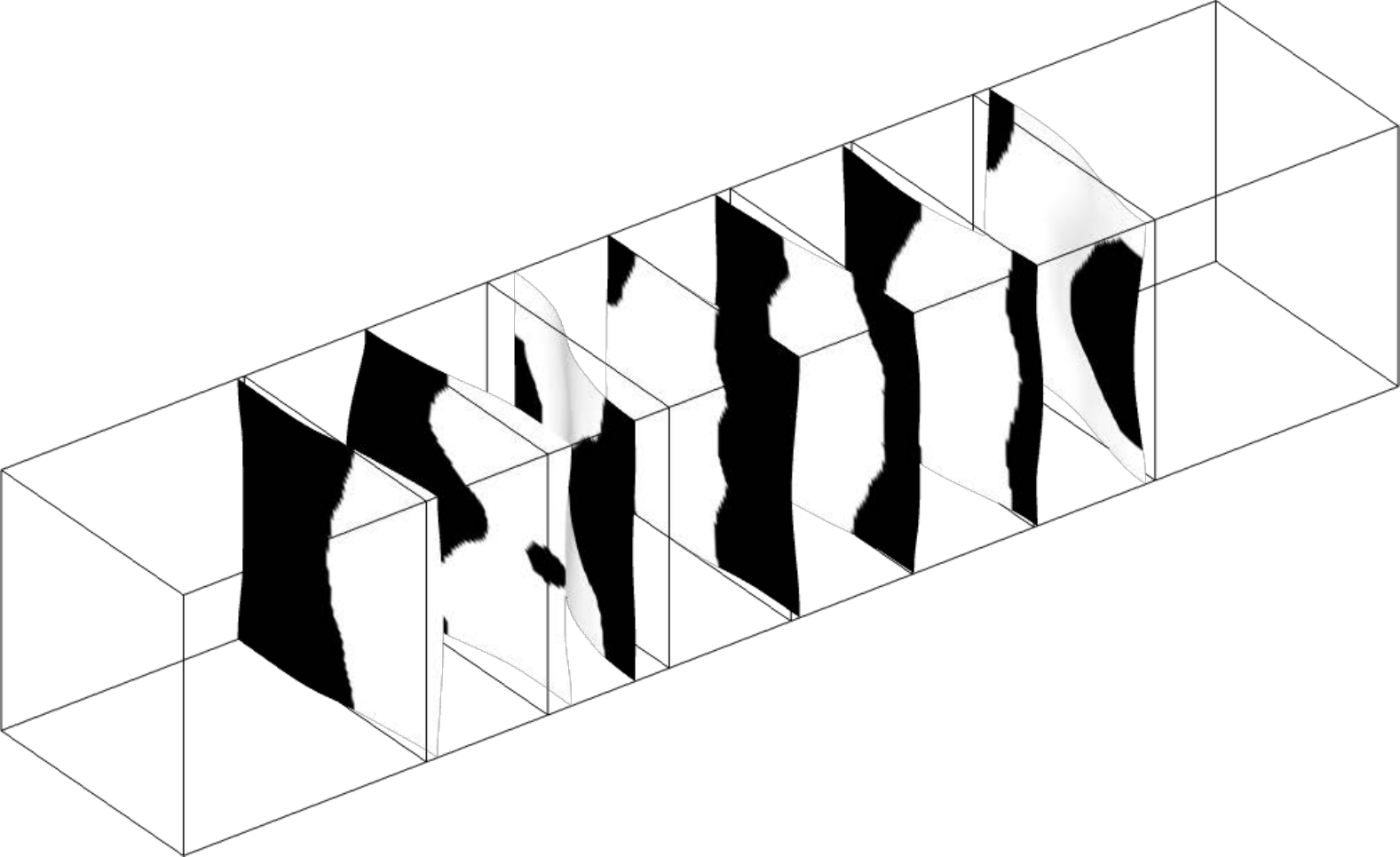}
        & \includegraphics[width=0.14\textwidth]{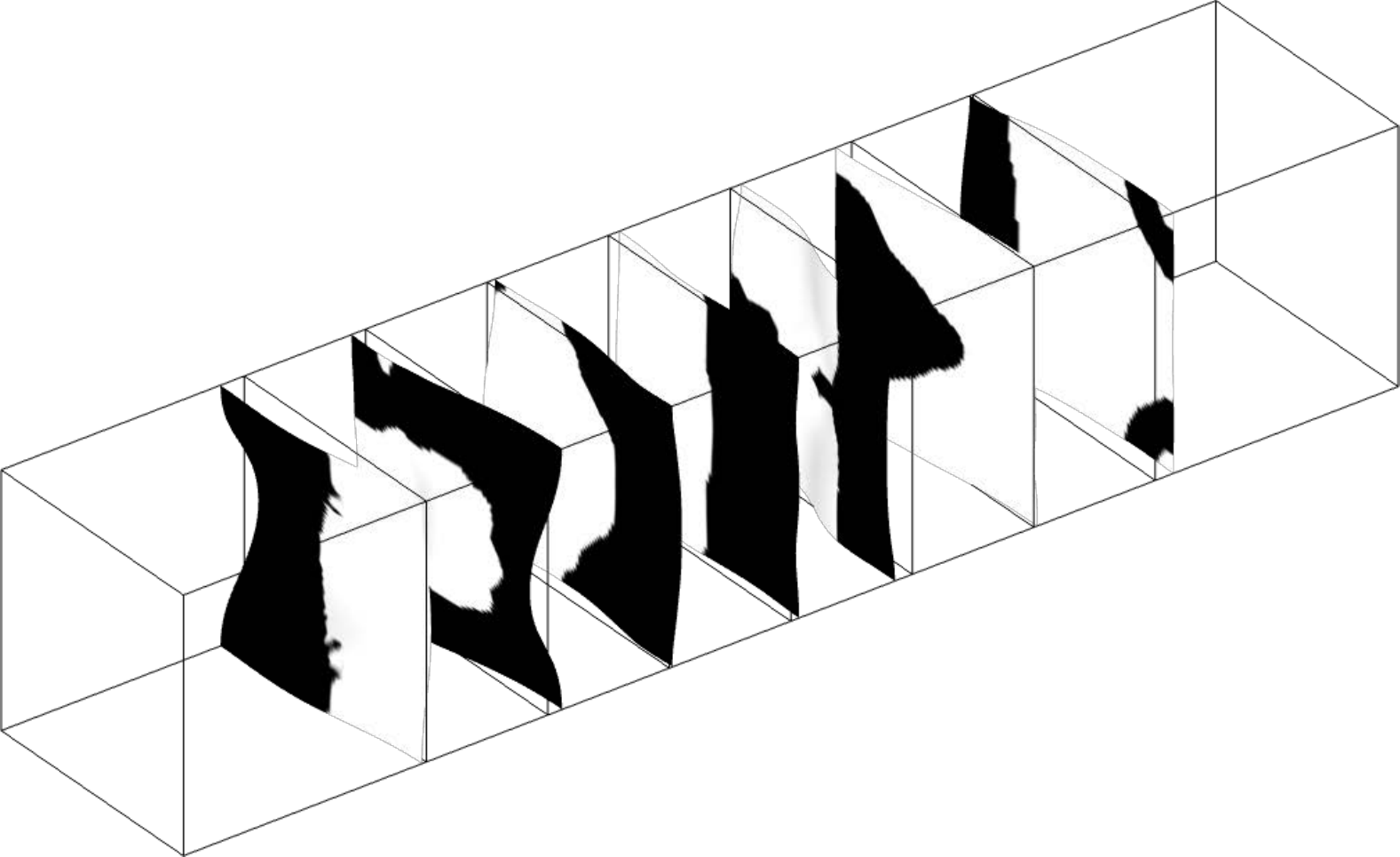}
        & \includegraphics[width=0.14\textwidth]{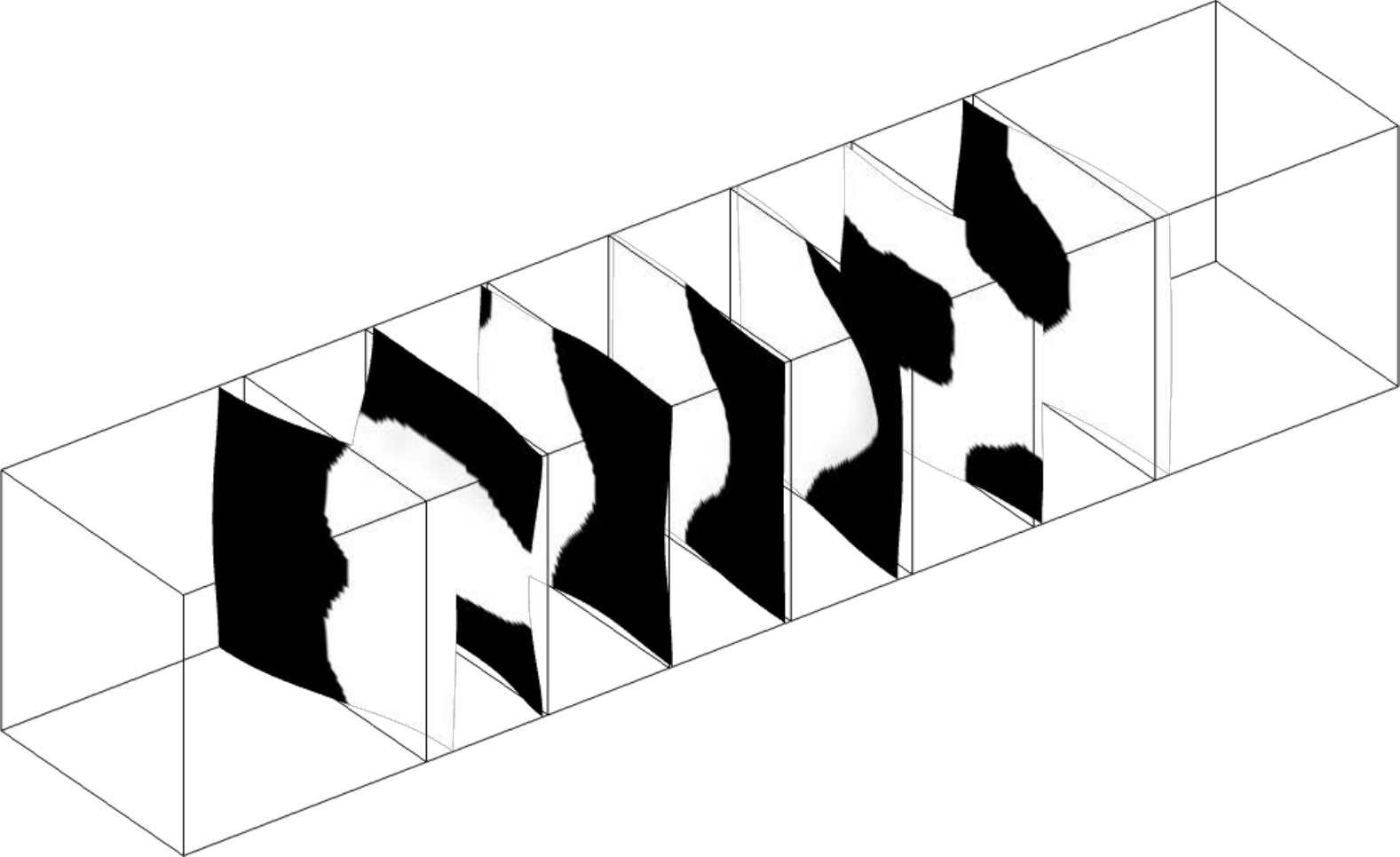} \\
   & $Pe = 1\times10^2$ & $Pe = 2\times10^2$ & $Pe = 3\times10^2$ & $Pe = 4\times10^2$ & $Pe = 5\times10^2$ \\     
  \midrule
  $Pe = 1\times10^2$ & $\mathbf{0.0159}$ & $0.0192$ & $0.0195$ & $0.0201$ & $0.0243$ \\
  \midrule
  $Pe = 2\times10^2$ & $0.1040$ & $\mathbf{0.1018}$ & $0.1218$ & $0.1159$ & $0.1266$ \\
  \midrule
  $Pe = 3\times10^2$ & $0.2430$ & $0.2513$ & $\mathbf{0.2310}$ & $0.2458$ & $0.2552$ \\
  \midrule
  $Pe = 4\times10^2$ & $0.2970$ & $0.2984$ & $0.3197$ & $\mathbf{0.2901}$ & $0.3030$ \\
  \midrule
  $Pe = 5\times10^2$ & $0.3675$ & $0.3689$ & $0.3886$ & $0.3705$ & $\mathbf{0.3633}$ \\
  \bottomrule
\end{tabular}}
\subtable[]
{\begin{tabular}{l|ccccc}
  \toprule
        & \includegraphics[width=0.14\textwidth]{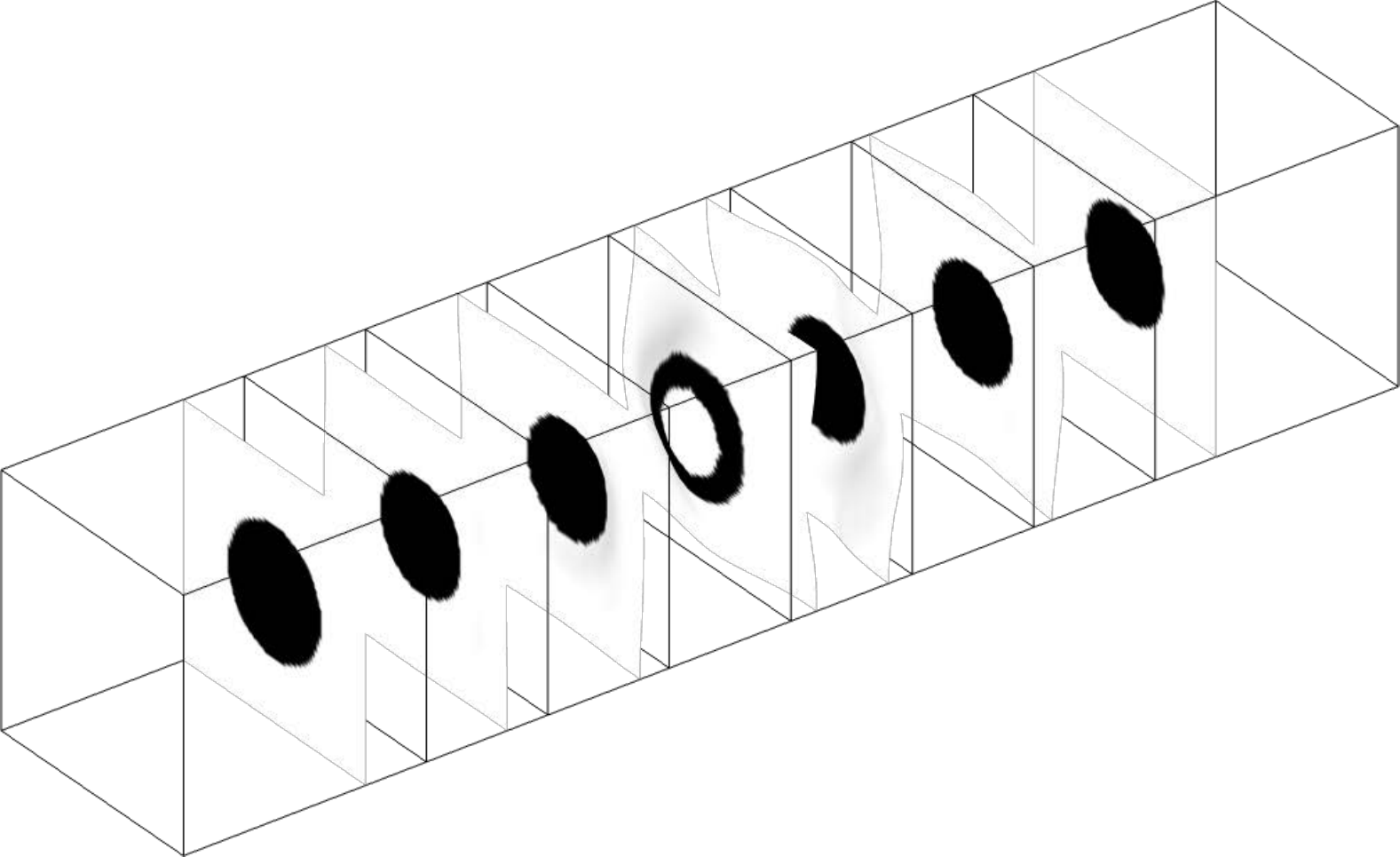}
        & \includegraphics[width=0.14\textwidth]{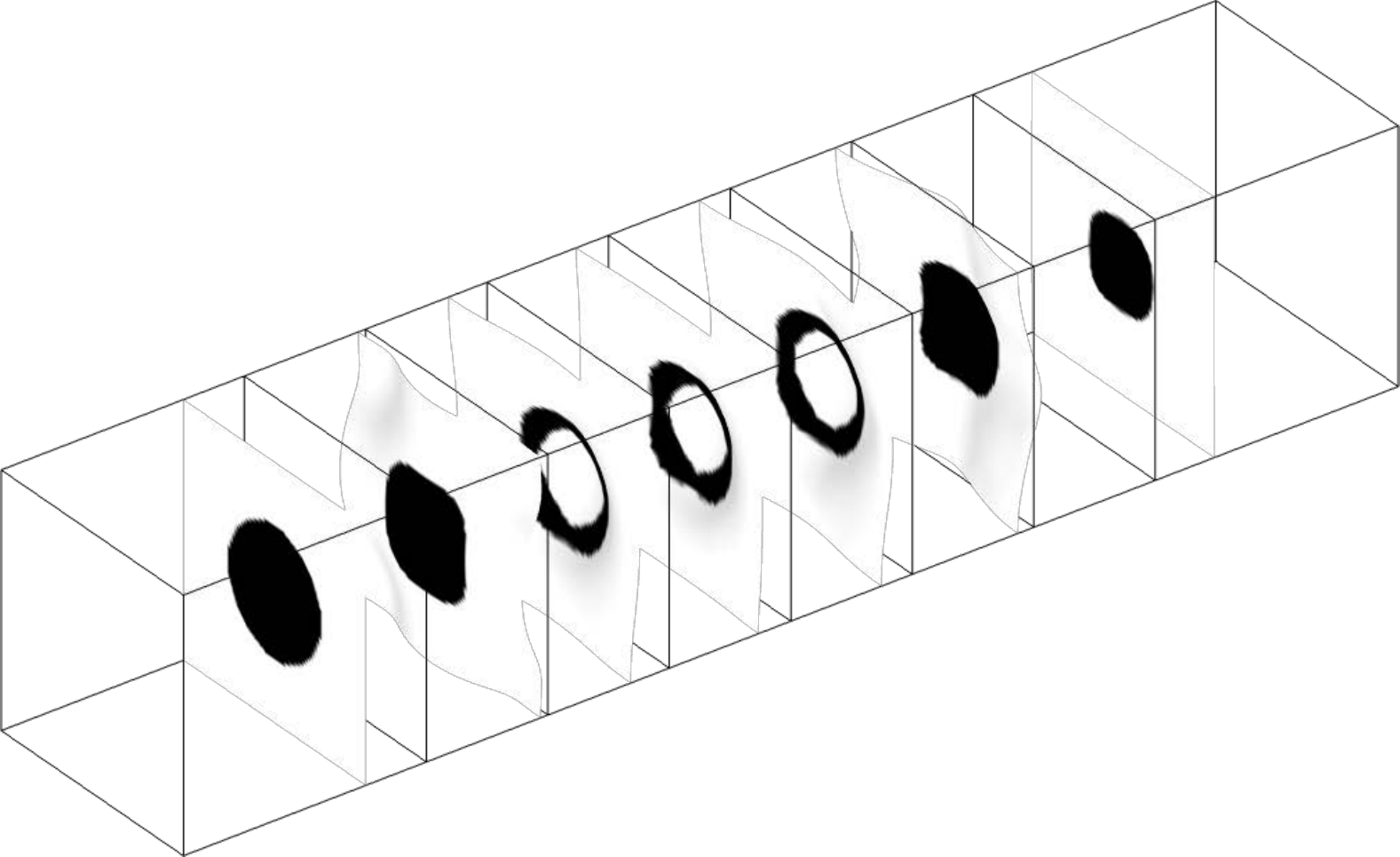}
        & \includegraphics[width=0.14\textwidth]{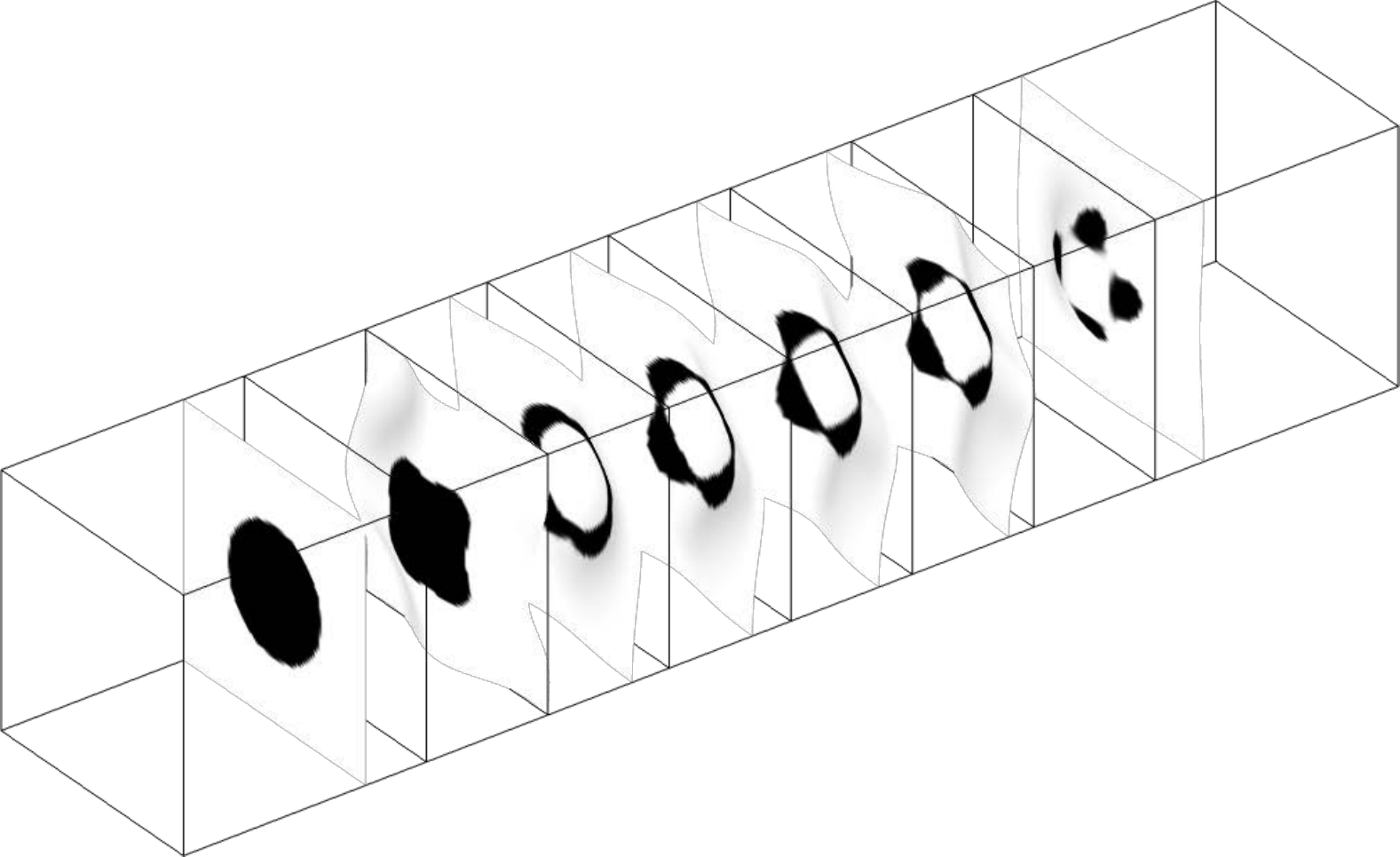}
        & \includegraphics[width=0.14\textwidth]{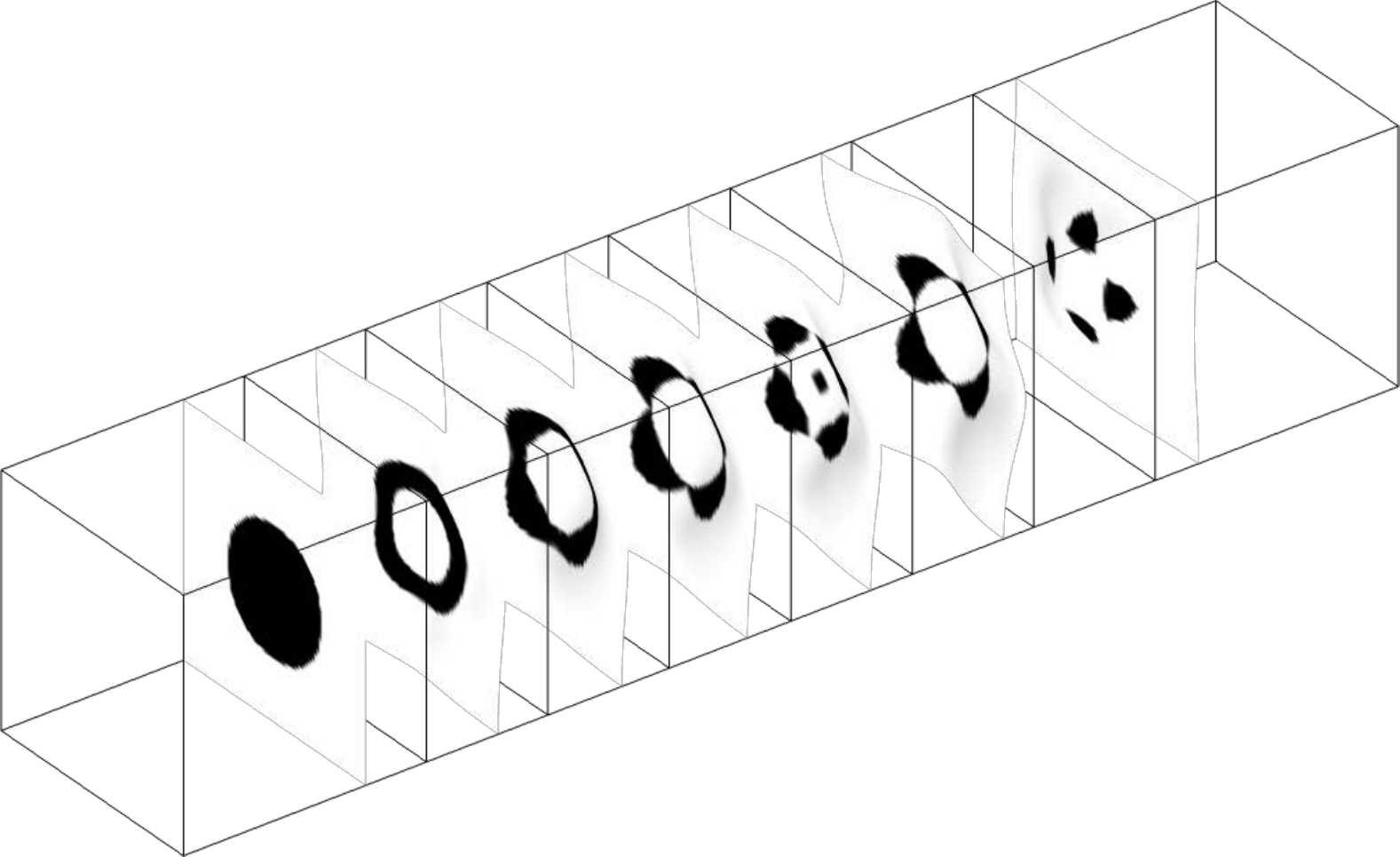}
        & \includegraphics[width=0.14\textwidth]{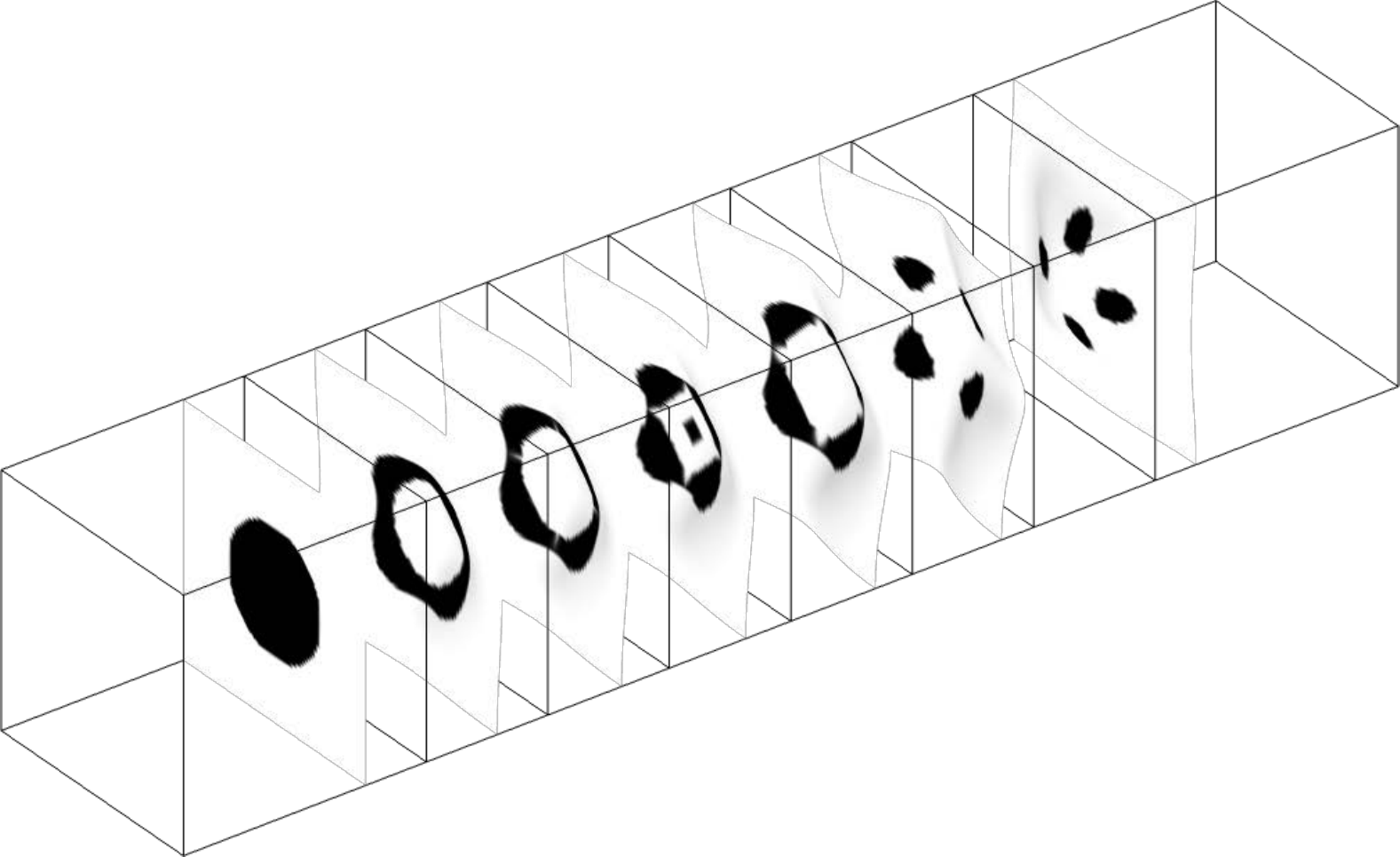} \\
   & $Pe = 1\times10^2$ & $Pe = 2\times10^2$ & $Pe = 3\times10^2$ & $Pe = 4\times10^2$ & $Pe = 5\times10^2$ \\     
  \midrule
  $Pe = 1\times10^2$ & $\mathbf{1.0359}$ & $1.0409$ & $1.0637$ & $1.0891$ & $1.1244$ \\
  \midrule
  $Pe = 2\times10^2$ & $1.4514$ & $\mathbf{1.4219}$ & $1.4296$ & $1.4363$ & $1.4655$ \\
  \midrule
  $Pe = 3\times10^2$ & $1.8390$ & $1.7712$ & $\mathbf{1.7443}$ & $1.7532$ & $1.7624$ \\
  \midrule
  $Pe = 4\times10^2$ & $2.1612$ & $2.0609$ & $2.0148$ & $\mathbf{1.9961}$ & $2.0065$ \\
  \midrule
  $Pe = 5\times10^2$ & $2.4251$ & $2.2994$ & $2.2389$ & $2.2159$ & $\mathbf{2.2088}$ \\
  \bottomrule
\end{tabular}}
\caption{Objective values for the obtained fiber bundles in Fig. \ref{fig:MassHeatTransferImmbeddedManifold7Pe} at different P\'{e}clet numbers, where the Reynolds number is remained as $1\times10^0$.}\label{tab:MassHeatTransferManifold3VolumeFlowOptimality}
\end{table}

\section{Conclusions}\label{sec:Conclusions}

Fiber bundle topology optimization for mass and heat transfer in surface and volume flow has been developed to find the optimized matching between the pattern of a surface structure and the implicit 2-manifold on which the surface pattern is defined. It can be regarded as topology optimization of surface structures for mass and heat transfer in fluid flow implemented on the variable design domains. Topology optimization for mass and heat transfer is thereby extended onto 2-manifolds with increased design freedom by including the design domains into the design space. Two sets of design variables are defined for the pattern of the surface structure and the implicit 2-manifold used to define the surface pattern, where the implicit 2-manifold is defined and evolved on the base manifold by using a differentiable homeomorphism. Two surface-PDE filters are used to regularize the design variables. The tangential gradient operator and the unit normal vector on the implicit 2-manifold are transformed based on the filtered design variable of the implicit 2-manifold. Transformed forms are hence derived for the weak forms of the surface-PDE filter of the implicit 2-manifold and the governing equations for mass and heat transfer in the surface and volume flow. Stabilized weak formulations for the governing equations of mass and heat transfer in the surface and volume flow have been provided for the linear finite elements based numerical solutions. Especially, the Laplace's equation is used to describe the deformation of the three dimensional domain of the volume flow, where the deformation is caused by the differentiable homeomorphism between the implicit 2-manifold and the base manifold. Continuous adjoint method has been used to analyze the fiber bundle topology optimization problems. In the numerical implementation, scaling factor based equivalent transformation of the constraints in the fiber bundle topology optimization problems is developed to scale the adjoint sensitivities of the constraints and ensure the robust satisfication of the constraints in the gradient based iterative procedures, by keeping the adjoint sensitivities of the constraints possess the same magnitude as that of the design objectives.

For the surface flow, fiber bundle topology optimization for mass and heat transfer has been developed to optimize the matching between the implicit 2-manifold and the pattern of the surface flow, where the mass and heat transfer processes in the surface flow are described by the surface Navier-Stokes equations, the surface convection-diffusion equation and the surface convective heat-transfer equation defined on the implicit 2-manifold. The material distribution method is used to implement topology evolution of the surface pattern, where an artificial Darcy friction force of the porous model is added to the surface Navier-Stokes equations. Numerical examples implemented on a series of curved surfaces obtained by deforming a flat surface into cylindrical surfaces have been presented to demonstrate fiber bundle topology optimization for mass and heat transfer in the surface flow, where the lengths of the inlet, the outlet and the wall boundaries are maintained to be unchanged during the deformation. The desired performance of the surface structure is set to achieve the anticipated distribution of the concentration at the outlet and minimize the thermal compliance in the surface flow for mass and heat transfer processes, respectively. The fiber bundles are derived with the topology of the zig-zag shaped curved-channel and the splitting-merging shaped curved-channel for the mass and heat transfer processes, respectively.

For the volume flow, fiber bundle topology optimization for mass and heat transfer has been developed to optimize the matching between the implicit 2-manifold defined on the cross-sections of the fluid channel and the pattern of the thin walls embedded in the fluid channel, where the mass and heat transfer processes in the volume flow are described by the Navier-Stokes equations, the convection-diffusion equation and the convective heat-transfer equation defined on the thee dimensional domain occupied by the fluid channel. The mixed boundary condition of the Navier-Stokes equations is used to implement fiber bundle topology optimization and it is interpolated by the material density used to represent the pattern of thin walls defined on the implicit 2-manifold. Numerical examples implemented on a series of cross-sections of a straight channel have been presented to demonstrate fiber bundle topology optimization for mass and heat transfer in the volume flow. The desired performance of thin walls is set to achieve the anticipated distribution of the concentration at the outlet of the fluid channel and minimize the thermal compliance in the volume flow for mass and heat transfer processes, respectively. Fiber bundles are derived with the topology of strengthening the secondary flow in the cross-sections of the channel to enhance the efficiencies of mass and heat transfer.

For both of the surface and volume flow, the variable amplitude of the implicit 2-manifold, the Reynolds number, the P\'{e}clet number, and the pressure drop or the dissipation power have been investigated in the numerical examples. The fiber bundle topology optimization problem can degenerate into the topology optimization problem for mass and heat transfer on a fixed 2-manifold, if the variable magnitude of the implicit 2-manifold is set to be zero. Increasing the value of the variable amplitude can enlarge the design space of the surface structures. However, the variable amplitude should be set reasonably to avoid its excessive value caused problems on numerical accuracy and divergence of the related finite element solution, because the non-zero value of the variable magnitude gives rise to the distortion of the mapped meshes on the implicit 2-manifold and that of the mapped meshes in the three dimensional domain. Convection of the surface and volume flow can be strengthened by setting reasonable values of the Reynolds number, the P\'{e}clet number, and the pressure drop or the dissipation power to improve the desired performance of the surface structures derived in fiber bundle topology optimization. Especially, the vortex based mixing mode appears on the topologically optimized fiber bundle for mass transfer in the surface flow, when the pressure drop is set to be large enough on the extended design space permitted by the variable amplitude of the implicit 2-manifold. 
Experimental verifications of the topologically optimized fiber bundles and the vortex based mixing mode for mass transfer in the surface flow will be implemented in the future research.

\section{Acknowledgements}\label{sec:Acknowledgements}

This research was supported by an EU2020 FET grant (TiSuMR, 737043), the DFG under grant KO 1883/20-1 Metacoils, funding within their framework of the German Excellence Intitiative under grant EXC 2082 "3D Matter Made to Order", and from the VirtMat initiative "Virtual Materials Design". The authors are also grateful to Prof. K. Svanberg of KTH for supplying the codes for the method of moving asymptotes.

\section{Appendix}\label{sec:AppendixMHM} 

This section provides the details for the adjoint analysis of the fiber bundle topology optimization problems in Eqs. \ref{equ:VarProToopSurfaceNSCD}, \ref{equ:VarProToopSurfaceNSCHM}, \ref{equ:VarProToopBulkNSCDMHT} and \ref{equ:VarProToopBulkNSCHTMHT}.

\subsection{Adjoint analysis for design objective in Eq. \ref{equ:VarProToopSurfaceNSCD}} \label{sec:AdjointAnalysisDesignObjectiveSurfaceCDMHM}

Based on the transformed design objective in Eq. \ref{equ:FurtherTransformedDesignObjectiveCD}, the variational formulations of the surface-PDE filters in Eqs. \ref{equ:VariationalFormulationPDEFilterBaseManifoldMHM} and \ref{equ:VariationalFormulationPDEFilterMHM}, the surface Navier-Stokes equations in Eq. \ref{equ:TransformedVariationalFormulationSurfaceNSEqusCD} and the surface convection-diffusion equation in Eq. \ref{equ:TransformedVariationalFormulationSurfaceCDEqu},  the augmented Lagrangian of the design objective in Eq. \ref{equ:VarProToopSurfaceNSCD} can be derived as
\begin{equation}\label{equ:AugmentedLagrangianMatchOptimizationCD}
\begin{split}
  \hat{J}_c = & \int_{l_{s,\Sigma}} \left( c - \bar{c} \right)^2 L^{\left( d_f \right)} \,\mathrm{d}l_{\partial\Sigma} \bigg/ \int_{l_{v,\Sigma}} \left( c_0 - \bar{c} \right)^2 L^{\left( d_f \right)} \,\mathrm{d}l_{\partial\Sigma} + \int_\Sigma \bigg[ \rho \left( \mathbf{u} \cdot \nabla_\Gamma^{\left(d_f\right)} \right) \mathbf{u} \cdot \mathbf{u}_a \\
  & + {\eta\over2} \left( \nabla_\Gamma^{\left(d_f\right)} \mathbf{u} + \nabla_\Gamma^{\left(d_f\right)} \mathbf{u}^\mathrm{T} \right) : \left( \nabla_\Gamma^{\left(d_f\right)} \mathbf{u}_a + \nabla_\Gamma^{\left(d_f\right)} \mathbf{u}_a^\mathrm{T} \right) - p \, \mathrm{div}_\Gamma^{\left( d_f \right)} \mathbf{u}_a - p_a \mathrm{div}_\Gamma^{\left( d_f \right)} \mathbf{u} \\
  & + \alpha \mathbf{u} \cdot \mathbf{u}_a + \lambda \mathbf{u}_a \cdot \mathbf{n}_\Gamma^{\left( d_f \right)} + \lambda_a \mathbf{u} \cdot \mathbf{n}_\Gamma^{\left( d_f \right)} \bigg] M^{\left( d_f \right)} \,\mathrm{d}\Sigma - \sum_{E_\Sigma\in\mathcal{E}_\Sigma} \int_{E_\Sigma} \tau_{BP,\Gamma}^{\left( d_f \right)} \nabla_\Gamma^{\left( d_f \right)} p \\
  & \cdot \nabla_\Gamma^{\left( d_f \right)} p_a M^{\left( d_f \right)} \,\mathrm{d}\Sigma + \int_\Gamma \left[ \left( \mathbf{u} \cdot \nabla_\Gamma^{\left(d_f\right)} c \right) c_a + D \nabla_\Gamma^{\left(d_f\right)} c \cdot \nabla_\Gamma^{\left(d_f\right)} c_a \right] M^{\left( d_f \right)} \,\mathrm{d}\Sigma \\
  & + \sum_{E_\Sigma\in\mathcal{E}_\Sigma} \int_{E_\Sigma} \tau_{PG,\Gamma}^{\left( d_f \right)} \left( \mathbf{u} \cdot \nabla_\Gamma^{\left( d_f \right)} c \right) \left( \mathbf{u} \cdot \nabla_\Gamma^{\left( d_f \right)} c_a \right) M^{\left( d_f \right)} \,\mathrm{d}\Sigma + \int_\Sigma f_{id,\Gamma} \\
  & \left( r_f^2 \nabla_\Gamma^{\left( d_f \right)} \gamma_f \cdot \nabla_\Gamma^{\left( d_f \right)} \gamma_{fa} + \gamma_f \gamma_{fa} - \gamma \gamma_{fa} \right) M^{\left( d_f \right)} + r_m^2 \nabla_\Sigma d_f \cdot \nabla_\Sigma d_{fa} \\
  & + d_f d_{fa} - A_d \left( d_m - {1\over2} \right) d_{fa} \,\mathrm{d}\Sigma, \\
\end{split}
\end{equation}
where $\hat{J}_c$ is the augmented lagrangian of $J_c$; and the adjoint variables satisfy
\begin{equation}\label{equ:ConstraintForAugmentedLagrangian}
  \left.\begin{split}
  & \mathbf{u}_a \in\left(\mathcal{H}\left(\Sigma\right)\right)^3 \\
  & p_a \in \mathcal{H}\left(\Sigma\right) \\
  & \lambda_a \in \mathcal{L}^2\left(\Sigma\right) \\
  & c_a \in \mathcal{H}\left(\Sigma\right) \\
  & \gamma_{fa} \in \mathcal{H}\left(\Sigma_D\right) \\
  & d_{fa} \in \mathcal{H}\left(\Sigma\right)
  \end{split}\right\}~\mathrm{with}~
  \left\{\begin{split}
  & \mathbf{u}_a = \mathbf{0}~ \mathrm{at} ~ \forall \mathbf{x}_\Sigma \in l_{v,\Sigma} \cup l_{v_0,\Sigma} \\
  & \lambda_a = 0~ \mathrm{at} ~ \forall \mathbf{x}_\Sigma \in l_{v,\Sigma} \cup l_{v_0,\Sigma} \\
  & c_a = 0~ \mathrm{at} ~ \forall \mathbf{x}_\Sigma \in l_{v,\Sigma} 
  \end{split}\right..
\end{equation}

Based on the transformed operators in Eqs. \ref{equ:TransformedTangentialOperatorMHM} and \ref{equ:TransformedDivergenceOperatorMHM} and their first order variationals in Eqs. \ref{equ:FirstOrderVarisForTangentialOperatorsMHM} and \ref{equ:FirstOrderVarisForDivergenceOperatorsMHM}, together with the first order variational of the 2-norm of a vector function
\begin{equation}\label{equ:FirstOrderVariOfVectorNormMHM}
\begin{split}
  \delta \left( \left\| \mathbf{f} \right\|_2 \right)^2
  = 2 \left\| \mathbf{f} \right\|_2 \delta \left\| \mathbf{f} \right\|_2 = \delta \mathbf{f}^2 = 2 \mathbf{f} \cdot \delta \mathbf{f} \Longrightarrow \delta \left\| \mathbf{f} \right\|_2
  = { \mathbf{f} \over \left\| \mathbf{f} \right\|_2 } \cdot \delta  \mathbf{f}
\end{split}
\end{equation}
with $\mathbf{f}$ representing the vector function, the first order variational of the augmented Lagrangian in Eq. \ref{equ:AugmentedLagrangianMatchOptimizationCD} can be derived as
\begin{equation}\label{equ:1stVarAugmentedLagrangianMatchOptimization}
\begin{split}
  \delta \hat{J}_c = & \int_{l_{s,\Sigma}} \left[ 2 \left( c - \bar{c} \right) L^{\left( d_f \right)} \delta c + \left( c - \bar{c} \right)^2 L^{\left( d_f, \delta d_f \right)} \right] \,\mathrm{d}l_{\partial\Sigma} \bigg/ \int_{l_{v,\Sigma}} \left( c_0 - \bar{c} \right)^2 L^{\left( d_f \right)} \,\mathrm{d}l_{\partial\Sigma} \\
  & - \int_{l_{s,\Sigma}} \left( c - \bar{c} \right)^2 L^{\left( d_f \right)} \, \mathrm{d}l_{\partial\Sigma} \int_{l_{v,\Sigma}} \left( c_0 - \bar{c} \right)^2 L^{\left( d_f, \delta d_f \right)} \,\mathrm{d}l_{\partial\Sigma} \Bigg/ \left( \int_{l_{v,\Sigma}} \left( c_0 - \bar{c} \right)^2 L^{\left( d_f \right)} \,\mathrm{d}l_{\partial\Sigma} \right)^2 \\
  & + \int_\Sigma \bigg[ \rho \left( \delta \mathbf{u} \cdot \nabla_\Gamma^{\left(d_f\right)} \right) \mathbf{u} \cdot \mathbf{u}_a + \rho \left( \mathbf{u} \cdot \nabla_\Gamma^{\left(d_f\right)} \right) \delta \mathbf{u} \cdot \mathbf{u}_a + \rho \left( \mathbf{u} \cdot \nabla_\Gamma^{\left(d_f, \delta d_f\right)} \right) \mathbf{u} \cdot \mathbf{u}_a \\
  & + {\eta\over2} \left( \nabla_\Gamma^{\left(d_f\right)} \delta \mathbf{u} + \nabla_\Gamma^{\left(d_f\right)} \delta \mathbf{u}^\mathrm{T} \right) : \left( \nabla_\Gamma^{\left(d_f\right)} \mathbf{u}_a + \nabla_\Gamma^{\left(d_f\right)} \mathbf{u}_a^\mathrm{T} \right) + {\eta\over2} \left( \nabla_\Gamma^{\left(d_f, \delta d_f\right)} \mathbf{u} + \nabla_\Gamma^{\left(d_f, \delta d_f\right)} \mathbf{u}^\mathrm{T} \right) \\
  & : \left( \nabla_\Gamma^{\left(d_f\right)} \mathbf{u}_a + \nabla_\Gamma^{\left(d_f\right)} \mathbf{u}_a^\mathrm{T} \right) + {\eta\over2} \left( \nabla_\Gamma^{\left(d_f\right)} \mathbf{u} + \nabla_\Gamma^{\left(d_f\right)} \mathbf{u}^\mathrm{T} \right) : \left( \nabla_\Gamma^{\left(d_f, \delta d_f\right)} \mathbf{u}_a + \nabla_\Gamma^{\left(d_f, \delta d_f\right)} \mathbf{u}_a^\mathrm{T} \right) \\
  & - \delta p \, \mathrm{div}_\Gamma^{\left( d_f \right)} \mathbf{u}_a - p \, \mathrm{div}_\Gamma^{\left( d_f, \delta d_f \right)} \mathbf{u}_a - p_a \mathrm{div}_\Gamma^{\left( d_f, \delta d_f \right)} \mathbf{u} - p_a \mathrm{div}_\Gamma^{\left( d_f \right)} \delta \mathbf{u} + \alpha \delta \mathbf{u} \cdot \mathbf{u}_a \\
  & + {\partial\alpha \over \partial \gamma_p} {\partial \gamma_p \over \partial \gamma_f} \mathbf{u} \cdot \mathbf{u}_a \delta \gamma_f + \left( \lambda \mathbf{u}_a + \lambda_a \mathbf{u} \right) \cdot \mathbf{n}_\Gamma^{\left( d_f, \delta d_f \right)} + \left( \delta \lambda \mathbf{u}_a + \lambda_a \delta \mathbf{u} \right) \cdot \mathbf{n}_\Gamma^{\left( d_f \right)} \bigg] M^{\left( d_f \right)} \\
  & + \bigg[ \rho \left( \mathbf{u} \cdot \nabla_\Gamma^{\left(d_f\right)} \right) \mathbf{u} \cdot \mathbf{u}_a + {\eta\over2} \left( \nabla_\Gamma^{\left(d_f\right)} \mathbf{u} + \nabla_\Gamma^{\left(d_f\right)} \mathbf{u}^\mathrm{T} \right) : \left( \nabla_\Gamma^{\left(d_f\right)} \mathbf{u}_a + \nabla_\Gamma^{\left(d_f\right)} \mathbf{u}_a^\mathrm{T} \right) \\
  & - p \, \mathrm{div}_\Gamma^{\left( d_f \right)} \mathbf{u}_a - p_a \mathrm{div}_\Gamma^{\left( d_f \right)} \mathbf{u} + \alpha \mathbf{u} \cdot \mathbf{u}_a + \lambda \mathbf{u}_a \cdot \mathbf{n}_\Gamma^{\left( d_f \right)} + \lambda_a \mathbf{u} \cdot \mathbf{n}_\Gamma^{\left( d_f \right)} \bigg] M^{\left( d_f, \delta d_f \right)} \,\mathrm{d}\Sigma \\
  & - \sum_{E_\Sigma\in\mathcal{E}_\Sigma} \int_{E_\Sigma} \tau_{BP,\Gamma}^{\left( d_f, \delta d_f \right)} \nabla_\Gamma^{\left( d_f \right)} p \cdot \nabla_\Gamma^{\left( d_f \right)} p_a M^{\left( d_f \right)} + \tau_{BP,\Gamma}^{\left( d_f \right)} \nabla_\Gamma^{\left( d_f \right)} \delta p \cdot \nabla_\Gamma^{\left( d_f \right)} p_a M^{\left( d_f \right)} \\
  & + \tau_{BP,\Gamma}^{\left( d_f \right)} \nabla_\Gamma^{\left( d_f, \delta d_f \right)} p \cdot \nabla_\Gamma^{\left( d_f \right)} p_a M^{\left( d_f \right)} + \tau_{BP,\Gamma}^{\left( d_f \right)} \nabla_\Gamma^{\left( d_f \right)} p \cdot \nabla_\Gamma^{\left( d_f, \delta d_f \right)} p_a M^{\left( d_f \right)} \\
  & + \tau_{BP,\Gamma}^{\left( d_f \right)} \nabla_\Gamma^{\left( d_f \right)} p \cdot \nabla_\Gamma^{\left( d_f \right)} p_a M^{\left( d_f, \delta d_f \right)} \,\mathrm{d}\Sigma + \int_\Sigma \bigg[ \left( \delta \mathbf{u} \cdot \nabla_\Gamma^{\left(d_f\right)} c \right) c_a + \left( \mathbf{u} \cdot \nabla_\Gamma^{\left(d_f\right)} \delta c \right) c_a \\
  & + \left( \mathbf{u} \cdot \nabla_\Gamma^{\left(d_f, \delta d_f\right)} c \right) c_a + D \nabla_\Gamma^{\left(d_f\right)} \delta c \cdot \nabla_\Gamma^{\left(d_f\right)} c_a + D \nabla_\Gamma^{\left(d_f, \delta d_f \right)} c \cdot \nabla_\Gamma^{\left(d_f\right)} c_a + D \nabla_\Gamma^{\left(d_f\right)} c \\
  & \cdot \nabla_\Gamma^{\left(d_f, \delta d_f\right)} c_a  \bigg] M^{\left( d_f \right)} + \left[ \left( \mathbf{u} \cdot \nabla_\Gamma^{\left(d_f\right)} c \right) c_a + D \nabla_\Gamma^{\left(d_f\right)} c \cdot \nabla_\Gamma^{\left(d_f\right)} c_a \right] M^{\left( d_f, \delta d_f \right)} \,\mathrm{d}\Sigma \\
  & + \sum_{E_\Sigma\in\mathcal{E}_\Sigma} \int_{E_\Sigma} \left( \tau_{PG,\Gamma}^{\left( d_f, \delta d_f \right)} + \tau_{PG,\Gamma}^{\left( d_f, \delta \mathbf{u} \right)} \right) \left( \mathbf{u} \cdot \nabla_\Gamma^{\left( d_f \right)} c \right) \left( \mathbf{u} \cdot \nabla_\Gamma^{\left( d_f \right)} c_a \right) M^{\left( d_f \right)} \\
  & + \tau_{PG,\Gamma}^{\left( d_f \right)} \left( \delta \mathbf{u} \cdot \nabla_\Gamma^{\left( d_f \right)} c + \mathbf{u} \cdot \nabla_\Gamma^{\left( d_f, \delta d_f \right)} c + \mathbf{u} \cdot \nabla_\Gamma^{\left( d_f \right)} \delta c \right) \left( \mathbf{u} \cdot \nabla_\Gamma^{\left( d_f \right)} c_a \right) M^{\left( d_f \right)} \\
  & + \tau_{PG,\Gamma}^{\left( d_f \right)} \left( \mathbf{u} \cdot \nabla_\Gamma^{\left( d_f \right)} c \right) \left( \delta \mathbf{u} \cdot \nabla_\Gamma^{\left( d_f \right)} c_a + \mathbf{u} \cdot \nabla_\Gamma^{\left( d_f, \delta d_f \right)} c_a \right) M^{\left( d_f \right)} + \tau_{PG,\Gamma}^{\left( d_f \right)} \left( \mathbf{u} \cdot \nabla_\Gamma^{\left( d_f \right)} c \right) \\
  & \left( \mathbf{u} \cdot \nabla_\Gamma^{\left( d_f \right)} c_a \right) M^{\left( d_f, \delta d_f \right)} \,\mathrm{d}\Sigma + \int_\Sigma f_{id,\Gamma} \bigg[ r_f^2 \nabla_\Gamma^{\left( d_f \right)} \delta \gamma_f \cdot \nabla_\Gamma^{\left( d_f \right)} \gamma_{fa} + \delta \gamma_f \gamma_{fa} - \delta \gamma \gamma_{fa} \\
  & + r_f^2 \left( \nabla_\Gamma^{\left( d_f, \delta d_f \right)} \gamma_f \cdot \nabla_\Gamma^{\left( d_f \right)} \gamma_{fa} + \nabla_\Gamma^{\left( d_f \right)} \gamma_f \cdot \nabla_\Gamma^{\left( d_f, \delta d_f \right)} \gamma_{fa} \right) \bigg] M^{\left( d_f \right)} + f_{id,\Gamma} \bigg( r_f^2 \nabla_\Gamma^{\left( d_f \right)} \gamma_f \\
  & \cdot \nabla_\Gamma^{\left( d_f \right)} \gamma_{fa} + \gamma_f \gamma_{fa} - \gamma \gamma_{fa} \bigg) M^{\left( d_f, \delta d_f \right)} + r_m^2 \nabla_\Sigma \delta d_f \cdot \nabla_\Sigma d_{fa} + \delta d_f d_{fa} - A_d \delta d_m d_{fa} \,\mathrm{d}\Sigma \\
\end{split}
\end{equation}
with the satisfication of the constraints in Eq. \ref{equ:ConstraintForAugmentedLagrangian}
and
\begin{equation}\label{equ:ConstraintForVariationalAugmentedLagrangian}
  \left.\begin{split}
  & \delta \mathbf{u} \in\left(\mathcal{H}\left(\Sigma\right)\right)^3 \\
  & \delta p \in \mathcal{H}\left(\Sigma\right) \\
  & \delta \lambda \in \mathcal{L}^2\left(\Sigma\right) \\
  & \delta c \in \mathcal{H}\left(\Sigma\right) \\
  & \delta \gamma_f \in \mathcal{H}\left(\Sigma_D\right) \\
  & \delta \gamma \in \mathcal{L}^2\left(\Sigma_D\right) \\
  & \delta d_f \in \mathcal{H}\left(\Sigma\right) \\
  & \delta d_m \in \mathcal{L}^2\left(\Sigma\right)
  \end{split}\right\} 
  ~ \mathrm{with} ~ 
  \left\{\begin{split}
  & \delta \mathbf{u} = \mathbf{0}~ \mathrm{at} ~ \forall \mathbf{x}_\Sigma \in l_{v,\Sigma} \cup l_{v_0,\Sigma} \\
  & \delta \lambda = 0~ \mathrm{at} ~ \forall \mathbf{x}_\Sigma \in l_{v,\Sigma} \cup l_{v_0,\Sigma} \\
  & \delta c = 0~ \mathrm{at} ~ \forall \mathbf{x}_\Sigma \in l_{v,\Sigma}
  \end{split}\right.,
\end{equation}
where $\delta$ is the first-order variational operator.

According to the Karush-Kuhn-Tucker conditions of the PDE constrained optimization problem \cite{HinzeSpringer2009}, the first order variational of the augmented Lagrangian to $c$ can be set to be zero as
\begin{equation}\label{equ:WeakAdjEquSCDEqu}
\begin{split}
  & \int_{l_{s,\Sigma}} 2 \left( c - \bar{c} \right) L^{\left( d_f \right)} \delta c \,\mathrm{d}l_{\partial\Sigma} \bigg/ \int_{l_{v,\Sigma}} \left( c_0 - \bar{c} \right)^2 L^{\left( d_f \right)} \,\mathrm{d}l_{\partial\Sigma} \\
  & + \int_\Sigma \left[ \left( \mathbf{u} \cdot \nabla_\Gamma^{\left(d_f\right)} \delta c \right) c_a + D \nabla_\Gamma^{\left(d_f\right)} \delta c \cdot \nabla_\Gamma^{\left(d_f\right)} c_a \right] M^{\left( d_f \right)} \,\mathrm{d}\Sigma \\
  & + \sum_{E_\Sigma\in\mathcal{E}_\Sigma} \int_{E_\Sigma} \tau_{PG,\Gamma}^{\left( d_f \right)} \left( \mathbf{u} \cdot \nabla_\Gamma^{\left( d_f \right)} \delta c \right) \left( \mathbf{u} \cdot \nabla_\Gamma^{\left( d_f \right)} c_a \right) M^{\left( d_f \right)} \,\mathrm{d}\Sigma = 0, \\
\end{split}
\end{equation}
the first order variational of the augmented Lagrangian to $\mathbf{u}$, $p$ and $\lambda$ can be set to be zero as
\begin{equation}\label{equ:WeakAdjEquSNSEqu}
\begin{split}
  & \int_\Sigma \bigg[ \rho \left( \delta \mathbf{u} \cdot \nabla_\Gamma^{\left(d_f\right)} \right) \mathbf{u} \cdot \mathbf{u}_a + \rho \left( \mathbf{u} \cdot \nabla_\Gamma^{\left(d_f\right)} \right) \delta \mathbf{u} \cdot \mathbf{u}_a + {\eta\over2} \left( \nabla_\Gamma^{\left(d_f\right)} \delta \mathbf{u} + \nabla_\Gamma^{\left(d_f\right)} \delta \mathbf{u}^\mathrm{T} \right) \\
  & : \left( \nabla_\Gamma^{\left(d_f\right)} \mathbf{u}_a + \nabla_\Gamma^{\left(d_f\right)} \mathbf{u}_a^\mathrm{T} \right) - \delta p \, \mathrm{div}_\Gamma^{\left( d_f \right)} \mathbf{u}_a - p_a \mathrm{div}_\Gamma^{\left( d_f \right)} \delta \mathbf{u} + \alpha \delta \mathbf{u} \cdot \mathbf{u}_a \\
  & + \left( \delta \lambda \mathbf{u}_a + \lambda_a \delta \mathbf{u} \right) \cdot \mathbf{n}_\Gamma^{\left( d_f \right)} + \left( \delta \mathbf{u} \cdot \nabla_\Gamma^{\left(d_f\right)} c \right) c_a \bigg] M^{\left( d_f \right)} \,\mathrm{d}\Sigma \\
  & + \sum_{E_\Sigma\in\mathcal{E}_\Sigma} \int_{E_\Sigma} \bigg[ - \tau_{BP,\Gamma}^{\left( d_f \right)} \nabla_\Gamma^{\left( d_f \right)} \delta p \cdot \nabla_\Gamma^{\left( d_f \right)} p_a + \tau_{PG,\Gamma}^{\left( d_f, \delta \mathbf{u} \right)} \left( \mathbf{u} \cdot \nabla_\Gamma^{\left( d_f \right)} c \right) \left( \mathbf{u} \cdot \nabla_\Gamma^{\left( d_f \right)} c_a \right) \\
  & + \tau_{PG,\Gamma}^{\left( d_f \right)} \left( \delta \mathbf{u} \cdot \nabla_\Gamma^{\left( d_f \right)} c \right) \left( \mathbf{u} \cdot \nabla_\Gamma^{\left( d_f \right)} c_a \right) + \tau_{PG,\Gamma}^{\left( d_f \right)} \left( \mathbf{u} \cdot \nabla_\Gamma^{\left( d_f \right)} c \right) \\
  & \left( \delta \mathbf{u} \cdot \nabla_\Gamma^{\left( d_f \right)} c_a \right) \bigg] M^{\left( d_f \right)} \,\mathrm{d}\Sigma = 0, \\
\end{split}
\end{equation}
the first order variational of the augmented Lagrangian to $\gamma_f$ can be set to be zero as
\begin{equation}\label{equ:WeakAdjEquSPDEFilterGa}
\begin{split}
  & \int_{\Sigma_D} \left( {\partial\alpha \over \partial \gamma_p} {\partial \gamma_p \over \partial \gamma_f} \mathbf{u} \cdot \mathbf{u}_a \delta \gamma_f + r_f^2 \nabla_\Gamma^{\left( d_f \right)} \delta \gamma_f \cdot \nabla_\Gamma^{\left( d_f \right)} \gamma_{fa} + \delta \gamma_f \gamma_{fa} \right) M^{\left( d_f \right)} \,\mathrm{d}\Sigma = 0, \\
\end{split}
\end{equation}
and the first order variational of the augmented Lagrangian to $d_f$ can be set to be zero as
\begin{equation}\label{equ:WeakAdjEquSPDEFilterDm}
\begin{split}
  & \int_{l_{s,\Sigma}} \left( c - \bar{c} \right)^2 L^{\left( d_f, \delta d_f \right)} \,\mathrm{d}l_{\partial\Sigma} \bigg/ \int_{l_{v,\Sigma}} \left( c_0 - \bar{c} \right)^2 L^{\left( d_f \right)} \,\mathrm{d}l_{\partial\Sigma} - \int_{l_{s,\Sigma}} \left( c - \bar{c} \right)^2 L^{\left( d_f \right)} \, \mathrm{d}l_{\partial\Sigma} \\
  & \int_{l_{v,\Sigma}} \left( c_0 - \bar{c} \right)^2 L^{\left( d_f, \delta d_f \right)} \,\mathrm{d}l_{\partial\Sigma} \Bigg/ \left( \int_{l_{v,\Sigma}} \left( c_0 - \bar{c} \right)^2 L^{\left( d_f \right)} \,\mathrm{d}l_{\partial\Sigma} \right)^2 + \int_\Sigma \bigg[ \rho \left( \mathbf{u} \cdot \nabla_\Gamma^{\left(d_f, \delta d_f\right)} \right) \mathbf{u} \\
  & \cdot \mathbf{u}_a + {\eta\over2} \left( \nabla_\Gamma^{\left(d_f, \delta d_f\right)} \mathbf{u} + \nabla_\Gamma^{\left(d_f, \delta d_f\right)} \mathbf{u}^\mathrm{T} \right) : \left( \nabla_\Gamma^{\left(d_f\right)} \mathbf{u}_a + \nabla_\Gamma^{\left(d_f\right)} \mathbf{u}_a^\mathrm{T} \right) \\
  & + {\eta\over2} \left( \nabla_\Gamma^{\left(d_f\right)} \mathbf{u} + \nabla_\Gamma^{\left(d_f\right)} \mathbf{u}^\mathrm{T} \right) : \left( \nabla_\Gamma^{\left(d_f, \delta d_f\right)} \mathbf{u}_a + \nabla_\Gamma^{\left(d_f, \delta d_f\right)} \mathbf{u}_a^\mathrm{T} \right) - p \, \mathrm{div}_\Gamma^{\left( d_f, \delta d_f \right)} \mathbf{u}_a \\
  & - p_a \mathrm{div}_\Gamma^{\left( d_f, \delta d_f \right)} \mathbf{u} + \left( \lambda \mathbf{u}_a + \lambda_a \mathbf{u} \right) \cdot \mathbf{n}_\Gamma^{\left( d_f, \delta d_f \right)} \bigg] M^{\left( d_f \right)} + \bigg[ \rho \left( \mathbf{u} \cdot \nabla_\Gamma^{\left(d_f\right)} \right) \mathbf{u} \cdot \mathbf{u}_a \\
  & + {\eta\over2} \left( \nabla_\Gamma^{\left(d_f\right)} \mathbf{u} + \nabla_\Gamma^{\left(d_f\right)} \mathbf{u}^\mathrm{T} \right) : \left( \nabla_\Gamma^{\left(d_f\right)} \mathbf{u}_a + \nabla_\Gamma^{\left(d_f\right)} \mathbf{u}_a^\mathrm{T} \right) - p \, \mathrm{div}_\Gamma^{\left( d_f \right)} \mathbf{u}_a \\
  & - p_a \mathrm{div}_\Gamma^{\left( d_f \right)} \mathbf{u} + \alpha \mathbf{u} \cdot \mathbf{u}_a + \lambda \mathbf{u}_a \cdot \mathbf{n}_\Gamma^{\left( d_f \right)} + \lambda_a \mathbf{u} \cdot \mathbf{n}_\Gamma^{\left( d_f \right)} \bigg] M^{\left( d_f, \delta d_f \right)} \\
  & + \bigg[ \left( \mathbf{u} \cdot \nabla_\Gamma^{\left(d_f, \delta d_f\right)} c \right) c_a + D \nabla_\Gamma^{\left(d_f, \delta d_f \right)} c \cdot \nabla_\Gamma^{\left(d_f\right)} c_a + D \nabla_\Gamma^{\left(d_f\right)} c \cdot \nabla_\Gamma^{\left(d_f, \delta d_f\right)} c_a  \bigg] M^{\left( d_f \right)} \\
  & + \left[ \left( \mathbf{u} \cdot \nabla_\Gamma^{\left(d_f\right)} c \right) c_a + D \nabla_\Gamma^{\left(d_f\right)} c \cdot \nabla_\Gamma^{\left(d_f\right)} c_a \right] M^{\left( d_f, \delta d_f \right)} \\
  & + f_{id,\Gamma} r_f^2 \left( \nabla_\Gamma^{\left( d_f, \delta d_f \right)} \gamma_f \cdot \nabla_\Gamma^{\left( d_f \right)} \gamma_{fa} + \nabla_\Gamma^{\left( d_f \right)} \gamma_f \cdot \nabla_\Gamma^{\left( d_f, \delta d_f \right)} \gamma_{fa} \right) M^{\left( d_f \right)} \\
  & + f_{id,\Gamma} \left( r_f^2 \nabla_\Gamma^{\left( d_f \right)} \gamma_f \cdot \nabla_\Gamma^{\left( d_f \right)} \gamma_{fa} + \gamma_f \gamma_{fa} - \gamma \gamma_{fa} \right) M^{\left( d_f, \delta d_f \right)} \\
  & + r_m^2 \nabla_\Sigma \delta d_f \cdot \nabla_\Sigma d_{fa} + \delta d_f d_{fa} \,\mathrm{d}\Sigma + \sum_{E_\Sigma\in\mathcal{E}_\Sigma} \int_{E_\Sigma} - \tau_{BP,\Gamma}^{\left( d_f, \delta d_f \right)} \nabla_\Gamma^{\left( d_f \right)} p \cdot \nabla_\Gamma^{\left( d_f \right)} p_a M^{\left( d_f \right)} \\
  & - \tau_{BP,\Gamma}^{\left( d_f \right)} \nabla_\Gamma^{\left( d_f, \delta d_f \right)} p \cdot \nabla_\Gamma^{\left( d_f \right)} p_a M^{\left( d_f \right)} - \tau_{BP,\Gamma}^{\left( d_f \right)} \nabla_\Gamma^{\left( d_f \right)} p \cdot \nabla_\Gamma^{\left( d_f, \delta d_f \right)} p_a M^{\left( d_f \right)} \\
  & - \tau_{BP,\Gamma}^{\left( d_f \right)} \nabla_\Gamma^{\left( d_f \right)} p \cdot \nabla_\Gamma^{\left( d_f \right)} p_a M^{\left( d_f, \delta d_f \right)} + \tau_{PG,\Gamma}^{\left( d_f, \delta d_f \right)} \left( \mathbf{u} \cdot \nabla_\Gamma^{\left( d_f \right)} c \right) \left( \mathbf{u} \cdot \nabla_\Gamma^{\left( d_f \right)} c_a \right) M^{\left( d_f \right)} \\
  & + \tau_{PG,\Gamma}^{\left( d_f \right)} \left( \mathbf{u} \cdot \nabla_\Gamma^{\left( d_f, \delta d_f \right)} c \right) \left( \mathbf{u} \cdot \nabla_\Gamma^{\left( d_f \right)} c_a \right) M^{\left( d_f \right)} + \tau_{PG,\Gamma}^{\left( d_f \right)} \left( \mathbf{u} \cdot \nabla_\Gamma^{\left( d_f \right)} c \right) \\
  & \left( \mathbf{u} \cdot \nabla_\Gamma^{\left( d_f, \delta d_f \right)} c_a \right) M^{\left( d_f \right)} + \tau_{PG,\Gamma}^{\left( d_f \right)} \left( \mathbf{u} \cdot \nabla_\Gamma^{\left( d_f \right)} c \right) \left( \mathbf{u} \cdot \nabla_\Gamma^{\left( d_f \right)} c_a \right) M^{\left( d_f, \delta d_f \right)} \,\mathrm{d}\Sigma = 0.
\end{split}
\end{equation}
The constraints in Eqs. \ref{equ:ConstraintForAugmentedLagrangian} and \ref{equ:ConstraintForVariationalAugmentedLagrangian} are imposed to Eqs. \ref{equ:WeakAdjEquSCDEqu}, \ref{equ:WeakAdjEquSNSEqu}, \ref{equ:WeakAdjEquSPDEFilterGa} and \ref{equ:WeakAdjEquSPDEFilterDm}. Then, the adjoint sensitivity of $J_c$ is derived as
\begin{equation}\label{equ:AdjSensitivityGaDmVariationalForm}
\begin{split}
\delta \hat{J}_c = - \int_{\Sigma_D} \gamma_{fa} \delta \gamma M^{\left( d_f \right)} \,\mathrm{d}\Sigma - \int_\Sigma A_d d_{fa} \delta d_m \,\mathrm{d}\Sigma.
\end{split}
\end{equation}

Without losing the arbitrariness of $\delta \mathbf{u}$, $\delta p$, $\delta \lambda$, $\delta c$, $\delta \gamma_f$, $\delta d_f$, $\delta \gamma$ and $\delta d_m$, one can set 
\begin{equation}
\left.\begin{split}
& \tilde{\mathbf{u}}_a = \delta \mathbf{u} \\
& \tilde{p}_a = \delta p \\
& \tilde{\lambda}_a = \delta \lambda \\
& \tilde{c}_a = \delta c \\
& \tilde{\gamma}_{fa} = \delta \gamma_f \\
& \tilde{d}_{fa} = \delta d_f \\
& \tilde{\gamma} = \delta \gamma \\
& \tilde{d}_m = \delta d_m 
\end{split}\right\}
~\mathrm{with}~
\left\{\begin{split}
& \forall \tilde{\mathbf{u}}_a \in \left(\mathcal{H}\left(\Sigma\right)\right)^3 \\
& \forall \tilde{p}_a \in \mathcal{H}\left(\Sigma\right)\\
& \forall \tilde{\lambda}_a \in \mathcal{L}^2\left(\Sigma\right) \\
& \forall \tilde{c}_a \in \mathcal{H}\left(\Sigma\right) \\
& \forall \tilde{\gamma}_{fa} \in \mathcal{H}\left(\Sigma_D\right) \\
& \forall \tilde{d}_{fa} \in \mathcal{H}\left(\Sigma\right)\\
& \forall \tilde{\gamma} \in \mathcal{L}^2\left(\Sigma_D\right) \\
& \forall \tilde{d}_m \in \mathcal{L}^2\left(\Sigma\right)
\end{split}\right.
\end{equation}
for Eqs. \ref{equ:WeakAdjEquSCDEqu}, \ref{equ:WeakAdjEquSNSEqu}, \ref{equ:WeakAdjEquSPDEFilterGa} and \ref{equ:WeakAdjEquSPDEFilterDm} to derive the adjoint system composed of Eqs. \ref{equ:WeakAdjEquSCDEquMHM}, \ref{equ:AdjSurfaceNavierStokesEqusJObjectiveMHM}, \ref{equ:AdjPDEFilterJObjectiveGaMHM} and \ref{equ:AdjPDEFilterJObjectiveDmMHM}.

\subsection{Adjoint analysis for constraint of pressure drop in Eq. \ref{equ:VarProToopSurfaceNSCD}} \label{sec:AdjointAnalysisPressureConstraintMHM}

Based on the variational formulations of the surface Navier-Stokes equations in Eq. \ref{equ:TransformedVariationalFormulationSurfaceNSEqusCD} and the surface-PDE filters in Eqs. \ref{equ:VariationalFormulationPDEFilterBaseManifoldMHM} and \ref{equ:VariationalFormulationPDEFilterMHM}, the augmented Lagrangian of the pressure drop $\Delta P$ in Eq. \ref{equ:TransformedPressureConstraintSurfaceNSCD} can be derived as
\begin{equation}\label{equ:AugmentedLagrangianPressureDrop}
\begin{split}
  \widehat{\Delta P} = & \int_{l_{v,\Sigma}} p L^{\left( d_f \right)} \,\mathrm{d}l_{\partial\Sigma} - \int_{l_{s,\Sigma}} p L^{\left( d_f \right)} \,\mathrm{d}l_{\partial\Sigma} \\
  & + \int_\Sigma \bigg[ \rho \left( \mathbf{u} \cdot \nabla_\Gamma^{\left(d_f\right)} \right) \mathbf{u} \cdot \mathbf{u}_a + {\eta\over2} \left( \nabla_\Gamma^{\left(d_f\right)} \mathbf{u} + \nabla_\Gamma^{\left(d_f\right)} \mathbf{u}^\mathrm{T} \right) : \left( \nabla_\Gamma^{\left(d_f\right)} \mathbf{u}_a + \nabla_\Gamma^{\left(d_f\right)} \mathbf{u}_a^\mathrm{T} \right) \\
  & - p \, \mathrm{div}_\Gamma^{\left( d_f \right)} \mathbf{u}_a - p_a \mathrm{div}_\Gamma^{\left( d_f \right)} \mathbf{u} + \alpha \mathbf{u} \cdot \mathbf{u}_a + \lambda \mathbf{u}_a \cdot \mathbf{n}_\Gamma^{\left( d_f \right)} + \lambda_a \mathbf{u} \cdot \mathbf{n}_\Gamma^{\left( d_f \right)} \\
  & + f_{id,\Gamma} \left( r_f^2 \nabla_\Gamma^{\left( d_f \right)} \gamma_f \cdot \nabla_\Gamma^{\left( d_f \right)} \gamma_{fa} + \gamma_f \gamma_{fa} - \gamma \gamma_{fa} \right) \bigg] M^{\left( d_f \right)} \\
  & + r_m^2 \nabla_\Sigma d_f \cdot \nabla_\Sigma d_{fa} + d_f d_{fa} - A_d \left( d_m - {1\over2} \right) d_{fa} \,\mathrm{d}\Sigma \\
  & - \sum_{E_\Sigma\in\mathcal{E}_\Sigma} \int_{E_\Sigma} \tau_{BP,\Gamma}^{\left( d_f \right)} \nabla_\Gamma^{\left( d_f \right)} p \cdot \nabla_\Gamma^{\left( d_f \right)} p_a M^{\left( d_f \right)} \,\mathrm{d}\Sigma, \\
\end{split}
\end{equation}
where $\widehat{\Delta P}$ is the augmented lagrangian of $\Delta P$; and the adjoint variables satisfy
\begin{equation}\label{equ:ConstraintForAugmentedLagrangianPressureDrop}
  \left.\begin{split}
  & \mathbf{u}_a \in\left(\mathcal{H}\left(\Sigma\right)\right)^3 \\
  & p_a \in \mathcal{H}\left(\Sigma\right) \\
  & \lambda_a \in \mathcal{L}^2\left(\Sigma\right) \\
  & \gamma_{fa} \in \mathcal{H}\left(\Sigma_D\right) \\
  & d_{fa} \in \mathcal{H}\left(\Sigma\right)
  \end{split}\right\}~\mathrm{with}~
  \left\{\begin{split}
  & \mathbf{u}_a = \mathbf{0} ~\mathrm{at}~ \forall \mathbf{x}_\Sigma \in l_{v,\Sigma} \cup l_{v_0,\Sigma} \\
  & \lambda_a = 0 ~\mathrm{at}~ \forall \mathbf{x}_\Sigma \in l_{v,\Sigma} \cup l_{v_0,\Sigma} 
  \end{split}\right..
\end{equation}
Based on the related results in Section \ref{sec:MethodologyFiberBundleTOOPTransferSurfaceFlows}, the first order variational of $\widehat{\Delta P}$ can be derived as
\begin{equation}\label{equ:1stVarAugmentedLagrangianConstrPressureDrop}
\begin{split}
  \delta \widehat{\Delta P} = & \int_{l_{v,\Sigma}} \delta p L^{\left( d_f \right)} + p L^{\left( d_f, \delta d_f \right)} \,\mathrm{d}l_{\partial\Sigma} - \int_{l_{s,\Sigma}} \delta p L^{\left( d_f \right)} + p L^{\left( d_f, \delta d_f \right)} \,\mathrm{d}l_{\partial\Sigma} \\
  & + \int_\Sigma \bigg[ \rho \left( \delta \mathbf{u} \cdot \nabla_\Gamma^{\left(d_f\right)} \right) \mathbf{u} \cdot \mathbf{u}_a + \rho \left( \mathbf{u} \cdot \nabla_\Gamma^{\left(d_f, \delta d_f \right)} \right) \mathbf{u} \cdot \mathbf{u}_a + \rho \left( \mathbf{u} \cdot \nabla_\Gamma^{\left(d_f\right)} \right) \delta \mathbf{u} \cdot \mathbf{u}_a  \\
  & + {\eta\over2} \left( \nabla_\Gamma^{\left(d_f\right)} \delta \mathbf{u} + \nabla_\Gamma^{\left(d_f\right)} \delta \mathbf{u}^\mathrm{T} \right) : \left( \nabla_\Gamma^{\left(d_f\right)} \mathbf{u}_a + \nabla_\Gamma^{\left(d_f\right)} \mathbf{u}_a^\mathrm{T} \right) \\
  & + {\eta\over2} \left( \nabla_\Gamma^{\left(d_f, \delta d_f\right)} \mathbf{u} + \nabla_\Gamma^{\left(d_f, \delta d_f\right)} \mathbf{u}^\mathrm{T} \right) : \left( \nabla_\Gamma^{\left(d_f\right)} \mathbf{u}_a + \nabla_\Gamma^{\left(d_f\right)} \mathbf{u}_a^\mathrm{T} \right) \\
  & + {\eta\over2} \left( \nabla_\Gamma^{\left(d_f\right)} \mathbf{u} + \nabla_\Gamma^{\left(d_f\right)} \mathbf{u}^\mathrm{T} \right) : \left( \nabla_\Gamma^{\left(d_f, \delta d_f\right)} \mathbf{u}_a + \nabla_\Gamma^{\left(d_f, \delta d_f\right)} \mathbf{u}_a^\mathrm{T} \right) - \delta p \, \mathrm{div}_\Gamma^{\left( d_f \right)} \mathbf{u}_a \\
  & - p \, \mathrm{div}_\Gamma^{\left( d_f, \delta d_f \right)} \mathbf{u}_a - p_a \mathrm{div}_\Gamma^{\left( d_f, \delta d_f \right)} \mathbf{u} - p_a \mathrm{div}_\Gamma^{\left( d_f \right)} \delta \mathbf{u} + \alpha \delta \mathbf{u} \cdot \mathbf{u}_a + {\partial\alpha\over\partial\gamma_p} {\partial\gamma_p\over\partial\gamma_f} \mathbf{u} \\
  & \cdot \mathbf{u}_a \delta \gamma_f + \delta \lambda \mathbf{u}_a \cdot \mathbf{n}_\Gamma^{\left( d_f \right)} + \lambda \mathbf{u}_a \cdot \mathbf{n}_\Gamma^{\left( d_f, \delta d_f \right)} + \lambda_a \delta \mathbf{u} \cdot \mathbf{n}_\Gamma^{\left( d_f \right)} + \lambda_a \mathbf{u} \cdot \mathbf{n}_\Gamma^{\left( d_f, \delta d_f \right)} \\
  & + f_{id,\Gamma} \bigg( r_f^2 \nabla_\Gamma^{\left( d_f \right)} \delta \gamma_f \cdot \nabla_\Gamma^{\left( d_f \right)} \gamma_{fa} + r_f^2 \nabla_\Gamma^{\left( d_f, \delta d_f \right)} \gamma_f \cdot \nabla_\Gamma^{\left( d_f \right)} \gamma_{fa} + r_f^2 \nabla_\Gamma^{\left( d_f \right)} \gamma_f \\
  & \cdot \nabla_\Gamma^{\left( d_f, \delta d_f \right)} \gamma_{fa} + \delta \gamma_f \gamma_{fa} - \delta \gamma \gamma_{fa} \bigg) \bigg] M^{\left( d_f \right)} + \bigg[ \rho \left( \mathbf{u} \cdot \nabla_\Gamma^{\left(d_f\right)} \right) \mathbf{u} \cdot \mathbf{u}_a \\
  & + {\eta\over2} \left( \nabla_\Gamma^{\left(d_f\right)} \mathbf{u} + \nabla_\Gamma^{\left(d_f\right)} \mathbf{u}^\mathrm{T} \right) : \left( \nabla_\Gamma^{\left(d_f\right)} \mathbf{u}_a + \nabla_\Gamma^{\left(d_f\right)} \mathbf{u}_a^\mathrm{T} \right) - p \, \mathrm{div}_\Gamma^{\left( d_f \right)} \mathbf{u}_a \\
  & - p_a \mathrm{div}_\Gamma^{\left( d_f \right)} \mathbf{u} + \alpha \mathbf{u} \cdot \mathbf{u}_a + \lambda \mathbf{u}_a \cdot \mathbf{n}_\Gamma^{\left( d_f \right)} + \lambda_a \mathbf{u} \cdot \mathbf{n}_\Gamma^{\left( d_f \right)} \\
  & + f_{id,\Gamma} \left( r_f^2 \nabla_\Gamma^{\left( d_f \right)} \gamma_f \cdot \nabla_\Gamma^{\left( d_f \right)} \gamma_{fa} + \gamma_f \gamma_{fa} - \gamma \gamma_{fa} \right) \bigg] M^{\left( d_f, \delta d_f \right)} + r_m^2 \nabla_\Sigma \delta d_f \\
  & \cdot \nabla_\Sigma d_{fa} + \delta d_f d_{fa} - A_d \delta d_m d_{fa} \,\mathrm{d}\Sigma - \sum_{E_\Sigma\in\mathcal{E}_\Sigma} \int_{E_\Sigma} \tau_{BP,\Gamma}^{\left( d_f, \delta d_f \right)} \nabla_\Gamma^{\left( d_f \right)} p \\
  & \cdot \nabla_\Gamma^{\left( d_f \right)} p_a M^{\left( d_f \right)} + \tau_{BP,\Gamma}^{\left( d_f \right)} \nabla_\Gamma^{\left( d_f \right)} \delta p \cdot \nabla_\Gamma^{\left( d_f \right)} p_a M^{\left( d_f \right)} + \tau_{BP,\Gamma}^{\left( d_f \right)} \nabla_\Gamma^{\left( d_f \right)} p \\
  & \cdot \nabla_\Gamma^{\left( d_f, \delta d_f \right)} p_a M^{\left( d_f \right)} + \tau_{BP,\Gamma}^{\left( d_f \right)} \nabla_\Gamma^{\left( d_f \right)} p \cdot \nabla_\Gamma^{\left( d_f \right)} p_a M^{\left( d_f, \delta d_f \right)} \,\mathrm{d}\Sigma \\
\end{split}
\end{equation}
with the satisfication of the constraints in Eq. \ref{equ:ConstraintForAugmentedLagrangianPressureDrop}
and 
\begin{equation}\label{equ:ConstraintForVariationalAugmentedLagrangianPressureDrop}
  \left.\begin{split}
  & \delta \mathbf{u} \in\left(\mathcal{H}\left(\Sigma\right)\right)^3 \\
  & \delta p \in \mathcal{H}\left(\Sigma\right) \\
  & \delta \lambda \in \mathcal{L}^2\left(\Sigma\right) \\
  & \delta \gamma_f \in \mathcal{H}\left(\Sigma_D\right) \\
  & \delta \gamma \in \mathcal{L}^2\left(\Sigma_D\right) \\
  & \delta d_f \in \mathcal{H}\left(\Sigma\right) \\
  & \delta d_m \in \mathcal{L}^2\left(\Sigma\right)
  \end{split}\right\} 
  ~ \mathrm{with} ~ 
  \left\{\begin{split}
  & \delta \mathbf{u} = \mathbf{0} ~\mathrm{at}~ \forall \mathbf{x} \in l_{v,\Sigma} \cup l_{v_0,\Sigma} \\
  & \delta \lambda = 0 ~\mathrm{at}~ \forall \mathbf{x} \in l_{v,\Sigma} \cup l_{v_0,\Sigma}
  \end{split}\right..
\end{equation}

According to the Karush-Kuhn-Tucker conditions of the PDE constrained optimization problem, the first order variational of the augmented Lagrangian to $\mathbf{u}$, $p$ and $\lambda$ can be set to be zero as
\begin{equation}\label{equ:WeakAdjEquSNSEquPressureDrop}
\begin{split}
  & \int_{l_{v,\Sigma}} \delta p L^{\left( d_f \right)} \,\mathrm{d}l_{\partial\Sigma} - \int_{l_{s,\Sigma}} \delta p L^{\left( d_f \right)} \,\mathrm{d}l_{\partial\Sigma} \\
  & + \int_\Sigma \bigg[ \rho \left( \delta \mathbf{u} \cdot \nabla_\Gamma^{\left(d_f\right)} \right) \mathbf{u} \cdot \mathbf{u}_a + \rho \left( \mathbf{u} \cdot \nabla_\Gamma^{\left(d_f\right)} \right) \delta \mathbf{u} \cdot \mathbf{u}_a + {\eta\over2} \left( \nabla_\Gamma^{\left(d_f\right)} \delta \mathbf{u} + \nabla_\Gamma^{\left(d_f\right)} \delta \mathbf{u}^\mathrm{T} \right) \\
  & : \left( \nabla_\Gamma^{\left(d_f\right)} \mathbf{u}_a + \nabla_\Gamma^{\left(d_f\right)} \mathbf{u}_a^\mathrm{T} \right) - \delta p \, \mathrm{div}_\Gamma^{\left( d_f \right)} \mathbf{u}_a - p_a \mathrm{div}_\Gamma^{\left( d_f \right)} \delta \mathbf{u} + \alpha \delta \mathbf{u} \cdot \mathbf{u}_a + \delta \lambda \mathbf{u}_a \cdot \mathbf{n}_\Gamma^{\left( d_f \right)} \\
  & + \lambda_a \delta \mathbf{u} \cdot \mathbf{n}_\Gamma^{\left( d_f \right)} \bigg] M^{\left( d_f \right)} \,\mathrm{d}\Sigma - \sum_{E_\Sigma\in\mathcal{E}_\Sigma} \int_{E_\Sigma} \tau_{BP,\Gamma}^{\left( d_f \right)} \nabla_\Gamma^{\left( d_f \right)} \delta p \cdot \nabla_\Gamma^{\left( d_f \right)} p_a M^{\left( d_f \right)} \,\mathrm{d}\Sigma = 0, \\
\end{split}
\end{equation}
the first order variational of the augmented Lagrangian to the variable $\gamma_f$ can be set to be zero as
\begin{equation}\label{equ:AdjEquAreaConstr1GaInformPressureDrop} 
\begin{split}
  & \int_{\Sigma_D} \left( {\partial\alpha\over\partial\gamma_p} {\partial\gamma_p\over\partial\gamma_f} \mathbf{u} \cdot \mathbf{u}_a \delta \gamma_f + r_f^2 \nabla_\Gamma^{\left( d_f \right)} \delta \gamma_f \cdot \nabla_\Gamma^{\left( d_f \right)} \gamma_{fa} + \delta \gamma_f \gamma_{fa} \right) M^{\left( d_f \right)} \,\mathrm{d}\Sigma = 0,
\end{split}
\end{equation}
and the first order variational of the augmented Lagrangian to the variable $d_f$ can be set to be zero as
\begin{equation}\label{equ:AdjEquAreaConstr1DmInformPressureDrop} 
\begin{split}
  & \int_{l_{v,\Sigma}} p L^{\left( d_f, \delta d_f \right)} \,\mathrm{d}l_{\partial\Sigma} - \int_{l_{s,\Sigma}} p L^{\left( d_f, \delta d_f \right)} \,\mathrm{d}l_{\partial\Sigma} + \int_\Sigma \bigg[ \rho \left( \mathbf{u} \cdot \nabla_\Gamma^{\left(d_f, \delta d_f \right)} \right) \mathbf{u} \cdot \mathbf{u}_a \\
  & + {\eta\over2} \bigg( \nabla_\Gamma^{\left(d_f, \delta d_f\right)} \mathbf{u} + \nabla_\Gamma^{\left(d_f, \delta d_f\right)} \mathbf{u}^\mathrm{T} \bigg) : \left( \nabla_\Gamma^{\left(d_f\right)} \mathbf{u}_a + \nabla_\Gamma^{\left(d_f\right)} \mathbf{u}_a^\mathrm{T} \right) + {\eta\over2} \left( \nabla_\Gamma^{\left(d_f\right)} \mathbf{u} + \nabla_\Gamma^{\left(d_f\right)} \mathbf{u}^\mathrm{T} \right) \\
  & : \bigg( \nabla_\Gamma^{\left(d_f, \delta d_f\right)} \mathbf{u}_a + \nabla_\Gamma^{\left(d_f, \delta d_f\right)} \mathbf{u}_a^\mathrm{T} \bigg) - p \, \mathrm{div}_\Gamma^{\left( d_f, \delta d_f \right)} \mathbf{u}_a - p_a \mathrm{div}_\Gamma^{\left( d_f, \delta d_f \right)} \mathbf{u} + \lambda \mathbf{u}_a \cdot \mathbf{n}_\Gamma^{\left( d_f, \delta d_f \right)} \\
  & + \lambda_a \mathbf{u} \cdot \mathbf{n}_\Gamma^{\left( d_f, \delta d_f \right)} + f_{id,\Gamma} \left( r_f^2 \nabla_\Gamma^{\left( d_f, \delta d_f \right)} \gamma_f \cdot \nabla_\Gamma^{\left( d_f \right)} \gamma_{fa} + r_f^2 \nabla_\Gamma^{\left( d_f \right)} \gamma_f \cdot \nabla_\Gamma^{\left( d_f, \delta d_f \right)} \gamma_{fa} \right) \bigg] \\
  & M^{\left( d_f \right)} + \bigg[ \rho \left( \mathbf{u} \cdot \nabla_\Gamma^{\left(d_f\right)} \right) \mathbf{u} \cdot \mathbf{u}_a + {\eta\over2} \left( \nabla_\Gamma^{\left(d_f\right)} \mathbf{u} + \nabla_\Gamma^{\left(d_f\right)} \mathbf{u}^\mathrm{T} \right) : \left( \nabla_\Gamma^{\left(d_f\right)} \mathbf{u}_a + \nabla_\Gamma^{\left(d_f\right)} \mathbf{u}_a^\mathrm{T} \right) \\
  & - p \, \mathrm{div}_\Gamma^{\left( d_f \right)} \mathbf{u}_a - p_a \mathrm{div}_\Gamma^{\left( d_f \right)} \mathbf{u} + \alpha \mathbf{u} \cdot \mathbf{u}_a + \lambda \mathbf{u}_a \cdot \mathbf{n}_\Gamma^{\left( d_f \right)} + \lambda_a \mathbf{u} \cdot \mathbf{n}_\Gamma^{\left( d_f \right)} \\
  & + f_{id,\Gamma} \left( r_f^2 \nabla_\Gamma^{\left( d_f \right)} \gamma_f \cdot \nabla_\Gamma^{\left( d_f \right)} \gamma_{fa} + \gamma_f \gamma_{fa} - \gamma \gamma_{fa} \right) \bigg] M^{\left( d_f, \delta d_f \right)} + r_m^2 \nabla_\Sigma \delta d_f \cdot \nabla_\Sigma d_{fa} \\
  & + \delta d_f d_{fa} \,\mathrm{d}\Sigma - \sum_{E_\Sigma\in\mathcal{E}_\Sigma} \int_{E_\Sigma} \tau_{BP,\Gamma}^{\left( d_f, \delta d_f \right)} \nabla_\Gamma^{\left( d_f \right)} p \cdot \nabla_\Gamma^{\left( d_f \right)} p_a M^{\left( d_f \right)} + \tau_{BP,\Gamma}^{\left( d_f \right)} \nabla_\Gamma^{\left( d_f \right)} p \\
  & \cdot \nabla_\Gamma^{\left( d_f, \delta d_f \right)} p_a M^{\left( d_f \right)} + \tau_{BP,\Gamma}^{\left( d_f \right)} \nabla_\Gamma^{\left( d_f \right)} p \cdot \nabla_\Gamma^{\left( d_f \right)} p_a M^{\left( d_f, \delta d_f \right)} \,\mathrm{d}\Sigma = 0. \\
\end{split}
\end{equation}
The constraints in Eqs. \ref{equ:ConstraintForAugmentedLagrangianPressureDrop} and \ref{equ:ConstraintForVariationalAugmentedLagrangianPressureDrop} are imposed to Eqs. \ref{equ:WeakAdjEquSNSEquPressureDrop}, \ref{equ:AdjEquAreaConstr1GaInformPressureDrop} and \ref{equ:AdjEquAreaConstr1DmInformPressureDrop}. Then, the adjoint sensitivity of $\Delta P$ is derived as
\begin{equation}\label{equ:AdjSensAreaConstr1Inform}
\begin{split}
  \delta \widehat{\Delta P} = - \int_{\Sigma_D} \delta \gamma \gamma_{fa} M^{\left( d_f \right)} \,\mathrm{d}\Sigma - \int_\Sigma A_d \delta d_m d_{fa} \,\mathrm{d}\Sigma.
\end{split}
\end{equation}

Without losing the arbitrariness of $\delta \mathbf{u}$, $\delta p$, $\delta \lambda$, $\delta \gamma_f$, $\delta d_f$, $\delta \gamma$ and $\delta d_m$, one can set
\begin{equation}
\left.\begin{split}
& \tilde{\mathbf{u}}_a = \delta \mathbf{u} \\
& \tilde{p}_a = \delta p \\
& \tilde{\lambda}_a = \delta \lambda \\
& \tilde{\gamma}_{fa} = \delta \gamma_f \\
& \tilde{d}_{fa} = \delta d_f \\
& \tilde{\gamma} = \delta \gamma \\
& \tilde{d}_m = \delta d_m 
\end{split}\right\}
~\mathrm{with}~
\left\{\begin{split}
& \forall \tilde{\mathbf{u}}_a \in \left(\mathcal{H}\left(\Sigma\right)\right)^3 \\
& \forall \tilde{p}_a \in \mathcal{H}\left(\Sigma\right)\\
& \forall \tilde{\lambda}_a \in \mathcal{L}^2\left(\Sigma\right) \\
& \forall \tilde{\gamma}_{fa} \in \mathcal{H}\left(\Sigma_D\right) \\
& \forall \tilde{d}_{fa} \in \mathcal{H}\left(\Sigma\right)\\
& \forall \tilde{\gamma} \in \mathcal{L}^2\left(\Sigma_D\right) \\
& \forall \tilde{d}_m \in \mathcal{L}^2\left(\Sigma\right)
\end{split}\right.
\end{equation}
for Eqs. \ref{equ:WeakAdjEquSNSEquPressureDrop}, \ref{equ:AdjEquAreaConstr1GaInformPressureDrop} and \ref{equ:AdjEquAreaConstr1DmInformPressureDrop} to derive the adjoint system composed of Eqs. \ref{equ:AdjEquSurfaceNSMHMPressureDrop}, \ref{equ:AdjPDEFilterPressureDropGaMHM} and \ref{equ:AdjPDEFilterJPressureDropDmMHM}.

\subsection{Adjoint analysis for design objective in Eq. \ref{equ:VarProToopSurfaceNSCHM}} \label{sec:AdjointAnalysisDesignObjectiveSurfaceCHTMHM}

Based on the transformed design objective in Eq. \ref{equ:TransformedDesignObjectiveSurfaceCHM}, the variational formulations of the surface-PDE filters in Eqs. \ref{equ:VariationalFormulationPDEFilterBaseManifoldMHM} and \ref{equ:VariationalFormulationPDEFilterMHM} and the surface Navier-Stokes equations in Eq. \ref{equ:TransformedVariationalFormulationSurfaceNSEqusHM} and the surface convective heat-transfer equation in Eq. \ref{equ:TransformedVariationalFormulationSurfaceCHMEqu}, the augmented Lagrangian of the design objective in Eq. \ref{equ:VarProToopSurfaceNSCHM} can be derived as
\begin{equation}\label{equ:AugmentedLagrangianMatchOptimizationCHM}
\begin{split}
  \hat{J}_T = & \int_\Sigma f_{id,\Gamma} k \nabla_\Gamma^{\left( d_f \right)} T \cdot \nabla_\Gamma^{\left( d_f \right)} T M^{\left( d_f \right)} + \bigg[ \rho \left( \mathbf{u} \cdot \nabla_\Gamma^{\left(d_f\right)} \right) \mathbf{u} \cdot \mathbf{u}_a + {\eta\over2} \left( \nabla_\Gamma^{\left(d_f\right)} \mathbf{u} + \nabla_\Gamma^{\left(d_f\right)} \mathbf{u}^\mathrm{T} \right) \\
  & : \left( \nabla_\Gamma^{\left(d_f\right)} \mathbf{u}_a + \nabla_\Gamma^{\left(d_f\right)} \mathbf{u}_a^\mathrm{T} \right) - p \, \mathrm{div}_\Gamma^{\left( d_f \right)} \mathbf{u}_a - p_a \, \mathrm{div}_\Gamma^{\left( d_f \right)} \mathbf{u} + \alpha \mathbf{u} \cdot \mathbf{u}_a + \lambda \mathbf{u}_a \cdot \mathbf{n}_\Gamma^{\left( d_f \right)} \\
  & + \lambda_a \mathbf{u} \cdot \mathbf{n}_\Gamma^{\left( d_f \right)} \bigg] M^{\left( d_f \right)} \,\mathrm{d}\Sigma - \sum_{E_\Sigma\in\mathcal{E}_\Sigma} \int_{E_\Sigma} \tau_{LS\mathbf{u},\Gamma}^{\left( d_f \right)} \left( \rho \mathbf{u} \cdot \nabla_\Gamma^{\left( d_f \right)} \mathbf{u} + \nabla_\Gamma^{\left( d_f \right)} p + \alpha \mathbf{u} \right) \\
  & \cdot \bigg( \rho \mathbf{u} \cdot \nabla_\Gamma^{\left( d_f \right)} \mathbf{u}_a + \nabla_\Gamma^{\left( d_f \right)} p_a \bigg) M^{\left( d_f \right)} + \tau_{LSp,\Gamma}^{\left( d_f \right)} \left( \rho \mathrm{div}_\Gamma^{\left( d_f \right)} \mathbf{u} \right) \left( \mathrm{div}_\Gamma^{\left( d_f \right)} \mathbf{u}_a \right) M^{\left( d_f \right)} \,\mathrm{d}\Sigma \\
  & + \int_\Sigma \bigg[ \bigg( \rho C_p \mathbf{u} \cdot \nabla_\Gamma^{\left(d_f\right)} T - Q \bigg) T_a + k \nabla_\Gamma^{\left(d_f\right)} T \cdot \nabla_\Gamma^{\left(d_f\right)} T_a \bigg] M^{\left( d_f \right)} \,\mathrm{d}\Sigma \\
  & + \sum_{E_\Sigma\in\mathcal{E}_\Sigma} \int_{E_\Sigma} \tau_{LST,\Gamma}^{\left( d_f \right)} \left( \rho C_p \mathbf{u} \cdot \nabla_\Gamma^{\left( d_f \right)} T - Q \right) \left( \rho C_p \mathbf{u} \cdot \nabla_\Gamma^{\left( d_f \right)} T_a \right) M^{\left( d_f \right)} \,\mathrm{d}\Sigma \\
  & + \int_\Sigma f_{id,\Gamma} \left( r_f^2 \nabla_\Gamma^{\left( d_f \right)} \gamma_f \cdot \nabla_\Gamma^{\left( d_f \right)} \gamma_{fa} + \gamma_f \gamma_{fa} - \gamma \gamma_{fa} \right) M^{\left( d_f \right)} \\
  & + r_m^2 \nabla_\Sigma d_f \cdot \nabla_\Sigma d_{fa} + d_f d_{fa} - A_d \left( d_m - {1\over2} \right) d_{fa} \,\mathrm{d}\Sigma, \\
\end{split}
\end{equation}
where $\hat{J_T}$ is the augmented lagrangian of $J_T$; and the adjoint variables satisfy
\begin{equation}\label{equ:ConstraintForAugmentedLagrangianCHM}
  \left.\begin{split}
  & \mathbf{u}_a \in\left(\mathcal{H}\left(\Sigma\right)\right)^3 \\
  & p_a \in \mathcal{H}\left(\Sigma\right) \\
  & \lambda_a \in \mathcal{L}^2\left(\Sigma\right) \\
  & T_a \in \mathcal{H}\left(\Sigma\right) \\
  & \gamma_{fa} \in \mathcal{H}\left(\Sigma_D\right) \\
  & d_{fa} \in \mathcal{H}\left(\Sigma\right)
  \end{split}\right\}~\mathrm{with}~
  \left\{\begin{split}
  & \mathbf{u}_a = \mathbf{0} ~ \mathrm{at} ~ \forall \mathbf{x} \in l_{v,\Sigma} \cup l_{v_0,\Sigma} \\
  & \lambda_a = 0 ~ \mathrm{at} ~ \forall \mathbf{x} \in l_{v,\Sigma} \cup l_{v_0,\Sigma} \\
  & T_a = 0 ~ \mathrm{at} ~ \forall \mathbf{x} \in l_{v,\Sigma} 
  \end{split}\right..
\end{equation}
The first order variational of the augmented Lagrangian in Eq. \ref{equ:AugmentedLagrangianMatchOptimizationCHM} can be derived as
\begin{equation}\label{equ:1stVariAugmentedLagrangianMatchOptimizationCHM}
\begin{split}
  \delta \hat{J}_T = & \int_\Sigma f_{id,\Gamma} {\partial k \over \partial \gamma_p} {\partial \gamma_p \over \partial \gamma_f} \nabla_\Gamma^{\left( d_f \right)} T \cdot \nabla_\Gamma^{\left( d_f \right)} T M^{\left( d_f \right)} \delta \gamma_f + 2 f_{id,\Gamma} k \nabla_\Gamma^{\left( d_f, \delta d_f \right)} T \cdot \nabla_\Gamma^{\left( d_f \right)} T M^{\left( d_f \right)} \\
  & + 2 f_{id,\Gamma} k \nabla_\Gamma^{\left( d_f \right)} \delta T \cdot \nabla_\Gamma^{\left( d_f \right)} T M^{\left( d_f \right)} + f_{id,\Gamma} k \nabla_\Gamma^{\left( d_f \right)} T \cdot \nabla_\Gamma^{\left( d_f \right)} T M^{\left( d_f, \delta d_f \right)} \\
  & + \bigg[ \rho \left( \delta \mathbf{u} \cdot \nabla_\Gamma^{\left(d_f\right)} \right) \mathbf{u} \cdot \mathbf{u}_a + \rho \left( \mathbf{u} \cdot \nabla_\Gamma^{\left(d_f, \delta d_f\right)} \right) \mathbf{u} \cdot \mathbf{u}_a + \rho \left( \mathbf{u} \cdot \nabla_\Gamma^{\left(d_f\right)} \right) \delta \mathbf{u} \cdot \mathbf{u}_a \\
  & + {\eta\over2} \left( \nabla_\Gamma^{\left(d_f\right)} \delta \mathbf{u} + \nabla_\Gamma^{\left(d_f\right)} \delta \mathbf{u}^\mathrm{T} \right) : \left( \nabla_\Gamma^{\left(d_f\right)} \mathbf{u}_a + \nabla_\Gamma^{\left(d_f\right)} \mathbf{u}_a^\mathrm{T} \right) \\
  & + {\eta\over2} \left( \nabla_\Gamma^{\left(d_f, \delta d_f\right)} \mathbf{u} + \nabla_\Gamma^{\left(d_f, \delta d_f\right)} \mathbf{u}^\mathrm{T} \right) : \left( \nabla_\Gamma^{\left(d_f\right)} \mathbf{u}_a + \nabla_\Gamma^{\left(d_f\right)} \mathbf{u}_a^\mathrm{T} \right) + {\eta\over2} \bigg( \nabla_\Gamma^{\left(d_f\right)} \mathbf{u} \\
  & + \nabla_\Gamma^{\left(d_f\right)} \mathbf{u}^\mathrm{T} \bigg) : \left( \nabla_\Gamma^{\left(d_f, \delta d_f\right)} \mathbf{u}_a + \nabla_\Gamma^{\left(d_f, \delta d_f\right)} \mathbf{u}_a^\mathrm{T} \right) - \delta p \, \mathrm{div}_\Gamma^{\left( d_f \right)} \mathbf{u}_a - p \, \mathrm{div}_\Gamma^{\left( d_f, \delta d_f \right)} \mathbf{u}_a \\
  & - p_a \mathrm{div}_\Gamma^{\left( d_f, \delta d_f \right)} \mathbf{u} - p_a \mathrm{div}_\Gamma^{\left( d_f \right)} \delta \mathbf{u} + {\partial \alpha \over \partial \gamma_p} {\partial \gamma_p \over \partial \gamma_f} \mathbf{u} \cdot \mathbf{u}_a \delta \gamma_f + \alpha \delta \mathbf{u} \cdot \mathbf{u}_a + \delta \lambda \mathbf{u}_a \cdot \mathbf{n}_\Gamma^{\left( d_f \right)} \\
  & + \lambda \mathbf{u}_a \cdot \mathbf{n}_\Gamma^{\left( d_f, \delta d_f \right)} + \lambda_a \delta \mathbf{u} \cdot \mathbf{n}_\Gamma^{\left( d_f \right)} + \lambda_a \mathbf{u} \cdot \mathbf{n}_\Gamma^{\left( d_f, \delta d_f \right)} \bigg] M^{\left( d_f \right)} + \bigg[ \rho \left( \mathbf{u} \cdot \nabla_\Gamma^{\left(d_f\right)} \right) \mathbf{u} \cdot \mathbf{u}_a \\
  & + {\eta\over2} \left( \nabla_\Gamma^{\left(d_f\right)} \mathbf{u} + \nabla_\Gamma^{\left(d_f\right)} \mathbf{u}^\mathrm{T} \right) : \left( \nabla_\Gamma^{\left(d_f\right)} \mathbf{u}_a + \nabla_\Gamma^{\left(d_f\right)} \mathbf{u}_a^\mathrm{T} \right) - p \, \mathrm{div}_\Gamma^{\left( d_f \right)} \mathbf{u}_a - p_a \mathrm{div}_\Gamma^{\left( d_f \right)} \mathbf{u} \\
  & + \alpha \mathbf{u} \cdot \mathbf{u}_a + \lambda \mathbf{u}_a \cdot \mathbf{n}_\Gamma^{\left( d_f \right)} + \lambda_a \mathbf{u} \cdot \mathbf{n}_\Gamma^{\left( d_f \right)} \bigg] M^{\left( d_f, \delta d_f \right)} \,\mathrm{d}\Sigma \\
  & - \sum_{E_\Sigma\in\mathcal{E}_\Sigma} \int_{E_\Sigma} \left( \tau_{LS\mathbf{u},\Gamma}^{\left( d_f, \delta d_f \right)} + \tau_{LS\mathbf{u},\Gamma}^{\left( d_f, \delta \mathbf{u} \right)} \right) \left( \rho \mathbf{u} \cdot \nabla_\Gamma^{\left( d_f \right)} \mathbf{u} + \nabla_\Gamma^{\left( d_f \right)} p + \alpha \mathbf{u} \right) \cdot \bigg( \rho \mathbf{u} \cdot \nabla_\Gamma^{\left( d_f \right)} \mathbf{u}_a \\
  & + \nabla_\Gamma^{\left( d_f \right)} p_a \bigg) M^{\left( d_f \right)} + \tau_{LS\mathbf{u},\Gamma}^{\left( d_f \right)} \bigg[ \rho \left( \delta \mathbf{u} \cdot \nabla_\Gamma^{\left( d_f \right)} \mathbf{u} + \mathbf{u} \cdot \nabla_\Gamma^{\left( d_f, \delta d_f \right)} \mathbf{u} + \mathbf{u} \cdot \nabla_\Gamma^{\left( d_f \right)} \delta \mathbf{u}\right) \\
  & + \nabla_\Gamma^{\left( d_f, \delta d_f \right)} p + \nabla_\Gamma^{\left( d_f \right)} \delta p + \alpha \delta \mathbf{u} + {\partial \alpha \over \partial \gamma_p} {\partial \gamma_p \over \partial \gamma_f} \mathbf{u} \delta \gamma_f \bigg] \cdot \left( \rho \mathbf{u} \cdot \nabla_\Gamma^{\left( d_f \right)} \mathbf{u}_a + \nabla_\Gamma^{\left( d_f \right)} p_a \right) M^{\left( d_f \right)} \\
  & + \tau_{LS\mathbf{u},\Gamma}^{\left( d_f \right)} \left( \rho \mathbf{u} \cdot \nabla_\Gamma^{\left( d_f \right)} \mathbf{u} + \nabla_\Gamma^{\left( d_f \right)} p + \alpha \mathbf{u} \right) \cdot \bigg[ \rho \left( \delta \mathbf{u} \cdot \nabla_\Gamma^{\left( d_f \right)} \mathbf{u}_a + \mathbf{u} \cdot \nabla_\Gamma^{\left( d_f, \delta d_f \right)} \mathbf{u}_a \right) \\
  & + \nabla_\Gamma^{\left( d_f, \delta d_f \right)} p_a \bigg] M^{\left( d_f \right)} + \tau_{LS\mathbf{u},\Gamma}^{\left( d_f \right)} \left( \rho \mathbf{u} \cdot \nabla_\Gamma^{\left( d_f \right)} \mathbf{u} + \nabla_\Gamma^{\left( d_f \right)} p + \alpha \mathbf{u} \right) \cdot \bigg( \rho \mathbf{u} \cdot \nabla_\Gamma^{\left( d_f \right)} \mathbf{u}_a \\
  & + \nabla_\Gamma^{\left( d_f \right)} p_a \bigg) M^{\left( d_f, \delta d_f \right)} + \left( \tau_{LSp,\Gamma}^{\left( d_f, \delta d_f \right)} + \tau_{LSp,\Gamma}^{\left( d_f, \delta \mathbf{u} \right)} \right) \left( \rho \mathrm{div}_\Gamma^{\left( d_f \right)} \mathbf{u} \right) \left( \mathrm{div}_\Gamma^{\left( d_f \right)} \mathbf{u}_a \right) M^{\left( d_f \right)} \\
  & + \tau_{LSp,\Gamma}^{\left( d_f \right)} \rho \left( \mathrm{div}_\Gamma^{\left( d_f, \delta d_f \right)} \mathbf{u} + \mathrm{div}_\Gamma^{\left( d_f \right)} \delta \mathbf{u} \right) \left( \mathrm{div}_\Gamma^{\left( d_f \right)} \mathbf{u}_a \right) M^{\left( d_f \right)} + \tau_{LSp,\Gamma}^{\left( d_f \right)} \left( \rho \mathrm{div}_\Gamma^{\left( d_f \right)} \mathbf{u} \right) \\
  & \left( \mathrm{div}_\Gamma^{\left( d_f, \delta d_f \right)} \mathbf{u}_a \right) M^{\left( d_f \right)} + \tau_{LSp,\Gamma}^{\left( d_f \right)} \left( \rho \mathrm{div}_\Gamma^{\left( d_f \right)} \mathbf{u} \right) \left( \mathrm{div}_\Gamma^{\left( d_f \right)} \mathbf{u}_a \right) M^{\left( d_f, \delta d_f \right)} \,\mathrm{d}\Sigma \\
  & + \int_\Sigma \bigg[ \rho C_p \left( \delta \mathbf{u} \cdot \nabla_\Gamma^{\left(d_f\right)} T + \mathbf{u} \cdot \nabla_\Gamma^{\left(d_f, \delta d_f\right)} T + \mathbf{u} \cdot \nabla_\Gamma^{\left(d_f\right)} \delta T \right) T_a + \rho {\partial C_p \over \partial \gamma_p} {\partial \gamma_p \over \partial \gamma_f} \mathbf{u} \cdot \nabla_\Gamma^{\left(d_f\right)} T \\
  & T_a \delta \gamma_f + {\partial k \over \partial \gamma_p} {\partial \gamma_p \over \partial \gamma_f} \nabla_\Gamma^{\left(d_f\right)} T \cdot \nabla_\Gamma^{\left(d_f\right)} T_a \delta \gamma_f + k \nabla_\Gamma^{\left(d_f, \delta d_f\right)} T \cdot \nabla_\Gamma^{\left(d_f\right)} T_a + k \nabla_\Gamma^{\left(d_f\right)} \delta T \cdot \nabla_\Gamma^{\left(d_f\right)} T_a \\
  & + k \nabla_\Gamma^{\left(d_f\right)} T \cdot \nabla_\Gamma^{\left(d_f, \delta d_f\right)} T_a \bigg] M^{\left( d_f \right)} + \left[ \left( \rho C_p \mathbf{u} \cdot \nabla_\Gamma^{\left(d_f\right)} T - Q \right) T_a + k \nabla_\Gamma^{\left(d_f\right)} T \cdot \nabla_\Gamma^{\left(d_f\right)} T_a \right] \\
  & M^{\left( d_f, \delta d_f \right)} \,\mathrm{d}\Sigma \\
  & + \sum_{E_\Sigma\in\mathcal{E}_\Sigma} \int_{E_\Sigma} \left( \tau_{LST,\Gamma}^{\left( d_f, \delta d_f \right)} + \tau_{LST,\Gamma}^{\left( d_f, \delta \mathbf{u} \right)} \right) \left( \rho C_p \mathbf{u} \cdot \nabla_\Gamma^{\left( d_f \right)} T - Q \right) \left( \rho C_p \mathbf{u} \cdot \nabla_\Gamma^{\left( d_f \right)} T_a \right) M^{\left( d_f \right)} \\
\end{split}
\end{equation}
\begin{equation*}
\begin{split}
  & + \tau_{LST,\Gamma}^{\left( d_f \right)} \rho C_p \left( \delta \mathbf{u} \cdot \nabla_\Gamma^{\left( d_f \right)} T + \mathbf{u} \cdot \nabla_\Gamma^{\left( d_f, \delta d_f \right)} T + \mathbf{u} \cdot \nabla_\Gamma^{\left( d_f \right)} \delta T \right) \left( \rho C_p \mathbf{u} \cdot \nabla_\Gamma^{\left( d_f \right)} T_a \right) M^{\left( d_f \right)} \\
  & + \tau_{LST,\Gamma}^{\left( d_f \right)} \left( \rho C_p \mathbf{u} \cdot \nabla_\Gamma^{\left( d_f \right)} T - Q \right) \rho C_p \left( \delta \mathbf{u} \cdot \nabla_\Gamma^{\left( d_f \right)} T_a + \mathbf{u} \cdot \nabla_\Gamma^{\left( d_f, \delta d_f \right)} T_a \right) M^{\left( d_f \right)} \\
  & + \tau_{LST,\Gamma}^{\left( d_f \right)} \left( \rho C_p \mathbf{u} \cdot \nabla_\Gamma^{\left( d_f \right)} T - Q \right) \left( \rho C_p \mathbf{u} \cdot \nabla_\Gamma^{\left( d_f \right)} T_a \right) M^{\left( d_f, \delta d_f \right)} + \Bigg( {\partial \tau_{LST,\Gamma}^{\left( d_f \right)} \over \partial C_p} {\partial C_p \over \partial \gamma_p} \\
  & + {\partial \tau_{LST,\Gamma}^{\left( d_f \right)} \over \partial k} {\partial k \over \partial \gamma_p} \Bigg) {\partial \gamma_p \over \partial \gamma_f} \left( \rho C_p \mathbf{u} \cdot \nabla_\Gamma^{\left( d_f \right)} T - Q \right) \left( \rho C_p \mathbf{u} \cdot \nabla_\Gamma^{\left( d_f \right)} T_a \right) M^{\left( d_f \right)} \delta \gamma_f \\
  & + \tau_{LST,\Gamma}^{\left( d_f \right)} \rho {\partial C_p \over \partial \gamma_p} {\partial \gamma_p \over \partial \gamma_f} \bigg[ \left( \mathbf{u} \cdot \nabla_\Gamma^{\left( d_f \right)} T \right) \left( \rho C_p \mathbf{u} \cdot \nabla_\Gamma^{\left( d_f \right)} T_a \right) \\
  & + \left( \rho C_p \mathbf{u} \cdot \nabla_\Gamma^{\left( d_f \right)} T - Q \right) \left( \mathbf{u} \cdot \nabla_\Gamma^{\left( d_f \right)} T_a \right) \bigg] M^{\left( d_f \right)} \delta \gamma_f \,\mathrm{d}\Sigma \\
  & + \int_\Sigma f_{id,\Gamma} \bigg[ r_f^2 \bigg( \nabla_\Gamma^{\left( d_f \right)} \delta \gamma_f \cdot \nabla_\Gamma^{\left( d_f \right)} \gamma_{fa} + \nabla_\Gamma^{\left( d_f, \delta d_f \right)} \gamma_f \cdot \nabla_\Gamma^{\left( d_f \right)} \gamma_{fa} + \nabla_\Gamma^{\left( d_f \right)} \gamma_f \cdot \nabla_\Gamma^{\left( d_f, \delta d_f \right)} \gamma_{fa} \bigg) \\
  & + \delta \gamma_f \gamma_{fa} - \delta \gamma \gamma_{fa} \bigg] M^{\left( d_f \right)} + f_{id,\Gamma} \left( r_f^2 \nabla_\Gamma^{\left( d_f \right)} \gamma_f \cdot \nabla_\Gamma^{\left( d_f \right)} \gamma_{fa} + \gamma_f \gamma_{fa} - \gamma \gamma_{fa} \right) M^{\left( d_f, \delta d_f \right)} \\
  & + r_m^2 \nabla_\Sigma \delta d_f \cdot \nabla_\Sigma d_{fa} + \delta d_f d_{fa} - A_d \delta d_m d_{fa} \,\mathrm{d}\Sigma \\
\end{split}
\end{equation*}
with the satisfication of the constraints in Eq. \ref{equ:ConstraintForAugmentedLagrangianCHM}
and
\begin{equation}\label{equ:ConstraintForVariationalAugmentedLagrangianCHM}
  \left.\begin{split}
  & \delta \mathbf{u} \in\left(\mathcal{H}\left(\Sigma\right)\right)^3 \\
  & \delta p \in \mathcal{H}\left(\Sigma\right) \\
  & \delta \lambda \in \mathcal{L}^2\left(\Sigma\right) \\
  & \delta T \in \mathcal{H}\left(\Sigma\right) \\
  & \delta \gamma_f \in \mathcal{H}\left(\Sigma_D\right) \\
  & \delta \gamma \in \mathcal{L}^2\left(\Sigma_D\right) \\
  & \delta d_f \in \mathcal{H}\left(\Sigma\right) \\
  & \delta d_m \in \mathcal{L}^2\left(\Sigma\right)
  \end{split}\right\} 
  ~ \mathrm{with} ~ 
  \left\{\begin{split}
  & \delta \mathbf{u} = \mathbf{0} ~ \mathrm{at} ~ \forall \mathbf{x}_\Sigma \in l_{v,\Sigma} \cup l_{v_0,\Sigma} \\
  & \delta \lambda = 0 ~ \mathrm{at} ~ \forall \mathbf{x}_\Sigma \in l_{v,\Sigma} \cup l_{v_0,\Sigma} \\
  & \delta T = 0 ~ \mathrm{at} ~ \forall \mathbf{x}_\Sigma \in l_{v,\Sigma}
  \end{split}\right..
\end{equation}

According to the Karush-Kuhn-Tucker conditions of the PDE constrained optimization problem \cite{HinzeSpringer2009}, the first order variational of the augmented Lagrangian to $T$ can be set to be zero as
\begin{equation}\label{equ:WeakAdjEquSCHMEqu}
\begin{split}
  & \int_\Sigma \left( 2 f_{id,\Gamma} k \nabla_\Gamma^{\left( d_f \right)} T \cdot \nabla_\Gamma^{\left( d_f \right)} \delta T + \rho C_p \mathbf{u} \cdot \nabla_\Gamma^{\left(d_f\right)} \delta T T_a + k \nabla_\Gamma^{\left(d_f\right)} \delta T \cdot \nabla_\Gamma^{\left(d_f\right)} T_a \right) M^{\left( d_f \right)} \,\mathrm{d}\Sigma \\
  & + \sum_{E_\Sigma\in\mathcal{E}_\Sigma} \int_{E_\Sigma} \tau_{LST,\Gamma}^{\left( d_f \right)} \left( \rho C_p \mathbf{u} \cdot \nabla_\Gamma^{\left( d_f \right)} \delta T \right) \left( \rho C_p \mathbf{u} \cdot \nabla_\Gamma^{\left( d_f \right)} T_a \right) M^{\left( d_f \right)} \,\mathrm{d}\Sigma = 0, \\
\end{split}
\end{equation}
the first order variational of the augmented Lagrangian to $\mathbf{u}$, $p$ and $\lambda$ can be set to be zero as
\begin{equation}\label{equ:WeakAdjEquSNSEquCHM} 
\begin{split}
  & \int_\Sigma \bigg[ \rho \left( \delta \mathbf{u} \cdot \nabla_\Gamma^{\left(d_f\right)} \mathbf{u} + \mathbf{u} \cdot \nabla_\Gamma^{\left(d_f\right)} \delta \mathbf{u} \right) \cdot \mathbf{u}_a + {\eta\over2} \left( \nabla_\Gamma^{\left(d_f\right)} \delta \mathbf{u} + \nabla_\Gamma^{\left(d_f\right)} \delta \mathbf{u}^\mathrm{T} \right) : \left( \nabla_\Gamma^{\left(d_f\right)} \mathbf{u}_a + \nabla_\Gamma^{\left(d_f\right)} \mathbf{u}_a^\mathrm{T} \right) \\
  & - \delta p \, \mathrm{div}_\Gamma^{\left( d_f \right)} \mathbf{u}_a - p_a \mathrm{div}_\Gamma^{\left( d_f \right)} \delta \mathbf{u} + \alpha \delta \mathbf{u} \cdot \mathbf{u}_a + \delta \lambda \mathbf{u}_a \cdot \mathbf{n}_\Gamma^{\left( d_f \right)} + \lambda_a \delta \mathbf{u} \cdot \mathbf{n}_\Gamma^{\left( d_f \right)} + \rho C_p \delta \mathbf{u} \cdot \nabla_\Gamma^{\left(d_f\right)} T T_a  \bigg] \\
  & M^{\left( d_f \right)} \,\mathrm{d}\Sigma - \sum_{E_\Sigma\in\mathcal{E}_\Sigma} \int_{E_\Sigma} \bigg\{ \tau_{LS\mathbf{u},\Gamma}^{\left( d_f, \delta \mathbf{u} \right)} \left( \rho \mathbf{u} \cdot \nabla_\Gamma^{\left( d_f \right)} \mathbf{u} + \nabla_\Gamma^{\left( d_f \right)} p + \alpha \mathbf{u} \right) \cdot \left( \rho \mathbf{u} \cdot \nabla_\Gamma^{\left( d_f \right)} \mathbf{u}_a + \nabla_\Gamma^{\left( d_f \right)} p_a \right) \\
  & + \tau_{LS\mathbf{u},\Gamma}^{\left( d_f \right)} \left[ \rho \left( \delta \mathbf{u} \cdot \nabla_\Gamma^{\left( d_f \right)} \mathbf{u} + \mathbf{u} \cdot \nabla_\Gamma^{\left( d_f \right)} \delta \mathbf{u}\right) + \nabla_\Gamma^{\left( d_f \right)} \delta p + \alpha \delta \mathbf{u} \right] \cdot \left( \rho \mathbf{u} \cdot \nabla_\Gamma^{\left( d_f \right)} \mathbf{u}_a + \nabla_\Gamma^{\left( d_f \right)} p_a \right) \\
  & + \tau_{LS\mathbf{u},\Gamma}^{\left( d_f \right)} \left( \rho \mathbf{u} \cdot \nabla_\Gamma^{\left( d_f \right)} \mathbf{u} + \nabla_\Gamma^{\left( d_f \right)} p + \alpha \mathbf{u} \right) \cdot \left( \rho \delta \mathbf{u} \cdot \nabla_\Gamma^{\left( d_f \right)} \mathbf{u}_a \right) + \tau_{LSp,\Gamma}^{\left( d_f, \delta \mathbf{u} \right)} \left( \rho \mathrm{div}_\Gamma^{\left( d_f \right)} \mathbf{u} \right) \\
  & \left( \mathrm{div}_\Gamma^{\left( d_f \right)} \mathbf{u}_a \right) + \tau_{LSp,\Gamma}^{\left( d_f \right)} \left( \rho \mathrm{div}_\Gamma^{\left( d_f \right)} \delta \mathbf{u} \right) \left( \mathrm{div}_\Gamma^{\left( d_f \right)} \mathbf{u}_a \right) - \tau_{LST,\Gamma}^{\left( d_f, \delta \mathbf{u} \right)} \left( \rho C_p \mathbf{u} \cdot \nabla_\Gamma^{\left( d_f \right)} T - Q \right) \\
  & \left( \rho C_p \mathbf{u} \cdot \nabla_\Gamma^{\left( d_f \right)} T_a \right) - \tau_{LST,\Gamma}^{\left( d_f \right)} \left( \rho C_p \delta \mathbf{u} \cdot \nabla_\Gamma^{\left( d_f \right)} T \right) \left( \rho C_p \mathbf{u} \cdot \nabla_\Gamma^{\left( d_f \right)} T_a \right) \\
  & - \tau_{LST,\Gamma}^{\left( d_f \right)} \left( \rho C_p \mathbf{u} \cdot \nabla_\Gamma^{\left( d_f \right)} T - Q \right) \left( \rho C_p \delta \mathbf{u} \cdot \nabla_\Gamma^{\left( d_f \right)} T_a \right) \bigg\} M^{\left( d_f \right)} \,\mathrm{d}\Sigma = 0, \\
\end{split}
\end{equation}
the first order variational of the augmented Lagrangian to $\gamma_f$ can be set to be zero as
\begin{equation}\label{equ:WeakAdjEquSPDEFilterGaCHM} 
\begin{split}
  & \int_{\Sigma_D} \bigg( {\partial \alpha \over \partial \gamma_p} {\partial \gamma_p \over \partial \gamma_f} \mathbf{u} \cdot \mathbf{u}_a \delta \gamma_f + {\partial k \over \partial \gamma_p} {\partial \gamma_p \over \partial \gamma_f} \nabla_\Gamma^{\left( d_f \right)} T \cdot \nabla_\Gamma^{\left( d_f \right)} T \delta \gamma_f + \rho {\partial C_p \over \partial \gamma_p} {\partial \gamma_p \over \partial \gamma_f} \mathbf{u} \cdot \nabla_\Gamma^{\left(d_f\right)} T T_a \delta \gamma_f \\
  & + {\partial k \over \partial \gamma_p} {\partial \gamma_p \over \partial \gamma_f} \nabla_\Gamma^{\left(d_f\right)} T \cdot \nabla_\Gamma^{\left(d_f\right)} T_a \delta \gamma_f + r_f^2 \nabla_\Gamma^{\left( d_f \right)} \delta \gamma_f \cdot \nabla_\Gamma^{\left( d_f \right)} \gamma_{fa} + \delta \gamma_f \gamma_{fa} \bigg) M^{\left( d_f \right)} \,\mathrm{d}\Sigma \\
  & - \sum_{E_\Sigma\in\mathcal{E}_\Sigma} \int_{E_\Sigma\cap\Sigma_D} \Bigg\{ \tau_{LS\mathbf{u},\Gamma}^{\left( d_f \right)} {\partial \alpha \over \partial \gamma_p} {\partial \gamma_p \over \partial \gamma_f} \mathbf{u} \cdot \left( \rho \mathbf{u} \cdot \nabla_\Gamma^{\left( d_f \right)} \mathbf{u}_a + \nabla_\Gamma^{\left( d_f \right)} p_a \right) - \Bigg( {\partial \tau_{LST,\Gamma}^{\left( d_f \right)} \over \partial C_p} {\partial C_p \over \partial \gamma_p} \\
  & + {\partial \tau_{LST,\Gamma}^{\left( d_f \right)} \over \partial k} {\partial k \over \partial \gamma_p} \Bigg) {\partial \gamma_p \over \partial \gamma_f} \left( \rho C_p \mathbf{u} \cdot \nabla_\Gamma^{\left( d_f \right)} T - Q \right) \left( \rho C_p \mathbf{u} \cdot \nabla_\Gamma^{\left( d_f \right)} T_a \right) - \tau_{LST,\Gamma}^{\left( d_f \right)} \rho {\partial C_p \over \partial \gamma_p} {\partial \gamma_p \over \partial \gamma_f} \\
  & \bigg[ \left( \mathbf{u} \cdot \nabla_\Gamma^{\left( d_f \right)} T \right) \left( \rho C_p \mathbf{u} \cdot \nabla_\Gamma^{\left( d_f \right)} T_a \right) + \left( \rho C_p \mathbf{u} \cdot \nabla_\Gamma^{\left( d_f \right)} T - Q \right) \\
  & \left( \mathbf{u} \cdot \nabla_\Gamma^{\left( d_f \right)} T_a \right) \bigg] \Bigg\} \delta \gamma_f M^{\left( d_f \right)} \,\mathrm{d}\Sigma = 0, \\
\end{split}
\end{equation}
and the first order variational of the augmented Lagrangian to $d_f$ can be set to be zero as
\begin{equation}\label{equ:WeakAdjEquSPDEFilterDmObjCHM} 
\begin{split}
  & \int_\Sigma \left( 2 f_{id,\Gamma} k \nabla_\Gamma^{\left( d_f, \delta d_f \right)} T \cdot \nabla_\Gamma^{\left( d_f \right)} T \right) M^{\left( d_f \right)} + f_{id,\Gamma} k \nabla_\Gamma^{\left( d_f \right)} T \cdot \nabla_\Gamma^{\left( d_f \right)} T M^{\left( d_f, \delta d_f \right)} \\
  & + \bigg[ \rho \left( \mathbf{u} \cdot \nabla_\Gamma^{\left(d_f, \delta d_f\right)} \right) \mathbf{u} \cdot \mathbf{u}_a + {\eta\over2} \left( \nabla_\Gamma^{\left(d_f, \delta d_f\right)} \mathbf{u} + \nabla_\Gamma^{\left(d_f, \delta d_f\right)} \mathbf{u}^\mathrm{T} \right) : \left( \nabla_\Gamma^{\left(d_f\right)} \mathbf{u}_a + \nabla_\Gamma^{\left(d_f\right)} \mathbf{u}_a^\mathrm{T} \right) \\
  & + {\eta\over2} \left( \nabla_\Gamma^{\left(d_f\right)} \mathbf{u} + \nabla_\Gamma^{\left(d_f\right)} \mathbf{u}^\mathrm{T} \right) : \left( \nabla_\Gamma^{\left(d_f, \delta d_f\right)} \mathbf{u}_a + \nabla_\Gamma^{\left(d_f, \delta d_f\right)} \mathbf{u}_a^\mathrm{T} \right) - p \, \mathrm{div}_\Gamma^{\left( d_f, \delta d_f \right)} \mathbf{u}_a \\
  & - p_a \mathrm{div}_\Gamma^{\left( d_f, \delta d_f \right)} \mathbf{u} + \lambda \mathbf{u}_a \cdot \mathbf{n}_\Gamma^{\left( d_f, \delta d_f \right)} + \lambda_a \mathbf{u} \cdot \mathbf{n}_\Gamma^{\left( d_f, \delta d_f \right)} \bigg] M^{\left( d_f \right)} + \bigg[ \rho \left( \mathbf{u} \cdot \nabla_\Gamma^{\left(d_f\right)} \right) \mathbf{u} \\
  & \cdot \mathbf{u}_a + {\eta\over2} \left( \nabla_\Gamma^{\left(d_f\right)} \mathbf{u} + \nabla_\Gamma^{\left(d_f\right)} \mathbf{u}^\mathrm{T} \right) : \left( \nabla_\Gamma^{\left(d_f\right)} \mathbf{u}_a + \nabla_\Gamma^{\left(d_f\right)} \mathbf{u}_a^\mathrm{T} \right) - p \, \mathrm{div}_\Gamma^{\left( d_f \right)} \mathbf{u}_a \\
  & - p_a \, \mathrm{div}_\Gamma^{\left( d_f \right)} \mathbf{u} + \alpha \mathbf{u} \cdot \mathbf{u}_a + \lambda \mathbf{u}_a \cdot \mathbf{n}_\Gamma^{\left( d_f \right)} + \lambda_a \mathbf{u} \cdot \mathbf{n}_\Gamma^{\left( d_f \right)} \bigg] M^{\left( d_f, \delta d_f \right)} \\
  & + \bigg[ \left( \rho C_p \mathbf{u} \cdot \nabla_\Gamma^{\left(d_f, \delta d_f\right)} T \right) T_a + k \nabla_\Gamma^{\left(d_f, \delta d_f\right)} T \cdot \nabla_\Gamma^{\left(d_f\right)} T_a + k \nabla_\Gamma^{\left(d_f\right)} T \cdot \nabla_\Gamma^{\left(d_f, \delta d_f\right)} T_a \bigg] \\
  & M^{\left( d_f \right)} + \left[ \left( \rho C_p \mathbf{u} \cdot \nabla_\Gamma^{\left(d_f\right)} T - Q \right) T_a + k \nabla_\Gamma^{\left(d_f\right)} T \cdot \nabla_\Gamma^{\left(d_f\right)} T_a \right] M^{\left( d_f, \delta d_f \right)} \\
  & + f_{id,\Gamma} r_f^2 \left( \nabla_\Gamma^{\left( d_f, \delta d_f \right)} \gamma_f \cdot \nabla_\Gamma^{\left( d_f \right)} \gamma_{fa} + \nabla_\Gamma^{\left( d_f \right)} \gamma_f \cdot \nabla_\Gamma^{\left( d_f, \delta d_f \right)} \gamma_{fa} \right) M^{\left( d_f \right)} \\
  & + f_{id,\Gamma} \left( r_f^2 \nabla_\Gamma^{\left( d_f \right)} \gamma_f \cdot \nabla_\Gamma^{\left( d_f \right)} \gamma_{fa} + \gamma_f \gamma_{fa} - \gamma \gamma_{fa} \right) M^{\left( d_f, \delta d_f \right)} \\
  & + r_m^2 \nabla_\Sigma \delta d_f \cdot \nabla_\Sigma d_{fa} + \delta d_f d_{fa} \,\mathrm{d}\Sigma \\
  & - \sum_{E_\Sigma\in\mathcal{E}_\Sigma} \int_{E_\Sigma} \tau_{LS\mathbf{u},\Gamma}^{\left( d_f, \delta d_f \right)} \left( \rho \mathbf{u} \cdot \nabla_\Gamma^{\left( d_f \right)} \mathbf{u} + \nabla_\Gamma^{\left( d_f \right)} p + \alpha \mathbf{u} \right) \cdot \left( \rho \mathbf{u} \cdot \nabla_\Gamma^{\left( d_f \right)} \mathbf{u}_a + \nabla_\Gamma^{\left( d_f \right)} p_a \right) \\
  & M^{\left( d_f \right)} + \tau_{LS\mathbf{u},\Gamma}^{\left( d_f \right)} \left( \rho \mathbf{u} \cdot \nabla_\Gamma^{\left( d_f, \delta d_f \right)} \mathbf{u} + \nabla_\Gamma^{\left( d_f, \delta d_f \right)} p \right) \cdot \left( \rho \mathbf{u} \cdot \nabla_\Gamma^{\left( d_f \right)} \mathbf{u}_a + \nabla_\Gamma^{\left( d_f \right)} p_a \right) M^{\left( d_f \right)} \\
  & + \tau_{LS\mathbf{u},\Gamma}^{\left( d_f \right)} \left( \rho \mathbf{u} \cdot \nabla_\Gamma^{\left( d_f \right)} \mathbf{u} + \nabla_\Gamma^{\left( d_f \right)} p + \alpha \mathbf{u} \right) \cdot \left[ \rho \mathbf{u} \cdot \nabla_\Gamma^{\left( d_f, \delta d_f \right)} \mathbf{u}_a + \nabla_\Gamma^{\left( d_f, \delta d_f \right)} p_a \right] M^{\left( d_f \right)} \\
  & + \tau_{LS\mathbf{u},\Gamma}^{\left( d_f \right)} \left( \rho \mathbf{u} \cdot \nabla_\Gamma^{\left( d_f \right)} \mathbf{u} + \nabla_\Gamma^{\left( d_f \right)} p + \alpha \mathbf{u} \right) \cdot \left( \rho \mathbf{u} \cdot \nabla_\Gamma^{\left( d_f \right)} \mathbf{u}_a + \nabla_\Gamma^{\left( d_f \right)} p_a \right) M^{\left( d_f, \delta d_f \right)} \\
  & + \tau_{LSp,\Gamma}^{\left( d_f, \delta d_f \right)} \left( \rho \mathrm{div}_\Gamma^{\left( d_f \right)} \mathbf{u} \right) \left( \mathrm{div}_\Gamma^{\left( d_f \right)} \mathbf{u}_a \right) M^{\left( d_f \right)} + \tau_{LSp,\Gamma}^{\left( d_f \right)} \left( \rho \mathrm{div}_\Gamma^{\left( d_f, \delta d_f \right)} \mathbf{u} \right) \left( \mathrm{div}_\Gamma^{\left( d_f \right)} \mathbf{u}_a \right) \\
  & M^{\left( d_f \right)} + \tau_{LSp,\Gamma}^{\left( d_f \right)} \left( \rho \mathrm{div}_\Gamma^{\left( d_f \right)} \mathbf{u} \right) \left( \mathrm{div}_\Gamma^{\left( d_f, \delta d_f \right)} \mathbf{u}_a \right) M^{\left( d_f \right)} + \tau_{LSp,\Gamma}^{\left( d_f \right)} \left( \rho \mathrm{div}_\Gamma^{\left( d_f \right)} \mathbf{u} \right) \\
  & \left( \mathrm{div}_\Gamma^{\left( d_f \right)} \mathbf{u}_a \right) M^{\left( d_f, \delta d_f \right)} - \tau_{LST,\Gamma}^{\left( d_f, \delta d_f \right)} \left( \rho C_p \mathbf{u} \cdot \nabla_\Gamma^{\left( d_f \right)} T - Q \right) \left( \rho C_p \mathbf{u} \cdot \nabla_\Gamma^{\left( d_f \right)} T_a \right) M^{\left( d_f \right)} \\
  & - \tau_{LST,\Gamma}^{\left( d_f \right)} \left( \rho C_p \mathbf{u} \cdot \nabla_\Gamma^{\left( d_f, \delta d_f \right)} T \right) \left( \rho C_p \mathbf{u} \cdot \nabla_\Gamma^{\left( d_f \right)} T_a \right) M^{\left( d_f \right)} \\
  & - \tau_{LST,\Gamma}^{\left( d_f \right)} \left( \rho C_p \mathbf{u} \cdot \nabla_\Gamma^{\left( d_f \right)} T - Q \right) \left( \rho C_p \mathbf{u} \cdot \nabla_\Gamma^{\left( d_f, \delta d_f \right)} T_a \right) M^{\left( d_f \right)} \\
  & - \tau_{LST,\Gamma}^{\left( d_f \right)} \left( \rho C_p \mathbf{u} \cdot \nabla_\Gamma^{\left( d_f \right)} T - Q \right) \left( \rho C_p \mathbf{u} \cdot \nabla_\Gamma^{\left( d_f \right)} T_a \right) M^{\left( d_f, \delta d_f \right)} \,\mathrm{d}\Sigma = 0.
\end{split}
\end{equation}
The constraints in Eqs. \ref{equ:ConstraintForAugmentedLagrangianCHM} and \ref{equ:ConstraintForVariationalAugmentedLagrangianCHM} are imposed to Eqs. \ref{equ:WeakAdjEquSCHMEqu}, \ref{equ:WeakAdjEquSNSEquCHM}, \ref{equ:WeakAdjEquSPDEFilterGaCHM} and \ref{equ:WeakAdjEquSPDEFilterDmObjCHM}. Then, the adjoint sensitivity of $J_T$ is derived as
\begin{equation}\label{equ:AdjSensitivityGaDmVariationalFormObjCHM}
\begin{split}
\delta \hat{J}_T = - \int_{\Sigma_D} \gamma_{fa} \delta \gamma M^{\left( d_f \right)} \,\mathrm{d}\Sigma - \int_\Sigma A_d d_{fa} \delta d_m \,\mathrm{d}\Sigma.
\end{split}
\end{equation}

Without losing the arbitrariness of $\delta \mathbf{u}$, $\delta p$, $\delta \lambda$, $\delta T$, $\delta \gamma_f$, $\delta d_f$, $\delta \gamma$ and $\delta d_m$, one can set 
\begin{equation}
\left.\begin{split}
& \tilde{\mathbf{u}}_a = \delta \mathbf{u} \\
& \tilde{p}_a = \delta p \\
& \tilde{\lambda}_a = \delta \lambda \\
& \tilde{T}_a = \delta T \\
& \tilde{\gamma}_{fa} = \delta \gamma_f \\
& \tilde{d}_{fa} = \delta d_f \\
& \tilde{\gamma} = \delta \gamma \\
& \tilde{d}_m = \delta d_m 
\end{split}\right\}
~\mathrm{with}~
\left\{\begin{split}
& \forall \tilde{\mathbf{u}}_a \in \left(\mathcal{H}\left(\Sigma\right)\right)^3 \\
& \forall \tilde{p}_a \in \mathcal{H}\left(\Sigma\right)\\
& \forall \tilde{\lambda}_a \in \mathcal{L}^2\left(\Sigma\right) \\
& \forall \tilde{T}_a \in \mathcal{H}\left(\Sigma\right) \\
& \forall \tilde{\gamma}_{fa} \in \mathcal{H}\left(\Sigma_D\right) \\
& \forall \tilde{d}_{fa} \in \mathcal{H}\left(\Sigma\right)\\
& \forall \tilde{\gamma} \in \mathcal{L}^2\left(\Sigma_D\right) \\
& \forall \tilde{d}_m \in \mathcal{L}^2\left(\Sigma\right)
\end{split}\right.
\end{equation}
for Eqs. \ref{equ:WeakAdjEquSCHMEqu}, \ref{equ:WeakAdjEquSNSEquCHM}, \ref{equ:WeakAdjEquSPDEFilterGaCHM} and \ref{equ:WeakAdjEquSPDEFilterDmObjCHM} to derive the adjoint system composed of Eqs. \ref{equ:WeakAdjEquSCHMEquMHMObj}, \ref{equ:AdjSurfaceNavierStokesEqusJObjectiveCHMObj}, \ref{equ:AdjPDEFilterJObjectiveGaCHMObj} and \ref{equ:AdjPDEFilterJObjectiveDmCHMObj}.

\subsection{Adjoint analysis for constraint of dissipation power in Eq. \ref{equ:VarProToopSurfaceNSCHM}} \label{sec:AdjointAnalysisDissipationConstraintMHM}

Based on the variational formulations of the surface Navier-Stokes equations in Eq. \ref{equ:TransformedVariationalFormulationSurfaceNSEqusHM} and the surface-PDE filters in Eqs. \ref{equ:VariationalFormulationPDEFilterBaseManifoldMHM} and \ref{equ:VariationalFormulationPDEFilterMHM}, the augmented Lagrangian of the dissipation power $\Phi$ in Eq. \ref{equ:TransformedDissipationSurfaceNSCHM} can be derived as
\begin{equation}\label{equ:AugmentedLagrangianDissipationPower}
\begin{split}
  \hat{\Phi} = & \int_\Sigma \bigg[ {\eta\over2} \left( \nabla_\Gamma^{\left( d_f \right)} \mathbf{u} + \nabla_\Gamma^{\left( d_f \right)} \mathbf{u}^\mathrm{T} \right) : \left( \nabla_\Gamma^{\left( d_f \right)} \mathbf{u} + \nabla_\Gamma^{\left( d_f \right)} \mathbf{u}^\mathrm{T} \right) + \alpha \mathbf{u}^2 + \rho \left( \mathbf{u} \cdot \nabla_\Gamma^{\left(d_f\right)} \right) \mathbf{u} \cdot \mathbf{u}_a \\
  & + {\eta\over2} \left( \nabla_\Gamma^{\left(d_f\right)} \mathbf{u} + \nabla_\Gamma^{\left(d_f\right)} \mathbf{u}^\mathrm{T} \right) : \left( \nabla_\Gamma^{\left(d_f\right)} \mathbf{u}_a + \nabla_\Gamma^{\left(d_f\right)} \mathbf{u}_a^\mathrm{T} \right) - p \, \mathrm{div}_\Gamma^{\left( d_f \right)} \mathbf{u}_a - p_a \mathrm{div}_\Gamma^{\left( d_f \right)} \mathbf{u} + \alpha \mathbf{u} \\
  & \cdot \mathbf{u}_a + \lambda \mathbf{u}_a \cdot \mathbf{n}_\Gamma^{\left( d_f \right)} + \lambda_a \mathbf{u} \cdot \mathbf{n}_\Gamma^{\left( d_f \right)} + f_{id,\Gamma} \left( r_f^2 \nabla_\Gamma^{\left( d_f \right)} \gamma_f \cdot \nabla_\Gamma^{\left( d_f \right)} \gamma_{fa} + \gamma_f \gamma_{fa} - \gamma \gamma_{fa} \right) \bigg] \\
  & M^{\left( d_f \right)} + r_m^2 \nabla_\Sigma d_f \cdot \nabla_\Sigma d_{fa} + d_f d_{fa} - A_d \left( d_m - {1\over2} \right) d_{fa} \,\mathrm{d}\Sigma - \sum_{E_\Sigma\in\mathcal{E}_\Sigma} \int_{E_\Sigma} \tau_{LS\mathbf{u},\Gamma}^{\left( d_f \right)} \\
  & \bigg( \rho \mathbf{u} \cdot \nabla_\Gamma^{\left( d_f \right)} \mathbf{u} + \nabla_\Gamma^{\left( d_f \right)} p + \alpha \mathbf{u} \bigg) \cdot \left( \rho \mathbf{u} \cdot \nabla_\Gamma^{\left( d_f \right)} \mathbf{u}_a + \nabla_\Gamma^{\left( d_f \right)} p_a \right) M^{\left( d_f \right)} \\
  & + \tau_{LSp,\Gamma}^{\left( d_f \right)} \left( \rho \mathrm{div}_\Gamma^{\left( d_f \right)} \mathbf{u} \right) \left( \mathrm{div}_\Gamma^{\left( d_f \right)} \mathbf{u}_a \right) M^{\left( d_f \right)} \,\mathrm{d}\Sigma,
\end{split}
\end{equation}
where $\hat{\Phi}$ is the augmented lagrangian of $\Phi$; and the adjoint variables satisfy
\begin{equation}\label{equ:ConstraintForAugmentedLagrangianDissipationPower}
  \left.\begin{split}
  & \mathbf{u}_a \in\left(\mathcal{H}\left(\Sigma\right)\right)^3 \\
  & p_a \in \mathcal{H}\left(\Sigma\right) \\
  & \lambda_a \in \mathcal{L}^2\left(\Sigma\right) \\
  & \gamma_{fa} \in \mathcal{H}\left(\Sigma_D\right) \\
  & d_{fa} \in \mathcal{H}\left(\Sigma\right)
  \end{split}\right\}~\mathrm{with}~
  \left\{\begin{split}
  & \mathbf{u}_a = \mathbf{0} ~ \mathrm{at} ~ \forall \mathbf{x}_\Sigma \in l_{v,\Sigma} \cup l_{v_0,\Sigma} \\
  & \lambda_a = 0 ~ \mathrm{at} ~ \forall \mathbf{x}_\Sigma \in l_{v,\Sigma} \cup l_{v_0,\Sigma} 
  \end{split}\right..
\end{equation}
Based on the related results in Section \ref{sec:MethodologyFiberBundleTOOPTransferSurfaceFlows}, the first order variational of $\hat{\Phi}$ can be derived as
\begin{equation}\label{equ:1stVarAugmentedLagrangianConstrDissipationPower}
\begin{split}
  \delta \hat{\Phi} = & \int_\Sigma \bigg\{ \eta \left( \nabla_\Gamma^{\left( d_f \right)} \delta \mathbf{u} + \nabla_\Gamma^{\left( d_f \right)} \delta \mathbf{u}^\mathrm{T} \right) : \left( \nabla_\Gamma^{\left( d_f \right)} \mathbf{u} + \nabla_\Gamma^{\left( d_f \right)} \mathbf{u}^\mathrm{T} \right) + \eta \bigg( \nabla_\Gamma^{\left( d_f, \delta d_f \right)} \mathbf{u} + \nabla_\Gamma^{\left( d_f, \delta d_f \right)} \mathbf{u}^\mathrm{T} \bigg) \\
  & : \left( \nabla_\Gamma^{\left( d_f \right)} \mathbf{u} + \nabla_\Gamma^{\left( d_f \right)} \mathbf{u}^\mathrm{T} \right) + 2 \alpha \mathbf{u} \cdot \delta \mathbf{u} + {\partial \alpha \over \partial \gamma_p} {\partial \gamma_p \over \partial \gamma_f} \mathbf{u}^2 \delta \gamma_f + \rho \bigg( \delta \mathbf{u} \cdot \nabla_\Gamma^{\left(d_f\right)} \mathbf{u} + \mathbf{u} \cdot \nabla_\Gamma^{\left(d_f, \delta d_f\right)} \mathbf{u} \\
  & + \mathbf{u} \cdot \nabla_\Gamma^{\left(d_f\right)} \delta \mathbf{u} \bigg) \cdot \mathbf{u}_a + {\eta\over2} \bigg( \nabla_\Gamma^{\left(d_f\right)} \delta \mathbf{u} + \nabla_\Gamma^{\left(d_f\right)} \delta \mathbf{u}^\mathrm{T} \bigg) : \left( \nabla_\Gamma^{\left(d_f\right)} \mathbf{u}_a + \nabla_\Gamma^{\left(d_f\right)} \mathbf{u}_a^\mathrm{T} \right) \\
  & + {\eta\over2} \bigg( \nabla_\Gamma^{\left(d_f, \delta d_f\right)} \mathbf{u} + \nabla_\Gamma^{\left(d_f, \delta d_f\right)} \mathbf{u}^\mathrm{T} \bigg) : \bigg( \nabla_\Gamma^{\left(d_f\right)} \mathbf{u}_a + \nabla_\Gamma^{\left(d_f\right)} \mathbf{u}_a^\mathrm{T} \bigg) + {\eta\over2} \left( \nabla_\Gamma^{\left(d_f\right)} \mathbf{u} + \nabla_\Gamma^{\left(d_f\right)} \mathbf{u}^\mathrm{T} \right) \\
  & : \bigg( \nabla_\Gamma^{\left(d_f, \delta d_f\right)} \mathbf{u}_a + \nabla_\Gamma^{\left(d_f, \delta d_f\right)} \mathbf{u}_a^\mathrm{T} \bigg) - \delta p \, \mathrm{div}_\Gamma^{\left( d_f \right)} \mathbf{u}_a - p \, \mathrm{div}_\Gamma^{\left( d_f, \delta d_f \right)} \mathbf{u}_a - p_a \mathrm{div}_\Gamma^{\left( d_f, \delta d_f \right)} \mathbf{u} \\
  & - p_a \mathrm{div}_\Gamma^{\left( d_f \right)} \delta \mathbf{u} + {\partial \alpha \over \partial \gamma_p} {\partial \gamma_p \over \partial \gamma_f} \mathbf{u} \cdot \mathbf{u}_a \delta \gamma_f + \alpha \delta \mathbf{u} \cdot \mathbf{u}_a + \delta \lambda \mathbf{u}_a \cdot \mathbf{n}_\Gamma^{\left( d_f \right)} + \lambda \mathbf{u}_a \cdot \mathbf{n}_\Gamma^{\left( d_f, \delta d_f \right)} \\
  & + \lambda_a \delta \mathbf{u} \cdot \mathbf{n}_\Gamma^{\left( d_f \right)} + \lambda_a \mathbf{u} \cdot \mathbf{n}_\Gamma^{\left( d_f, \delta d_f \right)} + f_{id,\Gamma} \bigg[ r_f^2 \bigg( \nabla_\Gamma^{\left( d_f \right)} \delta \gamma_f \cdot \nabla_\Gamma^{\left( d_f \right)} \gamma_{fa} + \nabla_\Gamma^{\left( d_f, \delta d_f \right)} \gamma_f \\
  & \cdot \nabla_\Gamma^{\left( d_f \right)} \gamma_{fa} + \nabla_\Gamma^{\left( d_f \right)} \gamma_f \cdot \nabla_\Gamma^{\left( d_f, \delta d_f \right)} \gamma_{fa} \bigg) + \delta \gamma_f \gamma_{fa} - \delta \gamma \gamma_{fa} \bigg] \bigg\} M^{\left( d_f \right)} + \bigg[ {\eta\over2} \bigg( \nabla_\Gamma^{\left( d_f \right)} \mathbf{u} \\
  & + \nabla_\Gamma^{\left( d_f \right)} \mathbf{u}^\mathrm{T} \bigg) : \bigg( \nabla_\Gamma^{\left( d_f \right)} \mathbf{u} + \nabla_\Gamma^{\left( d_f \right)} \mathbf{u}^\mathrm{T} \bigg) + \alpha \mathbf{u}^2 + \rho \left( \mathbf{u} \cdot \nabla_\Gamma^{\left(d_f\right)} \right) \mathbf{u} \cdot \mathbf{u}_a + {\eta\over2} \bigg( \nabla_\Gamma^{\left(d_f\right)} \mathbf{u} \\
  & + \nabla_\Gamma^{\left(d_f\right)} \mathbf{u}^\mathrm{T} \bigg) : \bigg( \nabla_\Gamma^{\left(d_f\right)} \mathbf{u}_a + \nabla_\Gamma^{\left(d_f\right)} \mathbf{u}_a^\mathrm{T} \bigg) - p \, \mathrm{div}_\Gamma^{\left( d_f \right)} \mathbf{u}_a - p_a \mathrm{div}_\Gamma^{\left( d_f \right)} \mathbf{u} + \alpha \mathbf{u} \cdot \mathbf{u}_a + \lambda \mathbf{u}_a \\
  & \cdot \mathbf{n}_\Gamma^{\left( d_f \right)} + \lambda_a \mathbf{u} \cdot \mathbf{n}_\Gamma^{\left( d_f \right)} + f_{id,\Gamma} \cdot \left( r_f^2 \nabla_\Gamma^{\left( d_f \right)} \gamma_f \cdot \nabla_\Gamma^{\left( d_f \right)} \gamma_{fa} + \gamma_f \gamma_{fa} - \gamma \gamma_{fa} \right) \bigg] M^{\left( d_f, \delta d_f \right)} \\
  & + r_m^2 \nabla_\Sigma \delta d_f \cdot \nabla_\Sigma d_{fa} + \delta d_f d_{fa} - A_d \delta d_m d_{fa} \,\mathrm{d}\Sigma - \sum_{E_\Sigma\in\mathcal{E}_\Sigma} \int_{E_\Sigma} \bigg\{ \left( \tau_{LS\mathbf{u},\Gamma}^{\left( d_f, \delta d_f \right)} + \tau_{LS\mathbf{u},\Gamma}^{\left( d_f, \delta \mathbf{u} \right)} \right) \\
  & \left( \rho \mathbf{u} \cdot \nabla_\Gamma^{\left( d_f \right)} \mathbf{u} + \nabla_\Gamma^{\left( d_f \right)} p + \alpha \mathbf{u} \right) \cdot \bigg( \rho \mathbf{u} \cdot \nabla_\Gamma^{\left( d_f \right)} \mathbf{u}_a + \nabla_\Gamma^{\left( d_f \right)} p_a \bigg) + \tau_{LS\mathbf{u},\Gamma}^{\left( d_f \right)} \bigg[ \rho \bigg( \delta \mathbf{u} \cdot \nabla_\Gamma^{\left( d_f \right)} \mathbf{u} \\
  & + \mathbf{u} \cdot \nabla_\Gamma^{\left( d_f, \delta d_f \right)} \mathbf{u} + \mathbf{u} \cdot \nabla_\Gamma^{\left( d_f \right)} \delta \mathbf{u} \bigg) + \nabla_\Gamma^{\left( d_f, \delta d_f \right)} p + \nabla_\Gamma^{\left( d_f \right)} \delta p + {\partial \alpha \over \partial \gamma_p} {\partial \gamma_p \over \partial \gamma_f} \mathbf{u} \delta \gamma_f + \alpha \delta \mathbf{u} \bigg] \\
  & \cdot \left( \rho \mathbf{u} \cdot \nabla_\Gamma^{\left( d_f \right)} \mathbf{u}_a + \nabla_\Gamma^{\left( d_f \right)} p_a \right) + \tau_{LS\mathbf{u},\Gamma}^{\left( d_f \right)} \bigg( \rho \mathbf{u} \cdot \nabla_\Gamma^{\left( d_f \right)} \mathbf{u} + \nabla_\Gamma^{\left( d_f \right)} p + \alpha \mathbf{u} \bigg) \\
  & \cdot \bigg[ \rho \left( \delta \mathbf{u} \cdot \nabla_\Gamma^{\left( d_f \right)} \mathbf{u}_a + \mathbf{u} \cdot \nabla_\Gamma^{\left( d_f, \delta d_f \right)} \mathbf{u}_a \right) + \nabla_\Gamma^{\left( d_f, \delta d_f \right)} p_a \bigg] + \bigg( \tau_{LSp,\Gamma}^{\left( d_f, \delta d_f \right)} + \tau_{LSp,\Gamma}^{\left( d_f, \delta \mathbf{u} \right)} \bigg) \\
  & \left( \rho \mathrm{div}_\Gamma^{\left( d_f \right)} \mathbf{u} \right) \left( \mathrm{div}_\Gamma^{\left( d_f \right)} \mathbf{u}_a \right) + \tau_{LSp,\Gamma}^{\left( d_f \right)} \rho \left( \mathrm{div}_\Gamma^{\left( d_f, \delta d_f \right)} \mathbf{u} + \mathrm{div}_\Gamma^{\left( d_f \right)} \delta \mathbf{u} \right) \\
  & \left( \mathrm{div}_\Gamma^{\left( d_f \right)} \mathbf{u}_a \right) + \tau_{LSp,\Gamma}^{\left( d_f \right)} \left( \rho \mathrm{div}_\Gamma^{\left( d_f \right)} \mathbf{u} \right) \left( \mathrm{div}_\Gamma^{\left( d_f, \delta d_f \right)} \mathbf{u}_a \right) \bigg\} M^{\left( d_f \right)} \\
  & + \bigg\{ \tau_{LS\mathbf{u},\Gamma}^{\left( d_f \right)} \bigg( \rho \mathbf{u} \cdot \nabla_\Gamma^{\left( d_f \right)} \mathbf{u} + \nabla_\Gamma^{\left( d_f \right)} p + \alpha \mathbf{u} \bigg) \cdot \left( \rho \mathbf{u} \cdot \nabla_\Gamma^{\left( d_f \right)} \mathbf{u}_a + \nabla_\Gamma^{\left( d_f \right)} p_a \right) \\
  & + \tau_{LSp,\Gamma}^{\left( d_f \right)} \left( \rho \mathrm{div}_\Gamma^{\left( d_f \right)} \mathbf{u} \right) \left( \mathrm{div}_\Gamma^{\left( d_f \right)} \mathbf{u}_a \right) \bigg\} M^{\left( d_f, \delta d_f \right)} \,\mathrm{d}\Sigma
\end{split}
\end{equation}
with the satisfication of the constraints in Eq. \ref{equ:ConstraintForAugmentedLagrangianDissipationPower}
and 
\begin{equation}\label{equ:ConstraintForVariationalAugmentedLagrangianDissipationPower}
  \left.\begin{split}
  & \delta \mathbf{u} \in\left(\mathcal{H}\left(\Sigma\right)\right)^3 \\
  & \delta p \in \mathcal{H}\left(\Sigma\right) \\
  & \delta \lambda \in \mathcal{L}^2\left(\Sigma\right) \\
  & \delta \gamma_f \in \mathcal{H}\left(\Sigma_D\right) \\
  & \delta \gamma \in \mathcal{L}^2\left(\Sigma_D\right) \\
  & \delta d_f \in \mathcal{H}\left(\Sigma\right) \\
  & \delta d_m \in \mathcal{L}^2\left(\Sigma\right)
  \end{split}\right\} 
  ~ \mathrm{with} ~ 
  \left\{\begin{split}
  & \delta \mathbf{u} = \mathbf{0} ~ \mathrm{at} ~ \forall \mathbf{x} \in l_{v,\Sigma} \cup l_{v_0,\Sigma} \\
  & \delta \lambda = 0 ~ \mathrm{at} ~ \forall \mathbf{x} \in l_{v,\Sigma} \cup l_{v_0,\Sigma}
  \end{split}\right..
\end{equation}

According to the Karush-Kuhn-Tucker conditions of the PDE constrained optimization problem, the first order variational of the augmented Lagrangian to $\mathbf{u}$, $p$ and $\lambda$ can be set to be zero as
\begin{equation}\label{equ:WeakAdjEquSNSEquDissipationPower}
\begin{split}
  & \int_\Sigma \bigg\{ \eta \left( \nabla_\Gamma^{\left( d_f \right)} \delta \mathbf{u} + \nabla_\Gamma^{\left( d_f \right)} \delta \mathbf{u}^\mathrm{T} \right) : \left( \nabla_\Gamma^{\left( d_f \right)} \mathbf{u} + \nabla_\Gamma^{\left( d_f \right)} \mathbf{u}^\mathrm{T} \right) + 2 \alpha \mathbf{u} \cdot \delta \mathbf{u} + \rho \bigg[ \left( \delta \mathbf{u} \cdot \nabla_\Gamma^{\left(d_f\right)} \right) \mathbf{u} \\
  & + \left( \mathbf{u} \cdot \nabla_\Gamma^{\left(d_f\right)} \right) \delta \mathbf{u} \bigg] \cdot \mathbf{u}_a + {\eta\over2} \left( \nabla_\Gamma^{\left(d_f\right)} \delta \mathbf{u} + \nabla_\Gamma^{\left(d_f\right)} \delta \mathbf{u}^\mathrm{T} \right) : \left( \nabla_\Gamma^{\left(d_f\right)} \mathbf{u}_a + \nabla_\Gamma^{\left(d_f\right)} \mathbf{u}_a^\mathrm{T} \right) \\
  & - \delta p \, \mathrm{div}_\Gamma^{\left( d_f \right)} \mathbf{u}_a - p_a \mathrm{div}_\Gamma^{\left( d_f \right)} \delta \mathbf{u} + \alpha \delta \mathbf{u} \cdot \mathbf{u}_a + \delta \lambda \mathbf{u}_a \cdot \mathbf{n}_\Gamma^{\left( d_f \right)} + \lambda_a \delta \mathbf{u} \cdot \mathbf{n}_\Gamma^{\left( d_f \right)} \bigg\} M^{\left( d_f \right)} \,\mathrm{d}\Sigma \\
  & - \sum_{E_\Sigma\in\mathcal{E}_\Sigma} \int_{E_\Sigma} \bigg\{ \tau_{LS\mathbf{u},\Gamma}^{\left( d_f, \delta \mathbf{u} \right)} \left( \rho \mathbf{u} \cdot \nabla_\Gamma^{\left( d_f \right)} \mathbf{u} + \nabla_\Gamma^{\left( d_f \right)} p + \alpha \mathbf{u} \right) \cdot \left( \rho \mathbf{u} \cdot \nabla_\Gamma^{\left( d_f \right)} \mathbf{u}_a + \nabla_\Gamma^{\left( d_f \right)} p_a \right) \\
  & + \tau_{LS\mathbf{u},\Gamma}^{\left( d_f \right)} \bigg[ \rho \left( \delta \mathbf{u} \cdot \nabla_\Gamma^{\left( d_f \right)} \mathbf{u} + \mathbf{u} \cdot \nabla_\Gamma^{\left( d_f \right)} \delta \mathbf{u} \right) + \nabla_\Gamma^{\left( d_f \right)} \delta p + \alpha \delta \mathbf{u} \bigg] \cdot \left( \rho \mathbf{u} \cdot \nabla_\Gamma^{\left( d_f \right)} \mathbf{u}_a + \nabla_\Gamma^{\left( d_f \right)} p_a \right) \\
  & + \tau_{LS\mathbf{u},\Gamma}^{\left( d_f \right)} \left( \rho \mathbf{u} \cdot \nabla_\Gamma^{\left( d_f \right)} \mathbf{u} + \nabla_\Gamma^{\left( d_f \right)} p + \alpha \mathbf{u} \right) \cdot \left( \rho \delta \mathbf{u} \cdot \nabla_\Gamma^{\left( d_f \right)} \mathbf{u}_a \right) + \tau_{LSp,\Gamma}^{\left( d_f, \delta \mathbf{u} \right)} \left( \rho \mathrm{div}_\Gamma^{\left( d_f \right)} \mathbf{u} \right) \\
  & \left( \mathrm{div}_\Gamma^{\left( d_f \right)} \mathbf{u}_a \right) + \tau_{LSp,\Gamma}^{\left( d_f \right)} \left( \rho \mathrm{div}_\Gamma^{\left( d_f \right)} \delta \mathbf{u} \right) \left( \mathrm{div}_\Gamma^{\left( d_f \right)} \mathbf{u}_a \right) \bigg\} M^{\left( d_f \right)} \,\mathrm{d}\Sigma = 0, \\
\end{split}
\end{equation}
the first order variational of the augmented Lagrangian to the variable $\gamma_f$ can be set to be zero as
\begin{equation}\label{equ:AdjEquAreaConstr1GaInformDissipationPower} 
\begin{split}
  & \int_{\Sigma_D} \left( {\partial \alpha \over \partial \gamma_p} {\partial \gamma_p \over \partial \gamma_f} \mathbf{u}^2 \delta \gamma_f + {\partial \alpha \over \partial \gamma_p} {\partial \gamma_p \over \partial \gamma_f} \mathbf{u} \cdot \mathbf{u}_a \delta \gamma_f + r_f^2 \nabla_\Gamma^{\left( d_f \right)} \delta \gamma_f \cdot \nabla_\Gamma^{\left( d_f \right)} \gamma_{fa} + \delta \gamma_f \gamma_{fa} \right) M^{\left( d_f \right)} \,\mathrm{d}\Sigma \\
  & - \sum_{E_\Sigma\in\mathcal{E}_\Sigma} \int_{E_\Sigma\cap\Sigma_D} \tau_{LS\mathbf{u},\Gamma}^{\left( d_f \right)} {\partial \alpha \over \partial \gamma_p} {\partial \gamma_p \over \partial \gamma_f} \mathbf{u} \cdot \left( \rho \mathbf{u} \cdot \nabla_\Gamma^{\left( d_f \right)} \mathbf{u}_a + \nabla_\Gamma^{\left( d_f \right)} p_a \right) \delta \gamma_f M^{\left( d_f \right)} \,\mathrm{d}\Sigma = 0,
\end{split}
\end{equation}
and the first order variational of the augmented Lagrangian to the variable $d_f$ can be set to be zero as
\begin{equation}\label{equ:AdjEquAreaConstr1DmInformDissipationPower} 
\begin{split}
  & \int_\Sigma \bigg[ \eta \left( \nabla_\Gamma^{\left( d_f, \delta d_f \right)} \mathbf{u} + \nabla_\Gamma^{\left( d_f, \delta d_f \right)} \mathbf{u}^\mathrm{T} \right) : \left( \nabla_\Gamma^{\left( d_f \right)} \mathbf{u} + \nabla_\Gamma^{\left( d_f \right)} \mathbf{u}^\mathrm{T} \right) + \rho \left( \mathbf{u} \cdot \nabla_\Gamma^{\left(d_f, \delta d_f\right)} \right) \mathbf{u} \cdot \mathbf{u}_a \\
  & + {\eta\over2} \left( \nabla_\Gamma^{\left(d_f, \delta d_f\right)} \mathbf{u} + \nabla_\Gamma^{\left(d_f, \delta d_f\right)} \mathbf{u}^\mathrm{T} \right) : \left( \nabla_\Gamma^{\left(d_f\right)} \mathbf{u}_a + \nabla_\Gamma^{\left(d_f\right)} \mathbf{u}_a^\mathrm{T} \right) + {\eta\over2} \left( \nabla_\Gamma^{\left(d_f\right)} \mathbf{u} + \nabla_\Gamma^{\left(d_f\right)} \mathbf{u}^\mathrm{T} \right) \\
  & : \left( \nabla_\Gamma^{\left(d_f, \delta d_f\right)} \mathbf{u}_a + \nabla_\Gamma^{\left(d_f, \delta d_f\right)} \mathbf{u}_a^\mathrm{T} \right) - p \, \mathrm{div}_\Gamma^{\left( d_f, \delta d_f \right)} \mathbf{u}_a - p_a \mathrm{div}_\Gamma^{\left( d_f, \delta d_f \right)} \mathbf{u} + \lambda \mathbf{u}_a \cdot \mathbf{n}_\Gamma^{\left( d_f, \delta d_f \right)} \\
  & + \lambda_a \mathbf{u} \cdot \mathbf{n}_\Gamma^{\left( d_f, \delta d_f \right)} + f_{id,\Gamma} r_f^2 \left( \nabla_\Gamma^{\left( d_f, \delta d_f \right)} \gamma_f \cdot \nabla_\Gamma^{\left( d_f \right)} \gamma_{fa} + \nabla_\Gamma^{\left( d_f \right)} \gamma_f \cdot \nabla_\Gamma^{\left( d_f, \delta d_f \right)} \gamma_{fa} \right) \bigg] M^{\left( d_f \right)} \\
  & + \bigg[ {\eta\over2} \left( \nabla_\Gamma^{\left( d_f \right)} \mathbf{u} + \nabla_\Gamma^{\left( d_f \right)} \mathbf{u}^\mathrm{T} \right) : \left( \nabla_\Gamma^{\left( d_f \right)} \mathbf{u} + \nabla_\Gamma^{\left( d_f \right)} \mathbf{u}^\mathrm{T} \right) + \alpha \mathbf{u}^2 + \rho \left( \mathbf{u} \cdot \nabla_\Gamma^{\left(d_f\right)} \right) \mathbf{u} \cdot \mathbf{u}_a \\
  & + {\eta\over2} \left( \nabla_\Gamma^{\left(d_f\right)} \mathbf{u} + \nabla_\Gamma^{\left(d_f\right)} \mathbf{u}^\mathrm{T} \right) : \left( \nabla_\Gamma^{\left(d_f\right)} \mathbf{u}_a + \nabla_\Gamma^{\left(d_f\right)} \mathbf{u}_a^\mathrm{T} \right) - p \, \mathrm{div}_\Gamma^{\left( d_f \right)} \mathbf{u}_a - p_a \mathrm{div}_\Gamma^{\left( d_f \right)} \mathbf{u} \\
  & + \alpha \mathbf{u} \cdot \mathbf{u}_a + \lambda \mathbf{u}_a \cdot \mathbf{n}_\Gamma^{\left( d_f \right)} + \lambda_a \mathbf{u} \cdot \mathbf{n}_\Gamma^{\left( d_f \right)} + f_{id,\Gamma} \bigg( r_f^2 \nabla_\Gamma^{\left( d_f \right)} \gamma_f \cdot \nabla_\Gamma^{\left( d_f \right)} \gamma_{fa} + \gamma_f \gamma_{fa} \\
  & - \gamma \gamma_{fa} \bigg) \bigg] M^{\left( d_f, \delta d_f \right)} + r_m^2 \nabla_\Sigma \delta d_f \cdot \nabla_\Sigma d_{fa} + \delta d_f d_{fa} \,\mathrm{d}\Sigma \\
  & - \sum_{E_\Sigma\in\mathcal{E}_\Sigma} \int_{E_\Sigma} \bigg[ \tau_{LS\mathbf{u},\Gamma}^{\left( d_f, \delta d_f \right)} \left( \rho \mathbf{u} \cdot \nabla_\Gamma^{\left( d_f \right)} \mathbf{u} + \nabla_\Gamma^{\left( d_f \right)} p + \alpha \mathbf{u} \right) \cdot \left( \rho \mathbf{u} \cdot \nabla_\Gamma^{\left( d_f \right)} \mathbf{u}_a + \nabla_\Gamma^{\left( d_f \right)} p_a \right) \\
  & + \tau_{LS\mathbf{u},\Gamma}^{\left( d_f \right)}  \left( \rho \mathbf{u} \cdot \nabla_\Gamma^{\left( d_f, \delta d_f \right)} \mathbf{u} + \nabla_\Gamma^{\left( d_f, \delta d_f \right)} p \right) \cdot \left( \rho \mathbf{u} \cdot \nabla_\Gamma^{\left( d_f \right)} \mathbf{u}_a + \nabla_\Gamma^{\left( d_f \right)} p_a \right) \\
  & + \tau_{LS\mathbf{u},\Gamma}^{\left( d_f \right)} \left( \rho \mathbf{u} \cdot \nabla_\Gamma^{\left( d_f \right)} \mathbf{u} + \nabla_\Gamma^{\left( d_f \right)} p + \alpha \mathbf{u} \right) \cdot \left( \rho \mathbf{u} \cdot \nabla_\Gamma^{\left( d_f, \delta d_f \right)} \mathbf{u}_a + \nabla_\Gamma^{\left( d_f, \delta d_f \right)} p_a \right) \\
  & + \tau_{LSp,\Gamma}^{\left( d_f, \delta d_f \right)} \left( \rho \mathrm{div}_\Gamma^{\left( d_f \right)} \mathbf{u} \right) \left( \mathrm{div}_\Gamma^{\left( d_f \right)} \mathbf{u}_a \right) + \tau_{LSp,\Gamma}^{\left( d_f \right)} \left( \rho \mathrm{div}_\Gamma^{\left( d_f, \delta d_f \right)} \mathbf{u} \right) \left( \mathrm{div}_\Gamma^{\left( d_f \right)} \mathbf{u}_a \right) \\
  & + \tau_{LSp,\Gamma}^{\left( d_f \right)} \left( \rho \mathrm{div}_\Gamma^{\left( d_f \right)} \mathbf{u} \right) \left( \mathrm{div}_\Gamma^{\left( d_f, \delta d_f \right)} \mathbf{u}_a \right) \bigg] M^{\left( d_f \right)} + \bigg[ \tau_{LS\mathbf{u},\Gamma}^{\left( d_f \right)} \left( \rho \mathbf{u} \cdot \nabla_\Gamma^{\left( d_f \right)} \mathbf{u} + \nabla_\Gamma^{\left( d_f \right)} p + \alpha \mathbf{u} \right) \\
  & \cdot \left( \rho \mathbf{u} \cdot \nabla_\Gamma^{\left( d_f \right)} \mathbf{u}_a + \nabla_\Gamma^{\left( d_f \right)} p_a \right) + \tau_{LSp,\Gamma}^{\left( d_f \right)} \left( \rho \mathrm{div}_\Gamma^{\left( d_f \right)} \mathbf{u} \right) \left( \mathrm{div}_\Gamma^{\left( d_f \right)} \mathbf{u}_a \right) \bigg] M^{\left( d_f, \delta d_f \right)} \,\mathrm{d}\Sigma = 0. \\
\end{split}
\end{equation}
The constraints in Eqs. \ref{equ:ConstraintForAugmentedLagrangianDissipationPower} and \ref{equ:ConstraintForVariationalAugmentedLagrangianDissipationPower} are imposed to Eqs. \ref{equ:WeakAdjEquSNSEquDissipationPower}, \ref{equ:AdjEquAreaConstr1GaInformDissipationPower} and \ref{equ:AdjEquAreaConstr1DmInformDissipationPower}. Then, the adjoint sensitivity of $\Phi$ is derived as
\begin{equation}\label{equ:AdjSensAreaConstr1InformDissipationPower}
\begin{split}
  \delta \hat{\Phi} = - \int_{\Sigma_D} \delta \gamma \gamma_{fa} M^{\left( d_f \right)} \,\mathrm{d}\Sigma - \int_\Sigma A_d \delta d_m d_{fa} \,\mathrm{d}\Sigma.
\end{split}
\end{equation}

Without losing the arbitrariness of $\delta \mathbf{u}$, $\delta p$, $\delta \lambda$, $\delta \gamma_f$, $\delta d_f$, $\delta \gamma$ and $\delta d_m$, one can set 
\begin{equation}
\left.\begin{split}
& \tilde{\mathbf{u}}_a = \delta \mathbf{u} \\
& \tilde{p}_a = \delta p \\
& \tilde{\lambda}_a = \delta \lambda \\
& \tilde{\gamma}_{fa} = \delta \gamma_f \\
& \tilde{d}_{fa} = \delta d_f \\
& \tilde{\gamma} = \delta \gamma \\
& \tilde{d}_m = \delta d_m 
\end{split}\right\}
~\mathrm{with}~
\left\{\begin{split}
& \forall \tilde{\mathbf{u}}_a \in \left(\mathcal{H}\left(\Sigma\right)\right)^3 \\
& \forall \tilde{p}_a \in \mathcal{H}\left(\Sigma\right)\\
& \forall \tilde{\lambda}_a \in \mathcal{L}^2\left(\Sigma\right) \\
& \forall \tilde{\gamma}_{fa} \in \mathcal{H}\left(\Sigma_D\right) \\
& \forall \tilde{d}_{fa} \in \mathcal{H}\left(\Sigma\right)\\
& \forall \tilde{\gamma} \in \mathcal{L}^2\left(\Sigma_D\right) \\
& \forall \tilde{d}_m \in \mathcal{L}^2\left(\Sigma\right)
\end{split}\right.
\end{equation}
for Eqs. \ref{equ:WeakAdjEquSNSEquDissipationPower}, \ref{equ:AdjEquAreaConstr1GaInformDissipationPower} and \ref{equ:AdjEquAreaConstr1DmInformDissipationPower} to derive the adjoint system composed of Eqs. \ref{equ:AdjEquSurfaceNSMHMDissipationPower}, \ref{equ:AdjPDEFilterDissipationPowerGaMHM} and \ref{equ:AdjPDEFilterJDissipationPowerDmMHM}.

\subsection{Adjoint analysis for area constraint in Eq. \ref{equ:VarProToopSurfaceNSCHM}} \label{sec:AdjointAnalysisAreaConstraintMHM}

The adjoint sensitivity of the area fraction of the surface structure can be derived from that of the area of the surface structure and the area of the implicit 2-manifold. The adjoint analysis in this section is implemented for those two areas.

\subsubsection{Adjoint analysis for area of surface structure}\label{sec:AdjointAnalysisAreaFluidStructureMHM}

Based on the variational formulations of the surface-PDE filters in Eqs. \ref{equ:VariationalFormulationPDEFilterBaseManifoldMHM} and \ref{equ:VariationalFormulationPDEFilterMHM}, the augmented Lagrangian of the area of the surface structure can be derived as
\begin{equation}\label{equ:AugLagAreaFluidStructureMHM}
\begin{split}
  \widehat{s\left|\Gamma_D\right|} = & \int_\Sigma f_{id,\Gamma} \left( \gamma_p + r_f^2 \nabla_\Gamma^{\left( d_f \right)} \gamma_f \cdot \nabla_\Gamma^{\left( d_f \right)} \gamma_{fa} + \gamma_f \gamma_{fa} - \gamma \gamma_{fa} \right) M^{\left( d_f \right)} \\
  & + r_m^2 \nabla_\Sigma d_f \cdot \nabla_\Sigma d_{fa} + d_f d_{fa} - A_d \left( d_m - {1\over2} \right) d_{fa} \,\mathrm{d}\Sigma,
\end{split}
\end{equation}
where $\widehat{s\left|\Gamma_D\right|}$ is the augmented lagrangian of $s\left|\Gamma_D\right|$; and the adjoint variables satisfy
\begin{equation}\label{equ:ConstraintForAugmentedLagrangianAreaFluidStructureMHM}
  \left\{\begin{split}
  & \gamma_{fa} \in \mathcal{H}\left(\Sigma_D\right) \\
  & d_{fa} \in \mathcal{H}\left(\Sigma\right)
  \end{split}\right..
\end{equation}

Based on the related results in Section \ref{sec:MethodologyFiberBundleTOOPTransferSurfaceFlows}, the first order variational of $\widehat{s\left|\Gamma_D\right|}$ can be derived as
\begin{equation}\label{equ:1stVarAugmentedLagrangianConstrAreaFluidStructure}
\begin{split}
  \delta \widehat{s\left|\Gamma_D\right|} = & \int_\Sigma f_{id,\Gamma} \bigg[ {\partial \gamma_p \over \partial \gamma_f} \delta \gamma_f + r_f^2 \bigg( \nabla_\Gamma^{\left( d_f \right)} \delta \gamma_f \cdot \nabla_\Gamma^{\left( d_f \right)} \gamma_{fa} + \nabla_\Gamma^{\left( d_f, \delta d_f \right)} \gamma_f \\
  & \cdot \nabla_\Gamma^{\left( d_f \right)} \gamma_{fa} + \nabla_\Gamma^{\left( d_f \right)} \gamma_f \cdot \nabla_\Gamma^{\left( d_f, \delta d_f \right)} \gamma_{fa} \bigg) + \delta \gamma_f \gamma_{fa} - \delta \gamma \gamma_{fa} \bigg] M^{\left( d_f \right)} \\
  & + f_{id,\Gamma} \bigg( \gamma_p + r_f^2 \nabla_\Gamma^{\left( d_f \right)} \gamma_f \cdot \nabla_\Gamma^{\left( d_f \right)} \gamma_{fa} + \gamma_f \gamma_{fa} - \gamma \gamma_{fa} \bigg) M^{\left( d_f, \delta d_f \right)} \\
  & + r_m^2 \nabla_\Sigma \delta d_f \cdot \nabla_\Sigma d_{fa} + \delta d_f d_{fa} - A_d \delta d_m d_{fa} \,\mathrm{d}\Sigma
\end{split}
\end{equation}
with the satisfication of Eq. \ref{equ:ConstraintForAugmentedLagrangianAreaFluidStructureMHM}
and 
\begin{equation}\label{equ:ConstraintForVariationalAugmentedLagrangianAreaFluidStructure}
  \left\{\begin{split}
  & \delta \gamma_f \in \mathcal{H}\left(\Sigma_D\right) \\
  & \delta \gamma \in \mathcal{L}^2\left(\Sigma_D\right) \\
  & \delta d_f \in \mathcal{H}\left(\Sigma\right) \\
  & \delta d_m \in \mathcal{L}^2\left(\Sigma\right)
  \end{split}\right..
\end{equation}

According to the Karush-Kuhn-Tucker conditions of the PDE constrained optimization problem, the first order variational of the augmented Lagrangian to $\gamma_f$ can be set to be zero as
\begin{equation}\label{equ:WeakAdjEquSNSEquAreaFluidStructure}
\begin{split}
  & \int_{\Sigma_D} \left( {\partial \gamma_p \over \partial \gamma_f} \delta \gamma_f + r_f^2 \nabla_\Gamma^{\left( d_f \right)} \delta \gamma_f \cdot \nabla_\Gamma^{\left( d_f \right)} \gamma_{fa} + \delta \gamma_f \gamma_{fa} \right) M^{\left( d_f \right)} \,\mathrm{d}\Sigma = 0, \\
\end{split}
\end{equation}
and the first order variational of the augmented Lagrangian to the variable $d_f$ can be set to be zero as
\begin{equation}\label{equ:AdjEquAreaConstr1GaInformAreaFluidStructure} 
\begin{split}
  & \int_\Sigma f_{id,\Gamma} r_f^2 \left( \nabla_\Gamma^{\left( d_f, \delta d_f \right)} \gamma_f \cdot \nabla_\Gamma^{\left( d_f \right)} \gamma_{fa} + \nabla_\Gamma^{\left( d_f \right)} \gamma_f \cdot \nabla_\Gamma^{\left( d_f, \delta d_f \right)} \gamma_{fa} \right) M^{\left( d_f \right)} \\
  & + f_{id,\Gamma} \left( \gamma_p + r_f^2 \nabla_\Gamma^{\left( d_f \right)} \gamma_f \cdot \nabla_\Gamma^{\left( d_f \right)} \gamma_{fa} + \gamma_f \gamma_{fa} - \gamma \gamma_{fa} \right) M^{\left( d_f, \delta d_f \right)} \\
  & + r_m^2 \nabla_\Sigma \delta d_f \cdot \nabla_\Sigma d_{fa} + \delta d_f d_{fa} \,\mathrm{d}\Sigma = 0.
\end{split}
\end{equation}
The constraints in Eqs. \ref{equ:ConstraintForAugmentedLagrangianAreaFluidStructureMHM} and \ref{equ:ConstraintForVariationalAugmentedLagrangianAreaFluidStructure} are imposed to Eqs. \ref{equ:WeakAdjEquSNSEquAreaFluidStructure} and \ref{equ:AdjEquAreaConstr1GaInformAreaFluidStructure}. Then, the adjoint sensitivity of $s\left|\Gamma_D\right|$ is derived as
\begin{equation}\label{equ:AdjSensAreaConstr1InformAreaFluidStructure}
\begin{split}
  \delta \widehat{s\left|\Gamma_D\right|} = - \int_{\Sigma_D} \delta \gamma \gamma_{fa} M^{\left( d_f \right)} \,\mathrm{d}\Sigma - \int_\Sigma A_d \delta d_m d_{fa} \,\mathrm{d}\Sigma.
\end{split}
\end{equation}

Without losing the arbitrariness of $\delta \gamma_f$, $\delta d_f$, $\delta \gamma$ and $\delta d_m$, one can set 
\begin{equation}
\left.\begin{split}
& \tilde{\gamma}_{fa} = \delta \gamma_f \\
& \tilde{d}_{fa} = \delta d_f \\
& \tilde{\gamma} = \delta \gamma \\
& \tilde{d}_m = \delta d_m 
\end{split}\right\}
~\mathrm{with}~
\left\{\begin{split}
& \forall \tilde{\gamma}_{fa} \in \mathcal{H}\left(\Sigma_D\right) \\
& \forall \tilde{d}_{fa} \in \mathcal{H}\left(\Sigma\right)\\
& \forall \tilde{\gamma} \in \mathcal{L}^2\left(\Sigma_D\right) \\
& \forall \tilde{d}_m \in \mathcal{L}^2\left(\Sigma\right)
\end{split}\right.
\end{equation}
for Eqs. \ref{equ:WeakAdjEquSNSEquAreaFluidStructure} and \ref{equ:AdjEquAreaConstr1GaInformAreaFluidStructure} to derive the adjoint system composed of Eqs. \ref{equ:AdjSensitivityGaDmAreaFluidStrucConstrCHM}, \ref{AdjPDEFilterAreaFluidStrucGaMHM} and \ref{equ:AdjPDEFilterAreaFluidStrucDfMHM}.

\subsubsection{Adjoint analysis for area of implicit 2-manifold}\label{sec:AdjointAnalysisAreaImplicitManifoldMHM}

Based on the variational formulation of the surface-PDE filter in Eq. \ref{equ:VariationalFormulationPDEFilterBaseManifoldMHM}, the augmented Lagrangian of the area of the implicit 2-manifold can be derived as
\begin{equation}\label{equ:AugLagAreaImplicitManifoldMHM}
\begin{split}
  \widehat{\left|\Gamma_D\right|} = & \int_\Sigma f_{id,\Gamma} M^{\left( d_f \right)} + r_m^2 \nabla_\Sigma d_f \cdot \nabla_\Sigma d_{fa} + d_f d_{fa} - A_d \left( d_m - {1\over2} \right) d_{fa} \,\mathrm{d}\Sigma,
\end{split}
\end{equation}
where $\widehat{\left|\Gamma_D\right|}$ is the augmented lagrangian of $\left|\Gamma_D\right|$; and the adjoint variables satisfy
\begin{equation}\label{equ:ConstraintForAugmentedLagrangianAreaImplicitManifoldMHM}
  d_{fa} \in \mathcal{H}\left(\Sigma\right).
\end{equation}

Based on the related results in Section \ref{sec:MethodologyFiberBundleTOOPTransferSurfaceFlows}, the first order variational of $\widehat{\left|\Gamma_D\right|}$ can be derived as
\begin{equation}\label{equ:1stVarAugmentedLagrangianConstrAreaImplicitManifold}
\begin{split}
  \delta \widehat{\left|\Gamma_D\right|} = & \int_\Sigma f_{id,\Gamma} M^{\left( d_f, \delta d_f \right)} + r_m^2 \nabla_\Sigma \delta d_f \cdot \nabla_\Sigma d_{fa} + \delta d_f d_{fa} - A_d \delta d_m d_{fa} \,\mathrm{d}\Sigma
\end{split}
\end{equation}
with the satisfication of Eq. \ref{equ:ConstraintForAugmentedLagrangianAreaImplicitManifoldMHM}
and 
\begin{equation}\label{equ:ConstraintForVariationalAugmentedLagrangianAreaImplicitManifold}
  \left\{\begin{split}
  & \delta d_f \in \mathcal{H}\left(\Sigma\right) \\
  & \delta d_m \in \mathcal{L}^2\left(\Sigma\right)
  \end{split}\right..
\end{equation}

According to the Karush-Kuhn-Tucker conditions of the PDE constrained optimization problem, the first order variational of the augmented Lagrangian to $d_f$ can be set to be zero as
\begin{equation}\label{equ:AdjEquAreaConstr1GaInformAreaImplicitManifold} 
\begin{split}
  & \int_\Sigma f_{id,\Gamma} M^{\left( d_f, \delta d_f \right)} + r_m^2 \nabla_\Sigma \delta d_f \cdot \nabla_\Sigma d_{fa} + \delta d_f d_{fa} \,\mathrm{d}\Sigma = 0.
\end{split}
\end{equation}
The constraints in Eqs. \ref{equ:ConstraintForAugmentedLagrangianAreaImplicitManifoldMHM} and \ref{equ:ConstraintForVariationalAugmentedLagrangianAreaImplicitManifold} are imposed to Eq. \ref{equ:AdjEquAreaConstr1GaInformAreaImplicitManifold}. Then, the adjoint sensitivity of $\left|\Gamma_D\right|$ is derived as
\begin{equation}\label{equ:AdjSensAreaConstr1InformAreaImplicitManifold}
\begin{split}
  \delta \widehat{\left|\Gamma_D\right|} = \int_\Sigma - A_d \delta d_m d_{fa} \,\mathrm{d}\Sigma.
\end{split}
\end{equation}

Without losing the arbitrariness of $\delta d_f$ and $\delta d_m$, one can set 
\begin{equation}
\left.\begin{split}
& \tilde{d}_{fa} = \delta d_f \\
& \tilde{d}_m = \delta d_m 
\end{split}\right\}
~\mathrm{with}~
\left\{\begin{split}
& \forall \tilde{d}_{fa} \in \mathcal{H}\left(\Sigma\right)\\
& \forall \tilde{d}_m \in \mathcal{L}^2\left(\Sigma\right)
\end{split}\right.
\end{equation}
for Eqs. \ref{equ:AdjEquAreaConstr1GaInformAreaImplicitManifold} and \ref{equ:AdjSensAreaConstr1InformAreaImplicitManifold} to derive the adjoint system composed of Eqs. \ref{equ:AdjSensitivityGaDmAreaImplitManifConstrCHM} and \ref{equ:AdjPDEFilterAreaImplitManifDfMHM}.

\subsection{Adjoint analysis for design objective in Eq. \ref{equ:VarProToopBulkNSCDMHT}} \label{sec:AdjointAnalysisDesignObjectiveBulkCHTMHM}

Based on the transformed design objective in Eq. \ref{equ:FurtherTransformedDesignObjectiveBulkCD}, the variational formulations of the Laplace's equation in Eq. \ref{equ:WeakFormLaplacianBulkFlowMHT}, the surface-PDE filters in Eqs. \ref{equ:VariationalFormulationPDEFilterBaseManifoldMHM} and \ref{equ:VariationalFormulationPDEFilterMHM} and the Navier-Stokes equations in Eq. \ref{equ:TransformedVariationalFormulationNavierStokesEquBulkFlowMHT} and the convection-diffusion equation in Eq. \ref{equ:TransformedVariationalFormulationBulkCDEqu}, the augmented Lagrangian of the design objective in Eq. \ref{equ:VarProToopBulkNSCDMHT} can be derived as
\begin{equation}\label{equ:AugmentedLagrangianMatchOptimizationBulkMT}
\begin{split}
  \hat{J}_c = & \int_{\Sigma_{s,\Omega}} \left( c - \bar{c} \right)^2 \,\mathrm{d}\Sigma_{\partial\Omega} \bigg/ \int_{\Sigma_{v,\Omega}} \left( c_0 - \bar{c} \right)^2 \,\mathrm{d}\Sigma_{\partial\Omega} + \int_\Omega \bigg[ \rho \left( \mathbf{u} \cdot \nabla_{\mathbf{x}_\Xi}^{\left(\mathbf{s}\right)} \right) \mathbf{u} \cdot \mathbf{u}_a \\
  & + {\eta\over2} \left( \nabla_{\mathbf{x}_\Xi}^{\left(\mathbf{s}\right)} \mathbf{u} + \nabla_{\mathbf{x}_\Xi}^{\left(\mathbf{s}\right)} \mathbf{u}^\mathrm{T} \right) : \left( \nabla_{\mathbf{x}_\Xi}^{\left(\mathbf{s}\right)} \mathbf{u}_a + \nabla_{\mathbf{x}_\Xi}^{\left(\mathbf{s}\right)} \mathbf{u}_a^\mathrm{T} \right) - p\,\mathrm{div}_{\mathbf{x}_\Xi}^{\left(\mathbf{s}\right)} \mathbf{u}_a - p_a \mathrm{div}_{\mathbf{x}_\Xi}^{\left(\mathbf{s}\right)} \mathbf{u} \bigg] K^{\left( \mathbf{s} \right)} \,\mathrm{d}\Omega \\
  & + \sum_{E_\Omega\in\mathcal{E}_\Omega} \int_{E_\Omega} - \tau_{BP,\Xi}^{\left(\mathbf{s}\right)} \nabla_{\mathbf{x}_\Xi}^{\left(\mathbf{s}\right)} p \cdot \nabla_{\mathbf{x}_\Xi}^{\left(\mathbf{s}\right)} p_a K^{\left( \mathbf{s} \right)} \,\mathrm{d}\Omega + \int_\Sigma \alpha \mathbf{u} \cdot \mathbf{u}_a M^{\left( d_f \right)} \, \mathrm{d}\Sigma \\
  & + \int_\Omega \left[ \left( \mathbf{u} \cdot \nabla_{\mathbf{x}_\Xi}^{\left(\mathbf{s}\right)} c \right) c_a + D \nabla_{\mathbf{x}_\Xi}^{\left(\mathbf{s}\right)} c \cdot \nabla_{\mathbf{x}_\Xi}^{\left(\mathbf{s}\right)} c_a \right] K^{\left(\mathbf{s}\right)} \,\mathrm{d}\Omega + \sum_{E_\Omega\in\mathcal{E}_\Omega} \int_{E_\Omega} \tau_{PG,\Xi}^{\left(\mathbf{s}\right)} \left( \mathbf{u} \cdot \nabla_{\mathbf{x}_\Xi}^{\left(\mathbf{s}\right)} c \right) \\
  & \left( \mathbf{u} \cdot \nabla_{\mathbf{x}_\Xi}^{\left(\mathbf{s}\right)} c_a \right) K^{\left(\mathbf{s}\right)} \,\mathrm{d}\Omega + \int_\Omega - \nabla_{\mathbf{x}_\Omega} \mathbf{s} : \nabla_{\mathbf{x}_\Omega} \mathbf{s}_a \,\mathrm{d}\Omega + \int_\Sigma \left( \mathbf{s} - d_f \mathbf{n}_\Sigma \right) \cdot \boldsymbol{\lambda}_{\mathbf{s}a} \\
  & + \boldsymbol{\lambda}_\mathbf{s} \cdot \mathbf{s}_a \,\mathrm{d}\Sigma + \int_\Sigma \left( r_f^2 \nabla_\Gamma^{\left( d_f \right)} \gamma_f \cdot \nabla_\Gamma^{\left( d_f \right)} \gamma_{fa} + \gamma_f \gamma_{fa} - \gamma \gamma_{fa} \right) M^{\left( d_f \right)} \,\mathrm{d}\Sigma \\
  & + \int_\Sigma r_m^2 \nabla_\Sigma d_f \cdot \nabla_\Sigma d_{fa} + d_f d_{fa} - A_d \left( d_m - {1\over2} \right) d_{fa} \,\mathrm{d}\Sigma,
\end{split}
\end{equation}
where the adjoint variables satisfy
\begin{equation}\label{equ:ConstraintForAugmentedLagrangianBulkMT}
  \left.\begin{split}
  & \mathbf{u}_a \in\left(\mathcal{H}\left(\Omega\right)\right)^3 \\
  & p_a \in \mathcal{H}\left(\Omega\right) \\
  & c_a \in \mathcal{H}\left(\Omega\right) \\
  & \mathbf{s}_a \in \left(\mathcal{H}\left(\Omega\right)\right)^3 \\
  & \boldsymbol{\lambda}_{\mathbf{s}a} \in \left(\mathcal{H}^{-{1\over2}}\left(\Sigma\right)\right)^3 \\
  & \gamma_{fa} \in \mathcal{H}\left(\Sigma\right) \\
  & d_{fa} \in \mathcal{H}\left(\Sigma\right)
  \end{split}\right\}~\mathrm{with}~
  \left\{\begin{split}
  & \mathbf{u}_a = \mathbf{0}~ \mathrm{at} ~ \forall \mathbf{x}_\Omega \in \Sigma_{v,\Omega} \cup \Sigma_{v_0,\Omega} \\
  & c_a = 0~ \mathrm{at} ~ \forall \mathbf{x}_\Omega \in \Sigma_{v,\Omega} \\
  & \mathbf{s}_a = \mathbf{0} ~ \mathrm{at} ~ \forall \mathbf{x}_\Omega \in \Sigma_{v,\Omega} \cup \Sigma_{s,\Omega} 
  \end{split}\right..
\end{equation}
The first order variational of the augmented Lagrangian in Eq. \ref{equ:AugmentedLagrangianMatchOptimizationBulkMT} can be derived as
\begin{equation}\label{equ:1stVariAugmentedLagrangianMatchOptimizationBulkMT}
\begin{split}
  \delta \hat{J}_c = & \int_{\Sigma_{s,\Omega}} 2 \left( c - \bar{c} \right) \delta c \,\mathrm{d}\Sigma_{\partial\Omega} \bigg/ \int_{\Sigma_{v,\Omega}} \left( c_0 - \bar{c} \right)^2 \,\mathrm{d}\Sigma_{\partial\Omega} + \int_\Omega \Big[ \rho \left( \delta \mathbf{u} \cdot \nabla_{\mathbf{x}_\Xi}^{\left(\mathbf{s}\right)} \right) \mathbf{u} \cdot \mathbf{u}_a \\
  & + \rho \left( \mathbf{u} \cdot \nabla_{\mathbf{x}_\Xi}^{\left(\mathbf{s},\delta\mathbf{s}\right)} \right) \mathbf{u} \cdot \mathbf{u}_a + \rho \left( \mathbf{u} \cdot \nabla_{\mathbf{x}_\Xi}^{\left(\mathbf{s}\right)} \right) \delta \mathbf{u} \cdot \mathbf{u}_a + {\eta\over2} \left( \nabla_{\mathbf{x}_\Xi}^{\left(\mathbf{s}\right)} \delta \mathbf{u} + \nabla_{\mathbf{x}_\Xi}^{\left(\mathbf{s}\right)} \delta \mathbf{u}^\mathrm{T} \right) : \Big( \nabla_{\mathbf{x}_\Xi}^{\left(\mathbf{s}\right)} \mathbf{u}_a \\
  & + \nabla_{\mathbf{x}_\Xi}^{\left(\mathbf{s}\right)} \mathbf{u}_a^\mathrm{T} \Big) + {\eta\over2} \Big( \nabla_{\mathbf{x}_\Xi}^{\left(\mathbf{s}, \delta \mathbf{s}\right)} \mathbf{u} + \nabla_{\mathbf{x}_\Xi}^{\left(\mathbf{s}, \delta \mathbf{s}\right)} \mathbf{u}^\mathrm{T} \Big) : \left( \nabla_{\mathbf{x}_\Xi}^{\left(\mathbf{s}\right)} \mathbf{u}_a + \nabla_{\mathbf{x}_\Xi}^{\left(\mathbf{s}\right)} \mathbf{u}_a^\mathrm{T} \right) + {\eta\over2} \left( \nabla_{\mathbf{x}_\Xi}^{\left(\mathbf{s}\right)} \mathbf{u} + \nabla_{\mathbf{x}_\Xi}^{\left(\mathbf{s}\right)} \mathbf{u}^\mathrm{T} \right) \\
  & : \left( \nabla_{\mathbf{x}_\Xi}^{\left(\mathbf{s}, \delta \mathbf{s}\right)} \mathbf{u}_a + \nabla_{\mathbf{x}_\Xi}^{\left(\mathbf{s}, \delta \mathbf{s}\right)} \mathbf{u}_a^\mathrm{T} \right) - \delta p\,\mathrm{div}_{\mathbf{x}_\Xi}^{\left(\mathbf{s}\right)} \mathbf{u}_a - p\,\mathrm{div}_{\mathbf{x}_\Xi}^{\left(\mathbf{s}, \delta \mathbf{s}\right)} \mathbf{u}_a - p_a \mathrm{div}_{\mathbf{x}_\Xi}^{\left(\mathbf{s}, \delta \mathbf{s}\right)} \mathbf{u} - p_a \mathrm{div}_{\mathbf{x}_\Xi}^{\left(\mathbf{s}\right)} \delta \mathbf{u} \Big] \\
  & K^{\left( \mathbf{s} \right)} + \Big[ \rho \left( \mathbf{u} \cdot \nabla_{\mathbf{x}_\Xi}^{\left(\mathbf{s}\right)} \right) \mathbf{u} \cdot \mathbf{u}_a + {\eta\over2} \left( \nabla_{\mathbf{x}_\Xi}^{\left(\mathbf{s}\right)} \mathbf{u} + \nabla_{\mathbf{x}_\Xi}^{\left(\mathbf{s}\right)} \mathbf{u}^\mathrm{T} \right) : \left( \nabla_{\mathbf{x}_\Xi}^{\left(\mathbf{s}\right)} \mathbf{u}_a + \nabla_{\mathbf{x}_\Xi}^{\left(\mathbf{s}\right)} \mathbf{u}_a^\mathrm{T} \right) - p\,\mathrm{div}_{\mathbf{x}_\Xi}^{\left(\mathbf{s}\right)} \mathbf{u}_a \\
  & - p_a \mathrm{div}_{\mathbf{x}_\Xi}^{\left(\mathbf{s}\right)} \mathbf{u} \Big] K^{\left( \mathbf{s}, \delta \mathbf{s} \right)} \,\mathrm{d}\Omega + \sum_{E_\Omega\in\mathcal{E}_\Omega} \int_{E_\Omega} - \tau_{BP,\Xi}^{\left(\mathbf{s}, \delta \mathbf{s}\right)} \nabla_{\mathbf{x}_\Xi}^{\left(\mathbf{s}\right)} p \cdot \nabla_{\mathbf{x}_\Xi}^{\left(\mathbf{s}\right)} p_a K^{\left( \mathbf{s} \right)} - \tau_{BP,\Xi}^{\left(\mathbf{s}\right)} \nabla_{\mathbf{x}_\Xi}^{\left(\mathbf{s}, \delta \mathbf{s}\right)} p \\
  & \cdot \nabla_{\mathbf{x}_\Xi}^{\left(\mathbf{s}\right)} p_a K^{\left( \mathbf{s} \right)} - \tau_{BP,\Xi}^{\left(\mathbf{s}\right)} \nabla_{\mathbf{x}_\Xi}^{\left(\mathbf{s}\right)} \delta p \cdot \nabla_{\mathbf{x}_\Xi}^{\left(\mathbf{s}\right)} p_a K^{\left( \mathbf{s} \right)} - \tau_{BP,\Xi}^{\left(\mathbf{s}\right)} \nabla_{\mathbf{x}_\Xi}^{\left(\mathbf{s}\right)} p \cdot \nabla_{\mathbf{x}_\Xi}^{\left(\mathbf{s}, \delta \mathbf{s}\right)} p_a K^{\left( \mathbf{s} \right)} - \tau_{BP,\Xi}^{\left(\mathbf{s}\right)} \nabla_{\mathbf{x}_\Xi}^{\left(\mathbf{s}\right)} p \\
  & \cdot \nabla_{\mathbf{x}_\Xi}^{\left(\mathbf{s}\right)} p_a K^{\left( \mathbf{s}, \delta \mathbf{s} \right)} \,\mathrm{d}\Omega + \int_\Sigma {\partial\alpha\over\partial\gamma_p} {\partial\gamma_p\over\partial\gamma_f} \mathbf{u} \cdot \mathbf{u}_a M^{\left( d_f \right)} \delta \gamma_f + \alpha \delta \mathbf{u} \cdot \mathbf{u}_a M^{\left( d_f \right)} + \alpha \mathbf{u} \cdot \mathbf{u}_a \\
  & M^{\left( d_f, \delta d_f \right)} \, \mathrm{d}\Sigma + \int_\Omega \Big[ \left( \delta \mathbf{u} \cdot \nabla_{\mathbf{x}_\Xi}^{\left(\mathbf{s}\right)} c + \mathbf{u} \cdot \nabla_{\mathbf{x}_\Xi}^{\left(\mathbf{s},\delta \mathbf{s}\right)} c + \mathbf{u} \cdot \nabla_{\mathbf{x}_\Xi}^{\left(\mathbf{s}\right)} \delta c \right) c_a + D \Big( \nabla_{\mathbf{x}_\Xi}^{\left(\mathbf{s},\delta \mathbf{s}\right)} c \cdot \nabla_{\mathbf{x}_\Xi}^{\left(\mathbf{s}\right)} c_a \\
  & + \nabla_{\mathbf{x}_\Xi}^{\left(\mathbf{s}\right)} \delta c \cdot \nabla_{\mathbf{x}_\Xi}^{\left(\mathbf{s}\right)} c_a + \nabla_{\mathbf{x}_\Xi}^{\left(\mathbf{s}\right)} c \cdot \nabla_{\mathbf{x}_\Xi}^{\left(\mathbf{s}, \delta \mathbf{s}\right)} c_a \Big) \Big] K^{\left(\mathbf{s}\right)} + \left[ \left( \mathbf{u} \cdot \nabla_{\mathbf{x}_\Xi}^{\left(\mathbf{s}\right)} c \right) c_a + D \nabla_{\mathbf{x}_\Xi}^{\left(\mathbf{s}\right)} c \cdot \nabla_{\mathbf{x}_\Xi}^{\left(\mathbf{s}\right)} c_a \right] \\
  & K^{\left(\mathbf{s}, \delta \mathbf{s}\right)} \,\mathrm{d}\Omega + \sum_{E_\Omega\in\mathcal{E}_\Omega} \int_{E_\Omega} \tau_{PG,\Xi}^{\left(\mathbf{s}, \delta \mathbf{s}\right)} \left( \mathbf{u} \cdot \nabla_{\mathbf{x}_\Xi}^{\left(\mathbf{s}\right)} c \right) \left( \mathbf{u} \cdot \nabla_{\mathbf{x}_\Xi}^{\left(\mathbf{s}\right)} c_a \right) K^{\left(\mathbf{s}\right)} + \tau_{PG,\Xi}^{\left(\mathbf{s}, \delta \mathbf{u}\right)} \left( \mathbf{u} \cdot \nabla_{\mathbf{x}_\Xi}^{\left(\mathbf{s}\right)} c \right) \\
  & \left( \mathbf{u} \cdot \nabla_{\mathbf{x}_\Xi}^{\left(\mathbf{s}\right)} c_a \right) K^{\left(\mathbf{s}\right)} + \tau_{PG,\Xi}^{\left(\mathbf{s}\right)} \left( \delta \mathbf{u} \cdot \nabla_{\mathbf{x}_\Xi}^{\left(\mathbf{s}\right)} c \right) \left( \mathbf{u} \cdot \nabla_{\mathbf{x}_\Xi}^{\left(\mathbf{s}\right)} c_a \right) K^{\left(\mathbf{s}\right)} + \tau_{PG,\Xi}^{\left(\mathbf{s}\right)} \left( \mathbf{u} \cdot \nabla_{\mathbf{x}_\Xi}^{\left(\mathbf{s}, \delta\mathbf{s}\right)} c \right) \\
  & \left( \mathbf{u} \cdot \nabla_{\mathbf{x}_\Xi}^{\left(\mathbf{s}\right)} c_a \right) K^{\left(\mathbf{s}\right)} + \tau_{PG,\Xi}^{\left(\mathbf{s}\right)} \left( \mathbf{u} \cdot \nabla_{\mathbf{x}_\Xi}^{\left(\mathbf{s}\right)} \delta c \right) \left( \mathbf{u} \cdot \nabla_{\mathbf{x}_\Xi}^{\left(\mathbf{s}\right)} c_a \right) K^{\left(\mathbf{s}\right)} + \tau_{PG,\Xi}^{\left(\mathbf{s}\right)} \left( \mathbf{u} \cdot \nabla_{\mathbf{x}_\Xi}^{\left(\mathbf{s}\right)} c \right) \\
  & \left( \delta \mathbf{u} \cdot \nabla_{\mathbf{x}_\Xi}^{\left(\mathbf{s}\right)} c_a \right) K^{\left(\mathbf{s}\right)} + \tau_{PG,\Xi}^{\left(\mathbf{s}\right)} \left( \mathbf{u} \cdot \nabla_{\mathbf{x}_\Xi}^{\left(\mathbf{s}\right)} c \right) \left( \mathbf{u} \cdot \nabla_{\mathbf{x}_\Xi}^{\left(\mathbf{s}, \delta \mathbf{s}\right)} c_a \right) K^{\left(\mathbf{s}\right)} + \tau_{PG,\Xi}^{\left(\mathbf{s}\right)} \left( \mathbf{u} \cdot \nabla_{\mathbf{x}_\Xi}^{\left(\mathbf{s}\right)} c \right) \\
  & \left( \mathbf{u} \cdot \nabla_{\mathbf{x}_\Xi}^{\left(\mathbf{s}\right)} c_a \right) K^{\left(\mathbf{s},\delta \mathbf{s}\right)}  \,\mathrm{d}\Omega + \int_\Omega - \nabla_{\mathbf{x}_\Omega} \delta \mathbf{s} : \nabla_{\mathbf{x}_\Omega} \mathbf{s}_a \,\mathrm{d}\Omega + \int_\Sigma \left( \delta \mathbf{s} - \delta d_f \mathbf{n}_\Sigma \right) \cdot \boldsymbol{\lambda}_{\mathbf{s}a} \\
  & + \delta \boldsymbol{\lambda}_\mathbf{s} \cdot \mathbf{s}_a + \bigg[ r_f^2 \bigg( \nabla_\Gamma^{\left( d_f, \delta d_f \right)} \gamma_f \cdot \nabla_\Gamma^{\left( d_f \right)} \gamma_{fa} + \nabla_\Gamma^{\left( d_f \right)} \delta \gamma_f \cdot \nabla_\Gamma^{\left( d_f \right)} \gamma_{fa} + \nabla_\Gamma^{\left( d_f \right)} \gamma_f \\
  & \cdot \nabla_\Gamma^{\left( d_f, \delta d_f \right)} \gamma_{fa} \bigg) + \delta \gamma_f \gamma_{fa} - \delta \gamma \gamma_{fa} \bigg] M^{\left( d_f \right)} + \bigg( r_f^2 \nabla_\Gamma^{\left( d_f \right)} \gamma_f \cdot \nabla_\Gamma^{\left( d_f \right)} \gamma_{fa} + \gamma_f \gamma_{fa} \\
  & - \gamma \gamma_{fa} \bigg) M^{\left( d_f, \delta d_f \right)} + r_m^2 \nabla_\Sigma \delta d_f \cdot \nabla_\Sigma d_{fa} + \delta d_f d_{fa} - A_d \delta d_m d_{fa} \,\mathrm{d}\Sigma
\end{split}
\end{equation}
with the satisfication of the constraints in Eq. \ref{equ:ConstraintForAugmentedLagrangianBulkMT}
and
\begin{equation}\label{equ:ConstraintForVariationalAugmentedLagrangianObjBulkMT}
  \left.\begin{split}
  & \delta \mathbf{u} \in\left(\mathcal{H}\left(\Omega\right)\right)^3 \\
  & \delta p \in \mathcal{H}\left(\Omega\right) \\
  & \delta c \in \mathcal{H}\left(\Omega\right) \\
  & \delta \mathbf{s} \in \left(\mathcal{H}\left(\Omega\right)\right)^3 \\
  & \delta \boldsymbol{\lambda}_{\mathbf{s}} \in \left(\mathcal{H}^{{1\over2}}\left(\Sigma\right)\right)^3 \\
  & \delta \gamma_f \in \mathcal{H}\left(\Sigma\right) \\
  & \delta d_f \in \mathcal{H}\left(\Sigma\right)
  \end{split}\right\}~\mathrm{with}~
  \left\{\begin{split}
  & \delta \mathbf{u} = \mathbf{0}~ \mathrm{at} ~ \forall \mathbf{x}_\Omega \in \Sigma_{v,\Omega} \cup \Sigma_{v_0,\Omega} \\
  & \delta c = 0~ \mathrm{at} ~ \forall \mathbf{x}_\Omega \in \Sigma_{v,\Omega} \\
  & \delta \mathbf{s} = \mathbf{0} ~ \mathrm{at} ~ \forall \mathbf{x}_\Omega \in \Sigma_{v,\Omega} \cup \Sigma_{s,\Omega}
  \end{split}\right..
\end{equation}

According to the Karush-Kuhn-Tucker conditions of the PDE constrained optimization problem, the first order variational of the augmented Lagrangian to $c$ can be set to be zero as
\begin{equation}\label{equ:WeakAdjEquCDEquBulkMT}
\begin{split}
  & \int_{\Sigma_{s,\Omega}} 2 \left( c - \bar{c} \right) \delta c \,\mathrm{d}\Sigma_{\partial\Omega} \bigg/ \int_{\Sigma_{v,\Omega}} \left( c_0 - \bar{c} \right)^2 \,\mathrm{d}\Sigma_{\partial\Omega} \\
  & + \int_\Omega \left[ \left( \mathbf{u} \cdot \nabla_{\mathbf{x}_\Xi}^{\left(\mathbf{s}\right)} \delta c \right) c_a + D \nabla_{\mathbf{x}_\Xi}^{\left(\mathbf{s}\right)} \delta c \cdot \nabla_{\mathbf{x}_\Xi}^{\left(\mathbf{s}\right)} c_a \right] K^{\left(\mathbf{s}\right)} \,\mathrm{d}\Omega \\
  & + \sum_{E_\Omega\in\mathcal{E}_\Omega} \int_{E_\Omega} \tau_{PG,\Xi}^{\left(\mathbf{s}\right)} \left( \mathbf{u} \cdot \nabla_{\mathbf{x}_\Xi}^{\left(\mathbf{s}\right)} \delta c \right) \left( \mathbf{u} \cdot \nabla_{\mathbf{x}_\Xi}^{\left(\mathbf{s}\right)} c_a \right) K^{\left(\mathbf{s}\right)} \,\mathrm{d}\Omega = 0,
\end{split}
\end{equation}
the first order variational of the augmented Lagrangian to $\mathbf{u}$ and $p$ can be set to be zero as
\begin{equation}\label{equ:WeakAdjEquNSEquBulkMT}
\begin{split}
  & \int_\Omega \Big[ \rho \left( \delta \mathbf{u} \cdot \nabla_{\mathbf{x}_\Xi}^{\left(\mathbf{s}\right)} \right) \mathbf{u} \cdot \mathbf{u}_a + \rho \left( \mathbf{u} \cdot \nabla_{\mathbf{x}_\Xi}^{\left(\mathbf{s}\right)} \right) \delta \mathbf{u} \cdot \mathbf{u}_a + {\eta\over2} \left( \nabla_{\mathbf{x}_\Xi}^{\left(\mathbf{s}\right)} \delta \mathbf{u} + \nabla_{\mathbf{x}_\Xi}^{\left(\mathbf{s}\right)} \delta \mathbf{u}^\mathrm{T} \right) : \Big( \nabla_{\mathbf{x}_\Xi}^{\left(\mathbf{s}\right)} \mathbf{u}_a \\
  & + \nabla_{\mathbf{x}_\Xi}^{\left(\mathbf{s}\right)} \mathbf{u}_a^\mathrm{T} \Big) - \delta p\,\mathrm{div}_{\mathbf{x}_\Xi}^{\left(\mathbf{s}\right)} \mathbf{u}_a - p_a \mathrm{div}_{\mathbf{x}_\Xi}^{\left(\mathbf{s}\right)} \delta \mathbf{u} \Big] K^{\left( \mathbf{s} \right)} \,\mathrm{d}\Omega + \sum_{E_\Omega\in\mathcal{E}_\Omega} \int_{E_\Omega} - \tau_{BP,\Xi}^{\left(\mathbf{s}\right)} \nabla_{\mathbf{x}_\Xi}^{\left(\mathbf{s}\right)} \delta p \\
  & \cdot \nabla_{\mathbf{x}_\Xi}^{\left(\mathbf{s}\right)} p_a K^{\left( \mathbf{s} \right)} \,\mathrm{d}\Omega + \int_\Sigma \alpha \delta \mathbf{u} \cdot \mathbf{u}_a M^{\left( d_f \right)} \, \mathrm{d}\Sigma + \int_\Omega \left( \delta \mathbf{u} \cdot \nabla_{\mathbf{x}_\Xi}^{\left(\mathbf{s}\right)} c \right) c_a K^{\left(\mathbf{s}\right)} \,\mathrm{d}\Omega \\
  & + \sum_{E_\Omega\in\mathcal{E}_\Omega} \int_{E_\Omega} \tau_{PG,\Xi}^{\left(\mathbf{s}, \delta \mathbf{u}\right)} \left( \mathbf{u} \cdot \nabla_{\mathbf{x}_\Xi}^{\left(\mathbf{s}\right)} c \right) \left( \mathbf{u} \cdot \nabla_{\mathbf{x}_\Xi}^{\left(\mathbf{s}\right)} c_a \right) K^{\left(\mathbf{s}\right)} + \tau_{PG,\Xi}^{\left(\mathbf{s}\right)} \left( \delta \mathbf{u} \cdot \nabla_{\mathbf{x}_\Xi}^{\left(\mathbf{s}\right)} c \right) \\
  & \left( \mathbf{u} \cdot \nabla_{\mathbf{x}_\Xi}^{\left(\mathbf{s}\right)} c_a \right) K^{\left(\mathbf{s}\right)} + \tau_{PG,\Xi}^{\left(\mathbf{s}\right)} \left( \mathbf{u} \cdot \nabla_{\mathbf{x}_\Xi}^{\left(\mathbf{s}\right)} c \right) \left( \delta \mathbf{u} \cdot \nabla_{\mathbf{x}_\Xi}^{\left(\mathbf{s}\right)} c_a \right) K^{\left(\mathbf{s}\right)} \,\mathrm{d}\Omega = 0, \\
\end{split}
\end{equation}
the first order variational of the augmented Lagrangian to $\mathbf{s}$ and $\boldsymbol{\lambda}_\mathbf{s}$ can be set to be zero as
\begin{equation}\label{equ:WeakAdjEquHarmonicEquBulkMT}
\begin{split}
  & \int_\Omega \Big[ \rho \left( \mathbf{u} \cdot \nabla_{\mathbf{x}_\Xi}^{\left(\mathbf{s},\delta\mathbf{s}\right)} \right) \mathbf{u} \cdot \mathbf{u}_a + {\eta\over2} \left( \nabla_{\mathbf{x}_\Xi}^{\left(\mathbf{s}, \delta \mathbf{s}\right)} \mathbf{u} + \nabla_{\mathbf{x}_\Xi}^{\left(\mathbf{s}, \delta \mathbf{s}\right)} \mathbf{u}^\mathrm{T} \right) : \left( \nabla_{\mathbf{x}_\Xi}^{\left(\mathbf{s}\right)} \mathbf{u}_a + \nabla_{\mathbf{x}_\Xi}^{\left(\mathbf{s}\right)} \mathbf{u}_a^\mathrm{T} \right) + {\eta\over2} \\
  & \left( \nabla_{\mathbf{x}_\Xi}^{\left(\mathbf{s}\right)} \mathbf{u} + \nabla_{\mathbf{x}_\Xi}^{\left(\mathbf{s}\right)} \mathbf{u}^\mathrm{T} \right) : \left( \nabla_{\mathbf{x}_\Xi}^{\left(\mathbf{s}, \delta \mathbf{s}\right)} \mathbf{u}_a + \nabla_{\mathbf{x}_\Xi}^{\left(\mathbf{s}, \delta \mathbf{s}\right)} \mathbf{u}_a^\mathrm{T} \right) - p\,\mathrm{div}_{\mathbf{x}_\Xi}^{\left(\mathbf{s}, \delta \mathbf{s}\right)} \mathbf{u}_a - p_a \mathrm{div}_{\mathbf{x}_\Xi}^{\left(\mathbf{s}, \delta \mathbf{s}\right)} \mathbf{u} \Big] K^{\left( \mathbf{s} \right)} \\
  & + \Big[ \rho \left( \mathbf{u} \cdot \nabla_{\mathbf{x}_\Xi}^{\left(\mathbf{s}\right)} \right) \mathbf{u} \cdot \mathbf{u}_a + {\eta\over2} \left( \nabla_{\mathbf{x}_\Xi}^{\left(\mathbf{s}\right)} \mathbf{u} + \nabla_{\mathbf{x}_\Xi}^{\left(\mathbf{s}\right)} \mathbf{u}^\mathrm{T} \right) : \left( \nabla_{\mathbf{x}_\Xi}^{\left(\mathbf{s}\right)} \mathbf{u}_a + \nabla_{\mathbf{x}_\Xi}^{\left(\mathbf{s}\right)} \mathbf{u}_a^\mathrm{T} \right) - p\,\mathrm{div}_{\mathbf{x}_\Xi}^{\left(\mathbf{s}\right)} \mathbf{u}_a \\
  & - p_a \mathrm{div}_{\mathbf{x}_\Xi}^{\left(\mathbf{s}\right)} \mathbf{u} \Big] K^{\left( \mathbf{s}, \delta \mathbf{s} \right)} \,\mathrm{d}\Omega + \sum_{E_\Omega\in\mathcal{E}_\Omega} \int_{E_\Omega} - \tau_{BP,\Xi}^{\left(\mathbf{s}, \delta \mathbf{s}\right)} \nabla_{\mathbf{x}_\Xi}^{\left(\mathbf{s}\right)} p \cdot \nabla_{\mathbf{x}_\Xi}^{\left(\mathbf{s}\right)} p_a K^{\left( \mathbf{s} \right)} - \tau_{BP,\Xi}^{\left(\mathbf{s}\right)} \nabla_{\mathbf{x}_\Xi}^{\left(\mathbf{s}, \delta \mathbf{s}\right)} p \\
  & \cdot \nabla_{\mathbf{x}_\Xi}^{\left(\mathbf{s}\right)} p_a K^{\left( \mathbf{s} \right)} - \tau_{BP,\Xi}^{\left(\mathbf{s}\right)} \nabla_{\mathbf{x}_\Xi}^{\left(\mathbf{s}\right)} p \cdot \nabla_{\mathbf{x}_\Xi}^{\left(\mathbf{s}, \delta \mathbf{s}\right)} p_a K^{\left( \mathbf{s} \right)} - \tau_{BP,\Xi}^{\left(\mathbf{s}\right)} \nabla_{\mathbf{x}_\Xi}^{\left(\mathbf{s}\right)} p \cdot \nabla_{\mathbf{x}_\Xi}^{\left(\mathbf{s}\right)} p_a K^{\left( \mathbf{s}, \delta \mathbf{s} \right)} \,\mathrm{d}\Omega \\
  & + \int_\Omega \Big[ \left( \mathbf{u} \cdot \nabla_{\mathbf{x}_\Xi}^{\left(\mathbf{s},\delta \mathbf{s}\right)} c \right) c_a + D \left( \nabla_{\mathbf{x}_\Xi}^{\left(\mathbf{s},\delta \mathbf{s}\right)} c \cdot \nabla_{\mathbf{x}_\Xi}^{\left(\mathbf{s}\right)} c_a + \nabla_{\mathbf{x}_\Xi}^{\left(\mathbf{s}\right)} c \cdot \nabla_{\mathbf{x}_\Xi}^{\left(\mathbf{s}, \delta \mathbf{s}\right)} c_a \right) \Big] K^{\left(\mathbf{s}\right)} \\
  & + \left[ \left( \mathbf{u} \cdot \nabla_{\mathbf{x}_\Xi}^{\left(\mathbf{s}\right)} c \right) c_a + D \nabla_{\mathbf{x}_\Xi}^{\left(\mathbf{s}\right)} c \cdot \nabla_{\mathbf{x}_\Xi}^{\left(\mathbf{s}\right)} c_a \right] K^{\left(\mathbf{s}, \delta \mathbf{s}\right)} \,\mathrm{d}\Omega + \sum_{E_\Omega\in\mathcal{E}_\Omega} \int_{E_\Omega} \tau_{PG,\Xi}^{\left(\mathbf{s}, \delta \mathbf{s}\right)} \left( \mathbf{u} \cdot \nabla_{\mathbf{x}_\Xi}^{\left(\mathbf{s}\right)} c \right) \\
  & \left( \mathbf{u} \cdot \nabla_{\mathbf{x}_\Xi}^{\left(\mathbf{s}\right)} c_a \right) K^{\left(\mathbf{s}\right)} + \tau_{PG,\Xi}^{\left(\mathbf{s}\right)} \left( \mathbf{u} \cdot \nabla_{\mathbf{x}_\Xi}^{\left(\mathbf{s}, \delta\mathbf{s}\right)} c \right) \left( \mathbf{u} \cdot \nabla_{\mathbf{x}_\Xi}^{\left(\mathbf{s}\right)} c_a \right) K^{\left(\mathbf{s}\right)} + \tau_{PG,\Xi}^{\left(\mathbf{s}\right)} \left( \mathbf{u} \cdot \nabla_{\mathbf{x}_\Xi}^{\left(\mathbf{s}\right)} c \right) \\
  & \left( \mathbf{u} \cdot \nabla_{\mathbf{x}_\Xi}^{\left(\mathbf{s}, \delta \mathbf{s}\right)} c_a \right) K^{\left(\mathbf{s}\right)} + \tau_{PG,\Xi}^{\left(\mathbf{s}\right)} \left( \mathbf{u} \cdot \nabla_{\mathbf{x}_\Xi}^{\left(\mathbf{s}\right)} c \right) \left( \mathbf{u} \cdot \nabla_{\mathbf{x}_\Xi}^{\left(\mathbf{s}\right)} c_a \right) K^{\left(\mathbf{s},\delta \mathbf{s}\right)}  \,\mathrm{d}\Omega \\
  & + \int_\Omega - \nabla_{\mathbf{x}_\Omega} \delta \mathbf{s} : \nabla_{\mathbf{x}_\Omega} \mathbf{s}_a \,\mathrm{d}\Omega + \int_\Sigma \delta \mathbf{s} \cdot \boldsymbol{\lambda}_{\mathbf{s}a} + \delta \boldsymbol{\lambda}_\mathbf{s} \cdot \mathbf{s}_a \,\mathrm{d}\Sigma = 0, \\
\end{split}
\end{equation}
the first order variational of the augmented Lagrangian to $\gamma_f$ can be set to be zero as
\begin{equation}\label{equ:WeakAdjEquPDEFilterEquGafBulkMT}
\begin{split}
  & \int_\Sigma \left( {\partial\alpha\over\partial\gamma_p} {\partial\gamma_p\over\partial\gamma_f} \mathbf{u} \cdot \mathbf{u}_a \delta \gamma_f + r_f^2 \nabla_\Gamma^{\left( d_f \right)} \delta \gamma_f \cdot \nabla_\Gamma^{\left( d_f \right)} \gamma_{fa} + \delta \gamma_f \gamma_{fa} \right) M^{\left( d_f \right)} \,\mathrm{d}\Sigma = 0,
\end{split}
\end{equation}
and the first order variational of the augmented Lagrangian to $d_f$ can be set to be zero as
\begin{equation}\label{equ:WeakAdjEquPDEFilterEqudffBulkMT}
\begin{split}
  & \int_\Sigma r_f^2 \left( \nabla_\Gamma^{\left( d_f, \delta d_f \right)} \gamma_f \cdot \nabla_\Gamma^{\left( d_f \right)} \gamma_{fa} + \nabla_\Gamma^{\left( d_f \right)} \gamma_f \cdot \nabla_\Gamma^{\left( d_f, \delta d_f \right)} \gamma_{fa} \right) M^{\left( d_f \right)} \\
  & + \left( \alpha \mathbf{u} \cdot \mathbf{u}_a + r_f^2 \nabla_\Gamma^{\left( d_f \right)} \gamma_f \cdot \nabla_\Gamma^{\left( d_f \right)} \gamma_{fa} + \gamma_f \gamma_{fa} - \gamma \gamma_{fa} \right) M^{\left( d_f, \delta d_f \right)} \\
  & + r_m^2 \nabla_\Sigma \delta d_f \cdot \nabla_\Sigma d_{fa} + \delta d_f d_{fa} - \mathbf{n}_\Sigma \cdot \boldsymbol{\lambda}_{\mathbf{s}a} \delta d_f \,\mathrm{d}\Sigma = 0.
\end{split}
\end{equation}

The constraints in Eqs. \ref{equ:ConstraintForAugmentedLagrangianBulkMT} and \ref{equ:ConstraintForVariationalAugmentedLagrangianObjBulkMT} are imposed to Eqs. \ref{equ:WeakAdjEquCDEquBulkMT}, \ref{equ:WeakAdjEquNSEquBulkMT}, \ref{equ:WeakAdjEquHarmonicEquBulkMT}, \ref{equ:WeakAdjEquPDEFilterEquGafBulkMT} and \ref{equ:WeakAdjEquPDEFilterEqudffBulkMT}. Then, the adjoint sensitivity of $J_c$ is derived as
\begin{equation}\label{equ:AdjSensitivityGaDmVariationalFormBulkMT}
\begin{split}
\delta \hat{J}_c = \int_\Sigma - \gamma_{fa} \delta \gamma M^{\left( d_f \right)} - A_d d_{fa} \delta d_m \,\mathrm{d}\Sigma.
\end{split}
\end{equation}

Without losing the arbitrariness of $\delta \mathbf{u}$, $\delta p$, $\delta c$, $\delta \mathbf{s}$, $\delta \boldsymbol{\lambda}_{\mathbf{s}}$, $\delta \gamma_f$, $\delta d_f$, $\delta \gamma$ and $\delta d_m$, one can set 
\begin{equation}
\left.\begin{split}
& \tilde{\mathbf{u}}_a = \delta \mathbf{u} \\
& \tilde{p}_a = \delta p \\
& \tilde{c}_a = \delta c \\
& \tilde{\mathbf{s}}_a = \delta \mathbf{s} \\
& \tilde{\boldsymbol{\lambda}}_{\mathbf{s}a} = \delta \boldsymbol{\lambda}_{\mathbf{s}} \\
& \tilde{\gamma}_{fa} = \delta \gamma_f \\
& \tilde{d}_{fa} = \delta d_f \\
& \tilde{\gamma} = \delta \gamma \\
& \tilde{d}_m = \delta d_m 
\end{split}\right\}
~\mathrm{with}~
\left\{\begin{split}
& \forall \tilde{\mathbf{u}}_a \in \left(\mathcal{H}\left(\Omega\right)\right)^3 \\
& \forall \tilde{p}_a \in \mathcal{H}\left(\Omega\right)\\
& \forall \tilde{c}_a \in \mathcal{H}\left(\Omega\right) \\
& \forall \tilde{\mathbf{s}}_a \in \left(\mathcal{H}\left(\Omega\right)\right)^3 \\
& \forall \tilde{\boldsymbol{\lambda}}_{\mathbf{s}a} \in \left(\mathcal{H}^{{1\over2}}\left(\Sigma\right)\right)^3 \\
& \forall \tilde{\gamma}_{fa} \in \mathcal{H}\left(\Sigma\right) \\
& \forall \tilde{d}_{fa} \in \mathcal{H}\left(\Sigma\right)\\
& \forall \tilde{\gamma} \in \mathcal{L}^2\left(\Sigma\right) \\
& \forall \tilde{d}_m \in \mathcal{L}^2\left(\Sigma\right)
\end{split}\right.
\end{equation}
for Eqs. \ref{equ:WeakAdjEquCDEquBulkMT}, \ref{equ:WeakAdjEquNSEquBulkMT}, \ref{equ:WeakAdjEquHarmonicEquBulkMT}, \ref{equ:WeakAdjEquPDEFilterEquGafBulkMT} and \ref{equ:WeakAdjEquPDEFilterEqudffBulkMT} to derive the adjoint system composed of Eqs. \ref{equ:WeakAdjEquCDEquBulkMTCa}, \ref{equ:AdjBulkNavierStokesEqusJObjectiveMTUaPa}, \ref{equ:WeakAdjEquHarmonicEquBulkMTSa}, \ref{equ:AdjPDEFilterJObjectiveGafMHMGafa} and \ref{equ:AdjPDEFilterJObjectiveDmMHMDfa}.

\subsection{Adjoint analysis for constraint of pressure drop in Eq. \ref{equ:VarProToopBulkNSCDMHT}} \label{sec:AdjointAnalysisPressureDropBulkCHTMHMBulk}

Based on the transformed pressure drop in Eq. \ref{equ:TransformedPressureConstraintSurfaceNSCD}, the variational formulations of the Laplace's equation in Eq. \ref{equ:WeakFormLaplacianBulkFlowMHT}, the surface-PDE filters in Eqs. \ref{equ:VariationalFormulationPDEFilterBaseManifoldMHM} and \ref{equ:VariationalFormulationPDEFilterMHM} and the Navier-Stokes equations in Eq. \ref{equ:TransformedVariationalFormulationNavierStokesEquBulkFlowMHT}, the augmented Lagrangian of the pressure drop in Eq. \ref{equ:VarProToopBulkNSCDMHT} can be derived as
\begin{equation}\label{equ:AugmentedLagrangianConstraintBulkMTDP}
\begin{split}
  \widehat{\Delta P} = & \int_{\Sigma_{v,\Omega}} p \,\mathrm{d}\Sigma_{\partial\Omega} - \int_{\Sigma_{s,\Omega}} p \,\mathrm{d}\Sigma_{\partial\Omega} + \int_\Omega \bigg[ \rho \left( \mathbf{u} \cdot \nabla_{\mathbf{x}_\Xi}^{\left(\mathbf{s}\right)} \right) \mathbf{u} \cdot \mathbf{u}_a + {\eta\over2} \left( \nabla_{\mathbf{x}_\Xi}^{\left(\mathbf{s}\right)} \mathbf{u} + \nabla_{\mathbf{x}_\Xi}^{\left(\mathbf{s}\right)} \mathbf{u}^\mathrm{T} \right) \\
  & : \left( \nabla_{\mathbf{x}_\Xi}^{\left(\mathbf{s}\right)} \mathbf{u}_a + \nabla_{\mathbf{x}_\Xi}^{\left(\mathbf{s}\right)} \mathbf{u}_a^\mathrm{T} \right) - p\,\mathrm{div}_{\mathbf{x}_\Xi}^{\left(\mathbf{s}\right)} \mathbf{u}_a - p_a \mathrm{div}_{\mathbf{x}_\Xi}^{\left(\mathbf{s}\right)} \mathbf{u} \bigg] K^{\left( \mathbf{s} \right)} \,\mathrm{d}\Omega + \sum_{E_\Omega\in\mathcal{E}_\Omega} \int_{E_\Omega} - \tau_{BP,\Xi}^{\left(\mathbf{s}\right)} \nabla_{\mathbf{x}_\Xi}^{\left(\mathbf{s}\right)} p \\
  & \cdot \nabla_{\mathbf{x}_\Xi}^{\left(\mathbf{s}\right)} p_a K^{\left( \mathbf{s} \right)} \,\mathrm{d}\Omega + \int_\Sigma \alpha \mathbf{u} \cdot \mathbf{u}_a M^{\left( d_f \right)} \, \mathrm{d}\Sigma + \int_\Omega - \nabla_{\mathbf{x}_\Omega} \mathbf{s} : \nabla_{\mathbf{x}_\Omega} \mathbf{s}_a \,\mathrm{d}\Omega + \int_\Sigma \left( \mathbf{s} - d_f \mathbf{n}_\Sigma \right) \\
  & \cdot \boldsymbol{\lambda}_{\mathbf{s}a} + \boldsymbol{\lambda}_\mathbf{s} \cdot \mathbf{s}_a \,\mathrm{d}\Sigma + \int_\Sigma \left( r_f^2 \nabla_\Gamma^{\left( d_f \right)} \gamma_f \cdot \nabla_\Gamma^{\left( d_f \right)} \gamma_{fa} + \gamma_f \gamma_{fa} - \gamma \gamma_{fa} \right) M^{\left( d_f \right)} \,\mathrm{d}\Sigma \\
  & + \int_\Sigma r_m^2 \nabla_\Sigma d_f \cdot \nabla_\Sigma d_{fa} + d_f d_{fa} - A_d \left( d_m - {1\over2} \right) d_{fa} \,\mathrm{d}\Sigma,
\end{split}
\end{equation}
where the adjoint variables satisfy
\begin{equation}\label{equ:ConstraintForAugmentedLagrangianConstraintBulkMTDP}
  \left.\begin{split}
  & \mathbf{u}_a \in\left(\mathcal{H}\left(\Omega\right)\right)^3 \\
  & p_a \in \mathcal{H}\left(\Omega\right) \\
  & \mathbf{s}_a \in \left(\mathcal{H}\left(\Omega\right)\right)^3 \\
  & \boldsymbol{\lambda}_{\mathbf{s}a} \in \left(\mathcal{H}^{-{1\over2}}\left(\Sigma\right)\right)^3 \\
  & \gamma_{fa} \in \mathcal{H}\left(\Sigma\right) \\
  & d_{fa} \in \mathcal{H}\left(\Sigma\right)
  \end{split}\right\}~\mathrm{with}~
  \left\{\begin{split}
  & \mathbf{u}_a = \mathbf{0} ~ \mathrm{at} ~ \forall \mathbf{x}_\Omega \in \Sigma_{v,\Omega} \cup \Sigma_{v_0,\Omega} \\
  & \mathbf{s}_a = \mathbf{0} ~ \mathrm{at} ~ \forall \mathbf{x}_\Omega \in \Sigma_{v,\Omega} \cup \Sigma_{s,\Omega} 
  \end{split}\right..
\end{equation}
The first order variational of the augmented Lagrangian in Eq. \ref{equ:AugmentedLagrangianConstraintBulkMTDP} can be derived as
\begin{equation}\label{equ:1stVariAugmentedLagrangianConstraintBulkMTDP}
\begin{split}
  \delta \widehat{\Delta P} = & \int_{\Sigma_{v,\Omega}} \delta p \,\mathrm{d}\Sigma_{\partial\Omega} - \int_{\Sigma_{s,\Omega}} \delta p \,\mathrm{d}\Sigma_{\partial\Omega} + \int_\Omega \bigg[ \rho \left( \delta \mathbf{u} \cdot \nabla_{\mathbf{x}_\Xi}^{\left(\mathbf{s}\right)} \right) \mathbf{u} \cdot \mathbf{u}_a + \rho \left( \mathbf{u} \cdot \nabla_{\mathbf{x}_\Xi}^{\left(\mathbf{s}, \delta \mathbf{s}\right)} \right) \mathbf{u} \\
  & \cdot \mathbf{u}_a + \rho \left( \mathbf{u} \cdot \nabla_{\mathbf{x}_\Xi}^{\left(\mathbf{s}\right)} \right) \delta \mathbf{u} \cdot \mathbf{u}_a + {\eta\over2} \left( \nabla_{\mathbf{x}_\Xi}^{\left(\mathbf{s}, \delta \mathbf{s}\right)} \mathbf{u} + \nabla_{\mathbf{x}_\Xi}^{\left(\mathbf{s}, \delta \mathbf{s}\right)} \mathbf{u}^\mathrm{T} \right) : \left( \nabla_{\mathbf{x}_\Xi}^{\left(\mathbf{s}\right)} \mathbf{u}_a + \nabla_{\mathbf{x}_\Xi}^{\left(\mathbf{s}\right)} \mathbf{u}_a^\mathrm{T} \right) \\
  & + {\eta\over2} \left( \nabla_{\mathbf{x}_\Xi}^{\left(\mathbf{s}\right)} \delta \mathbf{u} + \nabla_{\mathbf{x}_\Xi}^{\left(\mathbf{s}\right)} \delta \mathbf{u}^\mathrm{T} \right) : \left( \nabla_{\mathbf{x}_\Xi}^{\left(\mathbf{s}\right)} \mathbf{u}_a + \nabla_{\mathbf{x}_\Xi}^{\left(\mathbf{s}\right)} \mathbf{u}_a^\mathrm{T} \right) + {\eta\over2} \left( \nabla_{\mathbf{x}_\Xi}^{\left(\mathbf{s}\right)} \mathbf{u} + \nabla_{\mathbf{x}_\Xi}^{\left(\mathbf{s}\right)} \mathbf{u}^\mathrm{T} \right) \\
  & : \left( \nabla_{\mathbf{x}_\Xi}^{\left(\mathbf{s}, \delta \mathbf{s}\right)} \mathbf{u}_a + \nabla_{\mathbf{x}_\Xi}^{\left(\mathbf{s}, \delta \mathbf{s}\right)} \mathbf{u}_a^\mathrm{T} \right) - \delta p\,\mathrm{div}_{\mathbf{x}_\Xi}^{\left(\mathbf{s}\right)} \mathbf{u}_a - p\,\mathrm{div}_{\mathbf{x}_\Xi}^{\left(\mathbf{s}, \delta \mathbf{s}\right)} \mathbf{u}_a - p_a \mathrm{div}_{\mathbf{x}_\Xi}^{\left(\mathbf{s}, \delta \mathbf{s}\right)} \mathbf{u} \\
  & - p_a \mathrm{div}_{\mathbf{x}_\Xi}^{\left(\mathbf{s}\right)} \delta \mathbf{u} \bigg] K^{\left( \mathbf{s} \right)} + \bigg[ \rho \left( \mathbf{u} \cdot \nabla_{\mathbf{x}_\Xi}^{\left(\mathbf{s}\right)} \right) \mathbf{u} \cdot \mathbf{u}_a + {\eta\over2} \left( \nabla_{\mathbf{x}_\Xi}^{\left(\mathbf{s}\right)} \mathbf{u} + \nabla_{\mathbf{x}_\Xi}^{\left(\mathbf{s}\right)} \mathbf{u}^\mathrm{T} \right) \\
  & : \left( \nabla_{\mathbf{x}_\Xi}^{\left(\mathbf{s}\right)} \mathbf{u}_a + \nabla_{\mathbf{x}_\Xi}^{\left(\mathbf{s}\right)} \mathbf{u}_a^\mathrm{T} \right) - p\,\mathrm{div}_{\mathbf{x}_\Xi}^{\left(\mathbf{s}\right)} \mathbf{u}_a - p_a \mathrm{div}_{\mathbf{x}_\Xi}^{\left(\mathbf{s}\right)} \mathbf{u} \bigg] K^{\left( \mathbf{s}, \delta \mathbf{s} \right)} - \nabla_{\mathbf{x}_\Omega} \delta \mathbf{s} : \nabla_{\mathbf{x}_\Omega} \mathbf{s}_a \,\mathrm{d}\Omega \\
  & + \sum_{E_\Omega\in\mathcal{E}_\Omega} \int_{E_\Omega} - \tau_{BP,\Xi}^{\left(\mathbf{s}, \delta \mathbf{s}\right)} \nabla_{\mathbf{x}_\Xi}^{\left(\mathbf{s}\right)} p \cdot \nabla_{\mathbf{x}_\Xi}^{\left(\mathbf{s}\right)} p_a K^{\left( \mathbf{s} \right)} - \tau_{BP,\Xi}^{\left(\mathbf{s}\right)} \nabla_{\mathbf{x}_\Xi}^{\left(\mathbf{s}, \delta \mathbf{s}\right)} p \cdot \nabla_{\mathbf{x}_\Xi}^{\left(\mathbf{s}\right)} p_a K^{\left( \mathbf{s} \right)} - \tau_{BP,\Xi}^{\left(\mathbf{s}\right)} \\
  & \nabla_{\mathbf{x}_\Xi}^{\left(\mathbf{s}\right)} \delta p \cdot \nabla_{\mathbf{x}_\Xi}^{\left(\mathbf{s}\right)} p_a K^{\left( \mathbf{s} \right)} - \tau_{BP,\Xi}^{\left(\mathbf{s}\right)} \nabla_{\mathbf{x}_\Xi}^{\left(\mathbf{s}\right)} p \cdot \nabla_{\mathbf{x}_\Xi}^{\left(\mathbf{s}, \delta \mathbf{s}\right)} p_a K^{\left( \mathbf{s} \right)} - \tau_{BP,\Xi}^{\left(\mathbf{s}\right)} \nabla_{\mathbf{x}_\Xi}^{\left(\mathbf{s}\right)} p \cdot \nabla_{\mathbf{x}_\Xi}^{\left(\mathbf{s}\right)} p_a \\
  & K^{\left( \mathbf{s}, \delta \mathbf{s} \right)} \,\mathrm{d}\Omega + \int_\Sigma {\partial\alpha\over\partial\gamma_p} {\partial\gamma_p\over\partial\gamma_f} \mathbf{u} \cdot \mathbf{u}_a M^{\left( d_f \right)} \delta \gamma_f + \alpha \delta \mathbf{u} \cdot \mathbf{u}_a M^{\left( d_f \right)} + \alpha \mathbf{u} \cdot \mathbf{u}_a M^{\left( d_f, \delta d_f \right)} \\
  & + \left( \delta \mathbf{s} - \delta d_f \mathbf{n}_\Sigma \right) \cdot \boldsymbol{\lambda}_{\mathbf{s}a} + \delta \boldsymbol{\lambda}_\mathbf{s} \cdot \mathbf{s}_a + \bigg[ r_f^2 \bigg( \nabla_\Gamma^{\left( d_f, \delta d_f \right)} \gamma_f \cdot \nabla_\Gamma^{\left( d_f \right)} \gamma_{fa} + \nabla_\Gamma^{\left( d_f \right)} \delta \gamma_f \\
  & \cdot \nabla_\Gamma^{\left( d_f \right)} \gamma_{fa} + \nabla_\Gamma^{\left( d_f \right)} \gamma_f \cdot \nabla_\Gamma^{\left( d_f, \delta d_f \right)} \gamma_{fa} \bigg) + \delta \gamma_f \gamma_{fa} - \delta \gamma \gamma_{fa} \bigg] M^{\left( d_f \right)} \\
  & + \left( r_f^2 \nabla_\Gamma^{\left( d_f \right)} \gamma_f \cdot \nabla_\Gamma^{\left( d_f \right)} \gamma_{fa} + \gamma_f \gamma_{fa} - \gamma \gamma_{fa} \right) M^{\left( d_f, \delta d_f \right)} \\
  & + r_m^2 \nabla_\Sigma \delta d_f \cdot \nabla_\Sigma d_{fa} + \delta d_f d_{fa} - A_d \delta d_m d_{fa} \,\mathrm{d}\Sigma
\end{split}
\end{equation}
with the satisfication of the constraints in Eq. \ref{equ:ConstraintForAugmentedLagrangianConstraintBulkMTDP}
and
\begin{equation}\label{equ:ConstraintForVariationalAugmentedLagrangianConstraintBulkMTDP}
  \left.\begin{split}
  & \delta \mathbf{u} \in\left(\mathcal{H}\left(\Omega\right)\right)^3 \\
  & \delta p \in \mathcal{H}\left(\Omega\right) \\
  & \delta \mathbf{s} \in \left(\mathcal{H}\left(\Omega\right)\right)^3 \\
  & \delta \boldsymbol{\lambda}_{\mathbf{s}} \in \left(\mathcal{H}^{{1\over2}}\left(\Sigma\right)\right)^3 \\
  & \delta \gamma_f \in \mathcal{H}\left(\Sigma\right) \\
  & \delta d_f \in \mathcal{H}\left(\Sigma\right)
  \end{split}\right\}~\mathrm{with}~
  \left\{\begin{split}
  & \delta \mathbf{u} = \mathbf{0}~ \mathrm{at} ~ \forall \mathbf{x}_\Omega \in \Sigma_{v,\Omega} \cup \Sigma_{v_0,\Omega} \\
  & \delta \mathbf{s} = \mathbf{0} ~ \mathrm{at} ~ \forall \mathbf{x}_\Omega \in \Sigma_{v,\Omega} \cup \Sigma_{s,\Omega}
  \end{split}\right..
\end{equation}

According to the Karush-Kuhn-Tucker conditions of the PDE constrained optimization problem, the first order variational of the augmented Lagrangian to $\mathbf{u}$ and $p$ can be set to be zero as
\begin{equation}\label{equ:WeakAdjEquNSEquBulkMTDP}
\begin{split}
  & \int_{\Sigma_{v,\Omega}} \delta p \,\mathrm{d}\Sigma_{\partial\Omega} - \int_{\Sigma_{s,\Omega}} \delta p \,\mathrm{d}\Sigma_{\partial\Omega} + \int_\Omega \Big[ \rho \left( \delta \mathbf{u} \cdot \nabla_{\mathbf{x}_\Xi}^{\left(\mathbf{s}\right)} \right) \mathbf{u} \cdot \mathbf{u}_a + \rho \left( \mathbf{u} \cdot \nabla_{\mathbf{x}_\Xi}^{\left(\mathbf{s}\right)} \right) \delta \mathbf{u} \cdot \mathbf{u}_a \\
  & + {\eta\over2} \left( \nabla_{\mathbf{x}_\Xi}^{\left(\mathbf{s}\right)} \delta \mathbf{u} + \nabla_{\mathbf{x}_\Xi}^{\left(\mathbf{s}\right)} \delta \mathbf{u}^\mathrm{T} \right) : \left( \nabla_{\mathbf{x}_\Xi}^{\left(\mathbf{s}\right)} \mathbf{u}_a + \nabla_{\mathbf{x}_\Xi}^{\left(\mathbf{s}\right)} \mathbf{u}_a^\mathrm{T} \right) - \delta p\,\mathrm{div}_{\mathbf{x}_\Xi}^{\left(\mathbf{s}\right)} \mathbf{u}_a - p_a \mathrm{div}_{\mathbf{x}_\Xi}^{\left(\mathbf{s}\right)} \delta \mathbf{u} \Big] K^{\left( \mathbf{s} \right)} \,\mathrm{d}\Omega \\
  & + \sum_{E_\Omega\in\mathcal{E}_\Omega} \int_{E_\Omega} - \tau_{BP,\Xi}^{\left(\mathbf{s}\right)} \nabla_{\mathbf{x}_\Xi}^{\left(\mathbf{s}\right)} \delta p \cdot \nabla_{\mathbf{x}_\Xi}^{\left(\mathbf{s}\right)} p_a K^{\left( \mathbf{s} \right)} \,\mathrm{d}\Omega + \int_\Sigma \alpha \delta \mathbf{u} \cdot \mathbf{u}_a M^{\left( d_f \right)} \, \mathrm{d}\Sigma = 0, \\
\end{split}
\end{equation}
the first order variational of the augmented Lagrangian to $\mathbf{s}$ and $\boldsymbol{\lambda}_\mathbf{s}$ can be set to be zero as
\begin{equation}\label{equ:WeakAdjEquHarmonicEquBulkMTDP}
\begin{split}
  & \int_\Omega \Big[ \rho \left( \mathbf{u} \cdot \nabla_{\mathbf{x}_\Xi}^{\left(\mathbf{s}, \delta \mathbf{s}\right)} \right) \mathbf{u} \cdot \mathbf{u}_a + {\eta\over2} \left( \nabla_{\mathbf{x}_\Xi}^{\left(\mathbf{s}, \delta \mathbf{s}\right)} \mathbf{u} + \nabla_{\mathbf{x}_\Xi}^{\left(\mathbf{s}, \delta \mathbf{s}\right)} \mathbf{u}^\mathrm{T} \right) : \left( \nabla_{\mathbf{x}_\Xi}^{\left(\mathbf{s}\right)} \mathbf{u}_a + \nabla_{\mathbf{x}_\Xi}^{\left(\mathbf{s}\right)} \mathbf{u}_a^\mathrm{T} \right) + {\eta\over2} \\
  & \left( \nabla_{\mathbf{x}_\Xi}^{\left(\mathbf{s}\right)} \mathbf{u} + \nabla_{\mathbf{x}_\Xi}^{\left(\mathbf{s}\right)} \mathbf{u}^\mathrm{T} \right) : \left( \nabla_{\mathbf{x}_\Xi}^{\left(\mathbf{s}, \delta \mathbf{s}\right)} \mathbf{u}_a + \nabla_{\mathbf{x}_\Xi}^{\left(\mathbf{s}, \delta \mathbf{s}\right)} \mathbf{u}_a^\mathrm{T} \right) - p\,\mathrm{div}_{\mathbf{x}_\Xi}^{\left(\mathbf{s}, \delta \mathbf{s}\right)} \mathbf{u}_a - p_a \mathrm{div}_{\mathbf{x}_\Xi}^{\left(\mathbf{s}, \delta \mathbf{s}\right)} \mathbf{u} \Big] K^{\left( \mathbf{s} \right)} \,\mathrm{d}\Omega \\
  & + \int_\Omega \Big[ \rho \left( \mathbf{u} \cdot \nabla_{\mathbf{x}_\Xi}^{\left(\mathbf{s}\right)} \right) \mathbf{u} \cdot \mathbf{u}_a + {\eta\over2} \left( \nabla_{\mathbf{x}_\Xi}^{\left(\mathbf{s}\right)} \mathbf{u} + \nabla_{\mathbf{x}_\Xi}^{\left(\mathbf{s}\right)} \mathbf{u}^\mathrm{T} \right) : \left( \nabla_{\mathbf{x}_\Xi}^{\left(\mathbf{s}\right)} \mathbf{u}_a + \nabla_{\mathbf{x}_\Xi}^{\left(\mathbf{s}\right)} \mathbf{u}_a^\mathrm{T} \right) - p\,\mathrm{div}_{\mathbf{x}_\Xi}^{\left(\mathbf{s}\right)} \mathbf{u}_a \\
  & - p_a \mathrm{div}_{\mathbf{x}_\Xi}^{\left(\mathbf{s}\right)} \mathbf{u} \Big] K^{\left( \mathbf{s}, \delta \mathbf{s} \right)} \,\mathrm{d}\Omega + \sum_{E_\Omega\in\mathcal{E}_\Omega} \int_{E_\Omega} - \tau_{BP,\Xi}^{\left(\mathbf{s}, \delta \mathbf{s}\right)} \nabla_{\mathbf{x}_\Xi}^{\left(\mathbf{s}\right)} p \cdot \nabla_{\mathbf{x}_\Xi}^{\left(\mathbf{s}\right)} p_a K^{\left( \mathbf{s} \right)} - \tau_{BP,\Xi}^{\left(\mathbf{s}\right)} \nabla_{\mathbf{x}_\Xi}^{\left(\mathbf{s}, \delta \mathbf{s}\right)} p \\
  & \cdot \nabla_{\mathbf{x}_\Xi}^{\left(\mathbf{s}\right)} p_a K^{\left( \mathbf{s} \right)} - \tau_{BP,\Xi}^{\left(\mathbf{s}\right)} \nabla_{\mathbf{x}_\Xi}^{\left(\mathbf{s}\right)} p \cdot \nabla_{\mathbf{x}_\Xi}^{\left(\mathbf{s}, \delta \mathbf{s}\right)} p_a K^{\left( \mathbf{s} \right)} - \tau_{BP,\Xi}^{\left(\mathbf{s}\right)} \nabla_{\mathbf{x}_\Xi}^{\left(\mathbf{s}\right)} p \cdot \nabla_{\mathbf{x}_\Xi}^{\left(\mathbf{s}\right)} p_a K^{\left( \mathbf{s}, \delta \mathbf{s} \right)} \,\mathrm{d}\Omega \\
  & + \int_\Omega - \nabla_{\mathbf{x}_\Omega} \delta \mathbf{s} : \nabla_{\mathbf{x}_\Omega} \mathbf{s}_a \,\mathrm{d}\Omega + \int_\Sigma \delta \mathbf{s} \cdot \boldsymbol{\lambda}_{\mathbf{s}a} + \delta \boldsymbol{\lambda}_\mathbf{s} \cdot \mathbf{s}_a \,\mathrm{d}\Sigma = 0, \\
\end{split}
\end{equation}
the first order variational of the augmented Lagrangian to $\gamma_f$ can be set to be zero as
\begin{equation}\label{equ:WeakAdjEquPDEFilterEquGafBulkMTDP}
\begin{split}
  & \int_\Sigma \left( {\partial\alpha\over\partial\gamma_p} {\partial\gamma_p\over\partial\gamma_f} \mathbf{u} \cdot \mathbf{u}_a \delta \gamma_f + r_f^2 \nabla_\Gamma^{\left( d_f \right)} \delta \gamma_f \cdot \nabla_\Gamma^{\left( d_f \right)} \gamma_{fa} + \delta \gamma_f \gamma_{fa} \right) M^{\left( d_f \right)} \,\mathrm{d}\Sigma = 0,
\end{split}
\end{equation}
and the first order variational of the augmented Lagrangian to $d_f$ can be set to be zero as
\begin{equation}\label{equ:WeakAdjEquPDEFilterEqudffBulkMTDP}
\begin{split}
  & \int_\Sigma r_f^2 \left( \nabla_\Gamma^{\left( d_f, \delta d_f \right)} \gamma_f \cdot \nabla_\Gamma^{\left( d_f \right)} \gamma_{fa} + \nabla_\Gamma^{\left( d_f \right)} \gamma_f \cdot \nabla_\Gamma^{\left( d_f, \delta d_f \right)} \gamma_{fa} \right) M^{\left( d_f \right)} \\
  & + \left( r_f^2 \nabla_\Gamma^{\left( d_f \right)} \gamma_f \cdot \nabla_\Gamma^{\left( d_f \right)} \gamma_{fa} + \gamma_f \gamma_{fa} - \gamma \gamma_{fa} + \alpha \mathbf{u} \cdot \mathbf{u}_a \right) M^{\left( d_f, \delta d_f \right)} \\
  & + r_m^2 \nabla_\Sigma \delta d_f \cdot \nabla_\Sigma d_{fa} + \delta d_f d_{fa} - \delta d_f \mathbf{n}_\Sigma \cdot \boldsymbol{\lambda}_{\mathbf{s}a} \,\mathrm{d}\Sigma = 0.
\end{split}
\end{equation}

The constraints in Eqs. \ref{equ:ConstraintForAugmentedLagrangianConstraintBulkMTDP} and \ref{equ:ConstraintForVariationalAugmentedLagrangianConstraintBulkMTDP} are imposed to Eqs. \ref{equ:WeakAdjEquNSEquBulkMTDP}, \ref{equ:WeakAdjEquHarmonicEquBulkMTDP}, \ref{equ:WeakAdjEquPDEFilterEquGafBulkMTDP} and \ref{equ:WeakAdjEquPDEFilterEqudffBulkMTDP}. Then, the adjoint sensitivity of $J_c$ is derived as
\begin{equation}\label{equ:AdjSensitivityGaDmVariationalFormBulkMTDP}
\begin{split}
\delta \widehat{\Delta P} = \int_\Sigma - \gamma_{fa} \delta \gamma M^{\left( d_f \right)} - A_d d_{fa} \delta d_m \,\mathrm{d}\Sigma.
\end{split}
\end{equation}

Without losing the arbitrariness of $\delta \mathbf{u}$, $\delta p$, $\delta \mathbf{s}$, $\delta \boldsymbol{\lambda}_{\mathbf{s}}$, $\delta \gamma_f$, $\delta d_f$, $\delta \gamma$ and $\delta d_m$, one can set 
\begin{equation}
\left.\begin{split}
& \tilde{\mathbf{u}}_a = \delta \mathbf{u} \\
& \tilde{p}_a = \delta p \\
& \tilde{\mathbf{s}}_a = \delta \mathbf{s} \\
& \tilde{\boldsymbol{\lambda}}_{\mathbf{s}a} = \delta \boldsymbol{\lambda}_{\mathbf{s}} \\
& \tilde{\gamma}_{fa} = \delta \gamma_f \\
& \tilde{d}_{fa} = \delta d_f \\
& \tilde{\gamma} = \delta \gamma \\
& \tilde{d}_m = \delta d_m 
\end{split}\right\}
~\mathrm{with}~
\left\{\begin{split}
& \forall \tilde{\mathbf{u}}_a \in \left(\mathcal{H}\left(\Omega\right)\right)^3 \\
& \forall \tilde{p}_a \in \mathcal{H}\left(\Omega\right)\\
& \forall \tilde{\mathbf{s}}_a \in \left(\mathcal{H}\left(\Omega\right)\right)^3 \\
& \forall \tilde{\boldsymbol{\lambda}}_{\mathbf{s}a} \in \left(\mathcal{H}^{{1\over2}}\left(\Sigma\right)\right)^3 \\
& \forall \tilde{\gamma}_{fa} \in \mathcal{H}\left(\Sigma\right) \\
& \forall \tilde{d}_{fa} \in \mathcal{H}\left(\Sigma\right)\\
& \forall \tilde{\gamma} \in \mathcal{L}^2\left(\Sigma\right) \\
& \forall \tilde{d}_m \in \mathcal{L}^2\left(\Sigma\right)
\end{split}\right.
\end{equation}
for Eqs. \ref{equ:WeakAdjEquNSEquBulkMTDP}, \ref{equ:WeakAdjEquHarmonicEquBulkMTDP}, \ref{equ:WeakAdjEquPDEFilterEquGafBulkMTDP} and \ref{equ:WeakAdjEquPDEFilterEqudffBulkMTDP} to derive the adjoint system composed of Eqs. \ref{equ:AdjEquSurfaceNSMHMPressureDropBulk}, \ref{equ:WeakAdjEquHarmonicEquBulkMTSaPressureDrop}, \ref{equ:AdjPDEFilterPressureDropGaMHMBulk} and \ref{equ:AdjPDEFilterJPressureDropDmMHMBulk}.

\subsection{Adjoint analysis for design objective in Eq. \ref{equ:VarProToopBulkNSCHTMHT}} \label{sec:AdjointAnalysisDesignObjectiveBulkCHT}

Based on the transformed design objective in Eq. \ref{equ:TransformedDesignObjectiveBulkCHM}, the variational formulations of the Laplace's equation in Eq. \ref{equ:WeakFormLaplacianBulkFlowMHT}, the surface-PDE filters in Eqs. \ref{equ:VariationalFormulationPDEFilterBaseManifoldMHM} and \ref{equ:VariationalFormulationPDEFilterMHM} and the Navier-Stokes equations in Eq. \ref{equ:TransformedVariationalFormulationBulkNSEqusHM} and the convective heat-transfer equation in Eq. \ref{equ:TransformedVariationalFormulationBulkCHMEqu}, the augmented Lagrangian of the design objective in Eq. \ref{equ:VarProToopBulkNSCHTMHT} can be derived as
\begin{equation}\label{equ:AugmentedLagrangianMatchOptimizationBulkHT}
\begin{split}
  \hat{J}_T = & \int_\Omega f_{id,\Xi}^{\left( \mathbf{s} \right)} k \nabla_{\mathbf{x}_\Xi}^{\left( \mathbf{s} \right)} T \cdot \nabla_{\mathbf{x}_\Xi}^{\left( \mathbf{s} \right)} T K^{\left( \mathbf{s} \right)} + \Big[ \rho \left( \mathbf{u} \cdot \nabla_{\mathbf{x}_\Xi}^{\left(\mathbf{s}\right)} \right) \mathbf{u} \cdot \mathbf{u}_a + {\eta\over2} \left( \nabla_{\mathbf{x}_\Xi}^{\left(\mathbf{s}\right)} \mathbf{u} + \nabla_{\mathbf{x}_\Xi}^{\left(\mathbf{s}\right)} \mathbf{u}^\mathrm{T} \right) : \Big( \nabla_{\mathbf{x}_\Xi}^{\left(\mathbf{s}\right)} \mathbf{u}_a \\
  & + \nabla_{\mathbf{x}_\Xi}^{\left(\mathbf{s}\right)} \mathbf{u}_a^\mathrm{T} \Big) - p\,\mathrm{div}_{\mathbf{x}_\Xi}^{\left(\mathbf{s}\right)} \mathbf{u}_a - p_a \mathrm{div}_{\mathbf{x}_\Xi}^{\left(\mathbf{s}\right)} \mathbf{u} \Big] K^{\left(\mathbf{s}\right)} \,\mathrm{d}\Omega - \sum_{E_\Omega\in\mathcal{E}_\Omega} \int_{E_\Omega} \Big[ \tau_{LS\mathbf{u},\Xi}^{\left(\mathbf{s}\right)} \Big( \rho \mathbf{u} \cdot \nabla_{\mathbf{x}_\Xi}^{\left(\mathbf{s}\right)} \mathbf{u} \\
  & + \nabla_{\mathbf{x}_\Xi}^{\left(\mathbf{s}\right)} p \Big) \cdot \left( \rho \mathbf{u} \cdot \nabla_{\mathbf{x}_\Xi}^{\left(\mathbf{s}\right)} \mathbf{u}_a + \nabla_{\mathbf{x}_\Xi}^{\left(\mathbf{s}\right)} p_a \right) + \tau_{LSp,\Xi}^{\left(\mathbf{s}\right)} \left( \rho \mathrm{div}_{\mathbf{x}_\Xi}^{\left(\mathbf{s}\right)} \mathbf{u} \right) \left( \mathrm{div}_{\mathbf{x}_\Xi}^{\left(\mathbf{s}\right)} \mathbf{u}_a \right) \Big] K^{\left(\mathbf{s}\right)} \,\mathrm{d}\Omega \\
  & + \int_\Sigma \alpha \mathbf{u} \cdot \mathbf{u}_a M^{\left(d_f\right)} \, \mathrm{d}\Sigma +  \int_\Omega \left[ \left( \rho C_p \mathbf{u} \cdot \nabla_{\mathbf{x}_\Xi}^{\left(\mathbf{s}\right)} T - Q \right) T_a + k \nabla_{\mathbf{x}_\Xi}^{\left(\mathbf{s}\right)} T \cdot \nabla_{\mathbf{x}_\Xi}^{\left(\mathbf{s}\right)} T_a \right] \\
  & K^{\left(\mathbf{s}\right)} \,\mathrm{d}\Omega + \sum_{E_\Omega\in\mathcal{E}_\Omega} \int_{E_\Omega} \tau_{LST,\Xi}^{\left(\mathbf{s}\right)} \left( \rho C_p \mathbf{u} \cdot \nabla_{\mathbf{x}_\Xi}^{\left(\mathbf{s}\right)} T - Q \right) \left( \rho C_p \mathbf{u} \cdot \nabla_{\mathbf{x}_\Xi}^{\left(\mathbf{s}\right)} T_a \right) \\
  & K^{\left(\mathbf{s}\right)} \,\mathrm{d}\Omega - \int_\Omega \nabla_{\mathbf{x}_\Omega} \mathbf{s} : \nabla_{\mathbf{x}_\Omega} \mathbf{s}_a \,\mathrm{d}\Omega + \int_\Sigma \left( \mathbf{s} - d_f \mathbf{n}_\Sigma \right) \cdot \boldsymbol{\lambda}_{\mathbf{s}a} + \boldsymbol{\lambda}_\mathbf{s} \cdot \mathbf{s}_a \,\mathrm{d}\Sigma \\
  & + \int_\Sigma \bigg( r_f^2 \nabla_\Gamma^{\left( d_f \right)} \gamma_f \cdot \nabla_\Gamma^{\left( d_f \right)} \gamma_{fa} + \gamma_f \gamma_{fa} - \gamma \gamma_{fa} \bigg) M^{\left( d_f \right)} \,\mathrm{d}\Sigma \\
  & + \int_\Sigma r_m^2 \nabla_\Sigma d_f \cdot \nabla_\Sigma d_{fa} + d_f d_{fa} - A_d \left( d_m - {1\over2} \right) d_{fa} \,\mathrm{d}\Sigma,
\end{split}
\end{equation}
where the adjoint variables satisfy
\begin{equation}\label{equ:ConstraintForAugmentedLagrangianBulkHT}
  \left.\begin{split}
  & \mathbf{u}_a \in\left(\mathcal{H}\left(\Omega\right)\right)^3 \\
  & p_a \in \mathcal{H}\left(\Omega\right) \\
  & T_a \in \mathcal{H}\left(\Omega\right) \\
  & \mathbf{s}_a \in \left(\mathcal{H}\left(\Omega\right)\right)^3 \\
  & \boldsymbol{\lambda}_{\mathbf{s}a} \in \left(\mathcal{H}^{-{1\over2}}\left(\Sigma\right)\right)^3 \\
  & \gamma_{fa} \in \mathcal{H}\left(\Sigma\right) \\
  & d_{fa} \in \mathcal{H}\left(\Sigma\right)
  \end{split}\right\}~\mathrm{with}~
  \left\{\begin{split}
  & \mathbf{u}_a = \mathbf{0}~ \mathrm{at} ~ \forall \mathbf{x}_\Omega \in \Sigma_{v,\Omega} \cup \Sigma_{v_0,\Omega} \\
  & T_a = 0~ \mathrm{at} ~ \forall \mathbf{x}_\Omega \in \Sigma_{v,\Omega} \\
  & \mathbf{s}_a = \mathbf{0} ~ \mathrm{at} ~ \forall \mathbf{x}_\Omega \in \Sigma_{v,\Omega} \cup \Sigma_{s,\Omega} 
  \end{split}\right..
\end{equation}
The first order variational of the augmented Lagrangian in Eq. \ref{equ:AugmentedLagrangianMatchOptimizationBulkHT} can be derived as
\begin{equation}\label{equ:1stVariAugmentedLagrangianMatchOptimizationBulkHT}
\begin{split}
  \delta \hat{J}_T = & \int_\Omega f_{id,\Xi}^{\left( \mathbf{s}, \delta \mathbf{s} \right)} k \nabla_{\mathbf{x}_\Xi}^{\left( \mathbf{s} \right)} T \cdot \nabla_{\mathbf{x}_\Xi}^{\left( \mathbf{s} \right)} T K^{\left( \mathbf{s} \right)} + 2 f_{id,\Xi}^{\left( \mathbf{s} \right)} k \nabla_{\mathbf{x}_\Xi}^{\left( \mathbf{s}, \delta \mathbf{s} \right)} T \cdot \nabla_{\mathbf{x}_\Xi}^{\left( \mathbf{s} \right)} T K^{\left( \mathbf{s} \right)} + 2 f_{id,\Xi}^{\left( \mathbf{s} \right)} k \nabla_{\mathbf{x}_\Xi}^{\left( \mathbf{s} \right)} \delta T \\
  & \cdot \nabla_{\mathbf{x}_\Xi}^{\left( \mathbf{s} \right)} T K^{\left( \mathbf{s} \right)} + f_{id,\Xi}^{\left( \mathbf{s} \right)} k \nabla_{\mathbf{x}_\Xi}^{\left( \mathbf{s} \right)} T \cdot \nabla_{\mathbf{x}_\Xi}^{\left( \mathbf{s} \right)} T K^{\left( \mathbf{s}, \delta \mathbf{s} \right)} + \Big[ \rho \left( \delta \mathbf{u} \cdot \nabla_{\mathbf{x}_\Xi}^{\left(\mathbf{s}\right)} \right) \mathbf{u} \cdot \mathbf{u}_a + \rho \left( \mathbf{u} \cdot \nabla_{\mathbf{x}_\Xi}^{\left(\mathbf{s}, \delta \mathbf{s}\right)} \right) \mathbf{u} \\
  & \cdot \mathbf{u}_a + \rho \left( \mathbf{u} \cdot \nabla_{\mathbf{x}_\Xi}^{\left(\mathbf{s}\right)} \right) \delta \mathbf{u} \cdot \mathbf{u}_a + {\eta\over2} \left( \nabla_{\mathbf{x}_\Xi}^{\left(\mathbf{s}\right)} \delta \mathbf{u} + \nabla_{\mathbf{x}_\Xi}^{\left(\mathbf{s}\right)} \delta \mathbf{u}^\mathrm{T} \right) : \left( \nabla_{\mathbf{x}_\Xi}^{\left(\mathbf{s}\right)} \mathbf{u}_a + \nabla_{\mathbf{x}_\Xi}^{\left(\mathbf{s}\right)} \mathbf{u}_a^\mathrm{T} \right) \\
  & + {\eta\over2} \left( \nabla_{\mathbf{x}_\Xi}^{\left(\mathbf{s}, \delta \mathbf{s}\right)} \mathbf{u} + \nabla_{\mathbf{x}_\Xi}^{\left(\mathbf{s}, \delta \mathbf{s}\right)} \mathbf{u}^\mathrm{T} \right) : \left( \nabla_{\mathbf{x}_\Xi}^{\left(\mathbf{s}\right)} \mathbf{u}_a + \nabla_{\mathbf{x}_\Xi}^{\left(\mathbf{s}\right)} \mathbf{u}_a^\mathrm{T} \right) + {\eta\over2} \left( \nabla_{\mathbf{x}_\Xi}^{\left(\mathbf{s}\right)} \mathbf{u} + \nabla_{\mathbf{x}_\Xi}^{\left(\mathbf{s}\right)} \mathbf{u}^\mathrm{T} \right) \\
  & : \left( \nabla_{\mathbf{x}_\Xi}^{\left(\mathbf{s}, \delta \mathbf{s}\right)} \mathbf{u}_a + \nabla_{\mathbf{x}_\Xi}^{\left(\mathbf{s}, \delta \mathbf{s}\right)} \mathbf{u}_a^\mathrm{T} \right) - \delta p\,\mathrm{div}_{\mathbf{x}_\Xi}^{\left(\mathbf{s}\right)} \mathbf{u}_a - p\,\mathrm{div}_{\mathbf{x}_\Xi}^{\left(\mathbf{s}, \delta \mathbf{s}\right)} \mathbf{u}_a - p_a \mathrm{div}_{\mathbf{x}_\Xi}^{\left(\mathbf{s}, \delta \mathbf{s}\right)} \mathbf{u} \\
  & - p_a \mathrm{div}_{\mathbf{x}_\Xi}^{\left(\mathbf{s}\right)} \delta \mathbf{u} \Big] K^{\left(\mathbf{s}\right)} + \Big[ \rho \left( \mathbf{u} \cdot \nabla_{\mathbf{x}_\Xi}^{\left(\mathbf{s}\right)} \right) \mathbf{u} \cdot \mathbf{u}_a + {\eta\over2} \left( \nabla_{\mathbf{x}_\Xi}^{\left(\mathbf{s}\right)} \mathbf{u} + \nabla_{\mathbf{x}_\Xi}^{\left(\mathbf{s}\right)} \mathbf{u}^\mathrm{T} \right) \\
  & : \left( \nabla_{\mathbf{x}_\Xi}^{\left(\mathbf{s}\right)} \mathbf{u}_a + \nabla_{\mathbf{x}_\Xi}^{\left(\mathbf{s}\right)} \mathbf{u}_a^\mathrm{T} \right) - p\,\mathrm{div}_{\mathbf{x}_\Xi}^{\left(\mathbf{s}\right)} \mathbf{u}_a - p_a \mathrm{div}_{\mathbf{x}_\Xi}^{\left(\mathbf{s} \right)} \mathbf{u} \Big] K^{\left(\mathbf{s}, \delta \mathbf{s}\right)} \,\mathrm{d}\Omega \\
  & - \sum_{E_\Omega\in\mathcal{E}_\Omega} \int_{E_\Omega} \Big[ \left( \tau_{LS\mathbf{u},\Xi}^{\left(\mathbf{s}, \delta \mathbf{s}\right)} + \tau_{LS\mathbf{u},\Xi}^{\left(\mathbf{s}, \delta \mathbf{u}\right)} \right) \left( \rho \mathbf{u} \cdot \nabla_{\mathbf{x}_\Xi}^{\left(\mathbf{s}\right)} \mathbf{u} + \nabla_{\mathbf{x}_\Xi}^{\left(\mathbf{s}\right)} p \right) \cdot \left( \rho \mathbf{u} \cdot \nabla_{\mathbf{x}_\Xi}^{\left(\mathbf{s}\right)} \mathbf{u}_a + \nabla_{\mathbf{x}_\Xi}^{\left(\mathbf{s}\right)} p_a \right) \\
  & + \tau_{LS\mathbf{u},\Xi}^{\left(\mathbf{s}\right)} \Big( \rho \delta \mathbf{u} \cdot \nabla_{\mathbf{x}_\Xi}^{\left(\mathbf{s} \right)} \mathbf{u} + \rho \mathbf{u} \cdot \nabla_{\mathbf{x}_\Xi}^{\left(\mathbf{s}, \delta \mathbf{s}\right)} \mathbf{u} + \rho \mathbf{u} \cdot \nabla_{\mathbf{x}_\Xi}^{\left(\mathbf{s}\right)} \delta \mathbf{u} + \nabla_{\mathbf{x}_\Xi}^{\left(\mathbf{s}, \delta \mathbf{s}\right)} p + \nabla_{\mathbf{x}_\Xi}^{\left(\mathbf{s}\right)} \delta p \Big) \\
  & \cdot \left( \rho \mathbf{u} \cdot \nabla_{\mathbf{x}_\Xi}^{\left(\mathbf{s}\right)} \mathbf{u}_a + \nabla_{\mathbf{x}_\Xi}^{\left(\mathbf{s}\right)} p_a \right) + \tau_{LS\mathbf{u},\Xi}^{\left(\mathbf{s}\right)} \left( \rho \mathbf{u} \cdot \nabla_{\mathbf{x}_\Xi}^{\left(\mathbf{s}\right)} \mathbf{u} + \nabla_{\mathbf{x}_\Xi}^{\left(\mathbf{s}\right)} p \right) \\
  & \cdot \left( \rho \delta \mathbf{u} \cdot \nabla_{\mathbf{x}_\Xi}^{\left(\mathbf{s}\right)} \mathbf{u}_a + \rho \mathbf{u} \cdot \nabla_{\mathbf{x}_\Xi}^{\left(\mathbf{s}, \delta \mathbf{s}\right)} \mathbf{u}_a + \nabla_{\mathbf{x}_\Xi}^{\left(\mathbf{s}, \delta \mathbf{s}\right)} p_a \right) + \left( \tau_{LSp,\Xi}^{\left(\mathbf{s}, \delta \mathbf{s}\right)} + \tau_{LSp,\Xi}^{\left(\mathbf{s}, \delta \mathbf{u}\right)} \right) \\
  & \left( \rho \mathrm{div}_{\mathbf{x}_\Xi}^{\left(\mathbf{s}\right)} \mathbf{u} \right) \left( \mathrm{div}_{\mathbf{x}_\Xi}^{\left(\mathbf{s}\right)} \mathbf{u}_a \right) + \tau_{LSp,\Xi}^{\left(\mathbf{s}\right)} \left( \rho \mathrm{div}_{\mathbf{x}_\Xi}^{\left(\mathbf{s}, \delta \mathbf{s}\right)} \mathbf{u} \right) \left( \mathrm{div}_{\mathbf{x}_\Xi}^{\left(\mathbf{s}\right)} \mathbf{u}_a \right) \\
  & + \tau_{LSp,\Xi}^{\left(\mathbf{s}\right)} \left( \rho \mathrm{div}_{\mathbf{x}_\Xi}^{\left(\mathbf{s}\right)} \delta \mathbf{u} \right) \left( \mathrm{div}_{\mathbf{x}_\Xi}^{\left(\mathbf{s}\right)} \mathbf{u}_a \right) + \tau_{LSp,\Xi}^{\left(\mathbf{s}\right)} \left( \rho \mathrm{div}_{\mathbf{x}_\Xi}^{\left(\mathbf{s}\right)} \mathbf{u} \right) \left( \mathrm{div}_{\mathbf{x}_\Xi}^{\left(\mathbf{s}, \delta \mathbf{s}\right)} \mathbf{u}_a \right) \Big] K^{\left(\mathbf{s}\right)} \\
  & + \Big[ \tau_{LS\mathbf{u},\Xi}^{\left(\mathbf{s}\right)} \left( \rho \mathbf{u} \cdot \nabla_{\mathbf{x}_\Xi}^{\left(\mathbf{s}\right)} \mathbf{u} + \nabla_{\mathbf{x}_\Xi}^{\left(\mathbf{s}\right)} p + \alpha \mathbf{u} \right) \cdot \Big( \rho \mathbf{u} \cdot \nabla_{\mathbf{x}_\Xi}^{\left(\mathbf{s}\right)} \mathbf{u}_a + \nabla_{\mathbf{x}_\Xi}^{\left(\mathbf{s}\right)} p_a \Big) \\
  & + \tau_{LSp,\Xi}^{\left(\mathbf{s}\right)} \left( \rho \mathrm{div}_{\mathbf{x}_\Xi}^{\left(\mathbf{s}\right)} \mathbf{u} \right) \left( \mathrm{div}_{\mathbf{x}_\Xi}^{\left(\mathbf{s}\right)} \mathbf{u}_a \right) \Big] K^{\left(\mathbf{s}, \delta \mathbf{s}\right)} \,\mathrm{d}\Omega \\
  & + \int_\Sigma {\partial\alpha\over\partial\gamma_p} {\partial\gamma_p\over\partial\gamma_f} \mathbf{u} \cdot \mathbf{u}_a M^{\left(d_f\right)} \delta \gamma_f + \alpha \delta \mathbf{u} \cdot \mathbf{u}_a M^{\left(d_f\right)} + \alpha \mathbf{u} \cdot \mathbf{u}_a M^{\left(d_f, \delta d_f\right)} \, \mathrm{d}\Sigma \\
  & +  \int_\Omega \Big[ \left( \rho C_p \delta \mathbf{u} \cdot \nabla_{\mathbf{x}_\Xi}^{\left(\mathbf{s}\right)} T + \rho C_p \mathbf{u} \cdot \nabla_{\mathbf{x}_\Xi}^{\left(\mathbf{s}, \delta \mathbf{s}\right)} T + \rho C_p \mathbf{u} \cdot \nabla_{\mathbf{x}_\Xi}^{\left(\mathbf{s}\right)} \delta T \right) T_a + k \nabla_{\mathbf{x}_\Xi}^{\left(\mathbf{s}, \delta \mathbf{s}\right)} T \\
  & \cdot \nabla_{\mathbf{x}_\Xi}^{\left(\mathbf{s}\right)} T_a + k \nabla_{\mathbf{x}_\Xi}^{\left(\mathbf{s}\right)} \delta T \cdot \nabla_{\mathbf{x}_\Xi}^{\left(\mathbf{s}\right)} T_a + k \nabla_{\mathbf{x}_\Xi}^{\left(\mathbf{s}\right)} T \cdot \nabla_{\mathbf{x}_\Xi}^{\left(\mathbf{s}, \delta \mathbf{s}\right)} T_a \Big] K^{\left(\mathbf{s}\right)} + \Big[ \left( \rho C_p \mathbf{u} \cdot \nabla_{\mathbf{x}_\Xi}^{\left(\mathbf{s}\right)} T - Q \right) T_a \\
  & + k \nabla_{\mathbf{x}_\Xi}^{\left(\mathbf{s}\right)} T \cdot \nabla_{\mathbf{x}_\Xi}^{\left(\mathbf{s}\right)} T_a \Big] K^{\left(\mathbf{s}, \delta \mathbf{s}\right)} \,\mathrm{d}\Omega + \sum_{E_\Omega\in\mathcal{E}_\Omega} \int_{E_\Omega} \left( \tau_{LST,\Xi}^{\left(\mathbf{s}, \delta \mathbf{s}\right)} + \tau_{LST,\Xi}^{\left(\mathbf{s}, \delta \mathbf{u}\right)} \right) \left( \rho C_p \mathbf{u} \cdot \nabla_{\mathbf{x}_\Xi}^{\left(\mathbf{s}\right)} T - Q \right) \\
  & \left( \rho C_p \mathbf{u} \cdot \nabla_{\mathbf{x}_\Xi}^{\left(\mathbf{s}\right)} T_a \right) K^{\left(\mathbf{s}\right)} + \tau_{LST,\Xi}^{\left(\mathbf{s}\right)} \left( \rho C_p \delta \mathbf{u} \cdot \nabla_{\mathbf{x}_\Xi}^{\left(\mathbf{s}\right)} T + \rho C_p \mathbf{u} \cdot \nabla_{\mathbf{x}_\Xi}^{\left(\mathbf{s}, \delta \mathbf{s}\right)} T + \rho C_p \mathbf{u} \cdot \nabla_{\mathbf{x}_\Xi}^{\left(\mathbf{s}\right)} \delta T \right) \\
  & \left( \rho C_p \mathbf{u} \cdot \nabla_{\mathbf{x}_\Xi}^{\left(\mathbf{s}\right)} T_a \right) K^{\left(\mathbf{s}\right)} + \tau_{LST,\Xi}^{\left(\mathbf{s}\right)} \left( \rho C_p \mathbf{u} \cdot \nabla_{\mathbf{x}_\Xi}^{\left(\mathbf{s}\right)} T - Q \right) \Big( \rho C_p \delta \mathbf{u} \cdot \nabla_{\mathbf{x}_\Xi}^{\left(\mathbf{s}\right)} T_a + \rho C_p \mathbf{u} \\
  & \cdot \nabla_{\mathbf{x}_\Xi}^{\left(\mathbf{s}, \delta \mathbf{s} \right)} T_a \Big) K^{\left(\mathbf{s}\right)} + \tau_{LST,\Xi}^{\left(\mathbf{s}\right)} \left( \rho C_p \mathbf{u} \cdot \nabla_{\mathbf{x}_\Xi}^{\left(\mathbf{s}\right)} T - Q \right) \left( \rho C_p \mathbf{u} \cdot \nabla_{\mathbf{x}_\Xi}^{\left(\mathbf{s}\right)} T_a \right) K^{\left(\mathbf{s}, \delta \mathbf{s}\right)} \,\mathrm{d}\Omega \\
  & - \int_\Omega \nabla_{\mathbf{x}_\Omega} \delta \mathbf{s} : \nabla_{\mathbf{x}_\Omega} \mathbf{s}_a \,\mathrm{d}\Omega + \int_\Sigma \left( \delta \mathbf{s} - \delta d_f \mathbf{n}_\Sigma \right) \cdot \boldsymbol{\lambda}_{\mathbf{s}a} + \delta \boldsymbol{\lambda}_\mathbf{s} \cdot \mathbf{s}_a \,\mathrm{d}\Sigma \\
  & + \int_\Sigma \bigg[ r_f^2 \bigg( \nabla_\Gamma^{\left( d_f, \delta d_f \right)} \gamma_f \cdot \nabla_\Gamma^{\left( d_f \right)} \gamma_{fa} + \nabla_\Gamma^{\left( d_f \right)} \delta \gamma_f \cdot \nabla_\Gamma^{\left( d_f \right)} \gamma_{fa} + \nabla_\Gamma^{\left( d_f \right)} \gamma_f \\
  & \cdot \nabla_\Gamma^{\left( d_f, \delta d_f \right)} \gamma_{fa} \bigg) + \delta \gamma_f \gamma_{fa} - \delta \gamma \gamma_{fa} \bigg] M^{\left( d_f \right)} + \bigg( r_f^2 \nabla_\Gamma^{\left( d_f \right)} \gamma_f \cdot \nabla_\Gamma^{\left( d_f \right)} \gamma_{fa} + \gamma_f \gamma_{fa} \\
  & - \gamma \gamma_{fa} \bigg) M^{\left( d_f, \delta d_f \right)} \,\mathrm{d}\Sigma + \int_\Sigma r_m^2 \nabla_\Sigma \delta d_f \cdot \nabla_\Sigma d_{fa} + \delta d_f d_{fa} - A_d \delta d_m d_{fa} \,\mathrm{d}\Sigma
\end{split}
\end{equation}
with the satisfication of the constraints in Eq. \ref{equ:ConstraintForAugmentedLagrangianBulkHT}
and
\begin{equation}\label{equ:ConstraintForVariationalAugmentedLagrangianObjBulkHT}
  \left.\begin{split}
  & \delta \mathbf{u} \in\left(\mathcal{H}\left(\Omega\right)\right)^3 \\
  & \delta p \in \mathcal{H}\left(\Omega\right) \\
  & \delta T \in \mathcal{H}\left(\Omega\right) \\
  & \delta \mathbf{s} \in \left(\mathcal{H}\left(\Omega\right)\right)^3 \\
  & \delta \boldsymbol{\lambda}_{\mathbf{s}} \in \left(\mathcal{H}^{{1\over2}}\left(\Sigma\right)\right)^3 \\
  & \delta \gamma_f \in \mathcal{H}\left(\Sigma\right) \\
  & \delta d_f \in \mathcal{H}\left(\Sigma\right)
  \end{split}\right\}~\mathrm{with}~
  \left\{\begin{split}
  & \delta \mathbf{u} = \mathbf{0}~ \mathrm{at} ~ \forall \mathbf{x}_\Omega \in \Sigma_{v,\Omega} \cup \Sigma_{v_0,\Omega} \\
  & \delta T = 0~ \mathrm{at} ~ \forall \mathbf{x}_\Omega \in \Sigma_{v,\Omega} \\
  & \delta \mathbf{s} = \mathbf{0} ~ \mathrm{at} ~ \forall \mathbf{x}_\Omega \in \Sigma_{v,\Omega} \cup \Sigma_{s,\Omega}
  \end{split}\right..
\end{equation}

According to the Karush-Kuhn-Tucker conditions of the PDE constrained optimization problem, the first order variational of the augmented Lagrangian to $T$ can be set to be zero as
\begin{equation}\label{equ:WeakAdjEquCDEquBulkHT}
\begin{split}
  & \int_\Omega \left[ 2 f_{id,\Xi}^{\left( \mathbf{s} \right)} k \nabla_{\mathbf{x}_\Xi}^{\left( \mathbf{s} \right)} \delta T \cdot \nabla_{\mathbf{x}_\Xi}^{\left( \mathbf{s} \right)} T + \left( \rho C_p \mathbf{u} \cdot \nabla_{\mathbf{x}_\Xi}^{\left(\mathbf{s}\right)} \delta T \right) T_a + k \nabla_{\mathbf{x}_\Xi}^{\left(\mathbf{s}\right)} \delta T \cdot \nabla_{\mathbf{x}_\Xi}^{\left(\mathbf{s}\right)} T_a \right] K^{\left(\mathbf{s}\right)} \,\mathrm{d}\Omega \\
  & + \sum_{E_\Omega\in\mathcal{E}_\Omega} \int_{E_\Omega} \tau_{LST,\Xi}^{\left(\mathbf{s}\right)} \left( \rho C_p \mathbf{u} \cdot \nabla_{\mathbf{x}_\Xi}^{\left(\mathbf{s}\right)} \delta T \right) \left( \rho C_p \mathbf{u} \cdot \nabla_{\mathbf{x}_\Xi}^{\left(\mathbf{s}\right)} T_a \right) K^{\left(\mathbf{s}\right)} \,\mathrm{d}\Omega = 0,
\end{split}
\end{equation}
the first order variational of the augmented Lagrangian to $\mathbf{u}$ and $p$ can be set to be zero as
\begin{equation}\label{equ:WeakAdjEquNSEquBulkHT} 
\begin{split}
  & \int_\Omega \Big[ \rho \left( \delta \mathbf{u} \cdot \nabla_{\mathbf{x}_\Xi}^{\left(\mathbf{s}\right)} \right) \mathbf{u} \cdot \mathbf{u}_a + \rho \left( \mathbf{u} \cdot \nabla_{\mathbf{x}_\Xi}^{\left(\mathbf{s}\right)} \right) \delta \mathbf{u} \cdot \mathbf{u}_a + {\eta\over2} \left( \nabla_{\mathbf{x}_\Xi}^{\left(\mathbf{s}\right)} \delta \mathbf{u} + \nabla_{\mathbf{x}_\Xi}^{\left(\mathbf{s}\right)} \delta \mathbf{u}^\mathrm{T} \right) \\
  & : \left( \nabla_{\mathbf{x}_\Xi}^{\left(\mathbf{s}\right)} \mathbf{u}_a + \nabla_{\mathbf{x}_\Xi}^{\left(\mathbf{s}\right)} \mathbf{u}_a^\mathrm{T} \right) - \delta p\,\mathrm{div}_{\mathbf{x}_\Xi}^{\left(\mathbf{s}\right)} \mathbf{u}_a - p_a \mathrm{div}_{\mathbf{x}_\Xi}^{\left(\mathbf{s}\right)} \delta \mathbf{u} + \left( \rho C_p \delta \mathbf{u} \cdot \nabla_{\mathbf{x}_\Xi}^{\left(\mathbf{s}\right)} T \right) T_a \Big] K^{\left(\mathbf{s}\right)} \,\mathrm{d}\Omega \\
  & - \sum_{E_\Omega\in\mathcal{E}_\Omega} \int_{E_\Omega} \Big[ \tau_{LS\mathbf{u},\Xi}^{\left(\mathbf{s}, \delta \mathbf{u}\right)} \left( \rho \mathbf{u} \cdot \nabla_{\mathbf{x}_\Xi}^{\left(\mathbf{s}\right)} \mathbf{u} + \nabla_{\mathbf{x}_\Xi}^{\left(\mathbf{s}\right)} p \right) \cdot \left( \rho \mathbf{u} \cdot \nabla_{\mathbf{x}_\Xi}^{\left(\mathbf{s}\right)} \mathbf{u}_a + \nabla_{\mathbf{x}_\Xi}^{\left(\mathbf{s}\right)} p_a \right) \\
  & + \tau_{LS\mathbf{u},\Xi}^{\left(\mathbf{s}\right)} \left( \rho \delta \mathbf{u} \cdot \nabla_{\mathbf{x}_\Xi}^{\left(\mathbf{s} \right)} \mathbf{u} + \rho \mathbf{u} \cdot \nabla_{\mathbf{x}_\Xi}^{\left(\mathbf{s}\right)} \delta \mathbf{u} + \nabla_{\mathbf{x}_\Xi}^{\left(\mathbf{s}\right)} \delta p \right) \cdot \left( \rho \mathbf{u} \cdot \nabla_{\mathbf{x}_\Xi}^{\left(\mathbf{s}\right)} \mathbf{u}_a + \nabla_{\mathbf{x}_\Xi}^{\left(\mathbf{s}\right)} p_a \right) \\
  & + \tau_{LS\mathbf{u},\Xi}^{\left(\mathbf{s}\right)} \left( \rho \mathbf{u} \cdot \nabla_{\mathbf{x}_\Xi}^{\left(\mathbf{s}\right)} \mathbf{u} + \nabla_{\mathbf{x}_\Xi}^{\left(\mathbf{s}\right)} p \right) \cdot \left( \rho \delta \mathbf{u} \cdot \nabla_{\mathbf{x}_\Xi}^{\left(\mathbf{s}\right)} \mathbf{u}_a \right) + \tau_{LSp,\Xi}^{\left(\mathbf{s}, \delta \mathbf{u}\right)} \left( \rho \mathrm{div}_{\mathbf{x}_\Xi}^{\left(\mathbf{s}\right)} \mathbf{u} \right) \left( \mathrm{div}_{\mathbf{x}_\Xi}^{\left(\mathbf{s}\right)} \mathbf{u}_a \right) \\
  & + \tau_{LSp,\Xi}^{\left(\mathbf{s}\right)} \left( \rho \mathrm{div}_{\mathbf{x}_\Xi}^{\left(\mathbf{s}\right)} \delta \mathbf{u} \right) \left( \mathrm{div}_{\mathbf{x}_\Xi}^{\left(\mathbf{s}\right)} \mathbf{u}_a \right) - \tau_{LST,\Xi}^{\left(\mathbf{s}, \delta \mathbf{u}\right)} \left( \rho C_p \mathbf{u} \cdot \nabla_{\mathbf{x}_\Xi}^{\left(\mathbf{s}\right)} T - Q \right) \left( \rho C_p \mathbf{u} \cdot \nabla_{\mathbf{x}_\Xi}^{\left(\mathbf{s}\right)} T_a \right) \\
  & - \tau_{LST,\Xi}^{\left(\mathbf{s}\right)} \left( \rho C_p \delta \mathbf{u} \cdot \nabla_{\mathbf{x}_\Xi}^{\left(\mathbf{s}\right)} T \right) \left( \rho C_p \mathbf{u} \cdot \nabla_{\mathbf{x}_\Xi}^{\left(\mathbf{s}\right)} T_a \right) - \tau_{LST,\Xi}^{\left(\mathbf{s}\right)} \left( \rho C_p \mathbf{u} \cdot \nabla_{\mathbf{x}_\Xi}^{\left(\mathbf{s}\right)} T - Q \right) \\
  & \cdot \left( \rho C_p \delta \mathbf{u} \cdot \nabla_{\mathbf{x}_\Xi}^{\left(\mathbf{s}\right)} T_a \right) \Big] K^{\left(\mathbf{s}\right)} \,\mathrm{d}\Omega + \int_\Sigma \alpha \delta \mathbf{u} \cdot \mathbf{u}_a M^{\left(d_f\right)} \, \mathrm{d}\Sigma = 0, \\
\end{split}
\end{equation}
the first order variational of the augmented Lagrangian to $\mathbf{s}$ and $\boldsymbol{\lambda}_\mathbf{s}$ can be set to be zero as 
\begin{equation}\label{equ:WeakAdjEquHarmonicEquBulkHT}
\begin{split}
  & \int_\Omega \Big[ f_{id,\Xi}^{\left( \mathbf{s}, \delta \mathbf{s} \right)} k \nabla_{\mathbf{x}_\Xi}^{\left( \mathbf{s} \right)} T \cdot \nabla_{\mathbf{x}_\Xi}^{\left( \mathbf{s} \right)} T + 2 f_{id,\Xi}^{\left( \mathbf{s} \right)} k \nabla_{\mathbf{x}_\Xi}^{\left( \mathbf{s}, \delta \mathbf{s} \right)} T \cdot \nabla_{\mathbf{x}_\Xi}^{\left( \mathbf{s} \right)} T + \rho \left( \mathbf{u} \cdot \nabla_{\mathbf{x}_\Xi}^{\left(\mathbf{s}, \delta \mathbf{s}\right)} \right) \mathbf{u} \cdot \mathbf{u}_a \\
  & + {\eta\over2} \left( \nabla_{\mathbf{x}_\Xi}^{\left(\mathbf{s}, \delta \mathbf{s}\right)} \mathbf{u} + \nabla_{\mathbf{x}_\Xi}^{\left(\mathbf{s}, \delta \mathbf{s}\right)} \mathbf{u}^\mathrm{T} \right) : \left( \nabla_{\mathbf{x}_\Xi}^{\left(\mathbf{s}\right)} \mathbf{u}_a + \nabla_{\mathbf{x}_\Xi}^{\left(\mathbf{s}\right)} \mathbf{u}_a^\mathrm{T} \right) + {\eta\over2} \left( \nabla_{\mathbf{x}_\Xi}^{\left(\mathbf{s}\right)} \mathbf{u} + \nabla_{\mathbf{x}_\Xi}^{\left(\mathbf{s}\right)} \mathbf{u}^\mathrm{T} \right) \\
  & : \Big( \nabla_{\mathbf{x}_\Xi}^{\left(\mathbf{s}, \delta \mathbf{s}\right)} \mathbf{u}_a + \nabla_{\mathbf{x}_\Xi}^{\left(\mathbf{s}, \delta \mathbf{s}\right)} \mathbf{u}_a^\mathrm{T} \Big) - p\,\mathrm{div}_{\mathbf{x}_\Xi}^{\left(\mathbf{s}, \delta \mathbf{s}\right)} \mathbf{u}_a - p_a \mathrm{div}_{\mathbf{x}_\Xi}^{\left(\mathbf{s}, \delta \mathbf{s}\right)} \mathbf{u} + \left( \rho C_p \mathbf{u} \cdot \nabla_{\mathbf{x}_\Xi}^{\left(\mathbf{s}, \delta \mathbf{s}\right)} T \right) T_a \\
  & + k \nabla_{\mathbf{x}_\Xi}^{\left(\mathbf{s}, \delta \mathbf{s}\right)} T \cdot \nabla_{\mathbf{x}_\Xi}^{\left(\mathbf{s}\right)} T_a + k \nabla_{\mathbf{x}_\Xi}^{\left(\mathbf{s}\right)} T \cdot \nabla_{\mathbf{x}_\Xi}^{\left(\mathbf{s}, \delta \mathbf{s}\right)} T_a \Big] K^{\left(\mathbf{s}\right)} + \Big[ f_{id,\Xi}^{\left( \mathbf{s} \right)} k \nabla_{\mathbf{x}_\Xi}^{\left( \mathbf{s} \right)} T \cdot \nabla_{\mathbf{x}_\Xi}^{\left( \mathbf{s} \right)} T \\
  & + \rho \left( \mathbf{u} \cdot \nabla_{\mathbf{x}_\Xi}^{\left(\mathbf{s}\right)} \right) \mathbf{u} \cdot \mathbf{u}_a + {\eta\over2} \Big( \nabla_{\mathbf{x}_\Xi}^{\left(\mathbf{s}\right)} \mathbf{u} + \nabla_{\mathbf{x}_\Xi}^{\left(\mathbf{s}\right)} \mathbf{u}^\mathrm{T} \Big) : \left( \nabla_{\mathbf{x}_\Xi}^{\left(\mathbf{s}\right)} \mathbf{u}_a + \nabla_{\mathbf{x}_\Xi}^{\left(\mathbf{s}\right)} \mathbf{u}_a^\mathrm{T} \right) - p\,\mathrm{div}_{\mathbf{x}_\Xi}^{\left(\mathbf{s}\right)} \mathbf{u}_a \\
  & - p_a \mathrm{div}_{\mathbf{x}_\Xi}^{\left(\mathbf{s} \right)} \mathbf{u} + \left( \rho C_p \mathbf{u} \cdot \nabla_{\mathbf{x}_\Xi}^{\left(\mathbf{s}\right)} T - Q \right) T_a + k \nabla_{\mathbf{x}_\Xi}^{\left(\mathbf{s}\right)} T \cdot \nabla_{\mathbf{x}_\Xi}^{\left(\mathbf{s}\right)} T_a \Big] K^{\left(\mathbf{s}, \delta \mathbf{s}\right)} \,\mathrm{d}\Omega \\
  & - \sum_{E_\Omega\in\mathcal{E}_\Omega} \int_{E_\Omega} \Big[ \tau_{LS\mathbf{u},\Xi}^{\left(\mathbf{s}, \delta \mathbf{s}\right)} \left( \rho \mathbf{u} \cdot \nabla_{\mathbf{x}_\Xi}^{\left(\mathbf{s}\right)} \mathbf{u} + \nabla_{\mathbf{x}_\Xi}^{\left(\mathbf{s}\right)} p \right) \cdot \left( \rho \mathbf{u} \cdot \nabla_{\mathbf{x}_\Xi}^{\left(\mathbf{s}\right)} \mathbf{u}_a + \nabla_{\mathbf{x}_\Xi}^{\left(\mathbf{s}\right)} p_a \right) \\
  & + \tau_{LS\mathbf{u},\Xi}^{\left(\mathbf{s}\right)} \left( \rho \mathbf{u} \cdot \nabla_{\mathbf{x}_\Xi}^{\left(\mathbf{s}, \delta \mathbf{s}\right)} \mathbf{u} + \nabla_{\mathbf{x}_\Xi}^{\left(\mathbf{s}, \delta \mathbf{s}\right)} p \right) \cdot \left( \rho \mathbf{u} \cdot \nabla_{\mathbf{x}_\Xi}^{\left(\mathbf{s}\right)} \mathbf{u}_a + \nabla_{\mathbf{x}_\Xi}^{\left(\mathbf{s}\right)} p_a \right) \\
  & + \tau_{LS\mathbf{u},\Xi}^{\left(\mathbf{s}\right)} \left( \rho \mathbf{u} \cdot \nabla_{\mathbf{x}_\Xi}^{\left(\mathbf{s}\right)} \mathbf{u} + \nabla_{\mathbf{x}_\Xi}^{\left(\mathbf{s}\right)} p \right) \cdot \left( \rho \mathbf{u} \cdot \nabla_{\mathbf{x}_\Xi}^{\left(\mathbf{s}, \delta \mathbf{s}\right)} \mathbf{u}_a + \nabla_{\mathbf{x}_\Xi}^{\left(\mathbf{s}, \delta \mathbf{s}\right)} p_a \right) \\
  & + \tau_{LSp,\Xi}^{\left(\mathbf{s}, \delta \mathbf{s}\right)} \left( \rho \mathrm{div}_{\mathbf{x}_\Xi}^{\left(\mathbf{s}\right)} \mathbf{u} \right) \left( \mathrm{div}_{\mathbf{x}_\Xi}^{\left(\mathbf{s}\right)} \mathbf{u}_a \right) + \tau_{LSp,\Xi}^{\left(\mathbf{s}\right)} \left( \rho \mathrm{div}_{\mathbf{x}_\Xi}^{\left(\mathbf{s}, \delta \mathbf{s}\right)} \mathbf{u} \right) \left( \mathrm{div}_{\mathbf{x}_\Xi}^{\left(\mathbf{s}\right)} \mathbf{u}_a \right) \\
  & + \tau_{LSp,\Xi}^{\left(\mathbf{s}\right)} \left( \rho \mathrm{div}_{\mathbf{x}_\Xi}^{\left(\mathbf{s}\right)} \mathbf{u} \right) \left( \mathrm{div}_{\mathbf{x}_\Xi}^{\left(\mathbf{s}, \delta \mathbf{s}\right)} \mathbf{u}_a \right) - \tau_{LST,\Xi}^{\left(\mathbf{s}, \delta \mathbf{s}\right)} \left( \rho C_p \mathbf{u} \cdot \nabla_{\mathbf{x}_\Xi}^{\left(\mathbf{s}\right)} T - Q \right) \\
  & \left( \rho C_p \mathbf{u} \cdot \nabla_{\mathbf{x}_\Xi}^{\left(\mathbf{s}\right)} T_a \right) - \tau_{LST,\Xi}^{\left(\mathbf{s}\right)} \left( \rho C_p \mathbf{u} \cdot \nabla_{\mathbf{x}_\Xi}^{\left(\mathbf{s}, \delta \mathbf{s}\right)} T \right) \left( \rho C_p \mathbf{u} \cdot \nabla_{\mathbf{x}_\Xi}^{\left(\mathbf{s}\right)} T_a \right) \\
  & - \tau_{LST,\Xi}^{\left(\mathbf{s}\right)} \left( \rho C_p \mathbf{u} \cdot \nabla_{\mathbf{x}_\Xi}^{\left(\mathbf{s}\right)} T - Q \right) \left( \rho C_p \mathbf{u} \cdot \nabla_{\mathbf{x}_\Xi}^{\left(\mathbf{s}, \delta \mathbf{s} \right)} T_a \right) \Big] K^{\left(\mathbf{s}\right)} \\
  & + \Big[ \tau_{LS\mathbf{u},\Xi}^{\left(\mathbf{s}\right)} \left( \rho \mathbf{u} \cdot \nabla_{\mathbf{x}_\Xi}^{\left(\mathbf{s}\right)} \mathbf{u} + \nabla_{\mathbf{x}_\Xi}^{\left(\mathbf{s}\right)} p \right) \cdot \left( \rho \mathbf{u} \cdot \nabla_{\mathbf{x}_\Xi}^{\left(\mathbf{s}\right)} \mathbf{u}_a + \nabla_{\mathbf{x}_\Xi}^{\left(\mathbf{s}\right)} p_a \right) \\
  & + \tau_{LSp,\Xi}^{\left(\mathbf{s}\right)} \left( \rho \mathrm{div}_{\mathbf{x}_\Xi}^{\left(\mathbf{s}\right)} \mathbf{u} \right) \left( \mathrm{div}_{\mathbf{x}_\Xi}^{\left(\mathbf{s}\right)} \mathbf{u}_a \right) - \tau_{LST,\Xi}^{\left(\mathbf{s}\right)} \left( \rho C_p \mathbf{u} \cdot \nabla_{\mathbf{x}_\Xi}^{\left(\mathbf{s}\right)} T - Q \right) \\
  & \left( \rho C_p \mathbf{u} \cdot \nabla_{\mathbf{x}_\Xi}^{\left(\mathbf{s}\right)} T_a \right) \Big] K^{\left(\mathbf{s}, \delta \mathbf{s}\right)} \,\mathrm{d}\Omega - \int_\Omega \nabla_{\mathbf{x}_\Omega} \delta \mathbf{s} : \nabla_{\mathbf{x}_\Omega} \mathbf{s}_a \,\mathrm{d}\Omega \\
  & + \int_\Sigma \delta \mathbf{s} \cdot \boldsymbol{\lambda}_{\mathbf{s}a} + \delta \boldsymbol{\lambda}_\mathbf{s} \cdot \mathbf{s}_a \,\mathrm{d}\Sigma = 0, \\
\end{split}
\end{equation} 
the first order variational of the augmented Lagrangian to $\gamma_f$ can be set to be zero as
\begin{equation}\label{equ:WeakAdjEquPDEFilterEquGafBulkHT}
\begin{split}
  & \int_\Sigma \left( {\partial\alpha\over\partial\gamma_p} {\partial\gamma_p\over\partial\gamma_f} \mathbf{u} \cdot \mathbf{u}_a \delta \gamma_f + r_f^2 \nabla_\Gamma^{\left( d_f \right)} \delta \gamma_f \cdot \nabla_\Gamma^{\left( d_f \right)} \gamma_{fa} + \delta \gamma_f \gamma_{fa} \right) M^{\left( d_f \right)} \,\mathrm{d}\Sigma = 0,
\end{split}
\end{equation}
and the first order variational of the augmented Lagrangian to $d_f$ can be set to be zero as
\begin{equation}\label{equ:WeakAdjEquPDEFilterEqudffBulkHT}
\begin{split}
  & \int_\Sigma r_f^2 \left( \nabla_\Gamma^{\left( d_f, \delta d_f \right)} \gamma_f \cdot \nabla_\Gamma^{\left( d_f \right)} \gamma_{fa} + \nabla_\Gamma^{\left( d_f \right)} \gamma_f \cdot \nabla_\Gamma^{\left( d_f, \delta d_f \right)} \gamma_{fa} \right) M^{\left( d_f \right)} \\
  & + \left( r_f^2 \nabla_\Gamma^{\left( d_f \right)} \gamma_f \cdot \nabla_\Gamma^{\left( d_f \right)} \gamma_{fa} + \gamma_f \gamma_{fa} - \gamma \gamma_{fa} + \alpha \mathbf{u} \cdot \mathbf{u}_a \right) M^{\left( d_f, \delta d_f \right)} \\
  & + r_m^2 \nabla_\Sigma \delta d_f \cdot \nabla_\Sigma d_{fa} + \delta d_f d_{fa} - \delta d_f \mathbf{n}_\Sigma \cdot \boldsymbol{\lambda}_{\mathbf{s}a} \,\mathrm{d}\Sigma = 0.
\end{split}
\end{equation}

The constraints in Eqs. \ref{equ:ConstraintForAugmentedLagrangianBulkHT} and \ref{equ:ConstraintForVariationalAugmentedLagrangianObjBulkHT} are imposed to Eqs. \ref{equ:WeakAdjEquCDEquBulkHT}, \ref{equ:WeakAdjEquNSEquBulkHT}, \ref{equ:WeakAdjEquHarmonicEquBulkHT}, \ref{equ:WeakAdjEquPDEFilterEquGafBulkHT} and \ref{equ:WeakAdjEquPDEFilterEqudffBulkHT}. Then, the adjoint sensitivity of $J_T$ is derived as
\begin{equation}\label{equ:AdjSensitivityGaDmVariationalFormBulkHT}
\begin{split}
\delta \hat{J}_T = \int_\Sigma - \gamma_{fa} \delta \gamma M^{\left( d_f \right)} - A_d d_{fa} \delta d_m \,\mathrm{d}\Sigma.
\end{split}
\end{equation}

Without losing the arbitrariness of $\delta \mathbf{u}$, $\delta p$, $\delta T$, $\delta \mathbf{s}$, $\delta \boldsymbol{\lambda}_{\mathbf{s}}$, $\delta \gamma_f$, $\delta d_f$, $\delta \gamma$ and $\delta d_m$, one can set 
\begin{equation}
\left.\begin{split}
& \tilde{\mathbf{u}}_a = \delta \mathbf{u} \\
& \tilde{p}_a = \delta p \\
& \tilde{T}_a = \delta T \\
& \tilde{\mathbf{s}}_a = \delta \mathbf{s} \\
& \tilde{\boldsymbol{\lambda}}_{\mathbf{s}a} = \delta \boldsymbol{\lambda}_{\mathbf{s}} \\
& \tilde{\gamma}_{fa} = \delta \gamma_f \\
& \tilde{d}_{fa} = \delta d_f \\
& \tilde{\gamma} = \delta \gamma \\
& \tilde{d}_m = \delta d_m 
\end{split}\right\}
~\mathrm{with}~
\left\{\begin{split}
& \forall \tilde{\mathbf{u}}_a \in \left(\mathcal{H}\left(\Omega\right)\right)^3 \\
& \forall \tilde{p}_a \in \mathcal{H}\left(\Omega\right)\\
& \forall \tilde{T}_a \in \mathcal{H}\left(\Omega\right) \\
& \forall \tilde{\mathbf{s}}_a \in \left(\mathcal{H}\left(\Omega\right)\right)^3 \\
& \forall \tilde{\boldsymbol{\lambda}}_{\mathbf{s}a} \in \left(\mathcal{H}^{{1\over2}}\left(\Sigma\right)\right)^3 \\
& \forall \tilde{\gamma}_{fa} \in \mathcal{H}\left(\Sigma\right) \\
& \forall \tilde{d}_{fa} \in \mathcal{H}\left(\Sigma\right)\\
& \forall \tilde{\gamma} \in \mathcal{L}^2\left(\Sigma\right) \\
& \forall \tilde{d}_m \in \mathcal{L}^2\left(\Sigma\right)
\end{split}\right.
\end{equation}
for Eqs. \ref{equ:WeakAdjEquCDEquBulkHT}, \ref{equ:WeakAdjEquNSEquBulkHT}, \ref{equ:WeakAdjEquHarmonicEquBulkHT}, \ref{equ:WeakAdjEquPDEFilterEquGafBulkHT} and \ref{equ:WeakAdjEquPDEFilterEqudffBulkHT} to derive the adjoint system composed of Eqs. \ref{equ:WeakAdjEquHTEquBulkMTTa}, \ref{equ:AdjBulkNavierStokesEqusJObjectiveHTUaPa}, \ref{equ:WeakAdjEquHarmonicEquBulkHTSa}, \ref{equ:AdjPDEFilterJObjectiveGafHTGafa} and \ref{equ:AdjPDEFilterJObjectiveDmHTDfa}.

\subsection{Adjoint analysis for constraint of pressure drop in Eq. \ref{equ:VarProToopBulkNSCHTMHT}} \label{sec:AdjointAnalysisDissipationConstraintBulkMHM}

Based on the transformed pressure drop in Eq. \ref{equ:TransformedPressureConstraintSurfaceNSCD}, the variational formulations of the Laplace's equation in Eq. \ref{equ:WeakFormLaplacianBulkFlowMHT}, the surface-PDE filters in Eqs. \ref{equ:VariationalFormulationPDEFilterBaseManifoldMHM} and \ref{equ:VariationalFormulationPDEFilterMHM} and the Navier-Stokes equations in Eq. \ref{equ:TransformedVariationalFormulationBulkNSEqusHM}, the augmented Lagrangian of the dissipation power in Eq. \ref{equ:VarProToopBulkNSCHTMHT} can be derived as
\begin{equation}\label{equ:AugmentedLagrangianMatchOptimizationBulkDissipationHT}
\begin{split}
  \widehat{\Delta P} = & \int_{\Sigma_{v,\Omega}} p \,\mathrm{d}\Sigma_{\partial\Omega} - \int_{\Sigma_{s,\Omega}} p \,\mathrm{d}\Sigma_{\partial\Omega} + \int_\Omega \Big[ \rho \left( \mathbf{u} \cdot \nabla_{\mathbf{x}_\Xi}^{\left(\mathbf{s}\right)} \right) \mathbf{u} \cdot \mathbf{u}_a \\
  & + {\eta\over2} \left( \nabla_{\mathbf{x}_\Xi}^{\left(\mathbf{s}\right)} \mathbf{u} + \nabla_{\mathbf{x}_\Xi}^{\left(\mathbf{s}\right)} \mathbf{u}^\mathrm{T} \right) : \left( \nabla_{\mathbf{x}_\Xi}^{\left(\mathbf{s}\right)} \mathbf{u}_a + \nabla_{\mathbf{x}_\Xi}^{\left(\mathbf{s}\right)} \mathbf{u}_a^\mathrm{T} \right) - p\,\mathrm{div}_{\mathbf{x}_\Xi}^{\left(\mathbf{s}\right)} \mathbf{u}_a - p_a \mathrm{div}_{\mathbf{x}_\Xi}^{\left(\mathbf{s}\right)} \mathbf{u} \Big] K^{\left(\mathbf{s}\right)} \,\mathrm{d}\Omega \\
  & - \sum_{E_\Omega\in\mathcal{E}_\Omega} \int_{E_\Omega} \Big[ \tau_{LS\mathbf{u},\Xi}^{\left(\mathbf{s}\right)} \left( \rho \mathbf{u} \cdot \nabla_{\mathbf{x}_\Xi}^{\left(\mathbf{s}\right)} \mathbf{u} + \nabla_{\mathbf{x}_\Xi}^{\left(\mathbf{s}\right)} p \right) \cdot \left( \rho \mathbf{u} \cdot \nabla_{\mathbf{x}_\Xi}^{\left(\mathbf{s}\right)} \mathbf{u}_a + \nabla_{\mathbf{x}_\Xi}^{\left(\mathbf{s}\right)} p_a \right) \\
  & + \tau_{LSp,\Xi}^{\left(\mathbf{s}\right)} \left( \rho \mathrm{div}_{\mathbf{x}_\Xi}^{\left(\mathbf{s}\right)} \mathbf{u} \right) \left( \mathrm{div}_{\mathbf{x}_\Xi}^{\left(\mathbf{s}\right)} \mathbf{u}_a \right) \Big] K^{\left(\mathbf{s}\right)} \,\mathrm{d}\Omega + \int_\Sigma \alpha \mathbf{u} \cdot \mathbf{u}_a M^{\left(d_f\right)} \, \mathrm{d}\Sigma \\
  & + \int_\Omega - \nabla_{\mathbf{x}_\Omega} \mathbf{s} : \nabla_{\mathbf{x}_\Omega} \mathbf{s}_a \,\mathrm{d}\Omega + \int_\Sigma \left( \mathbf{s} - d_f \mathbf{n}_\Sigma \right) \cdot \boldsymbol{\lambda}_{\mathbf{s}a} + \boldsymbol{\lambda}_\mathbf{s} \cdot \mathbf{s}_a \,\mathrm{d}\Sigma \\
  & + \int_\Sigma \left( r_f^2 \nabla_\Gamma^{\left( d_f \right)} \gamma_f \cdot \nabla_\Gamma^{\left( d_f \right)} \gamma_{fa} + \gamma_f \gamma_{fa} - \gamma \gamma_{fa} \right) M^{\left( d_f \right)} \,\mathrm{d}\Sigma \\
  & + \int_\Sigma r_m^2 \nabla_\Sigma d_f \cdot \nabla_\Sigma d_{fa} + d_f d_{fa} - A_d \left( d_m - {1\over2} \right) d_{fa} \,\mathrm{d}\Sigma,
\end{split}
\end{equation}
where the adjoint variables satisfy
\begin{equation}\label{equ:ConstraintForAugmentedLagrangianDissipationBulkHT}
  \left.\begin{split}
  & \mathbf{u}_a \in\left(\mathcal{H}\left(\Omega\right)\right)^3 \\
  & p_a \in \mathcal{H}\left(\Omega\right) \\
  & \mathbf{s}_a \in \left(\mathcal{H}\left(\Omega\right)\right)^3 \\
  & \boldsymbol{\lambda}_{\mathbf{s}a} \in \left(\mathcal{H}^{-{1\over2}}\left(\Sigma\right)\right)^3 \\
  & \gamma_{fa} \in \mathcal{H}\left(\Sigma\right) \\
  & d_{fa} \in \mathcal{H}\left(\Sigma\right)
  \end{split}\right\}~\mathrm{with}~
  \left\{\begin{split}
  & \mathbf{u}_a = \mathbf{0}~ \mathrm{at} ~ \forall \mathbf{x}_\Omega \in \Sigma_{v,\Omega} \cup \Sigma_{v_0,\Omega} \\
  & \mathbf{s}_a = \mathbf{0} ~ \mathrm{at} ~ \forall \mathbf{x}_\Omega \in \Sigma_{v,\Omega} \cup \Sigma_{s,\Omega} 
  \end{split}\right..
\end{equation}
The first order variational of the augmented Lagrangian in Eq. \ref{equ:AugmentedLagrangianMatchOptimizationBulkDissipationHT} can be derived as
\begin{equation}\label{equ:1stVariAugmentedLagrangianMatchOptimizationDissipationBulkHT}
\begin{split}
  \delta \widehat{\Delta P} = & \int_{\Sigma_{v,\Omega}} \delta p \,\mathrm{d}\Sigma_{\partial\Omega} - \int_{\Sigma_{s,\Omega}} \delta p \,\mathrm{d}\Sigma_{\partial\Omega} \\
   & + \int_\Omega \Big[ \rho \left( \delta \mathbf{u} \cdot \nabla_{\mathbf{x}_\Xi}^{\left(\mathbf{s}\right)} \right) \mathbf{u} \cdot \mathbf{u}_a + \rho \left( \mathbf{u} \cdot \nabla_{\mathbf{x}_\Xi}^{\left(\mathbf{s}, \delta \mathbf{s}\right)} \right) \mathbf{u} \cdot \mathbf{u}_a + \rho \left( \mathbf{u} \cdot \nabla_{\mathbf{x}_\Xi}^{\left(\mathbf{s}\right)} \right) \delta \mathbf{u} \cdot \mathbf{u}_a \\
  & + {\eta\over2} \left( \nabla_{\mathbf{x}_\Xi}^{\left(\mathbf{s}, \delta \mathbf{s}\right)} \mathbf{u} + \nabla_{\mathbf{x}_\Xi}^{\left(\mathbf{s}, \delta \mathbf{s}\right)} \mathbf{u}^\mathrm{T} \right) : \left( \nabla_{\mathbf{x}_\Xi}^{\left(\mathbf{s}\right)} \mathbf{u}_a + \nabla_{\mathbf{x}_\Xi}^{\left(\mathbf{s}\right)} \mathbf{u}_a^\mathrm{T} \right) + {\eta\over2} \left( \nabla_{\mathbf{x}_\Xi}^{\left(\mathbf{s}\right)} \delta \mathbf{u} + \nabla_{\mathbf{x}_\Xi}^{\left(\mathbf{s}\right)} \delta \mathbf{u}^\mathrm{T} \right) \\
  & : \left( \nabla_{\mathbf{x}_\Xi}^{\left(\mathbf{s}\right)} \mathbf{u}_a + \nabla_{\mathbf{x}_\Xi}^{\left(\mathbf{s}\right)} \mathbf{u}_a^\mathrm{T} \right) + {\eta\over2} \left( \nabla_{\mathbf{x}_\Xi}^{\left(\mathbf{s}\right)} \mathbf{u} + \nabla_{\mathbf{x}_\Xi}^{\left(\mathbf{s}\right)} \mathbf{u}^\mathrm{T} \right) : \left( \nabla_{\mathbf{x}_\Xi}^{\left(\mathbf{s}, \delta \mathbf{s}\right)} \mathbf{u}_a + \nabla_{\mathbf{x}_\Xi}^{\left(\mathbf{s}, \delta \mathbf{s}\right)} \mathbf{u}_a^\mathrm{T} \right) \\
  & - \delta p\,\mathrm{div}_{\mathbf{x}_\Xi}^{\left(\mathbf{s}\right)} \mathbf{u}_a - p\,\mathrm{div}_{\mathbf{x}_\Xi}^{\left(\mathbf{s}, \delta \mathbf{s}\right)} \mathbf{u}_a - p_a \mathrm{div}_{\mathbf{x}_\Xi}^{\left(\mathbf{s}, \delta \mathbf{s}\right)} \mathbf{u} - p_a \mathrm{div}_{\mathbf{x}_\Xi}^{\left(\mathbf{s}\right)} \delta \mathbf{u} \Big] K^{\left(\mathbf{s}\right)} \\
  & + \Big[ {\eta\over2} \left( \nabla_{\mathbf{x}_\Xi}^{\left( \mathbf{s} \right)} \mathbf{u} + \nabla_{\mathbf{x}_\Xi}^{\left( \mathbf{s} \right)} \mathbf{u}^\mathrm{T} \right) : \left( \nabla_{\mathbf{x}_\Xi}^{\left( \mathbf{s} \right)} \mathbf{u} + \nabla_{\mathbf{x}_\Xi}^{\left( \mathbf{s} \right)} \mathbf{u}^\mathrm{T} \right) + \alpha \mathbf{u}^2 + \rho \left( \mathbf{u} \cdot \nabla_{\mathbf{x}_\Xi}^{\left(\mathbf{s}\right)} \right) \mathbf{u} \cdot \mathbf{u}_a \\
  & + {\eta\over2} \left( \nabla_{\mathbf{x}_\Xi}^{\left(\mathbf{s}\right)} \mathbf{u} + \nabla_{\mathbf{x}_\Xi}^{\left(\mathbf{s}\right)} \mathbf{u}^\mathrm{T} \right) : \left( \nabla_{\mathbf{x}_\Xi}^{\left(\mathbf{s}\right)} \mathbf{u}_a + \nabla_{\mathbf{x}_\Xi}^{\left(\mathbf{s}\right)} \mathbf{u}_a^\mathrm{T} \right) - p\,\mathrm{div}_{\mathbf{x}_\Xi}^{\left(\mathbf{s}\right)} \mathbf{u}_a - p_a \mathrm{div}_{\mathbf{x}_\Xi}^{\left(\mathbf{s}\right)} \mathbf{u} \Big] K^{\left(\mathbf{s}, \delta \mathbf{s}\right)} \,\mathrm{d}\Omega \\
  & - \sum_{E_\Omega\in\mathcal{E}_\Omega} \int_{E_\Omega} \Big[ \left( \tau_{LS\mathbf{u},\Xi}^{\left(\mathbf{s}, \delta \mathbf{s}\right)} + \tau_{LS\mathbf{u},\Xi}^{\left(\mathbf{s}, \delta \mathbf{u}\right)} \right) \left( \rho \mathbf{u} \cdot \nabla_{\mathbf{x}_\Xi}^{\left(\mathbf{s}\right)} \mathbf{u} + \nabla_{\mathbf{x}_\Xi}^{\left(\mathbf{s}\right)} p \right) \cdot \left( \rho \mathbf{u} \cdot \nabla_{\mathbf{x}_\Xi}^{\left(\mathbf{s}\right)} \mathbf{u}_a + \nabla_{\mathbf{x}_\Xi}^{\left(\mathbf{s}\right)} p_a \right) \\
  & + \tau_{LS\mathbf{u},\Xi}^{\left(\mathbf{s}\right)} \left( \rho \delta \mathbf{u} \cdot \nabla_{\mathbf{x}_\Xi}^{\left(\mathbf{s}\right)} \mathbf{u} + \rho \mathbf{u} \cdot \nabla_{\mathbf{x}_\Xi}^{\left(\mathbf{s}, \delta \mathbf{s}\right)} \mathbf{u} + \rho \mathbf{u} \cdot \nabla_{\mathbf{x}_\Xi}^{\left(\mathbf{s}\right)} \delta\mathbf{u} + \nabla_{\mathbf{x}_\Xi}^{\left(\mathbf{s}, \delta \mathbf{s}\right)} p + \nabla_{\mathbf{x}_\Xi}^{\left(\mathbf{s}\right)} \delta p \right) \\
  & \cdot \left( \rho \mathbf{u} \cdot \nabla_{\mathbf{x}_\Xi}^{\left(\mathbf{s}\right)} \mathbf{u}_a + \nabla_{\mathbf{x}_\Xi}^{\left(\mathbf{s}\right)} p_a \right) + \tau_{LS\mathbf{u},\Xi}^{\left(\mathbf{s}\right)} \left( \rho \mathbf{u} \cdot \nabla_{\mathbf{x}_\Xi}^{\left(\mathbf{s}\right)} \mathbf{u} + \nabla_{\mathbf{x}_\Xi}^{\left(\mathbf{s}\right)} p \right) \cdot \Big( \rho \delta \mathbf{u} \cdot \nabla_{\mathbf{x}_\Xi}^{\left(\mathbf{s}\right)} \mathbf{u}_a + \rho \mathbf{u} \\
  & \cdot \nabla_{\mathbf{x}_\Xi}^{\left(\mathbf{s}, \delta \mathbf{s}\right)} \mathbf{u}_a + \nabla_{\mathbf{x}_\Xi}^{\left(\mathbf{s}, \delta \mathbf{s}\right)} p_a \Big) + \left( \tau_{LSp,\Xi}^{\left(\mathbf{s}, \delta \mathbf{s}\right)} + \tau_{LSp,\Xi}^{\left(\mathbf{s}, \delta \mathbf{u}\right)} \right) \left( \rho \mathrm{div}_{\mathbf{x}_\Xi}^{\left(\mathbf{s}\right)} \mathbf{u} \right) \left( \mathrm{div}_{\mathbf{x}_\Xi}^{\left(\mathbf{s}\right)} \mathbf{u}_a \right) + \tau_{LSp,\Xi}^{\left(\mathbf{s}\right)} \\
  & \left( \rho \mathrm{div}_{\mathbf{x}_\Xi}^{\left(\mathbf{s}, \delta \mathbf{s}\right)} \mathbf{u} \right) \left( \mathrm{div}_{\mathbf{x}_\Xi}^{\left(\mathbf{s}\right)} \mathbf{u}_a \right) + \tau_{LSp,\Xi}^{\left(\mathbf{s}\right)} \left( \rho \mathrm{div}_{\mathbf{x}_\Xi}^{\left(\mathbf{s}\right)} \delta \mathbf{u} \right) \left( \mathrm{div}_{\mathbf{x}_\Xi}^{\left(\mathbf{s}\right)} \mathbf{u}_a \right) + \tau_{LSp,\Xi}^{\left(\mathbf{s}\right)} \left( \rho \mathrm{div}_{\mathbf{x}_\Xi}^{\left(\mathbf{s}\right)} \mathbf{u} \right) \\
  & \left( \mathrm{div}_{\mathbf{x}_\Xi}^{\left(\mathbf{s}, \delta \mathbf{s}\right)} \mathbf{u}_a \right) \Big] K^{\left(\mathbf{s}\right)} + \Big[ \tau_{LS\mathbf{u},\Xi}^{\left(\mathbf{s}\right)} \left( \rho \mathbf{u} \cdot \nabla_{\mathbf{x}_\Xi}^{\left(\mathbf{s}\right)} \mathbf{u} + \nabla_{\mathbf{x}_\Xi}^{\left(\mathbf{s}\right)} p \right) \cdot \left( \rho \mathbf{u} \cdot \nabla_{\mathbf{x}_\Xi}^{\left(\mathbf{s}\right)} \mathbf{u}_a + \nabla_{\mathbf{x}_\Xi}^{\left(\mathbf{s}\right)} p_a \right) \\
  & + \tau_{LSp,\Xi}^{\left(\mathbf{s}\right)} \left( \rho \mathrm{div}_{\mathbf{x}_\Xi}^{\left(\mathbf{s}\right)} \mathbf{u} \right) \left( \mathrm{div}_{\mathbf{x}_\Xi}^{\left(\mathbf{s}\right)} \mathbf{u}_a \right) \Big] K^{\left(\mathbf{s}, \delta \mathbf{s}\right)} \,\mathrm{d}\Omega + \int_\Sigma {\partial \alpha \over \partial \gamma_p} {\partial \gamma_p \over \partial \gamma_f} \mathbf{u} \cdot \mathbf{u}_a M^{\left(d_f\right)} \delta \gamma_f \\
  & + \alpha \delta \mathbf{u} \cdot \mathbf{u}_a M^{\left(d_f\right)} + \alpha \mathbf{u} \cdot \mathbf{u}_a M^{\left(d_f, \delta d_f\right)} \, \mathrm{d}\Sigma + \int_\Omega - \nabla_{\mathbf{x}_\Omega} \delta \mathbf{s} : \nabla_{\mathbf{x}_\Omega} \mathbf{s}_a \,\mathrm{d}\Omega \\
  & + \int_\Sigma \left( \delta \mathbf{s} - \delta d_f \mathbf{n}_\Sigma \right) \cdot \boldsymbol{\lambda}_{\mathbf{s}a} + \delta \boldsymbol{\lambda}_\mathbf{s} \cdot \mathbf{s}_a \,\mathrm{d}\Sigma + \int_\Sigma \bigg[ r_f^2 \bigg( \nabla_\Gamma^{\left( d_f, \delta d_f \right)} \gamma_f \cdot \nabla_\Gamma^{\left( d_f \right)} \gamma_{fa} \\
  & + \nabla_\Gamma^{\left( d_f \right)} \delta \gamma_f \cdot \nabla_\Gamma^{\left( d_f \right)} \gamma_{fa} + \nabla_\Gamma^{\left( d_f \right)} \gamma_f \cdot \nabla_\Gamma^{\left( d_f, \delta d_f \right)} \gamma_{fa} \bigg) + \delta \gamma_f \gamma_{fa} - \delta \gamma \gamma_{fa} \bigg] \\
  & M^{\left( d_f \right)} + \bigg( r_f^2 \nabla_\Gamma^{\left( d_f \right)} \gamma_f \cdot \nabla_\Gamma^{\left( d_f \right)} \gamma_{fa} + \gamma_f \gamma_{fa} - \gamma \gamma_{fa} \bigg) M^{\left( d_f, \delta d_f \right)} \,\mathrm{d}\Sigma \\
  & + \int_\Sigma r_m^2 \nabla_\Sigma \delta d_f \cdot \nabla_\Sigma d_{fa} + \delta d_f d_{fa} - A_d \delta d_m d_{fa} \,\mathrm{d}\Sigma
\end{split}
\end{equation}
with the satisfication of the constraints in Eq. \ref{equ:ConstraintForAugmentedLagrangianDissipationBulkHT}
and
\begin{equation}\label{equ:ConstraintForVariationalAugmentedLagrangianDissipationBulkHT}
  \left.\begin{split}
  & \delta \mathbf{u} \in\left(\mathcal{H}\left(\Omega\right)\right)^3 \\
  & \delta p \in \mathcal{H}\left(\Omega\right) \\
  & \delta \mathbf{s} \in \left(\mathcal{H}\left(\Omega\right)\right)^3 \\
  & \delta \boldsymbol{\lambda}_{\mathbf{s}} \in \left(\mathcal{H}^{{1\over2}}\left(\Sigma\right)\right)^3 \\
  & \delta \gamma_f \in \mathcal{H}\left(\Sigma\right) \\
  & \delta d_f \in \mathcal{H}\left(\Sigma\right)
  \end{split}\right\}~\mathrm{with}~
  \left\{\begin{split}
  & \delta \mathbf{u} = \mathbf{0}~ \mathrm{at} ~ \forall \mathbf{x}_\Omega \in \Sigma_{v,\Omega} \cup \Sigma_{v_0,\Omega} \\
  & \delta \mathbf{s} = \mathbf{0} ~ \mathrm{at} ~ \forall \mathbf{x}_\Omega \in \Sigma_{v,\Omega} \cup \Sigma_{s,\Omega}
  \end{split}\right..
\end{equation}

According to the Karush-Kuhn-Tucker conditions of the PDE constrained optimization problem, the first order variational of the augmented Lagrangian to $\mathbf{u}$ and $p$ can be set to be zero as
\begin{equation}\label{equ:WeakAdjEquNSEquDissipationBulkHT} 
\begin{split}
  & \int_{\Sigma_{v,\Omega}} \delta p \,\mathrm{d}\Sigma_{\partial\Omega} - \int_{\Sigma_{s,\Omega}} \delta p \,\mathrm{d}\Sigma_{\partial\Omega} + \int_\Omega \Big[ \rho \left( \delta \mathbf{u} \cdot \nabla_{\mathbf{x}_\Xi}^{\left(\mathbf{s}\right)} \right) \mathbf{u} \cdot \mathbf{u}_a + \rho \left( \mathbf{u} \cdot \nabla_{\mathbf{x}_\Xi}^{\left(\mathbf{s}\right)} \right) \delta \mathbf{u} \cdot \mathbf{u}_a \\
  & + {\eta\over2} \left( \nabla_{\mathbf{x}_\Xi}^{\left(\mathbf{s}\right)} \delta \mathbf{u} + \nabla_{\mathbf{x}_\Xi}^{\left(\mathbf{s}\right)} \delta \mathbf{u}^\mathrm{T} \right) : \left( \nabla_{\mathbf{x}_\Xi}^{\left(\mathbf{s}\right)} \mathbf{u}_a + \nabla_{\mathbf{x}_\Xi}^{\left(\mathbf{s}\right)} \mathbf{u}_a^\mathrm{T} \right) - \delta p\,\mathrm{div}_{\mathbf{x}_\Xi}^{\left(\mathbf{s}\right)} \mathbf{u}_a - p_a \mathrm{div}_{\mathbf{x}_\Xi}^{\left(\mathbf{s}\right)} \delta \mathbf{u} \Big] K^{\left(\mathbf{s}\right)} \,\mathrm{d}\Omega \\
  & - \sum_{E_\Omega\in\mathcal{E}_\Omega} \int_{E_\Omega} \Big[ \tau_{LS\mathbf{u},\Xi}^{\left(\mathbf{s}, \delta \mathbf{u} \right)} \left( \rho \mathbf{u} \cdot \nabla_{\mathbf{x}_\Xi}^{\left(\mathbf{s}\right)} \mathbf{u} + \nabla_{\mathbf{x}_\Xi}^{\left(\mathbf{s}\right)} p \right) \cdot \left( \rho \mathbf{u} \cdot \nabla_{\mathbf{x}_\Xi}^{\left(\mathbf{s}\right)} \mathbf{u}_a + \nabla_{\mathbf{x}_\Xi}^{\left(\mathbf{s}\right)} p_a \right) \\
  & + \tau_{LS\mathbf{u},\Xi}^{\left(\mathbf{s}\right)} \left( \rho \delta \mathbf{u} \cdot \nabla_{\mathbf{x}_\Xi}^{\left(\mathbf{s}\right)} \mathbf{u} + \rho \mathbf{u} \cdot \nabla_{\mathbf{x}_\Xi}^{\left(\mathbf{s}\right)} \delta\mathbf{u} + \nabla_{\mathbf{x}_\Xi}^{\left(\mathbf{s}\right)} \delta p \right) \cdot \left( \rho \mathbf{u} \cdot \nabla_{\mathbf{x}_\Xi}^{\left(\mathbf{s}\right)} \mathbf{u}_a + \nabla_{\mathbf{x}_\Xi}^{\left(\mathbf{s}\right)} p_a \right) \\
  & + \tau_{LS\mathbf{u},\Xi}^{\left(\mathbf{s}\right)} \left( \rho \mathbf{u} \cdot \nabla_{\mathbf{x}_\Xi}^{\left(\mathbf{s}\right)} \mathbf{u} + \nabla_{\mathbf{x}_\Xi}^{\left(\mathbf{s}\right)} p \right) \cdot \left( \rho \delta \mathbf{u} \cdot \nabla_{\mathbf{x}_\Xi}^{\left(\mathbf{s}\right)} \mathbf{u}_a \right) + \tau_{LSp,\Xi}^{\left(\mathbf{s}, \delta \mathbf{u} \right)} \left( \rho \mathrm{div}_{\mathbf{x}_\Xi}^{\left(\mathbf{s}\right)} \mathbf{u} \right) \left( \mathrm{div}_{\mathbf{x}_\Xi}^{\left(\mathbf{s}\right)} \mathbf{u}_a \right) \\
  & + \tau_{LSp,\Xi}^{\left(\mathbf{s}\right)} \left( \rho \mathrm{div}_{\mathbf{x}_\Xi}^{\left(\mathbf{s}\right)} \delta \mathbf{u} \right) \left( \mathrm{div}_{\mathbf{x}_\Xi}^{\left(\mathbf{s}\right)} \mathbf{u}_a \right) \Big] K^{\left(\mathbf{s}\right)} \,\mathrm{d}\Omega + \int_\Sigma \alpha \delta \mathbf{u} \cdot \mathbf{u}_a M^{\left(d_f\right)} \, \mathrm{d}\Sigma = 0, \\
\end{split}
\end{equation}
the first order variational of the augmented Lagrangian to $\mathbf{s}$ and $\boldsymbol{\lambda}_\mathbf{s}$ can be set to be zero as 
\begin{equation}\label{equ:WeakAdjEquHarmonicEquDissipationBulkHT}
\begin{split}
  & \int_\Omega \Big[ \rho \left( \mathbf{u} \cdot \nabla_{\mathbf{x}_\Xi}^{\left(\mathbf{s}, \delta \mathbf{s}\right)} \right) \mathbf{u} \cdot \mathbf{u}_a + {\eta\over2} \Big( \nabla_{\mathbf{x}_\Xi}^{\left(\mathbf{s}, \delta \mathbf{s}\right)} \mathbf{u} + \nabla_{\mathbf{x}_\Xi}^{\left(\mathbf{s}, \delta \mathbf{s}\right)} \mathbf{u}^\mathrm{T} \Big) : \left( \nabla_{\mathbf{x}_\Xi}^{\left(\mathbf{s}\right)} \mathbf{u}_a + \nabla_{\mathbf{x}_\Xi}^{\left(\mathbf{s}\right)} \mathbf{u}_a^\mathrm{T} \right) \\
  & + {\eta\over2} \left( \nabla_{\mathbf{x}_\Xi}^{\left(\mathbf{s}\right)} \mathbf{u} + \nabla_{\mathbf{x}_\Xi}^{\left(\mathbf{s}\right)} \mathbf{u}^\mathrm{T} \right) : \left( \nabla_{\mathbf{x}_\Xi}^{\left(\mathbf{s}, \delta \mathbf{s}\right)} \mathbf{u}_a + \nabla_{\mathbf{x}_\Xi}^{\left(\mathbf{s}, \delta \mathbf{s}\right)} \mathbf{u}_a^\mathrm{T} \right) - p\,\mathrm{div}_{\mathbf{x}_\Xi}^{\left(\mathbf{s}, \delta \mathbf{s}\right)} \mathbf{u}_a - p_a \mathrm{div}_{\mathbf{x}_\Xi}^{\left(\mathbf{s}, \delta \mathbf{s}\right)} \mathbf{u} \Big] K^{\left(\mathbf{s}\right)} \\
  & + \Big[ {\eta\over2} \left( \nabla_{\mathbf{x}_\Xi}^{\left( \mathbf{s} \right)} \mathbf{u} + \nabla_{\mathbf{x}_\Xi}^{\left( \mathbf{s} \right)} \mathbf{u}^\mathrm{T} \right) : \left( \nabla_{\mathbf{x}_\Xi}^{\left( \mathbf{s} \right)} \mathbf{u} + \nabla_{\mathbf{x}_\Xi}^{\left( \mathbf{s} \right)} \mathbf{u}^\mathrm{T} \right) + \alpha \mathbf{u}^2 + \rho \left( \mathbf{u} \cdot \nabla_{\mathbf{x}_\Xi}^{\left(\mathbf{s}\right)} \right) \mathbf{u} \cdot \mathbf{u}_a \\
  & + {\eta\over2} \left( \nabla_{\mathbf{x}_\Xi}^{\left(\mathbf{s}\right)} \mathbf{u} + \nabla_{\mathbf{x}_\Xi}^{\left(\mathbf{s}\right)} \mathbf{u}^\mathrm{T} \right) : \left( \nabla_{\mathbf{x}_\Xi}^{\left(\mathbf{s}\right)} \mathbf{u}_a + \nabla_{\mathbf{x}_\Xi}^{\left(\mathbf{s}\right)} \mathbf{u}_a^\mathrm{T} \right) - p\,\mathrm{div}_{\mathbf{x}_\Xi}^{\left(\mathbf{s}\right)} \mathbf{u}_a - p_a \mathrm{div}_{\mathbf{x}_\Xi}^{\left(\mathbf{s}\right)} \mathbf{u} \Big] K^{\left(\mathbf{s}, \delta \mathbf{s}\right)} \\
  & - \nabla_{\mathbf{x}_\Omega} \delta \mathbf{s} : \nabla_{\mathbf{x}_\Omega} \mathbf{s}_a \,\mathrm{d}\Omega \\
  & - \sum_{E_\Omega\in\mathcal{E}_\Omega} \int_{E_\Omega} \Big[ \tau_{LS\mathbf{u},\Xi}^{\left(\mathbf{s}, \delta \mathbf{s}\right)} \left( \rho \mathbf{u} \cdot \nabla_{\mathbf{x}_\Xi}^{\left(\mathbf{s}\right)} \mathbf{u} + \nabla_{\mathbf{x}_\Xi}^{\left(\mathbf{s}\right)} p \right) \cdot \left( \rho \mathbf{u} \cdot \nabla_{\mathbf{x}_\Xi}^{\left(\mathbf{s}\right)} \mathbf{u}_a + \nabla_{\mathbf{x}_\Xi}^{\left(\mathbf{s}\right)} p_a \right) \\
  & + \tau_{LS\mathbf{u},\Xi}^{\left(\mathbf{s}\right)} \Big( \rho \mathbf{u} \cdot \nabla_{\mathbf{x}_\Xi}^{\left(\mathbf{s}, \delta \mathbf{s}\right)} \mathbf{u} + \nabla_{\mathbf{x}_\Xi}^{\left(\mathbf{s}, \delta \mathbf{s}\right)} p \Big) \cdot \left( \rho \mathbf{u} \cdot \nabla_{\mathbf{x}_\Xi}^{\left(\mathbf{s}\right)} \mathbf{u}_a + \nabla_{\mathbf{x}_\Xi}^{\left(\mathbf{s}\right)} p_a \right) \\
  & + \tau_{LS\mathbf{u},\Xi}^{\left(\mathbf{s}\right)} \left( \rho \mathbf{u} \cdot \nabla_{\mathbf{x}_\Xi}^{\left(\mathbf{s}\right)} \mathbf{u} + \nabla_{\mathbf{x}_\Xi}^{\left(\mathbf{s}\right)} p \right) \cdot \Big( \rho \mathbf{u} \cdot \nabla_{\mathbf{x}_\Xi}^{\left(\mathbf{s}, \delta \mathbf{s}\right)} \mathbf{u}_a + \nabla_{\mathbf{x}_\Xi}^{\left(\mathbf{s}, \delta \mathbf{s}\right)} p_a \Big) \\
  & + \tau_{LSp,\Xi}^{\left(\mathbf{s}, \delta \mathbf{s}\right)} \left( \rho \mathrm{div}_{\mathbf{x}_\Xi}^{\left(\mathbf{s}\right)} \mathbf{u} \right) \left( \mathrm{div}_{\mathbf{x}_\Xi}^{\left(\mathbf{s}\right)} \mathbf{u}_a \right) + \tau_{LSp,\Xi}^{\left(\mathbf{s}\right)} \left( \rho \mathrm{div}_{\mathbf{x}_\Xi}^{\left(\mathbf{s}, \delta \mathbf{s}\right)} \mathbf{u} \right) \left( \mathrm{div}_{\mathbf{x}_\Xi}^{\left(\mathbf{s}\right)} \mathbf{u}_a \right) \\
  & + \tau_{LSp,\Xi}^{\left(\mathbf{s}\right)} \left( \rho \mathrm{div}_{\mathbf{x}_\Xi}^{\left(\mathbf{s}\right)} \mathbf{u} \right) \left( \mathrm{div}_{\mathbf{x}_\Xi}^{\left(\mathbf{s}, \delta \mathbf{s}\right)} \mathbf{u}_a \right) \Big] K^{\left(\mathbf{s}\right)} + \Big[ \tau_{LS\mathbf{u},\Xi}^{\left(\mathbf{s}\right)} \left( \rho \mathbf{u} \cdot \nabla_{\mathbf{x}_\Xi}^{\left(\mathbf{s}\right)} \mathbf{u} + \nabla_{\mathbf{x}_\Xi}^{\left(\mathbf{s}\right)} p \right) \\
  & \cdot \left( \rho \mathbf{u} \cdot \nabla_{\mathbf{x}_\Xi}^{\left(\mathbf{s}\right)} \mathbf{u}_a + \nabla_{\mathbf{x}_\Xi}^{\left(\mathbf{s}\right)} p_a \right) + \tau_{LSp,\Xi}^{\left(\mathbf{s}\right)} \left( \rho \mathrm{div}_{\mathbf{x}_\Xi}^{\left(\mathbf{s}\right)} \mathbf{u} \right) \left( \mathrm{div}_{\mathbf{x}_\Xi}^{\left(\mathbf{s}\right)} \mathbf{u}_a \right) \Big] K^{\left(\mathbf{s}, \delta \mathbf{s}\right)} \,\mathrm{d}\Omega \\
  & + \int_\Sigma \delta \mathbf{s} \cdot \boldsymbol{\lambda}_{\mathbf{s}a} + \delta \boldsymbol{\lambda}_\mathbf{s} \cdot \mathbf{s}_a \,\mathrm{d}\Sigma = 0, \\
\end{split}
\end{equation} 
the first order variational of the augmented Lagrangian to $\gamma_f$ can be set to be zero as
\begin{equation}\label{equ:WeakAdjEquPDEFilterEquGafDissipationBulkHT}
\begin{split}
  & \int_\Sigma \left( {\partial\alpha\over\partial\gamma_p} {\partial\gamma_p\over\partial\gamma_f} \mathbf{u} \cdot \mathbf{u}_a \delta \gamma_f + r_f^2 \nabla_\Gamma^{\left( d_f \right)} \delta \gamma_f \cdot \nabla_\Gamma^{\left( d_f \right)} \gamma_{fa} + \delta \gamma_f \gamma_{fa} \right) M^{\left( d_f \right)} \,\mathrm{d}\Sigma = 0,
\end{split}
\end{equation}
and the first order variational of the augmented Lagrangian to $d_f$ can be set to be zero as
\begin{equation}\label{equ:WeakAdjEquPDEFilterEqudffDissipationBulkHT}
\begin{split}
  & \int_\Sigma r_f^2 \left( \nabla_\Gamma^{\left( d_f, \delta d_f \right)} \gamma_f \cdot \nabla_\Gamma^{\left( d_f \right)} \gamma_{fa} + \nabla_\Gamma^{\left( d_f \right)} \gamma_f \cdot \nabla_\Gamma^{\left( d_f, \delta d_f \right)} \gamma_{fa} \right) M^{\left( d_f \right)} \\
  & + \left( r_f^2 \nabla_\Gamma^{\left( d_f \right)} \gamma_f \cdot \nabla_\Gamma^{\left( d_f \right)} \gamma_{fa} + \gamma_f \gamma_{fa} - \gamma \gamma_{fa} + \alpha \mathbf{u} \cdot \mathbf{u}_a \right) M^{\left( d_f, \delta d_f \right)} \\
  & + r_m^2 \nabla_\Sigma \delta d_f \cdot \nabla_\Sigma d_{fa} + \delta d_f d_{fa} - \delta d_f \mathbf{n}_\Sigma \cdot \boldsymbol{\lambda}_{\mathbf{s}a} \,\mathrm{d}\Sigma = 0.
\end{split}
\end{equation}

The constraints in Eqs. \ref{equ:ConstraintForAugmentedLagrangianDissipationBulkHT} and \ref{equ:ConstraintForVariationalAugmentedLagrangianDissipationBulkHT} are imposed to Eqs. \ref{equ:WeakAdjEquNSEquDissipationBulkHT}, \ref{equ:WeakAdjEquHarmonicEquDissipationBulkHT}, \ref{equ:WeakAdjEquPDEFilterEquGafDissipationBulkHT} and \ref{equ:WeakAdjEquPDEFilterEqudffDissipationBulkHT}. Then, the adjoint sensitivity of $\Delta P$ is derived as
\begin{equation}\label{equ:AdjSensitivityGaDmVariationalFormDissipationBulkHT}
\begin{split}
\delta \widehat{\Delta P} = \int_\Sigma - \gamma_{fa} \delta \gamma M^{\left( d_f \right)} - A_d d_{fa} \delta d_m \,\mathrm{d}\Sigma.
\end{split}
\end{equation}

Without losing the arbitrariness of $\delta \mathbf{u}$, $\delta p$, $\delta \mathbf{s}$, $\delta \boldsymbol{\lambda}_{\mathbf{s}}$, $\delta \gamma_f$, $\delta d_f$, $\delta \gamma$ and $\delta d_m$, one can set 
\begin{equation}
\left.\begin{split}
& \tilde{\mathbf{u}}_a = \delta \mathbf{u} \\
& \tilde{p}_a = \delta p \\
& \tilde{\mathbf{s}}_a = \delta \mathbf{s} \\
& \tilde{\boldsymbol{\lambda}}_{\mathbf{s}a} = \delta \boldsymbol{\lambda}_{\mathbf{s}} \\
& \tilde{\gamma}_{fa} = \delta \gamma_f \\
& \tilde{d}_{fa} = \delta d_f \\
& \tilde{\gamma} = \delta \gamma \\
& \tilde{d}_m = \delta d_m 
\end{split}\right\}
~\mathrm{with}~
\left\{\begin{split}
& \forall \tilde{\mathbf{u}}_a \in \left(\mathcal{H}\left(\Omega\right)\right)^3 \\
& \forall \tilde{p}_a \in \mathcal{H}\left(\Omega\right)\\
& \forall \tilde{\mathbf{s}}_a \in \left(\mathcal{H}\left(\Omega\right)\right)^3 \\
& \forall \tilde{\boldsymbol{\lambda}}_{\mathbf{s}a} \in \left(\mathcal{H}^{{1\over2}}\left(\Sigma\right)\right)^3 \\
& \forall \tilde{\gamma}_{fa} \in \mathcal{H}\left(\Sigma\right) \\
& \forall \tilde{d}_{fa} \in \mathcal{H}\left(\Sigma\right)\\
& \forall \tilde{\gamma} \in \mathcal{L}^2\left(\Sigma\right) \\
& \forall \tilde{d}_m \in \mathcal{L}^2\left(\Sigma\right)
\end{split}\right.
\end{equation}
for Eqs. \ref{equ:WeakAdjEquNSEquDissipationBulkHT}, \ref{equ:WeakAdjEquHarmonicEquDissipationBulkHT}, \ref{equ:WeakAdjEquPDEFilterEquGafDissipationBulkHT} and \ref{equ:WeakAdjEquPDEFilterEqudffDissipationBulkHT} to derive the adjoint system composed of Eqs. \ref{equ:AdjEquSurfaceNSMHMDissipationBulk}, \ref{equ:WeakAdjEquHarmonicEquBulkHTSaDissipation}, \ref{equ:AdjPDEFilterDissipationGaHTBulk} and \ref{equ:AdjPDEFilterJDissipationDmHTBulk}.

\end{document}